\documentclass[10pt,english]{article} 
\usepackage{amsmath, amsfonts, amsthm, amssymb, amscd}
\usepackage[mathcal]{euscript}
\usepackage{mathrsfs}
\usepackage{a4wide}
\usepackage{graphicx}
\usepackage{ulem}  
\usepackage{subfig}
\usepackage{mathrsfs}
\usepackage{rotating}
\usepackage{verbatim}
\usepackage{slashed}
\usepackage{tensor}    
\usepackage{dsfont} 
\usepackage{extarrows} 
\usepackage{lineno}
\usepackage {bbm}
\usepackage{cancel,ulem}
\DeclareFontFamily{U}{BOONDOX-calo}{\skewchar\font=45 }
\DeclareFontShape{U}{BOONDOX-calo}{m}{n}{
<-> s*[1.05] BOONDOX-r-calo}{}
\DeclareFontShape{U}{BOONDOX-calo}{b}{n}{
<-> s*[1.05] BOONDOX-b-calo}{}
\DeclareMathAlphabet{\mathcalboondox}{U}{BOONDOX-calo}{m}{n}
\SetMathAlphabet{\mathcalboondox}{bold}{U}{BOONDOX-calo}{b}{n}
\DeclareMathAlphabet{\mathbcalboondox}{U}{BOONDOX-calo}{b}{n}

\usepackage{perpage}
\MakePerPage{footnote} 
\usepackage[most]{tcolorbox} 
\newtcolorbox{done}{%
enhanced, breakable, size=minimal, parbox=false, after={\par}, 
before upper={\indent}, colback=white, 
overlay = {\draw[line width=2pt, blue] (frame.north east) -|
([xshift=3mm]frame.east)|-(frame.south east);},
overlay first={\draw[line width=2pt, blue] (frame.north east) -|
([xshift=3mm]frame.south east);},
overlay middle={\draw[line width=2pt, blue] ([xshift=3mm]frame.north east) -- 
([xshift=3mm]frame.south east);},
overlay last={\draw[line width=2pt, blue] ([xshift=3mm]frame.north east)|-
(frame.south east);},
} % 
\newtcolorbox{plfok}{%
enhanced, breakable, size=minimal, parbox=false, after={\par}, 
before upper={\indent}, colback=white, 
overlay = {\draw[line width=2pt, blue, dashed] (frame.north east) -|
([xshift=3mm]frame.east)|-(frame.south east);},
overlay first={\draw[line width=2pt, blue, dashed] (frame.north east) -|
([xshift=3mm]frame.south east);},
overlay middle={\draw[line width=2pt, blue, dashed] ([xshift=3mm]frame.north east) -- 
([xshift=3mm]frame.south east);},
overlay last={\draw[line width=2pt, blue,dashed] ([xshift=3mm]frame.north east)|-
(frame.south east);},
} 
\usepackage{graphicx}
\makeatletter
\def\widebreve#1{\mathop{\vbox{\m@th\ialign{##\crcr\noalign{\kern\p@}%
\brevefill\crcr\noalign{\kern0.1\p@\nointerlineskip}%
$\hfil\displaystyle{#1} \hfil$\crcr}}} \limits}
\def\brevefill{$\m@th \setbox \z@\hbox{}%
\hfill\scalebox{0.6}{\rotatebox[origin=c]{90}{(}} \kern1pt $}
\makeatletter 
\newtheorem{theorem}{Theorem}[section]
\newtheorem{proposition}[theorem]{Proposition}
\newtheorem{lemma}[theorem]{Lemma}
\newtheorem{claim}[theorem]{Claim}

\newtheorem{corollary}[theorem]{Corollary}
\newtheorem{definition}[theorem]{Definition}

\newtheorem{remark}[theorem]{Remark}

\numberwithin{equation}{section}
\newcommand \epss {\eps_\star}

\newcommand \crochet {\mathbf X}
\newcommand \nouveauS {\mathcalboondox S}        
\newcommand \Fenergy {\mathcalboondox F}           % {\mathbf F} 
\newcommand \Eenergy {\mathcalboondox E}    % {\mathbf E} 
\newcommand \Time {\mathbf T}

\newcommand \E {\mathcal{E}} % for Euclidean
\newcommand \ME {\mathcal{EM}} % for merging- Euclidean
\newcommand \EM {\mathcal{EM}} % for merging- Euclidean
\renewcommand \H {\mathcal H} % for hyperbolpidal

\newcommand \M {\mathcal M} % for merging
\newcommand \Mcal {\mathcal M} % fusioned to \M

\newcommand \N {\mathcal N} % for null

\newcommand \Lcal {\mathcal L} % exponent for the energy on the light cone 

\newcommand \Mscr       {\mathscr M} 
\newcommand \Mext       {\Mscr^{\mathcal E}} % see explications below

\newcommand \MH 		{\Mscr^{\mathcal{H}}} 
\newcommand \MME 		{\Mscr^{\mathcal{EM}}}  
\newcommand \MM 		{\Mscr^{\mathcal{M}}} 
\newcommand \Mtran      {\Mscr^{\mathcal{M}}}  %fussioned to \MM

\newcommand \MMEfar 	{\Mscr^{\textbf{far}}} 
\newcommand \MMEnear  	{\Mscr^{\textbf{near}}} 
 
\newcommand \Mnear {\Mscr^\textbf{near}}
\newcommand \Mfar{\Mscr^\textbf{far}}

\newcommand \near {\textbf{near}}
\newcommand \far {\textbf{far}}

\newcommand \Lfrak {\mathfrak L}  % Lorentzian group
\newcommand \Yfrak {\mathfrak Y}  % 

\newcommand \Rfrak {\mathfrak R}  % Rotations
\newcommand \Tfrak {\mathfrak T}  % Partial derivatives
\newcommand \Pbf {\mathbf P} % vectors for estimates based on integration along curves
\newcommand \delH {\del^{\mathcal H}} 
\newcommand \delsH {\slashed \del^{\mathcal H}}
\newcommand \PsiH {{\Psi^{\mathcal H}}{}}  
\newcommand \PhiH {{\Phi^{\mathcal H}}{}}  

\newcommand {\usH} {\slashed{u}^{\mathcal{H}}}
\renewcommand {\TH} {T^{\mathcal{H}}{}}
\newcommand \delN {\del^{\mathcal{N}}}
\newcommand \delsN {\slashed \del^{\mathcal{N}}}
\newcommand \PsiN {{\Psi^{\mathcal N}{}}}  
\newcommand \PhiN {{\Phi^{\mathcal N}}{}}

\newcommand \HN {H^{\mathcal{N}}{}}
\newcommand \hN {h^{\mathcal{N}}}

\newcommand \TN {T^{\mathcal{N}}{}}
\newcommand \delE {\del^\mathcal{E}} 
\newcommand \delsE {\slashed \del^\E}

\newcommand \delM {\del^{\mathcal M}} 
\newcommand \delsM {\slashed \del^\M}

\newcommand \delsEH {\slashed {\del}^{\mathcal{EH}}}
\newcommand \delsME {\slashed {\del}^{\mathcal{EM}}} % this is modified form $^{\mathscr{ME}}$ to $^{\mathcal{ME}}$

\newcommand \rhoH {{\mathbf r}^{\mathcal{H}}}    % {\rho^{\mathcal{H}}}
\newcommand \rhoE {{\mathbf r}^{\mathcal{E}}}                % {\rho^{\mathcal{E}}}

\newcommand {\hs}{\slashed{h}}

\newcommand {\us}{\slashed{u}}

\newcommand \Sbf {\mathbf S}
\newcommand {\Abb}{\mathbb{A}}

\newcommand {\Cbb}{\mathbb C}
\renewcommand {\Bbb}{\mathbb{B}}%\Bbb is occupied. Here renew-defined
\newcommand \Dbb{\mathbb{D}}

\newcommand \Fbb {\mathbb F}
\newcommand \Kbb{\mathbb K}

\newcommand \Pbb {\mathbb P}

\newcommand \Qbb {\mathbb Q}
\newcommand \Sbb {\mathbb S} 

\newcommand \SbbME{\mathbb S^{\mathcal{EM}}}
\newcommand {\Lbb} {\mathbb{L}}
\newcommand \Ibb	{\mathbb I}  	% interaction terms 

\newcommand \wR {\tensor[^{(w)}]{}{}  R}

\newcommand \la \langle
\newcommand \ra \rangle

\newcommand \init {\textbf{init}}

\newcommand \vecn {n} 
\newcommand \dive {\text{div} \hskip.05cm}

\newcommand \del \partial 
 \newcommand \delu \delH 
 
\newcommand \delEH {\del^{\mathcal{E \hskip-.06cm H}}} 

\newcommand \hb{\overline h}

\newcommand \uts   {{\slashed u}^\Ncal}
\newcommand \delus  \delsH
\newcommand \delts  \delsN

\def\blfootnote{\gdef\@thefnmark{} \@footnotetext}
\newcommand \Rwave {\tensor[^{(w)} ]{R}{^\star}}

\newcommand \vep \epsilon
\newcommand \Sch {S}

\newcommand \nub {{\color{red} \mathbf \nu}}
\newcommand \source{\textbf{sour}}

\newcommand \xti {\widetilde x}

\newcommand \Boxt {\widetilde \Box}

\newcommand \Yc{\Yfrak^{\text{com}}}

\newcommand \Dtrs  {D^{\mathcal{EM}}} 

\newcommand \Dext {D^{\mathcal E}}

\newcommand \Lscr{\mathscr{L}}

\newcommand {\ut}  {{u^E}}
\newcommand \BoxChapeau {\widehat \Box} 
\newcommand \Zt {Z}
\newcommand \Hcal {\mathcal H}

\newcommand \Hu {\underline{H}}

\newcommand {\gt}  {{g^E}}

\newcommand \RR{\mathbb{R}}
\newcommand \Tbf	{\mathbf{T}}
\newcommand \Abf	{\mathbf{A}}
\newcommand \Bbf	{\mathbf{B}}

\newcommand {\eps} \epsilon

\newcommand {\gb}{\overline{g}}

\let\oldmarginpar\marginpar
\renewcommand\marginpar[1]{\- \oldmarginpar[\raggedleft\footnotesize #1]%
{\raggedright\footnotesize #1}}

\newcommand \TJ T

\newcommand \gd g  
 
\newcommand \Rd  R 
\newcommand \Rbd R 
 
\newcommand \omegad  \omega 
\newcommand \nablabd \nabla  
\newcommand \Gammad \Gamma 
\newcommand \Gammab \Gamma 
\newcommand \Deltab \Delta 

\newcommand \coef \kappa

\newcommand \Ocal {\mathcal O}
\newcommand \Ncal {\mathcal N}

\newcommand \tdelME {\del^\Ncal}  
\newcommand \delME {\del^{\ME}}
\newcommand \tPhiME {\Phi^\Ncal{}}  
\newcommand \tPsiME {\Psi^\Ncal{}}

\renewcommand \ut {u^\Ncal} 

\renewcommand \gt {g^\Ncal{}}
\renewcommand \BoxChapeau {\widetilde \Box}

\renewcommand \Hu {H^\Hcal{}} 
 
\newcommand \ebf {\mathbf e} 
\newcommand \ebfME {\mathbf e^{\mathcal{EM}}} 

\renewcommand \nub {{\phantom{}}}

\newcommand \Gbf {\mathbf G}

\newcommand \Rbf {\mathbf R}
\newcommand \lbf {\mathbf l}

\newcommand \ord {\textbf{ord}}

\renewcommand \deg {\textbf{deg}}

\newcommand \Ext {\textbf{Ext}}
\newcommand \Int {\textbf{Int}}
\newcommand \Mink {\textbf{Mink}}
\newcommand \rank {\textbf{rank}}
\newcommand \gMink {g_\textbf{Mink}}
\renewcommand \Sch {\textbf{Sch}}

\newcommand \pertur {\textbf{pertur}}

\newcommand \Tbb {\mathbb T}

\newcommand \Mgood{\Mscr^{\textbf{good}}}
\newcommand \Mbad{\Mscr^{\textbf{bad}}}

\newcommand \rr {{rr}}

\newcommand \merging {\textbf{merging}}

\newcommand \err {\textbf{err}}

\newcommand \easy {\textbf{easy}}
\newcommand \hard {\textbf{hard}} 
\newcommand \super {\textbf{sub}} 

\newcommand \LOmega {Y_\text{rot}} % {\Xi}

\newcommand \exposant \rho
\newcommand \vecnnu \nu
\newcommand \notrelapse L

\newcommand \sour {\textbf{sour}}

\let\oldaligned\aligned
\def\aligned{\oldaligned\relax}
\begin{document}

\title{\bf Nonlinear stability of self-gravitating massive fields} 

%  Spelling:   LeFloch  or  LeFLOCH 

\author{\large Philippe G. LeFloch\footnote{Laboratoire Jacques-Louis Lions and Centre National de la Recherche Scientifique,
Sorbonne Universit\'e, 
4 Place Jussieu, 75252 Paris, France. Email: {\sl contact@philippelefloch.org}.
\newline $^\dag$ School of Mathematics and Statistics, Xi'an Jiaotong University, Xi'an, 710049 Shaanxi, People's Republic of China.
Email: {\sl yuemath@mail.xjtu.edu.cn}.
\newline
{\it Key words and phrases.} Self-gravitating massive field; wave-Klein-Gordon system; global nonlinear stability; Euclidean--hyperboloidal foliation; weighted Sobolev space; nonlinear hierarchy.
%
% 2000{\it \ AMS Subject Classification.}  Primary: Secondary:
% \hskip.cm Corresponding author: Y. Ma.  
%\hfill Version:~October~2022
} 
\, and Yue Ma$^\dag$} 

\date{}

\maketitle  

%-----------------------------------------------------------------------------------------------------------------------------------

\begin{abstract} 
We consider the global evolution problem for Einstein's field equations in the near-Minkowski regime and study the long-time dynamics of a massive scalar field evolving under its own gravitational field. We establish the existence of a globally hyperbolic Cauchy development associated with any initial data set that is sufficiently close to a data set in Minkowski spacetime. In addition to applying to massive fields, our theory allows us to cover metrics with slow decay in space. 
The strategy of proof, proposed here and referred to as the Euclidean-Hyperboloidal Foliation Method, applies, more generally, to nonlinear systems of coupled wave and Klein-Gordon equations. It is based on a spacetime foliation defined by merging together asymptotically Euclidean hypersurfaces (covering spacelike infinity) and asymptotically hyperboloidal hypersurfaces (covering timelike infinity). A transition domain (reaching null infinity) limited by two asymptotic light cones is introduced in order to realize this merging. On the one hand, we exhibit a boost-rotation hierarchy property (as we call it) which is associated with Minkowski's Killing fields and is enjoyed by commutators of curved wave operators and, on the other hand, we exhibit a metric hierarchy property (as we call it) enjoyed by components of Einstein's field equations in frames associated with our Euclidean-hyperboloidal foliation. The core of the argument is, on the one hand, the derivation of novel integral and pointwise estimates which lead us to almost sharp decay properties (at timelike, null, and spacelike infinity) and, on the other hand, the control of the (quasi-linear and semi-linear) coupling between the geometric and matter parts of the Einstein equations.  
\end{abstract}

%======================================================================================
% 
%\vskip.cm
%
{
\small 
\setcounter{tocdepth}{1} 
\tableofcontents
}  

\

\section{Introduction}   
\label{sectionN-1}

\subsection{The global evolution problem for self-gravitating matter}

\paragraph{Stability of vacuum spacetimes near Minkowski space.}

We are interested in the global evolution problem for Einstein's field equations of general relativity 
A {\sl local-in-time} existence theorem for the vacuum evolution problem was established first by Choquet-Bruhat in 1952 and, later on, extended to a broad set of self-gravitating matter fields; see \cite{CBG, YCB} and the references therein. 
The {\sl global} nonlinear stability problem for Minkowski spacetime was solved only later on. 
The Cauchy problem of interest is formulated by prescribing an initial data set on a spacelike hypersurface that is asymptotically Euclidean and sufficiently close to an (empty) hypersurface in Minkowski spacetime. 
For a class of suitably small data, Christodoulou and Klainerman~\cite{CK} first solved this problem via a gauge-invariant method, while solutions with lower decay in space were constructed next by Bieri~\cite{Bieri}. Later on, Lindblad and Rodnianski gave a different proof in wave coordinates \cite{LR2}. In the latter work, global coordinate functions satisfying the wave equation in the unknown metric are introduced and a global existence for a nonlinear system of coupled wave equations is established. Important additional contributions  are also reviewed below~\cite{BieriZipser, HintzVasy1, HintzVasy2}. All of these results easily extend to {\sl massless} matter fields. 

%-------------------------------------------------

\paragraph{New methodology for massive matter fields.}

The global dynamics of self-gravitating {\sl massive matter fields} has received far less attention, even in the regime of small perturbations near Minkowski spacetime, and only very limited results are available in the literature (reviewed below). Our objective in the present paper is to present a new methodology to tackle the global evolution problem for the Einstein-massive scalar field system. We expect our method to be useful also to study other classes of nonlinear wave systems that are not scaling invariant.  

The proposed strategy is an extension of the {\sl Hyperboloidal foliation method} which was at the core of our earlier work \cite{PLF-YM-one, PLF-YM-two} and allowed us to treat the restricted class of initial data sets coinciding with Schwarzschild data outside a spatially compact region. 
In the present paper we remove this restriction entirely and we assume a Schwarzschild-like decay only. Interestingly, our method is robust enough to allow for a broad range of asymptotic behaviors at spacelike infinity but, in order to limit the length of the present text, we postpone this additional analysis to follow-up work.
We refer to our new approach as the {\sl Euclidean-hyperboloidal foliation method}, as it relies on a merging of a family of asymptotically hyperboloidal hypersurfaces and a family {  of} asymptotically Euclidean hypersurfaces. 

While our project came under completion we learned that Ionescu and Pausader simultaneously solved the same problem by a completely different methodology, based on the notion of resonances; see \cite{IP3}. In fact, a different class of initial data sets is covered therein as far as the functional norms and the spatial decay are concerned (specially since the regularity in \cite{IP3} is stated in a weighted Fourier variable). In Section~\ref{section-related} below, important additional contributions are also reviewed~\cite{Bigorgne,Bigorgne2,FJS,FJS3,LTay,Smulevici,Wang,Wang2}, together with other related results. 

%------------------------------ 

\paragraph{Numerical study of asymptotically flat spacetimes.} 

Preliminary investigations of asymptotic flat spacetimes led physicists to suggest a {\sl nonlinear instability mechanism} in the evolution of self-gravitating massive fields, even for arbitrarily small perturbations. Namely,  the so-called family of ``oscillating soliton stars''~\cite{FLP,SS1991} seemed to provide a potential candidate for instabilities that would develop during the evolution of massive matter governed by the Einstein equations. However, after several controversies the most recent numerical developments (cf.~\cite{OCP}) have led to the definite {\sl conjecture} that, in asymptotically flat spacetimes, massive fields should be globally nonlinearly stable. 
Advanced numerical methods (including mesh refinement and high-order accuracy) were necessary (cf.~\cite{OCP} and the references therein) and, in the long-time evolution of arbitrarily small perturbations of oscillating soliton stars, the following nonlinear mechanism was observed. 
In a first phase of the evolution, the matter tends to collapse and thus evolves toward the formation of a black hole.
However, in an intermediate phase of the evolution, and below a certain threshold in the mass amplitude, the collapse phenomena significantly slows down. Eventually, the third and final phase of the evolution is reached and a strong dissipation mechanism dominates.  It was thus conjectured by physicists that dispersion effects should be dominant in the long-time evolution of self-gravitating massive matter fields in the small perturbation and asymptotically flat regime. The present work provides a rigorous proof of this conjecture. 

%---------------------------------------------

\paragraph{Asymptotically Anti-de Sitter spacetimes.}  

It is worth mentioning that, in asymptotically Anti-de Sitter (AdS) spacetimes, it was observed numerically that 
the evolution of generic (and  arbitrarily small) initial perturbations always leads to the formation of black holes. 
In such a spacetime, matter is confined (i.e.~timelike geodesics reach the AdS boundary in a finite proper time) and cannot disperse.  The effect of gravity remains dominant during all of the evolution, unavoidably leading to the formation of black holes; cf.~Bizon et al.~\cite{Bizon0}. Interestingly, this instability phenomenon in asymptotically AdS spacetimes was rigorously established by Moschidis \cite{Moschidis1} for the spherically symmetric evolution of massless Vlasov fields. 

%--------------------------------------------------------------------------------------------------------------------------------------------------

\subsection{Background on Einstein's field equations}

\paragraph{Einstein equations.}

Throughout, we are interested in four-dimen\-sional spacetimes $(\M, g)$ where $\M$ is 
the topological manifold $\M \simeq [0, + \infty) \times \RR^3$,
and
$g$ is a Lorentzian metric with signature $(-, +, +, +)$. The  Levi-Civita connection of the metric $g$ is denoted by $\nabla_g= \nabla$ from which we can determine the (Riemann, Ricci, scalar) curvature of the spacetime. 
The standard theory of gravity is based on the Hilbert-Einstein action 
defined as the integral of the scalar curvature  $\Rbf_g$ of the metric $g$, that is, 
\begin{equation} \label{eq:action1}
\int_\Mscr \Big( \Rbf_g + 16 \pi \, L[\phi, g] \Big) \, dV_g,
\end{equation}
where $dV_g$ denotes the canonical volume form on $(\Mscr, g)$
(and the Lagrangian $L[\phi, g]$ is discussed below). 
It is well-known that critical points of this action satisfy {\sl Einstein's field equations} 
\begin{equation} \label{eq 1 einstein-massif}
\Gbf= 8\pi \, \Tbf \quad \text{ in } (\Mscr,g), 
\end{equation}
in which the components of Einstein's curvature tensor
$\Gbf$ are $G_{\alpha\beta} := R_{\alpha\beta} - {\Rbf_g \over 2} \, g_{\alpha\beta}$, 
and $R_{\alpha\beta}$ denotes the Ricci curvature of the metric $g_{\alpha\beta}$. 
Throughout, Greek indices describe $0, 1,2,3$ and we use the standard convention of 
implicit summation over repeated indices, as well as 
raising and lowering indices with respect to the metric $g_{\alpha\beta}$ and its inverse denoted by $g^{\alpha\beta}$. 

%---------------------------------

\paragraph{Massive scalar field.}

The Lagrangian $16 \pi \, L[\phi, g]$ in \eqref{eq:action1}
describes the matter content of the spacetime and allows us to determine the {\sl energy-momentum tensor} 
\begin{equation}
T_{\alpha\beta} = T_{\alpha\beta}[\phi,g] := - 2 \, {\delta L \over \, \delta g^{\alpha\beta}} [\phi,g]  + g_{\alpha\beta} \, L[\phi,g].
\end{equation}
In view of the twice-contracted Bianchi identities $\nabla^\alpha R_{\alpha\beta} = {1 \over 2} \nabla_\beta \Rbf$, the Einstein tensor is checked to be divergence-free, that is,
$\nabla^\alpha G_{\alpha\beta} = 0$ 
and, consequently, the following {\sl matter evolution equations} hold 
\begin{equation} \label{Eq1-15bis}
\nabla^\alpha T_{\alpha\beta} =0 \quad \text{ in } (\Mscr,g).
\end{equation}
We are interested here in {\sl massive scalar fields} $\phi: \Mscr \mapsto \RR$ with energy-momentum tensor 
\begin{equation} \label{eq:Talphabeta}
T_{\alpha\beta} := \nabla_\alpha \phi \nabla_\beta \phi - \Big( {1 \over 2} \nabla_\gamma \phi \nabla^\gamma \phi + U(\phi) \Big) g_{\alpha\beta},
\end{equation}
in which the potential $U=U(\phi)$ is a prescribed function depending on the nature of the matter  
under consideration.

%-----------------------------------------

\paragraph{Nonlinear Klein-Gordon equation.}

From \eqref{Eq1-15bis}-\eqref{eq:Talphabeta},
we see that the field $\phi$ satisfies a {\sl nonlinear Klein-Gordon equation} associated with the unknown curved metric $g$:
\begin{equation} \label{eq-KGG}
\Box_g \phi - U'(\phi) = 0 \quad \text{ in } { (\Mscr,g).} 
\end{equation}
Throughout, we assume that
\begin{equation} \label{eq:Uofphi}
U(\phi) = {1\over 2} c^2  \phi^2 + \Ocal(\phi^3)
\end{equation}
for some constant $c>0$, referred to as the {\sl mass} of the scalar field. For instance, with the choice $U(\phi)={c^2 \over 2} \phi^2$,  \eqref{eq-KGG} is nothing but the Klein-Gordon equation $\Box_g \phi - c^2 \phi = 0$, which would be {\sl linear} for a known metric $g$. 
For suitable initial data, the equation \eqref{eq-KGG} is expected to uniquely determine the evolution of the matter.
Our challenge is precisely to study the nonlinear coupling problem when the metric $g$ itself is one of the unknowns and solves Einstein equations with suitably prescribed initial data. 

%-------------------------------------------------------------

\paragraph{Initial value problem.}

The formulation of the initial value problem for the Einstein equations requires to prescribe 
the intrinsic and extrinsic geometry of the initial hypersurface, that is, 
its induced metric and second fundamental form, together with the matter variables on this hypersurface,  that is, 
the scalar field and its Lie derivative in the timelike direction (normal to the hypersurface). 
Importantly, such an initial data cannot be chosen arbitrarily but should satisfy certain constraints of Gauss-Codazzi-type.  
When a foliation by spacelike hypersurfaces is chosen and a suitable gauge choice is made, the Einstein equations decompose into constraint equations 
\begin{equation} \label{equa-constraints}
G_{00} = 8\pi \, T_{00}, \qquad 
G_{0a} = 8\pi \, T_{0a}, 
\end{equation}
and evolution equations $G_{ab} = 8\pi \, T_{ab}$ (with $a,b=1,2,3$).
The four equations in \eqref{equa-constraints} involve the induced metric and second fundamental form 
associated with the initial hypersurface $\big\{ t=1\big\}$ (as well as the matter data) and therefore are interpreted as restrictions on the choice of the initial data sets. 

%----------------------------------------------------------------------------------------------------------- 

\paragraph{Vacuum spacetimes.}

As far as the global evolution problem in $3+1$ dimension is concerned, 
it is the class of {\sl vacuum} Einstein spacetimes that has received most attention by mathematicians
in the past twenty five years. 
The subject was initiated in 1993 when the global nonlinear stability of Minkowski spacetime was established by Christo\-doulou and Klainerman in a breakthrough and very influential work \cite{CK}. The method of proof introduced therein is fully geometric in nature and  relies on a clever use of Killing fields of Minkowski spacetime in order to define suitably weighted Sobolev norms and on 
a decomposition of the Einstein equations in a so-called null frame. Next, Bieri succeeded to weaken the asymptotic 
decay assumptions on initial data required in \cite{CK};  cf.~\cite{Bieri} and the monograph \cite{BieriZipser} written together with Zipser. For a yet weaker decay conditions, see Shen~\cite{Shen}. 

Next, Lindblad and Rodnianski \cite{LR1,LR2} discovered an alternative proof,
which is technically simpler and provides somewhat less control on the asymptotics of the solutions. (For instance, Penrose's peeling estimates were established in \cite{CK}.) 
{   
Their approach relies on a decomposition of the Einstein equations in wave coordinates and, on the other hand, takes its roots in pioneering work by Klainerman~\cite{Klainerman85-add, Klainerman86-add}
on the global existence problem for nonlinear wave equations and the intricate role of the Lorentz vector fields and the null structure. }
Much more recently, another successful method was discovered by Hintz and Vasy \cite{HintzVasy1,HintzVasy2} who treat a broad class of asymptotically Schwarzschild-type spacetimes. 

All of the methods above apply to vacuum spacetimes only or, more generally, to massless fields since the scaling vector field of Minkowski spacetime plays an essential role in their arguments. 
In contrast, massive fields are governed by Klein-Gordon equations which are {\sl not invariant} under scaling. Namely, Minkowski's scaling vector field does not commute with the Klein-Gordon operator and, therefore, cannot be used in defining weighted Sobolev norms and energy estimates. If this field is suppressed in the Sobolev norms introduced in \cite{CK,LR2}, then the corresponding estimates
are much too lax in order to provide a global (time, space) control of the dispersion and decay of solutions.

%------------------------------------------------------------------------------------------------------------------------------------

\subsection{Outline of the contribution in this paper} 

\paragraph{The Euclidean-Hyperboloidal Foliation Method.}

Our main contribution is the resolution of the global evolution problem for {massive} matter fields. In the course of this paper we also introduce a novel method which applies to nonlinear wave equations and does not require the use of Minkowski's scaling field. We refer to this method as the {\sl Euclidean-Hyperboloidal Foliation Method}: it can be regarded as a generalization  of the Hyperboloidal Foliation Method introduced earlier by the authors~\cite{PLF-YM-book,PLF-YM-CRAS,PLF-YM-one, PLF-YM-two}, {   which took its roots in pioneering work on the Klein-Gordon equation by Klainerman~\cite{Klainerman85}.} 
Our methodology allowed us to tackle a large class of nonlinear systems of coupled wave-Klein-Gordon equations, but was 
restricted to compactly supported solutions (or solutions prescribed in the exterior of a light cone region).  Then, 
we relied on a foliation of the interior of a light cone in Minkowski spacetime by spacelike hyperboloids, and established sharp pointwise and energy estimates while coping with the nonlinear coupling taking place between wave and Klein-Gordon equations. 
Our method used hierarchy properties and sharp estimates, which we precisely generalize in the present paper.  
The strategy adopted therein, as well as the one in the present paper, have the definite advantage of relying solely on translations, Lorentz boosts, and spatial rotations, rather than on the full family of (conformal) Killing fields of Minkowski spacetime.

\paragraph{Reference data and decay at spacelike infinity.}

With the Hyperboloidal Foliation Method, we solved the global nonlinear stability problem when the Einstein equations are coupled to a massive scalar field and are expressed in wave gauge. Only the restricted class of initial data sets coinciding with Schwarzschild data outside a spatially compact domain was treated in \cite{PLF-YM-two}. In the present paper, which was first distributed as~\cite{PLF-YM-2017,PLF-YM-2017-unpub},
we thus provide an extension and { cover} a broad class of initial data sets, while relying on a spacetime foliation based on glueing together asymptotically Euclidean hypersurfaces and asymptotically hyperboloidal hyper\-surfaces. Importantly, in order to describe the behavior at spacelike infinity 
we introduce the notion of reference spacetime metric. 
Our method applies to the classes of initial data sets enjoying slow decay conditions which were constructed in LeFloch and Nguyen~\cite{PLF-TCN} and Le~Floch and LeFloch~\cite{BLF-PLF-short,BLF-PLF-long}; therein, a notion of seed data sets is introduced and Einstein’s constraint equations are solved in suitably localized domain at infinity. Further results based on the Euclidean-hyperboloidal foliation method are found in \cite{Ma} for problems in $1+1$ dimensions and in \cite{PLF-YM-SecondPart,PLF-YM-companion} for the Einstein-massive field system under various decay conditions beyond those treated in the present paper. Our method also applies to generalization of the Einstein equations, namely the f(R) theory of gravity \cite{PLF-YM-grav,PLF-YM-comp,PLF-YM-long}. 

%----------------------------------------------------- 

\paragraph{Three spacetime domains.} 

We thus introduce a decomposition of the spacetime into ``interior'' and ``exterior'' regions, in which different foliations are used and which are merged together within a ``transition'' region. The precise definition will be given only later in Section~\ref{secti-42}. 

\begin{itemize}

\item {\bf Asymptotically hyperboloidal domain.} In our approach, a domain $\MH \subset \Mscr$ is foliated by spacelike hypersurfaces which are (truncated) hyperboloids in Minkowski spacetime $\RR^{3+1}$ and coincides with the future of a truncated asymptotically hyperboloidal initial hypersurface. This domain includes timelike infinity. 

\item {\bf Asymptotically Euclidean domain.} A domain $\Mext \subset \Mscr$ is foliated by asymptotically Euclidean hypersurfaces of $\Mscr$, which are flat spacelike hypersurfaces in Minkowski spacetime $\RR^{3+1}$.  This domain includes spacelike infinity. 

\item {\bf Merging domain.} These two domains are glued together by introducing a transition around the light cone 
(defined from the origin in our coordinate system), in which the geometry of the foliation changes drastically from being hyperboloidal to being Euclidean. Often, our estimates will be derived simultaneously in the merging and Euclidean domains, denoted by $\MME$. 

\end{itemize}

\noindent Furthermore, in analyzing the decay of solutions it will be important to distinguish between several frames of vector fields (and the associated high-order differential operators). The Cartesian frame $\del_\alpha$, the semi-hyperboloidal frame $\delH_\alpha$, as well as the semi-null frame 
$\delN_\alpha$ will be used at various stages of our analysis\footnote{We use the terminology ``semi-null'' for a frame containing three directions tangent to the null cone, as well as  ``semi-hyperboloidal'' for a frame containing three directions tangent to the hyperboloids of the foliation.}.

\begin{itemize} 

\item In the asymptotically hyperboloidal domain, we mainly rely on the {\bf semi-hyperboloidal frame} (SHF) 
denoted by $\delH = (\del_t, \delsH_a)$, consisting of $\delH_0 = \del_t$ together with three vector fields tangent to the foliation \begin{equation}
\delH_a  = \delsH_a = {x_a \over t} \del_t + \del_a.
\end{equation}

\item In the asymptotically Euclidean-merging domain, we mainly rely on the {\bf semi-null frame} (SNF)
denoted by $(\tdelME_\alpha) = (\del_t, \delsN_a)$, consisting of $\delN_0 = \del_t$ together with three vector fields tangent to the null cone 
\begin{equation}
\delN_a = \delsN_a = {x^a \over r} \del_t + \del_a. 
\end{equation}

\end{itemize} 

%------------------------------------------------------- 

Our global foliation allows us to properly connect (across our transition domain) the estimates enjoyed by the solution in the exterior to the ones in the interior. 
Since the interior of the light cone was already analyzed 
in \cite{PLF-YM-book,PLF-YM-CRAS,PLF-YM-one, PLF-YM-two} most of our technical analysis in the present paper 
will focus on the global dynamics in the exterior and transition domains. 
%

%----------------------------------------------------------------------

\paragraph{Wave-Klein-Gordon formulation.}

The equations under consideration are geometric in nature, and it is essential to fix the degrees of gauge freedom before tackling a nonlinear stability problem. Here, we require the existence of global coordinate functions $x^\alpha: \Mscr \mapsto \RR$ satisfying 
the wave gauge conditions ($\alpha=0,1,2,3$)
\begin{equation} \label{eq:wcooE}
\Box_g x^\alpha = 0. 
\end{equation}
In such a gauge, we have a nonlinear system of second-order partial differential equations, supplemented with second-order constraints. The main unknowns of the problem are the metric coefficients $g_{\alpha\beta}$ in the chosen coordinates, together with the scalar matter field $\phi$. It is well-known that the constraints are preserved during the time evolution (cf., for instance, \cite{YCB}).
More precisely, the Einstein equations \eqref{eq 1 einstein-massif} for a massive field 
$\phi$ satisfying \eqref{eq:Talphabeta} 
take the form of a nonlinear system of ten wave equations 
for the metric components $g_{\alpha\beta}$ coupled to a Klein-Gordon equation for the scalar field $\phi$: 
\begin{equation} \label{MainPDE-limit}
\aligned
\BoxChapeau_g g_{\alpha\beta} = \Fbb_{\alpha\beta}(g, g;\del g,\del g) 
-16\pi \, \big( \del_{\alpha}\phi\del_{\beta}\phi + U(\phi)g_{\alpha\beta} \big),
\qquad
\BoxChapeau_g \phi  - U'(\phi) = \, 0, 
\endaligned
\end{equation}
where  
$\BoxChapeau_{\gd} := \gd^{\alpha'\beta'} \del_{\alpha'} \del_{\beta'}$, referred to as the modified wave operator, takes the effects of the wave gauge into account. 
The system \eqref{MainPDE-limit} is also supplemented  with the wave gauge constraints 
\begin{equation} \label{eq:gamnul3}
\aligned
{ \Gamma^{\lambda}} & =   g^{\alpha\beta} \Gamma_{\alpha\beta}^\lambda = 0, 
\qquad
\Gamma_{\alpha \beta}^{\lambda}
= {1 \over 2} \,  g^{\lambda \lambda'} \big(\del_\alpha g_{\beta \lambda'}
+ \del_\beta g_{\alpha \lambda'} - \del_{\lambda'} g_{\alpha \beta} \big),
\endaligned
\end{equation}
together with Einstein's Hamiltonian and momentum constraints. We refer to~\cite{YCB} for this standard formulation.

%-----------------------------------------

\paragraph{Challenges overcome in this paper.} 

A major difficulty arising with the Einstein equations is coping with the (quasi-linear and semi-linear) coupling between the geometry and the matter terms, which potentially could lead to a blow-up phenomenon preventing global existence. To proceed, we deal with the following main issues. 

\begin{itemize}

\item {  
The Einstein equations do not satisfy the standard null condition, and the {\sl quasi-null tensorial structure} of the field equations (as we call it) plays an essential role in our proof and requires understanding the tensorial properties  of the Einstein equations in the wave gauge condition in the Euclidean-hyperboloidal foliation under consideration, and identifying suitable cancellation properties. This structure of the Einstein equations for global existence was discovered first by Lindblad and Rodnianski~\cite{LR1}, and later studied in~Alinhac~\cite{Alinhac-book} and following works. (See also Section~\ref{section-related}, below.) 
}

\item According to the positive mass theorem, physically admissible initial data (with the exception of Minkowski spacetime itself) must have a non-trivial tail at spacelike infinity. A typical behavior is the $1/r$ decay of Schwarzschild spacetime is assumed, while our strategy applies { to more general behaviors since we handle a much slower tail with decay $r^{-\lambda}$; see \eqref{equa-31-12-20} and \eqref{equa-new-conditions-hstar}. 
} 

\item We formulate a bootstrap argument and establish energy estimates of sufficiently high-order satisfied by the metric and the scalar matter field. Our bootstrap involves a suitable hierarchy of energy and pointwise estimates, 
and distinguishes between low- and high-order derivatives of the solutions. 

\item Throughout, the role of the boosts and rotation fields must be singled out in order to exhibit the necessary hierarchy between the equations. Hence, we carefully keep track of the boosts and rotations in our decompositions and estimates when necessary.

\item It turns out that sharp estimates on certain components of the metric as well as on the scalar field are necessary, and this part of the analysis is based on suitable integration formulas and techniques, and decompositions of the wave and Klein-Gordon operators  

\end{itemize}

\noindent  A detailed outline will be presented at the end of Section~\ref{sectionN-2} after stating our main results
in Theorem~\ref{theo:main1-geometric} (Euclidean foliation) and Theorem~\ref{thm main-PDE} (Euclidean-hyperboloidal foliation).

%--------------------------------------------------------------------------------------------------------------------------

\subsection{Related works} 
\label{section-related}

{  

\paragraph{Null conditions.}

The importance of the null condition was discovered in~\cite{Ch1986, Klainerman80, Klainerman86-add} and further investigated in \cite{Alinhac-bookfirst,Hormander}. Further major work was subsequently  done by Klainerman and Rodnianski \cite{KRimproved} as well as
Alinhac \cite{Alinhac} and Lindblad~\cite{Lindblad-global}. Concerning the quasi-null (or weak null) conditions, the discovery by Lindblad and Rodnianski~\cite{LR1} elucidated an ``apparent instability'' met with the wave gauge which was pointed out earlier by Choquet-Bruhat~\cite{CBG-instable}. Furthermore, the failure of a weak null structure for the Einstein equations was recognized first by Christodoulou and Klainerman~\cite{CK} who, for this reason, devised a fully geometric method. 

}

\paragraph{Hyperboloidal foliations.} 

{
The use of hyperbolic hypersurfaces for the global analysis of solutions to the Klein-Gordon equation was introduced in 1985 by Klainerman \cite{Klainerman85} and reexamined by H\"ormander in his textbook~\cite{Hormander}. 
}
This strategy was advocated and further developed by Tataru~\cite{Tataru96} for problems involving semi-linear and nonlinear wave equations. Matching different foliations (for Dirac or Klein-Gordon equations) in one space dimension was done by Candy and Lindblad \cite{CandyLindblad-1} and by Lindblad et al. \cite{Lindblad-et-al}. 

In the context of general relativity, in \cite{Friedrich81, Friedrich83} Friedrich studied hyperboloidal foliations of Einstein spacetimes and established global existence results for the Cauchy problem associated with the conformal vacuum field equations. Hyperboloidal foliations can also be constructed in a geometric manner by generalizing Christo\-doulou-Klainerman's method. In this direction, Wang \cite{Wang} gave an independent proof of the restricted theorem in \cite{PLF-YM-two}. In recent years, Wong \cite{WWYW1,WWYW2} also further investigated hyperboloidal foliations for wave equations. Moreover, such foliations for the Einstein equations were constructed numerically by Moncrief and Rinne \cite{MoncriefRinne}, and their work opened the way to an active domain of research in numerical relativity, as extensively pursued by Zenginoglu~\cite{Zenginoglu} and followers. For even more recent developments (established after this paper was first posted), we also refer the reader to Huneau and Stingo \cite{HuneauStingo}  (wave equations on a product space) and Ifrim and Stingo \cite{IfrimStingo} (almost global well-posedness), and Kauffman and Lindblad \cite{KauffmanLindblad} (massless Maxwell-Einstein model).

%------------------------------------------

\paragraph{Kinetic matter and perspectives.}

The Euclidean-Hyperboloidal Foliation Method should be relevant also for the study of the coupling of the Einstein equations with 
{\sl kinetic equations}, such as the Vlasov equation (cf. \cite{Rendall} for an introduction). In this direction, Fajman, Joudioux, and Smulevici \cite{FJS,FJS3} have analyzed the global existence problem for a class of relativistic transport equations and their coupling to wave equations and, by building upon the work \cite{PLF-YM-two} together with a new vector field technique \cite{FJS,Smulevici}, have established the stability of Minkowski spacetime for the Einstein-Vlasov system \cite{FJS3} 
for initial data sets coinciding with vacuum Schwarzschild data outside a spatially compact domain. 
Such a stability result for the Einstein-Vlasov system was also independently
proven by Lindblad and Taylor \cite{LTay} by a completely different method.  
Further progress on analyzing global solutions to the Vlasov equation and the nonlinear stability of Minkowski spacetimes was established in Bigorgne et al. \cite{Bigorgne,Bigorgne2}. 
The application to kinetic equations is worth being further investigated in these new directions and it would be desirable to extend our 
Euclidean-Hyperboloidal Method to the Einstein-Vlasov equations and, more generally, to the Einstein-Boltzmann equations.

%-------------------------

\paragraph{Further works on Klein-Gordon models.}

As already mentioned in the beginning of this introduction, an alternative proof is independently proposed by Ionescu and Pausader \cite{IP-two}, based on the technique of resonances developed by Shatah, Masmoudi, Germain, and followers; see~\cite{Shatah10} and \cite{B-Germain,DIPP,Germain,PS13}. 
Concerning the global existence problem for the Einstein-massive scalar { field system} 
with non-compact data, we also mention a research project by Wang \cite{Wang2} based on the fully geometric approach \cite{CK,Wang}.
With this strategy, 
LeFloch-Ma-Wang's model which couples together a wave equation and a Klein-Gordon equation (and was proposed independently in \cite{PLF-YM-one,Wang}) was successfully revisited by Ionescu and Pausader \cite{IP-two} for a class of non-compact matter fields. Nonlinear Klein-Gordon equations,
especially when they are posed on curved spacetimes, have been the subject of extensive research in the past two decades. 
A vast literature is available in this topic and we refer the interested reader to our former review in the introduction of \cite{PLF-YM-two}, as well as \cite{Bachelot88,Bachelot94,Hormander,Katayama12a,Katayama12b, Klainerman85,Shatah85} and the references cited therein. 

{ 
After this paper was distributed~\cite{PLF-YM-2017} (Dec. 2017), further major advances on the massive Maxwell Klein-Gordon system were established by Klainerman, Wang, and Yang~\cite{KlainermanWangYang} (Jan. 2018), and Fang, Wang, and Yang~\cite{FQY} (Feb. 2019): therein, the authors address the issue of non-compactly supported data for massive scalar fields by purely geometric, commuting vector-field techniques. 

}

%============================================================================================

\paragraph{Notation.} 

We use the notation $A \lesssim B$ when $A, B$ are positive functions satisfying $A \leq C \, B$ for some irrelevant constant $C>0$ ---which may be a universal constant (depending only upon the space dimension or the nonlinear wave system under consideration)
or, more generally, may depend upon the maximum order of differentiation under consideration below (denoted by $N$).    
All the relevant constants until Section~\ref{sectionN-9} (included) depend upon the foliation-defining function $\xi$ and the order of differentiation $N$, only. 
In the remaining sections, the constants also depend upon the system under consideration. 
In addition, we also use the notation $A \simeq B$ when both  $A \lesssim B$ and $B \lesssim A$ hold true. 
Furthermore, we write $A\ll B$ when $A \leq c_0(N) \, B$ for some small numerical constant  $c_0(N)>0$  
which we fix once for all throughout this paper. 

We display our main notation in the following table:  
$$
\aligned
& \aligned 
& g^\star = g_\Mink + h^\star
&& \text{ reference spacetime metric } 
&& \eqref{equa-gstar} 
% \text{Definition~\ref{def-seed}}
\\
& \Mscr_s = \MH_s \cup \MM_s \cup \Mext_s
&& \text{ spacelike slices}
&& \text{\eqref{eq:617def}}
\\
& \Time(s,r)
&& \text{ global time function}
&& \text{\eqref{eq5-05-05-2020}} 
\\
& \xi(s,r) 
&& \text{ foliation coefficient}
&& \text{\eqref{equa-def-xi}}
\\  %\\
& \zeta(s,r) 
&& \text{ energy coefficient}
&& \text{\eqref{equa-defzeta}}
\\  
& \delH_0 = \del_t, \quad \delH_a = \frac{x^a}{t} \del_t + \del_a
&& \text{ semi-hyperboloidal frame (SHF)}
&& \text{\eqref{eq:semihf}} 
\\
& \delN_0 = \del_t, \quad \delN_a = \delsN = {x^a \over r} \del_t + \del_a
&& \text{ semi-null frame (SNF)}
&& \text{\eqref{eq=nulllframedef}}
\\
& \delEH_s = ( \del_s T) \del_t,
\quad
% \delEH_a =
\delsEH_a = \del_a + (x^a /r) (\del_r \Time) \del_t  
&& \text{ Euclidean--hyperboloidal frame (EHF)}
&& \text{\eqref{equation-87}}
\\
& \crochet(s,r) 
&& \text{ energy weight }
&& \text{\eqref{eq:weight}}  
\\ 
\endaligned
\\
\endaligned 
$$

%=================================================================================== 

\section{Main results of global nonlinear stability} 
\label{sectionN-2}

\subsection{Admissible differential operators and weighted norms} 

\paragraph{Organization of this section.}

This section is devoted to the presentation of our main result concerning the nonlinear stability of self-gravitating scalar fields. We begin with some notation and then introduce our notions of asymptotic decay and the class of initial data sets under consideration. In~Theorem~\ref{theo:main1-geometric} we formulate our main existence and stability statement which encompasses a broad class of initial data sets and, next, in Section~\ref{sec-outlining} we outline the organization of the rest of this paper. 
Later on, in Section~\ref{secti-42} (cf.~Theorem~\ref{thm main-PDE}) 
we will give a nonlinear stability statement based on the Euclidean-hyperboloidal foliation which will provide additional quantitative estimates. 

%----------------------------------------------------------------------------------------

\paragraph{Vector fields and differential operators.}

We are interested in solving the initial value problem for the Einstein-scalar field system when the initial data set is a perturbation of a  spacelike hypersurface in Minkowski spacetime. 
The  Minkowski metric in standard Cartesian coordinates reads $\gMink = - dt^2 + \sum_{a=1,2,3} (dx^a)^2$. 
The restriction of this metric on a hypersurface of constant time $t$ coincides with the Euclidean metric $(\delta_{ab})$
and we also write 
\begin{equation}
\gMink = -dt^2 +  g_{\Mink, ab} dx^a dx^b 
= -dt^2 + \delta_{ab} dx^a dx^b. 
\end{equation} 
The future of the initial hypersurface $t=1$ is denoted by 
\begin{equation}
\RR_+^{3+1} := \big\{ t \geq 1, \, x \in \RR^3 \big\}.
\end{equation} 
Below, we also make use of the spatial radial vector field $(x^a/r)\del_a$, where the radius $r> 0$ is defined by $r^2 := \sum_{a=1,2,3} (x^a)^2$ for $x \neq 0$.

Minkowski spacetime admits three sets of {\sl Killing fields}. 
\begin{itemize}

\item[(1)] The  {\bf spacetime translations} generated by the coordinate vector fields $\del_\alpha$ ($\alpha=0,1,2,3$). 

\item[(2)]  The {\bf Lorentz boosts} generated by the vector fields 
$L_a := x_a \del_t  + t \, \del_a$ ($a=1,2,3$).  

\item[(3)] The {\bf spatial rotations} generated by the vector fields 
$\Omega_{ab} := x_a \del_b - x_b \del_a$ ($a, b=1,2,3$).
\end{itemize} 
\noindent 
We refer to $\del_\alpha, L_a, \Omega_{ab}$ as the {\bf admissible fields} which, importantly,  commute with the wave and Klein-Gordon operators in Minkowski spacetime, namely 
$\big[Y,  \Box_{\gMink} - c^2 \big] = 0$ for all admissible fields $Y \in \big\{ \del_\alpha, L_a, \Omega_{ab} \big\}$. In other words, for any  function $\phi$ 
$$
\Box_{\gMink} \phi - c^2 \phi = f \quad \text{ implies}  \quad \Box_{\gMink} Y \phi- c^2 Y \phi = Y f. 
$$ 
On the other hand, the {\bf scaling field} 
$S := t \, \del_t + r \, \del_r$
does not (even conformally) commute with the Klein-Gordon operator (while it conformally commutes with the wave operator). The admissible fields together with the scaling field are called conformal fields.

%----------------------------------------------------------- 

In defining our high-order norms, we use combinations of admissible/conformal vector fields.   
Importantly, we can reduce attention (as we prove it in Section~\ref{sectionN-5}) to ``ordered'' operators, a notion 
defined as follows. Unless specified differently we work with ordered operators.

\begin{definition} An operator $Z=\del^I L^J \Omega^K$ is called an {\bf ordered admissible operator.} 
To such an operator, one associates its {\bf order, degree,} and {\bf rank} by  
\begin{equation}
\ord(Z) = |I|+|J| + |K|, 
\qquad 
\deg(Z) = |I|, 
\qquad 
\rank(Z) = |J| + |K|,
\quad 
\text{ when } Z = \del^I L^J \Omega^K.
\end{equation}
An operator $\Gamma = \del^IL^J\Omega^KS^l$ is called an {\bf ordered conformal operator}. Its {\bf order}, {\bf degree}, and {\bf rank} are defined similarly:
$$
\ord(\Gamma) = |I|+|J| + |K| + l, 
\qquad 
\deg(\Gamma) = |I|, 
\qquad 
\rank(\Gamma) = |J| + |K| + l.
$$
\end{definition}

For convenience, we also introduce the Japanese bracket
$\la y \ra := \sqrt{ 1+ |y|^2}$ for all real $y$. 
Then, for any function $u=u(t,x)$ we define
\begin{equation}
|u|_N :=  \max_{\ord{Z}\leq N} |Z u|,
\qquad 
|u|^S_N := \max_{\ord(\Gamma)\leq N} |\Gamma u|, 
\end{equation}
where the first maximum is over all ordered admissible operators 
and the second one is over all ordered conformal operators. Furthermore, we define the {\sl pointwise
high-order norms} 
\begin{equation}\label{eq-notation-norms} 
\|u\|_{\Omega, \lambda,N} := \sup_{\Omega} 
% \la r-t\ra^{\eta} 
\la r+t\ra^{\lambda} |u|_N,
\qquad\qquad
\| u \|^S_{\Omega, \lambda,N} := \sup_{\Omega} 
% \la r-t\ra^{\eta} 
\la r+t\ra^{\lambda}  |u|^S_N, 
\end{equation}
%%----------------------- 
where $\Omega\subset \RR^{1+3}_+$ is a subset. When $\Omega = \RR^{1+3}_+$, {\sl 
we omit the subscript} $\Omega$. 
We point out that the norms involving the scaling field will not be propagated in time, but will only serve to construct classes of initial data sets (cf.~Section~\ref{section-use-S}).

In order to state minimal restrictions on our data, we may also localize the norms with respect to the \textsl{outgoing light cone} 
\begin{equation} \label{equa-lightcone}  
\qquad 
\Lscr:= \big\{ r = t-1 \big\} \subset \RR_+^{3+1}.
\end{equation} 
Its constant-$t$ slices are denoted by $\Lscr_t$.
In addition, a parameter $\ell \in (0,1/2]$ being fixed once for all, we introduce the {\sl near-light cone} domain
\begin{equation} \label{eq1-06-07-2021}
\Mscr^\near_{\ell}  := \Big\{  t \geq 2, \quad t-1 \leq r \leq \frac{t}{1-\ell} \Big\}, 
\end{equation} 
where, in agreement with our notation\footnote{This domain will be used in formulating our sign condition \eqref{equa-bending}, and 
is nothing but $\Mscr^{\near}_{\ell,[2,+\infty)}$ in the sense introduced later in Section~\ref{secti-10-1}.} in the next section, 
we restrict attention  to $t \geq 2$. 

%------------------------------------------------------------------------------------------------------------------------

\subsection{Initial data sets} 

\paragraph{Formulation of the Cauchy problem}

For the formulation of the initial value problem associated with the Einstein equations, we refer to the textbook~\cite{YCB}. Here, we outline the formulation in our setup. 
Let $M$ be a $3+1$ dimensional spacetime (that is, a time-oriented Lorentzian manifold possibly with boundary) equipped with a global foliation $M\simeq \RR_+\times \Sigma$ by spacelike hypersurfaces 
and a scalar field $\phi$. A globally defined time function $t: M \mapsto \RR_+$ is provided 
such that, for each time $t_1$, the hypersurface $\Sigma_{t_1} := t^{-1}(t_1) \simeq \Sigma$. In our setup, we have $\Sigma \simeq \RR^3$ and we denote by $t_0=1$ our initial time. 

\begin{definition} \label{def1-05-01-2022}
1. 
An initial data set is a quadruple $(g_0,k_0,\phi_0,\phi_1)$ where $g_0,k_0$ are symmetric two-tensors and $\phi_0, \phi_1$ are scalar fields defined on $\RR^3$ such that the following Einstein's constraint equations hold:
\begin{equation} \label{eq1-26-12-2021}
\aligned
R_0 - |k_0|^2 + \textbf{Tr}(k_0)^2 
& = 16 \pi \Big(
{1 \over 2} (\phi_1)^2 + {1 \over 2}  |\nabla_{g_0} \phi_0 |^2 + U(\phi_0) 
\Big),  
\\
\textbf{Tr}(k_0)  - \textbf{Div} \big( k_0)
& = 8 \pi \, \phi_1 {\nabla_{g_0}} \phi_0. 
\endaligned
\end{equation} 

2. A Cauchy hypersurface in $M$ is an embedding $i:\RR^3\mapsto\Sigma_{t_0}\subset M$ together with an initial data set  $(g_0,k_0,\phi_0,\phi_1)$ such that
$$
\aligned
&i^*(g) = g_0,\quad  i^*(\phi) = \phi_0, \quad i^*(\vecnnu \phi) = \phi_1,
\\
&\vecnnu \text{ is the normal vector to }\Sigma_{t_0},
\\
&k_0\text{ is the second fundamental form of } \Sigma_{t_0}. 
\endaligned
$$
3. If $(M,g,\phi)$ satisfies the Einstein-massive scalar field system \eqref{eq 1 einstein-massif} with the energy-momentum \eqref{eq:Talphabeta}, then $(M,g,\phi)$ is called a {\bf Cauchy development} of $(g_0,k_0,\phi_0,\phi_1)$. 
\end{definition}

For instance, Minkowski spacetime is a solution provided all of the data $(g_0 - \delta), k_0, \phi_0,\phi_1$ are taken to vanish. On the other hand,  when the data $(g_0-\delta),k_0,\phi_0,\phi_1$ are sufficiently small (in a sense to be specified), one may expect that a global development exists that remains globally close to Minkowski spacetime. This is the problem of nonlinear stability of  Minkowski spacetime  we address in the present paper. 

%-----------------------------------

\paragraph{Decomposition of the initial data.} 

We thus consider an initial metric $g_0$ that is sufficiently close the Euclidean metric and a second fundamental form $k_0$ that is sufficiently small, in a sense that we now explain.  Let us introduce the following decomposition  
\begin{equation}
g_{0ab} = \delta_{ab} + h^\star_{0ab} + u_{0ab}, 
\qquad 
k_{0 ab} = k^{\star}_{0ab} + l_{0ab}, 
\qquad a, b=1,2,3. 
\end{equation}
We aim at covering a variety of asymptotic behaviors and, at this juncture, it is convenient to introduce the following terminology. 

\begin{itemize}

\item The part $h^\star_{0}$ is referred to as the {\bf initial reference} and will be assumed to be small in a (weighted, high-order) pointwise norm. 

\item The part $u_{0}$ is referred to as the {\bf initial perturbation} and will be assumed to be small in  (weighted, high-order) energy norm. 

\end{itemize} 

An example of a such decomposition is provided by the construction in Lindblad and Rodnianski~\cite{LR1}, where the initial data is decomposed as the sum of a finite-energy perturbation plus an (asymptotically) Schwarzschild metric outside of a compact set (where the Schwarzschild mass is also assumed to be small). 
Observe that the Schwarzschild metric would be of infinite energy in the context of \cite{LR1}. 

We emphasize that the two parts will be treated differently.  Indeed, our strategy in the present work consists of propagating the contributions $h^\star_0$ and $u_0$ by using different arguments of analysis. 
The contribution $h^\star_{0ab}$ should be thought of as an Ansatz metric ---although it is more than that in the sense that 
$u_{0}$ is controlled by energy bounds, only, and need not be pointwise small in comparison to $h^\star_{0}$. 
We describe first the properties required on $h^\star_0$, and we then introduce the class of perturbations $u_0$ treated in the present paper.

%---------------------------------------------------------------------------------------------------------------------------

\subsection{The notion of reference spacetime}

\paragraph{Proposed notion.}

The following notation makes sense for any metric $g^\star = g_\Mink + h^\star$ defined in $\RR^{1+3}$. We also introduce the {\sl reduced Ricci curvature} associated with $h^\star$ (which coincides with the Ricci curvature in the wave gauge of interest) to be (as further analyzed in \eqref{equa-the-system})
\begin{equation}
\aligned
\wR^\star_{\alpha\beta} 
& := R^\star_{\alpha\beta}
- \frac{1}{2} \big(\del_\alpha\Gamma^\star_\beta + \del_\beta\Gamma^\star_\alpha\big) 
- \frac{1}{2}
\Big( g^{\star\delta \delta'} \del_{\delta} g^\star_{\alpha \beta} \Gamma^\star_{\delta'} - \Gamma^\star_\alpha \Gamma^\star_\beta\Big)
\\
& =: - \gd^{\star \alpha'\beta'} \del_{\alpha'} \del_{\beta'} g^\star_{\alpha\beta}
+ \frac{1}{2} \Fbb_{\alpha\beta}(g^\star, g^\star;\del h^{\star}, \del h^{\star}), 
\endaligned
\end{equation}
in which {
 $\Gamma^{\star \gamma} := g^{\star\alpha\beta} \Gamma_{\alpha\beta}^{\star\gamma}$} are the contracted Christoffel symbols in the global coordinate chart under consideration. We begin with the following definition which encompasses 
solutions that behave like the Schwarzschild metric at spacelike infinity, that is $1/r$. It is convenient to introduce the interior and the exterior of the light cone, defined by 
$$
\Ext := \{r\geq t-1\},\qquad \Int := \{r\leq t-1\}.
$$
{ 
A class of examples of metrics satisfying our decay conditions below is given by small perturbations of the Schwarzschild metric (suitably glued with the Minkowski metric). Our conditions below will be applied in the statement of our main theorem and, in the second part of this paper, will be weakened significantly; cf.~Section~\ref{sec1-23-05-2021}, where we will define two classes of reference metrics, referred to as Class A and Class B}. 
In the definition below, we rely on the notation \eqref{eq-notation-norms}
{ and we fix a small parameter $\varsigma$.}

\begin{definition} 
\label{def-seed-basic}
Fix some integer $N \geq 1$. 
Given parameters $\epss> 0$ and 
$\lambda \in (1/2,1]$,  
a Lorentzian metric 
\begin{equation} \label{equa-gstar} 
g^\star = g_\Mink + h^\star
\end{equation} 
defined on $M\simeq \RR^{1+3}_+$ is said to be
a $(\lambda, \epss,N)$-admissible 
{\bf reference metric} if the following conditions hold: 
\begin{equation}
\label{eq1-14-01-2022}
\aligned
&\text{Asymptotically Minkowski:} &&
{\aligned
& \max_{m=0,1,2} \| \del^m h^{\star}\|_{\lambda+ m,N+2-m}  
\lesssim \epss.
\\ 
\endaligned
}
\\ 
&\text{Almost Ricci flat in the exterior:}   
&&\max_{m=0,1}  
\| \del^m \wR^\star \|_{\Ext, {2+2\lambda}+m,N-m}  
\lesssim \epss^2. 
\\ 
&\text{Almost Ricci flat in the interior:}   
&& 
\max_{m=0,1} 
\| \del^m \wR^\star \|_{\Int,{2 +\lambda}+m,N-m} \lesssim \epss.
\\
&  \text{Approximate wave gauge:}
&& 
\max_\gamma |  \Gamma^{\star\gamma}  |_{N} \lesssim  \epss \,  \la r + t\ra^{-1} (1 + (r-t)_+) ^{-1-\varsigma}. 
\endaligned
\end{equation} 
One also says that $(M,g^{\star})$ is a {\bf reference spacetime}.
\end{definition}

%----------------------------------------------------------------------------------------------------------------------

\paragraph{Decay conditions.}

\begin{itemize} 

\item[] {\bf Metric decay.} The first condition in \eqref{eq1-14-01-2022}
guarantee that the reference metric is asymptotically flat. 
The restriction $\lambda > 1/2$ is natural since it is required for the ADM mass to be well-defined and finite. This condition  also ensures that the first-order derivatives $\del h^{\star}_{\alpha\beta}$ are globally square-integrable in $\RR^3$ on spacelike hypersurfaces. On the other hand, the upper bound $\lambda =1$ is also natural and is achieved by Schwarzschild-type solutions.

\item[] {\bf Curvature decay.} The second and third conditions in \eqref{eq1-14-01-2022} guarantee that $g^{\star}$ is an ``approximate solution'' to the vacuum Einstein equations: the linear part $\Box h^{\star}$ vanishes or enjoys the high decay $\la r+t\ra^{-2-2\lambda}$ (while remaining nonlinear terms automatically enjoy better decay by homogeneity). Obviously, by taking the (exact) Schwarzschild metric in the exterior domain, the second condition is trivially satisfied (see also below). 

\end{itemize}

%------------------------------------------------------------------

\paragraph{Light bending property. }

We introduce a property enjoyed by the Schwarzschild solution, namely 
 a sign condition enjoyed by a key metric component which we refer to as the {\sl light-bending property}. 
Heuristically, this condition means that most of the ``mass'' is initially contained in a sufficiently large ball. 
Specifically, we formulate this condition in terms of the {\bf incoming null component} 
\begin{equation} \label{eq1-03-01-2022}
g^\star(\lbf, \lbf) \quad \text{ with } 
\lbf := \del_t - (x^a/r)\del_a,
\end{equation} 
which we require to be strictly positive in a neighborhood of the light cone. 

\begin{definition} 
\label{def:basic12}   
A $(\lambda, \epss,N)$-admissible reference metric $g^\star$ is said to satisfy the \textbf{light-bending property} provided its incoming null component satisfies 
\begin{equation} 
\label{equa-bending} 
4\epss \leq \inf_{\Mscr^\near_{\ell}}  r \,  g^\star(\lbf, \lbf), 
\end{equation} 
in which $\Mscr^\near_{\ell}$ is defined in \eqref{eq1-06-07-2021}. 
A reference spacetime satisfying the light-bending property is called a {\bf light-bending reference spacetime}. 
\end{definition}

The condition \eqref{equa-bending}  guarantees that the light cone bends towards the center compared with the standard light cone. For instance the Schwarzschild metric with positive mass satisfies this condition (see below).  
Heuristically, the matter is attracting light rays, and our condition is reminiscent of the red-shift property enjoyed by the Schwarzschild metric. 

{

We point out the following features. 

\begin{itemize}

\item A reference enjoying the light-bending condition is much easier to deal with, and consequently we can cover a reasonably broad class of metrics and, simultaneously, achieves the proof of existence in a paper of moderate length (as opposed to a book). 

\item Heuristically, the stability of the light-bending property, as we establish in the present work, can be 
interpreted as follows: 
the property of attraction of matter toward the ``center'' 
 remains stable under small {\sl massive} perturbations, which is expected from a physical standpoint. 

\item Global existence without imposing this condition will be the subject of future work by the authors. 
However, we emphasize that an additional technique which {\sl supplements} (but does not supersede) 
the tools presented in this paper 
is necessary.  Again, our strategy in order to keep a reasonable length to the present article. 

\end{itemize}

}

%---------------------------------------------------------------------------------------------
\begin{figure}
\hskip6.cm
\begin{tikzpicture}[
scale=1.,
axis/.style={very thick, ->, >=stealth'},
important line/.style={thick},
dashed line/.style={dashed, thick},
every node/.style={color=black,}
]
\draw[loosely dashed] (0,0) -- (4.3,4.3); 
\draw[domain=0:3,blue,smooth] plot ({\x},{\x+\x^2/6}); 
\node[below] at (4.6,4.) {Minkowski};
\node[below] at (4.6,3.5) {light cone};
\node[below] at (1,3.) {spacetime};
\node[below] at (1,2.5) {light cone};

\draw[thick,->] (2.2,3) -- (1.2,4.2);
\node[below] at (1,5.) {Minkowski null vector $\lbf$};

\end{tikzpicture}

% \caption{\label{fig:f}{\bf }}
\end{figure}

%-------------------------------------------------------------------------------------------------------------------------------------------

\paragraph{Wave gauge condition.}

The reference metric need not satisfy the wave gauge condition. However a mild initial restriction is still assumed. 

\begin{definition}
The pair consisting of the 
induced metric and second fundamental form of the 
slice $t=1$ of a $(\lambda,N,\epss)$-admissible, light-bending, reference spacetime metric $g^\star$ 
is called a  $(\lambda,\kappa, N,\epss)$-admissible,  light-bending, {\bf initial reference}
and 
is denoted by  
$(g_0^{\star},k_0^{\star})$,  
provided the following \textbf{approximate wave gauge condition}  
with decay exponent $\kappa \in (1/2, 1]$ holds: 
\begin{equation}\label{eq2-04-11-2022}
\|\la r\ra^{\kappa + |I|}\del^I\Gamma^{\star\gamma}\big|_{\Sigma_{t_0}}\|_{L^2(\RR^3)}\leq \epss,\quad |I|\leq N. 
\end{equation}
\end{definition}

%-------------------------------------------

\paragraph{ An example based on the Schwarzschild solution.}

In wave coordinates the Schwarzschild metric reads 
\begin{subequations}
\label{equa-description-Schwarz-merging}
\begin{equation} \label{eq Sch-wave}
\aligned
g_{\Sch,00} & =   - \frac{r-m}{r+m},
\qquad g_{\Sch,0a} = 0,
\qquad 
g_{\Sch, ab} = \frac{r+m}{r-m} \omega_a \omega_b + \frac{(r+m)^2}{r^2}(\delta_{ab} - \omega_a \omega_b). 
\endaligned
\end{equation}
We introduce a smooth cut-off function $\chi^\star(r)$ that vanishes for $r\leq 1/2$ and is identically $1$ for all $r\geq 3/4$. Then for $0<m \leq 1/4$ (say, as we are interested only in small mass coefficients) we set
\begin{equation} \label{equa-defineMS} 
g_\merging^\star = \gMink + \chi^\star (r) \, \chi^\star(r/(t-1)) (g_\Sch - \gMink), 
\qquad 
t \geq 1. 
\end{equation} 
Observe that the approximate wave gauge condition \eqref{eq2-04-11-2022} is obviously satisfied. 
This expression coincides with $\gMink$ within the cone $r/(t-1)< 1/2$ and coincides with $g_\Sch$ in the cone exterior $r/(t-1)\geq 3/4$ which {\sl contains} the light cone $\Lscr$.  
This example satisfies our conditions, and 
here, the constant $\epss$ is the Schwarzschild mass $m>0$. More precisely, the Ricci flat condition holds true due to the fact that in the domain $\big\{ r > t-1 \big\}$ the reduced Ricci curvature $^{(w)}R^\star \equiv 0$ vanishes,  while in $\{r\leq t-1\}$, the homogeneity of $g^{\star}_{\merging}$ one has $|\wR^{\star} |_N\lesssim \epss \la t+r \ra^{-3}$ which satisfies the desired assumption.
Furthermore, for the light-bending property, a direct calculation shows that the light cone coefficient 
\begin{equation} \label{equa-Sch-bending} 
r \, g_{\merging}^{\star}(\lbf,\lbf) = 4m + \Ocal(1/r),  
\end{equation}
so that the light-bending property is satisfied. 
\end{subequations}
This construction is essentially the same as the decomposition in \cite{LR1}, where $g^{\star}_{\merging}$ takes the role of $g^0 := g_{\Mink} + h^0$ (with the notation therein). 

%%---------------------------------- 

\subsection{Admissible initial data and statement of the main theorem}

\paragraph{Nonlinear stability statement.}  

We are in a position to state our main global existence result, which establishes that any admissible perturbation 
generates a global-in-time solution to the Einstein equations. In other words, we find that
a set of pointwise and energy estimates are sufficient to prevent the formation of a singularity (such as a black hole)
and lead to a global solution that asymptotically converges to the Minkowski solution. The notion of ``global'' existence here 
is understood in a geometric way. At this junction, we recall that a future causally geodesically complete spacetime, by definition, has the property that every affinely parameterized geodesic (of null or timelike type) can be extended toward the future (for all values of its affine parameter).

To begin with we introduce the notion of admissible initial data set.

\begin{definition} \label{def1-03-01-2022}
An {\bf admissible initial data set} $\big( g_0,k_0,\phi_0,\phi_1\big)$ with parameters $(\lambda,\kappa,\epss, N,\mu,\eps)$ 
consists of two symmetric two-tensors $g_0,k_0$ and two scalar fields $\phi_0,\phi_1$ defined on $\RR^3$ and 
satisfying the following conditions. 

\begin{itemize}

\item Einstein's constraint equations \eqref{eq1-26-12-2021} are satisfied. 

\item There exists a decomposition into a sum 
\begin{equation} \label{eq1-06-01-2022}
g_{0ab} =  g^{\star}_{0ab} + u_{0ab},
\qquad\quad 
k_{0ab} = k^{\star}_{0ab} + l_{0ab}, 
\end{equation}
where  $(g_0^{\star},k_0^{\star})$  is a light-bending, $(\lambda,\kappa,\epss,N)$-admissible initial reference.

\item Moreover, the perturbation $(u_0, l_0,\phi_0,\phi_1)$ has finite energy in the sense 
that
% \begin{subequations} \label{eq4-06-01-2022}
\begin{equation} \label{eq2a-26-12-2021}
\big\|\la r\ra^{\kappa + |I|} \del^I \del u_0 \big\|_{L^2(\RR^3)} 
+ \big\|\la r\ra^{\kappa+|I|} \del^I l_0 \big\|_{L^2(\RR^3)}\leq \vep, 
\qquad |I|\leq N, 
\end{equation}
\begin{equation} \label{eq2b-26-12-2021}
\big\|\la r\ra^{\mu + N} \del^I \del \phi_0 \big\|_{L^2(\RR^3)}
+ \|\la r\ra^{\mu + N} \del^I \phi_0 \|_{L^2(\RR^3)}
+ \big\|\la r\ra^{\mu+N} \del^I \phi_1 \big\|_{L^2(\RR^3)}\leq \vep,
\qquad |I|\leq N.
\end{equation}
% \end{subequations}
\end{itemize} 
\end{definition}

We are ready to state our main result of nonlinear stability. 

\begin{theorem}[Nonlinear stability of self-gravitating massive fields]
\label{theo:main1-geometric}  
Fix some sufficiently large integer $N$ ($N=20$ being sufficient) and consider an admissible light-bending initial data $(g_0,k_0,\phi_0,\phi_1)$ with
parameters $(\lambda,\kappa,\epss,N,\mu,\eps)$ satisfying 
\begin{equation} \label{eq1-20-01-2022}
\kappa\in(1/2,1), 
\qquad 
\lambda \in (1/2,1], 
\qquad 
\mu\in(3/4,1),
\qquad 
\kappa\leq \mu. 
\end{equation} 
%%% 
Then there exists a small constant $c_0>0$ (determined by the Einstein system) such that for all 
\begin{equation}
\label{equa-all-conditions-exponents}
\aligned 
&  
N \, (1-\lambda) \leq c_0 \,\min(\kappa-1/2, \mu-3/4), 
\qquad
\qquad
\eps< c_0 \epss <c_0, 
\endaligned
\end{equation}
the maximal globally hyperbolic Cauchy development of $(g_0,k_0,\phi_0,\phi_1)$ 
associated with the Einstein-massive field system \eqref{eq 1 einstein-massif} 
is future causally geodesically complete, and asymptotically approaches Minkowski spacetime in all (timelike, null, spacetime) directions. Moreover,  this solution satisfies the light-bending property for all times and 
(in a suitable sense) 
remains close to the reference spacetime $(\RR_+^{3+1}, g^\star)$.  
\end{theorem}

The following observations are in order. 

\begin{itemize} 

\item {\bf Vacuum spacetimes.} While our main focus is on matter spacetimes, our formulation also allows us to recover results available in the literature for {\sl vacuum} spacetimes. The stability result in \cite{LR2} is also included in our theorem, by taking the scalar field to vanish identically and the initial reference to be Schwarzschild outside a compact set. Recall that \cite{LR2} was the first proof of the nonlinear stability of Minkowski spacetime {\sl in wave coordinates.} 

\item {\bf Our earlier result.} Theorem~\ref {theo:main1-geometric} applies to initial data sets considered in \cite{PLF-YM-book} which are taken therein to coincide with the Schwarzschild metric outside a compact set, and the reference metric to coincide with Schwarzschild outside a light cone). Recall that \cite{PLF-YM-book} together with the geometric proof in \cite{Wang} were the first results in the literature concerning the nonlinear stability of massive scalar fields.

\end{itemize}

%--------------------------------------------------------

\paragraph{Conditions on the reference metric and the perturbation.}

The decay and regularity assumptions made on the reference and the perturbation are made for simplicity in the presentation of this paper and will be revisited in the companion papers \cite{PLF-YM-SecondPart}--\cite{PLF-YM-long}. 

As explained in Section~\ref{sec1-23-05-2021} below, in our proof we find it convenient to introduce two classes of reference metrics, referred to as Class A and Class B, and to establish our main theorem under weakened conditions on the reference metrics beyond the ones presented in Definition~\ref{def-seed-basic}. The statement of these weaker conditions is involved and is thus postponed 
to Section~\ref{sec1-23-05-2021}. 
The conditions in Definition~\ref{def-seed-basic} encompass primarily metrics that have near-Schwarzschild behavior. In view of the decay properties established by Lindblad~\cite{Lindblad-3} 
for vacuum spacetimes in wave gauge, our method applies with a reference metric chosen, for instance, to be an exact solution to Einstein's vacuum equations (or a suitably approximate solution).
 
{ 
Interestingly, our slow decay conditions on the metric
at spacelike infinity are essentially the same to the ones in 
Bieri~\cite{Bieri}, who treated the class of vacuum spacetimes { but required only the existence of a few derivatives on the solutions (namely, worked at the $H^3$ regularity level).}
 More precisely, when the matter field is suppressed, our conditions on the metric can be slightly further relaxed so that our decay conditions coincide with the ones in~\cite{Bieri}.  
 Vacuum or matter evolution problems are very different in terms of scope and objectives. 
On the other hand, 
 Ionescu and Pausader \cite{IP3} and the present paper (distributed in 2017 as~\cite{PLF-YM-2017,PLF-YM-2017-unpub})
solve the same matter evolution problem\footnote{Indeed, these results were announced independently in ArXiv:1911.10652 and ArXiv:1712.10045, respectively.}. 
  While 
 we introduce function spaces in the physical spacetime and we impose slow decay on the metric but comparatively faster decay on the matter field, the authors \cite{IP3} develop a method in Fourier space and impose comparatively 
 more decay for the metric but less decay on the matter field.  
}  

\begin{theorem}[Nonlinear stability of self-gravitating massive fields. A generalization]
\label{theo:main1-geometric-AB}  
The statement in Theorem~\ref{theo:main1-geometric} remains valid if the conditions on the reference metric are weakened and replaced by the conditions defining Class A and Class B metrics  in \eqref{equa-paramet-repeat}--\eqref{equa-bending-repeat} below. 
\end{theorem}

%------------------------------------------------------------------------------------------------------------------------------------------------------ 

\subsection{Outline of this paper}
\label{sec-outlining}

The rest of this paper is organized in two main parts, as follows. 

\paragraph{Part I. The Euclidean--hyperboloidal foliation method.}

The proposed framework is based on a foliation that is asymptotically Euclidean in the vicinity of spacelike infinity while we cover timelike infinity with slices that are asymptotically hyperboloidal in the vicinity of null infinity. It is relevant for dealing with coupled nonlinear systems of wave and Klein-Gordon equations. This part contains several contributions of independent interest and is organized as follows. 

\begin{itemize}

\item {\bf Foliation, vector fields, and energy estimates} (Section~\ref{sectionN-3}).  

Our first task is to describe the foliation by asymptotically Euclidean or hyperboloidal slices, which is done via the introduction of a  coefficient $\xi=\xi(s,r)$ that interpolates between the ``interior domain'' in which $\xi= 1$ (for $r <\rhoH(s)$, a time-dependent radius) 
and an ``exterior domain''  { in which $\xi = 0$} (for $r >\rhoE(s)$, a larger radius).  The two foliations are merged across a transition  associated with the interval $[\rhoH(s), \rhoE(s)]$. Our parametrization is based on a time variable, denoted by $s$, which is connected to the standard Cartesian time function (that is, the variable $t$) and coincides with the standard hyperbolic time $\sqrt{t^2 - r^2}$ within the interior domain and is of the order of $\sqrt{t}$ in the exterior.  
Technical properties on the geometry of the foliation are provided in 
Lemmas~\ref{lem1-05-05-2020} and~\ref{lem1-03-06-2021}, which will later be supplemented with Lemma~\ref{lem 0 d-e-I}. 
We also introduce several  frames of interest associated with the proposed foliation, namely 
the ``semi-hyperboloidal frame'' (SHF) denoted by $\delH$ in \eqref{eq:semihf}, 
and the ``semi-null frame''  (SNF) $\delN$ in \eqref{eq=nulllframedef}. These vector fields are used to define
high-order operators and differentiate with the evolution equations of interest; they are also used to decompose tensor fields such as the metric.  
In addition, in Section~\ref{sectionN-3} for wave or Klein-Gordon equations on a curved spacetime we also
derive several versions of the weighted energy estimate associated with the proposed foliation; cf.~Proposition~\ref{prop 1 energy-curved}, as well as in Proposition~\ref{prop energy-ici-exterior} (Euclidean-merging domain) and Proposition~\ref{prop energy-ici-interior} (hyperboloidal domain).

%-------------------------

\item {\bf Functional inequalities} (Section~\ref{sectionN-4}).

Next, we establish new weighted Sobolev, Poincar\'e, and Hardy inequalities which are adapted to our Euclidean--hyperboloidal foliation. 
We begin with sup-norm Sobolev inequalities in the hyperboloidal domain in Proposition~\ref{prop:glol-Soin} (which involves the boost vectors) and next for the Euclidean-merging domain in Proposition~\ref{pro204-11-2}, which involves a {\sl weight function} associated with the distance to the light cone. 
Observe that our inequality in the exterior domain does not refer to values of the function in the interior domain.
A weighted Hardy inequality for the Euclidean-hyperboloidal foliation is derived in Proposition~\ref{lem1-hardy}, which will be necessary in order to control {\sl undifferentiated} terms such as metric coefficients. 
We supplement these results with two more inequalities of Poincar\'e-type, first for 
the Euclidean-merging domain in Proposition~\ref{propo-Poincare-ext} and, next,  
for the hyperboloidal domain in Proposition~\ref{prop poincare-int}. A feature of the latter inequality is that it involves 
a boundary term (cf. Lemma~\ref{lemma-4109}), controlled later by the energy in the Euclidean-merging domain. 

%-------------------------------

\item {\bf Calculus rules} (Sections~\ref{sectionN-5} and~\ref{sectionN-7}). 

All the necessary calculus rules enjoyed by the vector fields and operators 
of interest are provided next and represent key technical ingredients of our theory. 
For the convenience of the reader, most of the proofs are postponed to the Appendix.  
This includes ordering properties in Proposition~\ref{prop--fund-order}, 
allowing us to work with ordered admissible operators to $\del^I L^J \Omega^K$, as well 
as commutator estimates which will be used in order to commute vectors fields with differential operators.  
Here, we introduce the important notation $|u|_{p,k}$ in \eqref{eq1 notation} which helps
us to keep track of, both,
the total order of differentiation (which we call the {\sl order,}  
denoted by $p$) and the total number of boosts or rotations (which we call the {\sl rank,} denoted by $k \leq p$). 
At this stage we use a key {\sl hierarchy structure} enjoyed by quasi-linear commutators; cf. Proposition~\ref{lm 2 dmpo-cmm-H}. 
More refined calculus rules are next established in the Euclidean-merging domain in Section~\ref{sectionN-7}. We introduce the near/far decomposition. The above hierarchy property for quasi-linear commutators is revisited  in the Euclidean-merging domain  in Proposition~\ref{prop1-12-02-2020}, while our estimates for any wave field in terms of its energy density are summarized in Proposition~\ref{prop1-10-06-2021}.  

%-----------------------------------------------------------

\item {\bf Integral and Sobolev estimates} (Section~\ref{sectionN-9}). 

The above functional inequalities and calculus rules are then applied in order to derive  
estimates concerning $L^2$ weighted norms of solutions to wave or Klein-Gordon equations. We begin by writing down 
fundamental energy-based $L^2$ estimates for wave and Klein-Gordon fields in Proposition~\ref{prop 1 L2-be}. Observe that we state here inequalities that involve the Lorentzian boost or spatial rotations denoted by $\LOmega$. 
%%%%%%% {  \sout{ Also the $L^2$ norm we write the consequence of our Hardy-Poincar\'e inequality for high-order derivatives in Proposition~\ref{eq3-15-05-2020}.} }
%
We continue with consequences of our Sobolev inequalities, and in Proposition~\ref{lem 2 d-e-I} we state the 
Sobolev decay for wave fields in the Euclidean-merging domain. The proof is rather involved and, importantly, distinguishes between the near/far light cone domains of the spacetime. Finally, we analyze the pointwise decay of Klein-Gordon fields in Proposition~\ref{lem 2 d-e-II}. Observe that our estimates take a weight with exponent $\eta \in (1/2,1)$ into account. 

%--------------------------------------------------------------

\item {\bf Pointwise decay of wave fields} (Section~\ref{section-8-added}). 

Our next task is deriving estimates controlling the pointwise behavior of wave fields and their derivatives. 
Here, we rely on Kirchhoff's formula { and establish} sharp estimates for solutions to the wave equation under assumptions about the contribution from the source. In Proposition~\ref{Linfini wave}, we distinguish between sub-critical, critical, and super-critical regimes, and 
we prove estimates with various decay behaviors in terms of the radial distance and the distance to the light cone. 
The control of the Hessian for the wave equation is next derived, at arbitrary order, in Propositions~\ref{prop1-22-05-2020} 
and~\ref{propo2-22-05-2020}. The proof distinguishes between the near/far light cone regions and relies on two different decompositions of the wave operator in Lemma~\ref{lem1-06-12-2020} (near the light cone) 
and Lemma~\ref{eq1-07-12-2020} (away from the light cone). 
We provide a gradient estimate for wave fields and, specifically, in Proposition~\ref{prop1-23-07-2020} we control 
the weighted derivative $\crochet^\rho |(\del_t-\del_r) (ru)|(t,x)$ in the Euclidean-merging domain. The proof requires an analysis of the geometry of the characteristic curve associated with a suitable decomposition of the wave operator in a curved spacetime.

%--------------------------------------------------------------

\item {\bf Pointwise decay of Klein-Gordon fields }(Section~\ref{section-9-added}).

Finally, we derive sharp decay properties of Klein-Gordon fields and their derivatives. We deal with the quasi-linear commutators in the Euclidean-merging domain; see Proposition~\ref{prop1-28-03-2021}. 
The proof is based on yet another decomposition of the Klein-Gordon operator in Lemma~\ref{lem1-14-03-2021}. 
We conclude the presentation of tools relevant to the Euclidean-Hyperboloidal Method, and 
last we state pointwise decay of Klein-Gordon fields in Proposition~\ref{lem 1 d-KG-e}.

%--------------------------------------------------------------

\end{itemize} 

%---------------------------------------------------------------------------

\paragraph{Part II. Global nonlinear stability of self-gravitating massive fields.}

Next, having presented all of our technical tools, we turn our attention to the Einstein equations. 

\begin{itemize}

\item {\bf PDEs formulation of the global existence theory} (Section~\ref{section-new-11}). 

The main existence statement in Section~2 is reformulated in wave gauge.
Here, the connection between, geometric components and PDEs components is discussed. 
Observe also that we treat the interior and the exterior of the light cone successively, and our strategy provides more regularity in the exterior of the light cone. 
Our estimates also involve the geometric weight denoted by $\zeta$, which allows us to 
distinguish between the interior and exterior domains of our foliation.

\item {\bf Structure of Einstein's field equations} (Section~\ref{sectionN-12}).

Next, we decompose the Einstein-matter system in a form that is adapted to the Euclidean-hyper\-boloidal foliation 
and we analyze the nonlinear structure of the Einstein-massive scalar field equations. In particular, we exploit the wave gauge conditions
and separate certain components of the metric.  
Indeed, the wave gauge conditions play a central role in our proof. 
{
This is true in our derivation of, both, energy and pointwise estimates,} and 
eventually 
provides us with a control on components of the metric in presence of nonlinearities that do not obey the null condition.

%-----------------------------------------------

\item {\bf Strategy of proof and consequences of the energy estimates} (Section~\ref{sec1-23-05-2021}).

We then state our bootstrap assumptions and reformulate our assumptions concerning the initial set and  reference metric. 
This section serves as the backbone for the rest of the proof and, throughout our analysis, we refer to the conditions 
in Section~\ref{sec1-23-05-2021}, especially for the range of the decay exponents. 
We derive direct bounds provided by the energy functional, including 
direct consequences of the weighted Poincar\'e inequality in Proposition~\ref{eq3-15-05-2020} and 
of the generalized Sobolev inequality in Proposition~\ref{lem 2 d-e-I}.

\item {\bf Commutator and Hessian estimates for the metric perturbation} (Section~\ref{section---12}).

We consider the metric perturbation and establish estimates that are localized near or away from the light cone. We use the calculus rules in
Section~\ref{section----63} and rely on the boost-rotation hierarchy enjoyed by quasi-linear commutators, established in Proposition~\ref{prop1-12-02-2020}.  See the statement concerning the commutators in Proposition~\ref{Proposition12.1}. 
Next, we analyze the Hessian for the wave equation, as stated in Proposition~\ref{prop1-11-08-2021}.
Our arguments here use Propositions~\ref{prop1-22-05-2020} and~\ref{propo2-22-05-2020} and, again, we distinguish between the near-light cone and far-light cone regions.

\item {\bf Near-Schwarzschild decay of the null metric component} (Section~\ref{section---13}).

This is a key section of this paper where we establish that null metric component has a `near-Schwarzschild' decay; see~Proposition~\ref{prop1-14-08-2021}. In addition, the same technique allows us to derive the light-bending condition,
stated in Proposition~\ref{prop1-24-04-2021} below. 
We decompose the spacetime domain into two sub-domains, referred to as the ``bad'' and ``good'' regions (defined in \eqref{equa-def-good-bad} below). In the bad region, which is a (thick) neighborhood of the light cone (covering points up to a distance $\sqrt{t}$) we 
integrate toward the light cone from the good region.
On the other hand, in the good region we apply on Kirchhoff's formula and we integrate from the initial data, by taking the properties of the 
source terms into account. Here, we make use of the assumed decay of the reference metric and the contribution of the initial perturbation.

\item {\bf Sharp decay for good metric components} (Section~\ref{sec--gradient-good}).

We then estimate the gradient and Hessian of the good metric components, by
applying the method presented in Section~\ref{section----83}. 
We aim at applying the weighted pointwise estimate in Proposition~\ref{prop1-23-07-2020}. 
Our observation is as follows. 
In the general decomposition \eqref{eq 1 13-01-2019} given below, we observe that the right-hand side contains 
a problematic term, namely the quasi-null term $\Pbb$ which may not enjoy integrable decay,
By virtue of the tensorial structure given in \eqref{eq1 05-juillet-2019}, the quasi-null terms in the evolution equations of the {\sl good components} of the solution $u$ 
are actually null terms and, consequently, enjoy sufficient decay.

\item {\bf Pointwise estimate for metric components at low order} (Section~\ref{section---15}). 

We next control general components of the metric at lower order of differentiation, as stated in Proposition~\ref{section-15-1}.
The near-Schwarzschild decay (see \eqref{eq1-29-03-2021} and \eqref{eq3-29-03-2021}) is deduced from Proposition~\ref{Linfini wave}) 
and {  is} essential in our proof in order to deal
% { \sout{with}} 
with massive matter fields.
On the other, for massless fields a weaker estimate (stated in \eqref{eq1-09-05-2021} and \eqref{eq19-23-04-2021}) would be sufficient.

\item {\bf  Improved energy estimates } (Sections~\ref{section-16} and~\ref{sec-closing-bootstrap}). 

Finally, we close the bootstrap argument by establishing improved energy estimates at the highest-order of differentiation, both, first for general metric components (cf.~Proposition~\ref{proposition-section16-metric})
and then for the Klein-Gordon field (cf.~Proposition~\ref{proposition-section17-matter}). The boost-rotation hierarchy made evident in our earlier estimates is the key ingredient of this final step of the { proof}. 

\item {\bf Asymptotically hyperboloidal domain} (Section~\ref{section-18}). 

We complete our analysis with an overview of the method that was introduced earlier for dealing with the asymptotically hyperboloidal domain. 
We investigate the structure of the nonlinearities of the Einstein equations in the light cone region, 
we state the boost-rotation hierarchy, and we present the remaining arguments concerning the bootstrap argument. 

\end{itemize}

\noindent In addition, some technical material is postponed to the appendix:   
properties of the weight functions (Section~\ref{appendix-A}); 
proof of a Sobolev inequality (Section~\ref{appendix-B}); 
proof of calculus rules (Section~\ref{appendix-calculus}); 
proof of pointwise decay properties of wave fields (Section~\ref{appenDD});  
method of characteristics (Section~\ref{Annex-section-8}); 
passage from geometric to PDEs initial data (Section~\ref{sec1-07-01-2022});
estimates of sub-critical nonlinearities (Section~\ref{appendix-EEE}).

%===================================================================================

\part{The Euclidean--hyperboloidal foliation method} 

\section{The proposed framework}   
\label{sectionN-3} 

\subsection{Geometry of the foliation} 
\label{sec-geometry}

\paragraph{The time function.}

The {\bf Euclidean--hyperboloidal spacetime foliation} is given on a manifold $\Mscr$ covered by a single coordinate chart $(t,x)= (t,x^a)$ with $t \geq 1$ and $x \in \RR^3$. We denote by $r^2 = |x|^2 = \sum_{a=1}^3|x^a|^2$ and $g_{\Mink} := -dt^2 + \sum_a(dx^a)^2$ the Minkowski metric.
We concentrate on $\RR^{1+3}_+ := \{(t,x)\in\RR^{1+3},t\geq 1\}$ with is the future of the initial hypersurface $\{t=1\}$. Let $\chi: \RR \mapsto [0,1]$ be a cut-off function, that is, by definition a smooth function satisfying  
\begin{equation} \label{eq3-04-05-2020} 
\chi(x) = 
\begin{cases}
0,   & x \leq 0,
\\
1,   & x> 1, 
\end{cases}
\hskip1.cm
\quad
\chi^{(m)}(x)  > 0 \quad \text{ for } x \in (0,1/2) \text{ and } m= 0,1,2,3. 
\end{equation} 
Defining the {\sl hyperboloidal and Euclidean radii} at a time $s$, as we call them, by 
\begin{equation} \label{equa-rhoHrhoE}
\rhoH(s) := {1 \over 2} (s^2 -1), \qquad \rhoE(s) := {1 \over 2} (s^2 +1), 
\end{equation} 
Then, the function $\xi$ referred to as the {\bf foliation coefficient} and defined by 
\begin{equation} \label{equa-def-xi} 
\xi(s,r) := 1-\chi(r- \rhoH(s)) 
= \begin{cases}
1, \quad & r <\rhoH(s),
\\
0, \quad & r > \rhoE(s),
\end{cases}
\end{equation}
provides us with a cut-off function that ``selects'' the hyperboloidal domain. 
We next define the {\bf Euclidean--hyperboloidal time function} $T = \Time(s,r)$ by solving the ordinary differential equation 
\begin{equation} \label{eq5-05-05-2020}
\del_r \Time(s,r) = \frac{r \, \xi(s,r)}{(s^2+r^2)^{1/2}},
\qquad 
\Time(s,0) = s. 
\end{equation}
The following observations are immediate (while further properties enjoyed by the time function will be derived in Section~\ref{sec333}).  

\begin{lemma}[The Euclidean--hyperboloidal time function]
\label{lem1-05-05-2020}
The time function $\Time^\E=\Time^\E(s)$ defined by \eqref{eq5-05-05-2020} enjoys the following properties: 
\begin{equation} \label{eq1-05-05-2020}
\Time(s,r) = \begin{cases} 
(s^2+r^2)^{1/2},   \quad &  r \leq \rhoH(s) \qquad \text{ (hyperboloidal domain),} 
\\
r + 1 = (s^2+1)/2, \quad &  r = \rhoH(s),
\\
\Time^\E(s) &  r \geq \rhoE(s) \qquad \text{ (Euclidean domain),}
\end{cases}
\end{equation} 
while 
\begin{subequations} 
\begin{equation} \label{eq2-05-05-2020}
\aligned
& 0\leq \del_r \Time(s,r)<1 \quad 
&& \text{ (slices of constant $s$ are spacelike)},
\\ 
& 0 < \del_r\Time(s,r)<1 \quad 
&& \text{ when $0<r \leq \rhoH(s)$,}  
\endaligned
\end{equation}
and
\begin{equation}
|\del_r \, \del_r \Time(s,r)| \lesssim 1. 
\end{equation}
\end{subequations} 
\end{lemma}

%-------------------------------------------------------------------------------------

\paragraph{Spacetime foliation and decomposition.}
We define $\Mscr_s := \{(t,x^a)\in\Mscr\,/\, t = \Time(s,r) \}$. This forms a one-parameter family of space-like, asymptotically Euclidean hypersurfaces. The future of the initial surface $\{t=1\}$ is foliated decomposed as 
$$
\{t\geq 1\} = \Mscr^{\init}\cup \bigcup_{s\geq 2}\Mscr_s, 
$$
where $\Mscr^{\init} := \{(t,x)\,/\, 1\leq t\leq { {\bf T}}(2,r)\}$\footnote{This region is of less importance since we are principally interested in the asymptotic behavior at large times.}.  Each slice $\Mscr_s$ is decomposed into three parts\footnote{Strictly speaking, this is not a partition since we have defined each set to be a closed set containing its boundary.}:
\label{equa-nosconditions}
\begin{equation} \label{eq:617def}
\aligned
\MH_s :& =   \big \{ t = \Time(s,|x|), \quad \quad
|x| \leq \rhoH(s) 
\big\}
\qquad 
&& \text{asymptotically hyperboloidal,}
\\
\MM_s :& =   \big\{t = \Time(s,|x| ), \quad \, 
\rhoH(s)\leq  |x| \leq \rhoE(s) 
\big\}
\qquad 
&&  \text{merging (or transition),}
\\
\Mext_s :& =   \big\{ t=\Time(s), \quad \qquad\quad
\rhoE(s) \leq |x| 
\big\}
&& \text{asymptotically Euclidean}.
\endaligned
\end{equation}
Then $\Mscr_s = \MH_s \cup \Mtran_s \cup \Mext_s$ and thanks to \eqref{eq1-05-05-2020},
$$
\aligned
\MH_s & =  \Mscr_s\cap \big\{r \leq t-1 \big\} = \big\{(t,x)\in \Mscr_s \, / \, r \leq \rhoH(s) \big\}.
%\\
%\MM_s & =  \big\{(t,x)\in\Mscr_s \, / \, \rhoH(s) \leq r \leq \rhoE(s) \big\},
%\qquad
%&&
%\Mext_s =  \big\{(t,x)\in\Mscr_s \, / \, r\geq \rhoE(s) \big\}
\endaligned
$$
In addition, we write 
$
\MME_s := \Mext_s \cup \MM_s. 
$
Moreover, for any parameter value $s_1 > s_0$ we set 
$$
\Mscr_{[s_0,s_1]} : =   { \big\{ \Time(s_0,r)\leq t\leq \Time(s_1,r) \big\} = \bigcup_{s_0\leq s\leq s_1} \Mscr_s,}
\qquad \qquad 
\Mscr_{[s_0, + \infty)} := \bigcup_{s\geq s_0} \Mscr_s, 
$$
while 
$\Mscr^\H_{[s_0, s_1])}$, $\Mscr^\H_{[s_0, + \infty)}$, etc. are defined similarly. 

\begin{lemma} 
\label{lem1-03-06-2021}
There exists a function $c=c(s)\in (0,1)$ such that the radial variable $r$ in each of the three domains satisfies 
\begin{subequations}
\begin{equation} \label{eq 1 lem 2 position}
r = |x| \in  \,
\begin{cases}
\hskip1.cm [0, t-1],  
& \MH_{[s_0, + \infty)}, 
\\ 
[t-1,t -  c(s)],  
\qquad & \MM_{[s_0, + \infty)}, 
\\
[t - c(s), +\infty),   
& \Mext_{[s_0, + \infty)}. 
\end{cases}
\end{equation}
Furthermore, there exist universal constants $K_1, K_2>0$ such that
\begin{equation} \label{eq4-08-05-2020}
K_1\leq (1/s^2) \, \Time(s,r) \leq K_2 \qquad  
\text{ in } \MME_{[s_0, + \infty)}. 
\end{equation}
\end{subequations} 
\end{lemma} 

\begin{proof} The first case in \eqref{eq 1 lem 2 position} was already pointed out in \eqref{eq1-05-05-2020}. For the remaining two cases, $\Time(s, \rhoE(s))-\rhoE(s)$ is controlled as follows. We observe that the function
$q(s,r) := r - \Time(s,r)$ satisfies 
$\del_rq(s,r) = 1 - \del_r\Time(s,r)$, therefore $0 < \del_r q(s,r)<1$ for all $\rhoH(s)\leq r<\rhoE(s).$
From $q(s, \rhoH(s)) = {-1}$, we deduce that 
$$
\rhoE(s) - \Time(s, \rhoE(s)) = q(s, \rhoE(s)) = - 1 + \int_{\rhoH(s)}^{\rhoE(s)} \big(1 - \del_r \Time(s,r) \big) \, dr, 
$$
so that
$
-1< \rhoE(s) - \Time(s, \rhoE(s)) <0. 
$
When $(t,x) \in\MM_s$ with $t = \Time(s,r)$, the function $r \mapsto q(s,r)$ is strictly increasing, hence 
$
-1 = q(s, \rhoH(s))\leq r-t \leq q(s, \rhoE(s)) < 0. 
$ 
This establishes the last two cases in \eqref{eq 1 lem 2 position}. 
Next, we deal with \eqref{eq4-08-05-2020} as follows. 
In $\MM_{[s_0, + \infty)}$,  \eqref{eq 1 lem 2 position} shows that $\Time(s,r)-1\leq r<\Time(s,r)$, therefore
$ r<\Time(s,r)\leq r+1$; in the merging domain we find 
$$
(1/2) (s^2-1) = \rhoH(s)\leq r \leq \rhoE(s) = (1/2) (s^2+1),
$$
so \eqref{eq4-08-05-2020} holds in $\MM_{[s_0, + \infty)}$. Finally, within the Euclidean domain we simply have $\Time(s,r) = \Time(s, \rhoE(s))$ which does not depend on $r$. 
\end{proof}

%------------------------------------------------------------------ 

\paragraph{Parameterizations of the hypersurfaces.} 

It is natural to use the parameterization $(t,x)$ for the description of $\Mscr_{[s_0,s_1]}$ or $\Mscr_{[s_0, + \infty)}$. However, after introducing the time function { $ \Time$} we can also use the parameterization $(s,x)$ given by solving 
$(t,x) = (\Time(s,x),x)$.  
Since { $\Time$} is a strictly increasing in $s$, the change of parameter 
$(s,x) \mapsto (t,x) = (\Time(s,r),x)$ is a smooth and global diffeomorphism in $\Mscr{[s_0, + \infty)}$ with Jacobian matrix 
$$
\left(
\begin{array}{cc}
\del_s \Time &\del_sx
\\
\del_x \Time &\del_x x 
\end{array}
\right)
=
\left(
\begin{array}{cc}
\del_s \Time &0
\\
(x^a/r) \, \del_r \Time & I
\end{array}
\right).
$$
We will also use the volume element identity $dtdx = J \, dsdx$, which involves the Jacobian defined by  
$J(s,x) := \del_s\Time(s,x)$. Then, we state the following estimate for $J$ and 
the proof is given  
in Appendix~\ref{appendix-A}.

\begin{lemma}
\label{lem-Jacobian-bounds} 
The Jacobian associated with the Euclidean--hyperboloidal foliation  
satisfies the upper and lower bounds   
$$ 
J 
\leq
\begin{cases}
{s \over \Time}= s \, (s^2 +r^2)^{-1/2} \quad      & \text{ in } \MH_s,
\\
\xi s \, (s^2 +r^2)^{-1/2} + (1- \xi) \, 2s
\quad    & \text{ in } \MM_s, 
\\
2s \quad  & \text{ in } \Mext_s,  
\end{cases}
\qquad \quad  
J 
\geq
\begin{cases}
{s \over \Time} = s \, (s^2 +r^2)^{-1/2} \quad      & \text{ in } \MH_s,
\\
\xi \, s \, (s^2 + r^2)^{-1/2} + (1- \xi) 3s/5
\quad      & \text{ in } \MM_s, 
\\
3s/5 \quad      & \text{ in } \Mext_s. 
\end{cases}
$$ 
\end{lemma}

The future-oriented normal to the spacelike hypersurfaces $\Mscr_s$ (with respect to the Euclidean metric in $\RR^4$) reads (with $a=1,2,3$) 
\begin{subequations}
\label{normal-Ms-equa1-2-june}
\begin{equation} \label{normal-Ms}
n_s =  \frac{\big(1, - (x^a/r)\del_r \Time\big)}{\sqrt{1+|\del_r \Time|^2}} 
= \big( (1+\xi^2(s,r))r^2 +s^2 \big)^{-1/2} \Big((s^2 + r^2)^{1/2},-x^a\xi(s,r) \Big), 
\end{equation}
while the surface element (with respect to the Euclidean metric) is 
\begin{equation} \label{equa1-2-june}
d\sigma_s = (1+|\del_r \Time|^2) \, dx = \big( s^2 +r^2(1+\xi(s,r)^2)\big)^{1/2} (s^2 + r^2)^{-1/2} \, dx
\end{equation}
\end{subequations}
and, in particular,
\begin{equation} \label{eq1-08-05-2020}
n_sd\sigma_s = (1, - (x^a/r)\del_r \Time) \, dx = (1, - \del_a \Time) \, dx = \Big(1,\frac{-\xi(s,r)x^a}{(s^2+r^2)^{1/2}}\Big) \, dx.
\end{equation}

%-------------------------------------------

\paragraph{Frames of interest.}

In our approach, three choices of sets of vector fields are of interest. 

\begin{itemize}

\item {\bf The semi-hyperboloidal frame} (SHF)
\begin{equation} \label{eq:semihf}
\delH_0: = \del_t, 
\qquad\qquad
\delH_a = \delsH_a: = \frac{x^a}{t} \del_t + \del_a
\end{equation}
was already introduced by the authors in \cite{PLF-YM-book}. It is defined globally in $\Mscr_s$ and is relevant within the hyperboloidal domain. 
The transformations from this frame to the natural (Cartesian) frame $\del_\alpha$ (and vice-versa) are given by relations of the form $\delH_\alpha = \PhiH{}_\alpha^{\alpha'} \del_{\alpha'}$, while the inverse of $\PhiH$ is denoted by $\PsiH$, with 
$$
\big(\PhiH_{\alpha}^\beta  \big)
=
\left(
\aligned
&1 &&0 &&&0 &&&&0
\\
&x^1/t &&1 &&&0 &&&&0
\\
&x^2/t &&0 &&&1 &&&&0
\\
&x^3/t &&0 &&&0 &&&&1
\endaligned
\right),
\qquad
\qquad
\big(\PsiH_\alpha^\beta  \big) 
=
\left(
\aligned
&1 &&0 &&&0 &&&&0
\\
-&x^1/t &&1 &&&0 &&&&0
\\
-&x^2/t &&0 &&&1 &&&&0
\\
-&x^3/t &&0 &&&0 &&&&1
\endaligned
\right).
$$

The semi-hyperboloidal frame is the appropriate frame in the hyperboloidal domain in order to exhibit the
{ null and quasi-null} form structure of the nonlinearities of Einstein's field equations (as well as other nonlinear wave equations). 
It allows us to establish the relevant {\sl decay properties in the timelike and null directions.} Some of our arguments will involve a radial integration based on  
\begin{equation} \label{equa2---5mai}
\delsH_r := (x^a /r)\delsH_a.  
\end{equation}
For a two-tensor $T$ defined in $\Mscr_{[s_0, + \infty)}$, 
we use the notation 
$T = T^{\alpha\beta}\del_{\alpha}\otimes \del_{\beta}$ 
as well as,  in the semi-hyperboloidal frame,  
$T^{\alpha\beta} \del_{\alpha}\otimes \del_{\beta} = \TH^{\alpha\beta}\delH_{\alpha}\otimes\delH_{\beta}$
in which 
$\TH^{\alpha\beta} = \PsiH_{\alpha'}^\alpha \PsiH_{\beta'}^\beta T^{\alpha'\beta'}$. 

%------------------------------------------ 

\item {\bf The semi-null frame}  (SNF)
\begin{equation} \label{eq=nulllframedef}
\delN_0 : = \del_t, 
\qquad \qquad
\delN_a = \delsN_a:= {x^a \over r} \del_t + \del_a 
\end{equation}
is defined everywhere in $\Mscr_s$ {\sl except} on the center line $r=0$. 
The transformations between this frame and the natural frame are expressed by the identities $\delN_\alpha = \PhiN{}_\alpha^\beta \, \del_\beta$ and $\del_\alpha = \PsiN_{\alpha}^\beta \delN_\beta$, with 
$$
\big(\PhiN_{\alpha}^\beta \big)
= 
\left(
\begin{array}{cccc}
1 &0 &0 &0
\\
x^1/r& 1 &0 &0
\\
x^2/r&0  &1 &0
\\
x^3/r& 0 &0 &1
\end{array}
\right),
\qquad\qquad
\big(\PsiN_{\alpha}^\beta \big)
= 
\left(
\begin{array}{cccc}
1 &0 &0 &0
\\
-x^1/r& 1 &0 &0
\\
-x^2/r&0  &1 &0
\\
-x^3/r& 0 &0 &1
\end{array}
\right). 
$$
The semi-null frame is the most appropriate frame within the Euclidean-merging domain in order to exhibit the structure of the nonlinearities of the field equations and, in turn, to establish the  relevant {\sl 
decay properties in spatial and null directions}.
For a two-tensor $T$ defined in $\Mscr_{[s_0, + \infty)}\backslash\{r=0\}$, 
we use the notation 
$T = T^{\alpha\beta}\del_{\alpha}\otimes \del_{\beta}$ as well as, in the semi-hyperboloidal frame, 
$T^{\alpha\beta} \del_{\alpha}\otimes \del_{\beta} = \TN^{\alpha\beta}\delN_{\alpha}\otimes\delN_{\beta}$
in which 
$\TN^{\alpha\beta} = \PsiN_{\alpha'}^\alpha \PsiN_{\beta'}^\beta T^{\alpha'\beta'}$. 

%----------------------------------------------------------- 

\item {\bf The Euclidean--hyperboloidal frame} (EHF) is defined as 
\begin{equation} \label{equation-87} 
\aligned
\delEH_0 := \del_t, 
\qquad\qquad
\delEH_a & = \delsEH_a 
:=   \del_a + (x^a/r)\del_r \Time \, \del_t 
= \del_a + x^a\xi(s,r) (s^2 + r^2)^{-1/2} \del_t. 
\endaligned
\end{equation} 
By construction, $\delsEH_a$ are tangent vectors to the slices $\Mscr_s$. It is clear that $\delEH_a = \delH_a$ in $\MH_s$,
while $\delEH_a = \del_a$ in $\Mext_s$. The expressions of the vectors $\delEH_a$ are more involved in the merging $\MM_s$, where the vectors $\delEH_a$ interpolate between $\delH_a$ and $\del_a$. 
We also use the notation $\delE = (\del_\alpha)$ and $\delsE = (\del_a)$ for the restriction of $\delEH$ and $\delsEH$ in the Euclidean domain, while $\delM$ and $\delsM$ are the restriction of $\delEH$ and $\delsEH$ in the merging domain. 
We also define $\delME$ and $\delsME$ analogously. 
Some of our arguments will involve a radial integration along the hypersurfaces based on  
\begin{equation} \label{equa2---5mai-2} 
\delsEH_r := (x^a/r)\delsEH_a. 
\end{equation}

\end{itemize} 

%----------------------------------------------------------------------------------------------------------------------------------------------------- 

\subsection{Weighted energy estimates in Minkowski space}
\label{sec:energyrepeat}

\paragraph{The energy coefficient $\zeta$.}

The fundamental energy functional (stated shortly below) is going to involve another geometric weight, denoted by $\zeta = \zeta(t,x)$ and defined by 
\begin{equation} \label{equa-defzeta}
\zeta(s,r)^2 
: = 1 - \frac{r^2 \xi^2(s,r)}{s^2+r^2} = {s^2 \over s^2 +r^2} + (1- \xi^2(s,r)){r^2 \over s^2 +r^2}.
\end{equation}
It coincides with the weight $s/t = s /(s^2+r^2)^{1/2}$ in the hyperboloidal domain and reduces to $1$ in the Euclidean domain, and in turn will allow us to ``interpolate'' between the energy density induced on hyperboloids and the one induced on Euclidean slices. 
Moreover, a direct estimate on $\zeta$ is as follows.

\begin{lemma} 
\label{lme1-26-05-2021}
\label{Lem1-05-May-2020} 
In the Euclidean-merging domain, one has 
$$
\frac{|r-t|+1}{r} \lesssim \zeta^2 \leq\zeta\leq 1 \quad \text{ in } \MME.
$$
\end{lemma}

\begin{proof} Since $r\geq t-1$, we have $\frac{|r-t|}{r}\lesssim 1 = \zeta^2$  provided $r\geq \rhoE(s)$. So we only need to treat the merging domain $\MM_s$ in which $\rhoH(s) \leq r \leq \rhoE(s)$.
From \eqref{eq 1 lem 2 position}, we have already seen that $\frac{t-r}{r} \leq \frac{1}{r}$.
On the other hand, we observe that
$$
0\leq \zeta\leq \zeta^2 = \frac{s^2+(1-\xi^2(s,r)r^2)}{s^2+r^2} \leq 1
$$
and, since $\rhoH(s) \leq r \leq \rhoE(s)$, this leads us to $\zeta^2 \geq \frac{s^2}{s^2+r^2} \geq Cs^{-2} \geq Cr^{-1}$. 
\end{proof}

\begin{lemma}[Energy coefficient in the merging domain]
\label{lem1-22-05-2020}
In the merging domain the weight $\zeta$ is controled to the Jacobian, namely 
$$
K_1 \, \zeta^2 s\leq J\leq  K_2 \, \zeta^2 s \qquad \text{ in } \MM,  
$$ 
where $K_1,K_2$ are universal constants.
\end{lemma}

\begin{proof}
Comparing the inequality in Lemma~\ref{lem-Jacobian-bounds} (in the case $\MM_s$) with \eqref{equa-defzeta}, we see that 
$$
(1-\xi(s,r))s\leq (1-\xi^2(s,r))s\leq s\Big((1-\xi^2(s,r))\frac{r^2}{s^2+r^2} + \frac{s^2}{s^2+r^2} \Big) =  \zeta^2s.
$$
Then, recall that $Ks^2\leq r\leq K' s^2$ in $\MM_s$ with universal constants $K,K'$. From $r^2+s^2\lesssim s^4$ we deduce that
$\frac{1}{s^2+r^2}\lesssim \frac{s^4}{(s^2+r^2)^2}$
and, consequently,
$\frac{1}{(s^2+r^2)^{1/2}}\lesssim \frac{s^2}{s^2+r^2}$, finally leading us to
$
\xi(s,r) s \, (s^2+r^2)^{-1/2} \lesssim \zeta^2 s.
$
On the other hand, recalling the lower bound on $J$ established in Lemma~\ref{lem-Jacobian-bounds}, we observe that 
$$
(1-\xi^2(s,r))\frac{r^2}{s^2+r^2}\lesssim (1-\xi(s,r))
$$
and, in view of the property \eqref{eq4-08-05-2020} of the time function, we find
$\frac{s^2}{s^2+r^2}\lesssim 1-\xi(s,r)$ in the range $1-\xi(s,r) \geq 1/ r$.  
When $1-\xi(s,r) \leq 1/r$ we have $\xi(s,r) \geq 1-1/r\geq 1/2$ and therefore 
$s^2/ (s^2+r^2) \lesssim \xi(s,r) (s^2+r^2)^{-1/2}$. 
\end{proof}

%--------------------------

\paragraph{Weight function and energy estimate.}

We state first the basic energy estimate for a wave or Klein-Gordon equation in Minkowski spacetime and, next, explain how it carries over to a curved spacetime provided suitable geometric terms are taken into account. 
We first introduce a weight which reduces to a constant in the interior of the light cone and is essentially the distance to the light cone in the exterior domain.
To this end, we fix (once for all) 
{
a smooth and non-decreasing function $\aleph$ satisfying $\aleph(y) = 0$ for $y \leq { -2}$ and $\aleph(y) = y+2$ for $y \geq {-1}$,} and introduce the {\bf energy weight} 
\begin{equation} \label{eq:weight}  
\crochet := 1 + \aleph({r-t}). 
\end{equation} 
Recall that $\aleph'$ is non-negative. This choice is easier to work with, but an equivalent energy having a more geometric form can be based on the unknown metric $g$. 

Recall our notation $\Box  = \Box_{\gMink}  = - \del_t \, \del_t + \sum_{a=1,2,3} \del_a\del_a$ for the wave operator
in Minkowski spacetime. We treat simultaneously the wave and Klein-Gordon operators by assuming here that $c \geq 0$. Multiplying $\Box u - c^2 u$ by $- 2 \, \crochet^{2 \eta}  \del_t u$, we find the divergence identity 
$$
\aligned
& \del_tV^0_{\eta,c}[u] + \del_a V^a_{\eta,c}[u] - 2\eta \crochet^{-1} \aleph'({ r-t})(-1, x^a/r) \cdot V_{\eta,c}[u]
= 
- 2 \, \crochet^{2 \eta} \del_t u \big(\Box u - c^2u\big), 
\\
&
V_{\eta,c}[u]  
:= - \crochet^{2 \eta} \big(-|\del_t u|^2 - \sum_a|\del_au|^2 - c^2u^2, 2 \del_t u\del_au\big).
\endaligned
$$
Regarding now $\crochet=\crochet(s, r)$, 
we define our energy functional on each Euclidean--hyperboloidal slice $\Mscr_s$ as, { thanks to \eqref{eq1-08-05-2020}}, 
\begin{equation} \label{eq1-03-05-2020}
\aligned
\Eenergy_{\eta,c}(s,u) 
& := \int_{\Mscr_s}V_{\eta,c}[u] \cdot n_sd\sigma_s 
= \int_{\Mscr_s} 
\Big(|\del_tu|^2 + \sum_a|\del_au|^2 + \frac{2x^a\xi(s,r)}{(s^2+r^2)^{1/2}} \del_t u\del_a u + c^2u^2 \Big) \, 
\crochet(s, \cdot)^{2\eta} \, dx 
\\
& =  \int_{\Mscr_s} \Big(\zeta^2|\del_t u|^2 + \sum_a |\delsEH_au|^2 + c^2u^2 \Big) \,
\crochet(s, \cdot)^{2\eta} \, dx, 
\endaligned
\end{equation}
as well as in the equivalent form 
\begin{equation}  
\Eenergy_{\eta,c}(s,u) 
= \int_{\Mscr_s}
\Big(\zeta^2 \sum_a|\del_au|^2 +  \frac{\xi^2(s,r)}{s^2+r^2} \sum_{a<b} |\Omega_{ab}u|^2 
+ |\delsEH_r u|^2 + c^2 \, u^2 \Big) \, \crochet(s, \cdot)^{2\eta} \, dx,
\end{equation}
which involves the energy coefficient $\zeta$ as defined in \eqref{equa-defzeta}. 
Recall that this coefficient is non-trivial in the merging domain and in the hyperboloidal domain, only, and depends upon our choice of the foliation coefficient $\xi = \xi(s,r)$. 

%-------------------------------------------------------------------------------------------------------------------------------------------------  

\subsection{Energy functional on a curved space} 
\label{sec-energy-curved}

\paragraph{Energy identity.}

Consider the wave or Klein-Gordon equation with $c \geq 0$, defined on on a curved spacetime and with unknown $u$, 
\begin{equation} \label{eq:394-mod}
g^{\alpha\beta} \del_\alpha \del_\beta u - c^2 u = f, 
\end{equation}  
which is associated with a choice of metric 
$g^{\alpha\beta} =:  g_\Mink^{\alpha\beta} + H^{\alpha\beta}$ and right-hand side $f$. 
The energy-flux vector (with $a =1,2,3$) 
\begin{equation} \label{eq:39} 
V_{g, \eta,c}[u]
:=
-\crochet^{2 \eta} \Big(
g^{00} |\del_t u|^2 - g^{ab} \del_au \, \del_bu - c^2u^2, \ 2 g^{a\beta} \del_t u  \del_\beta u \Big),
\end{equation}
depends upon $g$ as well as the weight $\crochet^\eta$, and enjoys  the identity
$$ 
\aligned
& - 2 \, \crochet^{2 \eta} \del_t u \, \big(g^{\alpha\beta} \del_\alpha \del_\beta u -  c^2 \, u \big)
\\
& =   \dive  V_{g, \eta,c}[u] - 2\eta\crochet^{-1} \aleph'({ r-t})(-1,x^a/r) \cdot V_{g,\eta,c}[u]
- \crochet^{2\eta}\del_tH^{\alpha\beta}\del_{\alpha}u\del_{\beta}u 
+ 2 \, \crochet^{2\eta}\del_{\alpha}H^{\alpha\beta} \del_tu\del_{\beta}u.
\endaligned
$$
Hence, by setting   
\begin{equation} \label{eq7-08-05-2020}
\aligned
\Omega_{g, \eta,c}[u] 
& := 
-2\eta\crochet^{-1} \aleph'({ r-t}) (-1,x^a/r) \cdot V_{g,\eta,c}[u]
\\
& = 2 \, \eta\crochet^{2\eta-1} \aleph'({ r-t}) \, \big(g^{\N ab}\delsN_au\delsN_bu - H^{\N00} |\del_tu|^2 + c^2u^2\big),
\\
- G_{g, \eta}[u] 
& :=    \del_tH^{00} | \crochet^\eta \del_t u|^2 - \del_tH^{ab} \crochet^{2 \eta} \del_au\del_b u + 2 \crochet^{2 \eta}  \del_aH^{a\beta} \del_t u\del_{\beta} u, 
\endaligned
\end{equation}
we arrive at the \textbf{fundamental energy identity} 
\begin{equation} \label{eq 2 energy-curved-mod-0}
\aligned
-2 \crochet^{2 \eta} \, \del_tu f
= 
\dive  V_{g, \eta,c}[u] + \Omega_{g, \eta,c}[u] - G_{g, \eta}[u]. 
\endaligned
\end{equation}  
%
%-------------------------- 

\paragraph{Notation for the light cone energy.}

We introduce the {\bf light cone sections}
\begin{equation} \label{eq2-17-01-2021}
\Lscr_{[s_0,s_1]} = \big\{ (t,x)\in \RR^4 \, / \,  r=t-1 >0, \, \rhoH(s_0)\leq r \leq \rhoH(s_1) \big\}, 
\end{equation}
which is the locus ``between'' $\MH_{[s_0,s_1]}$ and $\MME_{[s_0,s_1]}$. Observe that the normal vector and the volume element of $\Lscr_{[s_0,s_1]}$ (with respect to the Euclidean metric) are 
$
n_{\Lscr} = \frac{\sqrt 2}{2}(1,-x^a/r)$
and $d\sigma_{\Lscr} = \sqrt{2}dx$. 
%\end{equation}
Furthermore, we have 
$$ 
V_{g,\eta,c}[u] \cdot n_{\Lscr}d\sigma_{\Lscr} = \Big(c^2u^2 + g^{\N ab}\delsN_au\delsN_bu - H^{\N00} |\del_t u|^2\Big) \, dx, 
$$
where $g^{\alpha\beta} = g_\Mink^{\alpha\beta} + H^{\alpha\beta}$  
and, finally, we define 
\begin{equation} \label{equa-Ebfgc}
\Eenergy_{g,c}^{\Lcal}(s_1,u;s_0) 
= \int_{\Lscr_{[s_0,s_1]}}V_{g, \eta,c}[u] \cdot n_{\Lscr}d\sigma_{\Lscr}
= \int \big(- \HN^{00} |\del_t u|^2 + g^{\N ab} \delsN_au\delsN_bu + c^2u^2 \big) \, dx,
\end{equation} 
where in the latter integral the domain of integration is defined by $t=r+1$ and $\rhoH(s_0)\leq r \leq \rhoH(s_1)$. 
Moreover, it is easily checked that 
\begin{equation} \label{eq8-27-03-2021}
\frac{d}{ds} \Eenergy_{g,c}^{\Lcal}(s,u;s_0) 
= s \int \big(- \HN^{00} |\del_t u|^2 + g^{\N ab} \delsN_au\delsN_bu + c^2u^2 \big)d\sigma,
\end{equation}
where the domain of integration is defined by $t=r+1$ and $r = \rhoH(s)$. 
The last two terms in \eqref{equa-Ebfgc} are non-negative whenever $g$ is sufficiently close to the Minkowski metric. 
However, they do not control the first term (due to the derivative in time).
Interestingly, the sign of $H^{\N00}$ is undetermined at this stage whereas 
it will play a key role as we will explain later on.

%-----------------------------------------------------------------------------------------------------------  

\paragraph{Energy on hypersurfaces.}

Next, we consider the integral  
\begin{equation}
\Eenergy_{g, \eta,c}(s,u) := \int_{\Mscr_s}V_{g, \eta,c}[u] \cdot n_s \, d\sigma_s.
\end{equation}
By integrating \eqref{eq 2 energy-curved-mod-0} over the domain limited by a slice of the foliation and the initial slice, and 
then using Stokes' formula together with \eqref{eq1-08-05-2020},  we arrive at the \textbf{fundamental energy identity} 
\begin{equation} \label{eq2-08-05-2020}
\Eenergy_{g, \eta,c}(s_1,u) - \Eenergy_{g, \eta,c}(s_0,u)
+ \int_{\Mscr_{[s_0,s_1]}} \!\!\!\!\!\!\!\!\big(\Omega_{g, \eta,c}[u] - G_{g, \eta}[u]\big) \, dxdt 
= -2 \int_{\Mscr_{[s_0,s_1]}} \!\!\!\!\!\!\!\!  \del_t uf \, \crochet^{2 \eta} dxdt.
\end{equation}
Under the change of variable $(t,x)\mapsto (s,x)$, by recalling the expression of the Jacobian $J(s,x) = \del_s\Time(s,x)$ we obtain
\begin{equation} \label{eq7-27-03-2021}
\aligned
& \Eenergy_{g, \eta,c}(s_1,u) - \Eenergy_{g, \eta,c}(s_0,u) + \int_{s_0}^{s_1} \int_{\Mscr_s} \big(\Omega_{g, \eta,c}[u] - G_{g, \eta}[u]\big) J dx ds
= - 2 \int_{s_0}^{s_1} \int_{\Mscr_s}   \del_t u \,  f \, J \,  \crochet^{2 \eta}  \, dxds.
\endaligned
\end{equation}

On one hand, by integrating \eqref{eq 2 energy-curved-mod-0} in $\MH_{[s_0,s_1]}$ and observing that 
$\Omega_{g,\eta,c}[u] = 0$ in 
$\MH_{[s_0,s_1]}$ and, therein, $J = s/\Time(s,r) = s/t$, we obtain 
\begin{equation}
\Eenergy_{g, \eta,c}^\H(s_1,u) - \Eenergy_{g, \eta,c}^\H(s_0,u) - \Eenergy_{g,c}^{\Lcal}(s_1,u;s_0) 
= \int_{s_0}^{s_1} \int_{\MH_s}  \big(G_{g, \eta}[u] - 2   \del_t u \,  f\big) \, (s/t) \, dxds, 
\end{equation} 
where $\Eenergy_{g,\eta,c}^\Hcal(s,u):= \int_{\Mscr^{\Hcal}_s}V_{g, \eta,c}[u] \cdot n_s \, d\sigma_s$.
On the other hand, by integrating  \eqref{eq 2 energy-curved-mod-0} in $\MME_{[s_0,s_1]}$ we find 
\begin{equation}
\aligned
& \Eenergy_{g, \eta,c}^{\ME}(s_1,u) - \Eenergy_{g, \eta,c}^{\ME}(s_0,u) + \Eenergy_{g,c}^{\Lcal}(s_1,u;s_0) + \int_{s_0}^{s_1} \int_{\Mscr^{\ME}_s} \big(\Omega_{g, \eta,c}[u] - G_{g, \eta}[u]\big) J dx ds
\\
& = -2 \int_{s_0}^{s_1} \int_{\MME_s}  \del_t u \,  f \, J \,  \crochet^{2 \eta}  \, dxds
\endaligned
\end{equation}
with\footnote{In $\Hcal_s$, the weight $\crochet$ is trivial, so the index $\eta$ is irrelevant and omitted.}
$\Eenergy_{g,\eta,c}^\ME(s,u) = \Eenergy_{g,c}^\ME(s,u) := \int_{\Mscr^{\ME}_s}V_{g, \eta,c}[u] \cdot n_s \, d\sigma_s$.
It remains to control the integral $\int_{\Mscr_s} \crochet^{2 \eta} \del_t u\, f J dx$ and, by recalling Lemma~\ref{lem-Jacobian-bounds},  we find 
$$
\aligned
\Big|\int_{\Mscr_s} \crochet^{2 \eta} \del_t u\, f J dx\Big| & \lesssim \int_{\MH_s} |(s/t)\del_t u | \, | f|dx + \int_{\MM_s} |\del_t u f|J\, dx
+ \int_{\Mext_s} | \crochet^\eta \del_t u | \, |  \crochet^\eta  s f|dx, 
\endaligned
$$
where we used $\crochet \lesssim 1$ in $\MH_s\cup\MM_s$ and Lemma~\ref{lem-Jacobian-bounds} in $\Mext_s$. We can then differentiate \eqref{eq7-27-03-2021} with respect to $s_1$. 

\begin{proposition}[Weighted energy estimate for the EH foliation on a curved spacetime]
\label{prop 1 energy-curved} 
Any solution to \eqref{eq:394-mod} (with sufficient regularity and fast decay at spacelike infinity) satisfies
$$
\aligned
&\frac{d}{ds} \Eenergy_{g,\eta,c}(s,u)
+ 2\eta \int_{\Mscr_s}  \big( g^{\N ab} \delsN_au\delsN_bu + c^2u^2
\big)  \aleph'({ r-t}) \crochet^{2\eta-1}\ Jdx
\\
&= \int_{\Mscr_s} \Big(G_{g, \eta}[u] + \eta\crochet^{2\eta-1} \aleph'({ r-t}) \HN^{00}  |\del_t u|^2   \Big)\ Jdx 
+ \int_{\Mscr_s}  |\del_t u f| \, \crochet^{2 \eta} \, Jdx, 
\endaligned
$$  
in which the latter integral can (for instance) be controlled by 
$$
\aligned
\int_{s_0}^{s_1} \Eenergy_{\eta,c}(s,u)^{1/2} \, \big(\| f\|_{L^2(\MH_s)}
+ \big\| s\zeta f\big\|_{L^2(\MM_s)}
+ \| s \, \crochet^\eta f\|_{L^2(\Mext_s)} \big) \, ds. 
\endaligned
$$ 
\end{proposition}

The following two results are established in a similar way. 

\begin{proposition}[Weighted energy estimate in the Euclidean-merging domain]
\label{prop energy-ici-exterior} 
Any solution $u: \MME_{[s_0,s_1]} \to \RR$ to the wave or Klein-Gordon equation \eqref{eq:394-mod} with  
right-hand side $f: \MME_{[s_0,s_1]} \to \RR$ satisfies  
$$ 
\aligned
&\frac{d}{ds} \Eenergy_{g,\eta,c}^{\ME}(s,u) + \frac{d}{ds} \Eenergy^{\Lcal}_{g, c}(s,u;s_0)  
+  2 \eta  \int_{\MME_s}\big( g^{\N ab} \delsN_au\delsN_bu +  c^2u^2\big)  \crochet^{2\eta-1} \aleph'({ r-t}) \, Jdx
\\
&=
\int_{\MME_s} \Big(G_{g, \eta}[u] + \eta\crochet^{2\eta-1} \aleph'({ r-t}) \HN^{00}  |\del_t u|^2   \Big) \, Jdxds 
+ \int_{\MME_s} \crochet^{2 \eta} \del_t u f \, Jdxds, 
\endaligned
$$
in which the latter integral is bounded by 
$
\int_{s_0}^{s_1} \, (\Eenergy_{\eta,c}^{\ME}(s,u))^{1/2} \, \big\|J \, \zeta^{-1} \crochet^{\eta} f\big\|_{L^2(\MME_s)} \, ds.
$
\end{proposition}

\begin{proposition}
[Weighted energy estimate in the asymptotically hyperboloidal domain]
\label{prop energy-ici-interior}
Any solution $u: \MH_{[s_0,s_1]} \mapsto \RR$ to the wave or Klein-Gordon equation \eqref{eq:394-mod} 
with right-hand side $f: \MH_{[s_0,s_1]} \mapsto \RR$ satisfies 
$$
\aligned
& \frac{d}{ds}\Eenergy^\H_{g,c}(s,u) - \frac{d}{ds}\Eenergy^{\Lcal}_{g,c}(s,u;s_0) 
=\int_{\MME_s} \crochet^{2 \eta} \del_t u f\ Jdx,
+  \int_{\MH_s}  G_{g, \eta}[u]  \, Jdx.  
\endaligned
$$ 
\end{proposition}

%==============================================================================================

\section{Sobolev, Hardy, and Poincar\'e inequalities} 
\label{sectionN-4}

\subsection{Sobolev inequalities on the Euclidean-hyperboloidal foliation} 

\paragraph{Preliminaries.}

We need to establish global Sobolev inequalities, separately in each of 
domains $\MH_s$ and $\MME_s$, 
when 
the functions under consideration, in general, are not compactly supported, nor can be approximated by functions compactly supported in each of these domains. It is important to do this without using the trace of the functions along the domain boundaries, as presented in this section. In fact, the role of smooth and compactly supported functions on $\RR^3$ is now played by functions which are compactly supported in a cone.

\paragraph{An inequality in a solid cone.}

We use here the notation $x=(x^a) \in \RR^3$ ($a=1, 2, 3$), together with  
$\RR^3_+ := \{x \in\RR^3 \, / \,  x^a\geq 0\}$
and consider functions that are defined in the domain $\RR^n_+$ and smoothly extendible  outside it.  
To any point $x \in \RR^3_+$ and any scale parameter $\rho>0$ we associate the cube
$
C_{\rho, x} := \big\{y\in\RR^3_+ \, / \, x^a\leq y^a\leq x^a + \rho \big\}.
$ 
Throughout, we restrict attention to sufficiently regular functions enjoying sufficiently fast decay at infinity. The proof of the following estimates is 
% 
%provided in the companion paper~\cite{PLF-YM-companion}. 
%}
postponed to Appendix~\ref{appendix-B}. 

\begin{lemma}[A sup-norm Sobolev inequality]
\label{lem 1 30 -10 -217}
Given any $\rho>0$ there exists a constant $C(\rho)>0$ such that for all functions $u: \RR^3_+ \mapsto \RR$, one has the  Sobolev inequality 
$$ 
|u(x)|
\lesssim C \, (\rho) \textstyle \sum_{|I| \leq 2} \| \del^I u\|_{L^2(C_{\rho, x})},
\qquad  
x \in \RR^3_+.
$$
\end{lemma}

%-----------------------

\paragraph{Inequalities in the exterior of a ball.} 

Next, we consider functions defined in the exterior of a ball, specifically either
$\Dtrs_s := \big\{x \in\RR^3 \, / \, |x| \geq {  (s^2 - 1)/2 = \rhoH(s)} \big\}$
or $\Dext_s  : = \big\{x \in\RR^3 \, / \, |x| \geq { (s^2+1)/2 = \rhoE(s)}\big\}$.  
We rely on the partial derivatives $\del_a$ and the rotations 
$\Omega_{ab} = x^a\del_b-x^b\del_a$ ($a,b=1,2,3$) and, from Lemma~\ref{lem 1 30 -10 -217}, we deduce the following result.

\begin{lemma}[Sup-norm Sobolev inequalities]
\label{pro2-10} 
\begin{subequations}
For any function $u=u(x)$ defined in $\Dtrs_s$ or $\Dext_s$, respectively, one has the  
following inequalities based on either the translation fields  
\begin{equation} \label{eq-dh1}
\sup_{\Dtrs_s}  |u|  \lesssim
\textstyle
\sum_{|I| \leq 2} \| \del^I u\|_{L^2(\Dtrs_s )}, 
\end{equation}
or the translation and rotation fields
\begin{equation} \label{eq-dh2}
\aligned
\sup_{\Dtrs_s} (1+ |x|) |u(x)| 
& \lesssim   \textstyle
\sum_{|I| + |J| \leq 2} \| \del^I\Omega^J u\|_{L^2(\Dtrs_s )}. 
\endaligned
\end{equation}
Here, the implied constants are independent of $s$ and the same inequalities hold with $\Dtrs_s$ replaced by $\Dext_s$. 
\end{subequations}
\end{lemma}

\begin{proof}  
1. The inequality \eqref{eq-dh1} is an immediate consequence of Lemma~\ref{lem 1 30 -10 -217} which was stated in a cube 
with a vertex located 
at an arbitrary point $x$. Given a point $x \in \Dext_s$, the cube $C_{\rho, x}$ is included in the exterior domain $\Dext_s$ and, consequently, \eqref{eq-dh1} (with the choice $\rho=1$) implies \eqref{eq-dh2} since $\| \del^I u\|_{L^2(C_{1, x})} \leq \| \del^I u\|_{L^2(\Dtrs_s )}$.

\vskip.3cm 

2.  Given any point $x_0\in \Dext_s$, we can apply a rotation and, without loss of generality, eventually assume that $x_0 = (r_0,0,0)$. We then consider the following domain described in standard spherical coordinates: 
$$
R_{x_0} := \big\{ r_0\leq r \leq r_0+1, \quad
0\leq \theta\leq \pi/6, \quad
\pi/3\leq \varphi\leq \pi/2 
\big\}.
$$ 
The case where $r_0$ is a priori bounded, say $r_0 \leq 1$, is a consequence of \eqref{eq-dh1}. 
So, from now on we assume that $r_0$ is bounded from below, say $r_0 \geq 1$. 
We consider the restriction of the function $u$ to the domain $R_{x_0}$, that is, 
$v_{x_0}( \rho, \theta, \varphi): = u(x^1, x^2, x^3)$
with  
$$
x^1 = (r_0 + \rho) \sin\varphi\, \cos\theta,
\qquad \quad
x^2 = (r_0 + \rho) \sin\varphi\, \sin\theta, 
\quad\qquad x^3 = (r_0 + \rho) \cos\varphi.
$$
We  express the partial derivative operators in the $(\theta, \varphi)$--variables 
in terms of the rotation fields and obtain the following identities 
(with uniformly bounded coefficients): 
$$
\aligned 
\del_{\theta} & =  \Omega_{12},
\qquad \qquad 
\del_\varphi   = - \sin \theta \, \Omega_{23} - \cos\theta \, \Omega_{13}, 
\qquad 
&& \del_\theta\del_\theta  = \Omega_{12} \Omega_{12},
\\
\del_\theta\del_\varphi 
& =  
- \sin \theta \, \Omega_{12} \Omega_{23} - \cos\theta \, \Omega_{12} \Omega_{13} - \cos\theta\Omega_{23} + \sin\theta\Omega_{13}, 
\\ 
\del_{\varphi} \del_{\varphi} 
& =   (\sin \theta)^2 \, \Omega_{23} \Omega_{23} + (\cos\theta)^2\Omega_{13}\Omega_{13} 
+  
(\sin\theta\cos\theta) (\Omega_{23} \Omega_{13} + \Omega_{13}\Omega_{23}). 
\endaligned
$$ 

By setting now  $\phi := \pi/2- \varphi$, the function $z_{x_0} (\rho, \theta, \phi) = v_{x_0} ( \rho, \theta, \varphi)$ is defined in the cube 
$\big\{ 0\leq \rho \leq 1, \quad 
0\leq \theta\leq \pi/6, \quad
0\leq \phi \leq \pi/6 \big\} \subset C_{1/2, 0}$ (since $\pi/6> 1/2$).
Therefore, in view of Lemma~\ref{lem 1 30 -10 -217} we have the Sobolev inequality 
\begin{equation} \label{eq 5 01-11-217}
|u(x_0)| = |z_{x_0}(0, 0, 0)|
\lesssim \textstyle
\sum_{|I| \leq 2} \| \del^Iz_{x_0} \|_{L^2(C_{1/2,0})}.
\end{equation}
Observe also that
$
| \del z_{x_0} | \lesssim | \del u| + \sum_{a\neq b} | \Omega_{ab} u|
$
and, more generally, 
$
| \del^{I'} z_{x_0} |
\lesssim
\sum_{|I| + |J| \leq 2} | \del^I\Omega^J u|$ for $|I'| \leq 2. 
$
On the other hand, by recalling that $r = r_0+\rho$ we find 
$$
\aligned
& \| \del^{I'} z_{x_0} \|_{L^2(C_{1/2,0})}^2 
= \int_{C_{1/2,0}} | \del^{I'} z_{x_0}(\rho, \theta, \phi)|^2 \,d{\rho} d\theta d\phi
\\
& \lesssim  
\sum_{|I| + |J| \leq 2} \int_{R_{x_0}} | \del^I\Omega^J u|^2 \,dr d\theta d\varphi
\lesssim
r_0^{-2} \sum_{|I| + |J| \leq 2} \int_{R_{x_0}} | \del^I\Omega^J u|^2 \,r^2 \sin\varphi \, dr d\theta d\varphi, 
\endaligned
$$
where we used $1\leq r/r_0\leq 2$ within $R_{x_0}$ as well as 
$\sqrt{3}/2 \leq \sin\varphi\leq 1$. We arrive at 
$$
\aligned
\| \del^{I'} z_{x_0} \|_{L^2(C_{1/2,0})}^2
& \lesssim  
r_0^{-2} \sum_{|I| + |J| \leq 2} \| \del^I\Omega^J u\|_{L^2(R_{x_0})}^2
\lesssim
(1+r_0)^{-2} \sum_{|I| + |J| \leq 2} \| \del^I\Omega^J u\|_{L^2(\Dext_s )}^2
\endaligned
$$
and, in combination with \eqref{eq 5 01-11-217}, the desired result follows for $r_0\geq 1$ and, therefore for all $r_0$.
\end{proof}

%-----------------------------------------------------------------------------------------------------------

\paragraph{Hyperboloidal domain.}

We rely here on the boosts $L_a = x^a\del_t+ t \, \del_a$ which are tangent to the hyperboloidal slices. 

\begin{proposition}[Sup-norm Sobolev inequality in the hyperboloidal domain]
\label{prop:glol-Soin}
For any function defined on a hypersurface $\MH_s$, the following estimate holds (in which $t^2 = s^2+ |x|^2$): 
$$
\sup_{\MH_s} t^{3/2} \, |u(t,x)|
\lesssim 
\sum_{|J| \leq 2} \| L^J u\|_{L^2(\MH_s)}
\simeq 
\sum_{m=0,1,2} \| t^m (\slashed \del^\H)^m  u\|_{L^2(\MH_s)}.   
$$
\end{proposition}

%----------------------------------------------------------------------------------------------------------

\begin{proof} It is equivalent to consider the restriction of a function $u$ to the hyperboloid $\MH_s$ with $|x| \leq \rhoH(s)$, that is, the function 
$v_s(x) := u((s^2 + r^2)^{1/2}, x)$. 
Then, we see that
$
\del_av_s = \delH_a u = t^{-1}L_a u = (s^2 +r^2)^{-1/2}L_a u.
$
Consider any point $x_0 \in \MH_s$ with $t_0 = \sqrt{s^2 + r_0^2}$ and, without loss of generality, assume that $x_0 = -3^{-1/2}(r_0,r_0,r_0)$. 
Our argument of proof has two parts. First of all, we consider the cube $C_{s/2, x_0} \subset \{|x| \leq { (s^2 - 1)/2 = \rhoH(s)} \}$. (Recall that $s \geq 2$ is assumed throughout.)  
In this cube we introduce the change of variable
$
y^a: = s^{-1}(x^a-x_0^a)
$
and consider the function 
$
w_{s, x_0}(y): = v_s(sy + x_0)$ for $y\in C_{1/2,0}$. 
We have 
$$
\aligned
\del_aw_{s, x_0} & =   s\del_av_s = \frac{s}{(s^2 + r^2)^{1/2}}L_au,
\qquad
\qquad
\del_b\del_aw_{s, x_0} & = \frac{s^2}{s^2 +r^2}L_bL_au -s^2x^b(s^2 +r^2)^{-3/2}L_au,
\endaligned
$$
thus,  for all $|I| \leq 2$ we obtain 
$
| \del^I w_{s, x_0} |
\lesssim
\sum_{|J| \leq 2} |L^J u|.
$
In view of Lemma~\ref{lem 1 30 -10 -217} we find 
$$
\aligned
|w_{s, x_0}(0)|^2 
\lesssim
\sum_{|I| \leq 2} \int_{C_{1/2,0}} | \del^Iw_{s, x_0} |^2 \, dy 
\lesssim 
s^{-3} \sum_{|J| \leq 2} \int_{C_{s/2,x_0}} \hskip-.3cm  |L^J u|^2 \, dx,
\endaligned
$$
which leads us to
$
|u(x_0)| 
\lesssim
s^{-3/2} \|L^J u\|_{L^2(\MH_s)}.
$
This inequality provides us with the desired conclusion in the range $r_0\leq \sqrt{3} s$, since $t_0 = \sqrt{s^2 +r_0^2} \leq 2s$. 

%----------------------------------

Next, in the range $r_0\geq \sqrt{3} s$ we consider the cone $C_{r_0/2, x_0}$ and introduce the function
$w_{x_0}(y) := v_s(r_0y+x_0)$ for $y\in C_{1/2,0}$. 
We have now 
$$
\aligned
\del_aw_{x_0} 
& =   \frac{r_0}{\sqrt{r^2 +s^2}}L_au, 
\qquad\qquad
\del_b\del_aw_{x_0} 
= \frac{r_0^2}{r^2 +s^2}L_bL_au - r_0^2x^b(s^2 +r^2)^{-3/2}L_au.
\endaligned
$$
In the cube $C_{r_0/2, x_0}$ we find $r \geq (1- \sqrt{3}/2) r_0$ and, therefore, 
$| \del^I w_{x_0} | \leq \sum_{|J| \leq 2} |L^Ju|$ for $|I| \leq 2$. 
Again, from Lemma~\ref{lem 1 30 -10 -217} we deduce 
$$
\aligned
|u(x_0)|^2 = |w_{x_0}(0)|^2 
& \lesssim  
\sum_{|J| \leq 2} \int_{C_{1/2,0}} | \del^Iw_{x_0} |^2 \, dy 
\lesssim 
r_0^{-3} \sum_{|J| \leq 2} \int_{C_{r_0/2,x_0}} |L^J u|^2 \, dx, 
\endaligned
$$
which leads us to $|u(x_0)| 
\lesssim
r_0^{-3/2} \sum_{|J| \leq 2} \|L^J u\|_{L^2(\MH_s)}
\lesssim t_0^{-3/2}\sum_{|J| \leq 2} \|L^J u\|_{L^2(\MH_s)}$ 
since, in this case, $4r_0^2\geq 3s^2 + 3r_0^2 = 3t_0^2$.
\end{proof}

%-----------------------------------------------------------------------------------------------------------

\paragraph{Euclidean-merging domain.} 

In the inequalities below, recall that $\delsME{}^I$ denotes any $|I|$-order operator determined from the fields $\{\delsME_a\}_{a=1,2,3}$, while $\delsE{}^I$ denotes any a $|I|$-order operator determined from the fields $\{\del_a\}_{a=1,2,3}$. Recall that $\crochet = 1 + \aleph({r-t})$ was introduced in \eqref{eq:weight}. 

\begin{proposition}[Weighted sup-norm Sobolev inequality in the Euclidean-merging domain]
\label{pro204-11-2}
Fix an exponent {$\eta \geq 0$} and set $C(\eta) := 1 + \eta + \eta^2$.
For all sufficiently regular functions defined in $\Mscr_{[s_0,s_1]}$ with $2 \leq s_0 \leq s \leq s_1$, one has  
\begin{subequations}
\begin{equation} \label{ineq 2 sobolev}
r  \crochet^\eta |u(t,x)| 
\lesssim 
C(\eta) \sum_{|I| + |J| \leq 2} \|\crochet^\eta  \delsME{}^I\Omega^J u\|_{L^2(\MME_s)}, 
\qquad (t,x)\in \MME_s, 
\end{equation}
\begin{equation} \label{ineq 1 sobolev}
r \crochet^\eta |u(t,x)| 
\lesssim
C(\eta) \sum_{|I| + |J| \leq 2} \|\crochet^\eta  \delsE{}^I\Omega^J u\|_{L^2( \Mext_s)},
\qquad 
(t,x)\in \Mext_s. 
\end{equation}
\end{subequations}
\end{proposition}

We establish a technical result first. We analyze the restriction to  $\MME_{s}$ of the weight function $\crochet$ in 
\eqref{eq:weight}, that is, $\omega_s(r) := 1 + \aleph(r-\Time(s,r))$ 
which satisfies $ \delsME_a\crochet  = \del_a\omega_s(r) =  \aleph'(r-t)(x^a/r)(1-\del_rT)$.  
We thus focus on the tangent derivatives of $\crochet^{\eta}$. 

\begin{lemma} 
\label{lem1-26-02-2020}
For any exponent {$\eta\geq 0$}, on $\MME_s$ one has 
$$ 
\big| \delsME ( \crochet^{\eta})\big| +  \big|\delsME \delsME  \crochet^{\eta} \big| 
\lesssim  C(\eta) \crochet^{\eta-1}.
$$
\end{lemma}

\begin{proof} 
Observe that
$
\delsME_a \crochet^{\eta} 
= \eta \crochet^{\eta-1} \aleph'(r-t)(x^a/r)(1-\del_rT)$  
and 
$$
\aligned
\delsME_b \delsME_a \crochet^{\eta} 
& = \eta (\eta-1) \crochet^{\eta-2} \aleph'(r-t)^2(x^a/r) (x^b/r) (1-\del_rT)^2
+  
\eta \crochet^{\eta-1} \, \aleph''( r-t )(x^a/r) (x^b/r) (1-\del_rT)^2 
\\
& \quad +  
\eta \crochet^{\eta-1} \aleph'( r-t ) \Big(   
(1/r)  \big( \delta_{ab} - (x^ax^b/r^2) \big) (1-\del_rT)
+ (x^a/r)(x^b/r) (-\del_r\del_rT) \Big). 
\endaligned
$$
We then observe that the functions $\aleph'$ and $\aleph''$ are bounded, while
both $\del_rT$ and $\del_r\del_r \Time$ are uniformly bounded (thanks to Lemma~\ref{lem1-05-05-2020}). 
\end{proof}

\begin{proof}[Proof of Proposition~\ref{pro204-11-2}] 
Consider the parameterization $(s,r)$ of $\Mscr_{[s_0,+\infty)}$ and recall that, in $\MME_s$, the function 
$s$ is constant and $t = \Time(s,r)$ with $r \geq \rhoH(s)$.
Consider the restriction of $ \crochet^{\eta} u$ to the slice $\MME_s$, that is, the function 
$
v_{s, \eta}(x) =  \crochet(s,r) ^{\eta} u(\Time(s,r), x). 
$
We have 
$$
\aligned
\del_a v_{s, \eta} (x)
& =  \delsME_a( \crochet(s,r)^{\eta} u(\Time(s,r),x)) 
=   \crochet^{\eta} \delsME_a u(t,x) +  \delsME_a \crochet^{\eta} u(t,x), 
\\ 
\del_b\del_a v_{s, \eta} (x)
& = \crochet^{\eta} \delsME_b\delsME_a u + \delsME_b \crochet^{\eta} \delsME_au(t,x)
+ \delsME_a \crochet^{\eta} \delsME_bu(t,x) + \delsME_b\delsME_a \crochet^{\eta} \ u(t,x), 
\endaligned
$$
while 
$$
\aligned
\Omega_{ab}v_{s, \eta}(x)
& 
=  \crochet^{\eta} \Omega_{ab} u(t,x),
&&
\Omega_{cd} \Omega_{ab} v_{s, \eta}(x) & =  \crochet^{\eta} \Omega_{cd} \Omega_{ab} u(t,x),
\\
\delsME_c \Omega_{ab}v_{s, \eta}(x) 
& 
=\delsME_c \crochet^{\eta} \Omega_{ab} u(t,x) +  \crochet^{\eta} \delsME_c\Omega_{ab} u(t,x),
\\ 
\Omega_{ab}\delsME_c v_{s,\eta}(x) 
&  
= \delsME_c\Omega_{ab}v_{s,\eta}(x) + \delta_c^b\delsME_a v_{s,\eta}(x) - \delta_c^a\delsME_b v_{s,\eta}(x). 
\endaligned
$$
In view of Lemma~\ref{lem1-26-02-2020}, the above bounds take the form  
$$
|\del^I\Omega^Jv_{s, \eta} |(x)\lesssim 
C(\eta) \crochet^{\eta} \hskip-.3cm  \sum_{|I'|+|J'| \leq 2} \big|\delsME{}^{I'} \Omega^{J'} u(t,x)\big|, \qquad |I|+|J| \leq 2.
$$
Then, applying Lemma~\ref{pro2-10} to the function $v_{s, \eta}$ we obtain
$$
\aligned
& \crochet(s,r)^{\eta} |u(\Time(s,r),x)| =  |v_{s, \eta}(x)| 
\lesssim
C(\eta) r^{-1} \hskip-.3cm \sum_{|I|+|J| \leq 2} \|\del^IL^Jv_{s, \eta} \|_{L^2(\Dtrs_s)}
\lesssim 
C(\eta) r^{-1} \hskip-.3cm  \sum_{|I|+|J| \leq 2} \| \crochet^{\eta} \delsME{}^I\Omega^J u\|_{L^2(\MME_s)}, 
\endaligned
$$
which is \eqref{ineq 2 sobolev}. The inequality \eqref{ineq 1 sobolev} is established in the same manner and we only need to 
observe that on $\delsME_a = \del_a$ in $\Mext_s$. 
\end{proof} 

%------------------------------------------------------------------------------------------------------  

\subsection{Hardy inequality on the Euclidean-hyperboloidal foliation}

In order to control the $L^2$ norm of the functions under consideration, we will proceed by estimating their first-order derivatives (thanks to a weighted energy functional) and then rely on the functional inequality presented now.  
Recall here our notation $\Fenergy_\eta := \Eenergy_\eta^{1/2}$ in \eqref{equa-defEF}. 

\begin{proposition}[Weighted Hardy inequality on the Euclidean-hyperboloidal foliation]
\label{lem1-hardy}
Fix some exponent {$\eta \geq 0$}. For any function $u$ defined in $\Mscr_{[s_0,s_1]}$ and sufficiently decaying at infinity, one has  
$$ 
\| r^{-1} \crochet^\eta u\|_{L^2(\Mscr_s)} 
\lesssim  
\| \crochet^\eta  \delsEH u\|_{L^2(\Mscr_s)}
\lesssim \Fenergy_\eta(s,u).
$$
\end{proposition}

\begin{proof}  
We write $v_s : = u(\Time(s,x),x)$ for the restriction of $u$ to the slice $\Mscr_s$, so that $\del_av_s = \delsEH_a u$
and, since
$1/r^2 = \del_a(x^a/r^2)$, we have\footnote{Throughout, the calculations are made with the parametrization $(s,x)$ in $\Mscr_s$, where $s$ is fixed.}
$$
\int_{\{\eps\leq |x|\}}r^{-2}(\omega_s^\eta v_s)^2 \, dx 
= \int_{\{\eps\leq |x|\}} \del_a\big(x^a(\omega_s^\eta v_s)^2/r^2 \big) \, dx
- 2 \int_{\{\eps\leq |x|\}}(x^a/r^2) \, \omega_s^\eta v_s (\del_a \omega_s^\eta v_s+ \omega_s^\eta\del_a v_s) \, dx, 
$$
where $\omega_s$ denotes the restriction of $\crochet$ to a slice.
Provided $u$ decays sufficiently fast at spacelike infinity, Stokes' formula implies 
$$
\int_{\{\eps\leq |x|\}} \del_a\big(x^a(\omega_s^\eta v_s)^2/r^2 \big) \, dx = \int_{|x| = \eps}(-x^a/r)_{a} \cdot (x^a(\omega_s^\eta v_s)^2/r^2 \big)_a d\sigma_{\eps}
= - \eps \int_{\Sbf^2}  (\omega_s^\eta v_s)^2d\sigma, 
$$
which tends to zero with $\eps\to 0$.
So we find
\begin{equation} \label{eq1-17-01-2021} 
\int_{\RR^3}r^{-2}(\omega_s^\eta v_s)^2 \, dx 
= - 2 \int_{\RR^3} (x^a/r)\del_a\omega_s^\eta \ r^{-1} \omega_s^\eta v_s^2 \, dx
- 2 \int_{\RR^3}(x^a/r)\omega_s^\eta \del_a v_s\ \big(r^{-1} \omega_s^\eta v_s\big) \, dx. 
\end{equation}
At this juncture, we recall that 
$
(x^a/r)\del_a \omega_s^\eta(r)  = \eta \crochet(s,r) ^{\eta-1} \aleph'(r-\Time(s,r)) (1-\del_r\Time(s,r))
$ 
and, thanks to the property $\del_r \Time\in [0, 1]$, we find $(x^a/r)\del_a \omega_s^\eta(r)\geq 0$. 
Consequently, \eqref{eq1-17-01-2021} leads us to
$$
\int_{\RR^3}r^{-2}(\omega_s^\eta v_s)^2 \, dx
\leq - 2\int_{\RR^3}(x^a/r)\omega_s^\eta \del_a v_s\ \big(r^{-1}  \omega_s^\eta v_s\big) \, dx, 
$$
and we obtain 
$\| r^{-1} \crochet^{\eta} u\|_{L^2(\Mscr_s)}^2
\lesssim \| r^{-1} \crochet^{\eta} u\|_{L^2(\Mscr_s)} \sum_a\|\crochet^{\eta} \delsEH_au\|_{L^2(\Mscr_s)}$. 
\end{proof}

%-------------------------------------------------------------------------------------------------------------------------------------------------

\subsection{Poincar\'e inequalities on the Euclidean-hyperboloidal foliation} 

\paragraph{Euclidean-merging domain.}

Estimates involving the $L^2$ norm of the wave field will be required later on, while the energy functional will only provide us with a control of first-order (or higher-order) derivatives. We cannot rely on a standard Hardy inequality since we are working in a domain with boundary and, especially, we cannot apply Stokes formula as is done in the proof of Hardy inequality. We refer to our inequality below as a Poincar\'e inequality since it relies on the
main idea behind the proof of the standard Poincar\'e inequality, i.e.~an integration from a boundary and an application of Fubini theorem. 

\begin{proposition}[Poincar\'e inequality in the Euclidean-merging domain.] 
\label{propo-Poincare-ext}
Fix an exponent {$\eta =1/2 +\delta$} with $\delta > 0$. 
For any function $u$ defined in $\MME_s = \{(t,x)\in\Mscr_s\, / \, |x|\geq \rhoH(s)\}$, one has  
\begin{equation} \label{Poincare-trex-sans-zeta}
\|  \crochet^{-1 + \eta}  u\|_{L^2(\MME_s)}
\lesssim
\big( 1+ \delta^{-1} \big) \|  \crochet^{\eta} \delsME u\|_{L^2(\MME_s)}
+ 
\|  r^{-1} \crochet^{\eta} u\|_{L^2(\MME_s)}, 
\end{equation}
\begin{equation} \label{Poincare-trex-zeta}
\|\crochet^{-1 + \eta}\zeta u\|_{L^2(\MME_s)}
\lesssim
(1+\delta^{-1})\|\crochet^{\eta}\zeta \delsME u\|_{L^2(\MME_s)}
+ 
\| r^{-1}\crochet^{\eta}\zeta u\|_{L^2(\MME_s)}. 
\end{equation}
\end{proposition}

Consequently, the contribution $\| r^{-1} \crochet^{\eta} u\|_{L^2(\MME_s)}$ in the right-hand side of the above inequality is controlled by Proposition~\ref{lem1-hardy} and we arrive at   
\begin{equation} \label{eq1-10-jan-2020}
\|  \crochet ^{-1 + \eta}  u\|_{L^2(\MME_s)} \lesssim
(1+\delta^{-1}) \, \Fenergy_{\eta}(s,u). 
\end{equation} 

\begin{proof}  
By replacing $\zeta$ by $1$ the first inequality can be deduced from the forthcoming proof of the second inequality. 

{\bf Step 1.} 
We work in spherical coordinates $(r, \omega) \in \RR^+ \times \Sbf^2$ defined in the hypersurface $\MME_s$. Let $\chi$ be a smooth cut-off function defined in $\RR$ with $\chi(y)=0$ for $y \leq 0$ and $\chi(y) = 1$ for $y \geq 1$. Given a function $u$ defined in the slab $\MME_{[s_0, s_1]}$, we set $u_s(x) := u(\Time(s,x),x)$ which is
the restriction of this function to the hypersurface $\MME_s$. 
We also introduce its exterior and interior parts 
$u^{\text{int}}$ and $u^{\text{ext}}$ by 
\begin{equation}
u_s^{\text{ext}}(r, \omega) := \chi(r/t-2)u_s(r, \omega), 
\qquad \qquad 
u_s^{\text{int}} := u_s - u_s^{\text{ext}}.
\end{equation}
By construction, $u_s^{\text{ext}}$ is supported outside the ball $r\geq2t$ where $\crochet\simeq r$ and { $\zeta =1$}, thus
\begin{equation} \label{eq 0 pr 16-09-2018}
\big\|\crochet^{-1 + \eta} \zeta \, u_s^{\text{ext}} \big\|_{L^2(\MME_s)}
\lesssim
\| r^{-1}\crochet^{\eta} \zeta \, u \|_{L^2(\MME_s)}.
\end{equation}
We thus focus our attention on the interior part $u_s^{\text{int}}$ and, for convenience, we set $v := u_s^{\text{int}}$. In view of 
\begin{equation} \label{eq 1 pr 16-09-2018}
\aligned
\del_r v(x) & =    -\chi'(r/t-2)t^{-1} u_s(x) + \chi(r/t-2)\del_r u_s(x) 
\\
& =    -\chi'(r/t-2)t^{-1} u(\Time(s,x),x) + \chi(r/t-2)\delsEH_r u(\Time(s,x),x)
\endaligned
\end{equation}
with $\delsEH_r = (x^a/r) \delsEH_a$, we deduce that 
\begin{equation} \label{eq 1' pr 16-09-2018}
|\del_r v(x)|
\lesssim
r^{-1} |u(\Time(s,x))| + | \delsEH_r u(\Time(s,x),x)|. 
\end{equation}
Observe that $\chi'(r/t-2)$ is supported in $2 \leq r/t\leq 3$ and in this region $r\sim t$. 

For any function $w$ defined in the interval $[t-1,3t]$ and vanishing at $3t$, we write 
$
w(r) = -\int_{r}^{3t} \del_rw(\rho) \, d\rho$
and
therefore 
\begin{equation} \label{eq 2' pr 16-09-2018}
\crochet(s,r) ^{2 \eta - 2}w(r) 
= - \la r- { \Time(s,r)}\ra^{2 \eta -2} \int_r^{3t} \del_r w(\rho) d\rho. 
\end{equation}
Using the average 
$
w(r) = \int_{\Sbf^2}v^2(r, \sigma)d\sigma$, 
we obtain 
\begin{equation} \label{eq 2'' pr 16-09-2018}
\int_{\rhoH(s)}^{3t} \crochet(s,r) ^{2\eta-2} w(r) \,r^2 dr = \|\crochet^{\eta-1} v\|_{L^2(\MME_s)}^2.
\end{equation}
On the other hand, from \eqref{eq 2' pr 16-09-2018}
and in view of $\del_rw(\rho) = 2\int_{\Sbf^2}v(\rho, \sigma)\del_rv(\rho, \sigma)d\sigma$, we deduce that  
$$
\aligned
& \int_{\rhoH(s)}^{3t} \crochet(s,r) ^{2-2\eta} \zeta^2 \, w(r) \,r^2 dr
= -2 \int_{\rhoH(s)}^{3t} \crochet(s,r) ^{-2+2 \eta} \zeta^2 \,  
\int_r^{3t} \Big(\int_{\Sbf^2}v(\rho, \sigma)\del_rv(\rho, \sigma) \,   d\sigma \Big) \, d\rho \, r^2 dr, 
\endaligned
$$
which gives us 
$$ 
\aligned
& \int_{\rhoH(s)}^{3t} \crochet(s,r) ^{2-2\eta} \zeta^2 \,  w(r) \,r^2 dr
\\
&
\leq  2 \int_{\rhoH(s)}^{3t} \crochet(s,r)^{-2+2 \eta} \zeta^2 \, 
\int_{\rhoH(s)}^{3t} \int_{\Sbf^2}
\frac{\mathbbm{1}_{\{\rho\geq r\}} \, \big| \crochet(s,\rho) ^{-1+\eta} \zeta \,  v(\rho, \sigma) \big| \, 
\big| \crochet(s,\rho) ^{\eta} \zeta \, \del_rv(\rho, \sigma) \big| \, \rho^2}{\rho^2\crochet(s,\rho)^{-1+2 \eta} \zeta^2(s,\rho) } \,d\rho d\sigma\, r^2 dr
\\
& =   \int_{\rhoH(s)}^{3t} \int_{\Sbf^2} 
\big|\crochet(s,\rho) ^{-1+\eta} \zeta \, v(\rho, \sigma) \big| \, 
\big|\crochet(s,\rho) ^{\eta} \zeta \, \del_r v(\rho, \sigma) \big|
\left(\int_{\rhoH(s)}^{3t} \frac{\mathbbm{1}_{\{\rho\geq r\}}r^2 \crochet(s,\rho)^{1 - 2 \eta} \zeta^2(s,r)}{\rho^2\crochet(s,r)^{2-2 \eta} \zeta^2(s,\rho)}dr\right) \rho^2d\rho d\sigma.
\endaligned
$$
We check the monotonicity of the function $r \mapsto r^2\zeta^2(s,r)$, by writing 
$$
\aligned
\del_r\big(r^2\zeta^2(s,r)\big)  &=  2r\zeta^2(s,r) + r^2\left(- \frac{2\xi(s,r)r^2\del_r\xi(s,r)}{s^2+r^2} - \frac{2\xi^2(s,r)r}{s^2+r^2} + \frac{2\xi^2(s,r)r^3}{(s^2+r^2)^2}\right),
\\
2r\zeta^2(s,r)  &=  2r - \frac{2\xi^2(s,r)r^3}{s^2+r^2} 
\endaligned
$$
and, with $\del_r\xi\leq 0$,
$$
\aligned
\del_r\big(r^2\zeta^2(s,r)\big) \geq& 2r - \frac{2\xi^2(s,r)r^3}{s^2+r^2} -  \frac{2\xi^2(s,r)r^3}{s^2+r^2} + \frac{2\xi^2(s,r)r^5}{(s^2+r^2)^2}
\\
 &= 2r\left(1-\frac{(2s^2r^2+r^4)\xi^2(s,r)}{(s^2+r^2)^2}\right)\geq 2r\left(1-\frac{2s^2r^2+r^4}{(r^2+s^2)^2}\right)\geq 0. 
\endaligned
$$
Here, we used $0\leq \xi(s,r)\leq 1$. This shows that $r^2\zeta^2(s,r)$ is increasing with respect to $r$.

\vskip.3cm

\noindent{\bf Step 2.} 
Now, with $\eta = 1/2+\delta$ (and $\delta>0$) we write  
$$
\aligned
&
\int_{\rhoH(s)}^{3t} \frac{\mathbbm{1}_{\{\rho\geq r\}}  r^2 \zeta^2(s,r)}{\rho^2 \zeta^2(s,\rho)}
\frac{\la\rho-\Time(s, \rho)\ra^{1 - 2 \eta}}{\crochet(s,r)^{2-2 \eta}}dr 
\\
& =   
\int_{\rhoH(s)}^{3t} \frac{\mathbbm{1}_{\{\rho\geq r\}} r^2 \zeta^2(s,r)}{\rho^2 \zeta^2(s,\rho)} 
\frac{\crochet(s,\rho) ^{1 - 2 \eta}}{\crochet(s,r) ^{2-2 \eta}}
\, 
\aleph'(r-\Time(s,r))(1 - \del_r\Time(s,r))
\, dr 
\\
& \quad + \int_{\rhoH(s)}^{3t} \frac{\mathbbm{1}_{\{\rho\geq r\}} r^2 \zeta^2(s,r)}{\rho^2 \zeta^2(s,\rho)} 
\frac{\crochet(s,\rho) ^{1 - 2 \eta}}{\crochet(s,r) ^{2-2 \eta}} 
\, 
\Big( 1 - \aleph'(r-\Time(s,r))(1 - \del_r\Time(s,r)) \Big)
\, dr 
=:T_1(s, \rho) +T_2(s, \rho). 
\endaligned
$$
To handle the term $T_1$, we apply the change of variable $y = \omega_s(r) = \crochet(s,r) $ 
(which is increasing with respect to $r$): 
$$
\aligned
T_1(s, \rho) 
& = \int_{\rhoH(s)}^{3t} \frac{\mathbbm{1}_{\{\rho\geq r\}}r^2 \zeta^2(s,r)}{\rho^2 \zeta^2(s,\rho)} 
\frac{\crochet(s,\rho) ^{1 - 2 \eta}}{\crochet(s,r) ^{2-2 \eta}}
\, 
\aleph'(r-\Time(s,r))(1 - \del_r\Time(s,r)) 
\, dr 
\\&
\leq 
\int_1^{\crochet(s,\rho) } \frac{\crochet(s,\rho) ^{1-2 \eta}}{y^{1 + (1-2 \eta)}}dy
=  \int_1^{\crochet(s,\rho) } \left(\frac{y}{\crochet(s, \rho)}\right)^{-1+ (2 \eta-1)}d\left(\frac{y}{\crochet(s, \rho)}\right)
\\
&
=   \int_{\crochet(s,\rho) ^{-1}}^1\!\!\!\!\!\! z^{-1+(2 \eta-1)}dz
=  (2\delta)^{-1} z^{2\delta} \Big|_{\crochet(s,\rho) ^{-1}}^1  \lesssim \delta^{-1}, 
\endaligned
$$
in which we used $2 \eta-1 = 2 \delta$. To handle $T_2$, we observe that $\del_r\Time(s,r) \equiv 0$ when $r\geq \rhoE(s) = \frac{s^2+1}{2}$ (i.e.  $(t,x)\in \Mext_s$).  On the other hand, $ \aleph'(\rho) \equiv 1$ for ${\rho\geq 0}$. Recalling
\eqref{eq 1 lem 2 position}, we have $-1\leq \rhoE(s) - \Time(s,\rhoE(s)) = c(s)\leq 0$.  
We also observe that $\Time(s,r) = \Time(s,\rhoE(s)) = T^{\E}(s)$ when $r\geq \rhoE(s)$.
Consequently, for all $r\geq \rhoE(s) + 1$ we have 
$$
r- \Time(s, r) = r - \Time(s,\rhoE(s)) \geq 1 + (\rhoE(s) - \Time(s,\rhoE(s)))  = 2 - c(s) \geq 0, 
$$
where we used \eqref{eq 1 lem 2 position}. Thus for all $r\geq \rhoE(s)+ 1$, we have 
$1 - \aleph'(r-\Time(s,r)+2)(1 - \del_r\Time(s,r)) \equiv 0$. On the other hand, when $\rhoH(s)\leq r \leq \rhoE(s) + 1$ 
and, observing that $r-\Time(s,r)$ is non-decreasing with respect to $r$ for $r\geq \rhoH(s)$, 
$$
0\geq r - \Time(s,r) \geq r - \Time(s,\rhoH(s)) 
= r - \rhoH(s) + \rhoH(s) - \Time(s,\rhoH(s)) \geq -1. 
$$
Then, using $1-2 \eta<0$ we have 
$$
T_2(s, \rho) = \int_{\rhoH(s)}^{\rhoE(s)+1} \frac{\mathbbm{1}_{\{\rho\geq r\}} r^2 \zeta^2(s,r)}{\rho^2 \zeta^2(s,\rho)}
\frac{\crochet(s,\rho) ^{1 - 2 \eta}}{\crochet(s,r) ^{2-2 \eta}}
\Big( 1 - \aleph'(r-\Time(s,r)+2)(1 - \del_r\Time(s,r)) \Big)
\, dr\lesssim 1. 
$$

\vskip.15cm

\noindent{\bf Step 3.} Returning to the identity derived in Step 1, it follows that 
$$ 
\aligned
& \int_{\rhoH(s)}^{3t}\crochet^{2\eta-2} \zeta^2(s,r) w(r) \,r^2 dr
\\
&
\lesssim
(1+\delta^{-1})\int_{\rhoH(s)}^{3t} \int_{\Sbf^2} |\crochet(s,\rho) ^{-1+\eta}\zeta(s,\rho) v(\rho, \sigma)
\, 
\la\rho-\Time(s, \rho)\ra^{\eta} \zeta(s,\rho) \del_rv(\rho, \sigma) \, |\rho^2d\rho d\sigma
\\
& =   (1+\delta^{-1})\|\crochet(s,r) ^{-1+\eta} \zeta(s,r) v\cdot \, \crochet(s,r) ^{\eta} \zeta(s,r) \del_rv\|_{L^1(\MME_s)}
\\
&
\lesssim
(1+\delta^{-1})\|\crochet(s,r) ^{-1+\eta}  \zeta v\|_{L^2(\MME_s)} \, \|\crochet(s,r) ^{\eta}  \zeta \, \del_rv\|_{L^2(\MME_s)}.
\endaligned
$$
In combination with \eqref{eq 2'' pr 16-09-2018} we obtain  
$
\|\crochet(s,r) ^{-1+\eta}  \zeta v\|_{L^2(\MME_s)} \lesssim 
(1+\delta^{-1}) \, \|\crochet(s,r) ^{\eta} \zeta \, \del_r v\|_{L^2(\MME_s)}
$
and, by recalling \eqref{eq 1' pr 16-09-2018}, 
$$
\aligned
& \|\crochet(s,r) ^{-1+\eta} \zeta u_s^{\text{int}} \|_{L^2(\MME_s)}
\lesssim
(1+\delta^{-1})\|\crochet(s,r) ^{\eta} \zeta \, \delsEH_r u\|_{L^2(\MME_s)} 
+ \|\crochet(s,r) ^{\eta}r^{-1} \zeta u\|_{L^2(\MME_s)}, 
\endaligned
$$
and we conclude by combining this result with \eqref{eq 0 pr 16-09-2018}.  
\end{proof}

%---------------------------------------------------------------------------------------------------------

\paragraph{Hyperboloidal domain.} 

We treat the interior domain, as follows. 

\begin{proposition}[Poincar\'e inequality in the hyperboloidal domain.] 
\label{prop poincare-int}  
For all functions $u$ defined in $\MH_{[s_0, s_1]}$, one has  
$$
\aligned
&  
\frac{d}{ds} \|u\|_{L^2(\MH_s)}^2 
\lesssim
\Bbf (\big( \rhoH(s),u \big) + \|u\|_{L^2(\MH_s)} \, \Fenergy_\eta(s,u), 
\qquad\qquad 
\Bbf(r,u) : =  \int_{\sigma\in \Sbf^2} u^2(r+1,r \sigma)r^2 \, d\sigma. 
\endaligned
$$
\end{proposition}

\begin{proof} We consider the vector field $V := (u^2,0, 0, 0)^T$ and compute $\dive  V= 2u\del_t u$. 
Integrating this identity over $\MH_{[s_0, s]}$ and applying Stokes' formula, we obtain 
$$
\aligned
& \int_{\MH_s}V\cdot \vecn d\sigma - \int_{\MH_{s_0}}V\cdot \vecn d\sigma + \int_{\Lscr_{[s_0,s]}} V\cdot\vecn d\sigma 
=\int_{\MH_{[s_0,s]}} \dive V dxdt 
= 2 \int_{s_0}^s\int_{\MH_s}(s/t)u\del_t u\, dx ds.
\endaligned
$$ 
Along the conical boundary $\Lscr_{[s_0,s]}$, we let $\vecn$ and $d\sigma$ be the  normal vector and volume element (with respect to Euclidean metric), respectively, so that  
$\vecn \, d\sigma = (-1,1) \, dx$.
For the integral along the conical boundary $\{r = t-1\}$, we observe that $s = (t^2 - r^2)^{1/2} = \sqrt{2r+1}$
and so $V\cdot \vecn d\sigma = u^2(r+1,x)$. 
In $\MH_s$, we have 
$\vecn d\sigma = (1,x^a/t) \, dx$
and, therefore, 
$$
\aligned
\|u\|_{L^2(\MH_s)}^2  & =    \|u\|_{L^2(\MH_{s_0})}^2
+ \int_{\rhoH(s_0)}^{\rhoH(s)} \!\!\int_{\sigma \in \Sbf^2} u^2(r+1,r \sigma)r^2d \sigma \, dr
+\int_{s_0}^s\int_{\MH_s}(s/t) \, \dive V \ dxds
\\
& =   \|u\|_{L^2(\MH_{s_0})}^2
+ \int_{\rhoH(s_0)}^{\rhoH(s)} \!\!\! \Bbf(r,u) r^2 \, dr
+\int_{s_0}^s\int_{\MH_s}(s/t) \, \dive V \, dxds.
\endaligned
$$
Finally, we differentiate the above identity with respect to $s$ and find 
$$ 
\aligned 
\frac{d}{ds} \|u\|_{L^2(\MH_s)}^2
= \Bbf (\big(\rhoH(s),u\big) + \int_{\MH_s}(s/t) \, \dive  V \, dx, 
\endaligned
$$ 
in which  
$\int_{\MH_s}(s/t) \, \dive Vdx = 2 \int_{\MH_s} u\ (s/t)\del_t u\ dx \leq 2 \, \|u\|_{L^2(\MH_s)} \, \Fenergy(s,u)$.  
\end{proof}

%-----------------------------------------------------------------------------------------------

\paragraph{A boundary term.}

The remaining term $\Bbf(r,u)$ in Proposition~\ref{prop poincare-int} is controlled by the energy in the Euclidean-merging domain, as follows.

\begin{lemma} 
\label{lemma-4109} 
Fix some exponent {$\eta = 1/2+\delta$} with $\delta>0$. 
For any function $u$ defined in $\MME_s$ one has 
$$
\Bbf(r,u) :=  
\int_{\sigma\in \Sbf^2} u^2(r+1,r\sigma)r^2 \, d\sigma
\lesssim 
(1+\delta^{-2}) \Eenergy_{\eta}^{\ME}(s,u)
\qquad \text{ at } r =  \rhoH(s).
$$
\end{lemma}

\begin{proof} If $v$ is a function defined in the half line $[t-1, + \infty)$ and vanishing at infinity, we have 
$
v(t-1) = -\int_{t-1}^{+\infty}  v'(\rho) \, d\rho
$
and, for $s\geq s_0$, we define  
$$
v_s(\rho) := \int_{\sigma \in \Sbf^2}(2+\rho-\Time(s, \rho))^{-2+2 \eta} |u(\Time(s,\rho), \rho\sigma)|^2 \rho^2 \, d\sigma.
$$ 
It then follows that 
$$
\aligned
v_s'(\rho)  
& =   (-2+2 \eta)\int_{\mathbb{S}^2}(2+\rho-\Time(s, \rho))^{-3+2 \eta}(1-\del_r\Time(s, \rho)) u^2 \rho^2 \, d\sigma
+ 2 \int_{\mathbb{S}^2}(2+\rho-\Time(s, \rho))^{-2+2 \eta} u\, \delsME_r u \rho^2 \, d\sigma 
\\
& \quad
+  2 \int_{\mathbb{S}^2}(2+\rho-\Time(s, \rho))^{-2+2 \eta} | u |^2 \ \rho d\sigma
=: T_1(s,\rho) + T_2(s, \rho) + T_3(s, \rho) 
\endaligned
$$
and therefore, with $t = T(\rhoH(s),s)$ and so $t-1 = \rhoH(s)$,  
$$
v_s(t-1) = \int_{\sigma\in \Sbf^2} |u(t, \rhoH(s)\sigma)|^2 \, d\sigma 
= - \int_{t-1}^{+\infty} v'(\rho) \, d\rho = -\int_{t-1}^{+\infty} 
\big( T_1 + T_2 + T_3\big)(s, \rho) \, d\rho. 
$$ 
We write $T_1 \leq \int_{\Sbf^2}(2+\rho + \Time(s, \rho))^{-2+2 \eta} u^2 \rho^2d\sigma$
and then 
$$
\aligned
\int_{t-1}^{+\infty}T_1(s, \rho) \, d\rho& \lesssim  \int_{t-1}^{+\infty}\int_{\sigma\in\Sbf^2} \, \big|(2+\rho-\Time(s,\rho))^{-1+\eta} u\big|^2\,\rho^2 \, d\sigma d\rho 
\lesssim \|\crochet^{-1+\eta}u\|_{L^2(\MME_s)}^2
\lesssim (1+\delta^{-2})\Eenergy_{\eta}(s,u), 
\endaligned
$$
thanks to \eqref{eq1-10-jan-2020}. 
For $T_2$, we have  
$$
\aligned
\Big|\int_{t-1}^{+\infty} T_2(s, \rho) \, d\rho\Big| 
& \leq  2 \int_{t-1}^{+\infty}  \int_{\Sbf^2} \big|(2+\rho-T)^{-1+\eta} u\,  (2+\rho-T)^{\eta} \delsME_r u\big|\ \rho^2 \, d\sigma d\rho
\\
& \lesssim  \|(2+r-t)^{\eta-1} u\|_{L^2(\MME_s)} \|(2+r-t)^{\eta} \delsME_r u\|_{L^2(\MME_s)}
\lesssim (1+\delta^{-1}) \Eenergy_{\eta}(s,u), 
\endaligned
$$
thanks to the Poincar\'e inequality \eqref{eq1-10-jan-2020}. For $T_3$, 
in view of Proposition~\ref{lem1-hardy} together with \eqref{eq1-10-jan-2020}, we find 
$$
\aligned
&\bigg|\int_{t-1}^{+\infty} T_3(s, \rho) \, d\rho\bigg| 
\leq  2 \int_{t-1}^{+\infty}  \int_{\Sbf^2} {\rho^{-1}}(2+\rho-\Time(s, \rho))^{-2+2 \eta} |u(\rho, \sigma)|^2 \ \rho^2 \, d\sigma d\rho
\\
& = 2 \int_{\MME_s}r^{-1}(2+r-t)^{-2+\eta} u^2dx
\lesssim \| \crochet^\eta r^{-1} u\|_{L^2(\MME_s)} \| \crochet^{-1+\eta} u\|_{L^2(\MME_s)} 
\lesssim  (1+\delta^{-1}) \, \Eenergy_{\eta}(s,u).\qedhere
\endaligned 
$$
\end{proof}

%-------------------------------------------------------------------------------------------

\paragraph{A final observation.}

Proposition~\ref{prop poincare-int} is not in a suitable form yet.  
We can use Proposition~\ref{prop poincare-int} together with Lemma~\ref{lemma-4109} and deduce  
(recalling again our notation $\Fenergy := \Eenergy^{1/2}$ in \eqref{equa-defEF}) 
\begin{equation} \label{eq1 prop poincare-int-copy}
\aligned
\|u\|_{L^2(\MH_s)} \frac{d}{ds} \|u\|_{L^2(\MH_s)}
\lesssim (1+\delta^{-2}) \, \Fenergy_{\eta}(s,u)^2
+ \|u\|_{L^2(\MH_s)} \, \Fenergy_\eta(s,u). 
\endaligned
\end{equation}
With the help of the technical lemma below, this inequality implies the final estimate  
\begin{equation} \label{eq2-26-01-2020}
\|u\|_{L^2(\MH_s)}
\lesssim
\|u\|_{L^2(\MH_{s_0})} + (1+\delta^{-2}) \, \Big( \sup_{s'\in[s_0,s]} \, \Fenergy_{\eta}(s',u) + \int_{s_0}^s \Fenergy_{\eta}(s',u) \, ds'\Big), 
\end{equation}
which we refer to as the {\bf hyperboloidal Poincar\'e inequality.}

\begin{lemma}
\label{ODE Gronwall-tech}
Let $P$ and $Q$ be non-negative functions defined in an interval $[s_\star,s^\star]$.  
Then, any solution $v \geq 0$ to the differential inequality
$
v(s) \, v'(s)\leq P(s)^2 + v(s)Q(s) 
$
satisfies 
$$
v(s)\leq v(s_\star) + \sup_{s'\in [s_\star,s]}\big( P(s')\big) + \int_{s_\star}^s(P+Q)(s') \, ds'.
$$
\end{lemma}

\begin{proof}
Let $I\subset[s_\star,s^\star]$ be the set where $v(s)>P(s)$, that is, 
$
I = \big\{ s \in[s_\star,s^\star] \, / \, v(s) > P(s) \text{ and } v(s) >0 \}.
$ 
Outside this set the conclusion is obvious. 
Since $I$ is an open subset of $[s_\star,s^{\star}]$ we can write 
$I = [s_\star,s^\star]\cap \bigcup_{i=1}^{+\infty} (s_i,S_i)$.
Given an arbitrary point $s\in [s_\star,s^\star]$ be a point in $I$, let us consider the corresponding
interval $(s_i,S_i)\ni s$. 
First of all, if $s_i\in (s_\star,s^\star]$ then, by continuity, $v(s_i) = P(s_i)\leq \sup_{s'\in[s_\star,s]}\{P(s')\}$ 
and we have 
$v'(s)\leq P(s) + Q(s)$ for all $s \in[s_i,S_i]$. 
By integrating this inequality over $[s_i,s]$  
we find 
$$
\aligned
v(s) 
& \leq  v(s_i) + \int_{s_i}^s(P(s') +Q(s'))ds' 
\\
& \leq  \sup_{s'\in [s_\star,s]}\big( P(s')\big) + \int_{s_i}^s(P+Q)(s')ds'
\leq 
\sup_{s'\in [s_\star,s]}\big( P(s')\big) + \int_{s_\star}^s(P+Q)(s')ds'.
\endaligned
$$
Second, if $s_i = s_\star$, we have $v(s_i) = v(s_\star)$ and we find 
$v(s) \leq  v(s_\star) + \int_{s_\star}^s(P(s') +Q(s'))ds'$.  
\end{proof}

%==============================================================================================

\section{Calculus rules with commuting vector fields}
\label{sectionN-5} 

\subsection{Fundamental ordering property} 
\label{sectn:cmm-ge}

\paragraph{Notation and terminology.}

In this section, we consider the spacetime domain $\Mscr_{[s_0, s_1]}$, and the set of vector fields $\Tfrak$, $\Lfrak$, and $\Rfrak$  of interest is follows. 
\begin{itemize}

\item {\bf Spacetime translation fields}: $\Tfrak := \big\{\del_\alpha \, / \, \alpha=0,1,2,3 \big\}$.

\item {\bf Lorentz boost fields:}  $\Lfrak := \big\{ L_a := x^a\del_t + t \, \del_a \, / \, a = 1,2,3 \big\}$.

\item {\bf Spatial rotation fields:} $\Rfrak:= \big\{\Omega_{ab} := x^a\del_b - x^b\del_a \, / \, 1 \leq a<b \leq 3\big\}$.

\end{itemize}
It will be convenient also to  denote by 
\begin{equation} \label{equa-notationLOmega} 
\LOmega  \text{ any of the fields } L_a \text{ or } \Omega_{ab}.
\end{equation}

All of these vector fields commute with, both, the (flat) wave operator $\Box=\Box_\Mink$ and the Klein-Gordon operator $\Box - c^2$. We thus work with the family of {\sl admissible vector fields}, also called the ``commuting family'',   
\begin{equation}
\Yc := \Tfrak\cup \Lfrak \cup \Rfrak. 
\end{equation}
If $\Yfrak = \{Y_1,Y_2, \ldots, Y_M\}$ is a family of operators and $I = (i_1,i_2, \ldots i_N)$ is an $N$-order multi-index (with $i_j \in \{1,2, \ldots, M \}$), then  $Y^I = Y_{i_1} Y_{i_2} \ldots  Y_{i_N}$ represents an arbitrary $N$-order differential operator in the family under consideration.

\begin{definition}
When $\Yfrak = \Yfrak_1\cup \Yfrak_2 \ldots \cup \Yfrak_Q$ consists of the union of disjoint sub-families $\Yfrak_k$, then  
an $N$-order operator $Y^I$ is said to be an {\bf operator of $\Yfrak$--type $(\tau_1, \tau_2, \ldots, \tau_Q)$} if, in the decomposition of $Y^I$, there are $\tau_k$ vectors belonging to $\Yfrak_k$. 
\end{definition}

For example, with respect to the translation and boost families $\Tfrak \cup \Lfrak$, the operator $\del_1\del_t L_1\del_2L_2$ is of $(\Tfrak \cup \Lfrak)$-type $(3,2)$. 
We will often focus our attention on the set $\Yc$ and be interested in estimates that take the {\sl number of boost and rotation fields} into account. By our definition above, the $\Yc$-type of a high-order operator is described by a
{\bf triple of integers} $(i,j,l)$, that is, such an operator $Y^I$ contains $i$ derivatives, $j$ boosts, and $l$ rotations 
with $|I| = i+ j + l$.
Given two integers $k \leq p$, it is convenient to introduce the notation 
\begin{equation}
\mathcal{I}_{p,k} = \big\{ I \, \big/ \, Y^I \text{ of $\Yc$--type} (i,j,l) \text{ with } i+j+l\leq p
\text{ and } j+l\leq k \big\}. 
\end{equation}
Equivalently, $\mathcal{I}_{p,k}$ consists of all operators of order $p$ containing at most $k$ boosts or rotations. 
We also introduce the (pointwise) expressions associated with a function $u$ (defined in a given domain of 
$\RR^{3+1}$): 
\begin{equation} \label{eq1 notation}
\aligned
|u|_{p,k} &:=  \max_{K\in \mathcal{I}_{p,k}} |Y^K u|,
\qquad \qquad &&|u|_p := \max_{0\leq k\leq p} |u|_{p,k},
\\
|\del u|_{p,k} &:= \max_{\alpha=0,1,2,3} |\del_\alpha u|_{p,k}, &&|\del u|_p := \max_{0\leq k\leq p} |\del u|_{p,k},
\\
|\del^m u| _{p,k} &:=\max_{|I| = m} |\del^I u|_{p,k}, &&|\del^m u|_p := \max_{0\leq k\leq p} |\del^m u|_{p,k},
\endaligned
\end{equation}
and, in addition, 
\begin{equation} \label{eq1-11-06-2021}
\aligned
|L u|_{p,k} : & =    \max_{a=1,2,3} |L_au|_{p,k}, 
\qquad 
\quad &&|L u|_p := \max_{0\leq k\leq p} |Lu|_{p,k}, 
\\
|\LOmega u|_{p,k} : & =    \max_{\LOmega \in \{ L_a,\Omega_{ab} \}} | \LOmega u|_{p,k}, 
\qquad 
\quad &&|\LOmega u|_p := \max_{0\leq k\leq p} |\LOmega u|_{p,k}, 
\endaligned
\end{equation}
where $\LOmega$ was introduced in \eqref{equa-notationLOmega}. 

For tensorial fields such as $T = T^{\alpha\beta} \del_\alpha \otimes\del_{\beta}$, we write similarly 
$
|T|_{p,k} := \max_{\alpha, \beta} |T^{\alpha\beta} |_{p,k}$
and 
$
|T|_p := \max_{0\leq k\leq p} |T|_{p,k}. 
$
We will also use a similar notation $|u|_{p,k} := \max_{\alpha, \beta} |u_{\alpha\beta} |_{p,k}$, etc. for tensors $u_{\alpha\beta}$ having lower indices. Several technical lemmas are stated below and allow us, in the rest of this paper, to always restrict our attention to ordered operators $\del^IL^J\Omega^K$ (in this order). {
It will sometimes be convenient to use the short-hand notation $|a,b|_p:= |a|_p + |b|_p$ for any functions or tensors $a,b$.
}

%-------------------------

\paragraph{Main result.}

A main conclusion of the present section is now stated. 

\begin{proposition}[Fundamental ordering property]
\label{prop--fund-order}
The following equivalence properties hold for any function $u$ defined in $\Mscr_{[s_0,s_1]}$:
$$ 
|u|_{p,k} \simeq \sum | \Zt u|,
\qquad\qquad 
|\del u|_{p,k} \simeq \sum |\del \Zt u|, 
\qquad\qquad 
|\del\del u|_{p,k} \simeq \sum |\del \del \Zt u|, 
$$ 
each sum being over all ordered admissible operators that satisfy $\ord(\Zt)\leq p$ and $\rank(\Zt) \leq k$, that is, 
the set of all 
$\Zt = \del^IL^J\Omega^K$ with $|I|+|J|+|K|\leq p$ and $|J|+|K|\leq k$.
\end{proposition}

The proof of this result consists of two parts. Lemmas~\ref{lem 1 depo mmu} to~\ref{lem 2 high-order} below 
show that the left-hand sides are bounded by the right-hand sides, while the reverse inequalities are the subject of Lemma~\ref{lem 1 notation} (but the first inequality is trivial).  

%--------------------------------------------------------------------------------------------------------------------------------------------------------- 

\subsection{Ordering and commutator lemmas}

\paragraph{Basic calculus rules.}

We begin with product and composition identities. 

\begin{lemma}[Generalized Leibniz identity]
\label{lem 1 Leibniz} 
Consider functions $u_k$ defined in $\Mscr_{[s_0,s_1]}$ for $k=1,2, \ldots,m$. 
Given any family $\Yfrak = \Yfrak_1\cup \ldots\cup \Yfrak_Q$, if $Y^I$ is an $N$-order operator of $\Yfrak $--type $(\tau_1, \tau_2, \ldots, \tau_Q)$, then the expression  
$Y^I(u_1 u_2 \ldots u_m)$ is a finite linear combination (with constant coefficients determined by $I$) of the terms
$Y^{I_1} u_1\,  Y^{I_2} u_2 \ldots Y^{I_m} u_m$, 
where, for $n=1, 2, \ldots, m$, each index $I_n$ is of $\Yfrak$--type $(\tau_{n1}, \tau_{n2}, \ldots, \tau_{nQ})$ and of  
total sum  
$$ 
(\tau_1, \tau_2, \ldots, \tau_Q) = \sum_{n=1}^m(\tau_{n1}, \tau_{n2}, \ldots, \tau_{nQ}).
$$
\end{lemma}

\begin{lemma}[Generalized F\`aa di Bruno rule]
\label{lem 1 faa}
Consider functions $u$ defined in $\Mscr_{[s_0,s_1]}$ together with a function $f$ defined in the range of $u$. Let $Y^I$ be an $N$-order operator of $\Yfrak$--type $(\tau_1, \tau_2, \ldots, \tau_Q)$
with respect to a family $\Yfrak = \Yfrak_1\cup\ldots\cup \Yfrak_Q$, and assume $N\geq 1$.   
Then, $Y^If(u)$ is a finite linear combination (with constant coefficients determined by $I$) of  
$f^{(k)}(u)Y^{I_1} uY^{I_2} u\ldots Y^{I_k} u$ with $1\leq k \leq |I|$, 
in which for each $n=1,2, \ldots, k$
the index $I_n$ is of $\Yfrak$--type $(\tau_{n_1}, \tau_{n_2}, \ldots, \tau_{n_Q})$
and of total sum   
$$ 
(\tau_1, \tau_2, \ldots, \tau_Q) = \sum_{n=1}^k (\tau_{n_1}, \tau_{n_2}, \ldots, \tau_{n_Q}).
$$
\end{lemma}

Motivated by the above two lemmas, we introduce the notation  
$$
\aligned
Y^I(u_1 u_2 \ldots u_m) 
\cong \hskip-.5cm
\sum_{I_1+I_2 +\ldots + I_m = I} \hskip-.7cm Y^{I_1} u_1 \, Y^{I_2} u_2 \ldots Y^{I_m} u_m, 
\qquad \qquad 
Y^I(f(u)) 
\cong 
\sum_{k=1}^{|I|}f^{(k)}(u) \hskip-.5cm \sum_{I_1 + \ldots + I_k = I} \hskip-.5cm Y^{I_1} u \, Y^{I_2} u \ldots Y^{I_k} u, 
\endaligned
$$
where the right-hand sides are finite linear combinations with constant coefficients. 
By convention, when a set of multi-indices is empty the sum is understood to be zero. 
More generally, we write $A \cong \sum_i A_i$ when an expression is decomposed linearly as a sum of terms $A_i$ with {\sl constant coefficients.}  

%--------------------------------------------------------------------------------------------------------------

\paragraph{Ordering lemmas.}

We now turn our attention to technical ordering lemmas for the family $\Yc$. 
We deal with operators $Y^I$ of $\Yc$--type $(i,j,k)$ with respect to the family $\Yc = \Tfrak \cup \Lfrak \cup \Rfrak$. The observations below were made first by the authors in \cite{PLF-YM-book}. 
The second lemma below takes the rotations into account, and our results are easily checked by induction (on $|I|$, $|J|$, $|K|$).

\begin{lemma}[Fundamental commutation relations. I]
\label{lem 1 depo mmu} 
For any multi-indices $I$ and $J$, the following identity holds:  
\begin{equation} \label{eq 1 comm}
[\del^I, L^J] \cong \sum_{|I'|=|I|{ \geq 1}\atop|J'|<|J|}\!\!\!  \del^{I'}L^{J'}. 
\end{equation}  
\end{lemma} 

\begin{lemma}[Ordering lemma. I] 
\label{lem 1 high-or} 
If $Y^I$ is of $\Yc$--type $(i,j,0)$ with respect to the family $\Tfrak \cup \Lfrak \cup \Rfrak$, 
then one can ensure the ordering 
$$
Y^K \cong \sum_{|I|=i\atop|J| \leq j} \del^I L^J,
$$
the implicit constant coefficients in the right-hand side being determined from the multi-index $I$.
\end{lemma}

\begin{lemma}[{ Fundamental commutation relations. II}]
\label{lem1-21-01-2020}
Given any three multi-indices $I,J,K$,
one has 
\begin{subequations}
\begin{equation} \label{eq 4' comm}
[L^J, \Omega^K] \cong \sum_{|J'|=|J|{ \geq 1}\atop|K'| < |K|}L^{J'} \Omega^{K'}, 
\qquad\qquad
[\del^I, \Omega^K] \cong \sum_{|I'| = |I|{ \geq 1}\atop |K'|<|K|}\del^{I'}\Omega^{K'}, 
\end{equation} 
\begin{equation} \label{eq 4 comm}
[ \del^I L^J, \Omega^{K} ]\cong \sum_{|I'|=|I|,|J'|=|J|\atop |K'|<|K|,{ |I|+|J|\geq 1}} \del^{I'}L^{J'} \Omega^{K'}.
\end{equation} 
\end{subequations}
\end{lemma} 

\begin{lemma}[Ordering lemma. II]
\label{lem 2 high-order}
With $\Yc = \Tfrak\cup\Lfrak\cup \Rfrak$, if $Y^L$ is a high-order operator of type $(i,j,k)$, 
then one can ensure the ordering $Y^L \cong \sum_{|I|=i,|J| \leq j\atop |K| \leq k} \del^I L^J \Omega^K$, 
the implicit constant coefficients in the right-hand side being determined from the multi-index $L$.
\end{lemma}

In particular, we arrive at 
\begin{equation} \label{equa2-2-juin}
|\del u|_{p,k} \lesssim \sum_{|I|+|J|+|K|\leq p \atop |J|+|K|\leq k} |\del \del^IL^J\Omega^K u|, 
\qquad \qquad
|\del\del u|_{p,k} \lesssim \sum_{|I|+|J|+|K|\leq p \atop |J|+|K|\leq k} |\del \del \del^IL^J\Omega^K u|. 
\end{equation} 
The above lemmas lead us to a proof of one set of inequalities in Proposition~\ref{prop--fund-order}, while the ``reverse'' inequalities arise as special cases of the following statement.

\begin{lemma}[Reverse inequality]
\label{lem 1 notation}
Consider functions $u$ defined in $\Mscr_{[s_0,s_1]}$.  
If $L$ is a multi-index of $\Yc$--type $(i+m,j,l)$ with $i+j+l\leq p$ and $j+l\leq k$, one has 
\begin{subequations}
\begin{equation} \label{eq1 lem 1 notation}
|Y^L u| \lesssim |\del^m u|_{p,k}.
\end{equation}
In particular, if $L'$ is a multi-index of $\Yc$--type $(i,j,l)$ with $i+j+l\leq p$ and $j+l\leq k$, then
\begin{equation} \label{eq9-14-03-2021}
|\del Y^{L'} u | \lesssim |\del u|_{p,k},
\qquad \qquad
|\del \del Y^{L'} u| \lesssim |\del\del u|_{p,k}.
\end{equation} 
\end{subequations} 
\end{lemma} 
The above results are checked by applying the ordering property in Lemma~\ref{lem 2 high-order}, then the commutation relations established in Lemmas~\ref{eq 1 comm} and~\ref{lem1-21-01-2020}. 
We omit the details but give a heuristic argument, as follows.
In the decomposition in Lemmas~\ref{lem 1 depo mmu} and~\ref{lem1-21-01-2020}, 
the degree (that is, the total order of partial derivatives) is preserved in each case; 
therefore, when we commute the partial derivatives arising in $Y^L$ we only produce terms with the {\sl same} degree.  

%-------------------------------------------------------------------------------------------------------------------------------------------------------- 

\subsection{Quasi-linear commutator properties}

We now turn our attention to the commutators $[Z,H^{\alpha\beta} \del_\alpha \del_{\beta}]$,
where $H^{\alpha\beta}$ is a two-tensor and $v$ is a function defined in $\Mscr_{[s_0, s_1]}$ and $Z$ an ordered operator. Such quasi-linear terms arise when dealing with the wave and Klein-Gordon operators on a curved background.  

\begin{lemma}[Estimates for linear commutators]
\label{prop-newpropo} 
For any indices satisfying $|I|+|J|+|K| \leq p$ and $|J|+|K| \leq k$ one has 
\begin{subequations}
\begin{equation} \label{eq2-31-01-2020}
|[\del^IL^J\Omega^K, \del] u | \lesssim |\del u|_{p-1,k-1},
\end{equation}
\begin{equation} \label{eq1-31-01-2020}
| [\del^IL^J\Omega^K, \del \del ] u | 
\lesssim |\del\del u|_{p-1,k-1} \lesssim |\del u|_{p,k-1}.
\end{equation}
\end{subequations}
\end{lemma} 

\begin{proof}
This estimate follows from several observations.  
If $I,J,K$ be multi-indices satisfying $|J| + |K|= k$, one has 
\begin{equation} \label{eq 1 lem 1 depo-cH}
\, [\del_{\alpha}, \del^I L^J \Omega^K] = \del^I([\del_{\alpha},L^J]\Omega^K\big) + \del^IL^J\big([\del_{\alpha},\Omega^K]\big)
\cong
\sum_{|J'| \leq|J|,|K'| \leq|K| \atop |J'| + |K'|<k}  \sum_\beta \del_\beta \del^IL^{J'} \Omega^{K'}, 
\end{equation}
which is a consequence of Lemmas~\ref{lem 1 depo mmu} and~\ref{lem1-21-01-2020}. Applying \eqref{eq 1 lem 1 depo-cH} twice, we arrive at a decomposition for second-order derivatives: 
\begin{equation} \label{eq 2 decompo-comm-H}
[\del_\alpha \del_{\beta}, \del^I L^J \Omega^K]
\cong
\sum_{|J'| \leq|J|,|K'| \leq|K| \atop |J'| + |K'|<k} \sum_{\gamma,\delta} \del_\gamma \del_{\delta} \del^I L^{J'} \Omega^{K'}.
\qedhere
\end{equation}
\end{proof}

We now arrive at one of our main observations. Recall that the notation $\LOmega$ was introduced in \eqref{equa-notationLOmega}. 
Observe that $H$ below is taken to be a real-valued function while, in the applications, it will be the components of a tensor (denoted by $H^{\alpha\beta}$ below). 

\begin{proposition}[Hierarchy structure for quasi-linear commutators]
\label{lm 2 dmpo-cmm-H}
Let $Z$ be an ordered operator with ${\ord(Z) = p}$ and $\rank(Z) = k$ and let $H,u$ be functions (defined on some domain of $\RR^{3+1}$). Then, one has 
\begin{equation} \label{eq8-14-03-2021}
|[Z,H]u|\lesssim  \sum_{p_1+p_2=p\atop k_1+k_2=k \text { with } k_1=p_1} | \LOmega H|_{p_1-1,p_1-1} |u|_{p_2,k_2}
+ \sum_{p_1+p_2=p\atop k_1+k_2=k} |\del H|_{p_1-1,k_1} |u|_{p_2,k_2}, 
\end{equation}
\begin{equation} \label{eq7-14-03-2021}
|[Z,H\del_{\alpha}\del_{\beta}]u|\lesssim |H| \, |\del\del u|_{p-1,k-1} 
+ \hskip-.7cm
\sum_{p_1+p_2=p\atop k_1+k_2=k \textbf{with } k_1=p_1} \hskip-.7cm
|\LOmega H|_{p_1-1,p_1-1} |\del\del u|_{p_2,k_2}
+ \hskip-.cm
\sum_{p_1+p_2=p\atop k_1+k_2=k}  \hskip-.cm  |\del H|_{p_1-1,k_1} |\del\del u|_{p_2,k_2}. 
\end{equation} 
\end{proposition}

Let us explain the interest of the above result.
If we disregard the terms containing the factor $\del H$ (which enjoy better $L^2$ and pointwise decay), 
the right-hand sides contain {\sl strictly fewer boosts or rotations} acting on the function. 
Consequently, the $L^2$ estimate of commutators will involve energy functionals at a {\sl lower rank}
and, in the derivation of our main energy estimates at high-order, this structure will be the key property allowing us 
to formulate an induction argument on the rank. A refinement of Proposition~\ref{lm 2 dmpo-cmm-H} will be established below in the Euclidean-merging domain; 
cf.~Proposition~\ref{prop1-12-02-2020}.

\begin{proof} Checking \eqref{eq8-14-03-2021} is immediate by observing that, in the first summation,
$k_1-1 \geq 0$ implies $k_1\geq 1$ and thus $p_2\leq p-1$, 
while, in the second summation, $p_1-1\geq k_1\geq 0$ implies $p_1\geq 1$ and thus $p_2\leq p-1$. Next, to deal with
\eqref{eq7-14-03-2021}, we perform a similar calculation and observe that $|H[Z,\del_{\alpha}\del_{\beta}]u|\lesssim |H||\del\del u|_{p-1,k-1}$ thanks to \eqref{eq1-31-01-2020} and \eqref{eq8-14-03-2021}.  
\end{proof}

%===================================================================================

\section{Calculus rules in the Euclidean-merging domain}    
\label{sectionN-7}

\subsection{Vector fields and high-order operators}  

\paragraph{The near/far decomposition.}

In this section we establish calculus rules involving vector fields and partial differential operators defined in the Euclidean-merging domain. Within our decompositions and estimates, in addition to the vector fields considered in the previous section we also now take into account  the {\sl null derivative fields} $\delsN_a$ associated with in the {\sl semi-null frame} \eqref{eq=nulllframedef}. 
Moreover, in order to fully analyze the decay of solutions it is important to distinguish between the behaviors ``near'' and ``far'' from the light cone. 
We thus introduce  the sub-domains 
\begin{equation} \label{equa-nearfardefinition}
\MMEnear_s  := \MME_s\cap \{  t-1 \leq r \leq 2t\}, \qquad\qquad
\MMEfar_s := \MME_s\cap \{r\geq 2t\}. 
\end{equation}
In this section, we are especially interested in deriving bounds on the first- and second-order terms
$|\delsN u|_{p,k}$, $|\del\delsN u|_{p,k}$, and $|\delsN\delsN u|_{p,k}$. It will be most technical to derive 
estimates in the following conical neighborhood of the light cone: 
\begin{equation} \label{equa-Mnear} 
\MMEnear_{[s_0,s_1]} := \MME_{[s_0,s_1]} \cap \big\{ t-1\leq r \leq 2t \big\}.
\end{equation}
On the other hand, the required decay properties within $\MMEfar_{[s_0,s_1]} :=\bigcup_{s \in [s_0,s_1]} \MMEfar_s$ will follow  easily from the Sobolev inequality (already stated in Proposition~\ref{pro204-11-2}) together with the inequality $r\lesssim \crochet$ valid far from the light cone.

%----------------------------

\paragraph{Statement of the main estimates.}

Our estimates involve both of the weights $\crochet$ and $\zeta$.  
The proofs are 
% given in the companion paper~\cite{PLF-YM-companion}. 
%}
postponed to Appendix~\ref{appendix-calculus}. 

\begin{proposition}
\label{prop1-02-02-2020}
Fix some exponent $\eta\geq 0$. 
\begin{subequations}
For any function $u$ defined in $\MMEnear_{[s_0, s_1]}$ one has 
\begin{equation} \label{eq2-03-02-2020} 
|\delsN u|_{p,k}
\lesssim  \sum_{\ord(Z)\leq p \atop \rank(Z)\leq k}\Big(|\delsN  Zu| + \frac{|r-t|+1}{r} |\del_tZ u|\Big),
\end{equation}
\begin{equation} \label{eq3-07-02-2020}
\crochet^{\eta} |\delsN u|_{p,k} 
\lesssim \crochet^{\eta} \sum_{\ord(Z)\leq p\atop \rank(Z)\leq k} \zeta \, \big|\del_t Z u\big|
+ \crochet^{\eta} \sum_{\ord(Z)\leq p\atop \rank(Z)\leq k} \big|\delsEH Zu\big|, 
\end{equation}
\end{subequations}
in which, as before, the summation is over all $Z = \del^IL^J\Omega^K$ in the specified range.  
\end{proposition}

\begin{proposition}
\label{prop1-07-02-2020}
\begin{subequations}
For any function $u$ defined in $\MMEnear_{[s_0, s_1]}$ one has 
\begin{equation} \label{eq1-06-02-2020}
\big|\del\delsN u\big|_{p,k} \lesssim  \frac{|r-t|}{r} | \del \del u|_{p,k} + r^{-1} |\del u|_{p+1,k+1},
\end{equation}
\begin{equation} \label{eq4-07-02-2020}
|\del\delsN u|_{p,k} \lesssim \Big(\frac{|r-t|}{r} \Big)^{1/2} \zeta \, | \del \del u|_{p,k} + r^{-1} |\del u|_{p+1,k+1}, 
\end{equation}
\begin{equation} \label{eq5-07-02-2020}
|\delsN \delsN u|_{p,k} \lesssim 
\frac{|r-t|^2}{r^2} | \del \del u|_{p,k} + \frac{1}{r} |\delsN u|_{p+1,k+1} +  \frac{|r-t|}{r^2} |\del u|_{p+1,k+1},
\end{equation}
\begin{equation} \label{eq6-07-02-2020}
|\delsN \delsN u|_{p,k} \lesssim 
\Big(\frac{|r-t|}{r} \Big)^{3/2} \zeta \, | \del \del u|_{p,k} + r^{-1} |\delsN u|_{p+1,k+1} + r^{-1} \Big(\frac{|r-t|}{r} \Big)^{1/2} \zeta \, |\del u|_{p+1,k+1}.
\end{equation}
\end{subequations}
\end{proposition} 

\begin{proposition} \label{prop1-10-02-2020}
\begin{subequations} \label{eq1-14-07-2021}
For any function $u$ defined in $\MMEnear_{[s_0, s_1]}$ one has 
\begin{equation} \label{eq2-10-02-2020}
|\LOmega u|_{p,k} \lesssim t \, | \delsN u|_{p,k} + { (} |r-t|{ +1)} \, | \del u|_{p,k},
\end{equation}
\begin{equation} \label{eq3-10-02-2020}
|\LOmega\LOmega u|_{p,k} \lesssim t \, | \delsN u|_{p+1,k+1} + { (} |r-t|{ +1)} \, | \del u|_{p+1,k+1},
\end{equation}
\begin{equation} \label{eq4-10-02-2020}
|\LOmega\LOmega u|_{p,k} \lesssim t|L u|_{p+1,k} \lesssim  t^2|\delsN u|_{p+1,k} + t{ (} |r-t|{+1)} \, | \del u|_{p+1,k},
\end{equation}
\end{subequations}
Moreover, when in $\Mfar_{[s_0,s_1]}$, one has
\begin{equation} \label{eq1-24-06-2021}
|\LOmega u|_{p,k}\lesssim r|\del u|_{p,k}.
\end{equation}
\end{proposition}

Moreover, by combining the bounds in $\Mnear_s$ and $\Mfar_s$, for any function $u$ defined in $\MME_{[s_0,s_1]}$ we find 
\begin{equation} \label{eq2-24-06-2021}
r^{-1} |\LOmega u|_{p,k}\lesssim |\delsN u|_{p,k} + \frac{|r-t|{ +1}}{r} |\del u|_{p,k}.
\end{equation}

%-----------------------------------------------------------------------------------------------------------

\paragraph{Homogeneous functions in the Euclidean-merging domain.}

For convenience, some terminology and technical results are introduced here. 

\begin{definition} 
A smooth function $f$ defined in the domain $\big\{ r \geq t/2, \, t > 0 \big\}$ is called {\bf exterior-homogeneous\footnote{When there is no ambiguity, we will simply write `homogeneous'.}
of degree $k$}
if it satisfies ($\Sbf^2 \subset \RR^3$ denoting the $2$-sphere)
$$
\aligned
f(\lambda t, \lambda x) & =   \lambda^{k} f(t, x),
& \lambda>0; 
\qquad \qquad
| \del^I f(t, \omega)|& \leq  C \, (I),   & \omega\in \Sbf^2, \quad 0 < t<2.
\endaligned
$$
\end{definition}

\begin{lemma}
\label{lem 1 homo-ext}
Let $f$ and $g$ be exterior-homogeneous functions of degree $m$ and $n$, respectively.  
\begin{itemize} 

\item When $m=n$, then $\alpha f + \beta g$ is exterior-homogeneous of degree $m$ (for any reals $\alpha,\beta$).

\item The product $fg$ is exterior-homogeneous of degree $(m+n)$.

\item $Z f$ is exterior-homogeneous of degree $m+k-p$ with $\ord(\Zt)= p$ and $\rank(\Zt) = k$ and, moreover, 
it holds $|Z f| \lesssim C(Z) r^{m+k-p}$ in $\MME_{[s_0, + \infty)}$.

\end{itemize} 
\end{lemma}

\begin{proof}
Only the third claim deserves a proof. We differentiate the identity $\lambda^m f(t,x) = f(\lambda t, \lambda x)$ with respect to $t$ and $x^a$ and obtain 
$
\del_\alpha f (\lambda t, \lambda x) = \lambda^{m-1} \del_\alpha f(t,x).
$
Clearly, the fields $L_a$ and $\Omega_{ab}$ preserve the degree of homogeneity while, under the action of $\del_\alpha$, the degree of homogeneity decreases by $1$.  Therefore, the desired result follows by induction. 
\end{proof}

%-------------------------------------

The weight function $|r-t|$ enjoys the following property. 

\begin{lemma} \label{lem1-31-05-2021}
In the region $\Mnear_{[s_0,+\infty)} := \{t-1 \leq r\leq 2t\} \cap \MME_{[s_0,+\infty)}$, the following estimate holds 
(with implied constants determined from the multi-indices $I,J$): 
\begin{equation} \label{eq1-31-05-2021}
\big|\del^IL^J (r-t)\big| \lesssim 
\begin{cases}
|r-t|,\quad  & |I| = 0,
\\
r^{-|I|+1}\leq|r-t|+1,\quad  & |I| \geq 1.
\end{cases}
\end{equation}
\end{lemma}

\begin{proof}
We observe that $L_a (r-t) = -(x^a/r)(r-t)$ where $-(x^a/r)$ is homogeneous of degree zero. By induction, we obtain 
$L^J(r-t) = \Gamma^J(r-t)$ where $\Gamma^J$ is a homogeneous function of degree zero. When $|I|\geq 1$, we find 
$\del_a(r-t) = (x^a/r)$ and 
$\del_t(r-t) = -1$, which are homogeneous of degree zero. The desired result is established by induction. 
\end{proof}

On the other hand, the weight function $|r-t|t^{-1}$ enjoys the following property.

\begin{lemma} \label{lem2-commu-ext}
In the near-light cone domain $\Mnear_{[s_0,+\infty)}$ (and for an implied constant depending upon by the multi-indices $I,J$), one has 
\begin{equation} \label{eq2-07-02-2020}
\big|\del^IL^J\big((r-t)t^{-1} \big)\big|
\lesssim
\begin{cases}
|r-t|t^{-1}, \quad & |I| = 0,
\\
t^{-|I|}\lesssim \frac{|r-t|+1}{r},     \quad & |I|\geq 1. 
\end{cases}
\end{equation}
\end{lemma}

\begin{proof} This follows by induction on $|J|$ that $L^J\big((r-t)t^{-1}\big) = \Gamma^J(r/t)^{\alpha(J)}(1-(r/t)^2)^{\beta(J)}$. Here, $\Gamma^J$ are homogeneous of degree zero and we proceed by induction on $|I|$. 
\end{proof}

%----------------------------------------------------------------------------------------------------------------------------

\subsection{Analysis of null multi-linear forms}  

\paragraph{Bilinear and trilinear forms.}

The classical null condition reads as follows. Let $T = T^{\alpha\beta}\del_{\alpha}\otimes\del_{\beta}$ and $H = H^{\alpha\beta\gamma}\del_{\alpha}\otimes \del_{\beta}\otimes \del_{\gamma}$ be  
bilinear and trilinear forms with constant coefficients, defined in some domain $\Omega\subset \RR^{1+3}$. Then, if for all $\zeta \in \RR^{1+3}$ satisfying $\xi_0^2 + \sum_a\xi_a^2 = 0$, one has
$$
T(\xi,\xi) = T^{\alpha\beta}\xi_{\alpha}\xi_{\beta} = 0,\qquad\qquad
H(\xi,\xi,\xi) = H^{\alpha\beta\gamma}\xi_{\alpha}\xi_{\beta}\xi_{\gamma} = 0,
$$ 
then the tensors $T$ and $H$ are called {\sl null multi-linear forms.}

\begin{proposition}
\label{proposition68}
Let $T, B$ be two-tensors fields and $H$ be a three-tensor field defined in $\MME_{[s_0, s_1]}$ and satisfying  the null condition. Assume that $|T|_{p,k}, |B|_{p,k}$, and $|H|_{p,k}$ are a priori bounded by a constant (depending on $p$)
and, for any two functions $u, v$, consider the maps 
$$
T(\del u, \del v) := T^{\alpha\beta} \del_\alpha u\del_{\beta}v, \qquad \quad
B(u, \del\del v):= B^{\alpha\beta} u\del_\alpha \del_{\beta}v, \qquad \quad
H^{\alpha\beta\gamma}(\del u, \del\del v) := H^{\alpha\beta\gamma} \del_{\gamma} u\del_\alpha \del_{\beta}v. 
$$
Then one has 
\begin{subequations}

\label{eq1-10-06-2021-all}
\begin{equation} \label{eq1-10-06-2021}
|T(\del u, \del v)|_{p,k} \lesssim |\del u|_{p_1,k_1} |\delsN v|_{p,k} + |\del u|_{p,k} |\delsN v|_{p_1,k_1}
+ | \delsN u|_{p,k} | \del v|_{p_1,k_1} +  | \delsN u|_{p_1,k_1} |  \del v|_{p,k},
\end{equation}
\begin{equation} \label{eq2-10-06-2021}
\aligned
|H(\del u, \del\del v)|_{p,k} 
& \lesssim |\del u|_{p_1,k_1} |\del\delsN v|_{p,k} + |\del u|_{p,k} |\del\delsN v|_{p_1,k_1} 
+ |\delsN u|_{p_1,k_1} |\del\del v|_{p,k} + |\delsN u|_{p,k} |\del\del v|_{p_1,k_1} 
\\
& \quad + r^{-1} \big(|\del u|_{p_1,k_1} |\del v|_{p,k} + |\del u|_{p,k} |\del v|_{p_1,k_1} \big), 
\endaligned
\end{equation}
\begin{equation} \label{eq3-10-06-2021}
|B(u, \del\del v)|_{p,k} \lesssim |u|_{p,k} |\del\delsN v|_{p_1,k_1} + |u|_{p_1,k_1} |\del\delsN v|_{p,k} 
+ r^{-1} \big(|u|_{p_1,k_1} |\del v|_{p,k} + |u|_{p,k} |\del v|_{p_1,k_1} \big), 
\end{equation}
\end{subequations}
in which $p_1 = [p/2] $ and $k_1=[k/2]$ (that is, the largest integers that do not exceed 
$p/2$ and $k/2$, respectively).
\end{proposition}

The above estimates will follow from the following observation. 

\begin{lemma}
Let $T = T^{\alpha\beta} \del_\alpha \otimes\del_{\beta}$ and $H = H^{\alpha\beta\gamma} \del_\alpha \otimes\del_{\beta} \otimes\del_{\gamma}$ be tensors defined in $\MME_{[s_0,s_1]}$ and satisfying the null condition. Then
one has 
\begin{equation}
\aligned
& T^{\Ncal 00} = 0, \qquad \quad
|T^{\Ncal \alpha\beta} |_{p,k} \lesssim |T|_{p,k},
\qquad\quad
H^{\Ncal 000} = 0, \qquad\quad
|H^{\Ncal \alpha\beta\gamma} |_{p,k} \lesssim |H|_{p,k}.
\endaligned
\end{equation}
\end{lemma}

\begin{proof} We only deal with the tensor $T$, since the arguments for the tensor $H$ are completely similar. 
Recall the transition relations 
$T^{\Ncal \alpha\beta} = \PsiN_{\alpha'}^\alpha  \PsiN_{\beta'}^\beta T^{\alpha'\beta'}$
and, especially,
$T^{\Ncal 00} = T^{\alpha\beta} \PsiN_{\alpha}^0\PsiN_{\beta}^0$. 
Observing that $\xi_\alpha = \PsiN_{\alpha}^0$ is a null vector, we find $\TN^{00} = 0$.
On the other hand, for any ordered operator $Z$ with $\ord(Z) \leq p$ and $\rank(Z) \leq k$, we have 
$$
Z\big(T^{\alpha\beta} \PsiN_{\alpha}^0\PsiN_{\beta}^0\big)
= 
\sum_{Z_1 \odot Z_2 = Z} 
Z_1 
T^{\alpha\beta}
Z_2  
\big(\PsiN_{\alpha}^0\PsiN_{\beta}^0\big),
$$
where $\PsiN_{\alpha}^0\PsiN_{\beta}^0$ is homogeneous of degree zero in the Euclidean-merging domain. We obtain
$
|Z_2 \big(\PsiN_{\alpha}^0\PsiN_{\beta}^0\big) | \lesssim 1
$
and thus 
$
\big| Z \big(T^{\alpha\beta} \PsiN_{\alpha}^0\PsiN_{\beta}^0\big)\big|
\lesssim  |T|_{p,k}.
$
In view of the first property in Proposition~\ref{prop--fund-order},
the estimate for $|T^{\Ncal \alpha\beta} |_{p,k}$ is established.
\end{proof}

%----------------------------------------------------------------------

\paragraph{Proof of Proposition~\ref{proposition68}.}

Let us consider an arbitrary two-tensor $T = T^{\alpha\beta} \del_\alpha \otimes\del_{\beta}$ defined in $\MME_{[s_0,s_1]}$ and, for any two functions $u,v$, let us introduce 
$$
\aligned
T(\del u, \del v)  & =   T^{\Ncal \alpha\beta} \delN_\alpha u\delN_{\beta}v 
= \sum_{(\alpha, \beta)\neq (0,0)} T^{\Ncal \alpha\beta} \delN_\alpha u\delN_{\beta}v.
\endaligned
$$
Then given any $Z$ with $\ord(Z) \leq p$ and $\rank(Z) \leq k$, we can write 
$$
\aligned
\big|Z(T(\del u,\del u))\big|
& 
\lesssim  \sum_{(\alpha,\beta)\neq(0,0)} \, 
\sum_{Z_1 \odot Z_2 \odot Z_3 = Z}
\big| Z_3 T^{\N\alpha\beta} \, \big|| Z_1 \delN_{\alpha}u|| Z_2 \delN_{\beta}v|
\lesssim  \sum_{(\alpha,\beta)\neq(0,0)\atop p_1+p_2\leq p,k_1+k_2\leq k} |\delN_{\alpha}u|_{p_1,k_1} |\delN_{\beta}v|_{p_2,k_2}.
\endaligned
$$
This leads to \eqref{eq1-10-06-2021}. Similarly, we have 
$$ 
H(\del u, \del\del v) =    
\sum_{(\alpha, \beta, \gamma) \neq (0,0,0)} \!\!\!\!
H^{\Ncal \alpha\beta\gamma} \delN_{\gamma} u\delN_\alpha \delN_{\beta}v
+ H^{\alpha\beta\gamma} \del_{\gamma} u\del_\alpha \big(\PsiN_{\beta}^{\beta'} \big)\delN_{\beta'}v, 
$$
where $\del_\alpha \big(\PsiN_{\beta}^{\beta'} \big)$ is homogeneous of degree $(-1)$ and bounded by $Cr^{-1}$, while 
$$
B(u, \del\del v) =    \sum_{(\alpha, \beta)\neq(0,0)} 
B^{\Ncal \alpha\beta} u\delN_\alpha \delN_{\beta}v + B^{\alpha\beta} u\del_\alpha \big(\PsiN_{\beta}^{\beta'} \big)\delN_{\beta'}v,
$$ 
and this allows us to derive all of the inequalities in \eqref{eq1-10-06-2021-all}. 

%------------------------------------------------------------------------------------------------------ 

\subsection{Boost-rotation hierarchy enjoyed by quasi-linear commutators}   
\label{section----63} 

In our derivation of energy estimates at arbitrary high-order, it will be essential to commute admissible fields with quasi-linear operators.
Recall that, according to our notation, a norm such as $ |\del\del u|_{p-1,k-1}$ below is non-vanishing only if $p\geq 1$ and 
$k \leq p$. Recall also that, near the light cone, the rotations are recovered from the boosts.

\begin{proposition}[Hierarchy property for quasi-linear commutators. Euclidean-merging domain] 
\label{prop1-12-02-2020}
{\bf 1. Estimate in the near-light cone domain:} for any function $u$ defined in  $\MMEnear_{[s_0, s_1]}$ 
and for any operator $Z$ with $\ord(Z) = p$ and $\rank(Z) = k$,  in the near-light cone domain one has
\begin{subequations}
\label{eq4-12-02-2020}
\begin{equation} \label{eq4a-12-02-2020}
\, \big|[Z,H^{\alpha\beta} \del_\alpha \del_{\beta}]u\big|
\lesssim  T^\textbf{hier} + T^\easy +T^{\textbf{super}}, 
\end{equation}
with 
\begin{equation} \label{eq4b-12-02-2020}
\aligned
T^\textbf{hier} & := \big(|\HN^{00} | + t^{-1} |r-t| |H|\big) \, |\del\del u|_{p-1,k-1} 
+ 
\sum_{p_1+p_2=p\atop p_1+k_2=k} \big(|L\HN^{00} |_{p_1-1} + t^{-1} |r-t| |L H|_{p_1{ -1}} \big) \, |\del\del u|_{p_2,k_2},
\\
T^\easy
& :=\sum_{p_1+p_2=p\atop k_1+k_2=k} \big(|\del \HN^{00} |_{p_1-1,k_1} + t^{-1} |r-t| |\del H|_{p_1-1,k_1} \big) \, |\del\del u|_{p_2,k_2}, 
\\ 
T^{\textbf{super}} & := t^{-1} |H||\del u|_p + t^{-1} \hskip-.3cm \sum_{0\leq p_1\leq p-1} |H|_{p_1+1} |\del u|_{p-p_1}. 
\endaligned
\end{equation}
\end{subequations} 
{\bf 2. Estimate in the Euclidean-merging domain:}
For any function $u$ defined in $\MME_{[s_0,s_1]}$, in the Euclidean-merging domain one has ($\LOmega$ below being defined in \eqref{equa-notationLOmega})
\begin{equation} \label{eq5-12-02-2020}
\aligned
|[Z,H^{\alpha\beta} \del_\alpha \del_{\beta}]u|
& \lesssim   
|H| \, | \del\del u|_{p-1,k-1} 
+ \hskip-.6cm
\sum_{p_1+p_2=p\atop p_1+k_2=k  \text{ with } k_1=p_1}  \hskip-.6cm 
| \LOmega H|_{p_1-1,p_1-1} |\del \del u|_{p_2,k_2}
+ \sum_{p_1+p_2=p\atop k_1+k_2=k} |\del H|_{p_1-1,k_1} |\del\del u|_{p_2,k_2}. 
\endaligned
\end{equation}
\end{proposition}

As already pointed out about Proposition~\ref{lm 2 dmpo-cmm-H}, the right-hand sides of \eqref{eq4-12-02-2020} and \eqref{eq5-12-02-2020} contain {\sl strictly fewer boosts or rotations} acting on $u$, or {\sl at least one partial derivative acting the metric $H$.} 
Namely, for instance in \eqref{eq5-12-02-2020} this property is clear for the first and last terms,
while in $ |\del \del u|_{p_2,k_2}$ one can not have $k_2=k$ due to the restriction $p_1+k_2=k$
with $p_1=0$ and therefore, by convention in this case,
$| \LOmega H|_{p_1-1,p_1-1}$ vanishes. 
This is the {\sl hierarchy property} enjoyed by quasi-linear commutators. 
For the proof, we observe that \eqref{eq5-12-02-2020} is a consequence of \eqref{eq7-14-03-2021}, so we focus our attention on deriving 
\eqref{eq4-12-02-2020} for which we state two technical lemmas first.

\begin{lemma} 
\label{lemma-611} 
For any ordered operator $Z = \del^IL^J\Omega^K$ with $\ord(Z)=p$ and $\rank(Z) = k$, 
in the domain $\Mscr_{[s_0, + \infty)}$ one has 
\begin{equation} \label{eq1-01-06-2021}
[Z,L_a] \cong \sum_{|I'| = p-k\atop |J'|+|K'| \leq k}\del^{I'} L^{J'} \Omega^{K'} 
\end{equation} 
and, more precisely, when $p=k\geq1$  
\begin{equation} \label{eq3-02-06-2021}
[Z,L_a] \cong \sum_{1 \leq  |J'|+|K'| \leq k} L^{J'}\Omega^{K'}.
\end{equation} 
\end{lemma}

%----------------------------------------------------------

\begin{lemma} 
\label{lem1-commu-ext}

\begin{subequations}
For any ordered operator $Z$ with $\ord(Z) = p$ and $\rank(Z) = k$, 
and any function $u$ defined in $\MMEnear_{[s_0,s_1]}$, in the near-light cone domain one has  
\begin{equation} \label{eq1-lem1-commu-ext}
|[Z, \del_t\delsN_a]u| =|[Z, \delsN_a\del_t]u| \lesssim  |1-r/t||\del\del u|_{p-1,k-1} + t^{-1} |\del u|_{p,k},
\end{equation}
\begin{equation} \label{eq2-lem1-commu-ext}
|[Z, \delsN_a\delsN_b]u| \lesssim |1-r/t||\del\del u|_{p-1,k-1} 
+ t^{-1} | \del u|_{p,k}. 
\end{equation}
\end{subequations}
\end{lemma}

%-------------------------------------------------------------------------------------------------------------

\paragraph{Proof of the hierarchy property \eqref{eq4-12-02-2020} in the near-light cone domain.}

We recall the semi-null decomposition 
\begin{equation} \label{decompo-H-ext}
H^{\alpha\beta} \del_\alpha \del_{\beta} u 
= \HN^{00} \del_t \, \del_t u + H^{\Ncal 0a} \del_t\delsN_a u + H^{\Ncal a0} \delsN_a\del_t u + H^{\Ncal ab} \delsN_a\delsN_bu + H^{\alpha\beta} \del_{\alpha}(\PsiN_{\beta}^{\beta'})\delN_{\beta'} u,
\end{equation}
which we use for the commutator decomposition 
\begin{equation} \label{eq1-Hessian-commu-ext}
\aligned
\,[Z,H^{\alpha\beta} \del_\alpha \del_{\beta}]u 
& =    [Z,\HN^{00} \del_t\del_t]u 
+ [Z, H^{\Ncal 0a} \del_t\delsN_a]u 
+ [Z, H^{\Ncal a0} \delsN_a\del_t]u 
\\
& \quad
+ [Z, H^{ \Ncal ab} \delsN_a\delsN_b]u
+[Z,H^{\alpha\beta} \del_{\alpha}(\PsiN_{\beta}^{\beta'})\delN_{\beta'}]u.
\endaligned
\end{equation}
Considering all of the terms in the right-hand side {\sl except} the last one, 
with $Z= \del^IL^J\Omega^K$
we have the identity
\begin{equation} \label{eq2-Hessian-commu-ext}
\aligned
\,[Z, H^{\Ncal \alpha\beta} \delN_\alpha \delN_{\beta}]u
& \cong  
\del^{I_1}L^{J_1} \Omega^{K_1} H^{\Ncal \alpha\beta} \del^{I_2}L^{J_2} \Omega^{K_2} \delN_\alpha \delN_{\beta} u 
\\
& \quad + L^{J_1'} \Omega^{K_1'} H^{\Ncal \alpha\beta} \del L^{J_2'} \Omega^{K_2'} \delN_\alpha \delN_{\beta} u 
+ H^{\Ncal\alpha\beta}[Z, \delN_\alpha \delN_{\beta}]u
\endaligned
\end{equation}
with $I_1+I_2=I$, $J_1+J_2=J_1'+J_2'=J$, and $K_1+K_2=K_1'+K_2'=K$, while 
$|I_1|\geq 1$ and $|J_1'|+|K_1'|\geq 1$.
Then, in $\MMEnear_{[s_0,s_1]}$ we have 
$$
\aligned
|[Z,\HN^{00} \del_t\del_t]u|
& \lesssim    |\del \HN^{00} |_{p_1-1,k_1} |\del\del u|_{p_2,k_2} + |L \HN^{00} |_{k_1'-1,k_1'-1} |\del\del u|_{p_2',k_2'} + |\HN^{00}  \big| \, | [Z, \del_t\del_t]u \big|, 
\endaligned
$$
where $p_1+p_2=  k_1'+p_2'=p$ 
and $k_1+k_2=k_1'+k_2'=k$.   
Recalling \eqref{eq1-31-01-2020}, we thus obtain 
\begin{equation} \label{eq1-11-02-2020}
\aligned
|[Z,\HN^{00} \del_t\del_t]u|
& \lesssim  |\HN^{00} | \, | \del\del u|_{p-1,k-1} 
+\sum_{p_1+p_2=p\atop k_1+k_2=k} |\del \HN^{00} |_{p_1-1,k_1} |\del\del u|_{p_2,k_2} 
+\sum_{p_1+p_2=p\atop p_1+k_2=k} |L\HN^{00} |_{p_1-1,p_1-1} |\del\del u|_{p_2,k_2}.
\endaligned
\end{equation}
For the second and third terms in the right-hand side of \eqref{eq1-Hessian-commu-ext}, we recall \eqref{eq2-Hessian-commu-ext}, \eqref{eq1-06-02-2020}, and \eqref{eq1-lem1-commu-ext}, which yield us  
\begin{subequations}
\label{eq1'-12-02-2020}\begin{equation}
\aligned
& |[Z, H^{\N0a}\del_t\delsN_a]u|
\lesssim  |H| \, |[ Z, \del\delsN ]u| 
+ \sum_{p_1+p_2=p\atop k_1+k_2=k} |\del H|_{p_1-1,k_1} |\del\delsN u|_{p_2,k_2} 
+ \sum_{p_1+p_2=p\atop p_1+k_2=k} |L H|_{p_1-1,p_1-1} |\del\delsN u|_{p_2,k_2}
\\
& \lesssim  t^{-1} |H|\big(|r-t| \, | \del\del u|_{p-1,k-1} + |\del u|_{p,k} \big) 
+ t^{-1} \sum_{p_1+p_2=p\atop k_1+k_2=k} |\del H|_{p_1-1,k_1} \Big(|r-t| \, | \del\del u|_{p_2,k_2} + |\del u|_{p_2+1,k_2+1} \Big)
\\
& \quad + t^{-1} \sum_{p_1+p_2=p\atop p_1+k_2=k} |L H|_{p_1-1,p_1-1} \Big(|r-t| \, | \del\del u|_{p_2,k_2} + |\del u |_{p_2+1,k_2+1} \Big)
\\
& \lesssim \frac{|r-t|}{t} |H||\del\del u|_{p-1,k-1} 
+ \sum_{p_1+p_2=p\atop k_1+k_2=k}\frac{|r-t|}{t} |\del H|_{p_1-1,k_1} |\del\del u|_{p_2,k_2} 
+ \sum_{p_1+p_2=p\atop p_1+k_2=k}\frac{|r-t|}{t} |LH|_{p_1-1,p_1-1} |\del\del u|_{p_2,k_2}
\\
& \quad + t^{-1} |H||\del u|_{p,k} 
+ t^{-1}\sum_{p_1+p_2=p\atop k_1+k_2=k} |\del H|_{p_1-1,k_1} |\del u|_{p_2+1,k_2+1}
+ t^{-1}\sum_{p_1+p_2=p\atop p_1+k_2=k} |LH|_{p_1-1,p_1-1} |\del u|_{p_2+1,k_2+1}, 
\endaligned
\end{equation}
which is controlled by 
\begin{equation} \label{eq1-12-02-2020}
\aligned
& \lesssim \frac{|r-t|}{t} |H||\del\del u|_{p-1,k-1} 
+ \sum_{p_1+p_2=p\atop k_1+k_2=k}\frac{|r-t|}{t} |\del H|_{p_1-1,k_1} |\del\del u|_{p_2,k_2} 
+ \sum_{p_1+p_2=p\atop p_1+k_2=k}\frac{|r-t|}{t} |LH|_{p_1-1,p_1-1} |\del\del u|_{p_2,k_2}
\\
& \quad + t^{-1} |H||\del u|_p + t^{-1}\sum_{0\leq p_1\leq p-1} |H|_{p_1+1} |\del u|_{p-p_1}.
\endaligned
\end{equation}
Observe that the latter sum is restricted to $p_1\leq p-1$.
\end{subequations}
Similarly, in view of \eqref{eq5-07-02-2020} and \eqref{eq2-lem1-commu-ext}, $|[Z, H^{\N ab} \delsN_a\delsN_b]u|$ is also bounded by the right-hand side of \eqref{eq1-12-02-2020}.
For the last term in the right-hand side of \eqref{eq1-Hessian-commu-ext}, we note that $\del\PhiN$ is homogeneous of degree $-1$. Then, we have 
$
H^{\alpha\beta} \del_{\alpha}(\PhiN_{\beta}^{\beta'})\delN_{\beta'} 
= r^{-1} H^{\alpha\beta} \lambda_{\alpha\beta}^{\gamma} \del_{\gamma} u, 
$
where $\lambda_{\alpha\beta}^{\gamma}$ are homogeneous of degree zero. 
Thanks to \eqref{eq2-31-01-2020}, we have 
\begin{equation} \label{eq3-12-02-2020}
\aligned
& \, \big|[Z,H^{\alpha\beta} 
\del_{\alpha}(\PhiN_{\beta}^{\beta'})\delN_{\beta'}]u \big|
\leq \big|[Z,t^{-1} \lambda_{\alpha\beta}^{\gamma}H^{\alpha\beta} \del_{\gamma}]u \big|
\\
& \lesssim   
\sum_{Z_1 \odot Z_2 = Z \atop \deg(Z_1) \geq 1} 
| Z_1(t^{-1} \lambda_{\alpha\beta}^{\gamma}H^{\alpha\beta})|
|Z_2 \del_{\gamma} u| 
+  \sum_{Z_1 \odot Z_2=Z\atop \deg(Z_1) = 0, \, \rank(Z_1) \geq 1}  
| Z_1(t^{-1} \lambda_{\alpha\beta}^{\gamma}H^{\alpha\beta})| |Z_2 \del_{\gamma} u|
+ t^{-1} |H | \, | [Z, \del_{\gamma}]u|  
\\
& \lesssim    t^{-1} |H| \, | \del u|_{p-1,k-1} + \sum_{p_1+p_2=p\atop k_1+k_2=k} |\del (t^{-1}H)|_{p_1-1,k_1} |\del u|_{p_2,k_2} + \sum_{p_1+p_2=p\atop p_1+k_2=k} |L (t^{-1}H)|_{p_1-1,p_1-1} |\del u|_{p_2,k_2}
\\
& \lesssim t^{-1} |H||\del u|_{p-1} 
+ t^{-1}\sum_{p_1+p_2=p} \Big(|H|_{p_1-1}+|\del H|_{p_1-1} + |LH|_{p_1-1}\Big) \, |\del u|_{p_2}
\\
& \lesssim t^{-1} |H||\del u|_p + t^{-1}\sum_{0\leq p_1\leq p-1} |H|_{p_1+1} |\del u|_{p-p_1-1}.
\endaligned
\end{equation}
Combining \eqref{eq1-11-02-2020}, \eqref{eq1-12-02-2020}, 
and \eqref{eq3-12-02-2020} together, we arrive at \eqref{eq4-12-02-2020}.

%------------------------------------------------------------------------------------------------------ 

\subsection{Main conclusion for the Euclidean-merging domain}

We now introduce the energy densities (defined in $\MME_{[s_0, + \infty)}$)
\begin{equation}
\aligned
\ebfME_{\eta,c}[u] 
& := \crochet^{2 \eta} \Big(\sum_{\alpha} |\zeta \, \del_\alpha u|^2 + \sum_a|\delsME_au|^2 +  c^2 u^2 \Big),
& 
\\
\ebf^{\ME,p}_{\eta,c}[u] & := \sum_{\ord(Z)\leq p} \ebfME_{\eta,c}[Z u],
\quad 
\qquad\qquad\qquad
&
\ebf^{\ME,p,k}_{\eta,c}[u] := \sum_{\ord(Z)\leq p\atop \rank(Z)\leq k} \ebf^{\ME}_{\eta,c}[Z u], 
\endaligned
\end{equation}
and formulate the properties \eqref{eq3-07-02-2020}, \eqref{eq1-06-02-2020}, and \eqref{eq5-07-02-2020} in terms of the energy densities, as follows. Recall that the weight $\zeta$ reduces to $1$ in the Euclidean domain and coincides with $s/t$ in the hyperboloidal domain. 

\begin{proposition}[Estimates of a wave field in terms of its energy densities]
\label{prop1-10-06-2021}
\begin{subequations}
For any function $u$ defined in $\MME_{[s_0, s_1]}$ one has
\begin{equation} \label{eq2-28-03-2021}
\crochet^\eta \zeta \, |\del u|_{p,k} + \crochet^\eta  |\delsN u|_{p,k} \lesssim \big(\ebf^{\E,p,k}_{\eta}[u]\big)^{1/2}
\quad \text{ in } \MME_{[s_0, s_1]}, 
\end{equation}
while near the light cone one controls second-order derivatives as follows: 
\begin{equation}
\crochet^\eta  \zeta \, | \del\delsN u|_{p,k} \lesssim \crochet^\eta  \frac{|r-t|}{t} |\del\del u|_{p,k} + t^{-1} \big(\ebf^{\ME,p+1,k+1}_{\eta}[u]\big)^{1/2}
\quad \text{ in } \MMEnear_{[s_0, s_1]}, 
\end{equation}
\begin{equation}
\crochet^\eta  |\delsN \delsN u|_{p,k} \lesssim \crochet^\eta  \Big(\frac{|r-t|}{t} \Big)^2|\del\del u|_{p,k} + t^{-1} \big(\ebf^{\ME,p+1,k+1}_{\eta}[u]\big)^{1/2}
\quad \text{ in } \MMEnear_{[s_0, s_1]}. 
\end{equation}
\end{subequations}
\end{proposition}

%=====================================================================================

\section{Integral and Sobolev estimates for wave and Klein-Gordon fields} 
\label{sectionN-9}

\subsection{Energy-based $L^2$ estimates in the Euclidean-merging domain} 

\paragraph{Immediate consequences.} 

We rely on the analysis in the previous section and we introduce the integrals associated with the densities $\ebf_{\eta,c}^{\ME,p,k}$, that is, 
for an arbitrary function 
\begin{equation} \label{eq6-03-01-2022}
\Eenergy^{\ME,p,k}_{\eta,c}(s,u) := \int_{\MME_s} \ebf^{\ME,p,k}_{\eta,c}[u] \, dx, 
\end{equation}
with $\Fenergy^{\ME,p,k}_{\eta,c}(s,u) := \Eenergy^{\ME,p,k}_{\eta,c}(s,u)^{1/2}$. For wave fields (which have $c=0$), we use the same notation with $c$ suppressed.  We will also us the analogous notation $\Eenergy^{\H,p,k}_{\eta,c}$ and $\Fenergy^{\H,p,k}_{\eta,c}$ for the hyperboloidal domain. 
First of for wave fields, it is immediate to rely on Proposition~\ref{prop1-10-06-2021} (together with \eqref{eq2-24-06-2021}) and for Klein-Gordon fields $v$, the energy functional provides us with the control 
$c \, \| \crochet^\eta \, v\|_{L^2(\MME_s)} \lesssim \Fenergy_{\eta,c}^{\ME}(s,v)$. We summarize these results as follows. 

\begin{proposition}[Energy-based $L^2$ estimates for wave and Klein-Gordon fields] 
\label{prop 1 L2-be}
For any $\eta\geq 0$, any wave field $u$ and Klein-Gordon field $v$ defined in $\MME_s$, and for all $k \leq p$ one has 
\begin{subequations}
\label{eq 1 prop 1 L2-be-0} 
\begin{equation} \label{eq 1 prop 1 L2-be}
\aligned
\| r^{-1} \crochet^\eta |\LOmega u|_{p,k} 
\|_{L^2(\MME_s)} 
& \lesssim  \Fenergy_\eta^{\ME,p,k}(s,u),
\\ 
\| \crochet^\eta \zeta \, |\del u|_{p,k} \|_{L^2(\MME_s)} + \| \crochet^\eta \, |\delsN u|_{p,k} \|_{L^2(\MME_s)} 
& \lesssim \Fenergy_\eta^{\ME,p,k}(s,u),
\endaligned
\end{equation}  
\begin{equation} \label{eq L2-v-repeat000}
c \, \| \crochet^\eta  |v|_{p,k} \|_{L^2(\MME_s)} 
\lesssim \Fenergy^{\ME,p,k}_{\eta,c}(s,v). 
\end{equation}
\end{subequations}
\end{proposition}

%---------------------------------------------------------------------------------------------------------------------------------------

\paragraph{$L^2$ estimate for $|u|_{p,k}$.} 

In order to apply Proposition~\ref{propo-Poincare-ext}, we need to bound the last term in its right-hand side. Let $Z=\del^IL^J\Omega^K$ with $\ord(Z) = |I|+|J|+|K|$ and  $\rank(Z) = |J|+|K|$.

\begin{itemize} 

\item {\bf ``At least one partial derivative''.} First, for all $\ord(Z) \leq p$ with $\rank(Z) \leq k$ and $|I|\geq 1$ we have 
$$
\aligned
\| r^{-1} \crochet^{\eta} Zu\|_{L^2(\MME_s)} 
& =  \| r^{-1} \crochet^{\eta} \del_{\alpha} \del^{I'}L^J\Omega^K u\|_{L^2(\MME_s)}
\\
& \lesssim \| \crochet^{\eta} \zeta \, \del_{\alpha} \del^{I'}L^J\Omega^K\|_{L^2(\MME_s)}
\lesssim \Fenergy_{\eta}^{\ME,p-1,k}(s,u), 
\endaligned
$$
thanks to the technical observations $\crochet \lesssim r \, \zeta^2$ made in Lemma~\ref{Lem1-05-May-2020}. Furthermore, in this case we observe that
\begin{equation} \label{eq6-17-08-2021}
\aligned
\| \crochet^{\eta} \zeta Zu\|_{L^2(\MME_s)} 
=  \| \crochet^{\eta} \zeta \del_{\alpha} \del^{I'}L^J\Omega^K u\|_{L^2(\MME_s)}
\lesssim \Fenergy_{\eta}^{\ME,p-1,k}(s,u).
\endaligned
\end{equation}

\item {\bf ``At least one boost''.} 
When $\ord(Z)  \leq p$ and $\rank(Z) \leq k$ with $|I| = 0$ and $|J|\geq 1$, we write
\begin{equation} \label{eq7-17-08-2021}
\aligned
& \| r^{-1} \crochet^{\eta} Z u\|_{L^2(\MME_s)} 
=  \| r^{-1} \crochet^{\eta}L_aL^{J'} \Omega^K u\|_{L^2(\MME_s)}
\\
& \lesssim  \| \crochet^{\eta} \delsN_a L^{J'} \Omega^K u\|_{L^2(\MME_s)} 
+  \| \crochet^{\eta} |r-t|r^{-1} \del_aL^{J'} \Omega^K u\|_{L^2(\MME_s)}
\lesssim \Fenergy_\eta^{\ME,p-1,k-1}(s,u).
\endaligned
\end{equation}

\item {\bf ``At least one rotation''.} 
When $|I| = |J| = 0$ and $1\leq |K|\leq k=p$, we obtain 
\begin{equation} \label{eq8-17-08-2021}
\aligned
& \| r^{-1} \crochet^{\eta} Z u\|_{L^2(\MME_s)}
=  \| r^{-1} \crochet^{\eta} \Omega_{ab} \Omega^{K'}u\|_{L^2(\MME_s)}
\\
& \lesssim  {\displaystyle \sum_{c}} \|t^{-1} \crochet^{\eta} L_c\Omega^{K'}u\|_{L^2(\Mnear_s)} 
+  {\displaystyle \sum_{c}} \| \crochet^{\eta} \del_c\Omega^{K'}u\|_{L^2(\Mfar_s)}
\lesssim \Fenergy_\eta^{\ME,p-1,k-1}(s,u).
\endaligned
\end{equation}

\item {\bf Partial conclusion.} We conclude that 
\begin{subequations} \label{eq2-15-05-2020-a}
\begin{equation} \label{eq2-15-05-2020-z}
\| r^{-1} \crochet^{\eta} Zu\|_{L^2(\MME_s)} 
\lesssim 
\begin{cases}
\Fenergy_\eta^{\ME, p-1,k}(s,u), 
&
1\leq \rank(Z) + 1\leq  \ord(Z) \leq p, \quad 0\leq \rank(Z)\leq k, 
\\
\Fenergy_{\eta}^{\ME,k-1}(s,u),\quad 
& 1\leq \ord(Z) = \rank(Z)\leq k. 
\end{cases} 
\end{equation}

\item {\bf No differentiation.} Finally, when $\ord(Z) = 0$, we rely on the weighted Hardy inequality in Proposition~\ref{lem1-hardy}
and obtain 
\begin{equation} \label{eq2-15-05-2020-c}
\| r^{-1} \crochet^{\eta} u\|_{L^2(\MME_s)} 
\lesssim \Fenergy_\eta^{0}(s,u)
\simeq \Fenergy_\eta^{\H, 0}(s,u) + \Fenergy_\eta^{\ME, 0}(s,u), 
\end{equation}
which involves the energy (at zero order) along the whole Euclidean--hyperboloidal slice. 

\end{subequations}

\end{itemize}

\noindent In combination with \eqref{eq2-15-05-2020-a} we thus obtain 
\begin{equation} \label{eq2-15-05-2020}
\| r^{-1} \crochet^{\eta} |u|_{p,k}\|_{L^2(\MME_s)} \lesssim \Fenergy_\eta^{\ME,p,k}(s,u) + \Fenergy_{\eta}^{0}(s,u), 
\end{equation}
and we conclude with the desired control of the right-hand side of the Poincar\'e inequality in Proposition~\ref{propo-Poincare-ext}.  We have arrive at the following result. 

\begin{proposition}[Hardy-Poincar\'e inequality for high-order derivatives] 
\label{eq3-15-05-2020}
For any $\eta= 1/2 +\delta$ with $\delta>0$ and any sufficiently decaying function $u$ defined in $\Mscr_{[s_0,s_1]}$ and 
for all $s \in [s_0, s_1]$ one has 
$$
\| \crochet^{-1 + \eta} |u|_{p,k}\|_{L^2(\MME_s)} 
\lesssim \big(1+\delta^{-1} \big) \, \Fenergy_\eta^{\ME,p,k}(s,u) + \Fenergy_{\eta}^{0}(s,u). 
$$
\end{proposition} 

We conclude with the following result.

\begin{proposition}
\label{17-08-2022-1}
For any $\eta= 1/2 +\delta$ with $\delta>0$ and any function $u$ defined in $\Mscr_{[s_0,s_1]}$ and 
for all $s \in [s_0, s_1]$ one has 
\begin{equation} \label{eq2-18-08-2021}
\| r^{-1} \crochet^{\eta} \zeta|\LOmega u|_{k-1}\|_{L^2(\MME_s)}\lesssim s^{-2} \, \Fenergy_{\eta}^{\ME,k}(s,u) 
+ \Fenergy_{\eta}^{\ME,k-1}(s,u)
\end{equation}
\begin{equation} \label{eq1-18-08-2021}
\| \crochet^{-1+\eta} \zeta|\LOmega u|_{k-1}\|_{L^2(\MME_s)}\lesssim  (1+\delta^{-1}) \, \Fenergy_{\eta}^{\ME,k}(s,u).
\end{equation}
\end{proposition}

\begin{proof}
We first derive \eqref{eq2-18-08-2021}. Let $Z$ be an ordered operator with $\ord(Z) = k-1$ and $\rank(Z) = j\leq k-1$. Then when $0\leq j<k-1$, $Z$ contains at least one partial derivative. Then we apply \eqref{eq6-17-08-2021} with $r^{-1}\lesssim t^{-1}\lesssim s^{-2}$. 
When $j=k-1$, we take $Z' = Z\LOmega$ and we are in the case \eqref{eq7-17-08-2021} and \eqref{eq8-17-08-2021}. So we conclude in view of \eqref{eq2-18-08-2021}. 
On the other hand, for the proof of \eqref{eq1-18-08-2021}, we consider $Z\LOmega u$ with $\ord(Z) \leq k-1$. When $\rank(Z)\leq \ord(Z)-1$, that is,
$Z = \del_{\alpha}Z'$ with $\ord(Z)\leq k-2$, we have 
$$
\|  \crochet^{-1 + \eta} \zeta Z\LOmega u\|_{L^2(\MME_s)}
\lesssim
\|  \crochet^{\eta} \zeta \del_{\alpha}Z'\LOmega u\|_{L^2(\MME_s)}
\lesssim \Fenergy_{\eta}^{\ME,k-1}(s,u).
$$
When $\rank(Z) = \ord(Z)$, we apply Proposition~\ref{propo-Poincare-ext} to $Z\LOmega u$:
$$
\|  \crochet^{-1 + \eta} Z\LOmega u\|_{L^2(\MME_s)}
\lesssim
\big( 1+ \delta^{-1} \big) \|  \crochet^{\eta} \delsME Z\LOmega u\|_{L^2(\MME_s)}
+ 
\|  r^{-1} \crochet^{\eta} Z\LOmega u\|_{L^2(\MME_s)}, 
$$
where the first term in the right-hand side is bounded by $\Fenergy_{\eta}^{\ME,k}(s,u)$, and the last term is bounded in view of
\eqref{eq2-15-05-2020-z} (second case). 
\end{proof}

%------------------------------------------------------------------------------------------------------------------------------------------------------------- 

\subsection{Weighted Sobolev decay for wave fields}
\label{sec333}

\paragraph{Objective.}

We are going to establish the following result. 

\begin{proposition}[Sobolev decay for wave fields in the Euclidean-merging domain] 
\label{lem 2 d-e-I}
For all $\eta \in [0,1)$ and all functions $u$, one has (with $k \leq p$)
\begin{subequations}
\begin{equation} \label{eq 1 lem 2 d-e-I}
\big\| r  \, \crochet^\eta \, |\del u|_{p,k} \big\|_{L^\infty(\MME_s)}  
+ \big\|  r^{1+ \eta} \, | \delsN  u |_{p,k} \big\|_{L^\infty(\MME_s)} 
\lesssim (1-\eta)^{-1} \, \Fenergy_\eta^{\ME,p+3, k+3}(s,u)
\end{equation} 
and, for $1/2 < \eta = 1/2+\delta < 1$,
\begin{equation} \label{eq 1 lem 2 d-e-I-facile}
\| r \, \crochet^{-1+\eta} |u|_{N-2}\|_{L^\infty(\MME_s)} 
\lesssim \delta^{-1} \, \Fenergy_\eta^{\ME,N}(s,u) + \Fenergy_{\eta}^{0}(s,u). 
\end{equation}
\end{subequations}
\end{proposition}

A direct application of Proposition~\ref{pro204-11-2} together with \eqref{eq2-28-03-2021} would {\sl not} provides us with a satisfactory decay bound, due to the weight $\zeta$ and the tangent derivatives $\delsME$. The interest of the above result is that, with a cost of a regularity loss, we manage to avoid the unpleasant objects such as $\zeta$ and $\delsME$.

%--------------------------

\paragraph{More about the weight functions.}

In order to establish Sobolev-type decay estimate in Euclidean-merging domain, we need to systematize the properties of the time function $\Time(s,r)$ and weight functions $\xi, \zeta$. First, we summarize the results of Lemmas~\ref{lem1-03-06-2021} to \ref{lem1-22-05-2020} in the following table. 
\begin{equation}
\aligned 
& 
\qquad & | & \qquad T \qquad && \qquad J \qquad & \qquad \zeta \qquad & \qquad \xi
\\
\hline
& \text{Domain } \MH \qquad & | & \qquad (s^2+r^2)^{1/2}  && \qquad s/t & \qquad s/t \qquad & \qquad 1
\\
\hline
& \text{Domain } \Mext  \qquad & | & \qquad \simeq s^2/2 && \qquad \simeq s & \qquad 1 \qquad & \qquad 0
\endaligned
\end{equation}
% Then we establish the following result.

\begin{lemma} \label{lem 0 d-e-I}
In the Euclidean-merging domain $\MME_s$, one has 
\begin{subequations}
\begin{equation} \label{eq 1 lem 0 d-e-I}
| \del_r\xi(s,r)| + | \del_r\del_r\xi(s,r)| 
\lesssim
(1-\xi(s,r))^{1/2}
\lesssim
\zeta(s,r), 
\end{equation} 
\begin{equation} \label{eq=zeta-bound} 
1 \lesssim s \, \zeta(s,r).
\end{equation}
\end{subequations}
\end{lemma} 

\begin{proof} 
These bounds are trivial in $\Mext_s$, and so we only need to derive them in $\Mtran_s$. We first check that  
$0\leq \del_r \xi(s,r) \lesssim \nub (1-\xi(s,r))^{1/2}$ when $\rhoH(s) \leq r \leq \rhoE(s)$, by relying on the properties  \eqref{eq3-04-05-2020} enjoyed by $\chi$. This follows from the 
observation that $0 \leq \chi' (x) \lesssim \chi(x)^{1/2}$ for $x \in [0,1]$. 
We only need to treat a neighborhood of the origin, say the interval $[0,1/2)$.
Indeed, given any $\eps_0>0$ let us consider the function $F(x) := \chi(x) - \eps \chi' (x)^2$ which, we claim, is non-negative. We find $F'(x) = \chi'(x) ( 1 -  2 \eps \chi''(x))$ which is positive for $\eps$ sufficiently small, hence $F$ is non-decreasing. Since $F(0)=0$, we find $F \geq 0$, that is, 
\begin{equation} \label{eq1-02-07-2021}
\chi(x)\geq \eps\chi'(x)^2
\qquad \text{in } [0,1].
\end{equation}
Next, we repeat the same argument but now with $\eps < (1/6) \| \chi'', \chi''''\|_\infty$ and by using the function 
$h_2 := \chi - \eps \chi''{}^2 $ which satisfies 
$$
h_2' = \chi' - 2 \eps \chi'' \chi''', 
\qquad 
h_2'' = \chi'' \big( 1 - 2 \eps \, \chi'''')  - 2 \eps \chi'''{}^2. 
$$
We claim that this function is convex, while it is vanishes together with its first derivative at the origin, and therefore $h_2 \geq 0$. 
Namely, this follows from the additional observation that, in the expression of $h_2''$, the latter term $2 \eps \chi'''{}^2$ is controlled by $\chi''$, as becomes clear by repeating the same argument to the function 
$$
h_3 := \chi''  - 3 \eps \chi'''{}^2 \geq 0, 
\qquad 
h_3' := \chi'''  \big( 1 - 6 \eps \chi'''' \big) \geq 0. 
$$ 
Finally, observe that the inequality \eqref{eq=zeta-bound} is equivalent to saying  
$
{1 \over s^2} \lesssim \frac{s^2}{s^2+r^2} \leq  \frac{s^2+(1-\xi(s,r))^2r^2}{s^2+r^2} = \zeta^2 
$
within the interval 
$
(s^2 - 1)/2 \leq r \leq (s^2 +1)/2.
$
Clearly, the latter holds since $\xi \in [0,1]$ and $s$ is bounded below by $\sqrt{2}$. 
\end{proof}

%---------------------------------------------------------------------------------------------------------------------------------------------------

\paragraph{Sobolev decay far from the light cone.}

The proof of the first statement \eqref{eq 1 lem 2 d-e-I} is established via several lemmas (below) which distinguish between ``far'' and ``near'' regimes, while the second statement \eqref{eq 1 lem 2 d-e-I-facile} is 
much easier (and proven below) 
in view of
Proposition~\ref{eq3-15-05-2020}
and the Sobolev inequality \eqref{ineq 2 sobolev}.   Observe that \eqref{eq 1 lem 2 d-e-I-facile} is {\sl not included} in \eqref{eq 1 lem 2 d-e-I},  since $|Zu|$ does not appear in the energy functional (for the wave equation).
Next, thanks to the ordering property in Proposition~\ref{prop--fund-order} and the Sobolev inequality in Proposition~\ref{pro204-11-2}, we can control the sup-norm far from the light cone directly. 

\begin{lemma} Fix $1/2<\eta<1$.  
For any function $u$ defined in $\MMEfar_s$ one has
\begin{equation} \label{equa-2938} 
\aligned
\| r^{1+\eta}(|\del u|_{p,k} + |\delsN u|_{p,k}) \|_{L^\infty(\MMEfar_s)} 
& \lesssim \Fenergy^{\E,p+2,k+2}_\eta(s,u), 
\\
\| r^\eta (|L u|_{p,k} + |\Omega u|_{p,k}) \|_{L^\infty(\MMEfar_s)} 
& \lesssim \Fenergy^{\E,p+2,k+2}_\eta(s,u).
\endaligned
\end{equation}
\end{lemma} 

\begin{proof} Far from the light cone we have $r/2 <  r-t$, so that the first inequality is immediate from the reordering property \eqref{equa2-2-juin} and the second Sobolev inequality in Proposition~\ref{pro204-11-2}. 
The second inequality is also direct after expanding  $Z L_a = \del^IL^J\Omega^K L_a$
(with $L_a = x^a\del_t  + t\del_a$) and using the homogeneity property. 
Namely, for all $|I|+|J| \leq p$ and $|J| \leq k$ we have 
$$
Z \big(r\big((x^a/r)\del_t u + (t/r)\del_au\big)\big) 
=
\sum_{I_1+I_2=I,J_1+J_2=J\atop K_1+K_2=K} 
\hskip-.3cm
\del^{I_1}L^{J_1} \Omega^{K_1} r\del^{I_2}L^{J_2} \Omega^{K_2} \big((x^a/r)\del_t u+ (t/r)\del_au\big) 
$$
and, by homogeneity of the coefficients,  
$
|Z L_au| \lesssim r|\del u|_{p,k}$, 
which leads us to $|L u|_{p,k} \lesssim r  |\del u|_{p,k}$. 
Applying \eqref{ineq 1 sobolev} in Proposition~\ref{pro204-11-2},
the bound on $|L u|_{p,k}$ is established. For $|\Omega u|_{p,k}$, we rely on
$\Omega_{ab} u = r\big((x^a/r)\del_b - (x^b/r)\del_a\big)$ and we proceed similarly. 
\end{proof}
%---------------------------------------------------- 

\paragraph{Sobolev decay near the light cone.} 

Next, we establish the following statement.

\begin{lemma}
\label{propo1-29-02-2020}
Fix $\eta \in [0,1)$. 
For any function $u$ defined in $\MMEnear_s$ one has
\begin{equation} \label{eq1-29-02-2020}
r \crochet^\eta |\del u|_{p,k} + r^{1+\eta} |\delsN u|_{p,k} \lesssim 
(1-\eta)^{-1} \, \Fenergy^{\ME,p+3,k+3}_{\eta}(s,u).
\end{equation}
\end{lemma} 

The proof is carried out in three steps. 
\begin{itemize}
\item In order to deal with the null derivatives we rely on the decomposition  
\begin{equation} \label{eq 1 d-e-I} 
\delsN_a u = (x^a/t) r^{-1}(t-r) \del_t u + t^{-1}L_a u. 
\end{equation}
Since in the first term the factor $\frac{t-r}{r}$ enjoys sufficient decay, we now focus our attention on the  term $t^{-1}L_a u$. 

\item We then consider the tangent radial derivative $\delsME_r := (x^a/r)\delsME_a$ (introduced in \eqref{equa2---5mai}). 
We establish that $\delsME_r L_a u$ enjoys {\sl sufficient decay} and, in fact, is controlled by $t^{-1} \crochet^{- \eta}$ (multiplied by $\Fenergy^{\ME,3}(s,u)$). 

\item We integrate $\delsME_r L_a u$ along radial directions from the boundary  
$\MME_s \cap \{r=2t\}$, since on this boundary the function $L_a u$ is known to have the desired decay, that is, ${(1+ t)^{- \eta}} \, \Fenergy^{\ME,2}(s,u)$. This is called {\sl integration toward the light cone}, and will be applied several times in the rest of this paper.

\end{itemize}  
For clarity, the proof of Lemma~\ref{propo1-29-02-2020} is divided in three steps, as follows.

\paragraph{Step I.} Bounds on $\delsME_r u$.

\begin{claim} \label{lem2-29-02-2020}
For any function $u$ defined in $\MME_{[s_0,s_1]}$ and any $|I| + |J| \leq2$, one has 
\begin{equation} \label{eq 1 lem 1 d-e-I}
\| \crochet^\eta  
\delsME{}^I\Omega^J \delsME_ru\|_{L^2(\MME_s)}
\lesssim \Fenergy_{\eta}^{\ME,2,2}(s,u) 
\end{equation}
and, by the Sobolev inequality \eqref{ineq 2 sobolev}, 
\begin{equation} \label{eq 2 lem 1 d-e-I}
\| r \, \crochet^{\eta}  \delsME_r u(t,x) \|_{L^\infty (\MME_s)}
\lesssim \Fenergy^{\ME,2,2}_{\eta}(s,u).
\end{equation}
\end{claim}

%----------------------------------------------  

We only need to check this result in the merging domain, since in the Euclidean domain $\delsME_a=\del_a$ and $\zeta=1$. Then the result is trivial.  Observe that $[\Omega_{ab}, \delsME_r] = 0$, so that 
$
\delsME{}^I \Omega^J \delsME_ru = \delsME{}^I \delsME_r\Omega^J u.
$
Then, we focus on the calculation of $\delsME_a \delsME_b$ and $\delsME_a \delsME_b \delsME_c$. 
Throughout our calculation, we use the parameterization $(s,x)$ and, for each fixed $s$, 
so $\delsME_a$ acts as an operator in the variable $x^a$.
By a direct calculation using $\delsME_a = \del_a + x^a\xi(s,r) (s^2 + r^2)^{-1/2} \del_t$ we find
$$
\aligned
\delsME_a \delsME_b u 
= 
& \frac{\xi(s,r)x^b}{(s^2 + r^2)^{1/2}} \delsME_a\del_t u + \delsME_a\del_b u 
+ \frac{\xi(s,r)((s^2 +r^2)\delta^{ab} - x^ax^b)}{(s^2 +r^2)^{3/2}} \del_t u
+ \frac{\del_r\xi(s,r)x^ax^b}{r(s^2 + r^2)^{1/2}} \del_t u.
\endaligned
$$ 
Since $0\leq \xi(s,r)\leq 1$, the first three terms are bounded by $| \delsME_a\del_\alpha u|$ or $\zeta | \del_t u|$, by recalling $1 \lesssim s \, \zeta(s,r)$ in \eqref{eq=zeta-bound}. 
Moreover, the fourth and last term (thanks to \eqref{eq 1 lem 0 d-e-I}) is also bounded by $\zeta \, | \del_t u|$. So by the expression of the energy, the $L^2$ norm of $\delsME_a\delsME_bu$ is bounded by the energy in the merging domain.

The estimate for $\delsME_a\delsME_b\delsME_c u$ is established similarly. A direct calculation shows that it is bounded by a linear combination (with constant coefficients) of the following terms:
$$
\delsME_a\del_\alpha \del_\beta u,
\qquad
s^{-1} \del_\alpha \del_\beta u,
\qquad 
s^{-1} \del_\alpha u,
\qquad
\del_r\xi(s,r)\del_\alpha \del_\beta u,
\qquad 
\del_r\del_r\xi(s,r)\del_\alpha u.
$$ 
Thanks to \eqref{eq1-03-05-2020}, \eqref{eq 1 lem 0 d-e-I}, and \eqref{eq=zeta-bound}, the $L^2$ norm of $\delsME_a\delsME_b\delsME_c u$ is bounded by the energy density in the transition domain. 
Finally, we observe that $\delsME{}^I\Omega^J \delsME_r u = \delsME{}^I\delsME_r\Omega^J u$
is a finite linear combination of $\delsME{}^{I'} \delsME_a\Omega^J u$ with homogeneous coefficients of degree $\leq 0$ and $|I'| \leq |I|$.  So \eqref{eq 1 lem 1 d-e-I} is established.
On the other hand, \eqref{eq 2 lem 1 d-e-I} is a combination of \eqref{eq 1 lem 1 d-e-I} and the Sobolev inequality in Proposition~\ref{pro204-11-2}.   

%-----------------------------------------------------------------------------------------------------------

\paragraph{Step II.}   Integration toward the light cone.

\begin{claim}
\label{claim-13-08-2020}
\begin{subequations}
For any exponent $\eta \in [0,1)$ and any function $u$ defined in $\MME_{[s_0, s_1]}$, on each slice $\MMEnear_s$ one has 
\begin{equation} \label{eq3-29-02-2020}
| r^{\eta}L_a u| \lesssim (1-\eta)^{-1} \, \Fenergy_{\eta}^{\ME,3,3}(s,u),
\end{equation}
\begin{equation} \label{eq4-29-02-2020}
| r \crochet^\eta  \del_\alpha u| \lesssim \Fenergy_{\eta}^{\ME,3,3}(s,u),
\end{equation}
\begin{equation} \label{eq1-01-03-2020}
| r^{1+\eta} \delsN_a u| \lesssim 
(1-\eta)^{-1} \, \Fenergy_{\eta}^{\ME,3,3}(s,u).
\end{equation}
\end{subequations}
\end{claim}

Namely, we rely on \eqref{eq 2 lem 1 d-e-I}. For \eqref{eq3-29-02-2020}, we replace $u$ by $L_au$ in \eqref{eq 2 lem 1 d-e-I} and obtain 
$$
|\delsME_r L_au| \lesssim r^{-1} \crochet^{-\eta} \, \Fenergy^{\ME,3,3}_{\eta}(s,u) 
\lesssim t^{-1} \crochet^{-\eta} \, \Fenergy_{\eta}^{\ME,3,3}(s,u).
$$
Let $(t,r)\in \Mnear_s$. We observe that $t = \Time(s,r)\simeq r \simeq s^2$, and with $w_{s,x}(\rho) := L_a u(\Time(s, \rho), \rho x/r)$ we obtain 
$$
w_{t,x}'(\rho) = (x^a/r)\delsME_a L_au(\Time(s, \rho), \rho x/r) = \delsME_r L_au(\Time(s, \rho), \rho x/r).
$$
Furthermore, observing that $(\Time(s),2 \, \Time(s)x/r)\in \{r=2t\}\subset \Mfar_s$, then by recalling \eqref{equa-2938}  
we obtain
$$
|w_{t,x}(2 \, \Time(s))| 
\lesssim r^{-\eta} \, \Fenergy_{\eta}^{\ME,2,2}(s,u)
\lesssim s^{-2\eta} \, \Fenergy_{\eta}^{\ME,2,2}(s,u).
$$
Then we write 
$
L_au (t,x) = w_{t,x}(r) = -\int_r^{2 \, \Time(s)}w_{t,x}'(\rho) \, d\rho + w_{t,x}(2 \, \Time(s)), 
$
which leads us to
$$
\aligned
|L_au(t,x)|  & \leq  s^{-2\eta} \, \Fenergy_{\eta}^{\ME,2,2}(s,u) + \int_r^{2 \, \Time(s)} |\delsME_r L_au(\Time(s,\rho), \rho)|d\rho
\\
& \lesssim   s^{-2\eta} \, \Fenergy_{\eta}^{\ME,2,2}(s,u) + \Fenergy_{\eta}^{\ME,3,3}(s,u) \int_r^{2 \, \Time(s)}\Time(s,\rho)^{-1}(2+\rho-\Time(s,\rho))^{-\eta} \, d\rho, 
\endaligned
$$
therefore
$$
\aligned
|L_au(t,x)| 
& \lesssim s^{-2\eta} \, \Fenergy_{\eta}^{\ME,2,2}(s,u) 
+ s^{-2} \, \Fenergy_{\eta}^{\ME,3,3}(s,u)\int_r^{2 \, \Time(s)}(2+\rho-\Time(s,\rho))^{-\eta}d(2 + \rho - \Time(s,\rho))
\\
&\quad + s^{-2} \, \Fenergy_{\eta}^{\ME,3,3}(s,u)\int_r^{2 \, \Time(s)}(2+\rho-\Time(s,\rho))^{-\eta}\del_r\Time(s,\rho) d\rho. 
\endaligned
$$
The first and second term are bounded as claimed in \eqref{eq3-29-02-2020}. For the last term, observe that $0\leq \del_r\Time(s,\rho)\leq 1$ and $\del_rT(r,\rho)\neq 0$ if and only if $\rhoH(s)\leq \rho\leq \rhoE(s)$. Then the last term does not vanish only if $r\leq \rhoE(s)$. In this case, we have s
$$
\aligned
&
s^{-2} \, \Fenergy_{\eta}^{\ME,3,3}(s,u)\int_r^{2 \, \Time(s)}(2+\rho-\Time(s,\rho))^{-\eta}\del_r\Time(s,\rho) d\rho
\\
& \lesssim s^{-2} \, \Fenergy_{\eta}^{\ME,3,3}(s,u)\int_r^{\rhoE(s)}(2+\rho-\Time(s,\rho))^{-\eta}\del_r\Time(s,\rho) d\rho
\lesssim t^{-1} \, \Fenergy_{\eta}^{\ME,3,3}(s,u), 
\endaligned
$$
where we used that $\rho - \Time(s,\rho)\geq -1$ when $\rhoH(s)\leq \rho\leq \rhoE(s)$
(thanks to Lemma~\ref{lem1-03-06-2021}) and $\rhoE(s)-\rhoH(s)=1$. In this case $|r-t|\leq 1$, and 
so \eqref{eq3-29-02-2020} is established.

%-----------------------------------------------------------------

Next, the derivation of \eqref{eq4-29-02-2020} is simpler. We observe that, thanks to \eqref{ineq 1 sobolev},
$
(1+t)\crochet^\eta |\del_\alpha u(t,x)| \lesssim \Fenergy_\eta^{\E,2,2}(s,u)$ in $\Mext_s$. 
So we only need to show the bound for $\MM_s$. We observe that 
$$
|\delsME_r\del_\alpha u| \lesssim \Fenergy_{\eta}^{\ME,3,3}(s,u) \, \Time(s,r)^{-1}(1+|r-\Time(s,r)|)^{-\eta} \leq s^{-2} \, \Fenergy^{\ME,3,3}(s,u)
$$
and we integrate in radial directions from $(t,x)$ to $(\Time(s, \rhoE(s)), \rhoE(s) x/r) \in \Mext_s\cap \Mtran_s$: 
$$
|\del_{\alpha}u(t,x)| \leq \int_r^{\rhoE(s)} |\delsME_r\del_\alpha u(\Time(s,\rho),\rho)|d\rho 
+ \del_{\alpha}u(\Time(s, \rhoE(s)), \rhoE(s) x/r).
$$
Observe that $\rhoE(s)-1 = \rhoH(s)\leq \rho\leq \rhoE(s)$, the desired bound is established.  
Finally, we handle \eqref{eq1-01-03-2020} by relying on the identity $\delsN_au = t^{-1}L_a u + (x^a/r)\frac{t-r}{t} \del_t u$ and applying \eqref{eq3-29-02-2020} and \eqref{eq4-29-02-2020}.

%-------------------------------------------------------

\paragraph{Step III.}

We finally establish the high-order version, as follows. In \eqref{eq4-29-02-2020} and \eqref{eq1-01-03-2020} we replace $u$ by $Zu$ with $\ord(Z)\leq p$ and $\rank(Z) \leq k$ and obtain 
$$
r \big(\crochet^\eta |\del_\alpha Z u| 
+ r^{\eta} |\delsN_a Z u|\big)
\lesssim 
(1-\eta)^{-1} \, \Fenergy_{\eta}^{\ME,p+3,k+3}(s,u),
$$
and by recalling the second property in Proposition~\ref{prop--fund-order} and \eqref{eq2-03-02-2020}, we arrive at \eqref{eq1-29-02-2020}.

%-----------------------------------

\paragraph{Completion of the proof of \eqref{eq 1 lem 2 d-e-I-facile}.}

We now complete the proof of Proposition~\ref{lem 2 d-e-I} and first establish the following estimate.

\begin{lemma}
\label{eq4-10-06-2021}
For all $1/2< \eta = 1/2+\delta$ and all functions $u$ defined in $\MMEnear_s$, one has\footnote{Here, $\delta$ may be large.} 
$$
\| r \, \crochet^{-1+\eta} |u|_{p,k} \|_{L^\infty(\Mext_s)} 
\lesssim 
(1+\delta^{-1}) \, \Fenergy^{\ME,p+2,k+2}_\eta(s,u) + \Fenergy^0_{\eta}(s,u).
$$
\end{lemma}

\begin{proof}
Let $Z$ be an ordered operator with $\ord(Z) \leq p$ and $\rank(Z) \leq k$. We replace $u$ in \eqref{ineq 1 sobolev} by $\crochet^{-1}Zu$. Then, observe that $\delsME_a = \del_a$  in $\Mext_s$ so that
$$
r \, \crochet^{\eta-1} |Zu(t,x) \, |\lesssim \sum_{|I|+|K|\leq 2} \|\crochet^{\eta}\del^I\Omega^K(\crochet^{-1}Zu)\|_{L^2(\Mext_s)},\qquad 
(t,x)\in \Mext_s.
$$
Recalling Lemma~\ref{lem1-26-02-2020}, the above bound leads to
$$
r \, \crochet^{\eta-1} |Zu(t,x) \, |\lesssim \sum_{|I|+|K|\leq 2} \|\crochet^{-1+\eta}\del^I\Omega^K Zu\|_{L^2(\Mext_s)},\qquad 
(t,x)\in \Mext_s.
$$
Then in view of Proposition~\ref{eq3-15-05-2020} we obtain the desired result.
\end{proof}

We can now derive \eqref{eq 1 lem 2 d-e-I-facile} in $\MM_s$. To this end we write \eqref{eq 2 lem 1 d-e-I} in the following form. 
If $Z$ denotes any ordered operator with $\ord(Z) \leq p$ and $\rank(Z) \leq k$, within $\MM_s$ we have 
$
|\delsME_rZu|\lesssim s^{-2} \, \Fenergy_{\eta}^{\ME,p+2,k+2}(s,u).
$
We integrate this bound toward the light cone $\{r=t-1\}$ and obtain 
$
t|Zu|\lesssim \Fenergy_{\eta}^{\ME,p+2,k+2}(s,u) + \Fenergy^0_{\eta}(s,u).
$
Finally, we observe that  $\crochet\lesssim 1$ in $\MM_s$, and we arrive at \eqref{eq 1 lem 2 d-e-I-facile}.

%-----------------------------------------------------------------------------------------------------------------------------------------

\subsection{Weighted Sobolev decay for Klein-Gordon fields}

We complete this section with an additional property. 

\begin{proposition}[Sobolev decay for Klein-Gordon fields in the Euclidean-merging domain] 
\label{lem 2 d-e-II}
Fix some $\eta \in [0,1)$. 
For any function $v$ one has  (with $k \leq p$) 
\begin{equation} \label{eq decay-v-repeat000-two}
c \, \| r \, \crochet^{\eta} |v|_{p,k} \|_{L^\infty (\MME_s)}
\lesssim
\Fenergy^{\ME,p+2,k+2}_{\eta,c}(s,v).
\end{equation}
\end{proposition}

\begin{proof} The derivation of \eqref{eq decay-v-repeat000-two} is straightforward in $\Mext_s$ by Proposition~\ref{pro204-11-2}, while a radial integration argument is required in order to cover the merging domain, as we now explain. 
We observe that
$
\|c\crochet^\eta |v|_{p,k} \|_{L^2(\Mext_s)} \lesssim \Fenergy^{\E,p+2,k+2}_{\eta,c}(s,v).
$
By Proposition~\ref{pro204-11-2}, for $(t,x)\in \Mext_s$ we have 
\begin{equation} \label{eq1-05-03-2020}
c \, r \, \crochet^\eta |v|_{p,k}(t,x)\lesssim \Fenergy^{\E,p+2,k+2}_{\eta,c}(s,v).
\end{equation}
For $(t,x)\in\MM_s$, we integrate on the hypersurface $\MME_s$ along radial directions and obtain 
$$
v(t,x) = v\big(T(\rhoE(s),s), \rhoE(s) x/r \big) - \int_{r}^{\rhoE(s)} \delsME_rv(\Time(s, \rho), \rho x/r) \, d\rho, 
$$
in which $(T(\rhoE(s),s), \rhoE(s)x/r)\in\Mext_s\cap\MM_s$ and $\rhoH(s) \leq r \leq \rhoE(s)$. Then, applying \eqref{eq1-05-03-2020} and \eqref{eq 2 lem 1 d-e-I} we obtain 
$$
\aligned
|v(t,x)|
& 
\lesssim  c^{-1}r^{-1} \crochet^{-\eta} \, \Fenergy^{\E,2,2}_{\eta,c}(s,v) + \Eenergy^{\ME,2,2}_{\eta,c} \int_{r}^{\rhoE(s)} \rho^{-1} \la\rho-t \ra^\eta \, d\rho
\\
& \lesssim  c^{-1}r^{-1} \crochet^{-\eta} \, \Fenergy^{\E,2,2}_{\eta,c}(s,v) + s^{-2} \, \Fenergy_{\eta,c}^{\ME,2,2}(s,v)
\lesssim  c^{-1}r^{-1} \crochet^{-\eta} \, \Fenergy^{\E,2,2}_{\eta,c}(s,v).
\endaligned
$$
For the second inequality we have used the fact that $r\simeq s^2$ in $\MM_s$, 
and for the last inequality we have used that $|\rho-t| \leq 1$ in $\MM_s$. We conclude that
$c \, r \, \crochet^\eta |v(t,x)| \lesssim \Fenergy^{\ME,2,2}_{\eta,c}(s,v)$ in $\MM_s$. 
Replacing $v$ by $Z v$ with $\ord(Z)\leq p$ and $\rank(Z)\leq k$ and applying Proposition~\ref{prop--fund-order}, we arrive at \eqref{eq1-05-03-2020} in $\MM_s$. Thus the inequality \eqref{eq decay-v-repeat000-two} is established.
\end{proof}   

%==============================================================================================

\section{Pointwise decay of wave fields and their derivatives} 
\label{section-8-added} 

\subsection{Pointwise estimates of wave fields}

\paragraph{Main statement.} \label{subsec1-07-01-2022}

We now derive several estimates concerning the decay of solutions to wave equations and their gradient and Hessian. 
It is convenient to introduce the following notation.
Given some data $f, u_0, u_1$ with sufficient regularity and decay (so that Kirchhoff's formula below makes sense), we 
consider the solution $u=u(t,x)$ to the Cauchy problem
\begin{equation} \label{eq7-28-12-2020}
\Box u = f, \qquad \qquad
u(1,x) = u_0(x),\qquad \qquad
\del_t u(1,x) = u_1(x), 
\qquad x \in \RR^3, 
\end{equation}
and use the short-hand notation 
\begin{equation}
\aligned 
& u = \Box^{-1}[u_0,u_1,f],
\qquad\qquad
\Box^{-1}_\init[u_0,u_1] := \Box^{-1}[u_0,u_1,0],
\qquad\qquad
\Box^{-1}_\source[f] := \Box^{-1}[0,0,f], 
\endaligned
\end{equation}
in which we find it convenient to distinguish between the contributions from the initial data and from the source. 
We then consider the effect of a (spacetime) decaying source, represented by the operator $\Box^{-1}_{\source}$.  
We revisit an earlier result that we established in \cite{PLF-YM-two} where the source was supported in the interior of a light cone --- a restriction we overcome in the present work.  
Throughout we denote by $\Lambda_{t,x} := \big\{ (\tau,y) \big/ \, t-\tau = |x-y|, \, 1\leq  \tau\leq t \big\}$ the truncated light cone associated with a point $(t,x)$. 

\begin{proposition}[Wave equation. Contribution from the source]
\label{Linfini wave}
The wave operator  $\Box^{-1}_\source$ on a function $f$ satisfying the decay conditions for some exponents $\alpha_1, \alpha_2, \alpha_3$, 
\begin{equation} \label{eq1-27-12-2020}
|f(\tau,y)| \lesssim C_1 \, 
\tau^{\alpha_1}(\tau + |y| )^{\alpha_2} \big( 1 + | \tau - |y| | \big)^{\alpha_3}, 
\qquad 
(\tau,y) \in \Lambda_{t,x}  
\end{equation}
enjoys the following properties:

\begin{subequations}

\noindent {\bf Case 1 (typical).} When $\alpha_1 = -1+\upsilon$ and $\alpha_2 = -1-\nu$ and $\alpha_3=-1+\mu$ 
for some  
$\upsilon + \mu < \nu$ and $0<\mu,\nu,\upsilon\leq 1/2$,
one has 
\begin{equation} \label{eq5-24-12-2020}
|\Box^{-1}_\source[f](t,x)|
\lesssim
C_1 \, \big(\upsilon^{-1} + \mu^{-1} + |\mu-\nu|^{-1}\big) \, |\upsilon + \mu-\nu|^{-1} (t+r)^{-1}.
\end{equation} 

\vskip.15cm

\noindent{\bf Case 2 (sub-critical).} When $\alpha_1=0$ and $\alpha_2 = -2-\nu$ and $\alpha_3 = -1+\mu$ 
for some $0 < \nu, \mu \leq 1/2$, one has 
\begin{equation} \label{Linfini wave ineq}
|\Box^{-1}_\source[f](t,x)|
\lesssim C_1 \, 
\begin{cases} 
\mu^{-1} |\mu-\nu|^{-1} (t+r)^{-1}t^{\mu-\nu},\qquad  
&\mu>\nu,
\\
\mu^{-1} (t+r)^{-1} \ln (t+1),\quad  & \mu=\nu, 
\\
\mu^{-1} |\mu-\nu|^{-1} (t+r)^{-1}, & \mu<\nu.
\end{cases} 
\end{equation} 

\vskip.15cm

\noindent{\bf Case 3 (critical).} When $\alpha_1 = 0$ and $\alpha_2 = -2$ and $\alpha_3= -1-\mu$ for some 
$\mu\in (0,1/2)$, one has  
\begin{equation} \label{eq1-10-01-2021}
|\Box^{-1}_\source[f](t,x)| \lesssim C_1\,\mu^{-1} (t+r)^{-1}\Big(1 + \crochet^{-\mu}\ln\Big(\frac{t}{\crochet}\Big)\Big).
\end{equation} 

\vskip.15cm

\noindent{\bf  Case 4 (super-critical).} When $\alpha_1 = 0$ and $\alpha_2 = -2+\nu$ and $\alpha_3 = -1-\mu$ for some 
$0<\nu< \mu < 1/2$, one has  
\begin{equation} \label{eq1-10-01-2021-case4}
|\Box^{-1}_\source[f](t,x)| \lesssim
C_1 \, \big(|\mu-\nu|^{-1} + \mu^{-1}\nu^{-1}\crochet^{-\mu}t^{\nu}\big) (t+r)^{-1}.
\end{equation} 
(Concerning the last two cases, recall that $\crochet \equiv 1$ when $r\leq t-1$). 
\end{subequations}
\end{proposition}

Kirchhoff's formula for the wave equation in $\RR^{3+1}$ with vanishing initial data at $\{t=1\}$ reads  
\begin{equation} \label{eq--source-Kirchhoff}
u(t,x) = \frac{1}{4\pi} \int_1^t\frac{1}{t- \tau} \int_{|y| =t- \tau}f(\tau,x-y) \, d\sigma(y) d\tau
\end{equation} 
and
(for any given point $(t,x)$ with $t \geq 1$, say)  involves an integration on the truncated cone $\Lambda_{t,x}$ based at the point $(t,x)$. 
With the change of variable $\lambda = \tau/t$ and $y' = y/t$, we find 
$$
\aligned
|u(t,x)| 
& \lesssim   
C_1 \int_1^t \int_{|y| = t- \tau} \tau^{\alpha_1}(\tau + |x-y| )^{\alpha_2} \big( 1 + | \tau - |x-y| | \big)^{\alpha_3} 
\, 
{d\sigma(y) d\tau \over t- \tau} 
\\ 
& =  C_1 \, t^{2+\alpha_1+\alpha_2+\alpha_3} \int_{1/t}^1\int_{|y'| = 1 - \lambda}
\lambda^{\alpha_1}(\lambda +  | y' - x/t | )^{\alpha_2}\Big( t^{-1} + | \lambda - | y' - x/t | \big| \Big)^{\alpha_3}
{d\sigma(y') \, d\lambda \over 1 - \lambda}.
\endaligned
$$
In order to establish Proposition~\ref{Linfini wave}, our task is to control this latter integral. It is necessary to distinguish between several cases and, at first, a simplest case is the following one. 

%----------------------------------------

\paragraph{Case of the center.}  

\begin{subequations}
When $x=0$ the inequality under consideration reduces to saying 
$$
\aligned
|u(t,0)| 
& \lesssim   C_1 \, t^{2+\alpha_1+\alpha_2+\alpha_3}
\int_{1/t}^1\int_{|y'| = 1 - \lambda}
\lambda^{\alpha_1}\Big( t^{-1} + |2\lambda -1|\Big)^{\alpha_3}
{d\sigma(y') \, d\lambda \over 1 - \lambda}
\\
& \lesssim  C_1 \, t^{2+\alpha_1+\alpha_2+\alpha_3}
\int_{1/t}^1\lambda^{\alpha_1}(t^{-1}+|2 \lambda-1|)^{\alpha_3}(1-\lambda)^{-1} \Big(\int_{|y'|=1-y}d\sigma(y')\Big) \, d\lambda
\\
& \lesssim C_1 \, t^{2+\alpha_1+\alpha_2+\alpha_3}  
\int_{1/t}^1\lambda^{\alpha_1}(t^{-1}+|2 \lambda-1|)^{\alpha_3}(1-\lambda) \, d\lambda.
\endaligned
$$
The integral in the right-hand side  has two (potential) singular points, namely $\lambda \to 0^+$ and $\lambda = 1/2$. In Case 1,  we have $\alpha_1 = -1+\upsilon>-1$ and $\alpha_3 = -1+\mu>-1$ and we find 
\begin{equation} \label{eq6-24-12-2020}
|u(t,0) \, |\lesssim 
C_1 \, (\upsilon^{-1} + \mu^{-1}) t^{-1+\upsilon+\mu-\nu}.
\end{equation}
In Case 2, we have $\alpha_1=0$ and $\alpha_3 = -1+\mu>-1$ and we find 
\begin{equation} \label{eq7-24-12-2020}
|u(t,0) \, |\lesssim 
C_1 \,  \mu^{-1} t^{-1+\mu-\nu}.
\end{equation}
In Case 3, we observe that
$$
\int_{1/t}^1(t^{-1} + |2\lambda-1|)^{-1-\mu}(1-\lambda) \, d\lambda
\lesssim \int_{1/t}^1(t^{-1} + |2\lambda-1|)^{-1-\mu}d\lambda\lesssim 
\mu^{-1}t^{\mu}
$$
and this leads us to
\begin{equation} \label{eq1-08-06-2021}
|u(t,0) \, |\lesssim C_1 \,  \mu^{-1} t^{-1}.
\end{equation}
Similarly for Case 4, we find  
\begin{equation}
|u(t,0) \, |\lesssim C_1 \,  \mu^{-1}t^{-1+\nu}.
\end{equation}
These estimates are stronger (in this very special case) than the ones we will derive in general, and this easily gives us the conclusion in Proposition~\ref{Linfini wave}. 

\end{subequations}

%----------------------------------------------------------------

\paragraph{Parametrization away from center.}

When $x\neq 0$, it is convenient to introduce an adapted parameterization. Without loss of generality we let $x = (r,0,0)$ and the sphere $\{|y| = 1 - \lambda\}$ is parameterized as follows.  
\begin{itemize}

\item $\theta$ denotes the angle from $(1,0,0)$ to $y$ with $0\leq \theta\leq \pi$, and 

\item $\phi$ denotes the angle from the plane passing by the points $(1,0,0)$ and $(0,1,0)$ to the plane passing by the points $y$ and $(1,0,0)$, in which $0\leq \phi\leq 2 \pi$. 

\end{itemize} 
We see that 
$y = (1 - \lambda)\big(\cos\theta, \sin\theta  \cos\phi, \sin\theta \sin\phi\big)$ while by elementary trigonometry we have 
$$
| y - x/t |^2 = (r/t)^2 + (1-\lambda)^2 - 2(r/t)(1-\lambda)\cos\theta
$$
and
$d\sigma(y) =(1 - \lambda)^2 \, \sin \theta \, d\theta d\phi$. It follows that 
\begin{equation}
\aligned
|u(t,x)|& \lesssim    
C_1 t^{2+\alpha_1+\alpha_2+\alpha_3} \int_{1/t}^1\int_0^{\pi}\int_0^{2 \pi}
\lambda^{\alpha_1}(\lambda+|y-x/t|)^{\alpha_2}\big(t^{-1}+\big|\lambda - |y-x/t|\big|\big)^{\alpha_3}
(1-\lambda)\sin\theta d\theta d\phi d\lambda
\\
& \lesssim  C_1 t^{2+\alpha_1+\alpha_2+\alpha_3}
\int_{1/t}^1(1-\lambda)\lambda^{\alpha_1}
\Big(
\int_0^{\pi}(\lambda+|y-x/t|)^{\alpha_2}\big(t^{-1}+\big|\lambda - |y-x/t|\big|\big)^{\alpha_3} \sin\theta d\theta \Big) \, d\lambda 
\\
&=:C_1 t^{2+\alpha_1+\alpha_2+\alpha_3} \int_{1/t}^1I(\lambda;t,r) \, d\lambda. 
\endaligned
\end{equation}
Before proceeding with our detailed analysis, we still need to simplify the expression $I(\lambda;t,r)$.
Let us define 
$
\rho =: |y - x/t|^2 = (r/t)^2 + (1-\lambda)^2 - 2(r/t)(1-\lambda)\cos\theta,
$
so that $d\rho = 2(r/t)(1-\lambda)\sin \theta d\theta$. With this notation the above expression simplifies, namely
\begin{equation} \label{Kirchhoff-source-basic}
I(\lambda;t,r) = \frac{t}{2r} \lambda^{\alpha_1}\int_{(r/t - (1-\lambda))^2}^{(r/t+(1-\lambda))^2}
(\lambda + \sqrt{\rho})^{\alpha_2} \big(t^{-1} + |\lambda - \sqrt{\rho} | \big)^{\alpha_3} \, d\rho.
\end{equation} 
The rest of the proof distinguishes between different regimes, depending whether the base point $(t,x)$ is inside, near, or outside the light cone; the proof is direct but requires very tedious calculations which are postponed to Appendix~\ref{appenDD}.
(The interested reader can also refer to \cite{PLF-YM-two} for the case when source is supported in the light cone.)  

%--------------------------------------------------------------------------------------------------------------------------------------------------------------

\subsection{Hessian estimates of wave fields} 
\label{secti-10-1}

\paragraph{Two decompositions.}

We now analyze the Hessian of a function and establish sharp control in terms of the wave operator. This requires us to decompose the domain $\MME_{[s_0,s_1]}$ in connection with the extended cone $\{r = (1-\ell)^{-1}t\}$ and, in addition to the  domain $\Mscr^{\near}_\ell$ introduced in \eqref{eq1-06-07-2021}, we set 
\begin{subequations}
\label{equa-defineMnear}
\begin{equation}
\aligned 
\Mnear_{\ell,[s_0,s_1]} :& = \MME_{[s_0,s_1]}\cap \Mnear_\ell \quad 
&& \text{ domain near the light cone,}
\\
\Mscr^{\far}_{\ell, [s_0,s_1]} :& =    \MME_{[s_0,s_1]} \setminus \Mscr^{\near}_{\ell,[s_0,s_1]} 
\quad 
&& \text{ domain far from the light cone.}
\endaligned
\end{equation} 
The decomposition $\MME_{[s_0, s_1]} = \MMEnear_{[s_0,s_1]} \cup \MMEfar_{[s_0,s_1]}$ used earlier in \eqref{equa-nearfardefinition} 
correspond to the special case where $\ell$ is replaced by $1/2$. 
For the associated slices, we use the notation 
\begin{equation}
\Mnear_{\ell,s} := \MME_s\cap\Mnear_\ell,
\qquad\qquad
\Mfar_{\ell,s}  := \MME_s\setminus\Mnear_{\ell,s}.
\end{equation} 
\end{subequations}

\begin{lemma}[Decomposition of the wave operator near the light cone]
\label{lem1-06-12-2020}
For any metric $g^{\alpha\beta} = g_\Mink^{\alpha\beta} + H^{\alpha\beta}$ and any function $u$ defined in $\MME_{[s_0, s_1]}$ one has\footnote{This decomposition is valid everywhere, but will be used mainly near the light cone (except for time-derivatives).} 
$$ 
\aligned
\Boxt_g  u 
& = \Theta[g] \, \del_t\del_t u + t^{-1}\Kbb_g^{\ME}[u],
\qquad
\qquad
&&
\big|\Kbb_g^{\ME}[u]\big| \lesssim |g||\del u|_{1,1}, 
\\
\Theta[g]
& := \big((r/t)^2-1\big) + H^{00} - 2\sum_a(x^a/t)H^{a0} + \sum_{a,b}(x^ax^b/t^2)H^{ab}.
\endaligned
$$
\end{lemma}

\begin{proof} We obtain the desired decomposition from the identities 
\begin{subequations} \label{eq1-17-07-2021}
\begin{equation} \label{decompo-at'-near}
\del_t\del_au = \del_a\del_t u = -(x^a/t)\del_t \, \del_t u + t^{-1} \big( \del_tL_a - \del_a\big)u,
\end{equation}
\begin{equation} \label{decompo-ab'-near}
\del_a\del_bu =  \frac{x^ax^b}{t^2} \del_t \, \del_t u + t^{-1} \Big(\del_aL_b - \delta_{ab} \del_t - (x^b/t)\del_tL_a + (x^b/t)\del_a \Big)u, 
\end{equation}
\end{subequations}
by a direct substitution into $\Boxt_g u = g^{\alpha\beta}\del_{\alpha}\del_{\beta}u$. Specifically, we find 
$$
\Kbb_g^{\ME}[u] :=  2 g^{a0} \big(\del_tL_a-\del_a \big) u 
+ g^{ab} \Big(\del_aL_b - \delta_{ab} \del_t - (x^b/t)\del_tL_a + (x^b/t)\del_a \Big) u.
\qedhere
$$
\end{proof}

\begin{lemma}[Decomposition of the wave operator away from the light cone]
\label{eq1-07-12-2020}
For any metric $g^{\alpha\beta} = g_\Mink^{\alpha\beta} + H^{\alpha\beta}$ and any function $u$ defined in $\MME_{[s_0, s_1]}$ one has\footnote{This decomposition is valid everywhere, but will be used away from the light cone.} 
$$ 
\aligned
\Boxt_g u 
& = \Theta^{tt}[g] \, \del_t\del_tu + \Theta^{rr}[g] \, \del_r\del_r u 
+\Dbb_g^{\ME}[u], 
\qquad \quad 
&& 
\big|\Dbb_g^{\ME}[u]\big| \lesssim \big(r^{-1} |g| + t^{-1} |H|\big) \, |\del u|_{1,1}, 
\\
\Theta^{tt}[g] & := -1 + H^{00} - 2(x^a/t)H^{a0}, 
\qquad
&& \Theta^{rr}[g] := 1+ \sum_{a,b}(x^ax^b/r^2)H^{ab}.
\endaligned
$$
\end{lemma}

\begin{proof}
In view of \eqref{decompo-at'-near} and the identity
$\del_a = (x^a/r)\del_r -\sum_{c\neq a}(x^c/r^2)\Omega_{ac}$, we obtain 
\begin{equation} \label{eq1-18-02-2020}
\aligned
\del_a\del_b  
&  =  (x^ax^b/r^2)\del_r\del_r 
- \sum_{c\neq b}(x^ax^c/r^3)\del_r\Omega_{bc} - \sum_{c\neq a}(x^bx^c/r^3)\del_r\Omega_{ac} 
+ \sum_{c\neq a,d\neq b}(x^cx^d/r^4)\Omega_{ac} \Omega_{bd}
+ r^{-1} \Lambda_{ab}^{\gamma} \del_{\gamma}
\\
& =:   (x^ax^b/r^2)\del_r\del_r  + r^{-1}\Dbb_{ab}, 
\endaligned
\end{equation} 
where $\Lambda_{ab}^{\gamma}$ are exterior-homogeneous of degree zero in the Euclidean-merging domain. It is clear that
$|\Dbb_{ab}[u]|\lesssim |\del u|_{1,1}$ and, moreover, with the notation in the lemma 
$$
\aligned
\Boxt_g u  & =  \big(g^{00} - 2(x^a/t)g^{a0}\big)\del_t\del_tu + \sum_{a,b}(x^ax^b/ r^2)g^{ab}\del_r\del_ru
+ r^{-1}g^{ab}\Dbb_{ab}[u] + 2t^{-1}g^{a0}\big(\del_tL_a - \del_a\big)u
\\
& =   \Theta^{tt}[g] \, \del_t\del_tu + \Theta^{rr}[g] \, \del_r\del_r u
+ r^{-1}g^{ab}\Dbb_{ab}[u] + 2t^{-1}g^{a0}\big(\del_tL_a - \del_a\big)u. 
\endaligned
$$
The desired result is established, by observing that that $g^{a0} = H^{a0}$, with 
$$
\hfill 
\Dbb_g^{\ME}[u] :=  r^{-1}g^{ab}\Dbb_{ab}[u] + 2t^{-1}g^{a0}\big(\del_tL_a - \del_a\big)u. 
\hfill 
\qedhere 
$$
\end{proof}

%------------------------------------------------------------------

\paragraph{Control near the light cone.} 

We begin our analysis near the light cone.

\begin{lemma}[Hessian for the wave equation near the light cone: zero-order]
\label{lem1-15-02-2020}
Consider a metric $g^{\alpha\beta} = g_\Mink^{\alpha\beta} + H^{\alpha\beta}$ defined in $\Mscr^{\near}_{\ell, [s_0,s_1]}$ and satisfying, for some $\eps_1 \ll 1$, 
\begin{equation} \label{eq2-28-12-2020}
|H|\leq \eps_1 \quad \text{ in } \Mscr^{\near}_{\ell, [s_0,s_1]}.
\end{equation}
Then, for any function $u$ defined in $\MME_{[s_0, s_1]}$ one has 
\begin{equation} \label{eq1-28-12-2020}
\frac{1+|r-t|}{r} |\del\del u|\lesssim |\Boxt_g u| + t^{-1} |\del u|_{1,1} +
|H^{\N00}\del_t\del_t u|.
\end{equation}
If, in addition,
\begin{equation} \label{eq1-15-02-2020}
\big|\HN^{00} \, \big|\leq \frac{1+|r-t|}{3r} \quad \text{ in } \Mscr^{\near}_{\ell, [s_0,s_1]}, 
\end{equation}
then actually one has 
\begin{equation} \label{eq3-15-02-2020}
\frac{1+|r-t|}{r} |\del\del u|\lesssim |\Boxt_g u| + t^{-1} |\del u|_{1,1}
\quad \text{ in } \Mscr^{\near}_{\ell, [s_0,s_1]}. 
\end{equation}
\end{lemma}

\begin{proof} Within $\Mscr^{\near}_{\ell,[s_0,s_1]} $, from \eqref{eq1-17-07-2021} we have $|\del_a\del_t u| + |\del_a\del_b u|\lesssim |\del_t\del_t u| + t^{-1} |\del u|_{1,1}$, so we can focus our attention on the component $\del_t\del_t u$ of $\del\del u$. We write 
$$
\aligned
g^{\N00} & = H^{\N00} = H^{00} - 2(x^a/r)H^{a0} + (x^ax^b/r^2)H^{ab},
\qquad\qquad
g^{ab} = \delta^{ab} + H^{ab},
\endaligned
$$
and Lemma~\ref{lem1-06-12-2020} gives us 
\begin{equation} \label{eq5-07-12-2020}
\aligned
\Boxt_g u + \frac{2(r+t)}{t^2}\del_t\del_t u
& = \Big(\frac{(2+r-t)(r+t)}{t^2} + 2\sum_a(x^a/r)\frac{t-r}{t}H^{a0} + \sum_{ab}(x^ax^b/r^2)\big((r/t)^2-1\big)H^{ab}\Big)\del_t\del_t u  
\\
& \quad + t^{-1}\Kbb_g^{\ME}[u] + H^{\N00}\del_t\del_t u.
\endaligned
\end{equation}
When \eqref{eq2-28-12-2020} holds, we observe that $2+r-t\geq 1+|r-t|$ in $\MME_{[s_0, s_1]}$ and we arrive at \eqref{eq1-28-12-2020}. The inequality \eqref{eq3-15-02-2020} then follows immediately.
\end{proof}

We have a similar conclusion at arbitrary order, as follows.  

\begin{proposition}[Hessian for the wave equation near the light cone: arbitrary order]
\label{prop1-22-05-2020}
Consider a metric $g^{\alpha\beta} = g_\Mink^{\alpha\beta} + H^{\alpha\beta}$ defined in $\Mscr^{\near}_{\ell, [s_0,s_1]}$ and satisfying \eqref{eq2-28-12-2020} for some $\eps_1 \ll 1$. 
Then, for any function $u$ defined in $\Mscr^{\near}_{\ell, [s_0,s_1]} $ one has 
\begin{equation} \label{eq8-04-06-2020}
\frac{1+|r-t|}{r} |\del\del u|_{p,k} \lesssim |\Boxt_g u|_{p,k} + t^{-1} |\del u|_{p+1,k+1} 
+ \sum_{\ord(Z) \leq p \atop \rank(Z) \leq k} \big| [Z,H^{\alpha\beta} \del_{\alpha} \del_{\beta}]u \big|
+ |H^{\N00} ||\del\del u|_{p,k}
\end{equation}
and, provided \eqref{eq1-15-02-2020} also holds, one has  
\begin{equation} \label{eq3-28-12-2020}
\frac{1+|r-t|}{r} |\del\del u|_{p,k} \lesssim |\Boxt_g u|_{p,k} + t^{-1} |\del u|_{p+1,k+1} 
+ \sum_{\ord(Z) \leq p \atop \rank(Z) \leq k}  \big| [Z,H^{\alpha\beta} \del_{\alpha} \del_{\beta}]u \big|. 
\end{equation}
The commutators in \eqref{eq8-04-06-2020} and \eqref{eq3-28-12-2020} are bounded in view of  \eqref{eq4-12-02-2020}.
\end{proposition}

\begin{proof}
Within $\Mscr^{\near}_{\ell, [s_0,s_1]}$, in the inequality \eqref{eq1-28-12-2020} we can replace $u$ by $Z u$ (with $\ord(Z) \leq p$ and $\rank(Z) \leq k$) and obtain
$$
\aligned
\frac{1+|r-t|}{t} |\del\del Z u|
& \lesssim  |\Boxt_g Z u| + t^{-1} |\del Z u|_{1,1} + |H^{\N00}\del_t\del_tZu|
\\
& \leq  | Z \Boxt_g u| + t^{-1} |\del u|_{p+1,k+1} 
+ |[Z,H^{\alpha\beta} \del_{\alpha} \del_{\beta}]u|
+ |H^{\N00} ||\del_t\del_t Z u|. 
\endaligned
$$
By recalling the second property in \eqref{equa2-2-juin} and the fact that $r\simeq t$ in $\Mscr^{\near}_{\ell, [s_0,s_1]}$, \eqref{eq8-04-06-2020} is established. On the other hand, \eqref{eq3-28-12-2020} is a
direct consequence of \eqref{eq8-04-06-2020} and \eqref{eq1-15-02-2020}.
\end{proof} 

%------------------------------------------------------------------------ 

\paragraph{Control away from the light cone.}

We now treat the domain $\Mscr^{\far}_{\ell,[s_0,s_1]}$ in which  $|r-t|\geq \ell r$. 

\begin{lemma}[Hessian for the wave equation away the light cone: zero-order]
\label{lem1-21-02-2020}
Consider a metric $g^{\alpha\beta} = g_\Mink^{\alpha\beta} + H^{\alpha\beta}$ defined in $\Mscr^{\far}_{\ell,[s_0,s_1]}$ and satisfying, for some $\eps_1\ll \ell$,  
\begin{equation} \label{eq7-15-02-2020}
|H| \leq \eps_1.
\end{equation} 
Then, for any function $u$ defined in $\Mscr^{\far}_{\ell,[s_0,s_1]}$ one has  
\begin{equation} \label{eq1-30-01-2021}
|\del\del u|\lesssim  (1 + t \, \crochet^{-1}) \, \big(|\Boxt_g u| + t^{-1} |\del u|_{1,1}\big).
\end{equation}
\end{lemma}

\begin{proof} Thanks to \eqref{lem1-21-02-2020}, in the domain $\Mfar_{\ell,[s_0,s_1]}$ we have $(r/t)|H|\ll \ell(r/t)^2\leq (r/t)^2-1$. Thanks to \eqref{eq7-15-02-2020},  the decomposition in Lemma~\ref{lem1-06-12-2020} together with $\big|\Kbb_g^{\ME}[u]\big| \lesssim |g||\del u|_{1,1}$ leads us to
\begin{equation} \label{eq3-07-12-2020}
|\del_t\del_t u|\lesssim t^2r^{-1}\crochet^{-1} |\Boxt_g u| 
+ r^{-1}(1+t \, \crochet^{-1}) \, |\del u|_{1,1}.
\end{equation}
On the other hand, in view of \eqref{eq7-15-02-2020} and Lemma~\ref{eq1-07-12-2020} together with $\big|\Dbb_g^{\ME}[u]\big| \lesssim \big(r^{-1} |g| + t^{-1} |H|\big) \, |\del u|_{1,1}$ therein,  we obtain 
$$
|\del_r\del_r u|\lesssim |\Boxt_g u| + \big(1+(r/t)|H|\big) \, |\del_t\del_t u| + r^{-1}\big(1 + (r/t)|H|\big) \, |\del u|_{1,1}.
$$
Substituting \eqref{eq3-07-12-2020} in the above inequality, we find 
$$
|\del_r\del_r u|\lesssim \Bigg(1 + \frac{t^2}{r \, \crochet} + \frac{t|H|}{\crochet}\Bigg) \, |\Boxt_g u|
+ r^{-1}\big(1+(r/t)|H|\big)\big(1 + t \, \crochet^{-1}\big) \, |\del u|_{1,1}, 
$$
while substituting the above inequality into \eqref{eq1-18-02-2020} gives us 
$$
|\del_a\del_b u|\lesssim \Bigg(1 + \frac{t^2}{r \, \crochet} + \frac{t|H|}{\crochet}\Bigg) \, |\Boxt_g u|
+ r^{-1}\big(1+(r/t)|H|\big)\big(1 + t \, \crochet^{-1}\big) \, |\del u|_{1,1}.
$$
In view of \eqref{eq3-07-12-2020}, from \eqref{decompo-at'-near} we finally deduce that  
$
|\del_t\del_a u|\lesssim t \, \crochet^{-1} |\Boxt_ g u| + \big(t^{-1} + \crochet^{-1}\big) \, |\del u|_{1,1} 
$, 
and we arrive at \eqref{eq1-30-01-2021}.
\end{proof}

\begin{proposition}[Hessian for the wave equation away from the light cone: arbitrary order]
\label{propo2-22-05-2020}
Under the assumption in Lemma~\ref{lem1-21-02-2020},  one has the pointwise Hessian inequality in $\Mfar_{\ell,[s_0,s_1]}$
(where the commutator is bounded thanks to \eqref{eq5-12-02-2020}): 
$$
\aligned
|\del\del u|_{p,k} & \lesssim   (1 + t \, \crochet^{-1}) \big(|\Boxt_g u|_{p,k} + t^{-1} |\del u|_{p+1,k+1}\big) 
+ \sum_{\ord(Z) \leq p\atop \rank(Z) \leq k} |[Z,H^{\alpha\beta} \del_\alpha \del_{\beta}]u|. 
\endaligned
$$
\end{proposition}

\begin{proof} We replace $u$ by $Z u$ in \eqref{eq1-30-01-2021} and obtain 
$$
\aligned
|\del\del Z u|
& \lesssim (1 + t \, \crochet^{-1})\big(|\Boxt_g Z u| +  t^{-1} |\del Z u|\big)
\\
& \lesssim   (1 + t \, \crochet^{-1})\big(| Z \Boxt_g u| + t^{-1} |\del u|_{p+1,k+1}\big)
+ |[Z, H^{\alpha\beta} \del_\alpha \del_{\beta}]u|
\\
& \lesssim   (1 + t \, \crochet^{-1})\big(|\Boxt_g u|_{p,k} +  t^{-1} |\del u|_{p+1,k+1}\big) + |[Z,H^{\alpha\beta} \del_\alpha \del_{\beta}]u|
\endaligned
$$
and, in view of \eqref{equa2-2-juin}, the desired bound is established. 
\end{proof} 

By taking into account the fact that $\crochet\geq \ell r$ in $\Mfar_{\ell,[s_0,s_1]}$, we conclude with the following result. 

\begin{corollary} With the notation in Proposition~\ref{propo2-22-05-2020}, from the inequality \eqref{eq1-30-01-2021} one has 
\begin{subequations}
\begin{equation} \label{eq3-21-02-2020}
|\del\del u|\lesssim \ell^{-1} |\Boxt_g u| + \ell^{-1}t^{-1} |\del u|_{1,1}, 
\end{equation}
while from Proposition~\ref{propo2-22-05-2020} one finds 
\begin{equation}
|\del\del u|_{p,k} 
\lesssim   \ell^{-1} \big(|\Boxt_g u|_{p,k} + t^{-1} |\del u|_{p+1,k+1}\big) 
+ \sum_{\ord(Z) \leq p\atop \rank(Z) \leq k} |[Z,H^{\alpha\beta} \del_\alpha \del_{\beta}]u|. 
\end{equation}
\end{subequations}
\end{corollary}

%-------------------------------------------------------------------------------------------------------------

\subsection{Gradient estimates of wave fields}
\label{section----83}

\paragraph{Operator of interest.}

We now consider the gradient of solutions to wave equations, and the technique now is more involved and relies on an integration along suitable characteristics.
We are going to rely on the weighted identity 
\begin{equation} \label{eq3'-17-01-2021}
-r(r-t+2)^\rho  \Box u = (\del_t+\del_r)\big((r-t+2)^\rho  \big(\del_t- \del_r \big)\big)(r u) - r (r-t+2)^\rho \sum_{a<b} \big(r^{-1} \Omega_{ab} \big)^2 u,
\end{equation}
in which we have used that $(r-t+2)$ commutes with $\del_t+\del_r$. We rewrite the semi-null decomposition \eqref{decompo-H-ext} as follows:
$$
\aligned
H^{\alpha\beta} \del_{\alpha} \del_{\beta}u 
& =  \HN^{00} \del_tu \del_tu + 2 \HN^{a0} \del_t\delsN_au + \HN^{ab} \delsN_a\delsN_b u + H^{\alpha\beta} \del_{\alpha} \big(\PsiN_{\beta}^{\beta'} \big)\delN_{\beta'}u
\\
&
=: \HN^{00} \del_t\del_t u + \slashed{H}^{\N}[u].
\endaligned
$$
The decomposition  
\begin{equation} \label{eq1-19-06-2020}
\aligned
r\del_t\del_t u 
& =  \frac{1}{4}(\del_t-\del_r)(\del_t-\del_r)(ru) - \frac{1}{4}r\Big((x^a/r)\delsN_a\Big)\Big((x^b/r)\delsN_b\Big)u 
+ x^a\del_t\delsN_au + \frac{1}{2}(\del_t-\del_r)u
\\
& =: 
\frac{1}{4}(\del_t-\del_r)(\del_t-\del_r)(ru) + X^{\N}[u]
\endaligned
\end{equation}
leads us to 
$$
-rH^{\alpha\beta} \del_{\alpha} \del_{\beta}u = -\frac{1}{4} \HN^{00}(\del_t-\del_r)^2(ru) - \HN^{00}X^{\N}[u] - r\slashed{H}^{\N}[u] 
$$
and, by including the weight function $(r-t+2)^\rho $, 
\begin{equation}
\aligned
-r(r-t+2)^\rho H^{\alpha\beta}\del_{\alpha}\del_{\beta}u 
& =  -\frac{1}{4}H^{\N 00}(\del_t-\del_r)\big((r-t+2)^\rho (\del_t-\del_r)(ru)\big)
- \frac{\rho}{2}H^{\N 00}(r-t+2)^{\rho-1}(\del_t - \del_r)(ru) 
\\
& \quad -  (r-t+2)^\rho \big(H^{\N 00}X^{\N}[u] + r\slashed{H}^{\N}[u]\big). 
\endaligned
\end{equation}
It is clear that
\begin{equation} \label{eq6-19-06-2020}
|X^{\N}[u]|\lesssim r|\del\delsN u| + |\del u|,
\qquad 
|\slashed{H}^{\N}[u]|\lesssim |H | \, | \del\delsN u| + r^{-1} |H | \, | \del u|. 
\end{equation}
Combining this observation with the decomposition of the wave operator in \eqref{eq3'-17-01-2021}, 
with an arbitrary $\rho$ we have 
\begin{equation}
\aligned
-r(r-t+2)^\rho \Boxt_g u 
=
&  \big((\del_t+\del_r) - \big(\HN^{00}/4\big)(\del_t-\del_r)\big)\big((r-t+2)^\rho (\del_t-\del_r)(ru)\big) 
\\
&
- \frac{\rho \HN^{00}}{2(r-t+2)}(r-t+2)^\rho (\del_t-\del_r)(ru)
\\
&  -  r(r-t+2)^\rho \sum_{a<b}(r^{-1} \Omega_{ab})^2u
- (r-t+2)^\rho \big(H^{\N 00}X^{\N}[u] + r\slashed{H}^{\N}[u]\big). 
\endaligned
\end{equation}

In conclusion, setting $\Pbf^{\N}_H := \del_t + \frac{4+\HN^{00}}{4-\HN^{00}} \del_r$, we arrive at our 
{\sl key decomposition within the Euclidean-merging domain}: 
\begin{equation} \label{eq1-05-08-2020}
\aligned
& \Big(\Pbf^N_H - \frac{2 \rho \HN^{00}}{(r-t+2)(4-\HN^{00})}\Big) 
\big((r-t+2)^\rho (\del_t-\del_r)(ru)\big)
\\
& = \frac{4r(r-t+2)^\rho}{4-\HN^{00}}\Big(-\Boxt_gu + \sum_{a<b}(r^{-1}\Omega_{ab}u)^2u + r^{-1}\HN^{00}X^{\N}[u] + \slashed{H}^{\N}[u]\Big). 
\endaligned
\end{equation}
This is an ordinary differential equation along the integral curves of the vector field $\Pbf^{\N}_H$, and we thus now study their global geometry.
Interestingly, the term $- \frac{2 \rho \HN^{00}}{(r-t+2)(4-\HN^{00})}$ 
will turn out to have a {\sl favorable sign} in our analysis, thanks to the assumption (made below)  $\HN^{00} \leq 0$.  

%----------------------------------------------------- 

\paragraph{Geometry of the characteristic curves.}

Denote by $\varphi_{t,x}(\tau) = (\tau, \varphi^a(\tau;t,x))\in \MME_{[s_0,s_1]}$ the integral curve of $\Pbf^{\N}_H$ satisfying $\varphi_{t,x} |_{\tau = t} = (t,x^a)$. For convenience we introduce
\begin{equation}
\Lscr_{\ell, [s_0,s_1]} := \big\{r = t/(1-\ell) \big\} \cap \MME_{[s_0,s_1]}. 
\end{equation}
The proof of the following statement is postponed to Appendix~\ref{Annex-section-8}.

\begin{lemma}[Geometry of the integral curves]\label{lem1-05-08-2020}
Consider a metric $g^{\alpha\beta} = g_\Mink^{\alpha\beta} + H^{\alpha\beta}$ defined in $\Mscr^{\near}_{\ell, [s_0,s_1]}$ and satisfying 
\begin{equation} \label{eq1'-10-01-2021}
|\HN^{00} |\ll 1,
\qquad
\HN^{00} \leq 0 \quad \text{ in } \Mscr^{\near}_{\ell, [s_0,s_1]}. 
\end{equation}
Given a point $(t,x)\in \Mscr^{\near}_{\ell, [s_0,s_1]}$,  consider the integral curve  $\varphi_{t,x}$ associated with  $\Pbf^{\N}_H$ and passing through that point.  Along this curve, there exists a unique time $t_0 \in [s_0, t]$ such that the following properties hold. 

\begin{itemize} 

\item[(1)] The integral curve remains in the near-light-cone domain, namely $\big\{\varphi_{t,x}(\tau) \, / \, t_0\leq \tau\leq t \big\} \subset \Mnear_{\ell,[s_0,s]}$. 

\item[(2)] The initial point $\varphi_{t,x}(t_0)$ lies on the hypersurface $\Mnear_{\ell,s_0}$ or on the cone $\Lscr_{\ell, [s_0,s_1]}$. 

\item[(3)] For each $s'\in[s_0,s]$, the curve $\varphi_{t,x}$ intersects $\Mnear_{\ell,s'}$ exactly once. 

\end{itemize} 
\end{lemma}

It remains to integrate \eqref{eq1-05-08-2020} (from $t_0$ to $t$) along the integral curve
and use  that the initial data is bounded in sup-norm in $\Mnear_{\ell,s_0} \cup \Lscr_{\ell, [s_0,s_1]}$,
while using \eqref{eq6-19-06-2020} for the control of the source term.  

\begin{proposition}[Weighted pointwise estimate in the Euclidean-merging domain] 
\label{prop1-23-07-2020}
Consider a metric $g^{\alpha\beta} = g_\Mink^{\alpha\beta} + H^{\alpha\beta}$ defined in $\Mscr^{\near}_{\ell, [s_0,s_1]}$ and satisfying \eqref{eq1'-10-01-2021}. 
Given any $\rho \geq 0$, for any function $u$ defined $\Mscr^{\near}_{\ell, [s_0,s_1]}$ one has
$$
\aligned
& \crochet^\rho |(\del_t-\del_r) (ru)|(t,x)
\\
& \lesssim
\sup_{\Omega^{\ell}_{s_0,s_1}} (r-t+2)^\rho \big( r \, |\del u| + |u|\big) 
+  \int_{t_0}^t  \crochet(\tau,r)^\rho \, r
\Big(
r^{-1} |\delsN u|_{1,1} +  |H | \, | \del\delsN u| + r^{-1} |H | \, | \del u| 
+ |\Boxt_g u| \Big)\big|_{\varphi_{t,x}(\tau)} d\tau, 
\endaligned
$$
in which the supremum is taken over the set $\Omega_{s_0,s_1}^{\ell} = \Lscr_{\ell, [s_0,s_1]}\cup \Mnear_{\ell,s_0}$. 
\end{proposition}

%==============================================================================

\section{Pointwise decay of Klein-Gordon fields and their derivatives} 
\label{section-9-added} 

\subsection{Quasi-linear commutators in the Euclidean-merging domain}

\paragraph{Main statement for this section.}

Our next result is based on the following observation.

\begin{lemma}[Decomposition of quasi-linear terms for Klein-Gordon fields]
\label{lem1-14-03-2021}
For every solution to $\Boxt_g\phi - c^2\phi = f$ and provided $|H|\ll 1$ within $\MME_{[s_0,s_1]}$, with the notation
$H^\rr : =(x^ax^b/r^2)H^{ab}$ one has 
$$ 
H^{\alpha\beta}\del_{\alpha}\del_{\beta}\phi 
 =  \frac{{ H^{\N00}}}{1+H^{\rr}} \del_t\del_t\phi +  \frac{2H^{a0}}{1+H^{\rr}}{ \delsN_a}\del_t\phi + \frac{H^\rr}{1+H^{\rr}} \big(c^2\phi + f\big)   + \frac{H^{ab} - H^{\rr}g_{\Mink}^{ab}}{r(1+H^{\rr})} \, \Dbb_{ab}[\phi]. 
$$  
\end{lemma}

\begin{proof} We begin with the decomposition 
\begin{subequations}
\begin{equation} \label{eq3-14-03-2021}
H^{\alpha\beta}\del_{\alpha}\del_{\beta} \phi 
= H^{00} \del_t\del_t\phi + 2H^{a0}\del_a \del_t\phi + H^{ab}\del_a\del_b\phi. 
\end{equation}
For the latter term we recall \eqref{eq1-18-02-2020} and obtain
\begin{equation} \label{eq5-14-03-2021}
H^{ab}\del_a\del_b \phi = H^{\rr}\del_r\del_r \phi + r^{-1}H^{ab}\Dbb_{ab}[\phi].
\end{equation}
On the other hand, we use the equation $\Boxt_g\phi - c^2\phi = f$ as follows. Recalling \eqref{eq1-18-02-2020}, we have
$$
\Boxt_g \phi = (-1+H^{00})\del_t\del_t \phi + (1+H^{\rr}){\del_r\del_r} \phi + 2H^{a0}\del_t\del_a\phi + r^{-1}g^{ab}\Dbb_{ab}[\phi], 
$$
which leads us to (thanks to the fact that $|H|\ll 1$)
$$
\del_r\del_r\phi = \frac{1-H^{00}}{1+H^{\rr}}\del_t\del_t\phi - \frac{2H^{a0}}{1+H^{\rr}}\del_t\del_a\phi + (1+H^{\rr})^{-1}\big(c^2\phi - r^{-1}g^{ab}\Dbb_{ab}[\phi] + f \big).
$$
By substituting this identity (together with \eqref{eq5-14-03-2021}) into \eqref{eq3-14-03-2021}  and recalling that { $H^{\N00} = H^{00} - 2(x^a/r)H^{a0} + H^{rr}$}, the proof is completed.
\end{subequations}
\end{proof}

Based on this identity, we establish the following result. 

\begin{proposition}[Estimate on quasi-linear commutator of Klein-Gordon equation]
\label{prop1-28-03-2021}
Under the condition of Lemma~\ref{lem1-14-03-2021} in which one takes
$f \equiv 0$ and 
\begin{equation}
\sum_{\circledast \in \{ \rr, 00, 0a \}} |H^\circledast|_{[p/2]}\ll 1, 
\end{equation}
the following estimate holds for all $Z$ with $\ord(Z)=p$ and $\rank(Z)=k$: 
\begin{subequations}
\begin{equation} \label{eq6-14-03-2021}
\aligned
|[Z,H^{\alpha\beta}\del_{\alpha}\del_{\beta}]\phi| 
& \lesssim  W_{p,k}^\hard  + W_{p,k}^\easy,  
\\
W_{p,k}^\hard 
&:=  \sum_{{\circledast} \in \{ \rr, 00, 0a \}} |H^\circledast||\del\del \phi|_{p-1,k-1} 
+ \sum_{k_1+p_2=p\atop k_1+k_2=k}  \sum_{\circledast \in \{ \rr, 00, 0a \}} 
| \LOmega  H^\circledast|_{k_1-1}\big(|\del\del \phi|_{p_2,k_2} +|\phi|_{p_2,k_2}\big)  
\\
W_{p,k}^\easy  :=  &{ \sum_{p_1+p_2=p\atop k_1+k_2=k} \Big( \frac{\la r-t\ra}{r}|\del H|_{p_1-1,k_1} +|\del H^{\N00}|_{p_1-1,k_1}\Big)\big(|\del\del\phi|_{p_2,k_2} + |\phi|_{p_2,k_2}\big)
}
\\
& + r^{-1} |H||\del \phi|_p 
+ r^{-1}\!\!\!\!\sum_{0\leq p_1\leq p-1}\!\!\!|H|_{p_1+1} |\del \phi|_{p-p_1}
 { + \sum_{{p_1+p_2+p_3=p\atop k_1+k_2+k_3=k}\atop{p_1> [p/2]}}|\del H^{rr}|_{p_1,k_1}|\del\del\phi|_{p_2,k_2}|H|_{p_3,k_3}}. 
\endaligned
\end{equation} 
\end{subequations}
\end{proposition}

%---------------------------------------------------------

\paragraph{Proof of the main statement.}

In order to prove this result, we establish two technical lemmas. 

\begin{lemma}
For each choice of ${ \circledast} \in \{ \rr, { \N00}, a0 \}$ and  under the condition 
$|H^\rr |_{[p/2]} +  |H^\circledast|_{[p/2]} \ll 1$ in $\MME_{[s_0,s_1]}$, one has 
\begin{equation} \label{eq2-11-07-2021}
\big|(1+H^{\rr})^{-1} H^\circledast\big|_{p,k}\lesssim 
|H^{\circledast}|_{p,k} + |H^{\rr} |_{p,k}.
\end{equation}
%\begin{equation}\label{eq1-26-05-2023}
% 
%\big|\del \big((1+H^{rr})^{-1}H^{\circledast}\big)\big|_{p,k}
%\lesssim |\del H^{\circledast}|_{p,k} 
%+ \sum_{p_1+p_2=p\atop k_1+k_2=k}
%\big(|H^{rr}|_{p_1,k_1}(|\del H^{\circledast}|_{p_2,k_2} + |\del H^{rr}|_{p_2,k_2})
% + |\del H^{rr}|_{p_1,k_1}|H^{\circledast}|_{p_2,k_2}\big).
%\end{equation}
\end{lemma}

\begin{proof} This is checked by a direct application of Lemma~\ref{lem 1 faa}. We introduce the function $f: (-1/2,1/2) \mapsto \RR$ by $f(\rho) := (1+\rho)^{-1}$. Then for all $\ord(Z) \leq p$ and $\rank(Z) \leq k$ we find 
$$
Z\big((1+H^{\rr})^{-1}\big) \cong \sum_{i=1}^pf^{(i)}(H^{\rr})\!\!\!\!\!\!\!\!\sum_{Z_1\odot\ldots\odot Z_i=Z}Z_1H^{\rr}Z_2H^{\rr}\ldots Z_iH^{\rr}.
$$
Thanks to the condition $|H^{\rr} |_{[p/2]}\ll 1$, this leads us to 
$
\big|(1+H^{\rr})^{-1} \, \big|_{p,k}\lesssim |H^{\rr} |_{p,k}.
$
Then we find 
$$
Z\big((1+H^{\rr})^{-1}H^\circledast\big) \cong \sum_{Z_1\odot Z_2=Z} Z_1\big((1+H^{\rr})^{-1}\big)Z_2H^\circledast.
$$
When $\ord{Z_1} \leq [p/2]$, the right-hand side is bounded by $|H^\circledast|_{p,k}$. When $\ord(Z_2)\leq [p/2]$, it is bounded by $|H^{\rr} |_{p,k}$ and \eqref{eq2-11-07-2021} the desired estimate is thus reached.  
%{ The proof of \eqref{eq1-26-05-2023} is similar, we omit the detail.}
\end{proof}

\begin{lemma}
For any function $\phi$ defined in $\MME_{[s_0,s_1]}$ and any ordered operator $Z$ with $\ord(Z)= p$
and $\rank(Z) = k$, one has 
\begin{equation} \label{eq8-11-07-2021}
\aligned
&\,[Z,\del_{\alpha}\Omega_c]\phi \cong \sum_{|I'|=|I|\atop |J'|+|K'|\leq k}\!\!\!\!\! \sum_\beta 
\del_{\beta}\del^{I'}L^{J'}\Omega^{K'},
%\\
%&{ 
%\,[Z,L_a\del_{\alpha}]\phi\cong  \sum_{|I'|=|I|\atop |J'|+|K'|\leq k}\!\!\!\!\! \sum_\beta 
%\del_{\beta}\del^{I'}L^{J'}\Omega^{K'}.
%}
\endaligned
\end{equation}
\end{lemma}

\begin{proof}
\begin{subequations}
%	{  These two estimates are established in the similar way. We only write the proof of the first in detail.}
For $Z = \del^IL^J\Omega^K$ we have 
\begin{equation} \label{eq4-11-07-2021}
[\del^IL^J\Omega^K,\del_{\alpha}\Omega_c]
= [\del^IL^J\Omega^K,\del_{\alpha}]\Omega_c + \del_{\alpha}\big([\del^IL^J\Omega^K,\Omega_c]\big).
\end{equation}
The first term in the right-hand side is decomposed via \eqref{eq 1 lem 1 depo-cH}. For the second term, we need to observe that
\begin{equation} \label{eq5-11-07-2021}
[\del^IL^J\Omega^K,\Omega_c] = \del^IL^J([\Omega^K,\Omega_c]) + [\del^IL^J,\Omega_c]\Omega^K.
\end{equation}
For the first term in the right-hand side of \eqref{eq5-11-07-2021}, we rely on the decomposition 
\begin{equation} \label{eq6-11-07-2021}
[\Omega^K,\Omega_c] \cong \sum_{|K'| = |K| \geq 1}\Omega^{K'}, 
\end{equation}
which can be checked by induction. Then we apply \eqref{eq6-11-07-2021} and \eqref{eq 4 comm} to \eqref{eq5-11-07-2021} and find
\begin{equation} \label{eq7-11-07-2021}
[\del^IL^J\Omega^K,\Omega_c]\cong \sum_{|K'|=|K| \geq 1}\del^IL^J\Omega^{K'} + \sum_{|I'|=|I|, |J'|=|J|\atop  |I|+|J|\geq 1}\del^{I'}L^{J'}\Omega^K. 
\end{equation}
Then we apply the above estimate, together with \eqref{eq 1 lem 1 depo-cH}, to the right-hand side of  \eqref{eq4-11-07-2021} and \eqref{eq8-11-07-2021} is thus established.
\end{subequations}
\end{proof}

\begin{proof}[Proof of Proposition~\ref{prop1-28-03-2021}] We commute $Z$ with the right-hand side of the identity in Lemma~\ref{lem1-14-03-2021}, 
and we then apply Proposition~\ref{lm 2 dmpo-cmm-H} { and Lemma~\ref{lem1-commu-ext}}. For controlling the first { two} terms in Lemma~\ref{lem1-14-03-2021},  we have 
$$
\aligned
\big|\big[Z,(1+H^{\rr})^{-1}H^\rr\big]\phi\big|& \lesssim  \sum_{k_1+p_2=p\atop k_1+k_2=k} |\LOmega H^\rr|_{k_1-1} |\phi|_{p_2,k_2}
+\sum_{p_1+p_2=p\atop k_1+k_2=k} |\del H^\rr|_{p_1-1,k_1} |\phi|_{p_2,k_2},
\\
\big|\big[Z,(1+H^{\rr})^{-1}H^{ \N00}\del_\alpha\del_t\big]\phi\big|
& \lesssim  | { H^*}||\del\del \phi|_{p-1,k-1} 
+ \hskip-.3cm 
\sum_{{k_1+p_2=p\atop k_1+k_2=k}} | \LOmega { H^*}|_{k_1-1} |\del\del \phi|_{p_2,k_2}
\\
&+\sum_{{p_1+p_2=p\atop k_1+k_2=k}}|\del  H^{\N00}|_{p_1-1,k_1} |\del\del\phi|_{p_2,k_2}
{
 + \sum_{{p_1+p_2+p_3=p\atop k_1+k_2+k_3=k}\atop{p_1> [p/2]}}|\del H^{rr}|_{p_1,k_1}|\del\del\phi|_{p_2,k_2}|H|_{p_3,k_3}}.
\endaligned
$$
{
 Here, we have $|H^*|_{p,k} := \max\{|H^{rr}|_{p,k},|H^{\N00}|_{p,k}, H^{a0}|_{p,k}\}$, and $|\LOmega H^*|_{p,k} := \max\{|\LOmega H^{rr}|_{p,k}, |\LOmega H^{\N00}|_{p,k}, |H^{a0}|_{p,k}\}$.}
 { 
 For the third term we rely on the calculations in \eqref{eq1'-12-02-2020} within $\Mnear_{[s_0,s_1]}$ and in \eqref{eq7-14-03-2021} within $\Mfar_{[s_0,s_1]}$, and write 
$$
|(1+H^{rr})^{-1}H^{a0}|_{p,k}\lesssim |H^*|_{p,k},\quad
\big|\LOmega\big((1+H^{rr})^{-1}H^{a0}\big)\big|_{p,k}\lesssim |\LOmega H^*|_{p,k},\quad
\big|\del\big((1+H^{rr})^{-1}H^{a0})\big|_{p,k}\lesssim |\del H|_{p,k}.
$$
}

For the last term in the identity of Lemma~\ref{lem1-14-03-2021}, we observe that, for all homogeneous coefficient $\Lambda$ of degree zero, 
\begin{equation} \label{eq3-11-07-2021}
\big|r^{-1} \lambda(1+H^{\rr})^{-1}H^{ab} \, \big|_{p,k}\lesssim r^{-1} |H|_{p,k}.
\end{equation}
We have 
$\Dbb_{ab} = \Lambda_{ab}^{\gamma c}\del_{\gamma}\Omega_c + \lambda_{ab}^{\gamma}\del_{\gamma}$,
where $\Lambda_{ab}^{\gamma c},\lambda_{ab}^{\gamma}$ are homogeneous functions of degree zero. Then $[Z,r^{-1}(1+H^{\rr})^{-1}H^{ab}\Dbb_{ab}]\phi$ is a finite linear combination of
$$
[Z, r^{-1} \lambda(1+H^{\rr})^{-1}H^{ab}\del_{\alpha}\Omega_c],\qquad\qquad 
[Z, r^{-1} \lambda(1+H^{\rr})^{-1}H^{ab}\del_{\alpha}].
$$
We only need to observe that 
\begin{equation} \label{eq3-12-07-2021}
\aligned
&\,[Z, r^{-1} \lambda(1+H^{\rr})^{-1}H^{ab}\del_{\alpha}\Omega_c]\phi 
\\
& =  \sum_{Z_1\odot Z_2=Z\atop \ord(Z_1)\geq 1}Z_1(r^{-1} \Lambda(1+H^{\rr})^{-1}H^{ab})Z_2\del_{\alpha}\Omega_c\phi 
+  r^{-1} \lambda(1+H^{\rr})^{-1}H^{ab}[Z,\del_{\alpha}\Omega_c]\phi
\\
& = \sum_{Z_1\odot Z_2=Z\atop \deg(Z_1)=0,\ord(Z_1)\geq 1}\!\!\!\!\!\!\!\!\!\!\!\!\!\!\!\!
L^{J_1}\Omega^{K_1}(r^{-1} \Lambda(1+H^{\rr})^{-1}H^{ab})Z_2\del_{\alpha}\Omega_c\phi 
+ \sum_{ Z_1\odot Z_2=Z\atop \deg(Z_1)\geq 1}\!\!\!\!\!
\del^{I_1}L^{J_1}\Omega^{K_1}(r^{-1} \Lambda(1+H^{\rr})^{-1}H^{ab})Z_2\del_{\alpha}\Omega_c\phi
\\
& \quad + r^{-1} \Lambda(1+H^{\rr})^{-1}H^{ab}[Z,\del_{\alpha}\Omega_c]\phi.
\endaligned
\end{equation}
Thanks to \eqref{eq3-11-07-2021}, the first term in the right-hand side is bounded by 
$$
r^{-1} |\LOmega H|_{k_1-1} |\del \phi|_{p_2+1,k_2+1} + r^{-1} |H| \, |\del\phi|_{p-1,k-1}
$$
with $k_1+p_2=p$ and $k_1+k_2=k$ and, obviously, we have $|H|_{0,0} = |H|$.

For the second term we distinguish between two cases. When there is at least one partial derivative distributed on $r^{-1}$, it is bounded by 
$$
r^{-2} |H|_{p_1-1,k_1} |\del\Omega \phi|_{p_2,k_2}\lesssim r^{-2} |H|_{p_1-1,k_1} |\del \phi|_{p_2+1,k_2+1}. 
$$
When there is no partial derivatives distributed on $r^{-1} \Lambda$, 
then there will be at least one acting on $(1+H^{\rr}) \, H^{ab}$. Then it is bounded by 
$$
r^{-1} |\del H|_{p_1-1,k_1} |\del\Omega\phi|_{p_2,k_2}\lesssim r^{-1} |\del H|_{p_1-1,k_1} |\del \phi|_{p_2+1,k_2+1}.
$$ We also recall that the last term in the right-hand side of \eqref{eq3-12-07-2021} is bounded by \eqref{eq8-11-07-2021}. Then we find 
$$ 
\big|[Z, r^{-1} \lambda(1+H^{\rr})^{-1}H^{ab}\del_{\alpha}\Omega_c]\phi \big|
\lesssim r^{-1} |H||\del \phi|_{p,k} 
+ r^{-1}\!\!\!\!\sum_{0\leq p_1\leq p-1}\!\!\!|H|_{p_1+1} |\del \phi|_{p-p_1} 
$$
and, in the same manner,
$$ 
\hfill 
\big|[Z, r^{-1} \lambda(1+H^{\rr})^{-1}H^{ab}\del_{\alpha}]\phi\big|\lesssim r^{-1} |H||\del \phi|_{p-1,k-1}
+  r^{-1}\!\!\!\!\sum_{0\leq p_1\leq p-1}\!\!\!|H|_{p_1+1} |\del \phi|_{p-p_1-1}. 
\hfill
$$ 
\qedhere
\end{proof}

%{ 
%On the other hand, we have the following decomposition near the light cone:
%\begin{lemma}[Decomposition of quasi-linear terms for Klein-Gordon fields near the light cone]
%For any function $\phi$ and metric $H^{\alpha\beta}$ defined in $\Mnear_{[s_0,s_1]}$, the following identity holds:
%$$
%\aligned
%H^{\alpha\beta}\del_{\alpha}\del_{\beta}\phi = &H^{\N00}\del_t\del_t\phi + (1-r/t)\Big(\frac{2x^a}{r}H^{a0} - \frac{x^ax^b}{r^2}(r/t+1)H^{ab}\Big)\del_t\del_t\phi
%\\
%&+t^{-1}\Big(2H^{a0}L_a\del_t + H^{ab}L_a\del_b - (x^a/t)H^{ab}L_b\del_t\Big)\phi.
%\endaligned
%$$
%\end{lemma}
%Based on this decomposition and follow a similar argument, we obtain
%\begin{proposition}
%...
%\end{proposition}
%}
%----------------------------------------------------------------------------------------------------------------------------------------------------------------------------------

\subsection{Decay of Klein-Gordon fields in the Euclidean-merging domain} 
\label{secti-20a}

Near the light cone, Proposition~\ref{lem 2 d-e-II} does not provide us  sufficient decay for our purpose. We take advantage of the Klein-Gordon structure and we {\sl control the mass term by the wave operator} and a source term. 

\begin{proposition}[Pointwise decay of Klein-Gordon fields]
\label{lem 1 d-KG-e}
Given any exponents $\eta, \lambda \in (0,1)$, for any  solution $v$ to $- \Box v + c^2 \, v = f$ defined in $\MME_{[s_0,s_1]}$ one has 
$$
{
c^2 \, |v|_{p,k} \lesssim 
\begin{cases}
r^{-2} \crochet^{1-\eta} \, \Fenergy_{\eta,c}^{\ME,p+4,k+4}(s,v) + |f|_{p,k}
&  \text{in }\Mnear_{[s_0,s_1]},
\\
r^{-1-\eta} \, \Fenergy_{\eta,c}^{\ME,p+2,k+2}(s,v)\quad 
& \text{in }\Mfar_{[s_0,s_1]}.
\end{cases}
}
$$
\end{proposition}

\begin{proof} In $\MMEnear_{[s_0,s_1]}$ we consider a solution $v$ to  
$- \Box v + c^2 \, v = f$ and start from the decomposition 
\begin{equation} \label{eq 1 d-KG-e}
\aligned
c^2 v 
& =   \big(r^2/t^2-1\big)\del_t \, \del_t v 
- t^{-1} \Big((2x^a/t)\del_tL_a - \sum_a\delsH_aL_a - (x^a/t)\delsH_a + \big(3+(r/t)^2 \big)\del_t\Big)v
+  f. 
\endaligned
\end{equation}
We write $c^2 \, |v| \lesssim t^{-1} |r-t| \, | \del\del v| + t^{-1} |\del v|_{1,1} + |f|$, and using this observation with $v$ replaced by $Zv$  
and recalling the ordering properties in Proposition~\ref{prop--fund-order}, 
we arrive at  
$$ 
c^2 \, |v|_{p,k} \lesssim t^{-1} |r-t| \, | \del v|_{p+1,k} + t^{-1} |\del v|_{p+1,k+1} + |f|_{p,k}. 
$$ 
Recalling the Sobolev inequality in Lemma~\ref{propo1-29-02-2020} 
together with the consequence \eqref{eq decay-v-repeat000-two} and 
substituting these bounds in the above inequality, we obtain 
$$
c^2 \, |v|_{p,k} \lesssim 
t^{-2} \crochet^{1-\eta} \, \Fenergy_{\eta,c}^{\ME,p+4,k+3}(s,v) + t^{-2} \crochet^{-\eta} \, \Fenergy_{\eta,c}^{\ME,p+4,k+4}(s,v) + |f|_{p,k}.
$$ 
This concludes the bound in $\Mnear_{[s_0,s_1]}$.
Finally, we again recall \eqref{eq decay-v-repeat000-two} which, in the far region $\MMEfar_s$, gives us the desired estimate.  
\end{proof}   

%=================================================================================

\

\part{The global nonlinear stability of self-gravitating massive fields}

\section{Global existence theory: PDEs formulation}
\label{section-new-11}

\subsection{Einstein's field equations in wave gauge}

\paragraph{Decomposition of the Ricci curvature.}

We now turn our attention to the Einstein equations and, before we can restate our main result in coordinates, we decompose these equations in the frames that are relevant for applying the Euclidean--hyperboloidal foliation framework presented in Part~I. We work in a global coordinate chart $(x^\alpha) =( t, x^a)$
and we introduce the contractions  
$\Gamma^\gamma := 
g^{\alpha \beta} \Gamma_{\alpha \beta}^\gamma$ and $ \Gamma_\alpha := 
g_{\alpha \beta} \Gamma^\beta$
of the corresponding Christoffel symbols. 
\begin{subequations}
\label{equa-the-system}
It is well-known that the Ricci curvature depends upon (up to) second-order derivatives of the metric $g$ and elementary (but tedious) calculations lead us to (cf.~for instance \cite{PLF-YM-two}) 
\begin{equation} \label{equa:sec8-01} 
2 \, R_{\alpha\beta}
= - g^{\mu\nu} \del_\mu \del_\nu g_{\alpha\beta} 
+  \Fbb_{\alpha\beta}(g,g;\del g, \del g) 
+ \big(\del_\alpha\Gamma_\beta + \del_\beta\Gamma_\alpha\big) 
+  W_{\alpha\beta},
\end{equation}
where
\begin{equation} \label{equa:sec8-02} 
\aligned
W_{\alpha\beta} 
& := g^{\delta \delta'} \del_{\delta} g_{\alpha \beta} \Gamma_{\delta'} - \Gamma_\alpha \Gamma_\beta,
\qquad
\Fbb_{\alpha\beta}(g,g;\del g, \del g) 
:= \Pbb_{\alpha\beta}(g,g;\del g, \del g)  + \Qbb_{\alpha\beta}(g,g;\del g, \del g). 
\endaligned
\end{equation}
Here, the quadratic nonlinearities $\Fbb_{\alpha\beta}$ are decomposed into {\bf  
quasi-null quadratic forms} (as we call them) 
\begin{equation} \label{equa:sec8-03} 
\Pbb_{\alpha\beta} (g,g;\del g, \del g) 
:= - \frac{1}{2} g^{\mu\mu'} g^{\nu\nu'} \del_\alpha g_{\mu\nu} \del_\beta g_{\mu'\nu'} + \frac{1}{4} g^{\mu\mu'} g^{\nu\nu'} \del_\alpha g_{\mu\mu'} \del_\beta g_{\nu\nu'}
\end{equation}
and {\bf null quadratic forms}
\begin{equation} \label{equa:sec8-04} 
\aligned
\Qbb_{\alpha\beta}(g,g;\del g, \del g) 
& := 
g^{\mu\mu'} g^{\nu\nu'} \del_\mu g_{\alpha\nu} \del_{\mu'} g_{\beta\nu'}
- g^{\mu\mu'} g^{\nu\nu'} \big(\del_\mu g_{\alpha\nu'} \del_\nu g_{\beta\mu'} - \del_\mu g_{\beta\mu'} \del_\nu g_{\alpha\nu'} \big)
\\
& \quad + g^{\mu\mu'} g^{\nu\nu'} \big(\del_\alpha g_{\mu\nu} \del_{\nu'} g_{\mu'\beta} - \del_\alpha g_{\mu'\beta} \del_{\nu'} g_{\mu\nu} \big)
+ \frac{1}{2} g^{\mu\mu'} g^{\nu\nu'} \big(\del_\alpha g_{\mu\beta} \del_{\mu'} g_{\nu\nu'} - \del_\alpha g_{\nu\nu'} \del_{\mu'} g_{\mu\beta} \big)
\\
& \quad + g^{\mu\mu'} g^{\nu\nu'} \big(\del_\beta g_{\mu\nu} \del_{\nu'} g_{\mu'\alpha} - \del_\beta g_{\mu'\alpha} \del_{\nu'} g_{\mu\nu} \big)
+ \frac{1}{2} g^{\mu\mu'} g^{\nu\nu'} \big(\del_\beta g_{\mu\alpha} \del_{\mu'} g_{\nu\nu'} - \del_\beta g_{\nu\nu'} \del_{\mu'} g_{\mu\alpha} \big). 
\endaligned
\end{equation} 
\end{subequations}
Since these expressions are quadratic in $(g,g)$ as well as in $(\del g, \del g)$, we can apply the standard polarization argument and define the corresponding symmetric bilinear forms.
More precisely, let $\mathbb{A}$ be a quadratic form (which can be $\Fbb$, $\Pbb$ or $\Qbb$), with a bit abuse of notation, we define
$$
\Abb(\del u,\del v) := \frac{1}{4}\big(\Abb(\del u+\del v,\del u+\del v) - \Abb(\del u- \del v,\del u -\del v)\big).
$$
Such a notation will be useful later on in this section when we expand a solution near a given reference metric. For instance, by definition we have 
\begin{equation} \label{equa-Pbb-four-arguments} 
\aligned
\Pbb_{\alpha\beta} (g,g';\del g'', \del g''') 
& := 
\frac{1}{4}\big(\Pbb_{\alpha\beta} (g,g';\del g'' + \del g''',\del g'' + \del g''') 
- \Pbb_{\alpha\beta} (g,g';\del g'' - \del g''',\del g'' - \del g''')\big) 
\\
&= 
{ 
 - {1 \over 4} g^{\mu\rho} g'^{\nu\sigma}( \del_\alpha g''_{\mu\nu} \del_\beta g'''_{\rho\sigma}  + \del_\alpha g'''_{\mu\nu} \del_\beta g''_{\rho\sigma})
+ \frac{1}{8} g^{\mu\rho} g'^{\nu\sigma} (\del_\alpha g''_{\mu\rho} \del_\beta g'''_{\nu\sigma} + \del_\alpha g'''_{\mu\rho} \del_\beta g''_{\nu\sigma}).
} 
\endaligned
\end{equation}
A similar notation is used for $\Qbb$ but also for other quadratic forms defined below such as $\Fbb$.

The following properties of the Ricci curvature  are central in the forthcoming analysis. First of all, its second-order part reads $- g^{\mu\nu} \del_\mu \del_\nu g_{\alpha\beta} +  \big(\del_\alpha\Gamma_\beta + \del_\beta\Gamma_\alpha\big)$, in which 
$- g^{\mu\nu} \del_\mu \del_\nu g_{\alpha\beta}$ is a nonlinear wave operator, 
since the metric $g$ has Lorentzian signature and, in fact, by our assumptions will remain close to the flat (Minkowski) metric. 
On the other hand, in presence of the additional terms
$\big(\del_\alpha\Gamma_\beta + \del_\beta\Gamma_\alpha\big)$ one would not be led to a hyperbolic operator, 
but hyperbololicity is achieved under the so-called {\bf wave gauge condition} 
\begin{equation} \label{eq2-27-05-2020} 
\Gamma^{\gamma} \equiv 0. 
\end{equation}
This condition is compatible with the evolution implied by the Einstein equations since, for instance as shown in \cite{YCB} Section 5.8, if the wave gauge condition holds on an initial Cauchy hypersurface, then it holds 
whenever the solution exists and remains sufficiently regular. These conditions also have important implications on the properties of quasi-null nonlinearities $\Pbb_{\alpha\beta}$, as we will show in the next section.

In view to the above observation, in order to deal with the Ricci curvature in coordinates satisfying the wave gauge conditions, 
it is convenient to introduce
\begin{equation} \label{eq1-16-07-2021}
\aligned
^{(w)}R_{\alpha\beta}: & =  R_{\alpha\beta} - \frac{1}{2} \Big(\del_{\alpha} \Gamma_{\beta} + \del_{\beta} \Gamma_{\alpha}) + W_{\alpha\beta} \Big)
=  - \frac{1}{2}g^{\mu\nu} \del_\mu \del_\nu g_{\alpha\beta}
+  \frac{1}{2} \Fbb_{\alpha\beta}(g,g;\del g, \del g). 
\endaligned
\end{equation}
By taking the wave gauge condition into account, this {\bf modified Ricci operator} $^{(w)}R_{\alpha\beta}$ provides us with a hyperbolic operator acting on each component $g_{\alpha\beta}$.  

%-----------------------------------------------------------------------------------------

\paragraph{Reference spacetime metric and perturbation.}

We are interested in solutions $g = g_{\alpha\beta}dx^\alpha dx^\beta $ that remain sufficiently close 
to a reference metric denoted by $g^\star  = g^\star_{\alpha\beta}dx^\alpha dx^\beta$, 
while the comparison with the Minkowski metric $g_\Mink = g_{\Mink,\alpha\beta}dx^\alpha dx^\beta$ 
will also be important in our analysis. Consequently, we will work with the 
following two decompositions: 
\begin{equation} \label{equa:sec8-11} 
g = g^\star  + u = g_\Mink + h^\star  + u, 
\end{equation}
in which $u = u_{\alpha\beta}dx^\alpha dx^\beta$ represents the perturbation of the reference metric $g^\star$,  
while $h^\star  + u$ denotes the perturbation of the flat solution.
In addition, denoting by $(g^{\alpha\beta})$ and $(g^{\star \alpha\beta})$ the inverse of $(g_{\alpha\beta})$ and $(g^\star_{\alpha\beta})$, respectively, we then {\sl define the new tensors} $u^{\alpha\beta}$ and $h^{\star \alpha\beta}$ by 
\begin{equation} \label{equa:sec8-13} 
h^{\star\alpha\beta} := g^{\star\alpha\beta} - g_{\Mink}^{\alpha\beta},\quad
\qquad
u^{\alpha\beta} := g^{\alpha\beta} - g_{\Mink}^{\alpha\beta} - h^{\star\alpha\beta} = g^{\alpha\beta} - g^{\star\alpha\beta}.
\end{equation}
Furthermore, the reduced wave operator is decomposed accordingly, namely 
\begin{equation} \label{equa:sec8-15} 
\aligned 
g^{\mu\nu} \del_\mu \del_\nu g_{\alpha\beta}  
& =  g^{\mu\nu} \del_\mu \del_\nu g^\star_{\alpha\beta}
+
g^{\mu\nu} \del_\mu \del_\nu u_{\alpha\beta} 
\\
& = g^{\star\mu\nu} \del_\mu \del_\nu g^\star_{\alpha\beta} 
+ g^{\mu\nu} \del_\mu \del_\nu u_{\alpha\beta}
+ u^{\mu\nu} \del_\mu \del_\nu g^\star_{\alpha\beta}. 
\endaligned
\end{equation}

\begin{lemma}[Raising indices] 
\label{lem-small}
Under the smallness condition $  | h^\star |_p + |u|_{[p/2]} \ll 1$, the inverse reference metric and perturbation satisfy 
$$
\max_{\alpha\beta} | h^{\star \alpha\beta} |_{p,k}
\lesssim | h^\star |_{p,k} = \max_{\alpha\beta} | h^\star_{\alpha\beta} |_{p,k}, 
\qquad 
\qquad 
\max_{\alpha\beta} | u^{\alpha\beta} |_{p,k}
\lesssim |u|_{p,k} = \max_{\alpha\beta} | u_{\alpha\beta} |_{p,k}.
$$  
\end{lemma}

\begin{proof} 
In view of the identity 
$
(g^{\star})^{-1} = (g_\Mink + h^{\star})^{-1} = g_\Mink + \sum_{m=1}^{+\infty} \big(- g_{\Mink} h^\star \big)^mg_{\Mink}
$, 
we obtain
{
$
h^{\star \alpha\beta} = \big((g^{\star})^{-1} - g_\Mink \big)_{\alpha\beta} 
= \big(\sum_{m=1}^{+\infty}(-g_{\Mink}h^\star)^mg_{\Mink}\big)_{\alpha\beta}.
$}
This series of functions is uniformly absolutely converging in the sup-norm up order $p$, that is, 
$$
\| (- g_{\Mink} h^\star)^m \|_{W^{p, +\infty}} \leq  c_{m,p} \| h^\star \|_{W^{p, +\infty}}^m = : c_{m,p} \xi^m 
$$
with $c_{m,p} \lesssim m^p$ 
and, in fact, is bounded by 
a series $\xi \mapsto \sum_m m^p  \xi^m$ whose radius of convergence is $1$. We can thus differentiate the series term by term. 
We can thus compute the derivatives of $h^{\star \alpha\beta}$ and apply the translation, boost, and rotation fields. 
In other words, applying $Z=\del^IL^J\Omega^K$ (with $|I|+|J| + |K|\leq p$ and $|J| + |K|\leq k$)
to each term in the series $\big(\sum_{m=1}^{+\infty}(- g_{\Mink}h^\star)^m\big)_{\alpha\beta}$ we find
$$
Z\big((- g_{\Mink}h^\star)^m\big) 
= (-1)^m  \hskip-.5cm
\sum_{|I_1|+\ldots |I_m| = |I|} 
\sum_{|J_1|+\ldots + |J_m| = |J|} 
\sum_{|K_1| +\ldots + |K_m| = |K|} 
\hskip-.cm
\prod_{i=1}^m g_{\Mink}^m\del^{I_i} L^{J_i} \Omega^{K_i} h^\star. 
$$ 
Each term is of degree $m \geq 1$ and is controlled by the product $m (m-1) \ldots (m-p) | h^\star |_{p,k} \, | h^\star |_{[p/2],k}^{m-1}$. Since the corresponding series in $m$ is converging when $| h^\star |_{[p/2],k} <1$ this yields the desired conclusion. For the second inequality we proceed similarly but from the identity $(g)^{-1} = (g^\star + u)^{-1} = \sum_{k=0}^{+\infty}\big((g^{\star})^{-1} u\big)^k (g^{\star})^{-1}$. 
Here, we require more information on $g^\star$, namely $  |h^{\star} |_p\ll 1$.
\end{proof}

\begin{remark} \label{rmk1-15-07-2021}
In the expression of $\Fbb_{\alpha\beta}$, the components with upper indices are undifferentiated. When we estimate such terms, we only need the smallness condition $|h^{\star} |_{p} + |u|_{[p/2]}\leq \eps_1\ll1$ in order to turn $|h^{\star\alpha\beta} |_{p,k}$ and $|u^{\alpha\beta} |_{p,k}$ into $|h^{\star} |_{p,k}$ and $|u|_{p,k}$, respectively (but we do not need $|\del h^{\star\alpha\beta} |_{p,k}$, nor derivatives beyond the ones in $|u|_{[p/2]}$). 
\end{remark}

%---------------------------------------------------------- 

\paragraph{Decomposition of the nonlinearities.}

Next, we return to \eqref{equa-the-system} and turn our attention to the nonlinear terms, which we further decompose as follows: 
\begin{subequations} \label{eq1-06-02-2022}
\begin{equation} \label{equa:sec8-18} 
\aligned
& \Fbb_{\alpha\beta}(g,g;\del g, \del g)
= \Fbb^\star_{\alpha\beta} [g^\star] + \Fbb^\star_{\alpha\beta} [u]
+ \Ibb^\star_{\alpha\beta} [u], 
\endaligned
\end{equation}
in which 
\begin{equation} \label{equa-b:sec8-18} 
\Fbb^\star_{\alpha\beta}[g^\star] := \Fbb_{\alpha\beta}(g^\star, g^\star;\del h^{\star}, \del h^{\star}),
\qquad
\Fbb^\star_{\alpha\beta} [u] := \Fbb_{\alpha\beta}(g^\star,g^\star;\del u, \del u) 
\end{equation}
and 
\begin{equation}
\Ibb^\star_{\alpha\beta}[u]:=  \Lbb^\star_{\alpha\beta}[u]   + \Bbb^\star_{\alpha\beta}[u] + \Cbb^\star_{\alpha\beta}[u], 
\end{equation}
where linear, bilinear, and cubic (and higher-order) interactions terms are defined, respectively, as  
\begin{equation} \label{eq4-06-01-2020}
\aligned
\Lbb^\star_{\alpha\beta}[u]
& := 
2 \, \Fbb_{\alpha\beta}(g^\star, g^\star; \del u, \del h^{\star})  
+  
{
\Fbb_{\alpha\beta}(u,g^\star;\del h^{\star}, \del h^{\star}) + \Fbb_{\alpha\beta}(g^\star,u;\del h^{\star}, \del h^{\star})}
\\
& = {
2 \, \Pbb_{\alpha\beta}(g^\star, g^\star; \del u, \del h^{\star}) 
+ 2\Qbb_{\alpha\beta}(g^\star, g^\star;\del u, \del h^{\star}) 
+ \Fbb_{\alpha\beta}(u,g^\star;\del h^{\star}, \del h^{\star}) 
+ \Fbb_{\alpha\beta}(g^\star,u;\del h^{\star}, \del h^{\star}) },
\\
\Bbb^\star_{\alpha\beta} [u]
& :=  
2 \, \Fbb_{\alpha\beta}(u,g^\star; \del u,\del h^{\star})  
+ 2 \, \Fbb_{\alpha\beta}(g^\star,u; \del u, \del h^{\star})  
+ \Fbb_{\alpha\beta}(u,u;\del h^{\star}, \del h^{\star}),
\\
\Cbb^\star_{\alpha\beta}  [u]
& :=   
{
\Fbb_{\alpha\beta}(u,g^\star;\del u, \del u) + \Fbb_{\alpha\beta}(g^\star,u;\del u, \del u)}  
+ 2 \, \Fbb(u,u; \del u, \del h^{\star})  
+ \Fbb_{\alpha\beta}(u,u;\del u, \del u).
\endaligned
\end{equation}
We also write
\begin{equation}\label{eq6-04-10-2022}
\aligned
&\Pbb^{\star}_{\alpha\beta}[u,v]:= \Pbb_{\alpha\beta}(g^{\star},g^{\star}; \del u,\del v),
\quad
&&\Qbb^{\star}_{\alpha\beta}[u,v]:= \Qbb_{\alpha\beta}(g^{\star},g^{\star}; \del u,\del v),
\\
&\Pbb^{\star}_{\alpha\beta}[u]:= \Pbb_{\alpha\beta}(g^{\star},g^{\star}; \del u,\del u),
\quad
&&\Qbb^{\star}_{\alpha\beta}[u]:= \Qbb_{\alpha\beta}(g^{\star},g^{\star}; \del u,\del u).
\endaligned
\end{equation}
\end{subequations}
We emphasize that, especially when $h^\star$ does not have a rapid decay at spacelike infinity, it might be better to think of $\Fbb_{\alpha\beta}(g^\star, g^\star;\del h^{\star}, \del u)$ (for instance) as being {\sl quadratic} in the two variables $(\del h^{\star}, \del u)$. Consequently, the modified Ricci curvature is decomposed as follows: 
\begin{equation} \label{equa-mainsys5}
\aligned
2 \, ^{(w)}R_{\alpha\beta}  
& =
2 \, R_{\alpha\beta} 
- \big(\del_\alpha\Gamma_\beta + \del_\beta\Gamma_\alpha\big) 
- W_{\alpha\beta} 
\\
& =  -\Boxt_g u_{\alpha\beta} 
- u^{\mu\nu} \del_\mu \del_\nu g^\star_{\alpha\beta}  
+ 2 \, \Rwave_{\alpha\beta}
+ \Fbb^{\star}_{\alpha\beta}[u] + \Ibb^{\star}_{\alpha\beta}[u], 
\endaligned
\end{equation}
in which the modified Ricci curvature of the reference spacetime metric reads 
\begin{equation} \label{eq1-08-01-2022}
2 \,  \Rwave_{\alpha\beta} 
= - g^{\star\mu\nu} \del_\mu \del_\nu g^\star_{\alpha\beta} + \Fbb^\star_{\alpha\beta}[g^\star]. 
\end{equation}  

%----------------------------------------------------------------------------------------------------------------------------------

\paragraph{Field equations in wave gauge.}

Finally, we formulate the Einstein equations in wave gauge by replacing the Ricci curvature by its modified version, that is, 
\begin{equation}
^{(w)}R_{\alpha\beta} = 8\pi T_{\alpha\beta} - 4\pi \, T \, g_{\alpha\beta}.
\end{equation}
In view of the expression \eqref{eq1-16-07-2021} of $^{(w)}R_{\alpha\beta}$, for the metric unknown we obtain 
\begin{equation} \label{eq1-30-11-2020} 
\Boxt_g g_{\alpha\beta} = \Fbb_{\alpha\beta}[g] - 8\pi \, (2 \, T_{\alpha\beta} - T \, g_{\alpha\beta})
\end{equation}
while the equations for the metric perturbation follow from \eqref{equa-mainsys5}:
\begin{equation} \label{eq 1 13-01-2019}
\aligned
\Boxt_g u_{\alpha\beta} 
& =  \Pbb^\star_{\alpha\beta}[u] + \Qbb^{\star}_{\alpha\beta}[u] 
- 8\pi \, (2 \, T_{\alpha\beta} - Tg_{\alpha\beta}) + 
\Big( \Ibb^{\star}_{\alpha\beta}[u] + 2 \, \Rwave_{\alpha\beta}
- u^{\mu\nu} \del_\mu \del_\nu g^\star_{\alpha\beta}  \Big). 
\endaligned
\end{equation} 
Moreover, the wave gauge condition $\Gamma^\alpha = 0$ reads
\begin{equation} \label{eq wave-condition 0}
g_{\beta \gamma} \del_{\alpha}g^{\alpha \beta} = \frac{1}{2}g_{\alpha \beta} \del_{\gamma}g^{\alpha \beta}, 
\end{equation}
and will be used in order to exhibit additional structure enjoyed by the quasi-null terms. 

%----------------------------------------------------------------------------------

\paragraph{Evolution equation for the matter.}

The evolution equation of the scalar field \eqref{eq-KGG} reduces into the following form after imposing the wave gauge condition,
\begin{equation} \label{eq10-15-05-2020}
\Boxt_g \phi - c^2 \phi = 0. 
\end{equation}
This equation is coupled to Einstein's field equations and, therefore, \eqref{eq 1 13-01-2019}--\eqref{eq10-15-05-2020}
is the main system of equations to be solved in terms of the geometric unknown $u_{\alpha\beta}$ 
and the matter field $\phi$. Our aim is to establish a global-in-time existence result 
by relying the properties of the nonlinearities derived in the present section and 
the methodology and the technical tools presented in Part~I.  

%-----------------------------------

\subsection{Statement based on the Euclidean-hyperboloidal foliation} 
\label{secti-42}

We consider the Cauchy problem associated with the Einstein-Klein-Gordon system consisting of the metric equations
\eqref{eq 1 13-01-2019} and the matter equation \eqref{eq10-15-05-2020} together with the wave gauge \eqref{eq wave-condition 0}. In other words, we now study the nonlinear wave-Klein-Gordon system
\begin{equation} \label{eq5-03-01-2022}
\aligned
&\Boxt_g u_{\alpha\beta} 
=  \Pbb^\star_{\alpha\beta}[u] + \Qbb^{\star}_{\alpha\beta}[u] 
- 8\pi \, \big( 2 \, T_{\alpha\beta} - Tg_{\alpha\beta} \big) 
+ \Big( \Ibb^{\star}_{\alpha\beta}[u] + 2 \, ^{(w)}R^{\star}_{\alpha\beta}
- u^{\mu\nu} \del_\mu \del_\nu g^\star_{\alpha\beta}  \Big), 
\\
&\Boxt_g \phi - c^2 \phi = 0,
\\
&g_{\beta \gamma} \del_{\alpha}g^{\alpha \beta} = \frac{1}{2}g_{\alpha \beta} \del_{\gamma}g^{\alpha \beta}, 
\\
&
u_{\alpha\beta}(1,x) = u_{0\alpha\beta}(x),\qquad \del_t u_{\alpha\beta}(1,x) = u_{1 \alpha\beta}(x),\qquad 
\phi(1,x) = \phi_0(x),\qquad \del_t\phi(1,x) = \overline\phi_1(x), 
\endaligned
\end{equation}
for given initial data $u_{0\alpha\beta}, u_{1 \alpha\beta}, \phi_0, \overline\phi_1$. Observe that 
our notation in $\overline\phi_1 = \del_t\phi(1,x)$
should be distinguished from our earlier notation $\phi_1 \neq \overline\phi_1$ since $\phi_1 = \vecnnu\phi$ involves the normal vector $\vecnnu$ to the initial slice (cf.~Definition \ref{def1-05-01-2022}), which, in general, does not coincides with $\del_t$.
Observe also that the system \eqref{eq5-03-01-2022} depends on a choice of reference spacetime metric $g^{\star}$ in $\RR^{1+3}_+$ satisfying the conditions 
\eqref{eq1-14-01-2022}. 
Concerning on the above initial data posed on the hypersurface $\{t=1\}$, we require smallness conditions in suitable 
weighted energy spaces, as we now discuss it. 
Then, the classical local existence theory for second-order hyperbolic systems applies, and 
\eqref{eq5-03-01-2022} possesses a unique local-in-time solution and our task will be to establish that the solution can be extended to all times. 

We recall that the fundamental weighed energy functional $\Eenergy_{\eta,c}(s,u)$ was introduced in \eqref{eq1-03-05-2020}, while high-order energies were analyzed after \eqref{eq6-03-01-2022}. The statement of our main result uses also the notation 
\begin{equation} \label{equa-defEF} 
\Eenergy_{c,\eta}^N(s,u): = \sum_{\ord(Z)\leq N} \Eenergy_{c,\eta}(s,Zu),\quad 
\qquad
\Fenergy_{c, \eta}^N(s,u) :=  \big(\Eenergy_{c, \eta}^N(s,u)\big)^{1/2}, 
\end{equation}
with a similar notation when the subscript $c$ is omitted, in which $Z$ denotes any ordered admissible high-order operator.
We are now in a position to supplement our main stability theorem with quantitative estimates, stated as follows.  

\begin{theorem}[Global existence theory. PDEs formulation] 
\label{thm main-PDE} 
Consider parameters $\kappa,\lambda,\mu$ satisfying \eqref{eq1-20-01-2022} and \eqref{equa-all-conditions-exponents}.
Let $g^{\star}$ be a reference spacetime metric defined in $\RR^{1+3}_+$ with regularity order ${ N\geq 20}$ and 
for a  sufficiently small $\epss$, and assume also that the light bending property \eqref{equa-bending} holds
 (as described in Definitions~\ref{def-seed-basic} and \ref{def:basic12}). 
Then there exists a small constant $\eps_0 > 0$ such that for all initial data satisfying  
\begin{equation} \label{eqergy-boundMas-initial} 
\aligned
\Fenergy_\kappa(s_0, Z u_{\alpha\beta} ) + \Fenergy_{c, \kappa}(s_0, Z \phi)
& \lesssim    
\eps \leq \eps_0, 
\quad &&   
\ord(Z) \leq N,
\endaligned
\end{equation} 
%
% \eqref{eq4-03-01-2022} holds, 
the Cauchy problem \eqref{eq5-03-01-2022} admits a global-in-time solution and the following estimates hold for all $s \geq s_0$
\begin{equation} \label{eqergy-boundMas} 
\aligned
\Fenergy_\kappa(s, Z u_{\alpha\beta} ) + {s^{-1} \, \Fenergy_{c, \kappa}(s, Z \phi) }
& \lesssim    
\eps \, s^{\delta},
\quad &&&& \ord(Z) \leq N-5,
\\
\Fenergy_{c, \kappa}(s, Z \phi)
& \lesssim    
\eps \, s^\delta,
\quad &&&& \ord(Z)  \leq N-7. 
\endaligned
\end{equation}  
\end{theorem}

Our estimates are stated with the energy on the Euclidean-hyperboloidal hypersurfaces, but also imply
estimates on the energy on the standard $t$=constant foliation. We emphasize that as stated in \eqref{eqergy-boundMas}, the energy of the metric has a slow growth of order $s^\delta$,
while the matter field has a slow growth of order $s^{1+ \delta}$,
except at the low order of differentiation where it also has a slow growth of order $s^\delta$. 
It is expected that assuming {stronger decay} on the initial data for the matter field would lead to a matter field growing at a smaller rate. 

%--------------------------------------------------------------------------------------------------------------------------

\subsection{From geometric to PDEs initial data} 
\label{Annexe-our-construction} 

\paragraph{Preliminary.}

Our main result stated earlier in Theorem \ref{theo:main1-geometric} is implied by Theorem \ref{thm main-PDE} above, as we now explain. In order to arrive at the Cauchy problem for the PDEs system as stated in \eqref{eq5-03-01-2022}, a choice of coordinates in the geometric formulation is required so that all geometric degrees of freedom are fixed and
the set of PDEs under consideration is locally well-posed. Let us explain how the PDEs initial data
\begin{equation}
g_{\alpha\beta}(t_0,\cdot),\quad \del_tg_{\alpha\beta}(t_0,\cdot), \quad \phi(t_0,\cdot),\quad \del_t\phi(t_0,\cdot)
\end{equation}
are determined from our geometric initial data $g_0, k_0, \phi_0,\phi_1$. If we formally 
count the available degrees of freedom, by taking into account that
$g_0$ and $k_0$ are symmetric two-tensors, the geometric initial data supplies us with $12$ scalar functions, namely\footnote{These components together with $\phi_0,\phi_1$ must also satisfy Einstein's constraint equations.}  $g_{0ab}$ and $k_{0ab}$. 
On the other hand, we need $20$ scalar functions in order to determine the PDEs initial data. 
At our disposal, we have four gauge conditions in spacetime which we can restrict to the initial slice. Yet, there are still four degrees of freedom unaccounted for. In other words, we must still impose four additional conditions on the choice of the PDEs initial data.

%------------------------------------------

\paragraph{Cauchy adapted frame and second fundamental form.}

The initial hypersurface is embedded in the spacetime via a map $i:\RR^3 \mapsto M$ and, associated with the foliation $M\simeq \RR_+ \times \Sigma$, we also have a global time function $t$ such that $i(\RR^3) = \{t_0\}\times\Sigma \simeq \Sigma_{t_0}$. Global coordinates $\{x^a\}_{a=1,2,3}$  are also assumed on $\Sigma\simeq \RR^3$, while $\{x^0=t,x^a\}$ represents a global coordinate chart on $\RR_+\times \Sigma$.
We denote by $\gb$ the restriction of the spacetime metric $g$ to the hypersurface $\Sigma_{t_0}$, and we 
introduce the vector field 
\begin{equation}
\vec{e}_0 := \del_t^{\perp} =: \del_t - \beta^a\del_a, 
\end{equation}
which is the orthogonal projection of $\del_t$ on the orthogonal space of the hypersurface $\Sigma_{t_0}$. Here, $\beta^a = \gb^{ab}g_{0b}$ is called the {\sl shift vector field}, and we finally also introduce the {\sl lapse function} $\notrelapse>0$ by
setting 
\begin{equation}
\notrelapse^2 := - g(\vec{e}_0,\vec{e}_0).
\end{equation} 
Consequently, we have 
$\vecnnu = \notrelapse^{-1}\vec{e}_0 = \notrelapse^{-1}\del_t^{\perp}$.  
We will refer to this frame $\{\vec{e}_0,\vec{e}_a:=\del_a\}$ as the {\sl Cauchy-adapted frame}, 
by following here the presentation in \cite[ Section 5]{YCB}. 

We begin by observing that
\begin{equation} \label{eq1-05-01-2022}
\del_t\phi|_{\Sigma_{t_0}} := \overline{\phi}_1 = \notrelapse\vecnnu \phi + \beta^c\del_c\phi.
\end{equation}
In addition, concerning the derivatives of the  metric in time-directions we recall that the
second fundamental form of a slice $\Sigma_t \subset M$ of the foliation
by definition
reads 
\begin{equation}
K_{ab} = - g(\nabla_a\vec{e}_b,\vecnnu\big)
= -\frac{1}{2\notrelapse}\Big(\vec{e}_0g_{ab} - g_{cb}\del_a\beta^c - g_{ac}\del_b\beta^c\Big),
\end{equation}
where $g_{\alpha\beta} =  g(\del_{\alpha},\del_{\beta})$ are the components in the natural frame $\{\del_0 = \del_t,\del_a\}$. This leads us to one of the standard ADM equation: 
\begin{equation} \label{eq3-16-12-2021}
\del_tg_{ab} = - 2\notrelapse K_{ab} + \beta^c\del_cg_{ab} + g_{cb}\del_a\beta^c + g_{ac}\del_b\beta^c, 
\end{equation}
which we are going to use on the initial slice. 

\paragraph{Wave gauge conditions.}

We now impose the wave gauge conditions which, for our purpose here, we restrict to $\Sigma_{t_0}$ and state in the so-called natural frame $\{\del_0 = \del_t,\del_a\}$, that is, 
\begin{equation}
\gd^{\alpha\beta} \big( 2 \del_\alpha \gd_{\beta\gamma} - \del_\gamma g_{\alpha\beta} \big) = 0,
\qquad \gamma=0, \ldots, 3. 
\end{equation}
These conditions are equivalent to saying 
\begin{equation} \label{eq2-16-12-2021}
\aligned
\del_tg_{00}  &=  (g^{00})^{-1}\big(-2g^{a0}\del_ag_{00} - 2g^{ab}\del_ag_{b0} + g^{ab}\del_tg_{ab}\big),
\\
\del_tg_{0a}  &=  (2g^{00})^{-1}\Big(
g^{\alpha\beta}\del_ag_{\alpha\beta} - 2g^{c0}\del_cg_{0a} - 2g^{0c}\del_tg_{ca} - 2g^{cb}\del_cg_{ba}
\Big), 
\qquad a=1,2,3. 
\endaligned
\end{equation}
For the PDEs initial data we impose $g_{ab}(t_0) = g_{0ab}$ and $\phi(t_0) = \phi_0$, but we still need to determine $g_{00}(t_0), g_{a0}(t_0), \del_t\phi(t_0)$, and $\del_tg_{\alpha\beta}(t_0)$. 
In the above equations, we also decompose the metric as 
$\gb_{ab} = g_{0ab} = \delta_{ab} + h_{0ab}$ and $K = k_0$, 
so that
$g_{00}$ and $g_{a0}$ are now the remaining degrees of freedom. Indeed, once these functions are chosen, all of the components of the metric will be fixed on the initial hypersurface while, in view of \eqref{eq1-05-01-2022} and \eqref{eq3-16-12-2021}, the time-derivatives $\del_t\phi, \del_t g_{ab}$ are also determined. Finally, by imposing \eqref{eq2-16-12-2021} the remaining time derivatives $\del_tg_{00}$ and $\del_tg_{a0}$ are also uniquely defined.

%----------------------------------

\paragraph{Expansion near the Euclidean geometry.}

The above expressions are quite involved. Since we work near the flat spacetime, we can expand $\del_tg_{\alpha\beta}$ and $\del_t\phi$ in power series with coefficients $g_{00}, g_{a0}, g_{0ab}, k_{0ab}$, as follows. For clarity in the presentation, we introduce 
\begin{equation}
g_{00} = : - 1+ M,
\qquad g_{a0} =: A_a,\qquad 
\gb_{ab} = g_{0ab} = \delta_{ab} + \overline h_{0ab}, 
\end{equation}
in which $|M| + |A| + |\hb|\ll 1$. 
We determine the tensor $h_{\alpha\beta} = \big( g_{\alpha\beta} - g_{\Mink,\alpha\beta} \big)|_{\Sigma_{t_0}}$ induced 
on the initial hypersurface by choosing 
\begin{equation}
h_{00} = M, \qquad h_{0a} = A_a, \qquad  h_{ab} = h_{0ab}. 
\end{equation}
Here, the lapse function and the shift vector are uniquely determined by choosing the values $M$ and $A_a$ and, more precisely, 
\begin{equation} \label{eq3-06-01-2022}
\beta^b = \gb^{ab}A_a,\qquad \notrelapse^2 = 1-M + \gb^{ab}A_aA_b.
\end{equation}

%-----------------------------

Similar calculations as the ones in the proof of Lemma \ref{lem-small} also yield us the following expansion for the perturbation of the inverse metric: 
\begin{equation}\label{eq-h-double}
h^{\alpha\beta} 
= - { {\mathbbm{ h}^{\alpha\beta}} + (g_{\Mink}^2h^2g_{\Mink})_{\alpha\beta}} + \Ocal_{\alpha\beta}(|h|^3), 
\end{equation}
where we define ${\mathbbm{h}}^{\alpha\beta} = 
\mathbbm{h}^{\alpha\beta}[h] := g_{\Mink}^{\alpha\alpha'}h_{\alpha'\beta'}g_{\Mink}^{\beta\beta'}$.

We can then expand the lapse and shift in the form 
$$
\aligned
&\beta^a = A_a + h^{ab}A_b = A_a - \sum_b{ \mathbbm{h}^{ab}}A_b +  \sum_b{  (g_{\Mink}^2h^2g_{\Mink})_{ab}}A_b + \sum_b\Ocal_{ab}(|h|^3)A_b,
\\
&\notrelapse = \big(1-M + \gb^{ab}A_aA_b\big)^{1/2} = 1 + \frac{1}{2}(-M + \gb^{ab}A_aA_b) -\frac{1}{4}\Big(
-M + \gb^{ab}A_aA_b \Big)^2 + \Ocal(|h|^3).
\endaligned
$$
In turn, substituting these relations into the ADM equation \eqref{eq3-16-12-2021}, \eqref{eq2-16-12-2021} and \eqref{eq1-05-01-2022}, we arrive at
\begin{equation} \label{eq4-16-12-2021}
\aligned
\del_tg_{ab}(t_0)   &=  -2k_{0ab} + \del_aA_b + \del_bA_a &&+ Q_{ab}[h,k_0]+ \Ocal(|h|^2)L_{ab}(\del_x h, k_0),
\\
\del_tg_{00}(t_0)  &=  2\sum_ak_{0aa}  &&+ Q_{a0}[h,k_0]+ \Ocal(|h|^2)L_{a0}(\del_x h, k_0),
\\
\del_t g_{a0}(t_0)  &=  \frac{1}{2}\del_a M - \frac{1}{2}\sum_b \del_ah_{0bb} + \sum_b\del_bh_{0ba}
&&+ Q_{00}[h,k_0] + \Ocal(|h|^2)L_{a0}(\del_x h, k_0), 
\\
\del_t\phi(t_0)  &= \phi_1 &&+ Q_{\phi}[h,\phi_0,\phi_1] + \Ocal(|h^2|)L_{\phi}(\del\phi_0,\phi_1), 
\endaligned
\end{equation} 
where the quadratic or cubic terms are $Q_{\alpha\beta},Q_{\phi}$ are bounded by $ |h| \Big( |k_0| + |\phi_1| + |\del_xh| + |\del\phi_0| \Big)$, and  the higher-order terms $\Ocal(|h|^2)L_{\alpha\beta}(\del_xh,k_0)$
and $\Ocal(|h^2|)L_{\phi}(\del\phi_0,\phi_1)$ are bounded by $|h^2| \Big( |k_0| + |\phi_1| + |\del_xh| + |\del\phi_0| \Big)$.

\paragraph{Expansion for the reference metric.}

 On the other hand, our geometric initial data enjoy the decomposition \eqref{eq1-06-01-2022}. Thus the PDEs initial data should also be decomposed as the sum of a reference plus a perturbation. To this purpose, we must first construct the PDEs initial reference by 
\begin{equation} \label{eq2-06-01-2022}
g^{\star}_{ab}(t_0) = g^{\star}_{0ab},\qquad g^{\star}_{a0} = A^{\star}_a,\qquad g^{\star}_{00} = -1+M^{\star}.
\end{equation}
For clarity, we introduce 
$$
w[g^{\star}]^{\gamma} := \Gamma^{\star\gamma} 
= g^{\star\alpha\beta}g^{\star\gamma\delta}\del_{\alpha}g^{\star}_{\beta\delta} 
- \frac{1}{2}g^{\star\alpha\beta}g^{\star\gamma\delta}\del_{\delta}g^{\star}_{\alpha\beta}.
$$
Recalling the { initial wave gauge conditions} satisfied by $g^{\star}$, i.e., \eqref{eq2-04-11-2022}, we have 
\begin{equation}\label{eq1-04-11-2022}
\|\la r\ra^{\kappa+ |I|} \del^I (w[g^{\star}]^{\gamma}) \big|_{\Sigma_{t_0}}\|_{L^2(\RR^3)} \leq \epss,
\qquad |I| \leq N. 
\end{equation}
We write
$
g^{\star\alpha\beta} \big( 2 \del_\alpha g^{\star}_{\beta\gamma} - \del_\gamma g^{\star}_{\alpha\beta} \big) = 2w[g^{\star}]_{\gamma}$ (with $\gamma=0, \ldots, 3$). 
and by repeating the derivation of \eqref{eq3-16-12-2021} and \eqref{eq2-16-12-2021}, we obtain
$$
\aligned
\del_tg^{\star}_{ab}  &=  - 2\notrelapse^{\star}k^{\star}_{ab} + \beta^{\star c}\del_cg^{\star}_{ab} + g^{\star}_{cb}\del_a\beta^{\star c} + g^{\star}_{ac}\del_b\beta^{\star c},
&&& \qquad a, b=1,2,3, 
\\
\del_tg^{\star}_{00}  &=  (g^{\star 00})^{-1}\big(-2g^{\star a0}\del_ag^{\star}_{00} - 2g^{\star ab}\del_ag^{\star}_{b0} + g^{\star ab}\del_tg^{\star}_{ab} { + 2w[g^{\star}]_0}\big),
\\
\del_tg^{\star}_{0a}  &=  (2g^{\star 00})^{-1}\big(g^{\star\alpha\beta}\del_ag^{\star}_{\alpha\beta} - 2g^{\star c0}\del_cg^{\star}_{0a} - 2g^{\star 0c}\del_tg^{\star}_{ca} - 2g^{\star cb}\del_cg^{\star}_{ba} { + 2w[g^{\star}]_a}\big), 
&&&\qquad a=1,2,3. 
\endaligned
$$
Here, $\beta^{\star b} = \gb^{\star ab}A^{\star}_a$ and 
$\notrelapse^{\star} = \sqrt{1-M^{\star} + \gb^{\star ab}A^{\star}_aA^{\star}_b}$, 
and $\overline{g}^{\star ab}$ denotes the inverse of $\bar{g}^{\star}_{ab}$. 
Our assumptions on the reference metric provide us with the bounds 
\begin{equation} \label{eq1-07-01-2022}
\la r \ra^{\lambda + m + |I|}\Big( |\del^m \del_x^I A^{\star} | + | \del^m \del_x^I M^{\star} | \Big)\lesssim \epss,\qquad |I|\leq N+2-m,
\quad m=0,1,2.
\end{equation}
As a matter of fact, the functions $(M^{\star},A^{\star}_a)$ are determined from the prescribed spacetime metric $(g^{\star},k^{\star})$. For example, in the case of the Schwarzschild asymptotics \eqref{equa-description-Schwarz-merging} we have 
\begin{equation}\label{equ-label-4nov-2022-1} 
M^{\star} = g^{\star}_{00} + 1 = \frac{2m}{r+m}, 
\qquad
A^{\star}_a = 0. 
\end{equation} 
%

%-----------------------------------------------------------------------------------------

For the reference metric we consider the expansion near the Minkowski metric, given by 
\begin{equation} 
\aligned
	\del_tg^{\star}_{ab}(t_0)   &=  -2k^{\star}_{0ab} + \del_aA^{\star}_b + \del_bA^{\star}_a &&+ Q_{ab}[h^{\star},k^{\star}_0]+ \Ocal(|h^{\star}|^3)L_{ab}(\del_x h^{\star}, k^{\star}_0),
\\
	\del_tg^{\star}_{00}(t_0)  &=  2\sum_ak^{\star}_{0aa} + 2w[g^{\star}]_0  
	&&
	+ 2\sum_{j=1}^{\infty}(-h^{\star}_{00})^jw[g^{\star}]_0
	+ Q_{a0}[h^{\star},k^{\star}_0]+ \Ocal(|h^{\star}|^3)L_{a0}(\del_x h^{\star}, k^{\star}_0),
\\
	\del_t g^{\star}_{a0}(t_0)  &=  \frac{1}{2}\del_a M^{\star} - \frac{1}{2}\sum_b \del_ah^{\star}_{0bb} + \sum_b\del_bh^{\star}_{0ba} + w[g^{\star}]_a
	&&+ \sum_{j=1}^{\infty}(-h^{\star}_{00})^jw[g^{\star}]_a + Q_{00}[h^{\star},k^{\star}_0] + \Ocal(|h^{\star}|^3)L_{a0}(\del_x h^{\star}, k^{\star}_0).
	\endaligned
\end{equation} 
We subtract the above expansion from \eqref{eq4-16-12-2021} and obtain the expression of $\del_t u_{0\alpha\beta}$. Observe that, at this stage, we still need to fix $u_{000}$ and $u_{0a0}$ arising in 
\begin{equation}
M = M^{\star} + u_{000},\qquad A_a = A^{\star}_a + u_{0a0}. 
\end{equation}
Since this is a consistent choice in view of our decay assumptions on the perturbation
we make the choice
\begin{equation}
M=M^\star, \qquad A = A^\star. 
\end{equation} 
In view of \eqref{eq4-16-12-2021}, having specified the components of the initial reference, the initial perturbation is thus determined explicitly from the geometric data.

%---------------------------------------------------------------------------------------------------------------

\subsection{Estimates for the PDEs initial data}
\label{section-use-S} 

\paragraph{The estimates on the initial data.}

It remains to translate our bounds \eqref{eq2a-26-12-2021}-\eqref{eq2b-26-12-2021}
on the geometric initial data into
bounds on the PDEs initial data. More precisely, we must check that our energy bounds at the initial time are 
guaranteed by our bounds on the initial data set. 
The proofs of the following two statements are postponed to Appendix \ref{sec1-07-01-2022}.

\begin{proposition}[Bounds on the spacetime PDEs components at the initial time. (I) $L^2$ norms]
\label{lem1-07-01-2022}
There exists a small positive constant $c_0$ such that the following property holds. 
Provided $|h^{\star} |_{N+2}\leq c_0$ and the initial perturbation bounds 
\eqref{eq2a-26-12-2021}-\eqref{eq2b-26-12-2021}
hold,
for  $\eps\leq c_0$, then the PDEs initial data satisfy 
\begin{subequations} \label{eq4-07-01-2022}
\begin{equation} \label{eq4a-07-01-2022}
\big\|\la r\ra^{\kappa+|I| } \del^I \del_a u_{0\alpha\beta} \big\|_{L^2(\RR^3)}
+ \big\|\la r\ra^{\kappa+|I| } \del^I  u_{1\alpha\beta} \big\|_{L^2(\RR^3)}
\leq \vep, 
\qquad |I|\leq N, 
\end{equation}
\begin{equation} \label{eq4b-07-01-2022}
\big\|\la r\ra^{\mu +{ N} } \del^I \del_a\phi_0 \big\|_{L^2(\RR^3)} 
+ \big\|\la r\ra^{\mu + { N} } \del^I \phi_0 \big\|_{L^2(\RR^3)} 
+\big\|\la r\ra^{\mu + { N}} \del^I \overline{\phi}_1\big\|_{L^2(\RR^3)}
\leq \vep,
\qquad |I|\leq N. 
\end{equation} 
\end{subequations}
\end{proposition}

In view of \eqref{eq4-07-01-2022}, by general arguments for second-order quasilinear hyperbolic systems we deduce that \eqref{eq5-03-01-2022} admits a unique local-in-time solution. 
In the following we prefer to 
write $|\del u|^S_{N} = \max_{\ord(\Gamma) \leq N} |\Gamma \del u|$ explicitly and, similarly, 
$|\del \phi|_{N} = \max_{\ord(Z) \leq N} |Z \del \phi|$. The proof of the following result is also postponed to Appendix \ref{sec1-07-01-2022}. 

\begin{proposition}[Bounds on the spacetime PDEs components at the initial time. (II) Energy norms]
\label{prop1-07-01-2022}
The local-in-time solution to \eqref{eq5-03-01-2022} restricted to the initial hypersurface $t_0 = 1$ enjoys  
the following bounds: 
\begin{equation} \label{eq5-07-01-2022}
\|\la r\ra^{\kappa}\Gamma \del u(1,\cdot)\|_{L^2(\RR^3)} + \|\la r\ra^{\mu}Z \del\phi(1,\cdot)\|_{L^2(\RR^3)} 
+ \|\la r\ra^{\mu}Z \phi(1,\cdot)\|_{L^2(\RR^3)}
\lesssim \eps
\end{equation}
for all ordered admissible operator $Z$ and all ordered conformal operator $\Gamma$ satisfying $\ord(\Gamma) \leq N$ and $\ord(Z)\leq N$.  
Consequently, 
on the hypersurface labelled $s_0 = 2$ and provided $\eps$ is sufficiently small so that the local-in-time solution extends to the domain $\Mscr^{\init}$ with 
\begin{equation} \label{eq2-02-02-2022}
\Fenergy_{g,\kappa}^N(s_0,u) + \Fenergy_{g,\mu,c}^N(s_0,\phi)\lesssim \eps.
\end{equation}
\end{proposition}

%------------------------------

\paragraph{The estimates on the spacetime contribution $u_{\init}$.}

Finally, we turn our attention to the issue of linearly propagating the initial perturbation. Recalling the notation in Section~\ref{subsec1-07-01-2022}, we consider 
\begin{equation}
\label{unit-21-octobre} 
u_{\init,\alpha\beta} = \Box^{-1}[u_{\alpha\beta}(1,\cdot),\del_t u_{\alpha\beta}(1,\cdot),0], 
\end{equation}
which is the spacetime contribution associated with the initial perturbation. The following estimate (whose proof is also postponed to~Appendix \ref{sec1-07-01-2022})
completes the discussion of the passage from geometric initial data to PDEs initial data. 

\begin{proposition} \label{prop1-12-01-2022}
For initial data satisfying \eqref{eq5-07-01-2022}, the solution \eqref{unit-21-octobre} satisfies the pointwise bound 
\begin{equation}
|u_{\init} |_{N-2}\lesssim \eps \, \la r+t\ra^{-1}.
\end{equation}
\end{proposition}

%=============================================================================================
\section{Einstein-matter system in the Euclidean-hyper\-boloidal foliation}
\label{sectionN-12}

\subsection{Nonlinearities of the Einstein equations}

\paragraph{Classification of the nonlinearities.} 

The right-hand side of \eqref{eq 1 13-01-2019} derived in Section~\ref{section-new-11} involves nonlinearities of a very different nature. 
\begin{itemize} 

\item The geometric nonlinearity $\Fbb^\star_{\alpha\beta}[u] = \Pbb_{\alpha\beta}^{\star}[u] + \Qbb_{\alpha\beta}^{\star}[u]$ and the matter source $8\pi \, (2 \, T_{\alpha\beta} - Tg_{\alpha\beta})$ come from the essential structure of the Einstein system and represent the most challenging contributions for our global-in-time analysis of the Cauchy problem. 

\item The interaction terms $\Ibb^{\star}_{\alpha\beta}[u]$ and $u^{\mu\nu}\del_{\mu}\del_{\nu}g^{\star}_{\alpha\beta}$
depend on the reference metric, and their control will depend on our assumptions on $h^\star$
(cf.~\eqref{equa-31-12-20}  or \eqref{equa-new-conditions-hstar} in~Section~\ref{sec1-23-05-2021} below). 

\item The Ricci curvature of the reference metric $\wR^{\star}_{\alpha\beta}$ will be controlled directly from our assumption that $g^{\star}$ is an approximate solution to Einstein equations.

\end{itemize} 

\paragraph{Interaction terms.}

We will control the interaction terms by using properties enjoyed by, both, the reference metric and the perturbation and we work under the basic condition
\begin{equation} \label{eq1-15-07-2021}
|h^{\star} |_p + |u|_{[p/2]} \leq \eps_1 \ll 1.
\end{equation}
Thanks to Lemma~\ref{lem-small} (together with Remark~\ref{rmk1-15-07-2021}), 
for $\Bbb^{\star}$ and $\Cbb^{\star}$ 
we have 
\begin{subequations}
\label{eq0-crossing} 
\begin{equation} \label{eq2-crossing} 
|\Bbb^{\star}_{\alpha\beta}[u] |_{p,k}
\lesssim  |\del h^\star |_p \sum_{p_1+p_2 = p} |u|_{p_1} |\del u|_{p_2} + |\del h^\star |_p^2 \sum_{p_1+p_2 = p} |u|_{p_1} |u|_{p_2},
\end{equation}
\begin{equation} \label{eq3-crossing}
|\Cbb^{\star}_{\alpha\beta}[u] |_{p,k}
\lesssim  \hskip-.3cm
\sum_{p_1+p_2+p_3 =p} \hskip-.3cm |u|_{p_1} |\del u|_{p_2} |\del u|_{p_3} 
+ |\del h^\star |_p  \hskip-.3cm  \sum_{p_1+p_2+p_3 =p} \hskip-.3cm |u|_{p_1} |u|_{p_2} |\del u|_{p_3} 
+ \hskip-.3cm \sum_{p_1+p_2 + p_3+p_4 = p} \hskip-.3cm  |\del u|_{p_1} |\del u|_{p_2} |u|_{p_3} |u|_{p_4}.
\end{equation}
On the other hand, when sufficient decay on $h^\star$ is available we can also bound $\Lbb^{\star}_{\alpha\beta}[u]$ in a similar way: 
\begin{equation} \label{eq1-crossing}
|\Lbb^{\star}_{\alpha\beta}[u] |_{p,k} \lesssim |\del h^\star |_p|\del u|_p + |\del h^\star |_p^2|u|_p.
\end{equation} 
\end{subequations}
More precisely, in the decomposition \eqref{eq4-06-01-2020}
of $\Lbb^\star_{\alpha\beta}[u]$ the first and second terms are bounded by the first and second terms in the right-hand side of \eqref{eq1-crossing}, respectively.
However, 
a more detailed estimate will be required to deal with $\Lbb^{\star}$ for reference metrics in Class B. As we will see in Section~\ref{section--125}, the above terms will enjoy integrable $L^2$ bounds.   
More precisely, we recall \eqref{eq4-06-01-2020} and \eqref{eq6-04-10-2022}, and make the following decomposition:
$$
\Lbb^{\star}_{\alpha\beta}[u] = 2\Pbb^{\star}_{\alpha\beta}[u,h^{\star}] 
+ 2\Qbb^{\star}_{\alpha\beta}[u,h^{\star}] 
+ \Fbb_{\alpha\beta}(u,g^\star;\del h^{\star}, \del h^{\star}) 
+ \Fbb_{\alpha\beta}(g^\star,u;\del h^{\star}, \del h^{\star}).
$$
In the above expression, the first two terms are critical, and  will be treated in the following general discussions about null and quasi-null terms.
 
%-----------------------------------------------

\paragraph{Estimates for null terms.} 

The rest of this section is devoted to a detailed analysis of null and quasi-null terms. 
We begin with a direct consequence of \eqref{eq1-10-06-2021} applied to each term in $\Qbb^{\star}_{\alpha\beta}[u,v]:= \Qbb_{\alpha\beta}(g^{\star},g^{\star}; \del u,\del v)$. We only need to observe that $g^{\star\mu\nu} = g^{\mu\nu}_{\Mink} + h^{\star\mu\nu}$ enjoys a null structure. For simplicity in the analysis,  we find it convenient to assume high regularity on the reference metric so that, in our final bootstrap estimates involving products of the reference metric and the metric perturbation, we will always control the reference metric in sup-norm and the solution in $L^2$ Sobolev norm. 

\begin{lemma}[{Estimates of null interaction terms}] 
\label{Null-Euclidean bilinear}
In the Euclidean-merging domain $\MME = \{r\geq t-1\}$, null forms are controlled by good derivatives and a contribution depending upon the reference metric:
\begin{equation} \label{equa-new-Qzero}
\aligned
|\Qbb^{\star}[u, v]|_{p,k} &:= \max_{\alpha,\beta} |\Qbb_{\alpha\beta}^\star[u, v] |_{p, k}
\\
& \lesssim  \sum_{p_1+p_2 = p\atop k_1+k_2=k} 
\Big( |\del u|_{p_1, k_1} |\delsN v |_{p_2, k_2}  + |\del v |_{p_1, k_1} |\delsN u|_{p_2, k_2}  \Big) 
+ | h^\star |_p  
\sum_{p_1+p_2=p\atop k_1+k_2=k} |\del u|_{p_1, k_1} |\del v |_{p_2, k_2} 
\endaligned
\end{equation} 
and, in particular, 
\begin{equation}\label{eq7-04-10-2022}
|\Qbb^{\star}[u]|_{p,k} := \max_{\alpha,\beta} |\Qbb_{\alpha\beta}^\star[u] |_{p, k}
\lesssim  \sum_{p_1+p_2 = p\atop k_1+k_2=k} 
|\del u|_{p_1, k_1} |\delsN u |_{p_2, k_2}  
+ | h^\star |_p  
\sum_{p_1+p_2=p\atop k_1+k_2=k} |\del u|_{p_1, k_1} |\del u |_{p_2, k_2}.
\end{equation}
\end{lemma}

%-------------------

\paragraph{Estimates for quasi-null terms.}

Although the quadratic form $\Pbb_{\alpha\beta}(g^\star,g^\star;\del u, \del v) $ cannot be written as a linear combination of ``good'' products  $(\delsN u) \, \del v$, yet we can control it by uncovering a suitable tensorial decomposition and taking the wave gauge into account. 
More precisely, we assume that $u$ (or more precisely $h$) satisfies the wave gauge (while $v$ satisfies this condition approximatively, only). 
In addition to the standard null forms, $\Pbb_{\alpha\beta}$ contains the term $\del \uts \del \slashed v^\Ncal$ which we are going to analyze in the null frame. Recalling the tensorial structure 
$\Pbb_{\alpha\beta}(g^{\star},g^{\star};\del u, \del v) = \Pbb(g^{\star},g^{\star};\del_\alpha u, \del_{\beta} v)$, 
we write 
\begin{equation} \label{eq1 05-juillet-2019}
\aligned
\Pbb _{\alpha\beta}^{\star\N}[u, v] 
:= 
\Pbb^\N _{\alpha\beta}(g^{\star},g^{\star}, \del u, \del v) 
:= \PhiN_{\alpha}^\gamma \PhiN_{\beta}^{\delta} \Pbb_{\gamma\delta}(g^{\star},g^{\star};\del u, \del v)
= \Pbb_{\alpha\beta}(g^{\star},g^{\star}; \delN_\alpha u, \delN_{\beta} v). 
\endaligned
\end{equation}
We also introduce the following (partial) norm of $\Pbb$ (with the component $\Pbb_{00}$ suppressed) 
\begin{equation}
|\slashed{\Pbb}^{\star\N} [u, v]| _{ p,k}
:= \max_{(\alpha, \beta)\neq(0,0)} |\Pbb_{\alpha\beta}[u, v]|_{p,k}, 
\end{equation}
which clearly satisfies 
\begin{equation} \label{eq2 05-juillet-2019}
|\slashed{\Pbb}^{\star\N}[u, v] |_{p,k}
\lesssim 
\sum_{p_1+p_2=p\atop  k_1+k_2=k} \Big(
|\del u|_{p_1,k_1} |\delsN v|_{p_2,k_2} +|\del v|_{p_1,k_1} |\delsN u|_{p_2,k_2} + | h^\star |_p |\del u|_{p_1} |\del v |_{p_2} 
\Big). 
\end{equation}
For the $(0,0)$-component of $\Pbb^{\star\N}[u, v]$, we observe that  
$
\del_t\PhiN^\alpha _{\beta} = 0
$
and, in view of the structure of $\Pbb_{\alpha\beta}^\star[u, v]$ in \eqref{equa-Pbb-four-arguments}, it is clear that 
the component $\Pbb^{\star\N}_{00}[u, v]$ is a linear combination of
$$
\aligned
A & :=  g^{\star\mu\mu'} g^{\star\nu\nu'} \del_t u_{\mu\mu'} \del_t v_{\nu\nu'} 
= \gt^{\star\mu\mu'} \gt^{\star\nu\nu'} \del_t \ut_{\mu\mu'} \del_t v^\Ncal_{\nu\nu'},  
\\
B & :=  g^{\star\mu\mu'}g^{\star\nu\nu'} \del_t u_{\mu\nu} \del_t v_{\mu'\nu'} 
= \gt^{\star\mu\mu'} \gt^{\star\nu\nu'} \del_t \ut_{\mu\nu} \del_t v^\Ncal_{\mu'\nu'}. 
\endaligned
$$
Dealing with the term $A$ is easy, since  
$$
\aligned
g^{\star\mu\mu'}g^{\star\nu\nu'} \del_t u_{\mu\mu'} \del_t v_{\nu\nu'} 
= \sum_{(\mu, \mu')\neq (0,0)\atop (\nu, \nu')\neq (0,0)} g_\Mink^{\Ncal \mu\mu'} g_\Mink^{\Ncal \nu\nu'} \del_t\ut_{\mu\mu'} \del_tv^\Ncal_{\nu\nu'}
+\Big(g_\Mink^{\Ncal \mu\mu'} h^{\star\Ncal \nu\nu'} +  h^{\star\Ncal \mu\mu'}g^{\star \nu\nu'} \Big)\del_t u_{\mu\mu'} \del_t v_{\nu\nu'}, 
\endaligned
$$
where we used $g_\Mink^{\Ncal 00} = 0$. Since   
$|g_\Mink^{\Ncal \mu\mu'} \, g_\Mink^{\Ncal \nu\nu'} \tdelME_t\ut_{\mu\mu'} \tdelME_t v^\Ncal_{\nu\nu'} |_{p,k}
\lesssim \sum_{p_1+p_2=p\atop { k_1+k_2=k}} |\del \uts|_{p_1,{ k_1}} |\del \slashed v^\Ncal |_{p_2,{ k_2}}$, it follows that  
\begin{equation} \label{eq3-04-12-2020}
\aligned
|A|_{p,k}  
& \lesssim  \sum_{p_1+p_2=p\atop k_1+k_2=k} |\del \uts|_{p_1,k_1} |\del \slashed v^\Ncal |_{p_2,k_2} 
+ \sum_{p_1+p_2+p_3=p} | h^\star |_{p_3} |\del u|_{p_1} |\del v |_{p_2}.
\endaligned
\end{equation}
We find it convenient to informally refer to the first sum in the right-hand side still as a quasi-null term: in some sense, it is the ``reduced'' form of the quasi-null terms. In fact the rest terms contained in $\Pbb$ are null terms or higher-order ones. 

Next, for the term $B$, we write
\begin{equation} \label{eq3-09-04-2020} 
B = \sum_{(\mu, \mu')\neq (0,0)\atop (\nu, \nu')\neq(0,0)} 
g_\Mink^{\Ncal \star \mu\mu'} \, g_\Mink^{\Ncal \star \nu\nu'} \del_t\ut_{\mu\nu} \del_tv^\Ncal_{\mu'\nu'} 
+ \Big(
g_\Mink^{\Ncal \mu\mu'} h^{\star\Ncal \nu\nu'} +  h^{\star\Ncal \mu\mu'}g^{\star \nu\nu'} \Big)
\del_t u_{\mu\nu} \del_t v_{\mu'\nu'}. 
\end{equation}
The first sum above contains two potentially critical terms arising for $(\mu, \nu) = (0,0)$ and $(\mu', \nu')\neq (0,0)$, as well 
as $(\mu, \nu)\neq(0,0)$ and $(\mu', \nu') = (0,0)$, respectively, that is,  
\begin{equation} \label{eq4-09-04-2020}
g_\Mink^{\Ncal 0c'} \, g_\Mink^{\Ncal 0d'} \del_t\ut_{00} \del_t v^\Ncal_{c'd'},
\qquad 
g_\Mink^{\Ncal c0} \, g_\Mink^{\Ncal d0} \del_t\ut_{cd} \del_t v^\Ncal_{00},
\end{equation}
while all of the remaining terms in the first sum of the right-hand side of \eqref{eq3-09-04-2020} are of the form $\del_t\uts \del_t \slashed v^\Ncal$.
It remains to analyze \eqref{eq4-09-04-2020}, as follows.

%------------------------------------------------------------------------ 

\paragraph{Second term in~\eqref{eq4-09-04-2020}.}

Concerning the metric perturbation $u$, we recall the wave gauge condition 
$g^{\alpha\beta} \del_{\alpha}h_{\beta\gamma} = \frac{1}{2}g^{\alpha\beta} \del_\gamma h_{\alpha\beta}$, 
which we express in the null frame as 
$$
\gt^{\alpha\beta} \tdelME_\alpha \hN_{\beta\gamma}
= \frac{1}{2} \gt^{\alpha\beta} \tdelME_\gamma\hN_{\alpha\beta}
+ \frac{1}{2}g^{\alpha\beta} \tdelME_\gamma\big(\tPsiME_{\alpha}^{\alpha'} \tPsiME_{\beta}^{\beta'} \big)\hN_{\alpha'\beta'}
- g^{\alpha\beta} \tPhiME_\gamma^{\gamma''} \del_\alpha \big(\tPsiME_{\beta}^{\beta'} \tPsiME_{\gamma''}^{\gamma'} \big)\hN_{\beta'\gamma'}. 
$$
Taking $\gamma = c= 1,2,3$ and recalling that $g_\Mink^{\Ncal 00} = 0$, the above identity gives us 
$$
\aligned
g_\Mink^{\Ncal 0b} \del_t \hN_{bc} 
& =      - g_\Mink^{\Ncal a\beta} \delts_a\hN_{\beta c} 
+ \frac{1}{2} \gt^{\alpha\beta} \delts_c\hN_{\alpha\beta}
\\
& \quad
- \hN{}^{\alpha\beta} \tdelME_\alpha \hN_{\beta c} + \frac{1}{2}g^{\alpha\beta} \delts_{c} \big(\tPsiME_{\alpha}^{\alpha'} \tPsiME_{\beta}^{\beta'} \big)\hN_{\alpha'\beta'}
- g^{\alpha\beta} \tPhiME_{c}^{\gamma''} \del_\alpha \big(\tPsiME_{\beta}^{\beta'} \tPsiME_{\gamma''}^{\gamma'} \big)\hN_{\beta'\gamma'}. 
\endaligned
$$
In view of Lemma~\ref{lem-small}, we have $|h^{\alpha\beta} | \lesssim |h|$, and therefore  
$|g_\Mink^{\Ncal 0b} \del_t\hN_{bc} | \lesssim |\delts h|+ r^{-1} |h| + r^{-1}| h|^2 + |h \, \del h|$ and, with the same lemma, we also find the higher-order version 
$$
|g_\Mink^{\Ncal 0b} \, \del_t\hN_{bc} |_{p,k} \lesssim |\delsN h|_{p,k} + r^{-1} |h|_{p,k} + \sum_{p_1+p_2 = p\atop k_1+k_2=k} \big( |\del h|_{p_1,k_1} |h|_{p_2,k_2} + r^{-1}|h|_{p_1,k_1} | h|_{p_2,k_2} \big). 
$$
Relying here on { $|h|_{[p/2]}\ll1$}, we obtain
\begin{equation}\label{eq1-04-10-2022}
|g_\Mink^{\Ncal 0b} \, \del_t\hN_{bc} |_{p,k} \lesssim |\delsN h|_{p,k} + r^{-1} |h|_{p,k} + \sum_{p_1+p_2 = p\atop k_1+k_2=k}|\del h|_{p_1,k_1} |h|_{p_2,k_2}. 
\end{equation}
After further decomposition, we thus obtain
$$
\aligned
|g_\Mink^{\Ncal 0b} \, \del_t \ut_{bc} |_{p,k}
& \lesssim   |\delsN u|_{p,k} + r^{-1} |u|_{p,k}   +|\del h^\star |_{p,k} + r^{-1} | h^\star |_{p,k}  
\\
&\quad + \sum_{p_1+p_2=p}|\del  u|_{p_1} |u|_{p_2} 
+ \sum_{p_1+p_2=p}\Big(| h^\star |_{p_1} |\del u|_{p_2} + |u|_{p_1} |\del h^\star |_{p_2} + | h^\star |_{p_1} |\del h^\star |_{p_2}\Big). 
\endaligned
$$
{
Modulo high-order contributions the term $g_\Mink^{\Ncal 0b} \del_t \ut_{bc}$ can therefore be treated like a {\sl  null derivative.} In turn, for the first term in~\eqref{eq4-09-04-2020} we conclude that 
\begin{equation}\label{eq3-04-10-2022}
|g_\Mink^{\Ncal c0} \, g_\Mink^{\Ncal d0} \del_t\ut_{cd} \del_t v^{\N}_{00} |_{p,k}
\lesssim 
\sum_{p_1+p_2=p\atop k_1+k_2=k} |\delsN u|_{p_1,k_1}|\del v|_{p_2,k_2}  + \sum_{p_1+p_2=p} \SbbME_{p_1}[u] \, |\del v|_{p_2}, 
\end{equation}
which involves the remainder $\SbbME_p[u]$ defined by  
\begin{equation}\label{eq2-04-10-2022}
\aligned
\SbbME_p[u]
&:=  
r^{-1} |u|_p   +|\del h^\star |_p + r^{-1} | h^\star |_p  
+ \sum_{p_1+p_2=p}|\del  u|_{p_1} |u|_{p_2} 
\\
& \quad
+ \sum_{p_1+p_2=p}\big(| h^\star |_{p_1} |\del u|_{p_2} + |u|_{p_1} |\del h^\star |_{p_2} + |h^{\star}|_{p_1}|\del h^{\star}|_{p_2}\big).
\endaligned
\end{equation}
From \eqref{eq3-04-12-2020} and \eqref{eq3-04-10-2022} we obtain (with $v=u$) the estimate \eqref{eq4-04-10-2022} for the diagonal case.}
 
%------------------------------------------------------------------------ 

\paragraph{First term in~\eqref{eq4-09-04-2020} and conclusion.}
The metric perturbation $v$ need not satisfy the wave condition exactly. So, we set
\begin{equation}\label{eq8-04-10-2022}
w[v]_{\gamma} := (\gMink + v)^{\alpha\beta} \del_{\alpha} v_{\beta\gamma} - \frac{1}{2} (\gMink + v)^{\alpha\beta} \del_\gamma v_{\alpha\beta},
\end{equation}
and rely on our suitable smallness and decay conditions. We have 
$$
\aligned
g_\Mink^{\Ncal 0b} \del_t v^\Ncal_{bc} 
& = \PhiN^{\gamma}_c w[v]_{\gamma}
- g_\Mink^{\Ncal a\beta} \delts_a v^\Ncal_{\beta c} 
+ \frac{1}{2} (\gMink + v)^{\alpha\beta} \delts_c v^\Ncal_{\alpha\beta}
\\
& \quad
- v^\Ncal{}^{\alpha\beta} \tdelME_\alpha v^\Ncal_{\beta c} 
+ \frac{1}{2} (\gMink + v)^{\alpha\beta} \delts_{c} \big(\tPsiME_{\alpha}^{\alpha'} \tPsiME_{\beta}^{\beta'} \big) v^\Ncal_{\alpha'\beta'}
- (\gMink + v)^{\alpha\beta} \tPhiME_{c}^{\gamma''} \del_\alpha \big(\tPsiME_{\beta}^{\beta'} \tPsiME_{\gamma''}^{\gamma'} \big)v^\Ncal_{\beta'\gamma'}. 
\endaligned 
$$
Then similar to \eqref{eq1-04-10-2022}, one has
\begin{equation}
\label{equa-12-aout-2022-3} 
|g_\Mink^{\Ncal 0b} \, \del_t v^{\Ncal}_{bc} |_{p,k} 
\lesssim \sum_\gamma |w[v]_\gamma|_{p,k}
+
|\delsN v |_{p,k} + r^{-1} | v|_{p,k} + \sum_{p_1+p_2 = p\atop k_1+k_2=k} |\del  v |_{p_1,k_1} | v |_{p_2,k_2}. 
\end{equation}
In turn, for the first term in \eqref{eq4-09-04-2020} we find  
$$
| g_\Mink^{\Ncal 0c'} \, g_\Mink^{\Ncal 0d'} \del_t\ut_{00} \del_t v^\Ncal_{c'd'}  |_{p,k}
\lesssim 
\sum_{p_1+p_2=p}  |\delts v |_{p_1} |\del u |_{p_2}
+ \sum_{p_1+p_2=p\atop { k_1+k_2=k}} \mathbb{W}^{\ME}_{p_1, k_1}[v] \, |\del u |_{p_2,k_2}, 
$$
which involves the remainder 
\begin{equation} \label{eq2-04-12-2020-v}
\aligned
\mathbb{W}^{\ME}_{p,k}[v]
&:=    
{ \sum_\gamma |w[v]_\gamma|_{p,k} 
}
+ \sum_{p_1+p_2 = p\atop k_1+k_2=k} |\del  v |_{p_1,k_1} | v |_{p_2,k_2}. 
\endaligned
\end{equation}

Finally, returning to the full expression \eqref{eq3-09-04-2020} and recall \eqref{eq3-04-10-2022}, we obtain 
\begin{equation} \label{eq1-12-04-2020}
\aligned 
|B|_{p,k} & \lesssim   
{\sum_{p_1+p_2=p\atop k_1+k_2=k} |\delts u|_{p_1,k_1} |\del v |_{p_2,k_2}} 
+ \sum_{p_1+p_2=p\atop k_1+k_2=k}  |\delts v |_{p_1,k_1} |\del u |_{p_2,k_2} 
\\
& \quad  
+ \hskip-.3cm  \sum_{p_1+p_2=p}\hskip-.3cm \SbbME_{p_1}[u] |\del v |_{p_2} 
+ \sum_{p_1+p_2=p\atop{k_1 + k_2 = k}} \mathbb{W}^{\EM}_{p_1,k_1}[v] \, |\del u |_{p_2,k_2}
+ \hskip-.3cm \sum_{p_1+p_2+p_3=p} \hskip-.3cm  | h^\star |_{p_3} |\del u|_{p_1} |\del v |_{p_2}. 
\endaligned
\end{equation}
It remains to {
combine \eqref{eq3-04-12-2020} with \eqref{eq1-12-04-2020}} and the desired conclusion is reached.

\begin{lemma}[Estimates of quasi-null interaction terms] 
\label{lem1-31-01-2021}
In the Euclidean-merging domain $\MME$, under the smallness condition $| h^\star |_p + |u|_{[p/2]} \ll 1$ and with $ \SbbME$ defined in \eqref{eq2-04-10-2022} , the quasi-null terms satisfy   
\begin{equation}\label{eq4-04-10-2022} 
\aligned
|\slashed{\Pbb}^{\star\N}[u] |_{p,k}
&
\lesssim \sum_{p_1+p_2=p\atop k_1+k_2=k} |\del u|_{p_1,k_1} |\delsN u |_{p_2,k_2} 
+ \sum_{p_1+p_2+p_3=p} | h^\star |_{p_3} |\del u|_{p_1} |\del u |_{p_2},
\\
|\Pbb_{00}^{\star \Ncal}[u] |_{p,k}
& \lesssim \sum_{p_1+p_2=p\atop k_1+k_2=k} |\del \us^{\Ncal}|_{p_1,k_1} |\del \us^{\N}|_{p_2,k_2} 
+ \hskip-.3cm  {\sum_{p_1+p_2=p} |\delsN u|_{p_1} |\del u |_{p_2}}  
+ \hskip-.3cm  \sum_{p_1+p_2=p}\hskip-.3cm \SbbME_{p_1}[u] |\del v |_{p_2} 
+ \hskip-.3cm \sum_{p_1+p_2+p_3=p} \hskip-.3cm  | h^\star |_{p_3} |\del u|_{p_1} |\del v |_{p_2}.
\endaligned
\end{equation}
{
If in addition $|v|_{[p/2]} \ll 1$ and with $\mathbb{W}^{\ME}$ defined in \eqref{eq2-04-12-2020-v}, 
\begin{equation}\label{eq5-04-10-2022} 
\aligned
|\slashed{\Pbb}^{\star\N}[u, v] |_{p,k}
&
\lesssim \sum_{p_1+p_2=p\atop k_1+k_2=k}\big(  |\del u|_{p_1,k_1} |\delts v |_{p_2,k_2} +  |\del v |_{p_1} |\delts u|_{p_2} \big) 
+ \sum_{p_1+p_2+p_3=p} | h^\star |_{p_3} |\del u|_{p_1} |\del v |_{p_2},
\\
|\Pbb_{00}^{\star \Ncal}[u, v] |_{p,k}
& \lesssim \sum_{p_1+p_2=p\atop k_1+k_2=k} |\del \uts|_{p_1,k_1} |\del \slashed v^\Ncal |_{p_2,k_2} 
+ \hskip-.3cm  {\sum_{p_1+p_2=p} |\delts u|_{p_1} |\del v |_{p_2}} 
+ \sum_{p_1+p_2=p}  |\delts v |_{p_1} |\del u |_{p_2} 
\\
& \quad  
+ \sum_{p_1+p_2=p\atop k_1 + k_2 = k} \mathbb{W}^{\EM}_{p_1,k_1}[v] \, |\del u |_{p_2,k_2}
+ \hskip-.3cm  \sum_{p_1+p_2=p}\hskip-.3cm \SbbME_{p_1}[u] |\del v |_{p_2} 
+ \hskip-.3cm \sum_{p_1+p_2+p_3=p} \hskip-.3cm  | h^\star |_{p_3} |\del u|_{p_1} |\del v |_{p_2}.
\endaligned
\end{equation}
}
\end{lemma}  

%-----------------------------------------------------------------------------------  

\subsection{Gradient and Hessian of the null component}

The component $\gt^{00}$ plays an essential role in the commutator estimates, but the 
bounds deduced from purely PDEs arguments are not sufficiently sharp
in order to handle such commutators. 
At this juncture, a key observation (first made by Lindblad and Rodnianski \cite{LR1} in the standard foliation) is that the wave gauge condition implies that the gradient of this component can be expressed in terms of ``good'' derivatives of other metric components. 
We return to the wave gauge condition
$$
g_{\beta\gamma} \del_{\alpha}h^{\alpha\beta} = \frac{1}{2}g_{\alpha\beta} \del_\gamma h^{\alpha\beta} 
{  - w[g]_{\gamma}}
$$ 
{ with, in fact, $w[g]_{\gamma} = 0$}
expressed in the semi-null frame, i.e. 
$$
g^{\N}_{\beta\gamma} \delN_\alpha h^{\N\alpha\beta} 
= \frac{1}{2} g^{\N}_{\alpha\beta} \delN_\gamma h^{\N\alpha\beta} 
+ \frac{1}{2}g_{\alpha'\beta'} \delN_\gamma\big({\PhiN_{\alpha}^{\alpha'} \PhiN_{\beta}^{\beta'} }\big)h^{\N\alpha\beta}
-h^{\N\alpha\beta} \del_{\alpha'} \big(\PhiN_{\alpha}^{\alpha'} \PhiN_{\beta}^{\beta'} \big)g_{\beta'\gamma'} {\PsiN}_\gamma^{\gamma'} { - \PhiN_{\gamma}^{\gamma'} w[g]_{\gamma'}}. 
$$
Taking $\gamma = c = 1,2,3$ we have 
\begin{equation} \label{eq5-04-12-2020}
\aligned
g^\Ncal_{\Mink, 0c} \del_th^{\N00} 
& =   - g^{\N}_{b c} \del_th^{\N 0b} - g^{\N}_{\beta c} \delts_ah^{\N a\beta} - \hN_{0c} \del_th^{\N00}  + \frac{1}{2} g^{\N}_{\alpha\beta} \delts_ch^{\N\alpha\beta} 
\\
& \quad + \frac{1}{2}g_{\alpha'\beta'} \delN_{ c}\big({\PhiN_{\alpha}^{\alpha'} \PhiN_{\beta}^{\beta'}} \big)h^{\N\alpha\beta}
-h^{\N\alpha\beta} \del_{\alpha'} \big(\tPhiME_{\alpha}^{\alpha'} \tPhiME_{\beta}^{\beta'} \big)g_{\beta'\gamma'} {\PsiN_c^{\gamma'}}  {- \PhiN_c^{\gamma'} w[g]_{\gamma'}}. 
\endaligned
\end{equation}
Recalling that $g_{\Mink, 0c}^{\N} = -(x^c/r)$ and $g_{\Mink, bc}^{\N} = -\frac{x^bx^c}{r^2} + \delta_{bc}$, therefore 
$(x^c/r)g_{\Mink,bc} = 0$, then 
multiplying \eqref{eq5-04-12-2020} by $(-x^c/r)$, and finally summing over $c$,  we find 
$$
\aligned
\del_th^{\N00} 
& = (x^c/r)\hN_{bc} \del_t h^{\N0b} -(x^c/r)\big(- g^{\N}_{\beta c} \delsN_ah^{\N a\beta} - \hN_{0c} \del_th^{\N00}  + \frac{1}{2} g^{\N}_{\alpha\beta} \delsN_ch^{\N \alpha\beta} 
\\
& \quad +  \frac{1}{2}g_{\alpha'\beta'} \delsN_c\big({\PhiN_{\alpha}^{\alpha'} \PhiN_{\beta}^{\beta'}} \big)\hN{}^{\alpha\beta}
-\hN{}^{\alpha\beta} \del_{\alpha'} \big(\tPhiME_{\alpha}^{\alpha'} \tPhiME_{\beta}^{\beta'} \big)g_{\beta'\gamma'} {  \PsiN_c^{\gamma'} }\big) { + \frac{x^c}{r}\PhiN^{\gamma'}_cw[g]_{\gamma'}}. 
\endaligned
$$
Recalling that $g^{\alpha\beta} = g_{\Mink}^{\alpha\beta} + h^{\alpha\beta}$, we arrive at the following estimate.

\begin{lemma}[Null component of the metric. I]
\label{lemma-12-04-2020} 
In the Euclidean-merging  domain $\MME$, the gradient of the null component of the metric satisfies 
$$ 
\aligned
|\del g^{\N00} |_{p,k}  \lesssim |\delsN h|_{p,k} + r^{-1} |h|_{p,k} + \sum_{p_1+p_2=p} |h|_{p_1} |\del h|_{p_2}. 
\endaligned
$$ 
while, for the metric perturbation,
$$|\del u^{\N00} |_{p,k} \lesssim |\delsN u|_{p,k} + r^{-1} |u|_{p,k} 
{  + \sum_\gamma |w^\star_\gamma|_{p,k}
+ \sum_{p_1+p_2=p\atop k_1+k_2=k}\big(|h^\star|_{p_1,k_1} |\del u|_{p_2,k_2} + |\del h^{\star}|_{p_1,k_1}|u|_{p_2,k_2} + |u|_{p_1,k_1}|\del u|_{p_2,k_2}\big)}.
$$  
\end{lemma} 

We will also need a bound on the second-order time-derivative $\del_t\del_t g^{\N00}$ and we first write 
$$
\aligned
g^{\N}_{\alpha\beta} 
& = g^{\N}_{\Mink,\alpha\beta} + \hN_{\alpha\beta} = \PhiN_{\alpha}^{\alpha'}\PhiN_{\beta}^{\beta'} \, g_{\Mink, \alpha'\beta'} + \hN_{\alpha\beta},
\qquad 
g^{\N\alpha\beta} 
= g_\Mink^{\N\alpha\beta} + h^{\N\alpha\beta} 
= \PsiN_{\alpha'}^\alpha \PsiN_{\beta'}^\beta g_\Mink^{\alpha'\beta'} + h^{\N\alpha\beta},
\endaligned
$$
where we recall that $\PhiN_{\alpha}^{\alpha'}$ and $\PsiN_{\alpha'}^\alpha $ are homogeneous functions {\sl
that do not depend on $t$}  (which is used in the calculation of \eqref{eq10-07-12-2020}, below) 
and $g_{\Mink,\alpha\beta}$ are constants. Thus we have 
$\del_t g^{\N}_{\alpha\beta} = \del_t \hN_{\alpha\beta}$
and $\del_t g^{\N\alpha\beta} = \del_t h^{\N\alpha\beta}$. 
We differentiate \eqref{eq5-04-12-2020} with respect to $\del_t$ and obtain 
\begin{equation} \label{eq10-07-12-2020}
\aligned
& g^{\N}_{\Mink,0c} \del_t\del_tg^{\N00} 
\\
& =  -  g^{\N}_{b c} \del_t\del_th^{\N0b} 
- g^{\N}_{\beta c} \del_t\delsN_ah^{\N a\beta} 
+ \frac{1}{2}g^{\N}_{\alpha\beta} \del_t\delsN_c\hN{}^{\alpha\beta}
-  \del_t\hN_{b c} \del_th^{\N 0b}
- \del_t\hN_{\beta c} \delsN_ah^{\N a\beta}+ \frac{1}{2}\del_t\hN_{\alpha\beta}\delsN_c\hN{}^{\alpha\beta}
\\
& \quad  - \del_t\big(\hN_{0c} \del_th^{\N00}  \big) 
+ \frac{1}{2}\del_t\big(g_{\alpha'\beta'} \tdelME_\gamma\big(\tPsiME_{\alpha}^{\alpha'} \tPsiME_{\beta}^{\beta'} \big)\hN{}^{\alpha\beta}\big)
- \del_t\big(\hN{}^{\alpha\beta} \del_{\alpha'} \big(\tPhiME_{\alpha}^{\alpha'} \tPhiME_{\beta}^{\beta'} \big)g_{\beta'\gamma'} \tPhiME_\gamma^{\gamma'}\big). 
\endaligned
\end{equation}
Similarly as in the derivation of Lemma~\ref{lemma-12-04-2020}, we multiply \eqref{eq10-07-12-2020} by $(-x^c/r)$ and sum up with respect to $c = 1,2,3$. In view of the identity $\frac{x^c}{r} g^\Ncal_{\Mink, bc} = 0$, we obtain 
$$
\aligned
\del_t\del_t h^{\N 00}
& = (x^c/r)\hN_{bc}\del_t\del_th^{\N0b} 
+ (x^c/r)g^{\N}_{\beta c}\del_t\delsN_ah^{\N a\beta} 
- \frac{x^c}{2r}g^{\N}_{\alpha\beta} \del_t\delsN_c\hN{}^{\alpha\beta}
\\
& \quad + (x^c/r)\del_t\hN_{b c} \del_th^{\N0b}
+ (x^c/r)\del_t\hN_{\beta c} \delsN_ah^{\N a\beta}
- \frac{x^c}{2r}\del_t\hN_{\alpha\beta}\delsN_c\hN{}^{\alpha\beta}
\\
& \quad + (x^c/r)\del_t\big(\hN_{0c} \del_th^{\N00}  \big)  
-(x^c/r) \frac{1}{2}\del_t\big(g_{\alpha'\beta'} \tdelME_\gamma\big(\tPsiME_{\alpha}^{\alpha'} \tPsiME_{\beta}^{\beta'} \big)\hN{}^{\alpha\beta}\big)
+ (x^c/r) \del_t\big(\hN{}^{\alpha\beta} \del_{\alpha'} \big(\tPhiME_{\alpha}^{\alpha'} \tPhiME_{\beta}^{\beta'} \big)g_{\beta'\gamma'} \tPhiME_\gamma^{\gamma'}\big),
\endaligned
$$
which leads us to the following estimate. 

\begin{lemma}[Null component of the metric. II]
\label{lem1-21-09-2022}
In the Euclidean-merging  domain $\MME$ the null component $h^{\N 00}$ satisfies  
\begin{subequations}
\begin{equation} 
|\del_t\del_t h^{\N 00} |
\lesssim |\del\delsN h| + r^{-1} \, (1+|h|) \, |\del h| + |\del h|^2 + |h||\del\del h|
\end{equation}
and, more generally, 
\begin{equation} \label{lem2-08-12-2020}
|\del_t\del_t h^{\N 00} |_{p,k}
\lesssim  |\del\delsN h|_{p,k} + r^{-1} | \del h|_{p,k} 
+  \hskip-.3cm  \sum_{p_1+p_2=p\atop k_1+k_2=k}   \hskip-.1cm   
\big( |h|_{p_1,k_1}    |\del\del h|_{p_2,k_2} 
+ |\del h|_{p_1,k_1} |\del h|_{p_2,k_2} \big)
+  r^{-1} \hskip-.3cm \sum_{p_1+p_2=p\atop k_1+k_2=k} |\del h|_{p_1,k_1} |h|_{p_2,k_2}.
\end{equation}
\end{subequations}
\end{lemma}

%====================================================================================================  

\section{Strategy of proof and consequences of the energy estimates}
\label{sec1-23-05-2021}

\subsection{Assumptions and bootstrap strategy}
\label{section-label-11-1}

\paragraph{Objective.}

We are now in a position to present our proof of nonlinear stability for the Einstein equations. In this section after stating our assumptions on the data and our bootstrap conditions, we begin with direct consequences of the energy estimates and (Sobolev, Hardy) functional inequalities. Recall that our global existence theory concerns the metric perturbation $u = (u_{\alpha\beta}) = g - g^\star$ (also referred to as the wave field) and the matter unknown $\phi$ (also referred to as the Klein-Gordon field).  We fix (once for all) a sufficiently large integer $N$ ($N =20$ being sufficient) which is determined only by the structure of the Einstein-matter system. 

From now on, we find it convenient to work under weaker assumptions in comparison to the ones first introduced in Section~2 and then restated in the PDE setup in Section~\ref{section-new-11}. Indeed, our analysis requires rather weak decay conditions, especially on the reference metric, but their formulation is quite involved and therefore is stated only in this Part~2. Our motivation for the organization of the proof is two-fold: first of all, many of our estimates are expected to remain relevant in future generalizations under low decay or regularity conditions; second, our presentation separates between various contributions of the reference metric, initial data, and solution components, and this helps understand the structure and the role of these contributions arising in the Einstein equations. 

Our estimates below cover the full range $\lambda \in (1/2,1]$ and so require low decay conditions. Although the regime $\lambda=1$ is included here, our estimates in this regime are not sharp in the form derived below and additional estimates are derived in the companion paper~\cite{PLF-YM-SecondPart}. 
Consequently, without loss of generality as far as the statement in Section~2 is concerned, we find it convenient to relax our assumptions and now assume $\lambda \in (1/2,1)$ {\sl in the following presentation} by ``loosing'' some decay from the assumptions in Section~2. 
Furthermore, our working assumptions on $h^\star$ now are much weaker and are presented in the following paragraphs as two sets of assumptions, namely the {\sl Class A and Class B metrics.} 

The decay conditions in Class A are stated in a pointwise manner and arise as conditions on perturbations of exact vacuum solutions. On the other hand, the decay conditions defining Class B metrics are typically realized when the reference metric is defined by a time evolution from a vacuum initial data set and, for instance, is an exact vacuum solution possibly expressed in an ``approximate'' wave gauge. 
The estimates required to cover Class B metrics are more involved and are presented separately when necessary.
 
%------------------------------------------

\paragraph{Conditions on the reference spacetime metric.} 

Given 
\begin{equation}\label{equa-paramet-repeat}
1/2 < \lambda  
< 1, 
\qquad  0<\theta< 1-\lambda, 
\qquad   
\kappa \in (1/2, 1), 
\end{equation} 
we consider a reference spacetime metric $(\RR^{3+1}_+, g^\star)$, understood in the following sense.  

\begin{itemize}
\item  
The following {\sl asymptotically Minkowski} conditions are assumed  
by $g^\star = \gMink + h^\star$. We distinguish between 
\begin{equation}\label{equa-31-12-20}    
\textbf{Class A:}
\quad
\begin{cases} 
\aligned
|h^{\star} |_{N+2} + \la r+t\ra|\del h^{\star} |_{N+1} + \la r+t\ra^2|\del\del h^{\star} |_{N}
& \lesssim \epss \la r+t\ra^{-\lambda},
\\
  
|  w^\star_\gamma  |_{N} &    \lesssim  \epss \,  \la r + t\ra^{-1}\crochet^{-1-\varsigma},
\endaligned
\end{cases}
\end{equation}
and 
\begin{equation}\label{equa-new-conditions-hstar}
\textbf{Class B:} \quad
\begin{cases}
\aligned
|h^\star |_{N+1} 
& \lesssim \epss \, \la r+t\ra^{-\lambda}, 
\\
|\del^{m} \del h^\star |_{N-m} 
& \lesssim \epss \, \la r+t\ra^{-1+\theta} \crochet^{-\kappa -m}, \quad 
&& m=0 \text{ and } 1, 
\\
|\del^m \slashed \del h^{\star} |_{N-m} 
& \lesssim \epss \, \la r+t\ra^{-1-\kappa}\crochet^{-m}, \quad 
&& m=0 \text{ and } 1, 
\\
|\del\hs^\star | & \lesssim \epss \la r+t\ra^{-1} \quad  && \text{in } \Mnear_{\ell,[s_0,\infty)}, 
\\
|  w^\star_\gamma  |_{N} & \lesssim  \epss \,  \la r + t\ra^{-1}\crochet^{-1-\varsigma},
&&  
\endaligned
\end{cases}
\end{equation} 
in which we have set $
w^\star_\gamma := w[h^{\star}]_{\gamma}
=
g^{\star \alpha\beta} \del_{\alpha} h^\star_{\beta\gamma} - \frac{1}{2} g^{\star \alpha\beta} \del_\gamma h^\star_{\alpha\beta}
$
and $\varsigma >0$ is a fixed parameter which can be taken to be infinity if the gauge condition holds exactly (that is, when $w^\star_\gamma=0$). 
Observe that the approximate wave gauge condition \eqref{eq2-04-11-2022} is then obviously satisfied. 
Here, the weight $\crochet$ is $1$ in the interior region and, in the exterior region, coincides with the distance to the light cone.

%------------------------------ 

\item The  {\sl radial and frame tame decay.} The following conditions are required for, both, Class A and Class B metrics. 
We introduce\footnote{The negative signs here come from raising the indices as explained in \eqref{equa-signe-a-noter}.}  the linear part of $h^{\star\alpha\beta}$ expressed by $h^{\star}_{\alpha\beta}$:
\begin{equation}\label{equa-signe-a-noter}
\Xi^{\star\alpha\beta} := -{ \mathbbm{h}^{\star\alpha\beta}},\quad
\qquad 
h^{\star\alpha\beta} = \Xi^{\star\alpha\beta} + \Abb^{\alpha\beta}[h^{\star}], 
\end{equation}
in which, using a matrix notation as in Lemma~\ref{lem-small}, 
\begin{equation}\label{eq1-07-05-2021}
\Abb^{\alpha\beta}[h] 
=  \big((h + g_{\Mink})^{-1} - g_{\Mink}\big)_{\alpha\beta} + { \mathbbm{h}^{\alpha\beta}} 
= { \Big(\sum_{k=2}^{+\infty}(-g_{\Mink}h)^k  g_{\Mink}\Big)_{\alpha\beta}}. 
\end{equation}

We also introduce the components 
\begin{equation}\label{eq2-03-01-2022}
\aligned
\Xi^{\star 00} & := - { \mathbbm{h}^{\star00}}, 
\qquad 
\Xi^{\star0a} := - { \mathbbm{h}^{\star0a}},
\qquad 
\Xi^{\star rr} := -(x^ax^b/r^2){ \mathbbm{h}^{\star  ab}}, 
\qquad
\\
\Xi^{\star\N00} & := - { \mathbbm{h}^{\star00}} + 2(x^a/r) { \mathbbm{h}^{\star0a}} - (x^ax^b/r^2){ \mathbbm{h}^{\star  ab}}, 
\endaligned
\end{equation} 
so 
$$
\Xi^{\star\N00} = \Xi^{\star 00} - 2(x^a/r)\Xi^{\star 0a} + \Xi^{\star rr}.
$$
By \eqref{equa-31-12-20}, and the fact that $|h^{\star} |\lesssim \epss\ll 1$,
\begin{equation}\label{eq3-03-01-2022}
|\Abb[h^{\star}]|_{N+2}\lesssim \epss^2 \, \la r+t\ra^{-2\lambda}.
\end{equation}
For these components, we require
\begin{equation}\label{eq2-29-03-2021-gstar}
\big| \Xi^{\star 00}\big|_N + \big|\Xi^{\star 0a}\big|_N + \big|\Xi^{\star\rr} \big|_{N} 
\lesssim \epss\la r+t\ra^{-1+\theta}
\qquad \text{ in }  \Mscr_{[s_0, + \infty)}. 
\end{equation}
Observe that this implies the same decay for the null component 
\begin{equation}\label{eq1-11-03-2021-gstar}
|\Xi^{\star\N00} |_{N} \lesssim \epss \la t+r\ra^{-1+\theta}
\qquad \text{ in }  \Mscr_{[s_0, + \infty)}. 
\end{equation}

%------------------------------ 

\item  The {\sl almost Ricci flat condition.} 
We also have a decay property on the Ricci curvature of $g^{\star}$ which expresses our assumption that $g^\star$ is an ``approximate solution'' to Einstein's vacuum equation. For metrics in either Class A or Class B
(in wave gauge with { $\kappa>1/2$}) we impose 
\begin{equation}\label{eq4-09-05-2021}
|\wR^{\star} |_{N} + \la r-t \ra|\del\wR^{\star} |_{N-1}
\lesssim
\begin{cases}
\epss^2 \, \la r+t\ra^{-2-2\lambda} \quad &\text{ in } \MME_{[s_0, + \infty)},
\\ 
\epss \la r+t\ra^{-2-\lambda} \quad &\text{ in } \MH_{[s_0, + \infty)}.
\end{cases}
\end{equation}  
In particular, we easily deduce from this pointwise condition \eqref{eq4-09-05-2021} we easily deduce the integral condition. 
\begin{equation}\label{eq1-21-05-2021}
\quad 
\aligned
\|\crochet^\kappa J{\zeta}^{-1} \, |\wR^\star_{\alpha\beta} |_N \|_{L^2(\MME_s)} 
& \leq
R^{\err}_{\star}(s), 
\qquad
\quad
R^\err_\star(s)\lesssim C_{R^{\star}}\epss^2 s^{-1-\delta}, 
\endaligned
\end{equation}
and in particular 
$\int_{s_0}^{+\infty} R^{\err}_{\star}(s') \, ds' \lesssim \delta^{-1} C_{R^{\star}} \epss^2$. 
% 

%------------------------------------- 

\item The {\sl light-bending property.} The following conditions are required for, both, Class A and Class B metrics. 
Recalling \eqref{eq1-03-01-2022} and \eqref{eq2-03-01-2022}, the light-bending condition \eqref{equa-bending} is written as 
$$
\inf_{\Mscr^\near_{\ell}} \big(-r\Xi^{\star\N00}\big)\geq 4\epss.
$$ 
Then combined with \eqref{eq3-03-01-2022} and recall the relation \eqref{equa-signe-a-noter}, the following bound holds, provided that $\epss\ll 1$:
\begin{equation}\label{equa-bending-repeat} 
\epss \leq \inf_{\Mscr^\near_{\ell}} \big(-rH^{\star\N00}\big).
\end{equation}
This sign condition can be derived from integral properties on the initial data set, by using Kirchhoff's formula and controlling quadratic terms within the Einstein equations.  

\end{itemize} 

{   
\noindent Our result announced in Section~2 is reached as follows. When $\lambda$ therein is strictly less than one, we can choose the above parameters such that $\eps \ll \theta \ll \delta \ll 1-\lambda$. 
When $\lambda$ equals $1$ in Section~2, it is obvious that all the conditions also hold for a smaller exponent and therefore the analysis in the forthcoming sections also applies; however, when $\lambda$ equals $1$ in the setup of Section~2 we can establish yet stronger estimates and we refer the reader to our companion paper~\cite{PLF-YM-SecondPart}. 
Interestingly, our calculations below are general since we carefully distinguish between different decay rates in each step of our analysis.}
%  

%-------------------------------------------

\paragraph{Conditions on initial data.}

  From now on,  {\sl  we fix the exponents $\kappa$ and $\mu$ since}
the factors $(1-\kappa)^{-1}$ (appearing for instance in \eqref{eq 1 lem 2 d-e-I} which is $(1-\eta)^{-1}$ with the notation therein) 
and $(1-\mu)^{-1}$ be implicitly arise in our estimates. In other words, the notation $\lesssim$ is used with 
constants tacitly depending upon 
$(1-\kappa)^{-1}$  and $(1-\mu)^{-1}$ (which are fixed once for all). 

\begin{itemize}   

\item In the hyperboloidal domain, we assume 
\begin{subequations}\label{eqs-int-00}
\begin{equation} 
\Fenergy_g^{\Hcal, N-5}(s_0,u) + s_0^{-1/2} \, \Fenergy_{g,c}^{\Hcal,N-5}(s_0, \phi) \leq C_0\eps \, s_0^{\delta},
\end{equation}
\begin{equation} 
\Fenergy_g^{\Hcal,N-7}(s_0,u) + \Fenergy_{g,c}^{\Hcal,N-7}(s_0, \phi) \leq C_0 \eps \, s_0^{\delta}. 
\end{equation}
\end{subequations}
\item In the Euclidean-merging domain, we assume 
\begin{subequations}\label{eqs-ext-00}

\begin{equation}\label{eq5-03-05-2020-00}
\Fenergy_{g,\kappa}^{\ME,N}(s_0,u) + s_0^{-1} \, \Fenergy_{g,\mu,c}^{\ME,N}(s, \phi) \leq C_0\eps \, s_0^{\delta},
\end{equation}
\begin{equation}\label{eq6-03-05-2020-00}
\Fenergy_{g,\kappa}^{\ME,N-5}(s_0,u) + \Fenergy_{g,\mu,c}^{\ME,N-5}(s_0, \phi) \leq C_0 \eps \, s_0^{\delta}. 
\end{equation}
\end{subequations}

\item The following decay on the linear development $u_\init$ of the initial data $(u_{0 \alpha\beta},u_{1 \alpha\beta})$ is a result of Proposition \ref{prop1-12-01-2022}.
\begin{equation}\label{eq3-09-05-2021}
|u_{\init, \alpha\beta} |_{N-4}\lesssim C_0\eps (t+r+1)^{-1+\theta}
\quad \text{ in }  \Mscr_{[s_0, + \infty)}.
\end{equation} 

\item Finally, the following light-bending condition 
is assumed:
\begin{equation}\label{eq3'-27-05-2020-initial}
\inf_{\Mscr^\near_{\ell, [s_0, + \infty)}} \big(- r \, u^{\N00}_\init - r \, \Xi^{\star\N00} \big) 
\geq \epss. 
\end{equation} 
\end{itemize}

%--------------------------------------------- 

\paragraph{Choosing the constants.} 

In the above conditions we have introduced the constants
$ 
N,\, \lambda, 
\, \epss,\, \ell,\, \kappa,\, \mu,\, C_1,\, \eps,\,\delta, 
$
which play different roles in the following analysis. Let us clarify the relations between them. We classify these constants into two groups and, specifically, we distinguish between the {\sl smallness parameters} $\epss,C_1,\eps$ and, on the other hand, the {\sl data and technical parameters.} Roughly speaking, the data and technical parameters determine the general class of initial data under consideration, as 
well as the energy spaces under consideration. These data and technical parameters are fixed, once for all, before developing the bootstrap argument.  In contrast, the smallness parameters describe the size of the initial data and will be determined later in the course of the bootstrap argument. 

We choose the constants as follows. 
\begin{itemize}

\item {\bf Reference parameters.} Our conditions\footnote{We recall that $A\ll B$ stands for $A \leq c_0(N) \, B$, where $c_0(N)>0$ is a small numerical constant.} on the reference metric $g^{\star}$ involve the exponents $(N,\lambda)$. The parameter $ \ell \in (0, 1/2]$ required in \eqref{equa-bending-repeat} is also fixed from now (and is arbitrary). We also fix a sufficiently large integer, say $N\geq 20$.

\item {\bf Energy parameters.} We require $\kappa\in (1/2,1)$ and $\mu\in (3/4,1)$, together with 
\begin{equation}\label{eq1-02-11-2021}
\theta \ll \min(\kappa-1/2,\mu -3/4).
\end{equation}

\item {\bf Technical parameter.} A small exponent denoted $\delta$ arises in our proof for technical reasons, and we require that  
\begin{equation}\label{eq2-02-11-2021}
\theta \ll  \delta < (2/9) \min(\kappa-1/2,\mu-3/4,\varsigma),
\qquad \delta\leq \ell,\qquad \delta\ll 1.
\end{equation} 
Clearly, the right-hand side of the first condition measures the ``criticality'' of the exponents $(\kappa,\mu)$. 

\item{\bf Smallness parameters.} 
Finally, the parameters $C_1,\eps, \epss$ are determined in the course of the bootstrap proof. For instance, we will see that the following condition is sufficient: 
\begin{equation}\label{eq3-02-11-2021} 
(\epss + C_1\eps)\lesssim \delta^6. 
\end{equation} 
\end{itemize}

\indent This completes the description of our choice of parameters and exponents. 

%--------------------------------------------------------

\paragraph{Bootstrap strategy.}

From the standard local existence theory, it is known that {the Einstein-Klein-Gordon system}
\eqref{eq5-03-01-2022} together with data prescribed on $\Mscr_{s_0}$ admits a local-in-time solution provided the initial data are sufficiently small and regular (say, when they belong to the Sobolev class $H^N(\Mscr_{s_0})$ with $N\geq 5$ at least). Our proof of global existence  relies on a bootstrap argument which addresses three issues. 
\begin{itemize} 

\item[1.] {\sl Blow-up criterion.} A sufficiently regular, local-in-time solution cannot approach its maximal time of existence, say $s^*$, at a time the energy at a sufficiently high order still remains bounded (with respect to the time variable). Namely, if this happens we can always extend this solution from $s^*-\alpha$ to $s^*+\alpha$, say, by applying the local-in-time existence theory for some sufficiently small $\alpha>0$ and this would contradict the fact that $s^*$ is maximal.

\item[2.] {\sl Continuity criterion.}
The  energy norms (as well as some other related expressions at lower-order of differentiation) associated with the regularity of the initial data then depend continuously upon the time variable, as long as a local-in-time solution exists. 

\item[3.] {\sl Improved bound criterion.}
Suppose that on a time interval $[s_0,s_1]$ the solution satisfies a set of inequalities containing 
(1) an energy bound at a sufficiently high-order of differentiation and, in addition, 
(2) other expressions (that is, functionals) of the solution. 
Suppose that we can prove that the same inequalities remain valid  
but as a {\sl stronger} set of inequalities with strictly smaller constants. In these circumstances,  
we deduce that the solution {\sl extends beyond} $s_1$.  

Indeed, this is so since if $[s_0,s_1]$ is the maximal interval on which the set of inequalities holds, then at the ``final'' time $s_1$ thanks to the continuous criterion, at least {\sl one of the inequalities} under consideration 
must become an {\sl equality.} However, in the case when we can prove that stronger inequalities holds on the {\sl same interval} then none of the inequalities should become an equality at $s_1$.  
This leads one to the conclusion that the set of inequalities does hold within the maximal time of existence $[s_0,s^*)$, 
which is excluded thanks to the blow-up criterion above, unless of course $s^* = +\infty$. 
\end{itemize}

%--------------------------------------------------------

\paragraph{Bootstrap assumptions.}

In earlier work on nonlinear wave equations, the set of inequalities under consideration, referred to as the {\sl bootstrap assumptions}
consist of energy and decay estimates, only. For the problem under consideration we work with a
somewhat non-standard set of bootstrap assumptions and distinguish between estimates at low- or high-order of differentiation, 
estimates within the hyperboloidal and Euclidean-merging domains, 
and a positivity condition near the light cone. 

\begin{itemize} 

\item In the hyperboloidal domain, for all $s \in [s_0, s_1]$ we assume 
\begin{subequations}\label{eqs-int}
\begin{equation}  \label{eqsa-int} 
\Fenergy^{\Hcal, N-5}(s,u) + s^{-1/2} \, \Fenergy_c^{\Hcal,N-5}(s, \phi) \leq (\epss+C_1\eps) \, s^{\delta},
\end{equation}
\begin{equation}  \label{eqsb-int}
\Fenergy^{\Hcal,N-7}(s,u) + \Fenergy_c^{\Hcal,N-7}(s, \phi) \leq (\epss+C_1\eps) \, s^{\delta}. 
\end{equation}
\end{subequations}

\item In the Euclidean-merging domain, for all $s \in [s_0, s_1]$ we assume 
\begin{subequations}\label{eqs1-14-01-2021}  
\begin{equation}\label{eq1-14-01-2021}
\Fenergy_{\kappa}^{\ME,N}(s,u) + s^{-1} \, \Fenergy_{\mu,c}^{\ME,N}(s, \phi) \leq(\epss+C_1\eps) \, s^{\delta},
\end{equation}
\begin{equation}\label{eq2-14-01-2021}
\Fenergy_{\kappa}^{\ME,N-5}(s,u) +\Fenergy_{\mu,c}^{\ME,N-5}(s, \phi) \leq(\epss+C_1\eps) \, s^{\delta}, 
\end{equation}
\end{subequations}
while, in the course of our analysis, we will also control the spacetime integral\footnote{which
we do not need to include in the set of bootstrap assumptions}
\begin{equation}\label{equa-new-spacetime-bound}
{\mathscr G}_\kappa^{\ME,p,k}(s_0, s,u)
:=
\sum_{\ord(Z)\leq p\atop \rank(Z)\leq k}\int_{s_0}^{s}
\int_{\MME_\tau}\crochet^{2\kappa-1} |\delsN Z u|^2 J \, dxd\tau
\leq (\epss+C_1\eps)^2 s^{2\delta}.
\end{equation}

%----------------

\item Near the light cone, we assume the light-bending condition
\begin{equation}\label{eq3-27-05-2020} 
\inf_{\Mscr^\near_{\ell, [s_0,s_1]}} (- \hN{}^{00} )\geq 0.
\end{equation}

\end{itemize} 

\noindent Observe that the inequalities \eqref{eqs-int} and \eqref{eqs1-14-01-2021} involve a (sufficiently large) constant $C_1>C_0$ (which will be chosen at the end of our bootstrap argument) as well as
{ the exponents $\kappa>1/2$ and $\mu >3/4$ }
which control the decay in spacelike directions. 
On the other hand, \eqref{eq3-27-05-2020} is neither an energy estimate nor a decay estimate; yet, it does make sense to include it as a bootstrap assumption.
Recall that we have decomposed  
\begin{equation}
\hN{}^{00} = h^{\star\N00} + u^{\Ncal 00}, 
\end{equation}
in which $h^{\star\N00}$ is continuous (at least) so that the left-hand side of \eqref{eq3-27-05-2020} 
is a continuous function of the time variable $s$, as long as the solution $(u_{\alpha\beta}, \phi)$ is 
(well-defined and) continuous at least. 

In some of our estimates we will distinguish between $h_{\alpha\beta}$ (with lower indices)  and $h^{\alpha\beta}$ and the notation 
$H^{\alpha\beta} := h^{\alpha\beta}$ will be used and, more specifically, 
in agreement with the notation in Section~\ref{sectn:cmm-ge}
we will write 
\begin{equation}\label{equa-notation-H} 
|H|_{p,k} := \max_{\alpha,\beta} |h^{\alpha\beta} |_{p,k}.  
\end{equation}

%------------------------------------------------------------------------------------

\paragraph{Main objective.}

We assume that $[s_0,s_1]$ is the maximal interval of time within which \eqref{eqs-int}, \eqref{eqs1-14-01-2021}, and \eqref{eq3-27-05-2020} hold so that, by continuity, one of these conditions is an equality at the end time $s_1$. 
Our objective is to establish the following {\bf  improved estimates} for all $s \in [s_0, s_1]$:
\begin{subequations}\label{eqs'-int}
\begin{equation}
\Fenergy^{\Hcal, N-5}(s,u) + s^{-1/2} \, \Fenergy_c^{\Hcal,N-5}(s, \phi) \leq \frac{1}{2}(\epss+C_1\eps) \, s^{\delta},
\end{equation}
\begin{equation}
\Fenergy^{\Hcal,N-7}(s,u) + \Fenergy_c^{\Hcal,N-7}(s, \phi) \leq \frac{1}{2}(\epss+C_1\eps) \, s^{\delta},
\end{equation}
\end{subequations}
\begin{subequations}\label{eqs'-ext}
\begin{equation}\label{eq5'-03-05-2020}
\Fenergy_{\kappa}^{\ME,N}(s,u) + s^{-1} \, \Fenergy_{\mu,c}^{\ME,N}(s, \phi) \leq\frac{1}{2}(\epss+C_1\eps) \, s^{\delta},
\end{equation}
\begin{equation}\label{eq6'-03-05-2020}
\Fenergy_{\kappa}^{\ME,N-5}(s,u) +\Fenergy_{\mu,c}^{\ME,N-5}(s, \phi) \leq \frac{1}{2}(\epss+C_1\eps) \, s^{\delta},
\end{equation}
\end{subequations}
\begin{equation}\label{eq3'-27-05-2020}
\inf_{\Mscr^\near_{\ell, [s_0,s_1]}} (- r \, \hN{}^{00} ) 
\geq  {1\over 2} \epss.
\end{equation}
The spacetime condition is unnecessary for metrics in Class A. 
It will follow that, at the time $s_1$, none of \eqref{eqs-int}, \eqref{eqs1-14-01-2021}, and \eqref{eq3-27-05-2020} can hold as an equality. Consequently, in view of the three bootstrap conditions discussed at the beginning of this section, the solution extends  indefinitely to all times $s \geq s_0$. This is our main task for the rest of this paper. 
\begin{itemize}

\item We focus our attention on the Euclidean-merging domain until Section~\ref{sec-closing-bootstrap}, while 
a direct generalization of our method in \cite{PLF-YM-one,PLF-YM-two} 
will suffice to cover the interior domain and will be presented in the final Section~\ref{section-18}. 

\item In the present section we begin, in the Euclidean-merging domain, the derivation of the improved bounds \eqref{eqs'-ext} and \eqref{eq3'-27-05-2020}. Accordingly, the arguments in the rest of this section are based on the bootstrap assumptions \eqref{eqs1-14-01-2021} and \eqref{eq3-27-05-2020}. 
One of our arguments below (when applying Proposition~\ref{eq3-15-05-2020}) will also require an estimate
on the zero-order norm $\Fenergy^{\Hcal,0}(s,u)$.  

\end{itemize}

%--------------------------------------------------------------------------------------------------------------------------------------------

\subsection{Basic $L^2$ and Sobolev estimates}
\label{subsec1-30-05-2020}

\paragraph{$L^2$ estimates.}

Relying mainly on the general integral estimates established in Proposition~\ref{prop 1 L2-be},  
we now derive $L^2$ estimates which are direct consequences of our bootstrap assumptions. 
We distinguish between the wave field $u$ (that is, any component of the metric perturbation) and the Klein-Gordon field $\phi$, 
as well as between estimates at low- or high-order of differentiation. 
We distinguish between the behavior of the fields as well as the behavior of their derivatives $\del$, $\del\del$, and $\delsN$.  At this stage, our estimates hold for, both, Class~A and Class~B.  

The following $L^2$ estimates for the wave fields and the Klein-Gordon fields 
are immediate from the bootstrap assumption \eqref{eq1-14-01-2021} and \eqref{eq2-14-01-2021} in the Euclidean-merging domain:
\begin{subequations}\label{eq7-03-05-2020}
\begin{equation}\label{eq7a-03-05-2020}
\| \crochet^\kappa \zeta \, | \del u|_N \|_{L^2(\MME_s)}
+
\| \crochet^\kappa |\delts u|_N\|_{L^2(\MME_s)} 
\lesssim (\epss+C_1\eps) \, s^{\delta}, 
\end{equation}
\begin{equation}\label{eq7b-03-05-2020}
\| \crochet^\mu \zeta \, |\del \phi|_p\|_{L^2(\MME_s)} + 
\| \crochet^\mu |\delts \phi|_p\|_{L^2(\MME_s)} +
\| \crochet^\mu |\phi|_p\|_{L^2(\MME_s)}
\lesssim
(\epss+C_1\eps) \, 
\begin{cases} 
s^{1+\delta}, \quad & p=N,
\\
s^{\delta}, & p=N-5,
\end{cases}
\end{equation}
\end{subequations}  

%--------------------------------------------------------------------------------------

\paragraph{Consequence of the weighted Poincar\'e inequality.}

Recalling our inequality in Proposition~\ref{eq3-15-05-2020} valid  for any 
{ $\eta>1/2$, with $\eta=\kappa \geq1/2 +\delta$, and $\kappa<1$,} we find 
\begin{equation}\label{eq1-12-05-2020}
\| \crochet^{-1+\kappa} |u|_{p,k} \|_{L^2(\MME_s)} 
\lesssim \delta^{-1} \, \Fenergy_{\kappa}^{\ME,p,k}(s,u) + \Fenergy_{\kappa}^{0}(s,u)
\lesssim \delta^{-1}  (\epss+C_1\eps)  \, s^{\delta},
\end{equation}
which provides us with a control of the metric components possibly without partial derivatives. 

%---------------------------------------------------------------------------

\paragraph{Pointwise decay of the metric components.} 

Basic sup-norm estimates are derived by relying on the generalized Sobolev estimate\footnote{We neglect the factor $1/(1-\kappa)$ since we have fixed $\kappa<1$.} 
\eqref{eq 1 lem 2 d-e-I} in Proposition~\ref{lem 2 d-e-I} in combination with the high-order bootstrap assumption \eqref{eq1-14-01-2021}. Namely, we control the wave fields at order $N-3$, as follows:  
\begin{equation}\label{eq10-02-05-2020}
r \, \crochet^\kappa \, |\del u|_{N-3} 
+ r^{1+\kappa} |\delts u|_{N-3} 
\lesssim (\epss+C_1\eps)  \, s^{\delta}
\quad \text{in }\MME_{[s_0,s_1]}.
\end{equation}
Combining this result with \eqref{equa-31-12-20} or \eqref{equa-new-conditions-hstar}, for  $\del h$ we obtain (recalling that $\theta\ll\delta$)
\begin{equation}\label{eq1-26-05-2021}
\crochet^{\min(\lambda,\kappa)} |\del h|_{N-3} + r^{\min(\lambda,\kappa)} |\delsN h|_{N-3}\lesssim (\epss + C_1\eps)r^{-1}s^{\delta}
\quad \text{in }\MME_{[s_0,s_1]}.
\end{equation}
We apply \eqref{eq 1 lem 2 d-e-I-facile} together with \eqref{eqsa-int}, \eqref{eqs1-14-01-2021} and obtain 
\begin{equation}\label{eq7-15-05-2020}
r \, \crochet^{\kappa-1} \,  |u|_{N-2} \lesssim \delta^{-1} \, (\epss+C_1\eps)  \, s^{\delta} 
\quad \text{in }\MME_{[s_0,s_1]}. 
\end{equation}
In view of $h_{\alpha\beta} = h^{\star}_{\alpha\beta} + u_{\alpha\beta}$ and recalling \eqref{equa-31-12-20} or \eqref{equa-new-conditions-hstar}, 
\begin{equation}\label{eq1-09-05-2021}
|h|_{N-2}\lesssim 
\delta^{-1} (\epss + C_1\eps)r^{-\min(\lambda,\kappa)}s^{\delta}
\quad \text{in }\MME_{[s_0,s_1]}.
\end{equation}

%--------------------------------------------------------

\paragraph{Pointwise decay of the Klein-Gordon field.}

Similarly as stated in \eqref{eq10-02-05-2020} (for a wave field),  the Sobolev decay inequality in 
Proposition~\ref{lem 2 d-e-I} and the bootstrap assumptions \eqref{eqs1-14-01-2021} provide us with sup-norm estimates for the Klein-Gordon field:
\begin{equation}\label{eq11a-02-05-2020}
\| r \, \crochet^\mu \, |\del \phi|_{p-3} \|_{L^\infty(\MME_s)}
+\| r^{1+\mu} |\delts \phi|_{p-3} \|_{L^\infty(\MME_s)}
\lesssim  (\epss+C_1\eps)  \, 
\begin{cases} 
s^{1+\delta}, \quad &p=N,
\\
s^{\delta}, \quad & p=N-5,
\end{cases}
\end{equation}
and, thanks to the consequence \eqref{eq decay-v-repeat000-two} of our generalized Sobolev inequality,
\begin{equation}\label{eq1-18-05-2020}
\| r \, \crochet^\mu \, |\phi|_{p-2} \|_{L^\infty(\MME_s)}
\lesssim  
(\epss+C_1\eps)  \, 
\begin{cases}
s^{1+\delta}, \quad
& p=N,
\\
s^{\delta}, & p=N-5.
\end{cases}
\end{equation}
However, within $\Mnear_s$ this is not sufficient for our purpose below and we can establish a stronger decay, as follows.

\begin{lemma} 
\label{lemma-111} 
Under the conditions stated in Section~\ref{section-label-11-1}, 
the Klein-Gordon field satisfies the pointwise bound 
$$ 
r \, \crochet^\mu \, |\phi|_{p-4} 
\lesssim 
(\epss+C_1\eps) \, \big(r^{-1}\crochet + r^{-\lambda}\big)
\begin{cases}
s^{1+2\delta},  \quad  
& p=N,
\\
s^{2\delta}, \quad 
&  p=N-5. 
\end{cases}
$$
\end{lemma} 

\begin{proof} We recall the previous statement \eqref {eq1-18-05-2020}. It remains to deal with the domain $\MMEnear_s$, and we need here the decay property near the light cone derived in Proposition~\ref{lem 1 d-KG-e}. We consider the Klein-Gordon equation 
$
g^{\alpha\beta} \del_{\alpha} \del_{\beta} \phi - c^2 \phi = 0
$
and with the notation in  Proposition~\ref{lem 1 d-KG-e}, we set 
$$
f = h^{\mu\nu} \del_{\mu} \del_{\nu} \phi = h^{\star\mu\nu} \del_{\mu} \del_{\nu} \phi + u^{\mu\nu} \del_{\mu}\del_{\nu}\phi.
$$ 
For the first term the right-hand side above, the decay condition on $h^{\star}$ in \eqref{equa-31-12-20}
is applied and  yields us 
$$
| h^{\star\mu\nu} \del_{\mu} \del_{\nu} \phi|_{p-4} \lesssim 
\epss C_1 \eps \, r^{-1-\lambda} \crochet^{-\mu}
\begin{cases} 
s^{1+\delta}, \quad & p = N,
\\
s^{\delta}, \quad & p=N-5.
\end{cases}
$$
For the second term $u^{\mu\nu} \del_{\mu}\del_{\nu}\phi$, by Lemma~\ref{lem-small} we have $|u^{\mu\nu} |_p\lesssim |u|_p$ and, by recalling the Sobolev decay \eqref{eq7-15-05-2020} and \eqref{eq11a-02-05-2020}, we find  
$$
|u^{\mu\nu} \del_{\mu} \del_{\nu} \phi|_{p-4} \lesssim |u|_{p-4} |\del\phi|_{p-3} \lesssim 
\delta^{-1}(\epss+C_1\eps)^2 r^{-2} \crochet^{1-\kappa - \mu}
\begin{cases} 
s^{1+2\delta},  & \quad p=N,
\\
s^{2\delta},   & \quad p=N-5. 
\end{cases}
$$ 
On the other hand, recalling \eqref{eqs1-14-01-2021} we have 
$$ 
r^{-2} \crochet^{1-\mu} \, \Fenergy_{\mu,c}^{\ME,p,k}(s,\phi)\lesssim 
(\epss+C_1\eps)  \, r^{-2}\crochet^{1-\mu}
\begin{cases} 
s^{1+\delta}, \quad &p=N,
\\
s^{\delta}, \quad &p=N-5.
\end{cases}
$$ 
We are thus in a position to apply  Proposition~\ref{lem 1 d-KG-e} and we arrive at the desired conclusion. 
\end{proof}  

%----------------------------------------------------------------------------------------------------

\subsection{Basic estimates for nonlinearities: energy norm} 
\label{sectionn33}

\paragraph{Improving the energy estimates.}
In order to improve the bootstrap bounds \eqref{eqs1-14-01-2021}, we are going to differentiate the wave equations \eqref{eq 1 13-01-2019} and the Klein-Gordon equation \eqref{eq10-15-05-2020} with respect to $Z=\del^IL^J\Omega^{K}$ 
(with $\ord(Z)=|I|+|J|+|K|\leq N$ or $\leq N-5$). We obtain 
\begin{subequations}
\label{eq11-15-05-2020-ab}
\begin{equation}\label{eq11-15-05-2020}
\aligned
\Boxt_g Z u 
&=   -[Z,h^{\mu\nu} \del_{\mu} \del_{\nu}]u_{\alpha\beta} + Z \big( \Pbb_{\alpha\beta}^{\star}[u]\big) + Z \big( \Qbb_{\alpha\beta}^{\star}[u] \big) 
\\
& - 8\pi \, Z \Big( 2 \, T_{\alpha\beta} - Tg_{\alpha\beta} \Big)
+ 
Z \Big( \Ibb^{\star}_{\alpha\beta}[u] + 2 \, \Rwave_{\alpha\beta}
- u^{\mu\nu} \del_{\mu} \del_{\nu} g^\star_{\alpha\beta}  \Big)
\endaligned
\end{equation}
and
\begin{equation}\label{eq12-15-05-2020}
\Boxt_g Z \phi - c^2Z\phi = -[Z, h^{\mu\nu} \del_{\mu} \del_{\nu}]\phi.
\end{equation}
\end{subequations}
In order to apply the energy estimate in Proposition~\ref{prop energy-ici-exterior}, we need to control the $L^2$ norm of the right-hand side of the associated equation and establish sufficient decay in time. This is our main task for the rest of this article. Specifically, in view of the energy estimate in Proposition~\ref{prop energy-ici-exterior} we need to control $\|J \, \zeta^{-1}  \crochet^{\kappa} |T|_N\|_{L^2(\MME_s)}$, where $T$ represents any of the terms in the right-hand sides of \eqref{eq11-15-05-2020-ab}. Thanks to the technical inequalities in Lemma~\ref{lem1-22-05-2020} we have 
\begin{equation}\label{eq1-17-08-2021}
\|J \, \zeta^{-1}  \crochet^{\kappa}  |T|_N\|_{L^2(\MME_s)}\lesssim \| s \, \crochet^{\kappa} \zeta |T|_N\|_{L^2(\MME_s)}
\end{equation} 
and, from now on, we focus on the right-hand side of \eqref{eq1-17-08-2021}. We treat first the comparatively easier terms, that is, the reference-perturbation interaction terms $\Ibb^{\star}[u]$, the term $u^{\mu\nu} \del_{\mu} \del_{\nu} h^\star_{\alpha\beta}$, 
and the source terms associated with the scalar field. On the other hand, the null terms, the quasi-null terms and the commutators require much more involved arguments and estimates, and this will be the subject of the following sections. 

%---------------------------------------

\paragraph{Linear terms and sub-critical nonlinearities.} 

In view of the expressions of the nonlinearities in \eqref{eq1-06-02-2022} and \eqref{equa-mainsys5}, it is natural to introduce the classification in which, as we will see, both $W^{\textbf{linear}}$ and $W^{\super}$ enjoy integrable $L^2$ bounds:
$$
\aligned
W^{\textbf{linear}}_{\alpha\beta} : &=   
{ 2}\Fbb_{\alpha\beta}(g^\star, g^\star;\del u, \del h^{\star}) - u^{\mu\nu} \, \del_{\mu} \del_{\nu} h^{\star}_{\alpha\beta},
\\
W^\super_{\alpha\beta} : &=  \Fbb_{\alpha\beta}(u,g^\star;\del h^{\star}, \del h^{\star}) + \Fbb_{\alpha\beta}(g^\star,u;\del h^{\star}, \del h^{\star})
+ \Bbb^\star_{\alpha\beta} [u] + \Cbb^{\star}_{\alpha\beta}[u]
- 8\pi \, \big( 2 \, T_{\alpha\beta} - T \, g_{\alpha\beta} \big)  + 2 \, \Rwave_{\alpha\beta}.
\endaligned 
$$
We begin with easier nonlinearities and postpone to Section~\ref{section--125} below the derivation of spacetime estimates for Class B metrics.  

\begin{lemma}[Linear terms and sub-nonlinearities] 
\label{lemma-112} 
Under the conditions stated in Section~\ref{section-label-11-1},  for all $s \in [s_0, s_1]$ one has 
\begin{equation}\label{eq2-02-06-2022}
\|J\,\zeta^{-1}\crochet^{\kappa} |W^{\textbf{linear}} |_{p,k}\|_{L^2(\MME_s)}
\lesssim \delta^{-1}(\epss+C_1\eps)^2s^{-1-\delta}
\qquad \text{ Class A,}
\end{equation}
while, for both Class A and Class B metrics, 
\begin{equation}\label{eq1-22-03-2021}
\|J \, \zeta^{-1}  \crochet^{\kappa} | W^\super |_{p,k}\|_{L^2(\MME_s)}
\lesssim  
\delta^{-1}(\epss + C_1\eps)^2s^{-1-\delta} { + R^{\err}_{\star}(s)}.
\end{equation}
\end{lemma}

\begin{proof} We substitute the basic $L^2$ and decay bounds in the corresponding expressions. We give here the proof of \eqref{eq2-02-06-2022} in Class A, while for supercritical terms, we refer to Appendix~\ref{appendix-EEE}. 
We recall \eqref{eq1-06-02-2022} and observe that $\Fbb^{\star}(g^{\star},g^{\star},\del h^{\star},\del u)$ involves null terms and quasi-null terms and that the latter were bounded in \eqref{eq5-04-10-2022}:
$$
|\Fbb^{\star}(g^{\star},g^{\star},\del h^{\star},\del u)|_{p,k}\lesssim 
\sum_{p_1+p_2=p\atop k_1+k_2=k} |\del h^{\star} |_{p_1,k_1} |\del u|_{p_2,k_2}
\lesssim \epss r^{-1-\lambda} |\del u|_{p,k}, 
$$
thanks to \eqref{equa-31-12-20}. Thanks to \eqref{eq1-17-08-2021}, we find 
$$
\aligned
\|J\zeta^{-1}\crochet^{\kappa} |\Fbb^{\star}(g^{\star},g^{\star},\del h^{\star},\del u)|_{p,k}\|_{L^2(\MME_s)}
& \lesssim  \epss s\| r^{-1-\lambda}\zeta \crochet^{\kappa} |\del u|_{p,k}\|_{L^2(\MME_s)}
\lesssim \epss s^{-1-2\lambda}\Fenergy_{\kappa}^{N,\ME}(s,u)^{1/2}
\\
& \lesssim  \epss  (\epss+C_1\eps)  s^{-1-2\lambda +\delta}\lesssim \delta^{-1}(\epss+C_1\eps)^2s^{-1-\delta}, 
\quad \text{Class A.}
\endaligned
$$
In turn, we apply the Hardy-Poincar\'e inequality in Proposition~\ref{eq3-15-05-2020} and, in view of \eqref{equa-31-12-20}, we have 
\begin{equation}\label{eq5-17-06-2020}
\|s\crochet^{\kappa} \zeta \, | u \, \del \del h^{\star} |_N\|_{L^2(\MME_s)} 
\lesssim \delta^{-1}\epss  (\epss+C_1\eps)  s^{-1-  2\lambda + \delta}\lesssim \delta^{-1}\epss  (\epss+C_1\eps)  s^{-2}, 
\quad \text{ Class A.}
\end{equation}
\end{proof}

The proof of the following two lemmas is easy and is postponed to Appendix~\ref{appendix-EEE}. 

\begin{lemma}[Sub-critical terms involving the reference metric]
\label{prop1-23-05-2021}
Under the conditions stated in Section~\ref{section-label-11-1}, 
for both Class A and Class B metrics
one has
$$
\aligned
& \|s\crochet^{\kappa}\zeta|\Fbb_{\alpha\beta}(u,g^\star;\del h^{\star}, \del h^{\star})|_N\|_{L^2(\MME_s)} + \|s\crochet^{\kappa}\zeta|\Bbb^{\star}_{\alpha\beta}[u]|_N \|_{L^2(\MME_s)}
+\|s\crochet^{\kappa}\zeta|\Cbb^{\star}_{\alpha\beta}[u]|_N \|_{L^2(\MME_s)} 
\\
&
\lesssim  
(\epss + C_1\eps)^2s^{-1-\delta}. 
\endaligned
$$
\end{lemma}

\begin{lemma}[Sub-critical terms involving the matter field]
\label{prop2-23-05-2021}
Under the conditions stated in Section~\ref{section-label-11-1}, 
for both Class A and Class B metrics 
one has 
$$ 
\| s \, \crochet^{\kappa} \zeta |2 \, T_{\alpha\beta} - Tg_{\alpha\beta} |_N\|_{L^2(\MME_s)} 
\lesssim  (\epss+C_1\eps)^2s^{-1-\delta}. 
$$
\end{lemma}

On the other hand, for $W^{\textbf{linear}}$ in Class B, the estimate is more involved and will be given in Section~\ref{section--125}. 

%------------------------------------------------------------------------------------------------------------------------------------

\subsection{Basic estimates for nonlinearities: pointwise norm}

\paragraph{Estimates for the Klein-Gordon field.}

For the matter interaction terms, thanks to \eqref{eq11a-02-05-2020}, \eqref{eq1-18-05-2020}, and Lemma~\ref{lemma-111} we have 
\begin{equation}\label{eq1-28-11-2020}
\sum_{\alpha, \beta} |2 \, T_{\alpha\beta} - ( T_{\gamma\gamma} g^{\gamma\gamma} ) \, g_{\alpha\beta} |_{N-3} 
=: | \Tbb(\phi)|_{N-3}
\lesssim (\epss+C_1\eps)^2
r^{-2}\crochet^{-2\mu}\big(r^{-1}\crochet + r^{-\lambda}\big)s^{1+3\delta} \quad \text{in }\MME_{[s_0,s_1]}.
\end{equation}
Here and from now on, we use the short-hand notation $\Tbb(\phi)$ for the matter term contributions $2 \, T_{\alpha\beta} -
( T_{\gamma\gamma} g^{\gamma\gamma} ) \, g_{\alpha\beta}$. 
For the source terms associated with the field $\phi$, we recall the expression of $T_{\alpha\beta}$ and establish  
the following bound based on \eqref{eq11a-02-05-2020} and Lemma~\ref{lemma-111}: 
$$
|\del\phi\del\phi|_{N-3}
\lesssim |\del\phi|_{N-3} |\del\phi|_{[(N-3)/2]}
\lesssim 
(\epss+C_1\eps)^2 \, r^{-2}\crochet^{-2\mu}\big(r^{-1}\crochet + r^{-\lambda}\big)s^{1+3\delta}.
$$
({ Here we require $N\geq 14$ in order to guarantee $[(p-3)/2]\leq N-9$.})
The bound on $|\phi^2|_{N-3}$ is similar and we omit the details. Using the fact that $|h^{\alpha\beta} |_{N-3}\lesssim 1$,  we obtain \eqref{eq1-28-11-2020}.
In the following analysis, since  $\mu \geq \kappa$, it is sometimes convenient to write the following weaker bound on $|\Tbb(\phi)|_{p,k}$
\begin{equation}\label{eq3-27-01-2021}
(r \, \crochet^\kappa)^2
|\Tbb(\phi)|_{p,k}\lesssim
(\epss+C_1\eps)^2
\big(r^{-1}\crochet + r^{-\lambda}\big)
\begin{cases} 
s^{1+3\delta},\quad & p\leq N-3,
\\
s^{3\delta},\quad & p\leq N-8.
\end{cases} 
\end{equation}

%------------------------------------------------------------

\paragraph{Bounds on null metric component.} 

The null component $g^{\N00}$ of the metric plays a special role in our analysis, and its decay is estimated in this section. First of all, in view of the consequence of the  wave gauge condition in Lemma~\ref{lemma-12-04-2020}, we have the decay property 
\begin{equation}\label{eq1-17-07-2020}
|\del g^{\N00} |_{N-3} \lesssim 
{ \delta^{-1}}\big(\epss + C_1\eps\big) r^{-1-\min(\lambda,\kappa)}s^{\delta}
\qquad \text{in }\MME_{[s_0,s_1]}, 
\end{equation}  
since { $0\leq \min(\lambda,\kappa)\leq 1-\delta/2$}.
Indeed, in the right-hand side 
the inequality in Lemma~\ref{lemma-12-04-2020} with $p=N-3$, namely 
$$ 
\aligned
|\del g^{\N00} |_{N-3} 
\lesssim  
& |\delsN h|_{N-3}  + r^{-1} |h|_{N-3} + \sum_{p_1+p_2 = N-3} |h|_{p_1} |\del h|_{p_2}.
\endaligned
$$ 
We need to substitute the Sobolev bounds \eqref{eq1-26-05-2021} on $|\del h|_{N-3}$ and $|\delts h|_{N-3}$, together with the bound on $|h|_{N-2}$ from \eqref{eq1-09-05-2021}. 

%------------------------------------------------------------

\paragraph{Estimates for the wave fields.}

We now proceed with estimates that are valid for Class A and Class B. Namely, we establish  
\begin{equation}
\label{eq1-04-12-2020} 
|\Ibb^{\star}_{\alpha\beta}[u]|_{N-4} + |u^{\mu\nu}\del_{\mu}\del_{\nu}g^{\star} |_{N-4} 
\lesssim
\begin{cases} 
\delta^{-1}(\epss + C_1\eps)^2 r^{-2-\lambda}\crochet^{-\kappa}s^{2\delta},\quad 
& \text{Class~A},
\\
\delta^{-1}(\epss + C_1\eps)^2 r^{-2+2\theta}\crochet^{-2\kappa}s^{\delta},\quad 
& \text{Class~B},
\end{cases}
\qquad \text{in }\MME_{[s_0,s_1]}.
\end{equation}
To this end, we substitute the bounds in Section~\ref{subsec1-30-05-2020} and, thanks to Lemma~\ref{lem-small}, we have 
\begin{equation}\label{eq11-04-06-2020}
|u^{\mu\nu} \del_{\mu} \del_{\nu}g^{\star}_{\alpha\beta} |_{N-3} 
\lesssim 
\begin{cases}
\delta^{-1}\epss  (\epss+C_1\eps)  \, r^{-3-\lambda} \crochet^{1-\kappa}s^{\delta},\qquad 
& \text{Class~A},
\\
\delta^{-1}\epss  (\epss+C_1\eps)  \, r^{-2+\theta} \crochet^{-2\kappa}s^{\delta},\qquad 
& \text{Class~B},
\end{cases}
\qquad \text{in } \MME_{[s_0,s_1]}.
\end{equation}
where the decay \eqref{eq7-15-05-2020} and the assumptions \eqref{equa-31-12-20} or \eqref{equa-new-conditions-hstar} are used.
By  \eqref{eq0-crossing} and \eqref{equa-31-12-20} or \eqref{equa-new-conditions-hstar}, together with \eqref{eq7-15-05-2020} and \eqref{eq10-02-05-2020} or \eqref{equa-new-conditions-hstar} we find 
\begin{equation}\label{eq4-04-06-2020}
\aligned
&|\Lbb^{\star}_{\alpha\beta}[u]|_{N-3} 
\lesssim
\begin{cases}
\epss (\epss+C_1\eps)  \, r^{-2-\lambda} \crochet^{-\kappa}s^{\delta}\quad 
& \text{Class~A},
\\
\epss(\epss + C_1\eps)\, r^{-2+2\theta}\crochet^{-2\kappa}s^{\delta}\quad 
&  \text{Class~B},
\end{cases}
\quad 
&&\quad \text{in }\MME_{[s_0,s_1]},
\\
&|\Bbb^{\star}_{\alpha\beta}[u]|_{N-3} \lesssim 
\begin{cases}
(\epss+C_1\eps)^2 r^{-3-\lambda} \crochet^{1-2 \kappa}s^{2 \delta}\quad 
& \text{Class~A},
\\
(\epss+C_1\eps)^2r^{-3+2\theta}\crochet^{1-3\kappa}s^{2\delta}\quad 
& \text{Class~B}.
\end{cases}
\quad 
&& \quad \text{in }\MME_{[s_0,s_1]},
\\
&|\Cbb^{\star}_{\alpha\beta}[u]|_{N-3} \lesssim 
\begin{cases}
(\epss+C_1\eps)^2r^{-3} \crochet^{1-3\kappa}s^{3\delta},\quad  
& \text{Class~A},
\\
(\epss+C_1\eps)^2r^{-3+\theta} \crochet^{1-3\kappa}s^{3\delta},\quad 
& \text{Class~B}, 
\end{cases}
\quad 
&& \quad \text{in }\MME_{[s_0,s_1]}.
\endaligned
\end{equation}

%---------------------------------------------------------------------------------------------------------------------------------------------------------------- 

\subsection{Spacetime estimates of $W^{\textbf{linear}}$ for Class B metrics}
\label{section--125}

\paragraph{Spacetime estimate associated with $u\del\del h^{\star}$.}

We are interested now in the spacetime integral 
\begin{equation} \label{eq1-22-08-2022-M}
 T^0_{p,k} := 
\int_{s_0}^{s}\|\crochet^{\kappa}\zeta |\del u|_{p,k}\|_{L^2(\MME_\tau)} \, \|\crochet^{\kappa}J\zeta^{-1} |u^{\N00}\del\del h^{\star} |_{p,k}\|_{L^2(\MME_\tau)} d\tau, 
\end{equation}
which is relevant in the derivation of our final energy estimate. We proceed by establishing various lemmas successively.

\begin{lemma}
For all functions $u$ defined in $\MME_{[s_0,s_1]}$, one has
\begin{equation}
\int_{s_0}^s\int_{\MME_\tau}\crochet^{2\kappa-1}|\delsN u|_{p,k}^2\,Jdxd\tau \lesssim 
\int_{s_0}^s\tau^{-1}\Eenergy_{\kappa}^{\ME,p,k}(\tau,u) d\tau + {\mathscr G}_\kappa^{\ME,p,k}(s_0,s,u).
\end{equation}
\end{lemma}

\begin{proof}
We distinguish between the near- and far- light-cone regions. 
In $\Mfar$, we have $\crochet^{-1}\lesssim t^{-1}$ and, therefore,
$$
\int_{s_0}^s\int_{\Mfar_\tau}\crochet^{2\kappa-1}|\delsN u|_{p,k}^2\,Jdxd\tau 
\lesssim \int_{s_0}^s\tau^{-1}\int_{\Mfar_{\tau}}\crochet^{2\kappa}\zeta^2|\delsN u|_{p,k}^2
\lesssim \int_{s_0}^s\tau^{-1}\Eenergy_{\kappa}^{\ME,p,k}(\tau,u) \, d\tau.
$$
In $\Mnear$, we rely on \eqref{eq2-03-02-2020} and observe that
$$
\aligned
&\int_{s_0}^s\int_{\Mnear_\tau}\crochet^{2\kappa-1}|\delsN u|_{p,k}^2\,Jdxd\tau
\\
& \lesssim  \sum_{\ord(Z)\leq p\atop \rank(Z)\leq k}
\int_{s_0}^s\int_{\Mnear_{\tau}}\crochet^{2\kappa-1}|\delsN Zu|_{p,k} dxd\tau + 
\sum_{\ord(Z)\leq p\atop \rank(Z)\leq k}
\int_{s_0}^s\int_{\Mnear_{\tau}}\crochet r^{-1}\, \crochet^{2\kappa-1}\zeta^2|\del u|_{p,k}dxd\tau
\\
& \lesssim  {\mathscr G}_\kappa^{\ME,p,k}(s_0,s,u) + \int_{s_0}^s\tau^{-1}\Eenergy_{\kappa}^{\ME,p,k}(\tau,u) d\tau.
\hskip3.cm\qedhere
\endaligned
$$
\end{proof}

\begin{lemma}[Spacetime estimate for Class B. I] 
\label{lemma--18-sept-22-000} 
Under the bootstrap assumptions, for metrics in Class B one has (for all $k \leq p \leq N$)
\begin{equation}\label{18-sept-22-000} 
T^0_{p,k} 
\lesssim   \delta^{-2} \, \epss(\epss+C_1\eps)^2 + \delta^{-1} \epss  \, {\mathscr G}_\kappa^{\ME,p,k}(s_0, s,u)
+\delta^{-1}\epss \int_{s_0}^s\tau^{-1}\Eenergy_{\kappa}^{\ME,p,k}(\tau,u) d\tau, 
\end{equation}
in which ${\mathscr G}_\kappa^{\ME,p,k}$ denotes the spacetime bulk energy \eqref{equa-new-spacetime-bound}. 
\end{lemma} 

%----------------------------------------------------------------------

In the semi-null frame we decompose $u^{\mu\nu} \del_\mu \del_\nu h^{\star}_{\alpha\beta}$ and obtain 
\begin{equation}
|u^{\mu\nu} \del_\mu \del_\nu h^{\star}_{\alpha\beta} |_{p,k}
\lesssim
|u^{\Ncal 00} \del_t \del_t h^{\star} |_{p,k}
+ 
| u^{\mu\nu} \del_\mu \delsN_\nu h^{\star} |_{p,k} + r^{-1} |u \del h^\star|_{p,k}
\end{equation}
and we then write $T^0_{p,k} \leq T^1_{p,k} + T^2_{p,k} + T^3_{p,k} + T^4_{p,k}$ with 
$$
\aligned 
T^1_{p,k} & :=  { \epss^{-1}} \int_{s_0}^{s}\tau^{1 + 3 \delta}
\|J\zeta^{-1} \crochet^{\kappa} |u^{\N00}\del\del h^{\star} |_{p,k}\|_{L^2(\MME_\tau)}^2 d\tau,
\quad
&& T^2_{p,k} := { \epss^{-1}}\int_{s_0}^{s}\tau^{1+3 \delta}\|J\zeta^{-1}  \crochet^{\kappa} | u \del \delsN h^{\star} |_{p,k}\|_{L^2(\MME_{\tau})}^2 d\tau,
\\
T^3_{p,k} & := { \epss^{-1}}\int_{s_0}^s \tau^{1+3 \delta}\|J\zeta^{-1} \la r \ra^{-1} \crochet^{\kappa} \, |u \del h^\star|_{p,k}\|_{L^2(\MME_{\tau})}^2d\tau, 
\quad
&& T^4_{p,k} := { \epss}\int_{s_0}^{s}\tau^{-1-3 \delta}\|\zeta\crochet^{\kappa} |\del u|_{p,k}\|_{L^2(\MME_{\tau})}^2d\tau. 
\endaligned
$$
We also recall that $J \simeq \zeta^2 s$ while $J \lesssim s$ and $\zeta \lesssim 1$, as well as $t \lesssim s^2$. First of all, recalling \eqref{eq1-14-01-2021}, the term $T^4$ is controlled as follows: 
$$
\aligned 
T^4_{p,k}
\lesssim { \epss}\int_{s_0}^s \tau^{-1-3 \delta}\Eenergy^{\ME,N}_{\kappa}(\tau,u) \, d\tau
\lesssim \delta^{-1} \epss (\epss+C_1\eps)^2, 
\endaligned
$$ 
where the bootstrap bound \eqref{eq1-14-01-2021} was used. To treat  the term $T^2$ we use our assumption 
$|\del \slashed \del h^{\star} |_{N} \lesssim \epss \, \la r\ra^{-1-\kappa}\crochet^{-1}$ and obtain
$$
\aligned 
T^2_{p,k}
&\lesssim
\epss \int_{s_0}^s  \tau^{1 + 3 \delta} \int_{\MME_\tau} J^2  
\zeta^{-2} \, \la r+t\ra^{-2-2\kappa} \big( \crochet^{-2+2\kappa} \, |u|_N^2 \big) \, dxd\tau
\\ 
&  
\lesssim \epss \, \int_{s_0}^s \tau^{-1 -4 \kappa +3\delta} \Big( \delta^{-1}  \Eenergy^{\ME,N}_{\kappa}(s,u) + \Eenergy_{\eta}^{0}(s,u) \Big) \, d\tau
\lesssim  \delta^{-1}  \epss \, (\epss+C_1\eps)^2,   
\endaligned
$$
thanks to Poincar\'e inequality in Proposition~\ref{eq3-15-05-2020}, namely with $\eta=\kappa >1/2$ 
$$
\| \crochet^{-1 + \eta} |u|_{N}\|_{L^2(\MME_\tau)}^2 
\lesssim \big(1+\delta^{-1} \big) \, \Eenergy_\eta^{\ME,N}(s,u) + \Eenergy_{\eta}^{0}(s,u). 
$$
For the term $T^3_{p,k}$ associated with $r^{-1} |u|_{p,k}$ we proceed similarly as above and write 
$$
\aligned
T^3_{p,k}
& = { \epss^{-1}} \int_{s_0}^s\tau^{1+3\delta}  \int_{\MME_\tau} J^2\zeta^{-2} \la r \ra^{-2} \crochet^{2\kappa} \, |u \del h^\star|_N^2 \, dxd\tau
\\
& 
\lesssim \epss  \int_{s_0}^s\tau^{1+3\delta}  \int_{\MME_\tau} J^2\zeta^{-2} \la r \ra^{-4 + 2 \theta} \crochet^{2-2\kappa} \, 
\big( \crochet^{-2+ 2\kappa} |u|_N^2\big) \, dxd\tau, 
\\
& 
\lesssim {   \epss 
\int_{s_0}^s \tau^{1+3\delta}  \int_{\MME_\tau}  J^2\zeta^{-2} \la r\ra^{-2-2\kappa+2\theta}
\Big( \delta^{-1}  \Eenergy^{\ME,N}_{\kappa}(s,u) + \Eenergy_{\eta}^{0}(s,u) \Big)dxd\tau}
\\
&  
\lesssim \epss \int_{s_0}^s \int_{\MME_\tau} \tau^{-1-4\kappa+4\theta+ 3\delta}
\Big( \delta^{-1}  \Eenergy^{\ME,N}_{\kappa}(s,u) + \Eenergy_{\eta}^{0}(s,u) \Big)
\lesssim   \delta^{-1} \epss   (\epss + C_1\eps)^2, 
\endaligned
$$
in which we again used Poincar\'e inequality in Proposition~\ref{eq3-15-05-2020}, namely with $\eta=\kappa >1/2$.

Dealing with the first term $T^1$ is more challenging. Thanks to our assumptions on $h^{\star}$ in \eqref{equa-new-conditions-hstar}, namely 
$|\del\del h^{\star} |_N\lesssim \epss \la r\ra^{-1+\theta}\crochet^{-1-\kappa}$ and then by applying the Hardy-Poincar\'e inequality in Proposition~\ref{propo-Poincare-ext}, with $\eta = \kappa + 1/2$ we have, 
$$ 
T^1_{p,k} 
\lesssim  \epss \int_{s_0}^s \tau^{-1+4\theta+3\delta} \|\crochet^{-1+1/2+\delta}\zeta |u^{\N00}|_{p,k} \|_{L^2(\MME_\tau)}^2 \, d\tau. 
$$
Then for any $\ord (Z) = p$ and $\rank(Z) = k$, 
$$
\aligned
&\epss \int_{s_0}^s \tau^{-1+4\theta+3\delta} \|\crochet^{-1+1/2+\delta}\zeta Zu^{\N00} \|_{L^2(\MME_\tau)}^2 \, d\tau
\\
& \lesssim   \delta^{-1} \epss  
\int_{s_0}^s  \tau^{-1+4\theta+3\delta} 
\Big(\|\crochet^{1/2+\delta}\zeta \delsME Zu^{\N00}\|_{L^2(\MME_\tau)}^2 
+ \|\crochet^{1/2+\delta}r^{-1}\zeta Zu^{\N00} \|_{L^2(\MME_\tau)}^2 \Big) \, d\tau
\\
& \simeq 
\delta^{-1} 
\epss  \int_{s_0}^s \tau^{-2+4\theta+3\delta}  \int_{\MME_\tau} \tau \zeta^2\crochet^{1+2\delta} |\del u^{\N00} |_{p,k}^2 \, dxd\tau
+   \delta^{-1} \epss   \int_{s_0}^s  \tau^{-1+4\theta+3\delta} \int_{\MME_\tau}\zeta^2\crochet^{1+2\delta}r^{-2} |u^{\N00} |^2_{p,k} \, dxd\tau
\\
&
=: T^{11}_{p,k} + {   T^{12}_{p,k}}. 
\endaligned
$$ 
%\\
{   
The term $T^{12}_{p,k}$ is comparatively easier and can be handled by techniques used so far, so we concentrate on the new aspects of the arguments. 
}
To treat $T_{11}$ we apply the wave gauge condition (cf.~Lemma \ref{lemma-12-04-2020}) which gives us 
$$
|\del u^{\N00} |_{p,k} \lesssim |\delsN u|_{p,k} + r^{-1} |u|_{p,k} 
+ \sum_\gamma |w^\star_\gamma|_{p,k}
+ {   
\sum_{p_1+p_2=p\atop k_1+k_2=k}\big(|h^\star|_{p_1,k_1} |\del u|_{p_2,k_2} + |\del h^{\star}|_{p_1,k_1}|u|_{p_2,k_2} + |u|_{p_1,k_1}|\del u|_{p_2,k_2}\big)},
$$
which allows us to decompose $T^{11}_{p,k} = T^{111}_{p,k} + T^{112}_{p,k} + T^{113}_{p,k} + T^{114}_{p,k}$ with obvious notation. Quadratic terms have much better decay, so we only treat the first three terms. The contribution $T^{111}$ associated with $\delsN u$ is bounded by 
$$ 
\aligned
T^{111}_{p,k}
&\lesssim  \delta^{-1} \epss  \int_{s_0}^s \tau^{-2+4\theta+3\delta} \int_{\MME_\tau} \crochet^{1+2\delta} |\delsN u|_{p,k}^2 (\tau\zeta^2) \, dx d\tau
\\
& \lesssim   \delta^{-1} \epss  
\int_{s_0}^s 
\tau^{-4(\kappa-1/2-\theta-\delta) + 3 \delta} 
\int_{\MME_\tau} \crochet^{2\kappa-1} |\delsN u|_{p,k}^2 J \, dxd\tau
  \lesssim
\delta^{-1} \epss   {\mathscr G}_\kappa^{\ME,p,k}(s_0, s,u) 
+ \delta^{-1} \epss \int_{s_0}^s\tau^{-1}\Eenergy_{\kappa}^{\ME,p,k}(\tau,u) d\tau
\endaligned
$$ 
in which the exponent $-4(\kappa-1/2-\theta-\delta)+3 \delta\leq 0$ has a favorable sign. 

%-----------------------------

Similarly, we write 
$$ 
\aligned
T^{112}_{p,k}
& \lesssim  \delta^{-1} \epss  \int_{s_0}^s \tau^{-1+4\theta+3\delta} \int_{\MME_\tau} \zeta^2 \crochet^{1+2\delta} r^{-2} |u|_N^2 \, dx d\tau
\lesssim   \delta^{-1} \epss 
\int_{s_0}^s \tau^{1-4\kappa +4\theta+7\delta} \int_{\MME_\tau}  \big( \crochet^{-2+2\kappa}  |u|_N^2 \big) \, dxd\tau
\\
&  
\lesssim  \delta^{-1}\epss
\int_{s_0}^s \tau^{1-4\kappa +4\theta+7\delta} \Big( \delta^{-1}  \Eenergy^{\ME,N}_{\kappa}(s,u) + \Eenergy_{\eta}^{0}(s,u) \Big) \, d\tau
\lesssim  \delta^{-2}  \epss \,   (\epss + C_1\eps)^2,
\endaligned
$$
and we note in passing that the second term $A_{12}$ in the decomposition of $A_1$ is treated by the exactly same argument. 

Finally, the wave gauge error $w^\star_\gamma$ is controlled by our assumptions on the reference metric, specifically  
$\sum_\gamma |w^\star_\gamma|_N^2 \lesssim \big( \epss \, r^{-1} \crochet^{- 1 - \varsigma}\big)^2$  
and we write    
\begin{equation} \label{equa-28-aout-2022}
\aligned
T^{113}_{p,k}
\lesssim 
\epss^3  \delta^{-1}\int_{s_0}^s \tau^{-1 - 4\varsigma +4\theta+9\delta} \int_{\MME_\tau} X^{-3-  \delta} 
\, dx d\tau
\lesssim \epss^3 \delta^{-2}. 
\endaligned
\end{equation}

%--------------------------------------------------------------

\paragraph{Spacetime estimates associated with the terms $\Fbb_{\alpha\beta}$.} 

In the terms
$
\Fbb_{\alpha\beta}(g^\star, g^\star;\del u, \del h^\star)
$
we need to pay attention to the contribution 
$\Qbb$ (thanks to Lemma~\ref{Null-Euclidean bilinear}) and $\Pbb$ (thanks to Lemma \ref{lem1-31-01-2021}).

\begin{lemma}[Spacetime estimate for Class B. II] 
\label{lemma--18-sept-22-01}  
Under the bootstrap assumptions,  for metrics in Class B one has ($k \leq p \leq N$)
\begin{equation}\label{18-sept-22-01} 
\aligned
&
\int_{s_0}^{s}\|\crochet^{\kappa} \zeta |\del u|_{p,k}\|_{L^2(\MME_{\tau})} 
\|J\zeta^{-1}\crochet^{\kappa} \Fbb_{\alpha\beta}(g^\star, g^\star;\del u, \del h^{\star})|_{p,k} \|_{L^2(\MME_{\tau})} \, d\tau
\\
& \lesssim   \epss \, (\epss+C_1\eps)^2 + \epss \,  {\mathscr G}_\kappa^{\ME,p,k}(s_0, s,u)
+ \epss \sum_{0\leq k_1\leq k}\int_{s_0}^{s} \tau^{-1+2k_1\theta} \Eenergy_{\kappa}^{\ME, p, k-k_1}(\tau,u) \, d\tau, 
\endaligned
\end{equation}
in which ${\mathscr G}_\kappa^{\ME,p,k}$ denotes the spacetime bulk energy \eqref{equa-new-spacetime-bound}. 
\end{lemma} 

We now give a proof of this lemma and we introduce
\begin{equation} \label{equa-28-aout-2022-02-0}
\aligned
A^0_{p,k} & 
:=  \sum_{p_1+p_2=p \atop k_1+k_2=k}
\int_{s_0}^{s}\|\crochet^{\kappa} \zeta|\del u|_{p,k}\|_{L^2(\MME_{\tau})}\|J\zeta^{-1}\crochet^{\kappa}|\del \uts|_{p_1,k_1} |\del \slashed h^\star{}^\Ncal|_{p_2,k_2}\|_{L^2(\MME_{\tau})}d\tau, 
\\
B^0_{p,k}& :=  
\int_{s_0}^{s} \|\crochet^{\kappa}\zeta|\del u|_{p,k}\|_{L^2(\MME_\tau)} \|J\zeta^{-1}\crochet^{\kappa}|w^\star_\gamma|_{p,k}|\del u|_{p,k}\|_{L^2(\MME_{\tau})} d\tau, 
\endaligned
\end{equation}
where $A^0_{p,k}$ and $B^0_{p,k}$ are discussed at the end of this section. 

Observing that third-order terms are comparatively easier and $A^0_{p,k}, B^0_{p,k}$ above already take into account the most challenging contribution in the quasi-null terms {and the wave-gauge terms}, we can now focus on each of the functions $a^j_{p,k}$ defined below and  
arising in our estimates, for which we study the corresponding spacetime integrals ($m=1,2$) 
\begin{equation}  
A^m_{p,k} := 
\int_{s_0}^{s}\|\crochet^{\kappa} \zeta |\del u|_{p,k}\|_{L^2(\MME_{\tau})}\|J\zeta^{-1}\crochet^{\kappa}a^m_{p,k}\|_{L^2(\MME_{\tau})}d\tau. 
\qquad 
\end{equation}
Namely, we are going to treat the remaining terms arising in \eqref{equa-new-Qzero} {and  \eqref{eq5-04-10-2022}} for the 
pair $(\del h^{\star}, \del u)$  
$$
a^1_{p,k} := \sum_{p_1+p_2 = p\atop k_1+k_2=k}  |\del u|_{p_1,k_1} |\delsN h^\star |_{p_2,k_2},
\qquad
a^2_{p,k} := \sum_{p_1+p_2 = p\atop k_1+k_2=k} |\del h^\star |_{p_1,k_1} |\delsN u|_{p_2,k_2}, 
\qquad 
$$  

%--------------------------------------------------------------

\noindent$\bullet$ {\bf Remaining contributions.}
$$ 
\aligned
A^1_{p,k}
& = \sum_{p_1+p_2=p\atop k_1+k_2=k}
\int_{s_0}^{s}\tau \| \crochet^{\kappa}\zeta|\del u|_N\|_{L^2(\MME_{\tau})}^2 \||\delsN h^\star|_N \|_{L^{\infty(\MME_{\tau})}} d\tau
\\
& \lesssim \epss \,  \int_{s_0}^{s}\tau^{-1-2\kappa}\Eenergy_{\kappa}^{\ME,N}(\tau,u) d\tau 
\lesssim \epss \, (\epss+C_1\eps)^2  \int_{s_0}^{s} \tau^{-1-2\kappa + 2 \delta} \, d\tau \lesssim \epss \, (\epss+C_1\eps)^2, 
\endaligned
$$
and next 
$$  
\aligned
A^2_{p,k}
& = \sum_{p_1+p_2=p\atop k_1+k_2=k}
\int_{s_0}^{s}\|\crochet^{\kappa}\zeta |\del u|_{p,k}\|_{L^2(\MME_{\tau})} 
\|J\zeta^{-1}\crochet^{\kappa}|\del h^\star |_{p_1, k_1} |\delsN u|_{p_2, k_2}\|_{L^2(\MME_{\tau})} d\tau
\\
& \lesssim   \epss \, \int_{s_0}^{s}\tau^{-3/2} \|\crochet^{\kappa}\zeta |\del u|_{p,k}\|_{L^2(\MME_{\tau})}^2 d\tau
+ \epss\,\int_{s_0}^s \tau^{3/2}\int_{\MME_{\tau}}J^2\zeta^{-2}\la r\ra^{-2+2\theta}|\delsN u|_{p,k}^2dxd\tau
\\
& \lesssim   \epss \,
\int_{s_0}^{s} \tau^{-3/2} \Eenergy_{\kappa}^{\ME,N}(\tau,u) d\tau 
+ 
\epss \, \int_{s_0}^{s}\tau^{-3/2+4\theta}\int_{\MME_\tau} J \crochet^{2\kappa-1}  |\delsN u|_{p,k}^2 \, dxd\tau
\\ 
& 
\lesssim \epss \, (\epss+C_1\eps)^2  \int_{s_0}^{s} \tau^{-3/2+2\delta} \, d\tau 
+ \epss \, {\mathscr G}_\kappa^{\ME,p,k}(s_0, s,u) 
+ \epss \int_{s_0}^s\tau^{-1}\Eenergy_{\kappa}^{\ME,p,k}(\tau,u) d\tau
\\
&\lesssim \epss \, (\epss+C_1\eps)^2 + \epss \,  {\mathscr G}_\kappa^{\ME,p,k}(s_0, s,u)
+\epss \int_{s_0}^s\tau^{-1}\Eenergy_{\kappa}^{\ME,p,k}(\tau,u) d\tau.
\endaligned
$$ 

%--------------------------------------------------------------

\noindent$\bullet$ {\bf Wave gauge error $B^0_{p,k}$.} The contribution $B^0_{p,k}$ that arises in \eqref{18-sept-22-01} comes from 
$
\sum_{p_1+p_2=p\atop k_1+k_2=k} \mathbb{W}_{p_1,k_1}[v] \, |\del u |_{p_2,k_2}, 
$
which is the one due to the remainder $ \mathbb{W}_{p, k}[h^{\star}]$ in \eqref{eq2-04-12-2020-v}, that is, from the approximate gauge term $\sum_\gamma |w^\star_\gamma|_{p,k}$. 
For this spacetime integral $B^0_{p,k}$ we observe that under the assumption ${\sum_\gamma |w^\star_\gamma|_N \lesssim \la r + t\ra^{-1}\crochet^{-1-\varsigma}}$
we can write 
$$
\aligned
B^0_{p,k}  &=  \int_{s_0}^{s} \|\crochet^{\kappa}\zeta|\del u|_{p,k}\|_{L^2(\MME_\tau)} \|J\zeta^{-1}\crochet^{\kappa}|w^\star_\gamma|_{p,k}|\del u|_{p,k}\|_{L^2(\MME_{\tau})} d\tau
  \lesssim  \epss \, 
\int_{s_0}^{s}\tau \Eenergy_{\kappa}^{\ME,p,k}(\tau,u)\|w_{\gamma}^{\star}\|_{L^{\infty}(\MME_{\tau})} d\tau, 
\endaligned
$$
which leads us to
\begin{equation}
B^0_{p,k}
\lesssim \epss \, \int_{s_0}^{s} {\tau^{-1}}\Eenergy^{\ME,p,k}_{\kappa}(\tau,u) \, d\tau.
\end{equation}
\noindent$\bullet$ {\bf Quasi-null terms.}
This term the most critical one. We first observe that in $\Mfar_{\ell,[s_0,s_1]}$, $\crochet^{-1}\lesssim \ell^{-1}\la r\ra^{-1}$. Then form \eqref{equa-new-conditions-hstar} the second and fourth bounds, 
\begin{equation}
|\del \hs^{\star}|_{p,k}\lesssim 
\begin{cases}
\ell^{-\theta} \epss \la r\ra^{-1},\quad &p=k=0,
\\
\epss \la r\ra^{-1+\theta},\quad &N\geq p\geq k\geq 1. 
\end{cases}
\end{equation}
Then we distinguish between two cases, as follows. 

1. Estimate for $p=k=0$. We have 
$$
\int_{s_0}^{s}\|\crochet^{\kappa}\zeta|\del u|_{p,k}\|_{L^2(\MME_{\tau})} 
\|J\zeta^{-1}\crochet^{\kappa} |\del \us|_{p,k}\del\hs^{\star}\|_{L^2(\MME_\tau)} d \tau
\lesssim \ell^{-1} \epss 
\int_{s_0}^s \tau^{-1}\Eenergy_{\kappa}^{\ME,p,k}(\tau,u) d\tau.
$$

2. Estimate for $k\geq 1$. In this case we have $p_1\leq p-1,k_1\leq k-1$ and so
$$
\aligned
&\int_{s_0}^{s}\|\crochet^{\kappa}\zeta|\del u|_{p,k}\|_{L^2(\MME_{\tau})} 
\|J\zeta^{-1}\crochet^{\kappa}|\del \us^{\Ncal}|_{p-1,k-1} |\del \slashed h^{\star\N}|_N\|_{L^2(\MME_{\tau})}d\tau
\\
& \lesssim  \epss\,\int_{s_0}^s \tau^{-1}\|\crochet^{\kappa}\zeta|\del u|_{p,k}\|_{L^2(\MME_\tau)}^2d\tau 
+ \epss\,\int_{s_0}^{s}\tau^{-1+2\theta}\|J\zeta^{-1}\crochet^{\kappa}|\del u|_{p-1,k-1}\|_{L^2(\MME_{\tau})}^2d\tau
\\
& \lesssim  \epss \sum_{0\leq k_1\leq k}\int_{s_0}^{s} \tau^{-1+2k_1\theta} \Eenergy_{\kappa}^{\ME, p, k-k_1}(\tau,u) \, d\tau.
\endaligned
$$

%-------------------------------------------------------

\paragraph{Conclusion.}

Combining Lemmas~\ref{lemma--18-sept-22-01} and \ref{lemma--18-sept-22-000}  together, we arrive at the following result. 

\begin{proposition}[Spacetime estimate for Class B. Conclusion] 
Under the bootstrap assumptions,  for metrics in Class B one has ($k \leq p \leq N$) 
\begin{equation}\label{eq1-05-10-2022}
\aligned
& \int_{s_0}^s \|\crochet^{2\kappa}|\del u|_{p,k}\|_{L^2(\MME_{\tau})} 
\|J\zeta^{-1}\crochet^{\kappa} W^{\textbf{linear}}|_{p,k}\|_{L^2(\MME_{\tau})}d\tau
\\
&
\lesssim \delta^{-2} \, \epss (\epss+C_1\eps)^2 + \delta^{-1} \epss  \, {\mathscr G}_\kappa^{\ME,p,k}(s_0, s,u)
+ 
\delta^{-1}\epss \sum_{0\leq k_1\leq k}\int_{s_0}^{s} \tau^{-1+2k_1\theta} \Eenergy_{\kappa}^{\ME, p, k-k_1}(\tau,u) \, d\tau.
\endaligned
\end{equation}
\end{proposition}

%============================================================================================

\section{Commutator and Hessian estimates for the metric perturbation}
\label{section---12} 

\subsection{Commutators of metric components}

We rely on the technique developed in Section~\ref{section----63} and on the hierarchy property enjoyed by quasi-linear commutators established in Proposition~\ref{prop1-12-02-2020}. Below, we also apply Propositions~\ref{prop1-22-05-2020} and~\ref{propo2-22-05-2020} in order to control the Hessian for the wave equation away from the light cone.  
 
\begin{proposition}[   Commutator estimates]
%  near the light cone and in the far region]
\label{Proposition12.1}
% {  \tt YM: This title is not good. This Proposition contains BOTH near and far region.}
%\\
Under the conditions stated in Section~\ref{section-label-11-1}, for all admissible $Z$ with $\ord(Z) = p \leq {N-4}$ and $\rank(Z) = k$, the commutators associated with the metric perturbation near the light cone satisfy
\begin{equation}\label{eq2-30-05-2020}
\aligned
|[Z,h^{\alpha\beta} \del_{\alpha} \del_{\beta}]u|
\lesssim
& \sum_{1\leq k_1 \leq k}|\hN{}^{00}|_{k_1}  |\del\del u|_{p-k_1,k-k_1} 
+ { \delta^{-1}}( \epss +C_1\eps) r^{-\min(\lambda,\kappa)}s^{\delta} \frac{\crochet}{r} |\del\del u|_{p,k}
\\
&  {  
+ { \delta^{-1}} (\epss + C_1\eps)^2 \, r^{-2-\min(\lambda,\kappa)}\crochet^{-\kappa}s^{2\delta}
}
\qquad 
\qquad \text{in } \Mnear_{ [s_0,s_1]},
\endaligned
\end{equation}
while 
\begin{equation}\label{eq12-04-06-2020}
\big|[Z, h^{\alpha\beta} \del_{\alpha} \del_{\beta}]u\big| 
\lesssim { \delta^{-1}}(\epss +C_1\eps) \, |\del\del u|_{p,k}
\qquad \text{in } \MME_{[s_0,s_1]}.   
\end{equation}
\end{proposition}

\begin{proof} 
In the notation of Proposition~\ref{prop1-12-02-2020} we have the decomposition $\HN^{00} = {\hN}{}^{00} = h^{\star\Ncal 00} + u^{\Ncal 00}$. Thanks to \eqref{eq1-26-05-2021}, \eqref{eq1-09-05-2021}, and \eqref{eq1-17-07-2020} and 
using the notation \eqref{equa-notation-H} (with upper indices), 
we have\footnote{These estimates hold at the regularity order $N-3$, but in the following we use them at the order $N-4$ only.} on each slice $\MME_s$
\begin{equation}\label{eq1-30-05-2020}
\aligned
|\HN^{00}|_{N-2} 
& \lesssim \epss \, r^{-\lambda} + { \delta^{-1}} (\epss+C_1\eps)  \frac{|r-t|}{r} \crochet^{-\kappa}s^{\delta},
&&
|\del \HN^{00}|_{N-3} 
\lesssim \delta^{-1}
\big(\epss + C_1\eps\big) r^{-1-\min(\lambda,\kappa)}s^{\delta},
\\ 
|H|_{N-3}
&  
\lesssim { \delta^{-1}}(\epss + C_1 \eps) r^{-\min(\lambda,\kappa)}s^{\delta},
\qquad \qquad 
&& 
|\del H|_{N-3}
\lesssim (\epss + C_1\eps)r^{-1}\crochet^{-\min(\lambda,\kappa)}s^{\delta}.
\endaligned
\end{equation}
The derivation of \eqref{eq12-04-06-2020} is comparatively easier, and follows if we directly substitute \eqref{eq1-30-05-2020} in \eqref{eq5-12-02-2020}. Therefore, in the rest of this proof we
restrict attention to the domain $\Mnear_{ [s_0,s_1]}$. 

Recalling the decomposition \eqref{eq4-12-02-2020}, we treat $T^\textbf{hier}$, $T^\easy$, and $T^{\textbf{super}}$,  successively.
For the first term in $T^\textbf{hier}$, we write 
$$
\aligned
\Big(|\HN^{00}| + \frac{|r-t|}{r}|H|\Big) \, |\del\del u|_{p-1,k-1} \lesssim  |h^{\N00}||\del\del u|_{p-1,k-1} 
+ { \delta^{-1}}(\epss + C_1\eps) r^{-\min(\lambda,\kappa)}s^{\delta} \frac{|r-t|}{r}|\del\del u|_{p-1,k-1}.
\endaligned
$$
The second term in $T^\textbf{hier}$ is handled in a similar way, namely 
$$
\frac{|r-t|}{t}|LH|_{p_1-1,p_1-1}|\del\del u|_{p_2,k_2} \lesssim \frac{|r-t|}{r}|H|_{p_1,k_1}|\del\del u|_{p_2,k_2}
\lesssim { \delta^{-1}}(\epss + C_1\eps) r^{-\min(\lambda,\kappa)}s^{\delta} \frac{|r-t|}{r}|\del\del u|_{p-1,k-1}.
$$
Here we used that $p_1-1\geq 0$ implies $p_2\leq p-1, k_2\leq k-1$. For the term $|L\HN^{00}|_{p_1-1,p_1-1}|\del\del u|_{p_2,k_2}$, we observe that 
the two conditions $p_1-1\geq 0$ and $p_1 + k_2 = k$ imply that $0\leq k_2 = k-p_1\leq k$, and therefore  
$$
|L\HN^{00}|_{p_1-1,p_1-1}|\del\del u|_{p_2,k_2} \lesssim |\hN{}^{00}|_{p_1,p_1}|\del\del u|_{p-p_1,k-p_1}, 
\qquad\quad 1\leq p_1\leq k. 
$$
This later term is bounded by the first term in the right-hand side of \eqref{eq2-30-05-2020}. 
For the term $T^{\bf easy}$, we use the second bound in \eqref{eq1-30-05-2020}and obtain
$$
|\del \HN^{00}|_{p_1-1,k_1}|\del\del u|_{p_2,k_2}
\lesssim  \delta^{-1}
(\epss + C_1\eps ) \, r^{-\min(\lambda,\kappa)}s^{\delta} \frac{\crochet}{r} \, |\del\del u|_{p,k}.
$$
Thanks to the last inequality in \eqref{eq1-30-05-2020} we also have
$$
\frac{|r-t|}{t}|\del H|_{p_1-1,k_1}|\del\del u|_{p_2,k_2}
\lesssim (\epss + C_1\eps)r^{-1}\crochet^{-\min(\lambda,\kappa)}s^{\delta}\frac{\crochet}{r} \, |\del\del u|_{p,k}.
$$
These expressions are also bounded by the second term in the right-hand side of \eqref{eq2-30-05-2020}.
The term $T^\textbf{super}$ contains a decreasing factor $t^{-1} \simeq r^{-1}$ (near the light cone) and, therefore, is
bounded by ${ \delta^{-1}}(\epss + C_1\eps)^2 \, r^{-2-\min(\lambda,\kappa)}\crochet^{-\kappa}s^{2\delta}$. Hence, \eqref{eq2-30-05-2020} is established. 
\end{proof}

%-------------------------------------------------------------------------------------------------------------------------------------------------

\subsection{Pointwise estimates for the Hessian of metric components}
\label{sec----17-2}

Next, we turn our attention to general second-order derivatives of the metric. 

\begin{proposition}\label{prop1-11-08-2021}
Under the conditions stated in Section~\ref{section-label-11-1}, 
% Under the bootstrap assumptions \eqref{eqs-int}-\eqref{eqs1-14-01-2021} together with the condition { $\min(\lambda,\kappa)\geq 1/2+\delta/2$ and $\delta^{-1}(\epss + C_1\eps)\ll 1$}, 
for all $p\leq N-4$ one has
\begin{equation}\label{eq3-29-01-2021}
\frac{\crochet}{r}|\del\del u|_{p,k} + |\del\delsN u|_{p,k}
\lesssim  (\epss+C_1\eps)  r^{-2}\crochet ^{-\kappa}s^{\delta} \, (r \, |h^{\N00}|_k) 
+  \big(\epss + C_1\eps\big) r^{-2}\crochet^{-\min(\lambda,\kappa)}s^{2\delta}
\qquad \text{ in } \Mnear_{ [s_0,s_1]}
\end{equation}
near the light cone, while far from the light cone
\begin{equation}\label{eq1-15-08-2021}
|\del\del u|_{N-4}\lesssim \ell^{-1} (\epss+C_1\eps)  t^{-1}r^{-1} \crochet^{-\min(\lambda,\kappa)}s^{2\delta}
\qquad 
\text{ in } \Mfar_{\ell, [s_0,s_1]}. 
\end{equation}
\end{proposition}

As a side remark, if we apply \eqref{eq1-09-05-2021} in the right-hand side of \eqref{eq3-29-01-2021} (and recall that { ${\delta^{-1}(\epss+C_1\eps)\ll 1}$}), we obtain the rough estimate (which is sufficient for most of following arguments)
\begin{equation}\label{eq10-04-06-2020}
\aligned
\frac{\crochet}{r}|\del\del u|_{N-4} + |\del\delsN u|_{N-4} \lesssim \big(\epss + C_1\eps\big) \, r^{-1-\min(\lambda,\kappa)}\crochet^{-\min(\lambda,\kappa)}s^{2\delta}
\qquad 
\text{ in } \Mnear_{[s_0,s_1]}.
\endaligned
\end{equation} 
We also deduce a simpler version of \eqref{eq1-15-08-2021} in which $\ell$ is replaced by $1$:
\begin{equation}\label{eq1-19-07-2020}
|\del\del u|_{N-4} \lesssim (\epss + C_1\eps)t^{-1}r^{-1}\crochet^{-\min(\lambda,\kappa)}s^{2\delta}
\qquad 
\text{ in } \Mscr^\far_{[s_0,s_1]}.
\end{equation}
The proof of Proposition~\ref{prop1-11-08-2021} is based on Propositions~\ref{prop1-22-05-2020} and~\ref{propo2-22-05-2020}, and it is convenient to distinguish between the proofs near the light cone and away from it, as follows. 
For \eqref{eq3-29-01-2021} we have the following preliminary result.

\begin{lemma}
\label{lemma-123} 
Under the conditions stated in Section~\ref{section-label-11-1}, 
% Under the bootstrap assumptions \eqref{eqs-int}-\eqref{eqs1-14-01-2021} and { suppose that $\min(\lambda,\kappa)\geq 1/2+\delta/2$ and $\delta^{-1}(\epss + C_1\eps)\lesssim 1$}, 
one has 
$$
% \begin{equation}\label{eq1-07-06-2020}
|\Boxt_g u|_{p,k} \lesssim \epss r^{-2-\lambda} +
\begin{cases}
(\epss+C_1\eps)  \, r^{-2}\crochet^{-\min(\lambda,\kappa)}s^{2\delta},\quad & \text{Class {A}},
\\
 (\epss+C_1\eps)  \, r^{-2}\crochet^{-\min(\lambda,\kappa)}s^{\delta}(r^{2\theta}+s^{\delta}),\quad & \text{Class {B}}.
\end{cases} 
\qquad \text{ in } \MME_{[s_0,s_1]},
\quad p\leq N-4, 
$$
\end{lemma}

\begin{proof}
Recall \eqref{eq 1 13-01-2019} together with \eqref{eq4-09-05-2021}, \eqref{eq1-04-12-2020} ({ with $\delta^{-1}(\epss + C_1\eps)\lesssim 1$}) and \eqref{eq3-27-01-2021} (with {$\min(\lambda,\kappa)\geq 1/2+\delta/2$}).  We only need to bound the quadratic forms $\Qbb^\star[u]$ and $\Pbb^\star[u]$ but, at this juncture, we do not use their structure 
and write straightforwardly 
\begin{equation}\label{eq5-04-06-2020}
|\del u \del u|_{N-3} \lesssim  (\epss+C_1\eps) ^2r^{-2} \crochet^{-2 \kappa}s^{2 \delta}
\qquad \text{ in } \MME_{[s_0,s_1]}. 
\end{equation}
This is the leading term, which provides us with the worst control. On the other hand, the Ricci contribution $^{(w)}R^{\star}_{\alpha\beta}$ is bounded by \eqref{eq4-09-05-2021}. 
\end{proof}

%----------------------------------------------------------------------------------

\begin{proof}[Proof of \eqref{eq3-29-01-2021}]
We rely on \eqref{eq1-06-02-2020} 
and \eqref{eq10-02-05-2020}. 
Assuming $p\leq N-4$, we find 
$$
|\del\delsN u|_{p,k}\lesssim \frac{|r-t|}{t}|\del\del u|_{p,k} + r^{-1}|\del u|_{p+1,k+1}
\lesssim \frac{|r-t|}{t}|\del\del u|_{p,k} +  (\epss+C_1\eps)  r^{-2}\crochet^{-\kappa}s^{\delta}, 
$$
and we can thus focus our attention on $\frac{|r-t|}{t}|\del\del u|_{p,k}$.
In order to apply the Hessian inequality in Proposition~\ref{prop1-22-05-2020}, we need to control the commutator as well as the wave operator source (arising as the contribution $|\Boxt_g u|_{p,k}$). Substituting the inequality in Lemma~\ref{lemma-123} together with
\eqref{eq2-30-05-2020} into \eqref{eq8-04-06-2020}, and
 observing that near the light cone, $r^{2\theta}\lesssim s^{\delta}$ thus the second case  in the estimate of Lemma \ref{lemma-123} reduces to the first case,  we find in the domain $\Mscr^\near_{ s}$ 
$$
\aligned
\frac{1+|r-t|}{r}|\del\del u|_{p,k} \lesssim 
& \sum_{1\leq k_1\leq k}|\hN{}^{00}|_{k_1}  |\del\del u|_{p-k_1,k-k_1} 
+ { \delta^{-1}}( \epss +C_1\eps) r^{-\min(\lambda,\kappa)}s^{\delta} \frac{\crochet}{r} |\del\del u|_{p,k}
+ |h^{\N00}||\del\del u|_{p,k}
\\
& \quad +   (\epss+C_1\eps)  r^{-2}\crochet^{-\kappa}s^{\delta} 
+ { \delta^{-1}} (\epss + C_1\eps)^2 
\, r^{-2}\crochet^{- \min(\lambda,\kappa)}s^{2\delta}
+ \epss r^{-2-\lambda}.
\endaligned
$$
Using that { $\delta^{-1}(\epss + C_1\eps)\ll 1$ and $\min(\lambda,\kappa)\geq 1/2 + \delta/2$},  we deduce that, near the light cone, 
$$
\frac{1+|r-t|}{r}|\del\del u|_{p,k} 
\lesssim\sum_{0\leq p_1\leq k}|\hN{}^{00}|_{p_1,p_1}  |\del\del u|_{p-p_1,k-p_1} 
+  \big(\epss + C_1\eps\big) r^{-2}\crochet^{-\min(\lambda,\kappa)}s^{2\delta}. 
$$
Recalling \eqref{eq10-02-05-2020}, we obtain \eqref{eq3-29-01-2021} for all $p\leq N-4$. 
\end{proof}

%-------------------------------------------------------- 

\begin{proof}[Proof of \eqref{eq1-15-08-2021}]
For the estimate far from the light cone, our argument now is based on Proposition~\ref{propo2-22-05-2020} and is comparatively simpler. In the domain $\Mfar_{\ell, [s_0,s_1]}$, one has $r\geq (1-\ell)^{-1}t$ implies $(1+t \, \crochet^{-1})\lesssim \ell^{-1}$. 
Next, substituting the inequality in Lemma~\ref{lemma-123} together with \eqref{eq12-04-06-2020} into the estimate in Proposition~\ref{propo2-22-05-2020} (with $p\leq N-4$), we obtain 
$$
\aligned
|\del\del u|_{p,k} 
& \lesssim (1+t \, \crochet^{-1})\big(  
|\Boxt_g u|_{p,k}
+ \epss r^{-2-\lambda} \big) 
\\
&   \quad  + (1+t \, \crochet^{-1})(\epss + C_1\eps)t^{-1}r^{-1}\crochet^{-\kappa}s^{\delta} 
+ { \delta^{-1}}(\epss + C_1\eps) \, |\del\del u|_{p,k}
\\
& \lesssim  {\delta^{-1}}(\epss + C_1\eps) \, |\del\del u|_{p,k} + 
\ell^{-1} (\epss + 
C_1\eps)
\begin{cases}
t^{-1}r^{-1}\crochet^{-\min(\lambda,\kappa)}s^{2\delta},\quad    & \text{Class {A}},
\\
t^{-1}r^{-1+2\theta}\crochet^{-\min(\lambda,\kappa)}s^{2\delta},\quad   & \text{Class {B}},
\end{cases}
\endaligned
$$ 
In view of our condition { $\delta^{-1}(\epss + C_1\eps)\ll 1$}, we arrive at
 \eqref{eq1-15-08-2021}. 
\end{proof}

%----------------------------------------------------------------------------------------------------------------------

\subsection{Application to the Hessian of $h^{\N00}$ near the light cone.}

Finally, we rely on \eqref{lem2-08-12-2020} and establish the following bound on $\del_t\del_t h^{\N00}$, which 
will play an essential role in the following analysis. 
We emphasize that the inequality below provides us with a {\sl sub-critical decay rate} near the light cone, in the sense that it is a decay faster than $1/r$. 

\begin{lemma} 
\label{eq7-08-12-2020-lem}
Under the conditions stated in Section~\ref{section-label-11-1},  as long as \eqref{eq10-04-06-2020} hold and using the wave gauge condition \eqref{eq2-27-05-2020} one has 
%\begin{equation}\label{eq7-08-12-2020}
$$
|\del_t\del_th^{\N00}|_{N-4}\lesssim (\epss + C_1\eps)r^{-1-\min(\lambda,\kappa)}\crochet^{-\min(\lambda,\kappa)}s^{2\delta}
\qquad \text{ in }\Mnear_{[s_0,s_1]}.
$$
\end{lemma}

\begin{proof} We recall that $h_{\alpha\beta} = h^{\star}_{\alpha\beta} + u_{\alpha\beta}$.  
Substituting the bounds \eqref{eq10-04-06-2020}, \eqref{eq1-26-05-2021}, and \eqref{eq1-09-05-2021} into \eqref{lem2-08-12-2020}, we obtain the statement in the lemma, provided  $\delta^{-1}(\epss + C_1\eps)\lesssim 1$. 
Concerning the estimate for the term $|h\del\del u|_{p,k}$, that is, (the third term in the right-hand side of \eqref{lem2-08-12-2020}), we apply \eqref{eq10-02-05-2020}, that is, 
$
|\del\del u|_{N-4}\lesssim |\del u|_{N-3}\lesssim  (\epss+C_1\eps)  r^{-1}\crochet^{-\kappa}s^{\delta}$.
\end{proof}

%======================================================================================

\section{Near-Schwarzschild decay of the null metric component}
\label{section---13} 

\subsection{Objective for this section}
\label{subsec1-02-09-2021}

\paragraph{Strategy.}

Our aim in this section is the derivation of the near-Schwarzschild decay of the null metric component, as stated below in Proposition~\ref{prop1-14-08-2021}, and in addition the derivation of the light-bending condition, as stated in Proposition~\ref{prop1-24-04-2021}. This second result will be proven by following the same arguments as the ones developed for the first result. The proof of Proposition~\ref{prop1-14-08-2021} relies on a decomposition of the spacetime domain into two sub-domains, referred to as the ``bad'' and ``good'' regions and defined in \eqref{equa-def-good-bad}. 
In the bad region, which is defined as a (thick) neighborhood of the light cone (covering points up to a distance $\sqrt{t}$) we 
integrate toward the light cone from the good domain. 
On the other hand, in the ``good'' region we rely on Kirchhoff's formula and we integrate from the initial data, by taking the properties of the 
source terms into account. 
Observe that Proposition~\ref{prop1-14-08-2021} relies on the assumed decay of the reference metric and the initial data of the perturbation.  We note in passing that this distance can be taken to be  
$(\epss+C_1\eps)^{1-c_3\delta}\sqrt{t}$, including in the critical regime treated in \cite{PLF-YM-SecondPart}. 
 
%------------------------------ 

\paragraph{A decomposition of the metric.} 

Recall that $h^{\alpha\beta} = g^{\alpha\beta} - g_{\Mink}^{\alpha\beta}$. 
We are going to rely on the decomposition 
\begin{equation}\label{eq1-19-04-2021}
h_{\alpha\beta} = h^{\star}_{\alpha\beta} + u_{\alpha\beta} = h^{\star}_{\alpha\beta} + u_{\init,\alpha\beta} + u_{\source, \alpha\beta},
\end{equation}
in which we distinguish between the contributions from the initial data and the source, namely 
$$
\aligned
&\Box u_{\init, \alpha\beta} = 0,
\qquad
&& u_{\init,\alpha\beta}(1,x) = u_{\alpha\beta}(1,x), \quad 
&&&& 
\del_t u_{\init,\alpha\beta}(1,x) = \del_t u_{\alpha\beta}(1,x),
\\
&\Box u_{\source,\alpha\beta} = \Box u_{\alpha\beta}, \quad 
&& u_{\source,\alpha\beta} = 0, 
\qquad
&&&& \del_t u_{\source,\alpha\beta}(1,x) = 0. 
\endaligned
$$
Then we recall \eqref{equa-signe-a-noter} and \eqref{eq1-07-05-2021} and, provided $|h|_{N-2}$ is sufficiently small, in view of the sup-norm bound \eqref{eq1-09-05-2021} we obtain 
\begin{equation}\label{eq4-19-04-2021}
|\Abb^{\alpha\beta}[h]|_{N-2}
\lesssim 
\delta^{-2}(\epss + C_1\eps)^2r^{-2\min(\lambda,\kappa)}s^{2\delta}. 
\end{equation}
Next, we introduce a decomposition of the null component 
$${ 
h^{\N00} = \PsiN_{\alpha}^0\PsiN_{\beta}^0 h^{\alpha\beta} = 
- \PsiN_{\alpha}^0\PsiN_{\beta}^0\mathbbm{h}^{\alpha\beta}[h]
+ \PsiN_{\alpha}^0\PsiN_{\beta}^0\Abb^{\alpha\beta}[h], } 
$$
as follows, which is our main decomposition of interest in this section\footnote{Here we abuse a bit the notation $u^{\N00}_{\init}$ because we are only interested in the linear contribution.}: 
\begin{equation}\label{eq2-19-04-2021}
\aligned
h^{\N00} & = { \Xi^{\star \N00}} + u^{\N00}_\init + h^{\N00}_{\pertur},  
\qquad 
&&{ \Xi^{\star\N00}}
:= - \PsiN_{\alpha}^0\PsiN_{\beta}^0{ \mathbbm{h}^{\star\alpha\beta}}, 
\\
u_{\init}^{\N00} 
& :=  - \PsiN_{\alpha}^0\PsiN_{\beta}^0{ g_{\Mink}^{\alpha\alpha'}u_{\init,\alpha'\beta'}g_{\Mink}^{\beta\beta'}},
\quad
\qquad 
&&
h^{\N00}_\pertur 
:=  - \PsiN_{\alpha}^0\PsiN_{\beta}^0{ g_{\Mink}^{\alpha\alpha'}u_{\sour,\alpha'\beta'}g_{\Mink}^{\beta\beta'}}
+ \PsiN_{\alpha}^0\PsiN_{\beta}^0\Abb^{\alpha\beta}[h].
\endaligned
\end{equation}
Clearly, we have
\begin{equation}\label{eq3-19-04-2021}
r \, |h^{\N00}|_{p,k}\lesssim r \, | { \Xi^{\star\N00}} |_{p,k} +r | u^{\N00}_\init |_{p,k} + r \, |u_{\source}|_{p,k} + (\epss + C_1\eps), 
\end{equation}
in which we used \eqref{eq4-19-04-2021} and our conditions { $\min(\lambda,\kappa)\geq 1/2+ \delta/2$ and $\delta^{-2}(\epss + C_1\eps)\lesssim 1$}. 

%--------------------------------

\paragraph{Main statement for this section.}

Our main task will be to establish the following result.

\begin{proposition}[Near-Schwarzschild decay of the null metric component]
\label{prop1-14-08-2021}
Under the conditions stated in Section~\ref{section-label-11-1},  
and using \eqref{eq1-11-03-2021-gstar} and \eqref{eq3-09-05-2021}, that is, 
\begin{equation}\label{eq15-23-04-2021}
\big|{ \Xi^{\star\N 00}} \, \big|_{N-4}\leq \epss \, r^{-1+\theta},
\qquad
 |u_{\init}^{\N00}|_{N-4}\leq  C_0\eps r^{-1+\theta}
\quad \text{in }\MME_{[s_0,s_1]}
\end{equation}
(with $ \theta \leq \delta/4$) and using \eqref{eq4-09-05-2021}, that is,
\begin{equation}\label{eq5-25-04-2021}
|^{(w)}R^{\star}_{\alpha\beta}|_{N-4}\lesssim \epss^2 r^{-2-3\delta}\crochet^{-1+\delta}
\quad \text{in }\MME_{[s_0,s_1]}, 
\end{equation}
one has 
\begin{equation}\label{eq19-23-04-2021}
|h^{\N00}|_{N-4}\lesssim  (\epss + C_1\eps)r^{-1+\theta}
\quad \text{ in } \MME_{[s_0, s_1]}.
\end{equation}
\end{proposition}

For convenience, we also introduce the domains 
\begin{equation}\label{equa-def-good-bad} 
\aligned
\Mgood_{[s_0,s_1]} 
& := \big\{ r\geq t-1 + (\epss + C_1\eps)^{1/2}t^{1/2} \big\}\cap \MME_{[s_0,s_1]},
\quad
\\
\Mbad_{[s_0,s_1]} 
& := \big\{ t-1\leq r\leq t-1 + (\epss + C_1\eps)^{1/2}t^{1/2} \big\}\cap \MME_{[s_0,s_1]}.
\endaligned
\end{equation}
We proceed by establishing first sufficient decay on the source contribution $|\Box u_{\alpha\beta}|_k$ (in \eqref{eq5-23-04-2021} and \eqref{eq6-23-04-2021} below) within the region $\Mgood_{[s_0,s_1]}$, 
next by applying Kirchhoff's formula (stated in Proposition~\ref{Linfini wave}) in order to handle the contribution $u_{\source}$ from the source term 
and establish \eqref{eq19-23-04-2021} in $\Mgood_{[s_0,s_1]}$. 
For the estimate in $\Mbad_{[s_0,s_1]}$, we proceed by a technique called ``integration toward the light cone'' which is introduced in Section~\ref{subsec1-25-04-2021}.  

%----------------------------------------------------------------------------------------- 

In addition, by applying \eqref{eq19-23-04-2021} to \eqref{eq3-29-01-2021} we obtain the {\sl improved pointwise estimate} for the components of the Hessian of the wave fields. 

\begin{corollary} The following improved pointwise estimates of the Hessian hold: 
\begin{subequations} \label{eqs2-07-09-2022-M}
\begin{equation}\label{eq1-09-08-2021}
|\del\del u|_{N-4} \lesssim (\epss + C_1\eps) r^{-1}\crochet^{-1-\min(\lambda,\kappa)} (r^{\theta} + s^{\delta})s^{\delta}
\quad\text{ in } \Mscr^{\near}_{[s_0,s_1]}, 
\end{equation}
\begin{equation}\label{eq2-09-08-2021}
|\del\delsN u|_{N-4}\lesssim (\epss + C_1\eps) r^{-2}\crochet^{-\min(\lambda,\kappa)}(r^{\theta} + s^{\delta})s^{\delta}
\quad\text{ in } \Mscr^{\near}_{[s_0,s_1]}.
\end{equation} 
\end{subequations} 
\end{corollary} 

%-----------------------------------------------------------------------------

\subsection{Control of the null component in the good region}
\label{subsec1-13-08-2021}

\paragraph{Wave operator contribution $|\Box u|_k$.} 

Recall that $|u|_k := \max_{\alpha,\beta}|u_{\alpha\beta}|_k$. We establish a preliminary result first. 

\begin{lemma}\label{lem1-25-04-2021}
For Class A and Class B metrics, under the conditions stated in Section~\ref{section-label-11-1},  
by using \eqref{eq5-25-04-2021} 
in $\Mgood_{[s_0,s]}\cap \Mnear_{[s_0,s]}$ with $s\in[s_0,s_1]$ 
one has 
\begin{equation}\label{eq5-23-04-2021}
\aligned
|\Box u|_k& \lesssim 
(\epss + C_1\eps)^{1-4\delta}r^{-2-3\delta}\crochet^{-1+\delta}\Abf_k(s)^2 
+ (\epss + C_1\eps)^{1-4\delta}r^{-2-3\delta}\crochet^{-1+\delta}\Abf_k(s) 
\\
& \quad +  \delta^{-1}(\epss + C_1\eps)^{2- 5 \delta} r^{-2-3\delta} \crochet^{-1+\delta},
\hskip6.cm 
k\leq N-4, 
\endaligned
\end{equation}
where $\Abf_k(s) := \sup_{\Mgood_{[s_0,s]}}\big( r \, |  u_{\source} |_k\big)$.  
On the other hand,  
in $ \Mfar_{[s_0,s]}$ one has 
\begin{equation}\label{eq6-23-04-2021}
|\Box u|_{N-4}\lesssim
\delta^{-1} (\epss + C_1\eps)^2r^{-2-3\delta}\crochet^{-1+\delta} + \delta^{-1}(\epss + C_1\eps)^2 t^{-1+(3/2)\delta}r^{-1-\min(\lambda,\kappa)}\crochet^{-1+(1-\min(\lambda,\kappa))}.
\end{equation}
\end{lemma}

\begin{proof} We need to establish pointwise bounds on the following quantities in $\Mgood_{[s_0,s_1]}$:
\begin{equation}\label{eq3-23-04-2021}
|h^{\mu\nu}\del_{\mu}\del_{\nu}u|_k,
\qquad |u^{\mu\nu}\del_{\mu}\del_{\nu}h^{\star}|_k,
\qquad 
|\Pbb^{\star}[u]|_k,
\qquad |\Qbb^{\star}[u]|_k,
\qquad |\Ibb^{\star}[u]|_k,
\qquad |\Tbb[\phi]|_k,
\qquad
|^{(w)}R^{\star}_{\alpha\beta}|_k.
\end{equation}
We handle the quadratic nonlinearities $\Pbb^{\star}[u]$ and $\Qbb^{\star}[u]$ without considering their (quasi-null or null) structure, and we directly rely on the rough bound \eqref{eq5-04-06-2020}. In 
$\Mgood_{[s_0,s_1]}\cap\Mnear_{[s_0,s_1]}$, since {$\min(\lambda,\kappa)\geq 1/2+(7/2)\delta$,} and $\crochet^{-1}\leq (\epss + C_1\eps)^{-1/2} r^{-1/2}$,  we find 
\begin{equation}\label{eq7-23-04-2021}
|\Pbb^{\star}[u]|_{N-4} + |\Qbb^{\star}[u]|_{N-4}
\lesssim (\epss + C_1\eps)^{2-4\delta} r^{-2-3\delta}\crochet^{-1+\delta}
\qquad 
\text{ in } \Mgood_{[s_0,s_1]} \cap\Mnear_{[s_0,s_1]}, 
\end{equation}
while 
in $\Mfar_{[s_0,s_1]}$, recalling  $r\lesssim \crochet$,  thanks to the pointwise bound \eqref{eq5-04-06-2020} again we get 
\begin{equation}\label{eq8-23-04-2021}
|\Pbb^{\star}[u]|_{N-4} + |\Qbb^{\star}[u]|_{N-4}
\lesssim (\epss + C_1\eps)^2 r^{-2-(\min(\lambda,\kappa)-\delta)}\crochet^{-1+(1-\min(\lambda,\kappa))}
\qquad \text{ in } 
\Mfar_{[s_0,s_1]}.
\end{equation}
Consequently, recalling the smallness condition $\epss + C_1\eps \leq  1$ 
and { $\min(\lambda,\kappa)\geq 2\delta$} we conclude that 
\begin{equation}\label{eq1-22-04-2021}
|\Pbb^{\star}[u]|_{N-4} + |\Qbb^{\star}[u]|_{N-4}
\lesssim (\epss + C_1\eps)^{2-4\delta} r^{-2-3\delta}\crochet^{-1+\delta}
\qquad 
\text{ in } \Mgood_{[s_0,s_1]}. 
\end{equation}
Recalling \eqref{eq1-04-12-2020}, 
 in $\Mgood_{[s_0, s_1]}$ we have 
\begin{equation}\label{eq4-23-04-2021}
\aligned
|\Ibb^{\star}[u]|_{N-4} + |u^{\mu\nu}\del_{\mu}\del_{\nu} h^{\star}|_{N-4}
& \lesssim  
\begin{cases}
\delta^{-1}(\epss + C_1\eps)^2r^{-2-(\min(\lambda,\kappa)-\delta)}\crochet^{-1+(1-\min(\lambda,\kappa))},\quad & \text{Class A},
\\
 \delta^{-1}(\epss + C_1\eps)^{2}r^{-2+2\theta}\crochet^{-2\kappa}s^{\delta},
\quad 
&  \text{Class B},
\end{cases}
\\
&\lesssim
\begin{cases}
\delta^{-1}(\epss+C_1\eps)^2r^{-2-3\delta}\crochet^{-1+\delta},\quad 
& \text{Class A},
\\
 \delta^{-1}(\epss+C_1\eps)^{2-4\delta}r^{-2-3\delta}\crochet^{-1+\delta},\quad 
&  \text{Class B}.
\end{cases} 
\endaligned
\end{equation}
Recalling \eqref{eq1-28-11-2020} and using { $\min(\lambda,\kappa)\geq 1/2+(9/2)\delta$, $\mu \geq 3/4 + (7/4)\delta$}, in $\MME_{[s_0, s_1]}$ we have 
\begin{equation}\label{eq9-23-04-2021}
\aligned
|\Tbb[\phi]|_{N-4}& \lesssim   (\epss+C_1\eps) ^2 r^{-2-(1/2 - (3/2)\delta)}\crochet^{-1+(2-2\mu)}
+  (\epss+C_1\eps) ^2 r^{-2 - (\lambda -1/2 - (3/2)\delta)}\crochet^{-1+(1-2\mu)} 
\\
& \lesssim   (\epss+C_1\eps) ^2 \, r^{-2-3\delta}\crochet^{-1+\delta}
 \qquad \text{in }\MME_{[s_0,s_1]}.
\endaligned
\end{equation}
The reduced Ricci curvature $|^{(w)}R^{\star}_{\alpha\beta}|_{N-4}$ is bounded by \eqref{eq5-25-04-2021}.

Now we have bounded all terms in \eqref{eq3-23-04-2021} except the first one. For this term, observe that in $\Mfar_{[s_0,s_1]}$ the bound is relatively trivial. Thanks to \eqref{eq1-19-07-2020} and \eqref{eq1-09-05-2021},
\begin{equation}\label{eq2-22-04-2021}
|h^{\mu\nu}\del_{\mu}\del_{\nu}u|_{N-4}\lesssim \delta^{-1}(\epss + C_1\eps)^2 t^{-1+(3/2)\delta}r^{-1-\min(\lambda,\kappa)}\crochet^{-1+(1-\min(\lambda,\kappa))}\quad \text{in }\Mfar_{[s_0,s_1]}
\end{equation}
In $\Mnear_{[s_0,s_1]}\cap\Mgood_{[s_0,s_1]}$, the estimate is more involved. Recall that
$$
h^{\mu\nu}\del_{\mu}\del_{\nu} u_{\alpha\beta} = h^{\N00}\del_t\del_t u_{\alpha\beta} 
+ \sum_{(\mu,\nu)\neq (0,0)}h^{\N\mu\nu}\delN_{\mu}\delN_{\nu}u_{\alpha\beta} + h^{\mu\nu}\del_{\mu}\big(\PsiN_{\nu}^{\nu'}\big)\delN_{\nu'}u_{\alpha\beta}.
$$ 
Observe that the last term has a decreasing factor $\del_{\mu}\big(\PsiN_{\nu}^{\nu'}\big)$ which is homogeneous of degree $(-1)$. Then we have 
\begin{equation}\label{eq3-22-04-2021}
|h^{\mu\nu}\del_{\mu}\del_{\nu} u_{\alpha\beta}|_k\lesssim |h^{\N00}\del_t\del_t u|_k 
+ | H \, \del\delsN u|_k + r^{-1}| H \, \del u|_{N-4}.
\end{equation}
The latter term contains a favorable factor $r^{-1}$ and is bounded by 
$
\delta^{-1}(\epss + C_1\eps)^2r^{-2-\min(\lambda,\kappa)}\crochet^{-\min(\lambda,\kappa)}s^{2\delta} 
$
thanks to  \eqref{eq10-02-05-2020} and \eqref{eq1-30-05-2020}. 
For the second term, 
recalling \eqref{eq3-19-04-2021} and \eqref{eq15-23-04-2021}, for $(t,x)\in \Mgood_{[s_0,s]}$
we have (provided  $C_0\leq C_1$) 
\begin{equation}\label{eq1-05-09-2021}
r|h^{\N00}|_k(t,x)\lesssim \Abf_k(s) + (\epss + C_1\eps)r^{\theta}. 
\end{equation}
We recall \eqref{eq1-30-05-2020}, \eqref{eq3-29-01-2021} and, for all $k\leq N-4$ and $(t,x)\in \Mnear_{[s_0,s]}$, we have 
\begin{equation}\label{eq7-25-04-2021}
\aligned
& 
| H \del\delsN u|_k(t,x)  \lesssim  | H |_{N-4}(t,x) \, |\del\delsN u|_k(t,x)
\\
& \lesssim \delta^{-1}(\epss + C_1\eps)
r^{-\min(\lambda,\kappa)}s^{\delta}
\big( (\epss+C_1\eps)  r^{-2}\crochet^{-\kappa}s^{\delta}(r|h^{\N00}|_k) + (\epss+C_1\eps) r^{-2}\crochet^{-\min(\lambda,\kappa)}s^{2\delta}\big)
\\
& \lesssim   (\epss+C_1\eps)  r^{-2-\min(\lambda,\kappa)}\crochet^{-\min(\lambda,\kappa)}s^{2\delta}\Abf_k(s) 
+  \delta^{-1}(\epss+C_1\eps)^2 r^{-2-\min(\lambda,\kappa)}\crochet^{-\min(\lambda,\kappa)}
s^{2\delta}(s^{\delta} + r^{\theta}), 
\endaligned
\end{equation}
where we used { $\delta^{-1}(\epss + C_1\eps)\lesssim 1$}. For the first term in the right-hand side of \eqref{eq3-22-04-2021},
recalling \eqref{eq3-29-01-2021} we write ($k\leq N-4$)
\begin{equation}\label{eq8-25-04-2021}
\aligned
|h^{\N00}\del\del u|_k& \lesssim  |h^{\N00}|_k|\del\del u|_k
\\
& \lesssim r^{-1}\big(\Abf_k(s) + (\epss + C_1\eps)r^{\theta}\big)
\big( (\epss+C_1\eps) r^{-1}\crochet^{-1-\kappa}s^{\delta}(r|h^{\N00}|_k) + (\epss+C_1\eps) r^{-1}\crochet^{-1-\min(\lambda,\kappa)}s^{2\delta}\big)
\\
& \lesssim    (\epss+C_1\eps)  r^{-2}\crochet^{-1-\min(\lambda,\kappa)}s^{\delta}\Abf_k(s)^2 
+ (\epss + C_1\eps)r^{-2}\crochet^{-1-\min(\lambda,\kappa)}s^{\delta}{ (s^{\delta} + r^{\theta})}\Abf_k(s)
\\
&\quad + (\epss + C_1\eps)^2r^{-2+\theta}\crochet^{-1-\min(\lambda,\kappa)}s^{\delta}(s^{\delta} + r^{\theta}).
\endaligned
\end{equation}
Then we conclude that
$$
\aligned
|h^{\mu\nu}\del_{\mu}\del_{\nu} u_{\alpha\beta}|_k
& \lesssim  (\epss+C_1\eps)  r^{-2}\crochet^{-1-\kappa}s^{\delta}\Abf_k(s)^2 
+(\epss + C_1\eps)r^{-2}\crochet^{-2\min(\lambda,\kappa)}s^{\delta}(s^{\delta} + r^{\theta})\Abf_k(s) 
\\
& \quad +\delta^{-1}(\epss + C_1\eps)^2r^{-2+\theta}\crochet^{-2\min(\lambda,\kappa)}s^{2\delta}(s^{\delta} + r^{\theta}). 
\endaligned
$$
The above bound {\sl would not} be sufficient for our purpose in $\Mnear_{[s_0,s_1]}$. However, since we are now interested in $\Mgood_{[s_0,s_1]}\cap\Mnear_{[s_0,s_1]}$ and by using $\crochet^{-1}\leq ( (\epss+C_1\eps)  r)^{-1/2}$, we obtain, { since $\min(\lambda,\kappa)\geq 1/2+ (9/2)\delta$ and $\min(\lambda,\kappa)\geq 6\delta$}, 
\begin{equation}\label{eq2-23-04-2021}
\aligned
|h^{\mu\nu}\del_{\mu}\del_{\nu} u_{\alpha\beta}|_k
& \lesssim (\epss+C_1\eps)^{1-4\delta} r^{-2-3\delta}\crochet^{-1+\delta}\Abf_{k}(s)^2 
+ (\epss + C_1\eps)^{1-4\delta}r^{-2-3\delta}\crochet^{-1+\delta}\Abf_k(s)
\\
& \quad + \delta^{-1}(\epss + C_1\eps)^{2{ -5\delta}} r^{-2- 3\delta}\crochet^{ -1+\delta}.
\endaligned
\end{equation}
Finally, \eqref{eq5-25-04-2021}, \eqref{eq1-22-04-2021}, \eqref{eq4-23-04-2021}, \eqref{eq9-23-04-2021}, and \eqref{eq2-23-04-2021} show \eqref{eq5-23-04-2021} in $\Mgood_{[s_0,s]}\cap \Mnear_{ [s_0,s]}$ with $s\in[s_0,s_1]$. The bound \eqref{eq6-23-04-2021} is 
now a consequence of \eqref{eq5-25-04-2021}, \eqref{eq8-23-04-2021}, \eqref{eq4-23-04-2021}, \eqref{eq9-23-04-2021}, and \eqref{eq2-22-04-2021}.
\end{proof}

%--------------------------------------------------------------------------------------

\paragraph{Application of Kirchhoff's formula.}

Now observe that $\Mgood_{[s_0,s]}$ is 
``past complete'', in the sense that any point $(t,x)\in \Mgood_{[s_0,s]}$
has the past light cone $\Lambda_{t,x} = \{(t',x')\in \MME_{[s_0,s]}, t-t' = |x-x'|\}\subset \Mgood_{[s_0,s]}$. Then we apply Proposition~\ref{Linfini wave}  (Cases 1 and 2) on each term in the right-hand side of \eqref{eq5-23-04-2021}.  Since  { $\min(\lambda,\kappa)\geq 1/2+(5/4)\delta$}, \eqref{eq6-23-04-2021}  and obtain 
\begin{equation}\label{eq13-23-04-2021}
r \, |u_{\source}|_k\lesssim \delta^{-2}(\epss + C_1\eps)^{1-4\delta}\Abf_k(s)^2
+ \delta^{-2}(\epss + C_1\eps)^{1-4\delta}\Abf_{k}(s) + \delta^{-3}(\epss + C_1\eps)^{2- 5\delta}
\qquad \text{ in } \Mgood_{[s_0,s]}. 
\end{equation}
Similarly, since  $\Mfar_{[s_0,s]}\cap\Mgood_{[s_0,s]}$ is also past complete, we obtain
\begin{equation}\label{eq12-23-04-2021}
r \, |u_{\source}|_k\lesssim \delta^{-3}(\epss + C_1\eps)^2\lesssim \epss + C_1\eps
\qquad \text{ in } \Mfar_{[s_0,s]}\cap\Mgood_{[s_0,s]}.
\end{equation}
Recalling \eqref{eq3-19-04-2021}, \eqref{eq15-23-04-2021}, and \eqref{eq13-23-04-2021}, and since  { $\delta^{-2}(\epss + C_1\eps)^{1-4\delta}\ll 1$}, we obtain 
\begin{equation}\label{eq17-23-04-2021}
\Abf_k(s)\lesssim  \delta^{-2}(\epss + C_1\eps)^{1-4\delta}\Abf_k(s)^2
+ \delta^{-3}(\epss + C_1\eps)^{2-5\delta}, 
\qquad k\leq N-4. 
\end{equation}
Clearly, the function $\Abf_k(s)$ are continuous functions of $s$. Denote by $\Mgood_{s_0} = \Mscr_{s_0}\cap\{t-1+(\epss + C_1\eps)^{1/2}t^{1/2}\leq r\}$, one has
$$
\Abf_k(s_0) = \sup_{\Mgood_{s_0}} \big( r \, | u_{\source}|_k\big) \lesssim \epss\ll 1.
$$
Let $[s_0,s^*]\subset [s_0,s_1]$ be the largest subset where $\Abf_k(s)\leq 1$. By continuity, $s^*>s_0$ and if $s^*<s_1$, we would have
$\Abf_k(s^*) = 1$.
Within $[s_0,s^*]$, \eqref{eq17-23-04-2021} leads us to, thanks to the smallness condition {$\delta^{-3}(\epss +C_1\eps)^{1-4\delta}\ll 1$}, 
\begin{equation}\label{equation-1326b} 
\Abf_k(s)\lesssim \delta^{-3}(\epss + C_1\eps)^{2-5 \delta}\ll 1.
\end{equation} 
This contradicts our assumption $\Abf_k(s^*) = 1$. So we conclude that $\Abf_k(s)\leq 1$ for all $s \in [s_0,s_1]$,
and  \eqref{eq17-23-04-2021} leads us to
\begin{equation}\label{eq14-23-04-2021}
\sup_{\Mgood_{[s_0,s_1]}} \big( r \, | u_{\source}|_{N-4} \big)
\lesssim  \delta^{-3}(\epss + C_1\eps)^{2-5\delta} \lesssim \epss + C_1\eps.
\end{equation}
Recalling \eqref{eq3-19-04-2021} and \eqref{eq15-23-04-2021} and { the condition $\delta^{-3}(\epss + C_1\eps)^{1-5\delta}\lesssim 1$}, we obtain the partial conclusion
$
| u_{\source}|_{N-4}\lesssim (\epss + C_1\eps)r^{-1}
$
$
\text{ in }\Mgood_{[s_0,s_1]}
$
which, thanks to \eqref{eq3-19-04-2021} and \eqref{eq15-23-04-2021}, leads us to 
\begin{equation}\label{eq18-23-04-2021}
|h^{\N00}|_{N-4}\lesssim (\epss + C_1\eps)r^{-1+\theta}
\qquad \text{in } \Mgood_{[s_0,s_1]}. 
\end{equation}

%----------------------------------------------------------------------------------------------

\subsection{Control of the null component in the bad region}
\label{subsec1-25-04-2021}

We are now in a position to estimate $|h^{\N00}|_{N-4}$ in the region $\Mbad_{[s_0,s_1]}$. Recall \eqref{eq1-17-07-2020}, namely 
$$
|\del_rZh^{\N00}|\leq \delta^{-1}(\epss+C_1\eps)r^{-1-\min(\lambda,\kappa)}s^{\delta},
\qquad
\quad \ord(Z)\leq N-4.
$$
We now perform an integration toward the light cone from the good region, as follows. 
Given any point $(t,x)\in \Mbad_{[s_0,s_1]}$ we consider a point $(t,\bar{x})\in\{r = t-1 + (\epss +C_1\eps)^{1/2}t^{1/2}\}\cap \MME_{[s_0,s]}$ with $\bar{x}/|\bar{x}| = x/|x|$ and we write 
$$
Zh^{\N00}(t,x) = Zh^{\N00}(t,\bar{x}) - \int_{|x|}^{|\bar{x}|}\del_rZh^{\N00}(t,\rho) \, d\rho. 
$$
This gives us {(since  $\min(\lambda,\kappa)\geq 1/2+\delta/2$ and $\delta^{-2}(\epss + C_1\eps)\lesssim 1$)},
\begin{equation}\label{equation1333}
\aligned
|Zh^{\N00}(t,x)|
& \lesssim |Zh^{\N00}(t,\bar {x})| + \delta^{-1}(\epss + C_1\eps)^{3/2}t^{-1/2-\min(\lambda,\kappa) + \delta/2}
\lesssim (\epss + C_1\eps)r^{-1+\theta} + (\epss + C_1\eps) r^{-1}.
\endaligned
\end{equation} 
Observe here that $r\simeq t$ in $\Mbad_{[s_0,s_1]}$. Combining the above bound with \eqref{eq18-23-04-2021}, the desired bound in $\MME_{[s_0, s_1]}$ is established. 

%----------------------------------------------------------------------------------------------------------------------------------

\subsection{Derivation of the light-bending property}

By applying the same technique of proof as for Proposition~\ref{prop1-14-08-2021}, we can also derive the sign condition \eqref{eq3'-27-05-2020}. 

\begin{proposition}[The light-bending property] 
\label{prop1-24-04-2021} 
Under the same conditions as in Proposition~\ref{prop1-14-08-2021} and, furthermore, by using  the
linear light-bending condition \eqref{eq3'-27-05-2020-initial}, restated here as 
\begin{equation}\label{eq9-25-04-2021} 
\epss  \leq - r \, (u_{\init}^{\N00}+\Xi^{\star\N00}) \leq \epss \, r^{\theta},
\end{equation}
% with {$\min(\lambda,\kappa)\geq 1/2+\delta/2$} and $ \theta\leq \delta/4$ and  { $(\epss+C_1\eps)^{2-5\delta}\ll \delta^3\epss$}, 
the null component of the metric satisfies 
\begin{equation}\label{eq1-24-04-2021}
r \, h^{\N00}
\leq -\epss/2<0 \qquad  \text{ in } \Mnear_{\ell,[s_0,s_1]}.
\end{equation}
\end{proposition}

\begin{proof} We follow the same strategy as in the proof of Proposition~\ref{prop1-14-08-2021}. In the good region, we have 
$$
\sup_{\Mgood_{[s_0,s_1]}} \big( r \, |u_{\source}| \big)
\lesssim  \delta^{-3}(\epss + C_1\eps)^{2- 5 \delta}, 
$$
which is \eqref{eq14-23-04-2021} at zero order. Then, in view of the decomposition \eqref{eq2-19-04-2021} and \eqref{eq9-25-04-2021} together with \eqref{eq4-19-04-2021} (thanks to  {$\min(\lambda,\kappa)\geq 1/2+\delta/2$}) we find 
$$
\epss - C\delta^{-3}(\epss + C_1\eps)^{2- 5 \delta} \leq -rh^{\N00}\leq \epss \, r^{\theta} + C\delta^{-3}(\epss + C_1\eps)^{2- 5 \delta}.
$$
Since { $ \delta^{-3}(\epss + C_1\eps)^{2- 5 \delta}\ll \epss$}, we obtain
$ 
-rh^{\N00} \geq 3\epss/4 $
$  \text{ in } \Mgood_{[s_0,s_1]}.
$
On the other hand, in $\Mbad_{[s_0,s_1]}$ and similarly to the analysis in Section~\ref{subsec1-25-04-2021}, we obtain 
the lower bound on $h^{\N00}$ by integration toward the light cone. 
Namely, given any point $(t,x)\in \Mbad_{[s_0,s_1]}$ we consider the
point $(t,\bar{x})\in\{r = t-1 + (\epss +C_1\eps)^{1/2}t^{1/2}\}\cap \MME_{[s_0,s]}$ with $\bar{x}/|\bar{x}| = x/|x|$ and write 
$
h^{\N00}(t,x) = h^{\N00}(t,\bar{x}) - \int_{|x|}^{|\bar{x}|}\del_rh^{\N00}(t,\rho) \, d\rho, 
$. 
{ In view of $\min(\lambda,\kappa)\geq 1/2+\delta/2$ and $\delta^{-1}\epss^{-1}(\epss + C_1\eps)^{1/2}\ll1$}, this gives us 
$$
rh^{\N00}(t,x)
\geq rh^{\N00}(t,\bar {x})
- \delta^{-1}r(\epss + C_1\eps)^{3/2}t^{-1/2-\min(\lambda,\kappa) + \delta/2}
\geq -\epss/2
$$
and the derivation of \eqref{eq1-24-04-2021} is completed.
\end{proof}

%=================================================================================

\section{Sharp decay for the gradient of the good metric components}  
\label{sec--gradient-good}

\subsection{Main statement for this section} 

Our objective now is to control the good metric components and, more precisely, their gradient and Hessian.   
To this end, we apply the gradient estimates for wave fields presented in Section~\ref{section----83}. 
We thus rely on the general decomposition \eqref{eq 1 13-01-2019} for Class A metrics and \eqref{eq1-30-11-2020} for Class B metrics. The right-hand side of \eqref{eq 1 13-01-2019} contains 
only one challenging term, namely the quasi-null term $\Pbb$ which may not enjoy integrable decay,
as is required for the weighted pointwise estimate in Proposition~\ref{prop1-23-07-2020}. However, thanks to the tensorial structure in \eqref{eq1 05-juillet-2019}, in the equations satisfied by the {\sl good components} of $u$ the relevant components of the quasi-null terms are actually {\sl null terms} and, consequently, enjoy sufficient decay for our argument. We will thus arrive at the following result which covers, both Class A and Class B metrics. 
Sections~\ref{subsec1-07-09-2022-M}--\ref{section---143} below are concerned with the proof for metrics in Class A,  while the proof for reference metrics in Class B is presented in Section~\ref{subsec1-21-09-2022}. 

\begin{proposition}[Estimates for the gradient of the good metric components]
\label{proposition14-1}
Assume that the reference metric belong to Class A. 
Under the conditions stated in Section~\ref{section-label-11-1} 
%Assume the bootstrap assumptions \eqref{eqs-int}--\eqref{eq3-27-05-2020} and { $0 \leq \epss + C_1\eps \leq \theta\leq \delta/2$ and  $\min(\lambda,\kappa)\geq 1/2+(5/2)\delta$, $\delta\leq 1/5$ and $\mu \geq 3/4+(7/4)\delta$}. 
and thanks to the near-Schwarzschild decay in \eqref{eq19-23-04-2021}, that is, 
\begin{equation}\label{eq1-05-05-2021}
| h^{\N00} |_{N-4}\leq (\epss + C_1\eps) (r+t)^{-1+\theta}
\qquad \text{ in } \Mnear_{\ell,[s_0,s_1]}, 
\end{equation} 
one has in the near-light cone domain $\Mnear_{\ell, [s_0,s_1]}$
\begin{subequations}\label{eq8-14-08-2021}
\begin{equation}\label{eq11-01-31-2021}
|\del \us^{\N}|_{N-4,k}\lesssim (\ell^{-\delta/2} + \delta^{-2})(\epss + C_1\eps) r^{-1+ k \theta}\crochet^{-1/2-\delta/2},
\qquad 0\leq k \leq N-4,
\end{equation}
\begin{equation}\label{eq8-24-03-2021}
|\del\del \us^{\N}|_{N-5,k}\lesssim (\ell^{-\delta}+\delta^{-2})(\epss + C_1\eps)r^{-1+k\theta}\crochet^{-1-\delta},
\qquad 0\leq k\leq N-5.
\end{equation}
\end{subequations}
In particular, the sharp decay rate $r^{-1}$ is achieved in the case $k=0$ when there is no boost or rotation. 
\end{proposition}
%-------------------------------------------------------------------------------------------------------------------------------------

\subsection{Linear contributions to the evolution} 
\label{subsec1-07-09-2022-M}

We begin by observing that the equations satisfied by the components  
$
u^{\N}_{\alpha\beta} = \PhiN_{\alpha}^{\alpha'}\PhiN_{\beta}^{\beta'}u_{\alpha'\beta'}
$
enjoy the decomposition 
\begin{equation}\label{eq4-23-07-2020}
\aligned
&\Boxt_gu^{\N}_{\alpha\beta} =  L_{1\alpha\beta} + L_{2\alpha\beta} + S_{1\alpha\beta},  
\quad
&& L_{1\alpha\beta} := u_{\alpha'\beta'}\Boxt_g\big(\PhiN_{\alpha}^{\alpha'}\PhiN_{\beta}^{\beta'}\big), 
\\
& L_{2\alpha\beta} := g^{\mu\nu}\del_{\mu}\big(\PhiN_{\alpha}^{\alpha'}\PhiN_{\beta}^{\beta'}\big)\del_{\nu}u_{\alpha'\beta'}, 
\qquad
&& S_{1\alpha\beta} := \PhiN_{\alpha}^{\alpha'}\PhiN_{\beta}^{\beta'}\Boxt_gu_{\alpha'\beta'}. 
\endaligned
\end{equation}
Differentiating \eqref{eq4-23-07-2020} with respect to $Z = \del^IL^J\Omega^K$ with $\ord(Z) \leq N-4$, we find
\begin{subequations}\label{eq5-14-08-2021}
\begin{equation}\label{eq10-25-07-2020}
\aligned
& \Boxt_g \big( Z u^{\N}_{\alpha\beta} \big) = Z L_{1\alpha\beta} + Z L_{2\alpha\beta} + Z S_{1\alpha\beta} + S_{2\alpha\beta}, 
\qquad 
&& S_{2\alpha\beta} = S_{2\alpha\beta}[u]
:= - [Z, h^{\mu\nu}\del_\mu\del_\nu] u^{\N}_{\alpha\beta},
\endaligned
\end{equation}
\begin{equation}\label{eq1-24-03-2021}
\Boxt_g(Z\del_t u^{\N}_{\alpha\beta}) = Z \del_t\big(L_{1\alpha\beta} + L_{2\alpha\beta} + S_{1\alpha\beta}\big) + S'_{2\alpha\beta},
\qquad 
S'_{2\alpha\beta} = S'_{2\alpha\beta}[u]
:= - [Z\del_t,h^{\mu\nu}\del_{\mu}\del_{\nu}]u^{\N}_{\alpha\beta}.
\end{equation}
\end{subequations}

\begin{proposition}
\label{proposition-label-144}
Assume that the reference metric belongs to Class A. 
Under the conditions stated in Section~\ref{section-label-11-1}, and under the decay condition \eqref{eq1-05-05-2021} 
% holds and { provided  $\min(\lambda,\kappa)\geq (1/2)+ (5/2)\delta$ and $\mu\geq 3/4 + (7/4)\delta$,} 
%
one has for all $\ord(Z)\leq N-4$ and $\rank(Z) = k\leq N-4$   
\begin{subequations}\label{eq6-14-08-2021}
\begin{equation}\label{eq9-31-01-2021}
\aligned
\crochet^{1/2+\delta/2}r|\Boxt_gZ\us^{\N}|
\lesssim  
(\epss + C_1\eps)r^{-1+\theta} \big(r \, \crochet^{1/2+\delta/2}|\del \us^{\N}|_{N-4,k-1}\big)
+ \delta^{-1} (\epss + C_1\eps) r^{-1-\delta}, 
\endaligned
\end{equation}
while for all $\ord(Z) \leq N-5$ and $\rank(Z) = k\leq N-5$ one has 
\begin{equation}\label{eq2-24-03-2021}
r \, \crochet^{1+\delta}|\Boxt_gZ\del_t \us^{\N}|
\lesssim (\epss + C_1\eps)r^{-1+\theta}\big(r \, \crochet^{\kappa}|\del\del \us^{\N}|_{N-5,k-1}\big)
+ \delta^{-1}(\epss + C_1\eps) r^{-1-\delta}.
\end{equation}
\end{subequations}
\end{proposition}

%---------------------------------------------

The term $L_1$ contains the decaying factor $\Boxt_g\big(\PhiN_{\alpha}^{\alpha'}\PhiN_{\beta}^{\beta'}\big)$ which is of order $r^{-2}$ (as is expected by homogeneity) and thus enjoys sufficient decay for our purpose. 
Indeed, using $|g^{\alpha\beta}|_{N-3}\lesssim 1$, 
in view of \eqref{eq10-02-05-2020} 
and \eqref{eq7-15-05-2020} and since { $\min(\lambda,\kappa)\geq 1/2+(5/2)\delta$} we have 
\begin{subequations}\label{eq7-14-08-2021}
\begin{equation}\label{eq2-25-07-2020} 
r \, \crochet^{1/2+\delta/2}|L_{1\alpha\beta}|_{N-4} 
\lesssim \delta^{-1} (\epss+C_1\eps)  \, r^{-1-\delta},
\end{equation}
\begin{equation}\label{eq5-24-03-2021}
r \, \crochet^{1+\delta}|\del L_{1\alpha\beta}|_{p,k}\lesssim \delta^{-1} (\epss+C_1\eps)  r^{-1-\delta}.
\end{equation}
\end{subequations}
Dealing with $L_2$ in \eqref{eq10-25-07-2020} is more involved and is the subject of the next lemma. 
Also, before we can apply Proposition~\ref{prop1-23-07-2020}, we will need to control the source term $S_1$
as well as the commutator $S_2$, and this will be the subject of Section~\ref{section---144}. 

\begin{lemma}
\label{lemma-6-octobre-2020} 
Under the conditions stated in Section~\ref{section-label-11-1},
%Provided  { $\min(\lambda,\kappa)\geq 1/2+ (5/2)\delta$ and $\delta^{-1}(\epss + C_1\eps)\lesssim 1$}, 
one has 
\begin{subequations}\label{eq3-14-08-2021}
\begin{equation}\label{eq3-25-07-2020}
\sum_{(\alpha,\beta)\neq (0,0)} 
r \, \crochet^{1/2+\delta/2} | L_{2\alpha\beta} |_{N-4}  \lesssim (\epss + C_1\vep) r^{-1-\delta}
\qquad \text{ in } \Mscr^{\near}_{[s_0,s_1]}, 
\end{equation}
\begin{equation}\label{eq6-24-03-2021}
\sum_{(\alpha,\beta)\neq (0,0)} 
r \, \crochet^{1+\delta}|\del L_{2\alpha\beta}|_{N-5}\lesssim (\epss + C_1\eps)r^{-1-\delta}
\qquad \text{ in } \Mscr^{\near}_{[s_0,s_1]}. 
\end{equation}
\end{subequations}
\end{lemma}

\begin{proof} {\bf Step 1.} Observe that we are interested in the components $(\alpha,\beta)\neq (0,0)$, only. 
A direct calculation shows that 
$$
\aligned
L_{2\alpha\beta}
& 
= (g_{\Mink}^{\mu\nu} + h^{\mu\nu}) \del_{\mu}\big(\PhiN_{\alpha}^{\alpha'}\PhiN_{\beta}^{\beta'}\big)\del_{\nu}u_{\alpha'\beta'}
= g_{\Mink}^{\N \mu\nu}\delN_{\mu}\big(\PhiN_{\alpha}^{\alpha'}\PhiN_{\beta}^{\beta'}\big)\delN_{\nu}u_{\alpha'\beta'}
+ h^{\mu\nu}\del_{\mu}\big(\PhiN_{\alpha}^{\alpha'}\PhiN_{\beta}^{\beta'}\big)\del_{\nu}u_{\alpha'\beta'}
\\
& = g_{\Mink}^{\N \mu 0}\delN_{\mu}\big(\PhiN_{\alpha}^{\alpha'}\PhiN_{\beta}^{\beta'}\big)\del_t u_{\alpha'\beta'}
+ g_{\Mink}^{\N \mu d}\delN_{\mu}\big(\PhiN_{\alpha}^{\alpha'}\PhiN_{\beta}^{\beta'}\big)\delsN_d u_{\alpha'\beta'}
+ h^{\mu\nu}\del_{\mu}\big(\PhiN_{\alpha}^{\alpha'}\PhiN_{\beta}^{\beta'}\big)\del_{\nu}u_{\alpha'\beta'}
\\
& =: A_1+A_2+A_3.
\endaligned
$$
The term $A_2$ contains good derivatives of $u$, while $A_3$ is quadratic and contains the {\sl decreasing factor} $\del_{\mu}\big(\PhiN_{\alpha}^{\alpha'}\PhiN_{\beta}^{\beta'}\big)$. Consequently, in view of \eqref{eq10-02-05-2020} and \eqref{eq1-30-05-2020}  we find 
$$
\aligned
|A_2| 
& \lesssim  r^{-1}|\delsN u|_{N-4}\lesssim  (\epss+C_1\eps)  \, r^{-2-\kappa}s^{\delta},
\\
|A_3| 
& \lesssim \delta^{-1}(\epss + C_1\eps)r^{-\min(\lambda,\kappa)}s^{\delta} \ r^{-1}\  (\epss+C_1\eps)  \, r^{-1}\crochet^{-\kappa}s^{\delta} 
\lesssim  (\epss+C_1\eps)  \, r^{-2-\min(\lambda,\kappa)}\crochet^{-\kappa} s^{2 \delta}, 
\endaligned
$$
where we used 
{$\delta^{-1}(\epss  + C_1\eps) \lesssim 1$}. 

It remains to show that, provided $(\alpha,\beta)\neq (0,0)$, the first term $A_1$ vanishes identically. In fact when $\alpha = 0$ and $\beta= b >0$, we obtain 
$$
\PhiN_{0}^{\alpha'}\PhiN_b^{\beta'} 
=
\begin{cases}
0, \quad           & \alpha' \neq 0 ,
\\
1,\quad         & \alpha' = 0, \, \beta'=b,
\\
x^{b}/r,\quad      & \alpha' = \beta' = 0.
\end{cases}  
$$
The only non-trivial case is $\alpha'=\beta'=0$ and, in this case, and recall that $g_{\Mink}^{00} = 0$,
$$
\aligned 
A_1 
& = g_{\Mink}^{\N \mu 0}\del_{\mu}^{\N}(x^b/r)\del_tu_{00} = g_{\Mink}^{\N c0}\delsN_c(x^b/r)\del_tu_{00} 
\\
& = -(x^c/r)\big(\del_c + (x^c/r)\del_t\big)(x^b/r)\del_tu_{00}
= -(x^c/r)\big(\delta_c^b/r - x^bx^c/r^3 \big)\del_tu_{00} = 0.
\endaligned
$$
Similarly, when $\alpha = a > 0,\beta = b>0$ we find 
$$
\PhiN_{a}^{\alpha'}\PhiN_b^{\beta'} = 
\begin{cases}
\text{constant},\quad    &\alpha',\beta'>0,
\\
x^a/r,\quad              & \alpha' = 0, \, \beta' = b,
\\
x^b/r,\quad            & \alpha' =a, \, \beta' = 0,
\\
x^ax^b/r^2,\quad & \alpha'=\beta'=0,
\end{cases}
$$
and in each case we find $g_{\Mink}^{\N \mu 0}\delN_{\mu}\big(\PhiN_a^{\alpha'}\PhiN_b^{\beta'}\big) = 0$. Now we conclude that
% \begin{equation}\label{eq1-14-08-2021}
$|L_{2\alpha\beta}|_{N-4}\lesssim  (\epss+C_1\eps)  r^{-2-\min(\lambda,\kappa)}s^{2\delta}$.
%\end{equation}
Using { $\min(\lambda,\kappa)\geq 1/2+(5/2)\delta$}, we obtain \eqref{eq3-25-07-2020}.

\vskip.3cm

\noindent{\bf Step 2.} The proof of \eqref{eq6-24-03-2021} is also direct. Recalling \eqref{eq10-02-05-2020} and \eqref{eq2-09-08-2021}, we have 
$$
|\del A_2|_{p,k} \lesssim  r^{-2}|\delsN u|_{N-5} + r^{-1}|\del\delsN u|_{N-5}\lesssim (\epss + C_1\eps)r^{-3}\crochet^{-\min(\lambda,\kappa)}
\,  s^{\delta}(s^{\delta} + r^{\theta}). 
$$
On the other hand, thanks to \eqref{eq10-02-05-2020} and \eqref{eq1-30-05-2020}  we have 
$$
\aligned
|\del A_3|_{p,k}
& \lesssim  r^{-1}\sum_{p_1+p_2=p\atop k_1+k_2=k}\big(|\del H |_{p_1,k_1}|\del u|_{p_2,k_2} + r^{-1}| H |_{p_1,k_1}|\del u|_{p_2,k_2} 
+ | H |_{p_1,k_1}|\del\del u|_{p_2,k_2}\big)
\\
& \lesssim  \delta^{-1}(\epss + C_1\eps)^2r^{-2-\min(\lambda,\kappa)}\crochet^{-\min(\lambda,\kappa)}s^{2\delta}. 
\endaligned
$$
{ Since $\min(\lambda,\kappa)\geq 1/2+ 2\delta$ and $\delta^{-1}(\epss + C_1\eps)\lesssim 1$},  we arrive at \eqref{eq6-24-03-2021}.
\end{proof}

%--------------------------------------------------------------------------------------------------------------------------------------------

\subsection{Contribution of the nonlinear sources} 
\label{section---144}

We focus first on the source $S_1$. Next, we will give the proof of Proposition~\ref{proposition-label-144} by including also the effect of the sources $S_{2\alpha\beta}$ and $S'_{2\alpha\beta}$. 

\begin{lemma}  Assume that the reference metric belong to Class A. 
Under the conditions stated in Section~\ref{section-label-11-1}
% Provided { $\delta^{-1}(\epss + C_1\eps)\lesssim 1$, $\min(\lambda,\kappa)\geq 1/2+(5/2)\delta$,  and $\mu \geq 3/4+(7/4)\delta$}, 
one has 
\begin{subequations}\label{eq4-14-08-2021}
\begin{equation}\label{eq1-26-07-2020}
\sum_{(\alpha,\beta)\neq(0,0)} 
r \, \crochet^{1/2+\delta/2} |S_{1\alpha\beta}|_{N-4}\lesssim (\epss + C_1\eps)r^{-1-\delta}
\qquad
\text{ in } \Mscr^{\near}_{[s_0,s_1]}, 
\end{equation}
\begin{equation}\label{eq7-24-03-2021}
\sum_{(\alpha,\beta)\neq(0,0)} 
r \, \crochet^{1+\delta}|\del_t S_{1\alpha\beta}|_{N-5}\lesssim (\epss + C_1\eps)r^{-1-\delta}
\qquad
\text{ in } \Mscr^{\near}_{[s_0,s_1]}. 
\end{equation}
\end{subequations}
\end{lemma}

\begin{proof}
Recalling \eqref{eq 1 13-01-2019} and using the tensorial structure \eqref{eq1 05-juillet-2019} of $\Pbb^{\star}[u]$, we write 
\begin{equation}\label{eq1-25-07-2020}
\aligned
S_{1\alpha\beta} 
= 
& \Pbb^{\star\N}_{\alpha\beta}[u] + \PsiN_{\alpha}^{\alpha'}\PsiN_{\beta}^{\beta'} \big(- u^{\mu\nu} \del_{\mu} \del_{\nu} g^\star_{\alpha'\beta'}
+ \Qbb^{\star}_{\alpha'\beta'}[u]  + \Ibb^{\star}_{\alpha'\beta'}[u] 
- 8\pi(2 \, T_{\alpha'\beta'} - Tg_{\alpha'\beta'})- 2 \, \Rwave_{\alpha'\beta'}\big).
\endaligned
\end{equation}
For all $(\alpha,\beta)\neq(0,0)$, by recalling \eqref{eq2 05-juillet-2019} in combination with \eqref{eq10-02-05-2020} and \eqref{equa-31-12-20} we obtain (an even stronger bound, for further application)
$$ 
r \, \crochet^{1+\delta}|\Pbb^{\star\N}_{\alpha\beta}[u]|_{N-4}  
\lesssim ( \epss+C_1\eps)^2r^{-1-\min(\lambda,\kappa)}\crochet^{1+\delta-\kappa} s^{2 \delta}
\lesssim ( \epss+C_1\eps )^2r^{-1-\delta}
$$
{ thanks to  $\min(\lambda,\kappa)\geq 1/2+(3/2)\delta$.} Similarly, the null terms are bounded thanks to \eqref{eq7-04-10-2022} combined with \eqref{eq10-02-05-2020} and \eqref{equa-31-12-20}:
$$  
r \, \crochet^{1+\delta}|\Qbb^{\star}_{\alpha'\beta'}[u]|_{N-4}\lesssim  (\epss+C_1\eps) ^2r^{-1-\delta}.
$$
The remaining terms in \eqref{eq1-25-07-2020} enjoy sufficient decay. 
In fact the reference-perturbation contributions $\Ibb^{\star}_{\alpha'\beta'}[u]$
together with $u^{\mu\nu} \del_{\mu} \del_{\nu} g^\star_{\alpha'\beta'}$ are bounded via \eqref{eq1-04-12-2020} { under the condition  $\delta^{-1} (\epss + C_1\eps) \lesssim 1$ and $\min(\lambda,\kappa)\geq 1/2+(3/2)\delta$:}
\begin{equation}\label{equa-28-aout-10}  
\aligned
&
r \, \crochet^{1+\delta}\big(|\Ibb^{\star}_{\alpha'\beta'}[u]|_{N-4} + |u^{\mu\nu} \del_{\mu} \del_{\nu} g^\star_{\alpha'\beta'}|_{N-4} \big)
\\
& \lesssim
\delta^{-1}(\epss + C_1\eps)^2 r^{-1-\lambda}\crochet^{1+\delta-\kappa}s^{2\delta}
   \lesssim (\epss + C_1\eps)r^{-1-\delta}.
\endaligned
\end{equation}
For matter source term, \eqref{eq1-28-11-2020}  leads us to, { in view of $\mu\geq 4/3+(7/4)\delta$ and $\min(\lambda,\kappa)\geq 1/2+(5/2)\delta$},
$$ 
\aligned
r \, \crochet^{1+\delta}|\Tbb(\phi)|_{N-3}
& \lesssim  (\epss + C_1\eps)^2\big(r^{-2}\crochet^{2-2\mu+\delta} + r^{-1-\lambda}\crochet^{1+\delta-2\mu}\big)s^{1+3\delta}
\lesssim  (\epss + C_1\eps)^2r^{-1-\delta}.
\endaligned
$$
Finally, for the curvature term $R^{\star}$ we apply \eqref{eq4-09-05-2021} which is assumed for Class A metrics.
\end{proof}

%-------------------------------------------------

\begin{proof}[Proof of Proposition~\ref{proposition-label-144}]
For these bounds we need to control $S_{2\mu\nu}[u]$ and $S'_{2\mu\nu}[u]$. We apply \eqref{eq2-30-05-2020} for $S_{2\mu\nu}[u] = -[Z,H^{\alpha\beta}\del_{\alpha}\del_{\beta}]\us_{\mu\nu}^{\N}$, which is possible since
$\us_{\mu\nu}^{\N}$ are linear combinations of $u_{\alpha\beta}$ with homogeneous coefficients of degree zero in $\MME_{[s_0, s_1]}$.
Hence, \eqref{eq2-30-05-2020} leads us to
\begin{equation}\label{eq3-24-03-2021}
|S_{2\mu\nu}[u]|
\lesssim \sum_{1\leq k_1 \leq k}|\hN{}^{00}|_{k_1} |\del\del \us^{\N}|_{p-k_1,k-k_1}
+ (\epss + C_1\eps) r^{-2-\min(\lambda,\kappa) + (3/2)\delta}\crochet^{-\min(\lambda,\kappa)},
\end{equation}
where the term $|\del\del u|_{p,k}$ in the second term of the right-hand side of \eqref{eq2-30-05-2020} is bounded by \eqref{eq1-09-08-2021}
which relies on \eqref{eq19-23-04-2021} or equivalently, \eqref{eq1-05-05-2021}) and the fact that { $\theta \leq \delta$}. For the last term in the right-hand side of \eqref{eq2-30-05-2020}, we used { $\delta^{-1}(\epss + C_1\eps)\lesssim 1$.} For $[Z\del_t,H^{\alpha\beta}\del_{\alpha} \del_\beta]$, fixing $\ord(Z)\leq N-5$, then $\ord(Z\del_t) = N-4$. { 
We have the identity
$$
[Z\del_t,H^{\alpha\beta}\del_{\alpha}\del_{\beta}]u = [Z,H^{\alpha\beta}\del_{\alpha}\del_{\beta}]\del_t u 
+ Z\big([\del_t,H^{\alpha\beta}\del_{\alpha}\del_{\beta}]u\big).
$$
}
{ For the first term, we apply \eqref{eq3-24-03-2021} and obtain (with $u$ now replaced by $\del_t u$ and $\ord(Z)\leq N-5$) 
}
\begin{equation}\label{eq1-06-09-2021}
|{[Z,H^{\alpha\beta}\del_{\alpha}\del_{\beta}]\del_t u}|\lesssim \sum_{1\leq   k_1\leq k}|\hN{}^{00}|_{k_1} |\del\del \del_t\us^{\N}|_{p-k_1,k-k_1}
+ (\epss + C_1\eps) r^{-2-\min(\lambda,\kappa) + (3/2)\delta}\crochet^{-\min(\lambda,\kappa)}.
\end{equation}
For the second term, we observe that
$$
[\del_t,H^{\alpha\beta}\del_{\alpha}\del_{\beta}]u = \del_tH^{\alpha\beta}\del_{\alpha}\del_{\beta}u
= \del_t H^{\N\alpha\beta}\delN_{\alpha}\delN_{\beta}u + \del_tH^{\alpha\beta}\del_{\alpha}\big(\PsiN^{\beta'}_{\beta}\big)\delN_{\beta'}u
$$
and, therefore, 
$$
\big|[\del_t,H^{\alpha\beta}\del_{\alpha}\del_{\beta}]u\big|_{N-5}
\lesssim |\del_t h^{\N00}\del_t\del_t u|_{N-5} + |\del H |_{N-5}|\del\delsN u|_{N-5} + r^{-1}| H \del u|_{N-5}.
$$
For the above three terms, we apply \eqref{eq1-17-07-2020}, \eqref{eq2-09-08-2021}, and \eqref{eq1-30-05-2020}
and obtain
\begin{equation}\label{eq2-06-09-2021}
\big|[\del_t,H^{\alpha\beta}\del_{\alpha}\del_{\beta}]u\big|_{N-5}\lesssim \delta^{-1}(\epss + C_1\eps)^2r^{-2-\min(\lambda,\kappa)+\delta}\crochet^{-\min(\lambda,\kappa)}.
\end{equation}
Then, based on \eqref{eq3-24-03-2021}, \eqref{eq1-06-09-2021}, and \eqref{eq2-06-09-2021} together with \eqref{eq1-05-05-2021} and thanks to { $\min(\lambda,\kappa)\geq 1/2 + (7/4)\delta$} and $\delta^{-1}(\epss + C_1\eps)\lesssim 1$, 
\begin{equation}\label{eq2-14-08-2021}
\aligned
&r \, \crochet^{1/2+\delta/2}|S_{2\alpha\beta}|
\lesssim (\epss + C_1\eps)r^{-1+\theta}\Big(r \, \crochet^{1/2+\delta/2}|\del \us^{\N}|_{N-4,k-1}\Big)
+ (\epss + C_1\eps) r^{-1-\delta},
\\
&r \, \crochet^{1+\delta}|S'_{2\alpha\beta}|\lesssim (\epss + C_1\eps)r^{-1+\theta}\big(r \, \crochet^{1+\delta}|\del\del_t \us^{\N}|_{N-5,k-1}\big)
+ (\epss + C_1\eps) r^{-1-\delta}.
\endaligned
\end{equation}
Now we substitute the above bound together with \eqref{eq7-14-08-2021},  \eqref{eq3-14-08-2021} and \eqref{eq4-14-08-2021} into  \eqref{eq10-25-07-2020}, then \eqref{eq6-14-08-2021} are established.
\end{proof}

%------------------------------------------------------------------------------------------------------

\subsection{Conclusion for metrics in Class A} 
\label{section---143}

\paragraph{Controlling the gradient.}

We are now in a position to conclude the derivation of \eqref{eq11-01-31-2021}, as follows. With the notation in Proposition~\ref{prop1-23-07-2020}, we observe that in view of \eqref{eq10-02-05-2020} expressed on the relevant cone $r = t(1-\ell)$ or 
$r-t\simeq \ell \, r$ 
$$
\crochet^{1/2+\delta/2}r |\del u|_{N-4}\lesssim  (\epss+C_1\eps)  \, \crochet^{1/2-\kappa+\delta/2}s^{\delta}\lesssim \ell^{-\delta/2}  (\epss+C_1\eps), 
$$
thanks to $\kappa \geq 1/2+\delta$. Thanks to \eqref{eq7-15-05-2020} we find 
$$
\crochet^{1/2+\delta/2}|u|_{N-4}\lesssim \delta^{-1} (\epss+C_1\eps)   r^{-1}\crochet^{3/2-\kappa+\delta/2}s^{\delta}
\lesssim  \delta^{-1} (\epss+C_1\eps).
$$
That is, { under the condition $\kappa  \geq1/2 + \delta$, $0<\ell<1$} and for ${\bf ord}(Z)\leq N-4$ one has 
$$
\sup_{\Omega_{s_0,s_1}^{\ell}} \crochet^{1/2+\delta/2}\big(  r \, |\del Z\us^{\N}| +  |Z\us^{\N}|\big) 
\lesssim \big(\ell^{-\delta/2} + \delta^{-1}\big) (\epss+C_1\eps) . 
$$
For $(t,x)\in \Mnear_{\ell, [s_0,s_1]}$ and $\ord(Z) = N-4$ with $\rank(Z) = k$, 
after observing that \eqref{eq1'-10-01-2021} are guaranteed by \eqref{eq3-27-05-2020} and 
\eqref{eq1-30-05-2020}, we obtain 
$$
\aligned
&\crochet^{1/2+\delta/2}|(\del_t-\del_r)(rZ\us^{\N})(t,x)|
\lesssim    (\ell^{-\delta/2} + \delta^{-1}) (\epss+C_1\eps) 
+ \int_{t_0}^t r\la r-\tau\ra^{1/2+\delta/2} |\Boxt_g Z\us^{\N}|_{\varphi_{t,x}(\tau)}d\tau
\\
&  +  \int_{t_0}^t 
\la r-\tau\ra^{1/2+\delta/2}\Big(|\delsN \us^{\N}|_{N-3} + r|H||\del\delsN Z\us^{\N}| + |H||\del \us^{\N}|_{N-4}\Big)\Big|_{\varphi_{t,x}(\tau)}d\tau
\\
& \lesssim   
(\ell^{-\delta/2} +  \delta^{-1}) (\epss+C_1\eps)  
+ \int_{t_0}^t(\epss + C_1\eps) r^{-1+\theta} \crochet^{1/2+\delta/2}r|\del \us^{\N}|_{N-4,k-1}\Big|_{\varphi_{t,x}(\tau)}d\tau 
+ \delta^{-1}(\epss + C_1\eps)\int_{t_0}^t \tau^{-1-\delta}d\tau
\\
& \lesssim    \big(\ell^{-\delta/2} +\delta^{-2}\big) (\epss + C_1\eps)  
+ (\epss + C_1\eps)\int_{t_0}^t\tau^{-1+\theta}\crochet^{1/2+\delta/2}r|\del \us^{\N}|_{N-4,k-1}\Big|_{\varphi_{t,x}(\tau)}d\tau. 
\endaligned
$$
Here, we used\footnote{Since $\us^{\N}$ is a finite linear combination of $u$ with homogeneous coefficients of degree zero, 
 the bounds can be applied.} 
\eqref{eq9-31-01-2021}, \eqref{eq1-30-05-2020}, \eqref{eq2-09-08-2021}, and the pointwise decay \eqref{eq10-02-05-2020} as well as the { condition $\min(\lambda,\kappa) \geq 1/2+2\delta$}.
On the other hand, observe that
$$
|Z u|\lesssim  \delta^{-1}  (\epss+C_1\eps)  r^{-1}\crochet^{1-\kappa}s^{\delta}
\lesssim \delta^{-1}  (\epss+C_1\eps)  r^{\delta/2 - \kappa}\lesssim  (\epss+C_1\eps) ,
$$
which we apply with $u$ replaced by $\us^{\N}$. 
This leads us to
\begin{equation}\label{eq1-29-11-2020}
\aligned
r\big|(\del_t-\del_r)Z \us^{\N}_{\alpha\beta} \, \big|
\lesssim  (\epss+C_1\eps)  + |(\del_t - \del_r)(r Z \us^{\N})|. 
\endaligned
\end{equation}
Now recalling \eqref{eq10-02-05-2020}, for all $\ord(Z)\leq N-4$ we have 
$
r \, |\delsN Z u|\lesssim   (\epss+C_1\eps) \, r^{-\kappa + \delta/2} \lesssim  (\epss+C_1\eps) .
$
Recalling the identities $2\del_t = (x^a/r)\delsN_a + (\del_t-\del_r)$
and $\del_a = \delsN_a - (x^a/r)\del_t$ and Lemma~\ref{lem 2 high-order},
we obtain in $\Mnear_{\ell,[s_0,s_1]}$: 
$$ 
\aligned
\crochet^{1/2+\delta/2}r |\del \us^{\N} |_{N-4,k}
& \lesssim 
\big(\ell^{-\delta/2} +\delta^{-2}\big) (\epss + C_1\eps)  
+ (\epss + C_1\eps)\int_{t_0}^t\tau^{-1+\theta} \crochet^{1/2+\delta/2}r|\del \us^{\N}|_{N-4,k-1}\Big|_{\varphi_{t,x}(\tau)}d\tau. 
\endaligned
$$
Finally, with the notation 
$\Bbf_k(t) := \sup_{\Mnear_{\ell,[s_0,s]}} \crochet^{1/2+\delta/2}r |\del \us^{\N} |_{N-4,k}$ 
(with $t = T^{\E}(s)$), 
the above estimate reads 
$$
\Bbf_k(t)\lesssim \big(\ell^{-\delta/2} +\delta^{-2}\big) (\epss + C_1\eps) 
+ (\epss + C_1\eps)\int_{t_0}^t\tau^{-1+\theta}\Bbf_{k-1}(\tau)d\tau, 
$$
in which the last term does not exist when $k=0$.  
By induction over $k$ varying from $k=0$ to $k=N-4$, we arrive at \eqref{eq11-01-31-2021}.

%------------------------------

\paragraph{Controlling the Hessian.}

It remains to establish \eqref{eq8-24-03-2021}. Again, we apply Proposition~\ref{prop1-23-07-2020} to \eqref{eq1-24-03-2021} with $\ord(Z)= N-5$ and $\rank(Z)=k\leq N-5$. First, thanks to \eqref{eq1-09-08-2021} {(with $\theta\leq \delta/2$)} and \eqref{eq10-02-05-2020} 
we have the following bounds for the initial data: 
\begin{equation}
\sup_{\Omega^{\ell}_{s_0,s_1}} \crochet^{1 + \delta}\big(  r \, |\del Z\del_t\us^{\N}| +  |Z\del_t\us^{\N}|\big) 
\lesssim \ell^{-\delta}(\epss + C_1\eps). 
\end{equation}
Observe that \eqref{eq1'-10-01-2021} are guaranteed by \eqref{eq1-24-04-2021}. We find
\begin{equation}\label{eq1-25-03-2021}
\aligned
&\crochet^{1+\delta} \, \big|(\del_t-\del_r)(rZ\del_t \us^{\N}_{\alpha\beta})(t,x)\big|
\\
& \lesssim    \ell^{-\delta}(\epss + C_1\eps) 
+ \int_{t_0}^t r \, \crochet^{1+\delta} |\Boxt_g Z\del_t u|_{\varphi_{t,x}(\tau)}d\tau
+ \int_{t_0}^t 
\crochet^{1+\delta}\Big(|\del\delsN u|_{N-4} + r|H||\del\delsN Z\del_t u| + |H||\del \del u|_{N-5}\Big)\Big|_{\varphi_{t,x}(\tau)}d\tau
\\
& \lesssim   
\ell^{-\delta}(\epss + C_1\eps) 
+ \int_{t_0}^t(\epss + C_1\eps) r^{-1+\theta} \crochet^{1+\delta}r|\del\del_t \us^{\N}|_{N-5,k-1}\Big|_{\varphi_{t,x}(\tau)}d\tau 
+ { \delta^{-1}}(\epss + C_1\eps)\int_{t_0}^t \tau^{-1-\delta}d\tau
\\
& \lesssim    \big(\ell^{-\delta} +\delta^{-2}\big) (\epss + C_1\eps)  
+ (\epss + C_1\eps)\int_{t_0}^t\tau^{-1+\theta} \crochet^{1+\delta}r|\del\del_t \us^{\N}|_{N-5,k-1}\Big|_{\varphi_{t,x}(\tau)}d\tau, 
\endaligned
\end{equation} 
where for the second and third terms in the third line we used \eqref{eq1-09-08-2021} and \eqref{eq2-09-08-2021} (with $\theta\leq \delta/2$). This is a consequence of \eqref{eq1-05-05-2021}  { under the condition $\min(\lambda,\kappa)\geq 1/2+ (3/2) \delta$.} 
Following a similar procedure, we also obtain
$$
\aligned
\crochet^{1+\delta}r|(\del_t-\del_r) Z\del_t\us^{\N}_{\alpha\beta}|
& \lesssim 
\crochet^{1+\delta}|(\del_t-\del_r)(rZ\del_t \us^{\N}_{\alpha\beta})| + \crochet^{1+\delta}|Z\del_t\us^{\N}_{\alpha\beta}|
\\
& \lesssim 
\crochet^{1+\delta}|(\del_t-\del_r)(rZ\del_t \us^{\N}_{\alpha\beta})| +  (\epss+C_1\eps) 
\endaligned
$$
and, thanks to \eqref{eq2-09-08-2021}, 
$$
\aligned
|\del_t Z\del_t\us^{\N}_{\alpha\beta}|& \lesssim  |(x^a/r)\delsN_a Z\del_t\us^{\N}_{\alpha\beta}| + |(\del_t-\del_r) Z\del_t\us^{\N}_{\alpha\beta}|
\lesssim |(\del_t-\del_r) Z\del_t\us^{\N}_{\alpha\beta}|
+ (\epss + C_1\eps)r^{-2+\delta}\crochet^{-\min(\lambda,\kappa)},
\\
|\del_a Z\del_t\us^{\N}_{\alpha\beta}|& \lesssim  |\delsN_aZ\del_t\us^{\N}_{\alpha\beta}| + |\del_t Z\del_t\us^{\N}_{\alpha\beta}|
\lesssim |(\del_t-\del_r) Z\del_t\us^{\N}_{\alpha\beta}| + (\epss + C_1\eps)r^{-2+\delta}\crochet^{-\min(\lambda,\kappa)}.
\endaligned
$$
Then, by Proposition~\ref{prop--fund-order} and \eqref{eq1-25-03-2021} we arrive at  
\begin{equation}\label{equa-3-avril-2021a}
\aligned
r \, \crochet^{1+\delta}|\del\del_t \us^{\N}_{\alpha\beta}(t,x)|_{N-5,k}& \lesssim   \big(\ell^{-\delta} +\delta^{-2}\big) (\epss + C_1\eps)  
+  (\epss + C_1\eps)\int_{t_0}^t\tau^{-1+\theta} \crochet^{1+\delta}r|\del\del_t \us^{\N}|_{N-5,k-1}\Big|_{\varphi_{t,x}(\tau)}d\tau. 
\endaligned
\end{equation}
We also have the inequality 
$$
\aligned
\crochet^{1+\delta}r|\del_{\alpha}\del_a u|_{N-5,k}
& \lesssim  \crochet^{1+\delta}r|\del\del_t u|_{N-5,k} + \crochet^{1+\delta}r|\del\delsN_a u|_{N-5,k}
+ \crochet^{1+\delta}|\del_t u|_{N-5,k}
\\
& \lesssim  (\epss + C_1\eps) + \crochet^{1+\delta}r|\del\del_t u|_{N-5,k}.
\endaligned
$$
Consequently, in terms of the function 
${\bf C}_k(t) := \sup_{\Mnear_{\ell,[s_0,s]}} \crochet^{1+\delta}r |\del\del \us^{\N} |_{N-5,k}$ 
(with $t = T^{\E}(s)$), 
the inequality \eqref{equa-3-avril-2021a} reads   
\begin{equation}
{\bf C}_k(t)\lesssim \big(\ell^{-\delta} +\delta^{-2}\big) (\epss + C_1\eps) 
+ (\epss + C_1\eps)\int_{t_0}^t\tau^{-1+\theta}{\bf C}_{k-1}(\tau)d\tau, 
\end{equation}
in which the last term does not exist when $k=0$.  
Proceeding by  induction on the integer $k$ varying from $k=0$ to $k=N-5$, we conclude that \eqref{eq8-24-03-2021} holds true. The proof of Proposition~\ref{proposition14-1} is complete. 

%--------------------------------------------------------------------------------------------------------

\subsection{Dealing with metrics in Class B}
\label{subsec1-21-09-2022}

The proof of Proposition~\ref{proposition14-1} for metrics in Class B follows the same arguments as the one above in this section, but we now directly study the metric equation \eqref{eq1-30-11-2020}. That is, we consider the equations satisfied by $h^{\N}_{\alpha\beta} = Z \PhiN_{\alpha}^{\alpha'}\PhiN_{\beta}^{\beta'}h_{\alpha'\beta'}$, stated now as 
\begin{subequations}\label{eqs1-07-09-2022-M}
\begin{equation}
\aligned
&\Boxt_gh^{\N}_{\alpha\beta} =  L_{1\alpha\beta} + L_{2\alpha\beta} + S_{1\alpha\beta},  
\quad
&& L_{1\alpha\beta} := h_{\alpha'\beta'}\Boxt_g\big(\PhiN_{\alpha}^{\alpha'}\PhiN_{\beta}^{\beta'}\big), 
\\
& L_{2\alpha\beta} := g^{\mu\nu}\del_{\mu}\big(\PhiN_{\alpha}^{\alpha'}\PhiN_{\beta}^{\beta'}\big)\del_{\nu}h_{\alpha'\beta'}, 
\qquad
&& S_{1\alpha\beta}[h] := \PhiN_{\alpha}^{\alpha'}\PhiN_{\beta}^{\beta'}\Boxt_gh_{\alpha'\beta'}, 
\endaligned
\end{equation}
so that 
\begin{equation}
\aligned
& \Boxt_g \big( Z h^{\N}_{\alpha\beta} \big) = Z L_{1\alpha\beta} + Z L_{2\alpha\beta} + Z S_{1\alpha\beta} + S_{2\alpha\beta}, 
\qquad  
&& S_{2\alpha\beta} = S_{2\alpha\beta}[h]
:= - [Z, h^{\mu\nu}\del_\mu\del_\nu] h^{\N}_{\alpha\beta},
\endaligned
\end{equation}
\begin{equation}
\Boxt_g(Z\del_t h^{\N}_{\alpha\beta}) = Z \del_t\big(L_{1\alpha\beta} + L_{2\alpha\beta} + S_{1\alpha\beta}\big) + S'_{2\alpha\beta},
\qquad 
S'_{2\alpha\beta} = S'_{2\alpha\beta}[h]
:= - [Z\del_t,h^{\mu\nu}\del_{\mu}\del_{\nu}]h^{\N}_{\alpha\beta}.
\end{equation}
\end{subequations}
We need first the following result which is an analogue of Proposition~\ref{proposition-label-144}. 

\begin{proposition}
\label{proposition-label-144-B}
Assume that the reference metric belongs to Class B. 
Under the conditions stated in Section~\ref{section-label-11-1} and under the near-Schwarzschild decay \eqref{eq1-05-05-2021} 
one has for all $\ord(Z)\leq N-4$ and $\rank(Z) = k\leq N-4$   
\begin{subequations}\label{eq6-14-08-2021-B}
\begin{equation}\label{eq9-31-01-2021-B}
\aligned
\crochet^{1/2+\delta/2}r|\Boxt_gZ\hs^{\N}|
\lesssim  
(\epss + C_1\eps)r^{-1+\theta} \big(r \, \crochet^{1/2+\delta/2}|\del \hs^{\N}|_{N-4,k-1}\big)
+ \delta^{-1} (\epss + C_1\eps) r^{-1-\delta}, 
\endaligned
\end{equation}
while for all $\ord(Z) \leq N-5$ and $\rank(Z) = k\leq N-5$  
\begin{equation}\label{eq2-24-03-2021-B}
r \, \crochet^{1+\delta}|\Boxt_gZ\del_t \hs^{\N}|
\lesssim (\epss + C_1\eps)r^{-1+\theta}\big(r \, \crochet^{\kappa}|\del\del \hs^{\N}|_{N-5,k-1}\big)
+ \delta^{-1}(\epss + C_1\eps) r^{-1-\delta}.
\end{equation}
\end{subequations}
\end{proposition}

For the proof, we need to control each term arising in \eqref{eqs1-07-09-2022-M}. First of all, we claim that the estimates \eqref{eq7-14-08-2021} and \eqref{eq3-14-08-2021} still hold in this case. Indeed, we only need to replace the bounds on $u$ by the bounds on $h$,  that is, \eqref{eq1-26-05-2021} and \eqref{eq1-09-05-2021}. For the estimate on $\del\delsN h$, we combine \eqref{eqs2-07-09-2022-M} and \eqref{equa-new-conditions-hstar}
and obtain
\begin{subequations}\label{eqs3-07-09-2022-M}
\begin{equation}\label{eq1-07-09-2022-M}
|\del\del h|_{N-4} \lesssim (\epss + C_1\eps) r^{-1}\crochet^{-1-\min(\lambda,\kappa)} (r^{\theta} + s^{\delta})s^{\delta} 
\quad\text{ in } \Mscr^{\near}_{[s_0,s_1]}, 
\end{equation}
\begin{equation}\label{eq2-07-09-2022-M}
|\del\delsN h|_{N-4}\lesssim (\epss + C_1\eps) r^{-1-\min(\lambda,\kappa)}\crochet^{-1}(r^{\theta} + s^{\delta})s^{\delta}
\quad\text{ in } \Mscr^{\near}_{[s_0,s_1]}.
\end{equation} 
\end{subequations}
We then focus our attention on the terms $S_1$ and $S_2$, whose treatment is somewhat less direct, as follows. 

\begin{lemma}
For metrics in Class B, one has
\begin{equation}\label{eq5-07-09-2022-M}
\sum_{(\alpha,\beta)\neq(0,0)}r \, \crochet^{1/2+\delta/2}|S_{1\alpha\beta}[h]|_{N-4} 
+ \sum_{(\alpha,\beta)\neq(0,0)}r \, \crochet^{1+\delta}|\del_t S_{1\alpha\beta}[h]|_{N-5} 
\lesssim (\epss+C_1\eps)r^{-1-\delta}
\quad\text{ in } \Mscr^{\near}_{[s_0,s_1]}.
\end{equation}
\end{lemma}

\begin{proof} As pointed out earlier, among the terms in the right-hand side of \eqref{eq1-30-11-2020}, the most critical ones are the quasi-null terms $\Pbb$ in the expression of the nonlinearity $\Fbb$, while the remaining terms enjoy sufficient pointwise decay, namely for any term other than $\Pbb_{\alpha\beta}$ we have 
\begin{equation}\label{eq3-07-09-2022-M}
r \, \crochet^{1/2+\delta/2}|T_{\alpha\beta}| + r \, \crochet^{1+\delta}|\del_t T_{\alpha\beta}|_{N-5}\lesssim (\epss+C_1\eps) r^{-1-\delta}. 
\end{equation} 
We check this claim as follows. The bound for $\Tbb_{\alpha\beta}$ stated in \eqref{eq1-28-11-2020} is sufficient. For the null term $\Qbb_{\alpha\beta}$ in \eqref{equa:sec8-04}, we have 
\begin{equation}\label{eq4-07-09-2022-M}
|\Qbb_{\alpha\beta}(g,g;\del g,\del g)|_{p,k} 
\lesssim \sum_{p_1+p_2 = p\atop k_1+k_2=k}|\delsN h|_{p_1,k_1}|\del h|_{p_2,k_2} 
+ |h|_p\sum_{p_1+p_2= p}|\del h|_{p_1}|\del h|_{p_2}, 
\end{equation} 
as is clear by proceeding similarly to the derivation of  \eqref{equa-new-Qzero}. We then substitute \eqref{eq1-26-05-2021} and \eqref{eq1-09-05-2021} and obtain
\begin{equation}
|\Qbb_{\alpha\beta}(g,g;\del g,\del g)|_{N-4}\lesssim (\epss+C_1\eps)^2r^{-2-\min(\lambda,\kappa)}\crochet^{-\min(\lambda,\kappa)}s^{3\delta},
\end{equation}
thanks to $\delta^{-1}(\epss+C_1\eps)\lesssim 1$, so that \eqref{eq3-07-09-2022-M} is also established for $\Qbb_{\alpha\beta}$. 

It thus remains to deal with the quasi-null terms $\Pbb_{\alpha\beta}$, and we use the following estimate (which is similar to the first inequality already established in Lemma \ref{lem1-31-01-2021}): 
\begin{equation}
|\slashed{\Pbb}(g,g;\del g,\del g)|_{p,k}\lesssim  \sum_{p_1+p_2 = p\atop k_1+k_2=k}|\delsN h|_{p_1,k_1}|\del h|_{p_2,k_2} 
+ |h|_p\sum_{p_1+p_2= p}|\del h|_{p_1}|\del h|_{p_2}. 
\end{equation}
Namely, this inequality is a consequence of the identity 
$
\PhiN_{\alpha}^\gamma \PhiN_{\beta}^{\delta} \Pbb_{\gamma\delta}(g,g;\del u, \del v)
= \Pbb_{\alpha\beta}(g,g; \delN_\alpha u, \delN_{\beta} v)
$
(based on the tensorial structure).
Thus, $\slashed{\Pbb}$ satisfies the same bound as $\Qbb$, and therefore \eqref{eq5-07-09-2022-M} is also established for all of the relevant nonlinearities.
\end{proof}

\begin{lemma} 
For metrics in Class B, in the near-light cone domain $\Mscr^{\near}_{[s_0,s_1]}$ one has 
\begin{equation}\label{eq6-07-09-2022-M}
\aligned
r \, \crochet^{1/2+\delta/2}|S_{2\alpha\beta}[h]|
&\lesssim (\epss + C_1\eps)r^{-1+\theta}\Big(r \, \crochet^{1/2+\delta/2}|\del \hs^{\N}|_{N-4,k-1}\Big)
+ (\epss + C_1\eps) r^{-1-\delta},
\\
r \, \crochet^{1+\delta}|S'_{2\alpha\beta}[h]|
& \lesssim (\epss + C_1\eps)r^{-1+\theta}\big(r \, \crochet^{1+\delta}|\del\del_t \hs^{\N}|_{N-5,k-1}\big)
+ (\epss + C_1\eps) r^{-1-\delta}.
\endaligned
\end{equation}
\end{lemma}

\begin{proof}
For this estimate we first establish \eqref{eq3-24-03-2021}. We thus need to revisit the proof of Proposition \ref{Proposition12.1}. In fact we only need to check the bound on $T^\textbf{super}$, since for the remaining terms, we have only used the bounds on $H$, i.e., \eqref{eq1-30-05-2020}. The quantity $u$ and its derivatives are preserved in the expressions. This leads to the fact that we can replace $u$ by $h$. However for $T^\textbf{super}$, we used the bounds on $u$ which might be invalid for $h$. We use \eqref{eq1-30-05-2020} in the expression of $T^\textbf{super}$, and obtain that it is bounded by $(\epss + C_1\eps)^2r^{-2-\min(\lambda,\kappa)}\crochet^{-\min(\lambda,\kappa)}s^{2\delta}$ in $\Mnear_{\ell,[s_0,s_1]}$. This guarantees the first estimate in \eqref{eq6-07-09-2022-M}.
For the second estimate, we repeat the argument for \eqref{eq2-14-08-2021}. In order to obtain the estimate on $[\del_t, H^{\mu\nu}\del_{\mu}\del_{\nu}]h$, we apply the argument for \eqref{eq2-06-09-2021}. Instead of \eqref{eq2-09-08-2021}, we apply \eqref{eq1-26-05-2021} and \eqref{eqs3-07-09-2022-M} for the bound in $\del\del h$ and $\del h$.
\end{proof}

%-------------------------------------

\begin{proof}[Proof of Proposition \ref{proposition14-1} for Class B]
We apply Proposition \ref{prop1-23-07-2020} together with the pointwise estimate \eqref{eq6-14-08-2021-B}. We observe that
$$
\sup_{\Omega_{s_0,s_1}^{\ell}} \crochet^{1/2+\delta/2}\big(  r \, |\del Z\hs^{\N}| +  |Z\hs^{\N}|\big) 
\lesssim \big(\ell^{-\delta/2} + \delta^{-1}\big) (\epss+C_1\eps), 
$$
$$
\sup_{\Omega^{\ell}_{s_0,s_1}} \crochet^{1 + \delta}\big(  r \, |\del Z\del_t\hs^{\N}| +  |Z\del_t\hs^{\N}|\big) 
\lesssim \ell^{-\delta}(\epss + C_1\eps).
$$
This is obtained by substituting \eqref{eq1-26-05-2021}, \eqref{eq1-09-05-2021} and \eqref{eqs3-07-09-2022-M} into the corresponding expressions and observing
 that $\crochet^{-1}\lesssim \ell^{-1}\la r\ra^{-1}$ on $\Lscr_{\ell, [s_0,s_1]}$. Then we follow the same line of arguments as in Section~\ref{section---143} but replace $u$ by $h$ (where we check that  \eqref{eq1-26-05-2021}, \eqref{eq1-09-05-2021} and \eqref{eqs3-07-09-2022-M} supply sufficient bounds on $|h|, |\del h|, |\delsN h|, |\del\del h|$ and $\del\delsN h$). The only essential difference is that, instead of controlling $|\del \us|$,  we control $|\del \hs|$ directly:
\begin{subequations}\label{eq8-14-08-2021-B}
\begin{equation}\label{eq11-01-31-2021-B}
|\del \hs^{\N}|_{N-4,k}\lesssim (\ell^{-\delta/2} + \delta^{-2})(\epss + C_1\eps) r^{-1+ k \theta}\crochet^{-1/2-\delta/2},
\qquad 0\leq k \leq N-4 \quad \text{in }\Mnear_{[s_0,s_1]},
\end{equation}
\begin{equation}\label{eq8-24-03-2021-B}
|\del\del \hs^{\N}|_{N-5,k}\lesssim (\ell^{-\delta}+\delta^{-2})(\epss + C_1\eps)r^{-1+k\theta}\crochet^{-1-\delta},
\qquad 0\leq k\leq N-5
 \qquad \text{in }\Mnear_{[s_0,s_1]}.
\end{equation}
\end{subequations}
Finally, in order to arrive at the desired estimate for the metric perturbation $\us$, we recall \eqref{equa-new-conditions-hstar} and, specifically, the fact that, by our assumption, $|\del_t\hs_\star^{\N}|\lesssim \epss t^{-1}$ has better decay. 
This completes the proof of Proposition \ref{proposition14-1}.
\end{proof}

%============================================================================== 

%=================================================================================

\section{Pointwise estimates for metric components at low order}
\label{section---15} 

\subsection{Objective}

We now consider {\sl general} components of the metric at lower order of differentiation (in comparison to the maximal order of differentiation). This section will be devoted to the proof of the following result. We emphasize that the near-Schwarz\-schild decay \eqref{eq1-29-03-2021} and \eqref{eq3-29-03-2021} below (which we will deduce from Proposition~\ref{Linfini wave}) are relevant only when dealing with massive matter fields, while for massless fields the pointwise decay \eqref{eq1-09-05-2021} 
and the estimate \eqref{eq19-23-04-2021} on $h^{\N00}$ would be sufficient. The following statement covers, both, Class A and Class B. 

\begin{proposition}[Pointwise estimate for general metric components]
\label{section-15-1} 
By using the estimates \eqref{eq8-14-08-2021} on the gradient and Hessian of the good metric components 
and under the conditions stated in Section~\ref{section-label-11-1}, 
% hold. 
% Then provided  { $\min(\lambda,\kappa) \geq 1/2+(9/2) \delta$ and $\mu\geq 3/4+(7/4) \delta$},  
the gradient and Hessian
of the metric satisfy 
\begin{equation}\label{eq5-25-03-2021}
\aligned
| { \del\hs^{\N}}|_{N-4,k} + |{ \del  \us^{\N}}|_{N-4,k}
& \lesssim  (\ell^{-\delta/2} + \delta^{-2})(\epss + C_1\eps) r^{-1+k\theta}\crochet^{-1/2-\delta/2}
\quad 
& \text{ in } \Mnear_{\ell, [s_0,s_1]}, 
\quad && 0\leq k\leq N-4,
\\
| { \del\del { \hs^{\N}}} |_{N-5,k} + | { \del\del { \us^{\N}}} |_{N-5,k} 
& \lesssim   (\ell^{-\delta} + \delta^{-2})(\epss + C_1\eps) r^{-1+k\theta}\crochet^{-1-\delta}
\quad 
& \text{ in } \Mnear_{\ell, [s_0,s_1]}, 
\quad && 0\leq k\leq N-5.
	\endaligned
\end{equation}
Moreover,
%  provided { $(N-4) \theta\leq \delta/2$ and $(\ell^{-1} + \delta^{-4}) \delta^{-1}\theta^{-1}(\epss + C_1\eps) \lesssim 1$}, 
the source terms $u_{\source,\alpha\beta}$ defined in \eqref{eq1-19-04-2021}
enjoys the near-Schwarz\-schild decay in the whole exterior domain:  
%
%\
%
%{\tt   Probablement ne change pas.}
%
%\
%
\begin{equation}\label{eq1-29-03-2021}
|u_{\source} |_k\lesssim  (\epss + C_1\eps) { r^{-1+{(3N+1)} \theta}}
\quad
\text{ in } \MME_{[s_0,s_1]}, 
\qquad 0\leq k\leq N-5. 
\end{equation}
\end{proposition}

We will also use estimates with upper indices, that is, $\del h^{\alpha\beta}$ below, and we recall our notation $H$ in \eqref{equa-notation-H}. 

\begin{corollary} 
In the whole exterior domain, when \eqref{eq2-29-03-2021-gstar} and \eqref{eq3-09-05-2021} hold, the lapse, orthogonal, and radial components of the metric enjoy the near-Schwarz\-schild decay 
\begin{equation}\label{eq3-29-03-2021}
| h^{00}, h^\rr, h^{0a} |_k
\lesssim (\epss + C_1\eps)  \, { r^{-1+{(3N+1)} \theta}}
\quad \text{ in } \MME_{[s_0,s_1]}, 
\qquad 0\leq k\leq N-5.
\end{equation}
\end{corollary}

{
The inequalities \eqref{eq5-25-03-2021} are a direct consequence of \eqref{eq8-14-08-2021}. We especially point out that when $k=0$ we rely on $|\del\hs^\star | \lesssim \epss \la r+t\ra^{-1}$ which is assumed in, both, Class A and Class B. 
}
{ 
The rest of this section is devoted to the proof of \eqref{eq1-29-03-2021} and \eqref{eq3-29-03-2021}. We emphasize that 
\eqref{eq1-29-03-2021} and \eqref{eq3-29-03-2021} will not be needed in Section~\ref{section-16}, thus we can use  \eqref{eq3-25-05-2023}
 in the present section.  
 }

%---------------------------------------------------------------------------------------------------------------------------------

\subsection{The near-Schwarz\-schild decay property (Class A)} \label{Sharpdecay-II}
\paragraph{Proof of \eqref{eq1-29-03-2021}.}
The proof of \eqref{eq1-29-03-2021} is a consequence of the general pointwise estimate for wave fields established earlier (in Proposition~\ref{Linfini wave})  and also uses our bound of $|\Box u_{\alpha\beta} |_k$. 
However, at this stage we need to control the terms listed in \eqref{eq3-23-04-2021} in the whole exterior domain $\MME_{[s_0, s_1]}$ rather than in the good domain $\Mgood_{[s_0,s_1]}$ (and our argument now should be compared with the one 
in Lemma~\ref{lem1-25-04-2021}). 

For the bound in $\Mnear_{\ell, [s_0,s_1]}$ we observe that thanks to \eqref{eq5-25-04-2021}, \eqref{eq1-04-12-2020}  and \eqref{eq9-23-04-2021}, 
the terms $\Ibb^{\star}[u]$, ${ u^{\mu\nu}\del_{\mu}\del_{\nu}h^{\star}}$,
 $\Tbb[\phi]$,  
and ${ ^{(w)}R^{\star}_{\alpha\beta}}$ are bounded { (provided $\min(\lambda,\kappa) \geq 1/2+(9/2) \delta$ and $\mu \geq 3/4 + (7/4) \delta$)} 
by
$
\delta^{-1}(\epss + C_1\eps)^2\, r^{-2-3\delta}\crochet^{-1+\delta}.
$ 
{
 To handle the null and quasi-null terms, we recall \eqref{eq7-04-10-2022} and \eqref{eq4-04-10-2022}. For the quadratic null and and high-order terms, we rely on the Sobolev pointwise estimates \eqref{eq10-02-05-2020}
 }
  together with the assumptions \eqref{equa-31-12-20} or \eqref{equa-new-conditions-hstar} (on $h^{\star}$). 
For quasi-null terms, we apply \eqref{eq5-25-03-2021} and find
\begin{equation}\label{eq126-04-2021}
|\Pbb^{\star}[u]|_{N-4, k} + |\Qbb^{\star}[u]|_{N-4,k}\lesssim (\ell^{-\delta} + \delta^{-4})(\epss + C_1\eps)^2 r^{-2+k\theta}\crochet^{-1-\delta},
\qquad 
\quad k\leq N-4.
\end{equation}

For the quasi-linear term, we recall the null decomposition \eqref{eq3-22-04-2021} and, in view of  { \eqref{eq3-25-05-2023}} and the sharp decay of the null component in \eqref{eq19-23-04-2021}, we obtain
$$
|h^{\N00}\del_t\del_t u|_{N-5, k}\lesssim (\ell^{-\delta} + \delta^{-2})(\epss + C_1\eps)^2 r^{-2+{(3N+1)} \theta}\crochet^{-1-\delta}.
$$

\

The Sobolev decay \eqref{eq1-09-05-2021} and the pointwise Hessian estimate \eqref{eq2-09-08-2021} lead us to 
$$
|h\del\delsN u|_{N-4}\lesssim \delta^{-1}(\epss + C_1\eps)^2r^{-2-\min(\lambda,\kappa)}\crochet^{-\min(\lambda,\kappa)}s^{2\delta}{ (r^{\theta} + s^{\delta})}.
$$
Finally, we have 
$$
r^{-1} |h\del u|_{N-4}\lesssim \delta^{-1}(\epss + C_1\eps)^2r^{-2-\min(\lambda,\kappa)}\crochet^{-\min(\lambda,\kappa)}s^{2\delta}.
$$
{ Thanks to the condition $\min(\lambda,\kappa) \geq 1/2 + (5/4) \delta$}, these inequalities lead us to
$$
|h^{\mu\nu}\del_{\mu}\del_{\nu}u_{\alpha\beta} |_k\lesssim (\ell^{-\delta} + \delta^{-2}) (\epss + C_1\eps)^2r^{-2+{(3N+1)} \theta}\crochet^{-1-\delta}
\qquad \text{in }\Mnear_{\ell,[s_0,s_1]}.
$$
So we conclude that in the near-light cone domain
\begin{equation}\label{eq2-09-05-2021}
\aligned
|\Box u_{\alpha\beta} |_k
& \lesssim \delta^{-1}(\epss + C_1\eps)^2 r^{-2-3\delta}\crochet^{-1+\delta} 
+ (\ell^{-\delta} + \delta^{-4}) (\epss + C_1\eps)^2r^{-2+{(3N+1)} \theta}\crochet^{-1-\delta}
\quad
\text{ in } \Mnear_{\ell,[s_0, s_1]}.
\endaligned
\end{equation}

Next for the domain $\Mfar_{\ell,[s_0,s_1]}$ far from the light cone, we also observe that  \eqref{eq4-23-04-2021}, \eqref{eq9-23-04-2021}, and \eqref{eq5-25-04-2021} control all terms in \eqref{eq3-23-04-2021} except the quasilinear term $h^{\mu\nu}\del_{\mu}\del_{\mu}u_{\alpha\beta}$,
the quasi-null terms $\Pbb^{\star}[u]$, and the null terms $\Qbb^{\star}[u]$. Furthermore,  recall that $\ell r\lesssim \crochet$ holds within $\Mfar_{\ell,[s_0,s_1]}$. Then \eqref{eq5-04-06-2020} and \eqref{eq1-15-08-2021} (together with \eqref{eq1-09-05-2021}) lead us to the following inequalities in $\Mfar_{\ell,[s_0,s_1]}$, { in view of  $\min(\lambda,\kappa) \geq 1/2+2\delta$}, 
\begin{equation}
\aligned
|\del u \del u|_{N-5}& \lesssim  \ell^{-4\delta} (\epss+C_1\eps)^2 r^{-2-3\delta}\crochet^{-1+\delta},
\\
|h^{\alpha\beta}\del_{\alpha}\del_{\beta} u|_{N-5}& \lesssim  \ell^{-1}{\delta^{-1}}(\epss + C_1\eps)^2t^{-1+{ (3/2)}\delta}r^{-1-\min(\lambda,\kappa)}
\crochet^{-1+(1-\min(\lambda,\kappa))}.
\endaligned
\end{equation}
Thus, $\MME_{[s_0, s_1]}$ and for all $k\leq N-5$, we arrive at a control of the wave operator  
\begin{equation}\label{eq1-23-09-2022}
\aligned
|\Box u_{\alpha\beta} |_k\lesssim  & \big(\ell^{-4\delta} + \delta^{-1}\big)(\epss + C_1\eps)^2 r^{-2-3\delta}\crochet^{-1+\delta} 
+ (\ell^{-\delta} + \delta^{-4}) (\epss + C_1\eps)^2r^{-2+{(3N+1)} \theta}\crochet^{-1-\delta}
\\
& \quad
+ \ell^{-1}{ \delta^{-1}}(\epss + C_1\eps)^2 t^{-1+{ (3/2)}\delta}r^{-1-\min(\lambda,\kappa)} \crochet^{-1+(1-\min(\lambda,\kappa))}.
\endaligned
\end{equation}
Consequently, by applying Proposition~\ref{Linfini wave} in Cases 1, 2, and 4, we deduce that  
$$
\aligned
|u_{\source} |_k
& \lesssim  \underbrace{\big(\ell^{-4\delta} + \delta^{-1}\big) \delta^{-2}(\epss + C_1\eps)^2r^{-1}}_{\text{Case 2 with } \mu = \delta, \nu = 3\delta} 
+\underbrace{(\ell^{-\delta} +\delta^{-4}) \delta^{-1}{(3N+1)^{-1}}\theta^{-1}(\epss + C_1\eps)^2r^{-1+{(3N+1)} \theta}}_{ 
\text{Case 4 with }\mu = \delta, \nu={(3N+1)} \theta,  {(3N+1)} \theta\leq \delta/2}
\\
& \quad
+ \underbrace{\ell^{-1}\delta^{-3}(\epss + C_1\eps)^2r^{-1}}_{\text{Case 1 with } \upsilon = (3/2) \delta\atop \mu = 1-\min(\lambda,\kappa), \nu = \min(\lambda,\kappa), \nu-\mu-\upsilon\geq \delta}
\endaligned
$$
Taking the smallness conditions ${(3N+1)}\theta\leq \delta/2$, 
and $\min(\lambda,\kappa) \geq 1/2 + (5/4) \delta$, 
as well as 
$(\ell^{-1} + \delta^{-4}) \delta^{-1}{(3N+1)^{-1}}\theta^{-1}(\epss + C_1\eps) \leq 1$ into account, 
we conclude that \eqref{eq1-29-03-2021} holds in the whole exterior domain. 

%-------------------------------------------------------

\paragraph{Proof of \eqref{eq3-29-03-2021}.} 

In view of \eqref{eq1-19-04-2021} and $h^{\alpha\beta} = -h_{\alpha\beta} + \Abb^{\alpha\beta}[h]$, we have 
$$
\aligned
h^{rr} 
& = \sum_{a,b}\frac{x^ax^b}{r^2}h^{ab} = -\frac{x^ax^b}{r^2}{ \mathbbm{h}^{ab}} + \sum_{a,b}(x^ax^b/r^2) \Abb^{ab}[h] 
\\
& = { \Xi^{\star rr}}  -\frac{x^ax^b}{r^2}u_{\init,ab} -\frac{x^ax^b}{r^2}u_{\source,{ ab}}  + \sum_{a,b}(x^ax^b/r^2) \Abb^{ab}[h], 
\endaligned
$$
{where we recall that $\Xi^{\star rr} = -(x^ax^b/r^2){\mathbbm{h}^{\star ab}}$.} Then by \eqref{eq2-29-03-2021-gstar}, \eqref{eq3-09-05-2021}, \eqref{eq1-29-03-2021} and \eqref{eq4-19-04-2021} with { $C_0\lesssim C_1$ and
$\min(\lambda,\kappa) \geq 1/2+\delta/2$ together with
$\delta^{-2}(\epss + C_1\eps) \lesssim 1$}, we arrive at the desired estimate for $h^{rr}$. The derivation of the estimates for $h^{a0}$ and $h^{00}$ is similar but simpler, and we omit the details. 
 
%------------------------------------------------------------------------------------------------

\subsection{Near-Schwarzschild decay property for Class B}

As before, we need to bound the terms listed in \eqref{eq3-23-04-2021}. We observe that the terms
$$
|h^{\mu\nu}\del_{\mu}\del_{\nu}u|_k,
\qquad 
|\Pbb^{\star}[u]|_k,
\qquad 
|\Qbb^{\star}[u]|_k,
\qquad 
|\Tbb[\phi]|_k,
\qquad
|^{(w)}R^{\star}_{\alpha\beta}|_k
$$
still enjoy the same bounds,  since all of the corresponding estimates are based on the pointwise Ricci bound 
\eqref{eq4-09-05-2021} 
together with 
 \eqref{eq1-09-05-2021}, \eqref{eq19-23-04-2021}, \eqref{eq9-23-04-2021}, and \eqref{eq3-22-04-2021}.  
 %          { \sout{ \eqref{eq7-26-03-2021}}}. 
 All of these estimates hold in both cases. Furthermore, from \eqref{eq4-04-06-2020} (Class B), the higher-order terms $\Bbb$ and $\Cbb$ still enjoy sufficient decay, i.e., provided that $\kappa \geq 1/2+\delta, 2\theta<\delta$, 
\begin{equation}
|\Bbb^{\star}[u]|_{N-3} + |\Cbb^{\star}[u]|_{N-3} \leq (\epss + C_1\eps)^2r^{-2-3\delta}\crochet^{-1+\delta} 
\qquad \text{in } \MME_{[s_0,s_1]}.
\end{equation} 
We only need to check $|u^{\mu\nu}\del_{\mu}\del_{\nu} h^{\star}|_k$ and $|\Lbb^{\star}[u]|_k$. When the reference belongs to Class B,  the pointwise bounds contained in \eqref{equa-new-conditions-hstar} still supply sufficient estimates on these two terms. However, the argument becomes more delicate. We first treat
$u^{\mu\nu}\del_{\mu}\del_{\nu}h^{\star}$ and establish the following estimate.

\begin{lemma}\label{lem1-22-09-2022}
When the reference metric belongs to Class B, one has
\begin{equation}\label{eq1-22-09-2022}
|u^{\mu\nu}\del_{\mu}\del_{\nu} h^{\star}|_{N-5}\lesssim (\epss+C_1\eps)^2r^{-2+2\theta}\crochet^{-1-\delta} + \delta^{-1}(\epss+C_1\eps)^2r^{-2-3\delta}\crochet^{-1+\delta}
\quad \text{in } \MME_{[s_0,s_1]}.
\end{equation}
\end{lemma}

\begin{proof}
This is still by the null decomposition  parallel to \eqref{eq3-22-04-2021}:
\begin{equation}\label{eq2-22-09-2022}
|u^{\mu\nu}\del_{\mu}\del_{\nu} h^{\star}|_{p,k}\lesssim |u^{\N00}\del_t\del_t h^{\star}|_{p,k} + |u\del\delsN h^{\star}|_{N-5} + r^{-1} |u\del h^{\star}|_{N-5}, 
\end{equation}
where by \eqref{equa-new-conditions-hstar} and \eqref{eq7-15-05-2020}, the second and third term are bounded by $\delta^{-1}(\epss + C_1\eps)^2\, r^{-2-3\delta}\crochet^{-1+\delta}$, provided that $\kappa\geq \frac{1}{2} + 2\delta, \theta<\delta$. For the first term, thanks to \eqref{equa-new-conditions-hstar} we have 
$
|u^{\N00}\del_t\del_th^{\star}|_{N-5}\lesssim \epss r^{-1+\theta}\crochet^{-1-\kappa}|u^{\N00}|.
$
Then we recall the decomposition \eqref{equa:sec8-13} together with \eqref{equa-signe-a-noter},
$$
u^{\N00} = h^{\N00} - \Xi^{\star\N00} - \PsiN^0_{\alpha}\PsiN^0_{\beta}\Abb^{\N00}[h^{\star}].
$$
Recalling \eqref{eq3-03-01-2022}, \eqref{eq1-11-03-2021-gstar} and \eqref{eq19-23-04-2021}, we obtain 
\begin{equation}
|u^{\N00}|_{N-4}\lesssim (\epss + C_1\eps) r^{-1+\theta}\quad \text{ in } \MME_{[s_0, s_1]}.
\end{equation}
Then we apply this bound for the first term in the right-hand side of \eqref{eq2-22-09-2022}, tand he desired bound is established.
\end{proof}

Similarly, we establish the following bound. 

\begin{lemma}
\label{lem-1DF9} 
When the reference metric belongs to Class B, one has
\begin{equation}
|\Lbb^{\star}[u]|_{N-5}\lesssim 
\begin{cases}
(\ell^{-\delta/2} + \delta^{-2}) (\epss + C_1\eps)^2r^{   -3+ 2\delta}\crochet^{ -\kappa}
 \quad& \text{in }\Mnear_{\ell,[s_0,s_1]},
\\
\ell^{-4\delta}(\epss+C_1\eps)^2r^{-2-3\delta}\crochet^{-1+\delta}
\quad& \text{in }\Mfar_{\ell,[s_0,s_1]}.
\end{cases}
\end{equation}
\end{lemma}

\begin{proof}
For the estimate in $\Mnear_{\ell,[s_0,s_1]}$, we recall \eqref{eq1-crossing} together with {    
%\sout{\eqref{eq7-26-03-2021}} 
\eqref{eq7-15-05-2020} and \eqref{eq10-02-05-2020} } (under the condition $\kappa\geq 1/2+\delta, \theta<\delta$). In $\Mfar_{\ell,[s_0,s_1]}$, we apply \eqref{equa-new-conditions-hstar} together with \eqref{eq10-02-05-2020} and \eqref{eq7-15-05-2020}, provided that $\kappa\geq 1/2+2\delta$.
\end{proof}

It remains to collect our bounds above together concerning the terms \eqref{eq3-23-04-2021} and we conclude that \eqref{eq1-23-09-2022} holds when the reference metric belongs to Class B. The remaining arguments are the same as the ones introduced for Class A, and \eqref{eq1-29-03-2021} and \eqref{eq3-29-03-2021} are then also established.

{
	\begin{remark} 
		For Class B, recalling  the bound on $|\delsN h^{\star}|$ in \eqref{equa-new-conditions-hstar}, and for Class A, recalling \eqref{equa-31-12-20}, we deduce that 
		\begin{equation}\label{eq1-02-10-2022}
			\aligned
			|\del h^{\star \N00} |_{N-4}& \lesssim  \delta^{-1}(\epss + C_1\eps)r^{-1}\crochet^{-1/2-\delta/2}
			\qquad &&\text{ in } \Mnear_{\ell,[s_0,s_1]}, 
			\\
			|\del\del h^{\star\N00} |_{N-5} & \lesssim  (\epss+C_1\eps)r^{-1}\crochet^{-1-\delta}
			\qquad &&\text{ in } \Mnear_{\ell,[s_0,s_1]}. 
			\endaligned
		\end{equation}
\end{remark}} 

%==============================================================================================

\section{Improved energy estimate for general metric components}
\label{section-16}

\subsection{Purpose of this section}

Our task now is to derive first several key estimates in (weighted, high-order) $L^2$ norm and then deduce a control of the energy of the metric components at the highest-order. We point out that, in the special case that the matter field would vanish identically, that is, if we would restrict attention to Einstein's vacuum equations (for which our main theorem is also new), then the estimate given next 
allows us to close the bootstrap argument and, in the vacuum, complete the proof of the main existence theorem. 
{ Except when stated otherwise below, 
all of our arguments here apply to, both, Class A and Class B metrics.}

\begin{proposition}[Improved energy estimate for the metric]
\label{proposition-section16-metric} 
Under the bootstrap conditions stated in Section~\ref{section-label-11-1},  the metric components satisfy 
\begin{equation}\label{eq11-06-10-2022}
\Fenergy_{\kappa}^{\ME,N}(s,u) 
\leq {C_1 \over 2} \, \eps \, s^\delta, 
\qquad s \in [s_0, s_1]. 
\end{equation} 
\end{proposition}

To establish this result, we consider the equation enjoyed by the metric coefficients 
\begin{equation}\label{eq2-20-03-2021}
\Boxt_g (Z u_{\alpha\beta}) 
= - [Z,h^{\mu\nu}\del_{\mu}\del_{\nu}]u_{\alpha\beta} + Z (\Boxt_gu_{\alpha\beta}), 
\end{equation}
and we control both the commutator term and the nonlinear source term in the right-hand side, in order to be in a position next to 
apply the weighted energy estimate in the Euclidean-merging domain, which was stated earlier in Proposition~\ref{prop energy-ici-exterior}. 

{   From now on we find it convenient to introduce the notation
\begin{equation} 
\nouveauS_{p,k}(s,u) := 
\|\crochet^{\kappa-1/2}J^{1/2}|\delsN u|_{p,k}\|_{L^2(\MME_s)}, 
\end{equation} 
which is a term that will be controlled at the end of our bootstrap argument only. A similar notation is used when the second index $k$ is suppressed.
}

%-------------------------------------------------------------------------------------------------------------------------------

\subsection{Sharp energy estimates for the commutators away from the light cone}

\paragraph{Statement for the commutators.}

We first treat the commutators, that is, the first term in the right-hand side of \eqref{eq2-20-03-2021}. As before, we 
distinguish between the near- and far-light cone regions. We also emphasize that our upper bound below involves 
a sum over energy norms with a {\sl different number} of boosts and spatial rotations. 
The proof will be based on Proposition~\ref{prop1-12-02-2020} (hierarchy property for commutators) and 
Propositions~\ref{prop1-22-05-2020} and~\ref{propo2-22-05-2020} (Hessian estimates for the wave equation). 

\begin{proposition}[Sharp energy estimate for the commutators]
\label{prop--eq1-27-03-2021}
Under the conditions stated in Section~\ref{section-label-11-1}, 
for all  $\ord(Z) = p$ and $\rank(Z) = k$ one has 
$$ 
\aligned
\|\crochet^{\kappa}J\zeta^{-1}[Z,h^{\alpha\beta}\del_{\alpha}\del_{\beta}]u\|_{L^2(\MME_s)}
& \lesssim
\delta^{-1}(\ell^{-1}+\delta^{-2})(\epss + C_1\eps) \sum_{0\leq k_1\leq k}s^{-1+2k_1\theta} \, \Fenergy_{\kappa}^{\ME,p,k-k_1}(s,u) 
\\
& \quad +  \ell^{-1}\delta^{-1}(\epss + C_1\eps) \|\crochet^{\kappa}J\zeta^{-1} |\Boxt_g u|_{p-1,k-1}\|_{L^2(\MME_s)}
\\
&\quad
+ (\epss+C_1\eps)s^{-1/2} { \nouveauS_{p,k}(s,u)}
+ \ell^{-1}\delta^{-1}(\epss + C_1\eps)^2s^{ -1-\delta}.
\endaligned
$$
\end{proposition}

%-------------------------------------------------------------

\paragraph{Argument away from the light cone.}

The bound away from the light cone is simpler and is derived first. As a preparation, we apply { the Hessian estimate in} Proposition~\ref{propo2-22-05-2020} and obtain
\begin{equation}\label{eq1-14-03-2021}
\aligned
\|\crochet^{\kappa}J\zeta^{-1}\, |\del\del u|_{p,k} \|_{L^2(\Mfar_{\ell,s})}
& \lesssim  
\sum_{\ord(Z') \leq p\atop \rank(Z') \leq k} \|\crochet^{\kappa}J\zeta^{-1}\, [Z',h^{\alpha\beta}\del_{\alpha}\del_{\beta}]u\|_{L^2(\Mfar_{\ell,s})} 
\\
& \quad + \ell^{-1}\|\crochet^{\kappa}J\zeta^{-1}\,|\Boxt_g u|_{p,k}\|_{L^2(\Mfar_{\ell,s})} + \ell^{-1}s^{-1} \, \Fenergy_{\kappa}^{\ME,p+1,k+1}(s,u).
\endaligned
\end{equation}
We turn our attention to the right-hand side of { {the hierarchy decomposition in}}
\eqref{eq5-12-02-2020} and we treat the second term therein first. Observe that 
$\crochet\geq \ell r\geq \ell t$ holds in $\Mfar_{\ell,s}$. We consider the case where $p_1\leq N-3$ and apply the third 
{ pointwise estimate} in \eqref{eq1-30-05-2020} (since $p_1\geq 1$ implies $p_2\leq p-1, k_2\leq k-1$) and for 
$A_1 := |\LOmega  h |_{p_1-1,p_1-1} |\del\del u|_{p_2,k_2}$ we find 
\begin{subequations}\label{eq2-17-08-2021}
\begin{equation}\label{eq2a-17-08-2021}
\aligned
\|\crochet^{\kappa}J\zeta^{-1}\, A_1  \|_{L^2(\Mfar_{\ell,s})}
\lesssim \delta^{-1} (\epss + C_1\eps) \|\crochet^{\kappa}J\zeta^{-1}\,|\del\del u|_{p-1,k-1}\|_{L^2(\Mfar_{\ell,s})}.
\endaligned
\end{equation}
When $p_1\geq N-2$, implying $ p_2\leq 2\leq N-4$, we obtain
%\marginpar{$N\geq 6$},
\begin{equation}\label{eq2b-17-08-2021}
\aligned
\|\crochet^{\kappa}J\zeta^{-1}\, A_1 \|_{L^2(\Mfar_{\ell,s})}
& \lesssim \|\crochet^{\kappa}J\zeta^{-1} |\LOmega h^{\star} |_{p_1-1,p_1-1} |\del\del u|_{p_2,k_2}\|_{L^2(\Mfar_{\ell,s})}
+ \|s \, \crochet^{\kappa}\zeta|\LOmega u|_{p_1-1,p_1-1} |\del\del u|_{p_2,k_2}\|_{L^2(\Mfar_{\ell,s})}
\\
& \lesssim \delta^{-1}(\epss + C_1\eps) \|\crochet^{\kappa}J\zeta^{-1} |\del\del u|_{p-1,k-1}\|_{L^2(\Mfar_{\ell,s})}
+ \ell^{-1}\delta^{-1}(\epss + C_1\eps)^2s^{-2}, 
\endaligned
\end{equation}
where for the second term we used  {{the Hessian estimate}} \eqref{eq1-15-08-2021} and the decomposition \eqref{equa:sec8-13}, Lemma~\ref{lem-small}, together with \eqref{eq1-12-05-2020} and  { $\min(\lambda,\kappa) \geq 1/2+(3/2) \delta$.} The first term in the right-hand side of \eqref{eq5-12-02-2020} is as in \eqref{eq2a-17-08-2021}, and we omit the details. The last term in the right-hand side of \eqref{eq5-12-02-2020} is easier thanks to 
the partial derivative acting on $H$. When $p_1\leq N-3$, we 
{
rely on the fourth inequality in \eqref{eq1-30-05-2020}}
and  for 
$A_2 := |\del  h |_{p_1-1,k_1} |\del\del u|_{p_2,k_2}$ 
we obtain
\begin{equation}\label{eq2c-17-08-2021}
\|\crochet^{\kappa}J\zeta^{-1}\, A_2 \|_{L^2(\Mfar_{\ell,s})}
\lesssim \ell^{-\delta/2}(\epss + C_1\eps)s^{-1} \, \Fenergy_{\kappa}^{\ME,p,k}(s,u).
\end{equation}
When $p_1\geq N-2\geq p-2\ge 1$. we also have $p_2\leq 2\leq p-1$, 
% \marginpar{$N\geq 6$} 
and therefore 
\begin{equation}\label{eq2d-17-08-2021}
\aligned
\|\crochet^{\kappa}J\zeta^{-1}\, A_2 \|_{L^2(\Mfar_{\ell,s})}
& \lesssim \|{ \crochet^{\kappa}J\zeta^{-1}\,| \del h^{\star}} |_{p_1-1,k_1} |\del\del u|_{p_2,k_2}\|_{L^2(\Mfar_{\ell,s})}
+ \|s \, \crochet^{\kappa} |\del u|_{p_1-1,k_1} |\del\del u|_{p_2,k_2}\|_{L^2(\Mfar_{\ell,s})}
\\
&
\lesssim  { \delta^{-1}(\epss + C_1\eps) \|\crochet^{\kappa}J\zeta^{-1} |\del\del u|_{p-1,k-1}\|_{L^2(\Mfar_{\ell,s})} } + \ell^{-\delta/2} (\epss + C_1\eps)s^{-1} \, \Fenergy_{\kappa}^{\ME,p,k}(s,u), 
\endaligned
\end{equation}
where \eqref{equa-31-12-20} and \eqref{eq10-02-05-2020} were used.
\end{subequations}
Now substituting \eqref{eq1-14-03-2021} into \eqref{eq2a-17-08-2021} and \eqref{eq2b-17-08-2021}, we conclude that, for all $\ord(Z)= p$ and all $\rank(Z)= k$ 
$$
\aligned
&
\|\crochet^{\kappa}J\zeta^{-1} 
[Z,h^{\alpha\beta}\del_{\alpha}\del_{\beta}]u
\|_{L^2(\Mfar_{\ell,s})}
\lesssim   \sum_{\ord(Z') \leq p-1\atop \rank(Z') \leq k-1}
\delta^{-1}(\epss + C_1\eps) \|\crochet^{\kappa}J\zeta^{-1} [Z',h^{\alpha\beta}\del_{\alpha}\del_{\beta}]u\|_{L^2(\Mfar_{\ell,s})}
\\
& + \ell^{-1}\delta^{-1}(\epss + C_1\eps) \|\crochet^{\kappa}J\zeta^{-1} |\Boxt_g u|_{p-1,k-1}\|_{L^2(\Mfar_{\ell,s})}
+ \ell^{-1}\delta^{-1}(\epss + C_1\eps)s^{-1} \, \Fenergy_{\kappa}^{\ME,p,k}(s,u)
+ \ell^{-1}\delta^{-1}(\epss + C_1\eps)^2s^{-2}.
\endaligned
$$
Summing over all operators having $\ord(Z) \leq p$ and $\rank(Z) \leq k$ and recalling the smallness condition { ${\delta^{-1}(\epss + C_1\eps) \ll 1}$}, we obtain 
\begin{equation}\label{eq1-21-03-2021}
\aligned
\!\!\!\!\sum_{\ord(Z) \leq p\atop \rank(Z) \leq k}\!\!\!\!
\|\crochet^{\kappa}J\zeta^{-1}\, 
[Z,h^{\alpha\beta}\del_{\alpha}\del_{\beta}]u
\|_{L^2(\Mfar_s)}
& \lesssim \ell^{-1}\delta^{-1}(\epss + C_1\eps) \|\crochet^{\kappa}J\zeta^{-1}\ |\Boxt_g u|_{p-1,k-1}\|_{L^2(\Mfar_s)}
\\
& \quad + \ell^{-1}\delta^{-1}(\epss + C_1\eps)s^{-1} \, \Fenergy_{\kappa}^{\ME,p,k}(s,u)
+ \ell^{-1}\delta^{-1}(\epss + C_1\eps)^2s^{-2}.
\endaligned
\end{equation}
{ Observe that the first term in the right-hand side vanishes when $k=0$. }

%-----------------------------------------------------------------------------------------------------------------------------------------

\subsection{Sharp energy estimates for the commutators near the light cone}
\label{sec--64}

\paragraph{Bounds on Hessian.} 

The control within $\Mnear_{\ell,s}$ is more involved, and we now rely on Propositions~\ref{prop1-12-02-2020} and~\ref{prop1-22-05-2020}. We analyze the Hessian first which is controlled by a direct application of \eqref{eq8-04-06-2020},
{
together with the pointwise bound \eqref{eq19-23-04-2021} for the null component $h^{\N00}$:}
\begin{equation}\label{eq3-18-08-2021}
\aligned
&
\|\crochet^{\kappa}J\zeta^{-1} \crochet t^{-1} |\del\del u|_{p,k}\|_{L^2(\Mnear_{\ell,s})}
\\
& \lesssim
\sum_{\ord(Z) \leq p\atop \rank(Z) \leq k}\|\crochet^{\kappa}J\zeta^{-1}[Z',h^{\alpha\beta}\del_{\alpha}\del_{\beta}]u\|_{L^2(\Mnear_{\ell,s})}
+ \|\crochet^{\kappa}J\zeta^{-1} |\Boxt_g u|_{p,k}\|_{L^2(\Mnear_{\ell,s})}
\\
& \quad + (\epss + C_1\eps)\textcolor{black}{\Big(}s^{-1+2\theta} \, \Fenergy_{\kappa}^{\ME,p+1,k}(s,u)
+ s^{-1} \, \Fenergy_{\kappa}^{\ME,p+1,k+1}(s,u)\textcolor{black}{\Big)}.
\endaligned
\end{equation}

%------------------

\paragraph*{Bounds on $T^{\bf hier}$.}

We rely on Proposition~\ref{prop1-12-02-2020}, and we successively control each term in the right-hand side of \eqref{eq4a-12-02-2020}. Consider any ordered operator $Z$ with $\ord(Z) = p$ and $\rank(Z) = k\leq p$.
We begin with the second term in $T^{\bf hier}$, and we observe that $p_1\geq 1$ implies $p_2\leq p-1$ and $k_2\leq k-1$. When $p_1\leq  N-4$, we apply \eqref{eq19-23-04-2021} and obtain
\begin{equation}\label{eq4-18-08-2021}
\|s \, \crochet^{\kappa}\zeta|L h^{\N00} |_{p_1-1,p_1-1} |\del\del u|_{p_2,k_2}\|_{L^2(\Mnear_{\ell,s})}
\lesssim (\epss + C_1\eps)s^{-1+2\theta} \, \Fenergy_{\kappa}^{\ME,p,k-1}(s,u).
\end{equation}
When $p_1\geq N-3$, therefore $p_2\leq 3\leq N-5$, 
% \marginpar{$N\geq 8$}, 
we have  the following decomposition:
\begin{equation}\label{eq5-17-08-2021}
\aligned
& \|\crochet^{\kappa} J\zeta^{-1} |L h^{\N00} |_{p_1-1,p_1-1} |\del\del u|_{p_2,k_2}\|_{L^2(\Mnear_{\ell,s})}
\\
& \lesssim s \|\crochet^{\kappa} |h^{\star\N00} |_{p_1}\zeta|\del\del u|_{p_2,k_2}\|_{L^2(\Mnear_{\ell,s})} 
+  \|\crochet^{\kappa} J\zeta^{-1} |L u^{ \N00}|_{p_1-1,p_1-1} |\del\del u|_{p_2,k_2}\|_{L^2(\Mnear_{\ell,s})}
=: G_1 + G_2. 
\endaligned
\end{equation} 
We use the following estimate on the null component $h^{\star\N00}$ of the reference:
\begin{equation}\label{eq3-17-08-2021}
|h^{\star\N00} |_N\lesssim \epss r^{-1+\theta},
\end{equation}
whose proof is postponed at the end of this Section~\ref{sec--64}. The first term in the right-hand side of \eqref{eq5-17-08-2021} is then bounded as follows:
\begin{equation}\label{eq7-18-08-2021}
{
G_1}
\lesssim 
(\epss + C_1\eps)s^{-1+2\theta} \, \Fenergy_{\kappa}^{\ME, p,k-1}(s,u).
\end{equation}
For the second one we rely 
{ 
on a calculation similar to the one in the proof of Lemma~\ref{lemma--18-sept-22-000}. 
}
Recalling \eqref{eq1-09-08-2021},  we have 
$$
G_2\lesssim (\epss + C_1\eps) s^{1+2\delta} \|\crochet^{-1-\min(\lambda,\kappa) + \kappa}\zeta r^{-1}|Lu^{\N00}|_{p_1-1}\|_{L^2(\Mnear_{\ell,s})}.
$$
where $2\theta\leq \delta$ is applied. Now let $Z$ be a admissible operator with $\ord(Z)\leq p_1-1$. Then by Proposition~\ref{propo-Poincare-ext} we have 
$$
\aligned
& 
(\epss + C_1\eps) s^{1+2\delta}\|\crochet^{-1-\min(\lambda,\kappa) + \kappa}\zeta r^{-1}ZLu^{\N00}\|_{L^2(\Mnear_{\ell,s})}
\\
& \lesssim \,  (\epss + C_1\eps) s^{1+2\delta}\|\crochet^{1/2+\delta}\zeta \del(r^{-1}ZLu^{\N00})\|_{L^2(\MME_s)}
 +  (\epss + C_1\eps) s^{1+2\delta}\|\crochet^{1/2+\delta}\zeta r^{-2}ZLu^{\N00}\|_{L^2(\MME_s)}
\\
& \lesssim \, (\epss + C_1\eps) s^{1+2\delta}\|\crochet^{1/2+\delta}\zeta r^{-1} |\del Lu^{\N00}|_{p_1-1}\|_{L^2(\MME_s)} 
+ (\epss + C_1\eps) s^{1+2\delta}\|\crochet^{1/2+\delta}\zeta r^{-2}ZLu^{\N00}\|_{L^2(\MME_s)}
\\
& =:\,  G_{21} + G_{22}.
\endaligned
$$
The term $G_{22}$ is easier to handle and, after recalling \eqref{eq2-24-06-2021}, we find 
$$
\aligned
 G_{22}
& \lesssim   (\epss + C_1\eps) s^{1+2\delta}\|\crochet^{1/2+\delta}\zeta r^{-1}|\delsN u|_{p_1-1}\|_{L^2(\MME_s)} 
+   (\epss + C_1\eps) s^{1+2\delta}\|\crochet^{3/2+\delta}\zeta r^{-2}|\del u|_{p_1-1}\|_{L^2(\MME_s)} 
\\
& \lesssim (\epss + C_1\eps) s^{1/2+2\delta}\|\crochet^{\kappa-1/2}s^{1/2}\zeta r^{-\kappa+\delta}|\delsN u|_{p_1-1}\|_{L^2(\MME_s)} 
+ (\epss + C_1\eps) s^{1+2\delta}\|\crochet^{\kappa}r^{-1/2+\delta-\kappa}\zeta|\del u|_{p_1-1}\|_{L^2(\MME_s)}
\\
& \lesssim (\epss + C_1\eps) s^{1/2+4\delta-2\kappa}\|\crochet^{\kappa-1/2}J^{1/2}|\delsN u|_{p_1-1}\|_{L^2(\MME_s)}
+ (\epss+C_1\eps)s^{-2\kappa+4\delta}\Fenergy_{\kappa}^{\ME,p,k}(s,u)
\\
&  
\lesssim  (\epss + C_1\eps)s^{-1/2}  { \nouveauS_{p_1-1}(s,u)}
+ (\epss + C_1\eps)s^{-1}\Fenergy_{\kappa}^{\ME,p,k}(s,u).
\endaligned
$$
The estimate on $G_{21}$ requires us to apply Lemma~\ref{lemma-12-04-2020} (the second conclusion), hence 
$$
\aligned
&  
\|\crochet^{1/2+\delta}\zeta r^{-1} |\del L u^{\N00}|_{p_1-1}\|_{L^2(\MME_s)}
\\
& \lesssim  \|\crochet^{1/2+\delta}r^{-1}\zeta|\delsN u|_{p_1}\|_{\MME_s} 
+ \|\crochet^{1/2+\delta}r^{-2}\zeta|Lu|_{p_1-1}\|_{L^2(\MME_s)} 
+ \|\crochet^{1/2+\delta}\zeta r^{-1} |w^{\star}|_{p_1}\|_{L^2(\MME_s)}
\\
& \quad
+\sum_{p_{11}+p_{12} = p_1}\|\crochet^{1/2+\delta}r^{-1}\zeta\big(|h^\star|_{p_{11}} |\del u|_{p_{12}} + |\del h^{\star}|_{p_{11}}|u|_{p_{12}} + |u|_{p_{11}}|\del u|_{p_{12}}\big)\|_{L^2(\MME_s)}.
\endaligned
$$
For the first term above, we observe that
$$
\aligned
(\epss+C_1\eps)s^{1+2\delta}\|\crochet^{1/2+\delta}r^{-1}\zeta|\delsN u|_{p_1}\|_{\MME_s} 
\lesssim  
(\epss+C_1\eps)s^{-1/2}  { \nouveauS_{p_1}(s,u)}
%%% 
\endaligned
$$
while for the second term, we recall \eqref{eq2-24-06-2021}: 
$$
\aligned
&(\epss + C_1\eps) s^{1+2\delta}\|\crochet^{1/2+\delta}r^{-2}\zeta|Lu|_{p_1-1}\|_{L^2(\MME_s)} 
\\
& \lesssim (\epss + C_1\eps) s^{1+2\delta}\|\crochet^{1/2+\delta}r^{-1}\zeta|\delsN u|_{p_1-1}\|_{L^2(\MME_s)} + (\epss + C_1\eps)s^{1+2\delta}\|\crochet^{3/2+\delta}r^{-2}\zeta|\del u|_{p_1-1}\|_{L^2(\MME_s)}
\\
& \lesssim 
 (\epss+C_1\eps)s^{-1/2}   { \nouveauS_{p_1}(s,u)}
+ (\epss+C_1\eps)s^{-1}\Fenergy_{\kappa}^{\ME,p_1,p_1}(s,u).
\endaligned
$$
The third term relies on our assumption on the reference. In fact recalling the {   
wave gauge condition assumed on $g^{\star}$, namely $|w^{\star}|_N\lesssim \crochet^{-1-\varsigma}r^{-1}$} we have 
$$
(\epss+C_1\eps)s^{1+2\delta}\|\crochet^{1/2+\delta}\zeta r^{-1} |w^{\star}|_{p_1}\|_{L^2(\MME_s)}
\lesssim 
   \delta^{-1}(\epss+C_1\eps)^2s^{-1-\delta}.
$$
The estimate on high-order terms are much easier, and we omit the details.

We thus conclude that
\begin{equation}\label{eq5-18-08-2021}
G_2\lesssim (\epss+C_1\eps)s^{-1}\Fenergy_{\kappa}^{\ME,p,k}(s,u) 
+ (\epss + C_1\eps)s^{{ -}1/2}   { \nouveauS_{p,k}(s,u)}
   + \delta^{-1}(\epss+C_1\eps)^2s^{-1-\delta}.
\end{equation}
where we have used that $|\delsN u|_{p_1}\lesssim |\delsN u|_{p,k}$ (since $p_1+k_2=k$, thus $p_1\leq k$).
% 

%======================================

For the term $\frac{|r-t|}{t} |L H |_{p_1-1,p_1-1} |\del\del u|_{p_2,k_2}$, we observe that $p_2\leq p-1$ and $k_2\leq k-1$. When $p_1\leq [N/2]\leq N-3$, we apply \eqref{eq1-30-05-2020}: 
\begin{equation}\label{eq6-18-08-2021}
\aligned
& \|\crochet^{\kappa}J\zeta^{-1} 
|L H |_{p_1-1,p_1-1} \crochet t^{-1} |\del\del u|_{p_2,k_2}\|_{L^2(\Mnear_{\ell,s})}
\\
& \lesssim \delta^{-1}(\epss + C_1\eps)s^{\delta-2\min(\lambda,\kappa)}\|\crochet^{\kappa}J\zeta^{-1} \crochet t^{-1} |\del\del u|_{p-1,k-1}\|_{L^2(\Mnear_{\ell,s})} 
\endaligned
\end{equation}
with a right-hand side bounded by \eqref{eq3-18-08-2021}. When $p_1\geq N-2\geq p-2$, we have $k_2\leq p_2\leq 2\leq p-1$ and so 
$$
\aligned
& \|\crochet^{\kappa}J\zeta^{-1} |L h |_{p_1-1,p_1-1}\crochet t^{-1} |\del\del u|_{p_2,k_2}\|_{L^2(\Mnear_{\ell,s})}
\\
& \lesssim
\|\crochet^{\kappa}J\zeta^{-1} |L h^{\star} |_{p_1-1,p_1-1}\crochet t^{-1} |\del\del u|_{p_2,k_2}\|_{L^2(\Mnear_{\ell,s})} 
+ \|\crochet^{\kappa}J\zeta^{-1} |L u|_{p_1-1,p_1-1}\crochet t^{-1} |\del\del u|_{p_2,k_2}\|_{L^2(\Mnear_{\ell,s})}.
\endaligned
$$
{{In view of \eqref{equa-31-12-20} (Class A) or \eqref{equa-new-conditions-hstar} (Class B)}}, the first term is also bounded as
\begin{equation}\label{eq9-18-08-2021}
\aligned
& \|\crochet^{\kappa}J\zeta^{-1} |L h^{\star} |_{p_1-1,p_1-1}\crochet t^{-1} |\del\del u|_{p_2,k_2}\|_{L^2(\Mnear_{\ell,s})} 
\\
& \lesssim
(\epss + C_1\eps)s^{\delta-2\min(\lambda,\kappa)}\|\crochet^{\kappa}J\zeta^{-1}\crochet t^{-1} |\del\del u|_{p-1,k-1}\|_{L^2(\Mnear_{\ell,s})}.
\endaligned
\end{equation}
For the second term we apply \eqref{eq10-04-06-2020} (namely, a rough bound therein, observing that $\Mnear_s\supset \Mnear_{\ell,s}$) and \eqref{eq1-18-08-2021}, and obtain (with { $\min(\lambda,\kappa) \geq 1/2+(3/4) \delta$})
\begin{equation}\label{eq10-18-08-2021}
\aligned
& \|\crochet^{\kappa}J\zeta^{-1} |L u|_{p_1-1,p_1-1}\crochet t^{-1} |\del\del u|_{p_2,k_2}\|_{L^2(\Mnear_{\ell,s})}
\\
& \lesssim 
(\epss + C_1\eps)s^{1+2\delta - 4\min(\lambda,\kappa)} \|\crochet^{\kappa-1} |\LOmega u|_{p_1-1,p_1-1}\|_{L^2(\Mnear_{\ell,s})}
\\
&
\lesssim \delta^{-1}(\epss + C_1\eps)s^{1+3\delta - 4\min(\lambda,\kappa)} \, \Fenergy_{\kappa}^{\ME,p_1,p_1}
  \lesssim \delta^{-1}s^{-1}(\epss + C_1\eps) \Fenergy_{\kappa}^{\ME,p,k}.
\endaligned
\end{equation}
The first term in $T^{\bf hier}$ is bounded as we have done for \eqref{eq4-18-08-2021} and \eqref{eq6-18-08-2021}, so we omit the details and write 
\begin{equation}\label{eq8-18-08-2021}
\aligned
& \| \crochet^{\kappa}J\zeta^{-1}\big(|\HN^{00} | + t^{-1} |r-t| |H|\big) \, |\del\del u|_{p-1,k-1} \|_{L^2(\Mnear_{\ell,s})}
\\
& \lesssim
(\epss + C_1\eps)s^{-1+2\theta} \, \Fenergy_{\kappa}^{\ME,p,k-1}(s,u) 
+ \delta^{-1}(\epss + C_1\eps)s^{\delta-2\min(\lambda,\kappa)}\|\crochet^{\kappa}J\zeta^{-1} \crochet t^{-1} |\del\del u|_{p-1,k-1}\|_{L^2(\Mnear_{\ell,s})}.
\endaligned
\end{equation} 
{{
Consequently, in view \eqref{eq4-18-08-2021} and \eqref{eq7-18-08-2021}--\eqref{eq8-18-08-2021}, we find 
}}
\begin{equation}\label{eq15-18-08-2021}
\aligned
\|\crochet^{\kappa}J\zeta^{-1}T^{\bf hier}\|_{L^2(\Mnear_{\ell,s})}
& \lesssim 
\delta^{-1}(\ell^{-\delta} + \delta^{-2})(\epss + C_1\eps) \sum_{0\leq k_1\leq k}s^{-1+2k_1\theta} \, \Fenergy_{\kappa}^{\ME,p,k-k_1}(s,u)
\\
& \quad + \delta^{-1}(\epss + C_1\eps)s^{\delta-2\min(\lambda,\kappa)}\|\crochet^{\kappa}J\zeta^{-1}\crochet t^{-1} |\del\del u|_{p-1,k-1}\|_{L^2(\Mnear_{\ell,s})}
\\
&\quad +  (\epss + C_1\eps)s^{ -1/2}  { \nouveauS_{p}(s,u)}    
+ \delta^{-1}(\epss+C_1\eps)^2s^{-1-\delta}.
\endaligned
\end{equation}

%---------------------------------------------------------------------

\paragraph*{Bounds on $T^{\bf easy}$ and $T^{\bf super}$.}

These are relatively easier to handle, since they contain a partial derivative acting on $H$ and a sufficiently decaying factor. We first treat $|\del H^{\N00} |_{p_1-1,k_1} |\del\del u|_{p_2,k_2}$. When $p_1-1\leq N-3$, we apply \eqref{eq1-17-07-2020} and obtain (since $p_1\geq1$ implies $p_2\leq p-1$)
$$
\|\crochet^{\kappa}J\zeta^{-1} |\del h^{\N00} |_{p_1-1,k_1} |\del\del u|_{p_2,k_2}\|_{L^2(\Mnear_{\ell,s})}
\lesssim
\delta^{-1} (\epss + C_1\eps)s^{-1-2\min(\lambda,\kappa)+\delta}\|\crochet^{\kappa}\zeta|\del\del u|_{p,k}\|_{L^2(\Mnear_{\ell,s})}.
$$
{ Provided $\min(\lambda,\kappa) \geq 1/2+\delta$,}
$$
\|\crochet^{\kappa}J\zeta^{-1} |\del h^{\N00} |_{p_1-1,k_1} |\del\del u|_{p_2,k_2}\|_{L^2(\Mnear_{\ell,s})}\lesssim \delta^{-1}(\epss + C_1\eps)^2s^{-2}.
$$
When $p_1-1\geq N-2$ which implies $k_2\leq p_2\leq 1\leq p-1$ we have 
$$
\aligned
& \|\crochet^{\kappa}J\zeta^{-1} |\del h^{\N00} |_{p_1-1,k_1} |\del\del u|_{p_2,k_2}\|_{L^2(\Mnear_{\ell,s})}
\\
& \lesssim
s\|\crochet^{\kappa}\zeta|\del h^{\star} |_{p_1-1,k_1} |\del\del u|_{p_2,k_2}\|_{L^2(\Mnear_{\ell,s})} 
+ s\|\crochet^{\kappa}\zeta|\del u^{ \N00}|_{p_1-1,k_1} |\del\del u|_{p_2,k_2}\|_{L^2(\Mnear_{\ell,s})}, 
\endaligned
$$
where 
{ by {\eqref{eq1-02-10-2022}}} 
the first term is bounded by 
$$
\delta^{-1}(\epss+C_1\eps)s^{-1}\|\crochet^{\kappa}J\zeta^{-1}|\del\del u|_{p-1,k-1}\|_{L^2(\Mnear_{\ell,s})}. 
$$
For the second term, we rely on 
 { 
 \eqref{eq10-02-05-2020} together with Lemma~\ref{lemma-12-04-2020} (the second estimate)}
  and obtain
{
\begin{equation}\label{eq12-18-08-2021}
\aligned
&s\|\crochet^{\kappa}\zeta|\del u^{ \N00}|_{p_1-1,k_1} |\del\del u|_{p_2,k_2}\|_{L^2(\Mnear_{\ell,s})}
\\
& \lesssim  (\epss + C_1\eps) s^{1+\delta}\|r^{-1}\zeta|\del u^{\N00}|_{p_1-1,k_1}\|_{L^2(\Mnear_s)}
\\
& \lesssim  (\epss + C_1\eps)s^{1+\delta}\|r^{-1}\zeta|\delsN u|_{p_1-1,k_1}\|_{L^2(\Mnear)}
+ (\epss + C_1\eps)s^{1+\delta}\|r^{-2}\zeta |u|_{p_1-1,k_1}\|_{L^2(\Mnear_s)}
\\
&+ (\epss + C_1\eps)s^{1+\delta}\|r^{-1}\zeta |w^{\star}|_{p_1-1,k_1}\|_{L^2(\Mnear_s)}
\\
&+ (\epss + C_1\eps) s^{-1+\delta}\sum_{p_{11}+p_{12}=p_1-1}\|r^{-1}\zeta(|\del h^{\star}|_{p_{11}}|u|_{p_{12}} + |h^{\star}|_{p_{11}}|\del u|_{p_{12}} + |u|_{p_{11}}|\del u|_{p_{12}})\|_{L^2(\Mnear_s)}
\\
& \lesssim  (\epss+C_1\eps)s^{-3/2+\delta}\|\crochet^{\kappa-1/2}J^{1/2}|\delsN u|_{p_1-1}\|_{L^2(\MME_s)} 
+ \delta^{-1}(\epss+C_1\eps)^2s^{-1-\delta},
\endaligned
\end{equation}
where we applied \eqref{eq1-12-05-2020}. 
{
These terms are bounded as we have done in \eqref{eq5-18-08-2021}} and we conclude that
\begin{equation}\label{eq13-18-08-2021}
\aligned
& \|\crochet^{\kappa}J\zeta^{-1} |\del { h^{ \N00}}|_{p_1-1,k_1} |\del\del u|_{p_2,k_2}\|_{L^2(\Mnear_{\ell,s})}
 \lesssim  \delta^{-1}(\epss+C_1\eps)^2s^{-1-\delta} + (\epss + C_1\eps)s^{-1/2}  { \nouveauS_{p_1-1}(s,u).}
\endaligned
\end{equation}
Since  {$\min(\lambda,\kappa) \geq 1/2+\delta$}, the second term in $T^{\bf easy}$ in \eqref{eq4b-12-02-2020} is bounded as
$$
\|\crochet^{\kappa}J\zeta^{-1}\crochet r^{-1} \, |\del H|_{p_1-1,k_1} |\del\del u|_{p_2,k_2}\|_{L^2(\Mnear_{\ell,s})}\lesssim(\epss + C_1\eps)^2s^{-2},
$$ 
and we may omit the details.  
Then we write the bound on $T^{\bf easy}$:
\begin{equation}\label{eq14-18-08-2021}
\aligned
&
\|\crochet^{\kappa}J\zeta^{-1}T^{\bf easy}\|_{L^2(\Mnear_{\ell, s})}
 \lesssim 
 (\ell^{-\delta} + \delta^{-2})(\epss + C_1\eps) \sum_{0\leq k_1\leq k}s^{-1+2k_1\theta} \, \Fenergy_{\kappa}^{\ME,p,k-k_1}(s,u)
 % }}  
 + (\epss + C_1\eps)s^{-1/2}  { \nouveauS_{p_1-1}(s,u)}
\\
&   + \delta^{-1}(\epss + C_1\eps)s^{\delta-2\min(\lambda,\kappa)}\|\crochet^{\kappa}J\zeta^{-1}\crochet t^{-1} |\del\del u|_{p-1,k-1}\|_{L^2(\Mnear_{\ell,s})}
+ \delta^{-1}(\epss + C_1\eps)^2s^{-2}.
\endaligned
\end{equation}
Finally, the terms contained in $T^{\bf super}$ are relatively trivial due to the additional decreasing factor $t^{-1}$ and, { in view of   $\min(\lambda,\kappa) \geq 1/2+\delta$}, we find 
\begin{equation}\label{eq16-18-08-2021}
\|\crochet^{\kappa}J\zeta^{-1}T^{\bf super}\|_{L^2(\Mnear_{\ell, s})}\lesssim \delta^{-1}(\epss + C_1\eps)^2 s^{-2}.
\end{equation}

%------------------------

\paragraph*{Closing the proof of  Proposition~\ref{prop--eq1-27-03-2021}.}

Recalling \eqref{eq15-18-08-2021}, \eqref{eq14-18-08-2021} and \eqref{eq16-18-08-2021}, and apply \eqref{eq3-18-08-2021},  for $\ord(Z) = p$ and $\rank(Z) = k$,
$$
\aligned
&
\|\crochet^{\kappa}J\zeta^{-1}[Z,h^{\alpha\beta}\del_{\alpha}\del_{\beta}]u\|_{\Mnear_{\ell,s}}
 \lesssim 
\delta^{-1}(\ell^{-\delta} + \delta^{-2})(\epss + C_1\eps) \sum_{0\leq k_1\leq k}s^{-1+2k_1\theta} \, \Fenergy_{\kappa}^{\ME,p,k-k_1}(s,u) 
\\
& \quad + \delta^{-1}(\epss + C_1\eps)s^{\delta-2\min(\lambda,\kappa)}\!\!\!\!\!\!
\sum_{\ord(Z') \leq p-1\atop \rank(Z') \leq k-1}\!\!\!\!\!\! 
\|\crochet^{\kappa}J\zeta^{-1}[Z',h^{\alpha\beta}\del_{\alpha}\del_{\beta}]u\|_{L^2(\Mnear_{\ell,s})}
\\
& \quad + \delta^{-1} (\epss + C_1\eps)s^{\delta-2\min(\lambda,\kappa)} \|\crochet^{\kappa}J\zeta^{-1} |\Boxt_g u|_{p-1,k-1}\|_{L^2(\Mnear_{\ell,s})}
   + (\epss + C_1\eps)s^{ -1/2} { \nouveauS_p(s,u)} 
 \\
 &
 \quad + \delta^{-1}(\epss + C_1\eps)^2s^{ -1-\delta}. 
\endaligned
$$
Similarly to the case in $\Mfar_{\ell,s}$, we sum up the above bound for all $\ord(Z) \leq p$ and $\rank(Z) \leq k$. Recalling  { $\delta^{-1}(\epss + C_1\eps) \ll 1$}, we obtain
$$ 
\aligned
&
\sum_{\ord(Z) \leq p\atop \rank(Z) \leq k}\|\crochet^{\kappa}J\zeta^{-1}[Z,h^{\alpha\beta}\del_{\alpha}\del_{\beta}]u\|_{\Mnear_{\ell,s}}
 \lesssim 
\delta^{-1}(\ell^{-\delta} + \delta^{-2})(\epss + C_1\eps) \sum_{0\leq k_1\leq k}s^{-1+2k_1\theta} \, \Fenergy_{\kappa}^{\ME,p,k-k_1}(s,u) 
\\
& \quad +  \delta^{-1}(\epss + C_1\eps)s^{\delta-2\min(\lambda,\kappa)} \|\crochet^{\kappa}J\zeta^{-1} |\Boxt_g u|_{p-1,k-1}\|_{L^2(\Mnear_{\ell,s})}
 + (\epss + C_1\eps)s^{ -1/2}   { \nouveauS_p(s,u)}  
\\
&\quad + \delta^{-1}(\epss + C_1\eps)^2s^{-1-\delta}. 
\endaligned
$$
Combining the estimate \eqref{eq1-21-03-2021} away from the light cone with our estimate above, we arrive at the desired conclusion in Proposition~\ref{prop--eq1-27-03-2021}. 

%----------------------------------------------------

\paragraph*{Proof of \eqref{eq3-17-08-2021}.} 

We recall our decomposition of the null component 
$$
h^{\star\N00} = -\PsiN_{\alpha}^0\PsiN_{\beta}^0h^{\star\alpha\beta} 
= -\sum_{\alpha,\beta}\PsiN_{\alpha}^0\PsiN_{\beta}^0 { \mathbbm{h}^{\alpha\beta}[h^{\star}]}  + \PsiN_{\alpha}^0\PsiN_{\beta}^0\Abb^{\alpha\beta}[h^{\star}] = \Xi^{\star\N00} + \PsiN_{\alpha}^0\PsiN_{\beta}^0\Abb^{\alpha\beta}[h^{\star}], 
$$
where the expression $\Abb^{\alpha\beta}$ is defined in \eqref{eq1-07-05-2021} (and consists of high-order terms obtained when expressing $h^{\star\alpha\beta}$ in terms of $h^{\star}_{\alpha\beta}$). Then, we
substitute \eqref{eq1-11-03-2021-gstar} and 
{ 
\eqref{equa-31-12-20} (Class A) 
or \eqref{equa-new-conditions-hstar} (Class B)
} into the above expression, 
and we obtain \eqref{eq3-17-08-2021} under our condition { $\min(\lambda,\kappa) \geq 1/2$}.

%-------------------------------------------------------------------------------------------------------------------------------------------------------------------

\subsection{Sharp energy estimates for source terms}

We begin with the second source term in \eqref{eq2-20-03-2021}. Our upper bound below involves a sum of energies associated with
boosts and spatial rotations up to rank $k$.
We consider successively each term in the decomposition
$\Boxt_g u_{\alpha\beta} = \Pbb^\star_{\alpha\beta}[u] + \Qbb^{\star}_{\alpha\beta}[u] + W^{\textbf{linear}}_{\alpha\beta} + W^\super_{\alpha\beta}$ 
of the right-hand side of \eqref{eq 1 13-01-2019}. For Class A, both $W^{\textbf{linear}}_{\alpha\beta}$ and $W^\super_{\alpha\beta}$ enjoy sufficient $L^2$ estimates (cf.~Lemma \ref{lemma-112}).  We now focus on estimating the null and quasi-null terms which is more involved as we need to distinguish between the near- and far-light cone regions, as follows. We emphasize again that the following estimates on $\Pbb^\star_{\alpha\beta}[u]$ and $\Qbb^{\star}_{\alpha\beta}[u]$ are valid for {\sl both} Class A and Class B. 

%----------------------------------------------------

\paragraph{Null semi-linear terms near the light cone.}

Recalling the basic decomposition of null terms derived in \eqref{eq7-04-10-2022} and using the decay property of the reference metric in~\eqref{equa-31-12-20} or \eqref{equa-new-conditions-hstar}, under the condition $p_1\leq [p/2]\leq N-4$ 
{
(and provided $N\geq 7$, say)}
we have 
\begin{equation}
\label{equation-8-sept-2021}
\aligned
|\Qbb^\star [u] |_{p,k}
& \lesssim  
\sum_{p_1+p_2 = p\atop k_1+k_2=k} 
|\del u|_{p_1,k_1} |\delsN u|_{p_2, k_2} + |\del u|_p|\delsN u|_{p_1, k_1} + \epss 
r^{-\lambda} |\del u|_{p,k} |\del u|_{p_1, k_1} 
\\
& \lesssim {  (\epss + C_1\eps) s^{\delta}r^{-1} \crochet^{-\kappa} }
|\delsN u|_{ p,k} 
+  (\epss+C_1\eps)  \, s^{\delta} r^{-1-\min(\lambda,\kappa)} |\del u|_{ p,k}
=: G_1 + G_2.
\endaligned 
\end{equation}
where we used the basic pointwise metric bound in { \eqref{eq10-02-05-2020}}. 
The second term $G_2$ 
is treated as follows. 
Recalling the inequality $J \lesssim \zeta^2 s$ { (from Lemma~\ref{lem1-22-05-2020})} and the fact that $s^2 \lesssim r$ in $\MME_s$ as well as the condition { $\min(\lambda,\kappa) \geq 1/2\geq (3/2) \delta$}, we arrive at 
$$ 
\| \crochet^{\kappa}J\zeta^{-1}\,  G_2 \|_{L^2(\Mnear_{\ell,s})}
\lesssim  (\epss+C_1\eps)  s^{-1-2\min(\lambda,\kappa)+\delta}\|\crochet^{\kappa}\zeta|\del u|_p\|_{L^2(\Mnear_{\ell,s})}\lesssim (\epss+C_1\eps)^2s^{-1-\delta}.
$$
On the other hand, for the first term $G_1$
we write 
$$
\aligned 
\| \crochet^{\kappa}J\zeta^{-1}\, G_1 \|_{L^2(\Mnear_{\ell,s})}
& \lesssim (\epss+C_1\eps) s^{1/2+\delta}\|J^{1/2}r^{-1}|\delsN u|\|_{L^2(\Mnear_s)}
\lesssim  
(\epss+C_1\eps) s^{-1/2}  { \nouveauS_{p,k}(s,u)}. 
\endaligned
$$
{ We conclude that, for both classes of metrics,}
\begin{equation}\label{eq1-01-10-2022}
\aligned
& 
\|\crochet^{\kappa}J\zeta^{-1} |\Qbb^\star [u] |_{p,k}\|_{L^2(\Mnear_{\ell,s})}
 \lesssim
  (\epss+C_1\eps) s^{-1/2}    { \nouveauS_{p,k}(s,u)}  
+ { (\epss+C_1\eps)^2s^{-1-\delta}}.
\endaligned
\end{equation}

%-------------------------------------------------------------------------------

\paragraph{Quasi-null terms near the light cone.}

By Lemma~\ref{lem1-31-01-2021} (with $u=v$) we have 
\begin{equation}\label{eq13-01-31-2021}
\aligned
& |\Pbb^{\star}[u]|_{p,k} \lesssim  |\Pbb_{00}^{\star}[u]|_{p,k} + |\slashed{\Pbb}^{\star}[u]|_{p,k}
\\
& \lesssim   \sum_{p_1+p_2=p\atop k_1+k_2=k} |\del \us^{\N} |_{p_1,k_1} |\del \us^{\N} |_{p_2,k_2}
+ \sum_{p_1+p_2=p}\Big(|\delts u|_{p_1} |\del u|_{p_2}
+ | \SbbME_{p_1}[u] | |\del u|_{p_2}  \Big) 
+ \sum_{p_1+p_2+p_3=p} |h^{\star} |_{p_3} |\del u|_{p_1} |\del u|_{p_2}, 
\endaligned
\end{equation}
and we consider the weighted $L^2$ norm of each term.  The first term in the right-hand side is the most challenging term. Recalling  \eqref{eq1-17-08-2021}, we only need to treat $\|s \, \crochet^{\kappa}\zeta|\del \us^{\N} |_{p_1,k_1} |\del \us^{\N} |_{p_2,k_2}\|_{L^2(\Mnear_{\ell,s})}$. 
Without loss of generality, we assume that $p_1\leq[p/2]\leq N-4$
% \marginpar{$N\geq 7$} 
and we apply \eqref{eq11-01-31-2021}: 
$$
\aligned
& \|s \, \crochet^{\kappa}\zeta \, | \del \us^{\N} |_{p_1,k_1} |\del \us^{\N} |_{p_2,k_2}\|_{L^2(\Mnear_{\ell,s})}
\lesssim  \||\del \us^{\N} |_{N-4,k_1} \,s \, \crochet^{\kappa}\zeta \, | \del \us^{\N} |_{p_2,k_2}\|_{L^2(\Mnear_{\ell,s})}
\\
& \lesssim  (\ell^{-\delta/2} + \delta^{-2})(\epss + C_1\eps)s^{-1+2k_1\theta} \|\zeta\crochet^{\kappa} |\del \us^{\N} |_{p_2,k_2}\|_{L^2(\Mnear_{\ell,s})} 
\\
&
\lesssim  (\ell^{-\delta/2} + \delta^{-2})(\epss + C_1\eps) s^{-1 + 2k_1\theta} \, \Fenergy_{\kappa}^{\ME,p,k-k_1}(s,u), 
\endaligned
$$
where $0\leq k_1\leq k$. The remaining terms in the right-hand side of \eqref{eq13-01-31-2021} are much easier, since they are either null quadratic terms (controlled in the previous paragraph) or cubic terms. 
({ Here, we need $\min(\lambda,\kappa) \geq 1/2 + 2\delta \geq (3/2) \delta$ and $\delta^{-1}(\epss + C_1\eps) \lesssim 1$}.)
We omit the details and conclude that
\begin{equation}
{ 
\|\crochet^{\kappa} J\zeta^{-1}|\Pbb^{\star}[u]|_{p,k}\|_{\Mnear_{\ell,s}}\lesssim (\ell^{-\delta/2}+\delta^{-2})(\epss + C_1\eps)s^{-1+2k_1\theta} \, \sum_{0\leq k_1\leq k} \Fenergy_{\kappa}^{\ME,p,k-k_1}(s,u) + (\epss+C_1\eps)^2s^{-1-\delta}.
}
\end{equation}

%----------------------------------------------------- 

\paragraph*{Quadratic semi-linear terms away from the light cone.}

It remains to derive the desired bound in $\Mfar_{\ell,s}$. We observe that $\crochet^{-1}\lesssim \ell^{-1}s^{-2}$ and, by Lemmas~\ref{Null-Euclidean bilinear} and~\ref{lem1-31-01-2021} (on null and quasi-null interactions) and \eqref{equa-31-12-20} or \eqref{equa-new-conditions-hstar} (bound on $h^{\star}$) we obtain 
$$
\aligned
G_3 
& := 
|\Qbb^{\star}[u]|_{p,k} + |\Pbb^{\star}[u]|_{p,k}
\lesssim  (1+|h^{\star} |_N)|\del u\del u|_N
\lesssim \sum_{p_1+p_2=p\atop k_1+k_2=k} |\del u|_{p_1,k_1} |\del u|_{p_2,k_2}
\\
& \lesssim |\del u|_{p,k} |\del u|_{[N/2]}\lesssim   (\epss+C_1\eps)  r^{-1}\crochet^{-\kappa}s^{\delta} |\del u|_{p,k}
\lesssim  \ell^{-\delta/2}   (\epss+C_1\eps)  r^{-1}\crochet^{-\kappa+\delta/2} |\del u|_{p,k}.
\endaligned
$$
We thus find 
$$
\aligned 
\|\crochet^{\kappa}J\zeta^{-1}\, G_3 \|_{L^2(\Mfar_{\ell,s})}
& \lesssim  \ell^{-\delta/2}  (\epss+C_1\eps) \| r^{-1}s \, \crochet^{\kappa}\zeta|\del u|_{p,k}\|_{L^2(\Mfar_{\ell,s})}
\lesssim   \ell^{-\delta/2}  (\epss+C_1\eps)  s^{-1} \, \Fenergy_{\kappa}^{\ME,p,k}(s,u).
\endaligned
$$
We thus conclude that
\begin{equation}\label{eq2-05-10-2022}
\aligned
&  
\|\crochet^{\kappa}J\zeta^{-1} |\Pbb^\star [u] |_{p,k}\|_{L^2(\MME_{\ell,s})}
+\|\crochet^{\kappa}J\zeta^{-1} |\Qbb^\star [u] |_{p,k}\|_{L^2(\MME_{\ell,s})}
\\
& \lesssim   (\ell^{-\delta/2}+\delta^{-2})(\epss + C_1\eps)s^{-1+2k_1\theta} \, \sum_{0\leq k_1\leq k} \Fenergy_{\kappa}^{\ME,p,k-k_1}(s,u) + (\epss+C_1\eps)^2s^{-1-\delta} + (\epss+C_1\eps) s^{-1/2}  { \nouveauS_{p,k}(s,u)}.
\endaligned
\end{equation}
At the end of this Subsection, we conclude the following estimate for Class A {by recalling Lemma~\ref{lemma-112}}. 
{  
The analysis of Class B metrics is postponed to Section~\ref{subsec1-03-10-2022}.}

\begin{proposition}[Sharp energy estimates for nonlinearities. { Class A metrics}]
When the reference metric belongs to Class A 
%%%% 
and for all $s \in [s_0, s_1]$ one has 
\begin{equation}\label{eq2-21-03-2021}
\aligned
 \|\crochet^{\kappa}J\zeta^{-1} |\Boxt_g u |_{p,k}\|_{L^2(\MME_s)}
& \lesssim   (\ell^{-\delta/2} + \delta^{-2})(\epss + C_1\eps) \sum_{0\leq k_1\leq k} s^{-1 + 2k_1\theta} \, \Fenergy_{\kappa}^{\ME,p,k-k_1}(s,u)
\\
& \quad 
 + (\epss+C_1\eps) s^{-1/2}   { \nouveauS_{p,k}(s,u)} 
 + \delta^{-1}(\epss + C_1\eps)^2s^{-1-\delta} + R^\err_\star(s), 
\endaligned
\end{equation}
where $R^\err_\star(s)$ denotes the Ricci upper bound in~\eqref{eq1-21-05-2021}. 
\end{proposition}

%------------------------------------------------------------------------------------------------------------

\subsection{Derivation of the sharp energy estimate: preliminaries}

To complete the derivation of our energy estimate, we need the following result.

\begin{lemma}[Comparison of energy functionals]
\label{lem1-28-03-2021} 
Under the conditions stated in Section~\ref{section-label-11-1}, { for both Class A and Class B}
% { Provided  $(\epss + C_1\eps) \ll 1$}, 
with $(\eta,c^*,w) = (\kappa,0,u)$ or $(\eta,c^*,w) = (\mu,c,\phi)$ one has 
$
\Eenergy_{g, \eta,c^*}^{\ME}(s,u) \geq (1/4) \, \Eenergy_{\eta,c^*}^{\ME}(s,u). 
$
\end{lemma}

\begin{proof} We rely on the identity 
$$
\aligned
\Eenergy_{\eta,c^*}^{\ME}(s,w) 
=\Eenergy_{g,\eta,c^*}^{\ME}(s,w)  
& + \int_{\MME_s}\crochet^{2\eta}h^{\N00} |\del_t w|^2 dx
- \int_{\MME_s}\crochet^{2\eta}h^{\N ab}\delsN_aw\delsN_bw
\\
&  -
\int_{\MME_s} \crochet^{2\eta}(2x^a/r) \Big(\frac{r\xi(s,r)}{(s^2+r^2)^{1/2}}-1\Big) \del_t w \big(h^{a0}\del_tw + h^{ab}\del_bw\big) \, dx.
\endaligned
$$
Concerning the second term in the right-hand side, we know that $h^{\N00}\leq 0$ in $\Mnear_{\ell,s}$. 
We then find 
$$
\int_{\MME_s}\crochet^{2\eta}h^{\N00} |\del_t w|^2 dx\leq  \int_{\Mfar_{\ell,s}}\crochet^{2\eta} h^{\N00} |\del_t w|^2dx
\lesssim (\epss + C_1\eps) \, \Eenergy_{\eta}^{\ME}(s,w) 
\lesssim (\epss + C_1\eps) \, \Eenergy_{\eta,c^*}^{\ME}(s,w), 
$$
where, for the second inequality, we used $\zeta\equiv 1$ in $\Mfar_{\ell,s}$. For the remaining terms, 
we recall \eqref{eq2-28-03-2021} together with 
$
|h|\lesssim (\epss + C_1\eps)r^{-\min(\lambda,\kappa)}s^{\delta}\leq (\epss + C_1\eps)r^{-1/2}, 
$
and  therefore 
$$
\Big|\int_{\MME_s}\crochet^{2\eta}h^{\N ab}\delsN_aw\delsN_bw dx \Big|\lesssim (\epss + C_1\eps) \, \Eenergy_{\eta,c^*}^{\ME}(s,w).
$$
Finally, we observe that $\big|\frac{r\xi(s,r)}{(s^2+r^2)^{1/2}}-1\big|\lesssim \zeta^2$, thus 
$$
\Big|\int_{\MME_s} \crochet^{2\eta}(2x^a/r) \Big(\frac{r\xi(s,r)}{(s^2+r^2)^{1/2}}-1\Big) \del_t w \big(h^{a0}\del_t w + h^{ab}\del_b w\big) \, dx\Big|
\lesssim (\epss + C_1\eps) \, \Eenergy_{\eta,c^*}^{\ME}(s,w).  
$$
Since $(\epss + C_1\eps) \ll 1$, we arrive at the desired conclusion.
\end{proof}

%-------------------------------------------------------------------------------

We also need the following inequality for ($\ord(Z) \leq p$ and $\rank(Z) \leq k$ and)
$(\eta,w) = (\kappa,u) $ or $(\mu,\phi)$:
\begin{equation}\label{eq9-27-03-2021}
\int_{\MME_s} |J G_{g, \eta}[Zw]| dx\lesssim
 (\epss+C_1\eps)   { \Big( \nouveauS_{0}(s,Zw)\Big)^2} 
+ 
(\epss + C_1\eps)s^{-1} \Eenergy_{\eta}^{\ME}(s,Zw) ,
\end{equation}
where $G_{g, \eta}[w]$ is defined in  \eqref{eq7-08-05-2020}.  
This can be checked from~\eqref{eq1-26-05-2021} and \eqref{eq1-17-07-2020} together with the following identity: 
$$
G_{g,\eta}[u] = \crochet^{2\eta}\Big(\del_th^{\N00} + 2\delsN_ah^{a0} - \frac{2x^b}{r}\delsN_ah^{ab}\Big)|\del_tu|^2 
+ 2\crochet^{2\eta}\delsN_ah^{ab}\del_tu\delsN_bu
-\crochet^{2\eta}\del_th^{ab}\delsN_au\delsN_bu.
$$

Recalling Proposition~\ref{prop energy-ici-exterior}, we also need the estimate 
$$
\aligned
& \int_{\MME_s}\eta \crochet^{2\eta-1} \aleph'(r-t) h^{\N 00}  |\del_t Zw|^2 \ Jdx
= \int_{\Mnear_{\ell,s}} + \int_{\Mfar_{\ell,s}} 
\eta\crochet^{2\eta-1} \aleph'(r-t) h^{\N 00}  |\del_t Zw|^2 \ Jdx 
\\
& \lesssim \int_{\Mfar_{\ell,s}} \crochet^{2\eta-1} \aleph'(r-t) h^{\N 00}  |\del_t Zw|^2 \ Jdx
\lesssim \ell^{-\delta}(\epss + C_1\eps)  s\int_{\Mfar_{\ell,s}}\crochet^{2\eta}r^{-1+\theta-\delta}\zeta^2|\del_t Zw|^2\ dx,
\endaligned
$$
where for the first inequality we applied \eqref{eq3-27-05-2020} and for the second inequality we used that $\ell r\lesssim \crochet$ in $\Mfar_{\ell,s}$. This leads us to, {for both Class A and Class B}, 
\begin{equation}\label{eq2-30-03-2021}
\eta\int_{\MME_s} \crochet^{2\eta-1} \aleph'(r-t) h^{\N 00}  |\del_t Zw|^2 \ Jdx\lesssim \ell^{-\delta}(\epss + C_1\eps)s^{-1} \Eenergy_{\eta}^{\ME}(s,Zw).
\end{equation}
%-----------------------------------------------------------------------------------------------------------

\subsection{Derivation of the sharp energy estimate for the metric components: Class A}

By combining Proposition~\ref{prop--eq1-27-03-2021}.
and \eqref{eq2-21-03-2021}, provided $\delta^{-1}\ell^{-1}(\epss + C_1\eps) \lesssim 1$, for 
all $\ord(Z) = p\leq N, \rank(Z)=k\leq p$ thanks to \eqref{eq2-20-03-2021}, { for Class A metrics only we find}
\begin{equation}\label{eq2-27-03-2021}
\aligned
& \|\crochet^{\kappa}J\zeta^{-1}\,\Boxt_g Z u_{\alpha\beta}\|_{L^2(\MME_s)}
\\
& \lesssim  \delta^{-1}(\ell^{-1} + \delta^{-2})(\epss + C_1\eps) \sum_{0\leq k_1\leq k} s^{-1 + 2k_1\theta} \, \Fenergy_{\kappa}^{\ME,p,k-k_1}(s,u)
+ (\epss+C_1\eps)s^{-1/2}   { \nouveauS_{p,k}(s,u)}  
\\
&+ \ell^{-1}\delta^{-1}(\epss + C_1\eps)^2s^{-1-\delta}
+ R^\err_\star(s). 
\endaligned
\end{equation}
Now we apply Proposition~\ref{prop energy-ici-exterior} on  \eqref{eq2-20-03-2021} with $(\eta,w,c^*) = (\kappa,u,0)$: 
\begin{equation}\label{eq3-27-03-2021}
\aligned
& \frac{d}{ds}\Eenergy_{g,\kappa}^{\ME}(s,Z u)  + \frac{d}{ds} \Eenergy^{\Lcal}_{g, c}(s,Z u;s_0)  
+  2 \kappa \int_{\MME_s}\big( g^{\N ab} \delsN_aZu \delsN_bZu 
\big) \crochet^{2\kappa-1} \aleph'(r-t) \ Jdx
\\
& \lesssim \big(\ell^{-\delta} + \delta^{-2}\big)(\epss + C_1\eps)s^{-1}\Eenergy_{\kappa}^{\ME}(s,Zu)
+ \big\|\zeta\crochet^{\kappa}\del_t Zu\|_{L^2(\MME_s)}\|\crochet^{\kappa}J\zeta^{-1} \Boxt_gZu \big\|_{L^2(\MME_s)}
\\
& \lesssim \delta^{-1}\big(\ell^{-1} + \delta^{-2}\big)(\epss + C_1\eps) 
\sum_{0\leq k_1\leq k} s^{-1+2k_1\theta} \, \Fenergy_{\kappa}^{\ME}(s,Zu) \,\Fenergy_{\kappa}^{\ME,p,k-k_1}(s,u) 
 + (\epss+C_1\eps)s^{-1/2}  { \nouveauS_{p,k}(s,u)}  
\, \Fenergy_{\kappa}^{\ME}(s,Zu)
\\
& \quad 
+ \ell^{-1}\delta^{-1}(\epss + C_1\eps)^2 s^{-1-\delta} \, \Fenergy_{\kappa}^{\ME}(s,Zu) 
+ R^\err_\star(s) \Fenergy_{\kappa}^{\ME}(s,Zu), 
\endaligned
\end{equation}
where for the first inequality we applied \eqref{eq9-27-03-2021} and \eqref{eq2-30-03-2021}. For the later one we used \eqref{eq2-27-03-2021}. 
Now let us focus our attention on the left-hand side. 
In the third term we observe that $\aleph'(r-t) \geq 0$ in $\MME_s$. Since $|h|\ll 1$ we have 
$
\sum_{a} |\delsN_a u|^2\lesssim g^{\N ab}\delsN_a u\delsN_b u
$
and this leads us to
$$
{ 0\leq \frac{d}{ds}{\mathscr G}_\kappa^{\ME,p,k}(s_0, s,u)}\lesssim \sum_{\ord(Z) \leq p\atop \rank(Z) \leq k}\int_{\MME_s}\crochet^{2\kappa -1} 
\aleph'(r-t) g^{\N ab}\delsN_a Zu\delsN_b Zu \, Jdx.
$$
We sum up \eqref{eq3-27-03-2021} for all $\ord(Z) \leq p\leq N$ and all $\rank(Z) \leq k\leq p$ and so, 
\begin{subequations}
\begin{equation}\label{eq2-28-05-2023}
\aligned
&\frac{d}{ds}\Eenergy_{g,\kappa}^{\ME,p,k}(s,u)  
+ \frac{d}{ds} \Eenergy^{\Lcal,p,k}_{g, c}(s,u;s_0)  
+ { \frac{d}{ds}{\mathscr G}_\kappa^{\ME,p,k}(s_0, s,u)}
\\
& \lesssim  \delta^{-1}\big(\ell^{-1} + \delta^{-2}\big)(\epss + C_1\eps) \sum_{0\leq k_1\leq k}
s^{-1+2k_1\theta} \, \Fenergy_{\kappa}^{\ME,p,k}(s,u) \Fenergy_{\kappa}^{\ME,p,k-k_1}(s,u)
\\
&\quad + (\epss+C_1\eps) { \frac{d}{ds}{\mathscr G}_\kappa^{\ME,p,k}(s_0, s,u)}
+   (\epss+C_1\eps) s^{-1} { \Eenergy_{\kappa}^{\ME,p,k}(s,Zu)}
\\
& \quad 
+ \ell^{-1}\delta^{-1}(\epss + C_1\eps)^2 s^{-1-\delta} \, \Fenergy_{\kappa}^{\ME,p,k}(s,u)
+ R^\err_\star(s) \Fenergy_{\kappa}^{\ME,p,k}(s,u).
\endaligned
\end{equation}
Here we need to point out that, thanks to \eqref{eq2-03-02-2020},
\begin{equation}\label{eq4-27-03-2021}
\aligned
\Big(  { \nouveauS_{p,k}(s,u)}  \Big)^2  
& \lesssim  
\sum_{\ord(Z)\leq p\atop \rank(Z)\leq k}\int_{\MME_s}\crochet^{2\kappa-1} |\delsN Z u|^2 J \, dx
 + s^{-1}\Eenergy_{\kappa}^{\ME,p,k}(s,u) 
\\
& \lesssim 
\frac{d}{ds} \Eenergy^{\Lcal,p,k}_{g, c}(s,u;s_0)+ s^{-1}\Eenergy_{\kappa}^{\ME,p,k}(s,u). 
 \endaligned
\end{equation}
\end{subequations}
Given that $\epss + C_1\eps$ sufficiently small, 
{ 
the second term in the right-hand-side cancels out thanks to the third term in the left-hand side.} Now, recalling \eqref{eq8-27-03-2021} 
{ (for the evolution of the energy contribution $\Eenergy^{\Lcal,p,k}_{g, c}$)}
and \eqref{eq3-27-05-2020} 
{ (that is, the light-bending condition)}
we have 
$\frac{d}{ds} \Eenergy^{\Lcal}_{g, c}(s,Z u;s_0) \geq 0$. In view of Lemma~\ref{lem1-28-03-2021} we have
\begin{equation}\label{eq5-27-03-2021}
\aligned
\frac{d}{ds} \, \Fenergy_{g,\kappa}^{\ME,p,k}(s,u) 
 \leq K_0 \sum_{0\leq k_1\leq k}
s^{-1+2k_1\theta} \, \Fenergy_{g,\kappa}^{\ME,p,k-k_1}(s,u)
+ C(N) \ell^{-1}\delta^{-1}(\epss + C_1\eps)^2 s^{-1-\delta} 
+ C(N) R^\err_\star(s), 
\endaligned
\end{equation}
where $C_{R^{\star}}$ was introduced in \eqref{eq1-21-05-2021}, $K_0 = C(N) \delta^{-1}(\ell^{-1} + \delta^{-2})(\epss + C_1\eps)$ and $C(N)$ is a constant determined by $N$. We also find 
\begin{equation}\label{eq6-27-03-2021}
\aligned
 \frac{d}{ds}\Eenergy^{\Lcal,p,k}_{g, c}(s, u;s_0) 
& \lesssim  \delta^{-1}\big(\ell^{-1} + \delta^{-2}\big)(\epss + C_1\eps) \sum_{0\leq k_1\leq k}
s^{-1+2k_1\theta} \, \Fenergy_{\kappa}^{\ME,p,k}(s,u) \Fenergy_{\kappa}^{\ME,p,k-k_1}(s,u)
\\
& \quad + \ell^{-1}\delta^{-1}(\epss + C_1\eps)^2 s^{-1-\delta} \, \Fenergy_{\kappa}^{\ME,p,k}(s,u)
+ R^\err_\star(s) \Fenergy_{\kappa}^{\ME,p,k}(s,u).
\endaligned
\end{equation} 

At this stage, we can focus our attention on \eqref{eq5-27-03-2021}. Gronwall inequality leads us to (for { $K_0\leq \delta/2$ and $s_0\geq 1$}) 
$$
\aligned
\Fenergy_{g,\kappa}^{\ME,p,k}(s,u)
& \leq  s^{K_0} \, \Fenergy_{g,\kappa}^{\ME,p,k}(s_0,u)
+ K_0s^{K_0}\sum_{1\leq k_1\leq k}\int_{s_0}^s {s'}^{-1+2k_1\theta-K_0} \, \Fenergy_{g,\kappa}^{\ME,p,k-k_1}(s',u)ds'
\\
& \quad  + C(N)\ell^{-1}\delta^{-2}s^{K_0}(\epss + C_1\eps)^2
+ C(N)s^{K_0}\int_{s_0}^sR^{\err}_{\star}(s'){s'}^{-K_0} \, ds'
\\
& \leq  s^{K_0} \, \Fenergy_{g,\kappa}^{\ME,p,k}(s_0,u)
+ K_0s^{K_0}\sum_{1\leq k_1\leq k}\int_{s_0}^s {s'}^{-1+2k_1\theta-K_0} \, \Fenergy_{g,\kappa}^{\ME,p,k-k_1}(s',u)ds'
 + C(N)(\ell^{-1}\delta^{-2} + C_{R^{\star}})(\epss + C_1\eps)^2, 
\endaligned
$$
where for the last inequality we used \eqref{eq1-21-05-2021}. Observe that the second term in the right-hand side does not exist when $k=0$. Fixing $p=N$ and $k=0$ and imposing the smallness condition {(recall that $K_0 = C(N)\delta^{-1}(\ell^{-1} + \delta^{-2})(\epss + C_1\eps)$)}
\begin{equation}\label{eq1-13-10-2021}
C_0\eps + C(N)(\ell^{-1}\delta^{-2}+C_{R^{\star}})(\epss + C_1\eps)^2 + \frac{K_0}{8\theta}  (\epss+C_1\eps) \leq (1/4) (\epss + C_1\eps), 
\end{equation} 
we arrive at 
$$
\Fenergy_{g,\kappa}^{\ME,N,0}(s,u)
\leq s^{K_0} \, \Fenergy_{g,\kappa}^{\ME,N,0}(s_0,u)  
+ C(N)(\ell^{-1}\delta^{-2} + C_{R^{\star}})(\epss + C_1\eps)^2 
\leq (1/4)  (\epss+C_1\eps)  s^{K_0}.
$$
Then, by induction we obtain
$
\Fenergy_{g,\kappa}^{\ME,N,k}(s,u) \leq (1/4)  (\epss+C_1\eps)  s^{K_0 + 2k\theta}
$
and finally by Lemma~\ref{lem1-28-03-2021} concerning the comparison of energy functionals, we arrive at the  conclusion 
\begin{equation}
\label{equation-1633} 
\Fenergy_{\kappa}^{\ME,N}(s,u) \leq 2 \, \Fenergy_{g,\kappa}^{\ME,N}(s,u) \leq (1/2)  (\epss+C_1\eps) s^{K_0 + 2N\theta}.
\end{equation}
When {  
$K_0 + 2N\theta\leq \delta$}, we have thus improved the energy bounds for $u$, as stated earlier in Proposition~\ref{proposition-section16-metric}. 
Furthermore, provided that $K_0\leq N\theta$,  we have 
\begin{equation}\label{eq1-28-05-2023}
 \int_{s_0}^{s} \Big( { \nouveauS_{p,k}(s,u)} \Big)^2 \, d\tau 
 = 
 \int_{s_0}^{s} { \|J^{1/2}\crochet^{\kappa-1/2}|\delsN u|_{p,k}\|_{L^2(\MME_\tau)}^2} \, d\tau
 \lesssim (\epss+C_1\eps)^2s^{2K_ + 4N\theta}
 \lesssim (\epss+C_1\eps)^2s^{6N\theta}.
\end{equation}
To see this, we recall \eqref{eq2-28-05-2023} with $(\epss+C_1\eps)$ sufficiently small, and write 
$$
\aligned
&\frac{d}{ds}\Eenergy_{g,\kappa}^{\ME,p,k}(s,u)  
+ \frac{d}{ds} \Eenergy^{\Lcal,p,k}_{g, c}(s,u;s_0)  
+ { \frac{d}{ds}{\mathscr G}_\kappa^{\ME,p,k}(s_0, s,u)}
\\
& \lesssim  \delta^{-1}\big(\ell^{-1} + \delta^{-2}\big)(\epss + C_1\eps) \sum_{0\leq k_1\leq k}
s^{-1+2k_1\theta} \, \Fenergy_{\kappa}^{\ME,p,k}(s,u) \Fenergy_{\kappa}^{\ME,p,k-k_1}(s,u)
+ R^\err_\star(s) \Fenergy_{\kappa}^{\ME,p,k}(s,u).
\endaligned
$$
Then we only need to integrate the above inequality on $[s_0,s]$ and substitute $\Fenergy_{g,\kappa}^{\ME,N,k}(s,u) \leq (1/4)  (\epss+C_1\eps)  s^{K_0 + 2k\theta}$ into the right-hand-side.
}

%--------------------------------------------------------------

\subsection{Derivation of the sharp energy estimate for the metric components: Class B}
\label{subsec1-03-10-2022}

The estimate in Class B is more involved, especially since \eqref{eq2-21-03-2021} is no longer valid. However, the only problematic term is $W^{\textbf{linear}}$, which was (partially) investigated in Section~\ref{section--125}. 
{
For clarity in the presentation, we
introduce the following classification of the terms arising in the right-hand side of \eqref{eq 1 13-01-2019}.
}
$$
\Boxt_g u_{\alpha\beta} = T_{\alpha\beta}[u] + W^{\textbf{linear}}_{\alpha\beta}[u], 
$$
where 
$$
\aligned
T_{\alpha\beta}[u] : &=  \Pbb^\star_{\alpha\beta}[u] + \Qbb^{\star}_{\alpha\beta}[u] + W^\super_{\alpha\beta},
\\
W^{\textbf{linear}}_{\alpha\beta}[u] : &=   \Fbb_{\alpha\beta}(g^\star, g^\star;\del h^{\star}, \del u) + \Fbb_{\alpha\beta}(g^\star, g^\star;\del u, \del h^{\star}) - u^{\mu\nu} \, \del_{\mu} \del_{\nu} h^{\star}_{\alpha\beta}.
\endaligned
$$
In view of  \eqref{eq1-22-03-2021}, {\eqref{eq1-01-10-2022},} and \eqref{eq2-05-10-2022}, we have 
\begin{equation}\label{eq3-05-10-2022}
\aligned 
\|\crochet^{\kappa}J\zeta^{-1}|T[u]|_{p,k}\|_{L^2(\MME_s)} 
& \lesssim 
(\ell^{-\delta/2} + \delta^{-2})(\epss + C_1\eps) \sum_{0\leq k_1\leq k} s^{-1 + 2k_1\theta} \, \Fenergy_{\kappa}^{\ME,p,k-k_1}(s,u)
\\
&\quad+ (\epss+C_1\eps) s^{-1/2} { \nouveauS_{p,k}(s,u)} 
  + \delta^{-1}(\epss + C_1\eps)^2s^{-1-\delta} + R^\err_\star(s).
\endaligned
\end{equation}
Substituting the above bound into Proposition~\ref{prop--eq1-27-03-2021}, we obtain
\begin{equation}\label{eq1-06-10-2022}
\aligned
&
\|\crochet^{\kappa}J\zeta^{-1}[Z,h^{\alpha\beta}\del_{\alpha}\del_{\beta}]u\|_{L^2(\MME_s)}
 \lesssim
\delta^{-1}(\ell^{-1}+\delta^{-2})(\epss + C_1\eps) \sum_{0\leq k_1\leq k}s^{-1+2k_1\theta} \, \Fenergy_{\kappa}^{\ME,p,k-k_1}(s,u) 
\\
&\quad + (\epss+C_1\eps) s^{-1/2}  { \nouveauS_{p,k}(s,u)} 
 + \|W^{\textbf{linear}}[u]\|_{L^2(\MME_s)} 
  + \ell^{-1}\delta^{-1}(\epss + C_1\eps)^2s^{-1-\delta} + R^\err_\star(s).
\endaligned
\end{equation}
Then we establish the following result.

\begin{lemma} 
For $\ord(Z) = p\leq N, \rank(Z) = k$, thanks to the condition 
\begin{equation}\label{eq4-06-10-2022} 
R^\err_\star(s)\lesssim C_{R^{\star}}\epss^2 s^{-1-\delta}, 
\end{equation}
one has 
\begin{equation}\label{eq3-06-10-2022}
\aligned
\int_{s_0}^s\int_{\MME_\tau}\crochet^{2\kappa}|\del_t Zu\, \Boxt_g Zu|J\ dxd\tau
& \lesssim  (\ell^{-1}\delta^{-2} + C_{R^{\star}})(\epss+C_1\eps)^3 + \delta^{-1}\epss {\mathscr G}_\kappa^{\ME,p,k}(s_0, s,u)
\\
& \quad
+ \delta^{-1}(\ell^{-1}+\delta^{-2}) \big(\epss + C_1\eps \big) \sum_{0\leq k_1\leq k}\int_{s_0}^s\tau^{-1+2k_1\theta} 
\Eenergy_{\kappa}^{\ME,p,k-k_1}(\tau,u) d\tau.
\endaligned
\end{equation}
\end{lemma}

\begin{proof}
In view of  \eqref{eq2-20-03-2021}, we have
\begin{equation}\label{eq2-06-10-2022}
\aligned
&
\int_{s_0}^s\int_{\MME_\tau}\crochet^{2\kappa}|\del_t Zu \,\Boxt_g Zu|J dxdt 
 \lesssim  \int_{s_0}^s\|\crochet^{\kappa}\zeta|\del u|_{p,k}\|_{L^2(\MME_{\tau})} 
\|J\zeta^{-1}\crochet^{\kappa}|\Boxt_g u|_{p,k}\|_{L^2(\MME_{\tau})}d\tau
\\
& \quad + \int_{s_0}^s\|\crochet^{\kappa}\zeta|\del u|_{p,k}\|_{L^2(\MME_\tau)} \big\|J\zeta^{-1}\crochet^{\kappa}|[Z,h^{\mu\nu}\del_{\mu}\del_{\nu}]u |_{p,k}\big\|_{L^2(\MME_{\tau})}d\tau.
\endaligned
\end{equation}
For the second term, we apply \eqref{eq1-06-10-2022} and obtain
$$
\aligned
& \int_{s_0}^s\|\crochet^{\kappa}\zeta|\del u|_{p,k}\|_{L^2(\MME_\tau)} \|J\zeta^{-1}\crochet^{\kappa}|[Z,h^{\mu\nu}\del_{\mu}\del_{\nu}]u |_{p,k}\|_{L^2(\MME_{\tau})}d\tau
\\ 
& \lesssim  \int_{s_0}^s \|\crochet^{\kappa}\zeta|\del u|_{p,k}\|_{L^2(\MME_{\tau})}
\|J\zeta^{-1}\crochet^{\kappa} W^{\textbf{linear}}|_{p,k}\|_{L^2(\MME_{\tau})}d\tau
\\
& \quad + \delta^{-1}(\ell^{-1}+\delta^{-2})(\epss + C_1\eps) \sum_{0\leq k_1\leq k}\int_{s_0}^s\tau^{-1+2k_1\theta} 
\|\crochet^{\kappa}\zeta|\del u|_{p,k}\|_{L^2(\MME_{\tau})} \Fenergy_{\kappa}^{\ME,p,k-k_1}(\tau,u) d\tau
\\
& {
\quad + (\epss+C_1\eps)\int_{s_0}^s s^{-1/2}{ \nouveauS_{p,k}(\tau,u)}\|\crochet^{\kappa}\zeta|\del u|_{p,k}\|_{L^2(\MME_\tau)} d\tau }
\\
& \quad + (\ell^{-1}\delta^{-1} + C_{R^{\star}})(\epss + C_1\eps)^2 \int_{s_0}^s\tau^{-1-\delta}\|\crochet^{\kappa}\zeta|\del u|_{p,k}\|_{L^2(\MME_{\tau})} 
\\
& \lesssim  \int_{s_0}^s \|\crochet^{\kappa}\zeta|\del u|_{p,k}\|_{L^2(\MME_{\tau})}
\|J\zeta^{-1}\crochet^{\kappa} W^{\textbf{linear}}|_{p,k}\|_{L^2(\MME_{\tau})}d\tau
+{ (\epss+C_1\eps){\mathscr G}_\kappa^{\ME,p,k}(s_0, s,u)}
\\
& \quad +  \delta^{-1}(\ell^{-1}+\delta^{-2})(\epss + C_1\eps) \sum_{0\leq k_1\leq k}\int_{s_0}^s\tau^{-1+2k_1\theta} 
\Eenergy_{\kappa}^{\ME,p,k-k_1}(\tau,u) d\tau 
+ (\ell^{-1}\delta^{-2} + C_{R^{\star}})(\epss + C_1\eps)^3,
\endaligned
$$
where for the last inequality we used  { \eqref{eq4-27-03-2021} (third term) and}  the fact
$$
\aligned
&\ell^{-1}\delta^{-1}(\epss + C_1\eps)^2\int_{s_0}^s\tau^{-1-\delta}\|\crochet^{\kappa}\zeta|\del u|_{p,k}\|_{L^2(\MME_{\tau})} d\tau
\\
& \lesssim  
\ell^{-1}\delta^{-1}\int_{s_0}^s(\epss+C_1\eps)^{3/2}\tau^{-1/2-\delta}\, 
(\epss+C_1\eps)^{1/2} \tau^{-1/2}\|\crochet^{\kappa}\zeta|\del u|_{p,k}\|_{L^2(\MME_{\tau})}d\tau
\\
& \lesssim  \ell^{-1}\delta^{-1}(\epss+C_1\eps)^3\int_{s_0}^s\tau^{-1-2\delta}d\tau 
+ \ell^{-1}\delta^{-1}(\epss+C_1\eps)\int_{s_0}^s \tau^{-1}\|\crochet^{\kappa}\zeta|\del u|_{p,k}\|_{L^2(\MME_{\tau})}^2d\tau
\\
& \lesssim  \ell^{-1}\delta^{-2}(\epss+C_1\eps)^3 + \ell^{-1}\delta^{-1}(\epss+C_1\eps)\int_{s_0}^s\tau^{-1}\Eenergy_{\kappa}^{\EM,p,k}(\tau,u) \, d\tau.
\endaligned
$$
Then we apply \eqref{eq1-05-10-2022}, and obtain
$$
\aligned
& \int_{s_0}^s\|\crochet^{\kappa}\zeta|\del u|_{p,k}\|_{L^2(\MME_\tau)} \|J\zeta^{-1}\crochet^{\kappa}|[Z,h^{\mu\nu}\del_{\mu}\del_{\nu}]u |_{p,k}\|_{L^2(\MME_{\tau})}d\tau
\\
& \lesssim (\ell^{-1}\delta^{-2} + C_{R^{\star}})(\epss+C_1\eps)^3 + \delta^{-1}{   (\epss+C_1\eps)} {\mathscr G}_\kappa^{\ME,p,k}(s_0, s,u)
\\
& \quad + \delta^{-1}(\ell^{-1}+\delta^{-2})(\epss + C_1\eps) \sum_{0\leq k_1\leq k}\int_{s_0}^s\tau^{-1+2k_1\theta} 
\Eenergy_{\kappa}^{\ME,p,k-k_1}(\tau,u) d\tau.
\endaligned
$$
For the first term in the right-hand side of \eqref{eq2-06-10-2022}, we apply directly \eqref{eq3-05-10-2022} together with \eqref{eq1-05-10-2022}.
\end{proof}

{ To the equation \eqref{eq2-20-03-2021} we apply the weighted energy equation derived in 
Proposition~\ref{prop energy-ici-exterior} 
with the choice of parameter $(\eta,w,c^*) = (\kappa,u,0)$.}
Again, by recalling \eqref{eq8-27-03-2021} and \eqref{eq3-27-05-2020}, we have 
$\frac{d}{ds} \Eenergy^{\Lcal}_{g, c}(s,Z u;s_0) \geq 0$. Furthermore, by \eqref{eq9-27-03-2021} and \eqref{eq2-30-03-2021},
$$
\aligned
&\frac{d}{ds} \Eenergy_{g,\kappa}^{\ME}(s,Zu) 
+  2 \kappa  \int_{\MME_s} g^{\N ab} \delsN_aZu\delsN_bZu \crochet^{2\kappa-1} \aleph'({ r-t}) \, Jdx
\\
&\lesssim 
(\ell^{-\delta} + \delta^{-2})(\epss + C_1\eps)s^{-1} \Eenergy_{\kappa}^{\ME}(s,Zu) 
+ \int_{\MME_\tau}\crochet^{2\kappa}|\del_t Zu\, \Boxt_g Zu|J\ dx.
\endaligned
$$
Integrating the above inequality on $[s_0,s]$ and apply \eqref{eq3-06-10-2022}, we obtain
\begin{equation}\label{eq5-06-10-2022}
\aligned
& \Eenergy_{g,\kappa}^{\ME}(s,Zu) + 2\kappa \int_{s_0}^s\int_{\MME_\tau} g^{\N ab} \delsN_aZu\delsN_bZu \crochet^{2\kappa-1} \aleph'({ r-t}) \, Jdxd\tau
\\
& \lesssim  \Eenergy_{g,\kappa}^{\ME}(s_0, Zu) + (\ell^{-1}\delta^{-2} + C_{R^{\star}})(\epss+C_1\eps)^3 + \delta^{-1}{   (\epss+C_1\eps)} {\mathscr G}_\kappa^{\ME,p,k}(s_0, s,u)
\\
& \quad + \delta^{-1}(\ell^{-1}+\delta^{-2})(\epss + C_1\eps) \sum_{0\leq k_1\leq k}\int_{s_0}^s\tau^{-1+2k_1\theta} 
\Eenergy_{\kappa}^{\ME,p,k-k_1}(\tau,u) d\tau, 
\endaligned
\end{equation}
and we focus first on the left-hand side. Recalling \eqref{eq:weight}, we observe that $\aleph'(r-t)\geq 1$ in $\MME_{[s_0,s_1]}$
and, by \eqref{eq1-30-05-2020}, we obtain 
$
g^{\N ab}\delsN_aZu\delsN_bZu\geq (1/2) \, |\delsN Zu|^2.
$
After recalling that $\kappa\geq 1/2$ this leads us to 
\begin{equation}\label{eq6-06-10-2022}
2 \kappa  \int_{\MME_s} g^{\N ab} \delsN_aZu\delsN_bZu \crochet^{2\kappa-1} \aleph'({ r-t}) \, Jdx\geq 
\frac{1}{2}\int_{\MME_s}\crochet^{2\kappa-1}|\delsN Zu|^2Jdx.
\end{equation}

Now we sum up \eqref{eq5-06-10-2022} for all $\ord(Z)\leq p$ and $\rank(Z)\leq k$. Recalling
\eqref{equa-new-spacetime-bound} and the above bound \eqref{eq6-06-10-2022}, we obtain {   (similarly as in  \eqref{eq5-27-03-2021})}
$$
\aligned
& 
\Eenergy_{g,\kappa}^{\ME,p,k}(s,u) + \frac{1}{2}{\mathscr G}_\kappa^{\ME,p,k}(s_0, s,u)
\\
& \leq \Eenergy_{g,\kappa}^{\ME,p,k}(s_0,u) 
+ C(N)(\ell^{-1}\delta^{-2} + C_{R^{\star}})(\epss+C_1\eps)^3 + C(N)\delta^{-1}{   (\epss+C_1\eps)} {\mathscr G}_\kappa^{\ME,p,k}(s_0, s,u)
\\
&\quad + C(N)\delta^{-1}(\ell^{-1}+\delta^{-2})(\epss + C_1\eps) \sum_{0\leq k_1\leq k}\int_{s_0}^s\tau^{-1+2k_1\theta} 
\Eenergy_{\kappa}^{\ME,p,k-k_1}(\tau,u) d\tau,
\endaligned
$$
where $C(N)$ is a { constant depending upon $N$. }
For {   $(\epss+C_1\eps)$} sufficiently small, the third term in the right-hand side can be absorbed by the second term in the left-hand side. Thanks to \eqref{eq5-03-05-2020-00}, we then have 
\begin{equation}\label{eq7-06-10-2022}
\aligned
\Eenergy_{g,\kappa}^{\ME,p,k}(s,u) { + \frac{1}{2} {\mathscr G}_\kappa^{\ME,p,k}(s_0, s,u)}
& \leq (C_0\eps)^2s_0^{2\delta}
+ C(N)(\ell^{-1}\delta^{-2} + C_{R^{\star}})(\epss+C_1\eps)^3 
\\
& \quad + C(N)\delta^{-1}(\ell^{-1}+\delta^{-2})(\epss + C_1\eps) \sum_{0\leq k_1\leq k}\int_{s_0}^s\tau^{-1+2k_1\theta} 
\Eenergy_{\kappa}^{\ME,p,k-k_1}(\tau,u) d\tau.
\endaligned
\end{equation}
Especially when $k=0$, we obtain 
\begin{equation}\label{eq8-06-10-2022}
\aligned
\Eenergy_{g,\kappa}^{\ME,p,0}(s,u) { + \frac{1}{2}{\mathscr G}_\kappa^{\ME,p,k}(s_0, s,u)}
& \leq (C_0\eps)^2s_0^{2\delta} 
+ C(N)(\ell^{-1}\delta^{-2} + C_{R^{\star}})(\epss+C_1\eps)^3 
\\
& \quad + C(N)\delta^{-1}(\ell^{-1}+\delta^{-2})(\epss + C_1\eps)\int_{s_0}^s\tau^{-1} 
\Eenergy_{\kappa}^{\ME,p,0}(\tau,u) d\tau.
\endaligned
\end{equation} 
We apply Gronwall inequality to \eqref{eq8-06-10-2022} and obtain
$$
\Eenergy_{g,\kappa}^{\ME,p,0}(s,u){ + \frac{1}{2}{\mathscr G}_\kappa^{\ME,p,0}(s_0, s,u)} \leq (C_0\eps)^2s_0^{2\delta} s^{K_0}
+ C(N)(\ell^{-1}\delta^{-2} + C_{R^{\star}})(\epss+C_1\eps)^3 s^{K_0}
$$
with $K_0 :=  C(N)\delta^{-1}(\ell^{-1}+\delta^{-2})(\epss + C_1\eps)$. 
{  
Assuming now that the constants are chosen so that } 
\begin{equation}\label{eq9-06-10-2022}
(C_0\eps)^2s_0^{2\delta} + C(N)(\ell^{-1}\delta^{-2} + C_{R^{\star}})(\epss+C_1\eps)^3 
+ \frac{K_0}{32\theta}(\epss+C_1\eps)^2 \leq \frac{1}{16}(\epss+C_1\eps)^2, 
\end{equation}
{  
we proceed by induction and conclude that}
$$
\aligned
&
\Eenergy_{g,\kappa}^{\ME,p,k}(s,u) { + \frac{1}{2}{\mathscr G}_\kappa^{\ME,p,k}(s_0, s,u)}
\\
& \leq  (C_0\eps)^2s_0^{2\delta} s^{K_0 + 2k\theta}
+ C(N)(\ell^{-1}\delta^{-2} + C_{R^{\star}})(\epss+C_1\eps)^3 s^{K_0+2k\theta} + \frac{K_0}{32\theta}(\epss+C_1\eps)^2 s^{K_0+2k\theta}
\\
& \leq  \frac{1}{16}(\epss+C_1\eps)^2s^{K_0+2k\theta}.
\endaligned
$$
Thus by Lemma~\ref{lem1-28-03-2021}, and provided $K_0+2N\theta\leq 2\delta$, we arrive at 
\begin{equation}\label{eq2-25-05-2023}
\Eenergy_{\kappa}^{\ME,N}(s,u) \leq 4\Eenergy_{g,\kappa}^{\ME,N}(s,u)\leq \frac{1}{4}(\epss+C_1\eps)^2s^{2\delta}, 
\end{equation}
which leads us to the conclusion \eqref{eq11-06-10-2022} stated on $\Fenergy_{\kappa}^{\ME,N}(s,u)$. {
 We point out that \eqref{eq1-28-05-2023} also holds in this case.}

%-------------------------------------------------------------------------------------------------------------------------------------------

\subsection{ Improved Sobolev pointwise decay}

{

To complete this section, we apply the improved energy estimate (deduced from \eqref{equation-1633} for Class~A metrics
 and \eqref{eq2-25-05-2023} for Class~B metrics), that is, 
\begin{equation}\label{eq1-25-05-2023}
\Eenergy_{g,\kappa}^{\ME,p,k}(s,u)\lesssim (\epss+C_1\eps)^2s^{K_0+2k\theta}, 
\end{equation}
together with the Sobolev inequality presented in Proposition~\ref{lem 2 d-e-I}. Provided that $\epss+C_1\eps\ll N$, we obtain
}
\begin{equation}\label{eq3-25-05-2023}
|\del u|_{N-3} + |\delsN u|_{N-3}\lesssim (\epss+C_1\eps)s^{3N\theta}, 
\end{equation}
{
which is our claim in Section~\ref{Sharpdecay-II}.
}

%==============================================================================================

\section{Improved energy estimate for the Klein-Gordon field} 
\label{sec-closing-bootstrap}

\subsection{Purpose of this section}

We now turn our attention to the Klein-Gordon component and handle the energy at the highest-order, as follows. 

\begin{proposition}[Improved energy estimates for the Klein-Gordon field]
\label{proposition-section17-matter} 
Under the bootstrap assumptions and conditions stated in Section~\ref{section-label-11-1}, 
the matter field satisfies 
$
\Fenergy_{\mu,c}^{\ME,N}(s,\phi) 
\leq  {C_1 \over 2} \, \eps \,  s^{1+\delta}
$
for all $s \in [s_0, s_1]$. 
\end{proposition}

%---------------------------------------------------------------------------------------------------------------

\subsection{Sharp energy estimates for the commutators}

\paragraph{Easy terms.}

We rely on Proposition~\ref{prop1-28-03-2021} which provides us with a control of the quasi-linear commutators. Among the terms in the right-hand side of \eqref{eq6-14-03-2021}, dealing with $W_{p,k}^\hard$ is the most challenging task. We find it convenient to treat first
$W_{p,k}^\easy$ and we claim that  
\begin{equation}\label{eq1-28-03-2021}
\aligned
\|\crochet^{\mu}J\zeta^{-1}\, W_{p,k}^\easy  \|_{L^2(\MME_s)}
& \lesssim
 (\epss+C_1\eps)s^{-1}\Fenergy_{\mu,c}^{\ME,p,k}(s,\phi) (\epss+C_1\eps)s^{-3/2+\delta}
  { \nouveauS_{p-1,k}(s,u)}
  %%% 
 +{   \delta^{-1}}(\epss + C_1\eps)^2s^{-1-\delta}. 
\endaligned
\end{equation}
{   Here, we only rely on the Sobolev decay. In the expression of $ W_{p,k}^\easy$, in the first sum the term involving $\frac{\la r-t\ra}{r}$ is a null term; the last term is of cubic in nature, and the second term is similar to the third term.}
Therefore it is sufficient to analyze the { following two terms:} 
$$
|H|_{p_1+1} |\del \phi|_{p-p_1}, \quad |\del H^{\N00}|_{p_1-1,k_1}|\del\del\phi|_{p_2,k_2} 
\quad  \text{ with } p_1+p_2=p,\quad k_1+k_2=k.
$$
{ For the first term,} on one hand, when $p_1\leq N-4$, we recall the third inequality in \eqref{eq1-30-05-2020} and, 
{ thanks to $\min(\lambda,\kappa) \geq 1/2+(3/2) \delta$}, 
$$
\aligned
\|\crochet^{\mu}J\zeta^{-1}\, r^{-1} |H|_{p_1+1} |\del\phi|_{p-p_1}\|_{L^2(\MME_s)}
& \lesssim  \delta^{-1}(\epss + C_1\eps)s^{-1-2\min(\lambda,\kappa)+\delta}\|\crochet^{\mu}\zeta |\del \phi|_{p-p_1}\|_{L^2(\MME_s)}
\\
& \lesssim \delta^{-1}(\epss + C_1\eps)^2s^{-2\min(\lambda,\kappa)+2\delta}
\lesssim \delta^{-1}(\epss + C_1\eps)^2s^{-1-\delta}. 
\endaligned
$$
% The first term in $W_{p,k}^\easy$ is and we omit the details.
On the other hand, when $p_1\geq N-3$, that is, 
$p-p_1\leq 3\leq N-10$, we apply Lemma~\ref{lemma-111} (lower-order) to the term $|\del \phi|_{p-p_1}$ and, 
{ since $\min(\lambda,\kappa,\mu) \geq 2\delta$}, 
$$
\aligned
& \|\crochet^{\mu}J\zeta^{-1}\, r^{-1} |H|_{p_1+1} |\del\phi|_{p-p_1}\|_{L^2(\MME_s)}
\\
& \lesssim  s^{-1}\|| h^{\star} |_{p_1+1}\crochet^{\mu}\zeta|\del\phi|_{p-p_1}\|_{L^2(\MME_s)}
+ s^{-1}\||u|_{p_1+1}\crochet^{\mu}\zeta|\del\phi|_{p-p_1}\|_{L^2(\MME_s)}
\\
& \lesssim  \epss s^{-1-2\lambda} \, \Fenergy^N_{\mu,c}(s,\phi) 
+ (\epss+C_1\eps) s^{-1+2\delta}\| r^{-2}\crochet^{2-\kappa}\, \crochet^{\kappa-1} |u|_N\|_{L^2(\MME_s)} 
\\
& \quad +  (\epss+C_1\eps) s^{-1+2\delta}\| r^{-1-\lambda}\crochet^{1-\kappa}\, \crochet^{\kappa-1} |u|_N\|_{L^2(\MME_s)}
  \lesssim  \delta^{-1} (\epss+C_1\eps)^2s^{-1-\delta}. 
\endaligned
$$

{   For the second term, we rely on the wave gauge condition. When $p_1-1\leq N-4$, \eqref{eq1-17-07-2020} yields us  
$$
\aligned
&
\|\crochet^{\mu}J\zeta^{-1}|\del H^{\N00}|_{p_1-1,k_1}|\del\del\phi|_{p_2,k_2}\|_{L^2(\MME_s)}
\lesssim
(\epss+C_1\eps)s^{1+\delta}\|r^{-1-\min(\lambda,\kappa)}\crochet^{\mu}\zeta|\del\del\phi|_{p,k}\|_{L^2(\MME_s)}
\\
& \lesssim \delta^{-1}(\epss+C_1\eps)C_1\eps s^{-1-2\min(\lambda,\kappa)+2\delta}\lesssim \delta^{-1}(\epss+C_1\eps)^2s^{-1-\delta}.
\endaligned
$$
When $p_1-1\geq N-3$, we closely follow the calculation in the proof of Lemma~\ref{lemma--18-sept-22-000} and obtain 
$$
\aligned
&\|\crochet^{\mu}J\zeta^{-1}|\del H^{\N00}|_{p_1-1,k_1}|\del\del\phi|_{p_2,k_2}\|_{L^2(\MME_s)}
\\
& \lesssim
\|\crochet^{\mu}J\zeta^{-1}|\del h^{\star\N00}|_{p_1-1,k_1}|\del\del\phi|_{p_2,k_2}\|_{L^2(\MME_s)}
+\|\crochet^{\mu}J\zeta^{-1}|\del u^{\N00}|_{p_1-1,k_1}|\del\del\phi|_{p_2,k_2}\|_{L^2(\MME_s)}
 =: G_1 + G_2.
\endaligned
$$
Here, thanks to \eqref{eq1-02-10-2022}, $G_1$ is bounded by 
$$
G_1\lesssim (\epss+C_1\eps) s^{-1}\Fenergy_{\mu,c}^{\ME,p,k-1}(s,\phi).
$$
For $G_2$, thanks to Lemma~\ref{lemma-111} (the lower-order case) we have 
$$
\aligned
\|\crochet^{\mu}J\zeta^{-1}|\del u^{\N00}|_{p_1-1,k_1}|\del\del\phi|_{p_2,k_2}\|_{L^2(\Mfar_s)}&\lesssim 
(\epss+C_1\eps)s^{1+2\delta}\|r^{-1-\kappa}\crochet^{\kappa}\zeta|\del u|_{p,k}\|_{L^2(\Mfar_s)}\lesssim (\epss + C_1\eps)s^{-1-\delta},
\endaligned
$$
while
$$
\aligned
&
\|\crochet^{\mu}J\zeta^{-1}|\del u^{\N00}|_{p_1-1,k_1}|\del\del\phi|_{p_2,k_2}\|_{L^2(\Mnear_s)}
\lesssim (\epss+C_1\eps)s^{1+2\delta}\|r^{-1}\zeta|\del u^{\N00}|_{p_1-1,k_1}\|_{L^2(\MME_s)}
\\
&\lesssim (\epss+C_1\eps)s^{-3/2+\delta}\|\crochet^{\kappa-1/2}J^{1/2}|\delsN u|_{p_1-1,k_1}\|_{L^2(\MME_s)} 
+ \delta^{-1}(\epss+C_1\eps)^2s^{-1-\delta}.
\endaligned
$$
Here for $\Mnear_s$, we rely on the computation made in \eqref{eq12-18-08-2021} (second line), and the desired result is obtained.
}
%---------------------------------------------------------------------------------------------------------------------------------------------

\paragraph{Challenging terms.}

Next, we treat $W_{p,k}^\hard$ in the right-hand side of \eqref{eq6-14-03-2021} as follows. 

\begin{proposition}\label{prop1-30-03-2021}
Under the condition \eqref{eq2-29-03-2021-gstar}, one has
$$
\|\crochet^{\mu}J\zeta^{-1} W_{p,k}^\hard \|_{L^2(\MME_s)}\lesssim { (\epss+C_1\eps)s^{-1+2(3N+1)\theta}\Fenergy_{\mu,c}^{\ME,p,k-1}(s,\phi) +}
\begin{cases}
0\quad & p\leq N-5, 
\\
\delta^{-1}(\epss + C_1\eps)^2, 
&  p\geq N-4.
\end{cases} 
$$
\end{proposition}

\begin{proof} 
{ We} consider the second term in $W_{p,k}^\hard$ in the right-hand side of \eqref{eq6-14-03-2021}. Observe that $k_1\geq 1$ implies $p_2\leq p-1$ and $k_2\leq k-1$. When $k_1-1\leq N-6$, we apply \eqref{eq3-29-03-2021} and obtain
\begin{equation}\label{eq1-02-09-2021}
\aligned
\|\crochet^{\mu}J\zeta^{-1} |\LOmega  h^\circledast|_{k_1-1} |\del\del \phi|_{p_2,k_2}\|_{L^2(\MME_s)} 
& \lesssim (\epss + C_1\eps) \|\crochet^{\mu}J\zeta^{-1}r^{-1+({ 3N+1}) \theta} |\del \phi|_{p_2+1,k_2}\|_{L^2(\MME_s)}  
\\
& \lesssim (\epss + C_1\eps)s^{-1+2({ 3N+1}) \theta} \, \Fenergy_{\mu,c}^{\ME,{ p,k-1}}(s,\phi). 
\endaligned
\end{equation}
The first term in $W_{p,k}^\hard$ in \eqref{eq6-14-03-2021} is bounded in the same manner, we omit the detail. When $k_1-1\geq N-5$, we have  
$p_2\leq p-(N-4) = 4-(N-p) \leq 4\leq N-11$, and we emphasize that {\sl this case occurs only when $p\geq N-4$}. At this junction, we need the following observation. Recalling the decomposition 
$
h^{\alpha\beta} = - { \mathbbm{h}[h^{\star}]^{\alpha\beta}} - {\mathbbm{h}[u]^{\alpha\beta}} + \Abb^{\alpha\beta}[h].
$
In
Section~\ref{subsec1-02-09-2021}, together with \eqref{eq1-07-05-2021} and \eqref{eq1-09-05-2021}, 
we have 
$$
|\Abb^{\alpha\beta}[h]|_{p,k}
\lesssim \sum_{p_1+p_2=p\atop k_1+k_2=k} |h|_{p_1,k_1} |h|_{p_2,k_2} 
\lesssim |h^{\star} |_N^2 + \big(|h^{\star} |_N + |u|_{N-2}\big)|u|_{p,k}
$$
and therefore 
$
|\Abb^{\alpha\beta}[h]|_{p,k}\lesssim \epss^2r^{-2\lambda} + (\epss + C_1\eps)r^{-\min(\lambda,\kappa)}s^{\delta} |u|_{p,k}.
$
In view of \eqref{eq2-29-03-2021-gstar}, { \eqref{eq3-29-03-2021}} and thanks to {$\min(\lambda,\kappa) \geq 1/2+\delta$}, we have 
$$
|\LOmega h^\circledast|_{p-1,k-1}\lesssim (\epss + C_1\eps)r^{-1+\theta} + |\LOmega u|_{ p-1,k-1}.
$$
Applying the above bound together with Lemma~\ref{lemma-111} and \eqref{eq1-12-05-2020} and  
{ using $\min(\lambda,\kappa) \geq 1/2+\delta$}, we obtain 
\begin{equation}
\aligned
& \|\crochet^{\mu}J\zeta^{-1} |\LOmega  h^\circledast|_{k_1-1} |\del\del \phi|_{p_2,k_2}\|_{L^2(\MME_s)} 
\\
& \lesssim (\epss + C_1\eps)s\|\crochet^{\mu}r^{-1+\theta}\zeta |\del \phi|_{p,k-1}\|_{L^2(\MME_s)} 
+ \|\crochet^{\mu}J\zeta^{-1} | \LOmega  u|_{k_1-1} |\del\del \phi|_{p_2,k_2}\|_{L^2(\MME_s)}
\\
& \lesssim \epss s^{-1+2\theta} \, \Fenergy_{\mu,c}^{\ME,p,k-1}(s,\phi)
+ (\epss + C_1\eps)s^{1+2\delta}\|(r^{-2}\crochet + r^{-1-\lambda}) \zeta |\LOmega u|_{p-1,k-1}\|_{L^2(\MME_s)}
\\
& \lesssim \epss s^{-1+2\theta} \, \Fenergy_{\mu,c}^{\ME,p,k-1}(s,\phi)
+ \delta^{-1}(\epss + C_1\eps)^2s^{1-2\min(\lambda,\kappa)+2\delta}
\lesssim \epss s^{-1+2\theta} \, \Fenergy_{\mu,c}^{\ME,p,k-1}(s,\phi) + \delta^{-1}(\epss + C_1\eps)^2.
\endaligned
\end{equation}
By combining the above bound together with \eqref{eq1-02-09-2021}, the proof of Proposition~\ref{prop1-30-03-2021} is completed. 
\end{proof}

%-------------------------------------------------------------------------------------------------------------------------------- 

\subsection{Derivation of the sharp energy estimate for the matter field}

We are finally in a position to complete the bootstrap argument. We apply the energy estimate in Proposition~\ref{prop energy-ici-exterior} to the equation
\begin{equation}
\Boxt_g Z\phi + c^2 Z \phi 
= -[Z,H^{\alpha\beta}\del_{\alpha}\del_{\beta}]\phi
\end{equation} 
with $\ord(Z) = p$ and $\rank(Z) = k$. 
Recalling Lemma~\ref{lem1-28-03-2021}, \eqref{eq9-27-03-2021}, and \eqref{eq2-30-03-2021} 
(with the choice $(\eta,w,c^*) = (\mu,\phi,c)$), we obtain
$$
\aligned
& \frac{d}{ds}\Eenergy_{g,\mu,c}^{\ME}(s,Z\phi) 
+ \frac{d}{ds} \Eenergy^{\Lcal}_{g, c}(s,Z \phi;s_0)  
+  2 \mu \int_{\MME_s}\big( g^{\N ab} \delsN_aZ\phi \delsN_bZ\phi +  c^2|Z\phi|^2\big) \crochet^{2\mu-1} \aleph'(r-t) \ Jdx 
\\
& \lesssim{ (\epss + C_1\eps)}s^{-1} \Eenergy_{\mu,c}^{\ME}(s,Z\phi) 
+ {   \delta^{-1}}(\epss + C_1\eps)^2s^{-1-\delta} \, \Fenergy_{\mu,c}^{\ME}(s,Z\phi)
\\
&\quad
+ (\epss+C_1\eps)s^{-3/2+\delta}\Fenergy_{\mu,c}^{\ME}(s,Z\phi)\,
 { \nouveauS_{p,k}(s,u)}
+ \Fenergy_{\mu,c}^{\ME}(s,Z\phi) \, \|\crochet^{\mu} J\zeta^{-1} W_{p,k}^\hard \|_{L^2(\MME_s)},  
\endaligned
$$ 
in which the second term involving $\Fenergy_{\mu,c}^{\ME}(s,Z\phi)$ comes from $ W_{p,k}^\easy$ in \eqref{eq1-28-03-2021}. 
The term $W_{p,k}^\hard$ is bounded by Proposition~\ref{prop1-30-03-2021}. 
Observe that by \eqref{eq8-27-03-2021} (evolution of the energy)
and \eqref{eq1-24-04-2021} (light-bending property), we have 
$\frac{d}{ds} \Eenergy^{\Lcal}_{g, c}(s,Z \phi;s_0) \geq 0$. 
Consequently, in view of the comparison property in Lemma~\ref{lem1-28-03-2021} we have 
\begin{equation}\label{eq3-30-03-2021}
\aligned
\frac{d}{ds} \, \Fenergy_{g,\mu,c}^{\ME}(s,Z\phi)
& \lesssim 
 { (\epss + C_1\eps)}s^{-1} \Fenergy_{\mu,c}^{\ME}(s,Z\phi) + {   \delta^{-1}}(\epss + C_1\eps)^2s^{-1-\delta}
\\
& \quad + (\epss+C_1\eps)s^{-3/2+\delta} \,  { \nouveauS_{p,k}(s,u)}
+ \|\crochet^{\mu} J\zeta^{-1} W_{p,k}^\hard \|_{L^2(\MME_s)},
\endaligned
\end{equation}
and 
\begin{equation}\label{eq4-30-03-2021}
\aligned
\frac{d}{ds} \Eenergy^{\Lcal}_{g, c}(s,Z \phi;s_0)
& \lesssim{ (\epss + C_1\eps)}s^{-1} \Eenergy_{\mu,c}^{\ME}(s,Z\phi) 
+ {   \delta^{-1}}(\epss + C_1\eps)^2s^{-1-\delta} \, \Fenergy_{\mu,c}^{\ME}(s,Z\phi)
\\
&\quad 
+ (\epss+C_1\eps)s^{-3/2+\delta}\Fenergy_{\mu,c}^{\ME}(s,Z\phi)\,
 { \nouveauS_{p,k}(s,u)}
 + \Fenergy_{\mu,c}^{\ME}(s,Z\phi) \, \|\crochet^{\mu} J\zeta^{-1} W_{p,k}^\hard \|_{L^2(\MME_s)}.
\endaligned
\end{equation}
For $p\leq N-5$, we apply Proposition~\ref{prop1-30-03-2021} (low-order case) and take the sum over $\ord(Z) \leq p\leq N-5$ and $\rank(Z) \leq k\leq p$:
\begin{equation}
\aligned
\frac{d}{ds} \, \Fenergy_{g,\mu,c}^{\ME,p,k}(s,\phi)
& \leq 
K_0 s^{-1} \Fenergy_{g,\mu,c}^{\ME,p,k}(s,\phi) 
 { + K_0s^{-1+2(3N+1)\theta}\Fenergy_{g,\mu,c}^{\ME,p,k-1}(s,\phi) }
 \\
 &
 \quad
+K_0s^{-3/2+\delta}  { \nouveauS_{p,k}(s,u)}
 +K(N){   \delta^{-1}}(\epss + C_1\eps)^2s^{-1-\delta},
\endaligned
\end{equation}
where $K_0 = K(N)x 
(\epss + C_1\eps)$ and $K(N)$ is a constant determined by $N$.
Furthermore,  the last term in the right-hand side does not exist when $k=0$. By Gronwall's inequality { and \eqref{eq1-28-05-2023},} we have 
\begin{equation}
\aligned
\Fenergy_{g,\mu,c}^{\ME,p,k}(s,\phi) 
& \leq  C_0\eps s^{K_0} + {   \delta^{-2}}K(N)s^{K_0+{ 3N\theta}}(\epss + C_1\eps)^2
+ K_0s^{K_0}
\int_{s_0}^s {s'}^{-1+2{ (3N+1)} \theta - K_0} \, \Fenergy_{g,\mu,c}^{\ME,p,k-1}(s',\phi) ds'.
\endaligned
\end{equation}
Fixing $p=N-5$ and $k=0$, under the condition 
\begin{equation}\label{condition-24-10-21} 
C_0\eps + {   \delta^{-2}}K(N)(\epss + C_1\eps)^2 + \theta^{-1} { (8N+1)(3N+1)^{-1}} K(N) (\epss+C_1\eps)^2 \leq (1/4)  (\epss+C_1\eps), 
\end{equation}
we obtain 
$\Fenergy_{g,\mu,c}^{\ME,N-5,0}(s,\phi)
\leq (1/4) (\epss+C_1\eps) s^{{ 3N\theta + }K_0}$
and therefore, by induction,
$$
\Fenergy_{g,\mu,c}^{\ME,N-5,k}(s,\phi)
\leq (1/4) (\epss+C_1\eps) s^{ 2k(3N+1)\theta + 3N\theta + K_0}, 
\qquad 
\quad 0\leq k\leq N-5.
$$
Using {$ (6N^2+5N) \theta \leq \delta/2$} and { $K_0\leq \delta/2$,} thanks to Lemma~\ref{lem1-28-03-2021} again (comparison property for  energies) 
this leads us to
\begin{equation}
\Fenergy_{\mu,c}^{\ME,N-5}(s,\phi)
\leq 2 \, \Fenergy_{g,\mu,c}^{\ME,N-5}(s,\phi) \leq (1/2) (\epss+C_1\eps) s^{ 2k(3N+1)\theta + 3N\theta + K_0}\leq  (1/2)  (\epss+C_1\eps) s^{\delta}, 
\end{equation}
which improves the lower-order energy bounds. For the higher-order estimates, taking \eqref{eq3-30-03-2021} together with (the high--order case of) Proposition~\ref{prop1-30-03-2021}, we find 
$$
\aligned
\frac{d}{ds} \, \Fenergy_{g,\mu,c}^{\ME,p,k}(s,\phi)
& \leq K_0 s^{-1} \, \Fenergy_{g,\mu,c}^{\ME,p,k}(s,\phi) 
 { + K_0s^{-1+2(3N+1)\theta}\Fenergy_{g,\mu,c}^{\ME,p,k-1}(s,\phi) 
}
\\
&
\quad 
+ K_0s^{-3/2+\delta}   { \nouveauS_{p,k}(s,u)}
 + K(N) \delta^{-1}(\epss + C_1\eps)^2
\endaligned
$$
with $ K_0 := K(N) 
(\epss + C_1\eps)$.
Next, Gronwall's lemma implies 
$$
\aligned
\Fenergy_{g,\mu,c}^{\ME,p,k}(s,\phi) \leq& C_0\eps s^{K_0}
+ K(N) \delta^{-1}(\epss + C_1\eps)^2s^{1+K_0}
 { + K_0s^{-1+2(3N+1)\theta}\Fenergy_{g,\mu,c}^{\ME,p,k-1}(s,\phi)}, 
\endaligned
$$
where the second term in the right-hand side does not exist when $k=0$. 
Fixing $p=N$ and $k=0$, and for 
\begin{equation}\label{equa-21-10-21b}
C_0\eps + K(N) \delta^{-1}(\epss+C_1\eps)^2 + { (1/4)} K_0  (\epss+C_1\eps) \leq (1/4)  (\epss+C_1\eps),
\end{equation}
we thus have 
$$
\Fenergy_{g,\mu,c}^{\ME,N,0}(s,\phi) \leq \big(C_0\eps +  (\epss + C_1\eps)^2\big)s\leq (1/4)  (\epss+C_1\eps) s^{1+K_0} 
$$
and, by induction,
$$
\Fenergy_{g,\mu,c}^{\ME,N,k}(s,\phi) \leq (1/4)  (\epss+C_1\eps) s^{1+2k{ (3N+1)}\theta+K_0},\qquad 0\leq k\leq N.
$$
{ Using { $(6N^2+2N)\theta\leq \delta/2$} and $K_0\leq \delta/2$}, we conclude that 
$$
\Fenergy_{\mu,c}^{\ME,N}(s,\phi) \leq 2 \, \Fenergy_{g,\mu,c}^{\ME,N}(s,\phi) \leq (1/2)  (\epss+C_1\eps) s^{1+2k{ (3N+1)}\theta+K_0} \leq (1/2) (\epss+C_1\eps) s^{1+\delta}, 
$$
which improves the high-order energy estimate for the Klein-Gordon component, as stated in Proposition~\ref{proposition-section17-matter}. This completes the bootstrap argument and, in turn, the proof of the nonlinear stability of self-gravitating massive fields stated in Theorem~\ref{theo:main1-geometric}.  

%==============================================================================================

\section{Bootstrap argument in the hyperboloidal domain} 
\label{section-18}

\subsection{Strategy in the hyperboloidal domain}

\paragraph{Aim.}

The analysis of the interior domain was the subject of earlier work by the authors, namely the paper \cite{PLF-YM-one} and the monograph \cite{PLF-YM-two}. It will sufficient here to refer the reader to this monograph while presenting the modifications that are required in order to apply our methodology in the hyperboloidal domain. In \cite{PLF-YM-two}, the exterior domain was not an ``arbitrary'' solution as we cover in the present work, but the metric was assumed to coincide with the Schwarzschild metric outside a light cone. 
In the present paper, we perturb about a reference metric rather than by keeping the metric to be exactly 
the Schwarzschild metric in the Euclidean domain. However, our estimates turn out to be strong enough in order to provide {\sl 
suitable decay estimates along the light cone}, that is, estimates that are are sufficiently strong in order 
to perform the same steps as in \cite{PLF-YM-two} with only minor changes. 
The novelty comes from boundary terms arising in several of our arguments, as explained below. Observe that we allow the solution to loose a few degrees of regularity within the hyperboloidal domain.

\paragraph{Bootstrap argument.} 

Recall that in \eqref{eqs-int}, for the hyperboloidal domain and for all $s \in [s_0, s_1]$, we assumed
\begin{subequations}\label{eqs-int-repeat}
\begin{equation}  \label{eqsa-int-repeat} 
\Fenergy^{\Hcal, N-5}(s,u) + s^{-1/2} \, \Fenergy_c^{\Hcal,N-5}(s, \phi) \leq  (\epss+C_1\eps) \, s^{\delta},
\end{equation}
\begin{equation}  \label{eqsb-int-repeat}
\Fenergy^{\Hcal,N-7}(s,u) + \Fenergy_c^{\Hcal,N-7}(s, \phi) \leq (\epss+C_1\eps) \, s^{\delta}. 
\end{equation}
\end{subequations}
We are interested in deriving \eqref{eqs'-int}, that is, 
\begin{subequations}\label{eqs'-int-repeat}
\begin{equation}
\Fenergy^{\Hcal, N-5}(s,u) + s^{-1/2} \, \Fenergy_c^{\Hcal,N-5}(s, \phi) \leq \frac{1}{2} (\epss+C_1\eps) \, s^{\delta},
\end{equation}
\begin{equation}
\Fenergy^{\Hcal,N-7}(s,u) + \Fenergy_c^{\Hcal,N-7}(s, \phi) \leq \frac{1}{2} (\epss+C_1\eps) \, s^{\delta}. 
\end{equation}
\end{subequations}
We present the modifications required in comparison to \cite{PLF-YM-one}, which are only due to contributions from the merging-Euclidian domain ``toward'' the hyperboloidal domain. These contributions arise both as boundary contributions in energy estimates and boundary contributions in pointwise estimates. In addition, the contribution from the Ricci curvature 
of the reference in, both, the energy and pointwise estimates is controlled thanks to \eqref{eq4-09-05-2021}.

%-------------------------------

\paragraph{Null and quasi null nonlinearities.}

As presented in \cite{PLF-YM-two}, the field equations in the semi-hyperboloidal frame and in wave gauge enjoy a null and quasi-null structure, we repeat here for the sake of comparison with the exterior domain. We emphasize that, in comparison with the structure in the exterior, an \textsl{additional term} arises in the interior. 

\begin{lemma}[Null interaction terms] 
\label{Null-Euclidean bilinear-hyper}
In the hyperboloidal domain $\Mscr^\H$, null forms are controlled by good derivatives and contributions depending upon
the reference metric: 
\begin{equation}
|\Qbb_{\alpha\beta}^\star[u] |_p
\lesssim  \sum_{p_1+p_2 = p} |\del u|_{p_1} \big( |\delsH u|_{p_2} + (s/t)^2 |\del u|_{p_2} \big)
+ | h^\star |_p\sum_{p_1+p_2=p} |\del u|_{p_1} |\del u|_{p_2}.
\end{equation}
\end{lemma}

For the quasi-null terms, it is convenient to also introduce 
\begin{equation}\label{eq2-04-12-2020-hyper}
\aligned
\Sbb_p^\H[u] 
& := t^{-1} |u|_{p_1} + \big(|\del h^\star |_{p_1} + t^{-1} | h^\star |_{p_1} \big)
+ \sum_{p_1+p_2=p_1} \big(|\del  u|_{p_1} |u|_{p_2} + |u|_{p_1} |u|_{p_2} \big)
\\
& \quad +   \sum_{p_1+p_2=p_1} \big(| h^\star |_{p_1} |\del u|_{p_2} + |u|_{p_1} |\del h^\star |_{p_2} + | h^\star |_{p_1} |\del h^\star |_{p_2} \big)
+  \sum_{p_1+p_2=p_1} 
\big(| h^\star |_{p_1} | u|_{p_2} + |u|_{p_1} | h^\star |_{p_2} + | h^\star |_{p_1} | h^\star |_{p_2} \big). 
\endaligned
\end{equation}

%---------------------------- 

\begin{lemma}[Quasi-null interaction terms] 
\label{eq3 05-juillet-2019-hyper}
In the hyperboloidal domain $\Mcal^\Hcal$ and under the smallness condition $| h^\star |_p + |u|_{[p/2]} \ll 1$,  the quasi-null terms satisfy (with $\Sbb_p^\H[u]$ defined in \eqref{eq2-04-12-2020-hyper})
$$
\aligned
|\slashed{\Pbb}^{\star\H} [u] |_p
& \lesssim \sum_{p_1+p_2=p} |\del u|_{p_1} \big(  |\delsH u|_{p_2} + (s/t)^2 |\del u|_{p_2} \big) 
+ \sum_{p_1+p_2+p_3=p} | h^\star |_{p_3} |\del u|_{p_1} |\del u|_{p_2},
\\
|\Pbb_{00}^{\star\H} [u] |_{p,k}
& \lesssim     \hskip-.3cm  \sum_{p_1+p_2=p}   \hskip-.3cm  \Big(|\del \usH |_{p_1} |\del \usH |_{p_2} 
+ \big( |\delsH u|_{p_1} + (s/t)^2 |\del u|_{p_1} \big)  |\del u|_{p_2}
\Big) 
+   \hskip-.3cm 
\sum_{p_1+p_2=p}  \hskip-.3cm  | \Sbb_p^\H[u] |_{p_1} |\del u|_{p_2}  
+  \hskip-.3cm  \sum_{p_1+p_2+p_3=p}   \hskip-.3cm  | h^\star |_{p_3} |\del u|_{p_1} |\del u|_{p_2}. 
\endaligned
$$
\end{lemma} 

%------------------------------------------- 

\paragraph{Boost-rotation hierarchy in the hyperboloidal domain.}   

The basic analysis of commutators $[Z,H^{\alpha\beta} \del_\alpha \del_{\beta}]$, including in the interior domain, was given 
in Lemma~\ref{eq2-31-01-2020}  and Proposition~\ref{lm 2 dmpo-cmm-H}. 
As done in Section~\ref{section----63} for the exterior domain, in our derivation of energy estimates at arbitrary high-order we will need a hierarchy property, discovered first in  \cite[Section 4]{PLF-YM-one} and 
\cite[Section 7.3]{PLF-YM-two} for the interior domain.

\begin{proposition}[Hierarchy property for quasi-linear commutators in the hyperboloidal domain] 
\label{prop1-12-02-2020-interior}
For any function $u$ defined in $\MH_{[s_0,s_1]}$ and for any operator $Z$ with $\ord(Z) = p$ and $\rank(Z) = k$ one has 
\begin{equation}\label{equa-hsgd57} 
\aligned
\, 
& |[Z,H^{\alpha\beta} \del_\alpha \del_{\beta}]u|
\lesssim  T^\textbf{hier}  + T^\easy +  T^{\textbf{super}}, 
\endaligned
\end{equation}
\begin{equation}
\aligned
T^\textbf{hier} & 
|\Hu^{00} | \, | \del\del u|_{p-1,k-1} +\sum_{k_1+p_2=p\atop k_1+k_2=k} |L \Hu^{00} |_{k_1-1,k_1-1} |\del\del u|_{p_2,k_2}
\\
T^\easy
& :=
\sum_{p_1+p_2=p\atop k_1+k_2=k} |\del \Hu^{00} |_{p_1-1,k_1} |\del\del u|_{p_2,k_2}
+ t^{-1} |H| \, | \del u|_{p} 
\\
T^{\textbf{super}} &
:= \sum_{p_1+p_2=p} |\delsH H|_{p_1-1} |\del u|_{p_2+1}
+ t^{-1} \!\!\!\!\sum_{p_1+p_2=p} \!\!\!\! |\del H|_{p_1-1} |\del u|_{p_2+1}.
\endaligned
\end{equation}
\end{proposition}

Let us consider the right-hand side of \eqref{equa-hsgd57}. The first and second terms require particular attention in our analysis, since 
only the gradient of the metric naturally enjoys good $L^2$ and $L^\infty$ decay. 
Moreover, we emphasize that, in the second term, we have $k_1\geq 1$ (since the case $k_1=0$ is precisely understood to vanish by our convention) thus $k_2 \leq k-1$. Consequently, these (first and second) key terms contain {\sl fewer boosts and rotations} in comparison to the terms in the  left-hand side of \eqref{equa-hsgd57}. This property will lead us to a {\sl hierarchy structure} for our bootstrap argument presented at the end of this paper.

%----------------------------------------------------------------------------------------------------------

\subsection{Boundary and exterior contributions} 

%------------------------------------------------------------------- 

\paragraph{Boundary contribution in the energy estimates.}

By integrating the local energy identity within two hypersurfaces of the hyperboloidal foliation, we obtain a boundary integral along the light cone. This fact was already pointed out in Section~\ref{sec-energy-curved} when, specifically, we derived a weighted energy estimate in the asymptotically hyperboloidal domain in Proposition~\ref{prop energy-ici-interior}. 
This boundary integral term reads 
$
\Eenergy^{\Lcal}_{g, c}(s,u;s_0) 
$
and, more precisely, for high-order energy estimates our analysis in the interior will require a bound on $\Eenergy^{\Lcal}_{g, c}(s,Z u;s_0)$ for $\ord(Z)$ up to some high order ($N$ or, in fact, slightly smaller).  One approach in order to control this term is the following argument.  Let us return to our conclusion \eqref{equation-1633}, namely 
\begin{equation}
\Fenergy_{\kappa}^{\ME,N}(s,u) \leq (1/2) (\epss+C_1\eps) s^{K_0 + 2N\theta}, 
\end{equation}
but let us now impose a stronger condition on $K_0$, that is, $K_0 \leq C \theta$ for a possibly large numerical constant $C$. Consequently, we find 
\begin{equation}
\Fenergy_{\kappa}^{\ME,N}(s,u) \leq (1/2) (\epss+C_1\eps) s^{C \, \theta}
\end{equation}
for some constant $C>0$. Next, returning to the inequality \eqref{eq6-27-03-2021} and, appyling with the same choice of constant $K_0$, leads us to a control of the boundary term   
\begin{equation}
\Eenergy^{\Lcal,N}_{g, c}(s, u;s_0) \leq (1/2) (\epss+C_1\eps) s^{C\theta}.
\end{equation} 
By returning to our consequence of the generalized Sobolev inequality we obtain the somewhat stronger statement involving $s^{-1+C \theta}$. 

%-----------------------------------------------------------

\paragraph{Boundary contribution in the weighted Hardy inequality.}

The Hardy inequality, established now by integration along a hyperboloidal hypersurface, includes a contribution from the boundary. We follow the steps in the proof of  \cite[Lemma 3.12]{PLF-YM-two} but instead of the contribution in $1/s$ due to the (outside) Schwarzschild metric, we have now the slightly weaker decay $s^{-1+C \theta}$. 
The new boundary term is now controlled as  
\begin{equation}
\int_{r= \rhoH(s)}r^{-1} |Z u|^2 \, d\sigma \lesssim  s^{-2+C \theta}, \qquad \ord(Z) \leq N-1, 
\end{equation} 
and the standard calculation leading to the Hardy inequality yields us 
\begin{equation}
\aligned
&
\| r^{-1} Zu \|^2_{L^2(\M^\H_s)}
& \lesssim \| r^{-1} Zu \|_{L^2(\M^\H_s)}  \|\del_r u_s \|_{L^2(\M^\H_s))} +  s^{-2+C \theta}
\endaligned
\end{equation} 
and, therefore, 
\begin{equation}
\| r^{-1} Zu \|_{L^2(\M^\H_s))} \lesssim \|\del_r Zu \|_{L^2(\M^\H_s))} + s^{-1+ C\theta/2}, \qquad \ord(Z) \leq N-1.
\end{equation}

%-----------------------------------------------------------

\paragraph{Exterior contribution from the Kirchhoff formula.}

Again when applying the Kirchhoff argument we must take the contribution along the light cone into account, which has a decay $s^{-1+C \theta}$.

%-----------------------------------------------------------

\paragraph{Boundary contribution for the Klein-Gordon equation.}

We refer to the technical inequality in \cite[Proposition 3.5 ]{PLF-YM-two} (first established in \cite{PLF-YM-one}, by integration along lines from the origin).  In this earlier work this technical estimate was used with a vanishing scalar field in the outside but now the scalar field is non-vanishing but yet controlled by Lemma~\ref{lemma-111}. That is, we can use that the Klein-Gordon field satisfies the pointwise bound 
\begin{equation}\label{eq9-15-05-2020} 
r \, \crochet^\mu \, |\phi|_{p-4} 
\lesssim 
(\epss+C_1\eps) \, \big(r^{-1}\crochet + r^{-\lambda}\big) s^{2\delta}, \quad   p=N-5.  
\end{equation}
However, in order to loose fewer derivatives we can improve the conclusion of Lemma~\ref{lemma-111} by 
returning to its proof and applying once more our argument based on Proposition~\ref{lem 1 d-KG-e}.

%===================================================================== 

\paragraph{Acknowledgments}

Part of this work was done when P.L.F was a visiting research fellow at the Courant Institute for Mathematical Sciences, New York University, and at the School of Mathematical Sciences, Fudan University, Shanghai. He was also supported by the  project Einstein-PPF from the Agence Nationale de la Recherche (ANR), and the MSCA Staff Exchange project Einstein-Waves (grant number 101131233) from the European Research Council (ERC). The work of Y.M. was supported by a Special Financial Grant from the China Postdoctoral Science Foundation under the grant number
2017T100732. 
 
%%================================================================================
 
\bibliography{references}

\addcontentsline{toc}{section}{References}

\pagestyle{plain}

%============================================================================================== 

\

\small

\appendix 

\section{Properties of the weight functions}
\label{appendix-A} 

\begin{proof}[Proof of Lemma~\ref{lem-Jacobian-bounds}] All of the following calculation are based on the parameterization $(s,x)$. We differentiate \eqref{eq5-05-05-2020} and obtain
\begin{equation}\label{eq1-03-02-2021}
\del_s\del_r \Time(s,r) = \del_r\del_s\Time(s,r) = \frac{{ s}r \, \del_r\xi(s,r)}{(s^2+r^2)^{1/2}} 
- \frac{sr\xi(s,r)}{(s^2+r^2)^{3/2}}, 
\end{equation}
from which we are going to evaluate $\del_s T$ by integration with respect to $r$.

\vskip.3cm

\noindent{\bf Hyperboloidal domain $0\leq r \leq \rhoH(s)$.} In this case, the foliation coefficient
$\xi$ is identically $1$ and \eqref{eq1-03-02-2021} reduces to  
$$
\del_r\del_s \Time(s,r) = - \frac{sr}{(s^2+r^2)^{3/2}}, 
\qquad r\leq \rho^{\H}(s). 
$$
We integrate this identity with respect to $r$, from the center $r=0$ at which one has $\Time(s,0) = s$ and thus $\del_s \Time(s,0) = 1$,
and find
$$
\aligned
J = \del_s\Time(s,r)
& = 1 - \int_0^r\frac{s\rho}{(s^2+\rho^2)^{3/2}} \, d\rho
= \frac{s}{(s^2+r^2)^{1/2}} = {s \over \Time(s,r)}.
\endaligned
$$
\vskip.15cm

\noindent{\bf Merging domain $\rhoH(s)\leq r \leq \rhoE(s)$.} In this case, the foliation coefficient $\xi$ is non-trivial and, by integrating \eqref{eq1-03-02-2021} from the boundary of the interior domain, we find
$$
\aligned
\del_s \Time(s,r)  & =  1 - \int_0^{\rhoH(s)} \frac{s\rho d\rho}{(s^2+\rho^2)^{3/2}}
- \int_{\rhoH(s)}^r\frac{s\rho\xi(s, \rho) \, d\rho}{(s^2+\rho^2)^{3/2}}
- s\int_{\rhoH(s)}^r\frac{\rho\del_\rho\xi(s, \rho) d\rho}{(s^2+\rho^2)^{1/2}}
\\
& = \frac{2s}{s^2+1} + s\int_{\rhoH(s)}^r\xi(s, \rho)\ d\left((s^2+\rho^2)^{-1/2} \right)
- s\int_{\rhoH(s)}^r\frac{\rho\del_\rho\xi(s, \rho) d\rho}{(s^2+\rho^2)^{1/2}}
\\
& =  \frac{2s}{s^2+1}
+ \left[\frac{s\xi(s, \rho)}{(s^2+\rho^2)^{1/2}} \right]_{\frac{s^2-1}{2}}^r -s\int_{\frac{s^2-1}{2}}^r\frac{\del_\rho\xi(s, \rho) \, d\rho}{(s^2+\rho^2)^{1/2}}
- s\int_{\frac{s^2-1}{2}}^r\frac{\rho\del_\rho\xi(s, \rho) d\rho}{(s^2+\rho^2)^{1/2}}, 
\endaligned
$$
and therefore 
\begin{equation}
\aligned
\del_s \Time(s,r) 
& \leq  \frac{2s}{s^2+1}
+ \left[\frac{s\xi(s, \rho)}{(s^2+\rho^2)^{1/2}} \right]_{\frac{s^2-1}{2}}^r
- s\int_{\frac{s^2-1}{2}}^r\frac{\rho \del_\rho\xi(s, \rho) \, d\rho}{{ (}s^2+\rho^2{)^{1/2}}}
- s\int_{\frac{s^2-1}{2}}^r\frac{\rho\del_\rho\xi(s, \rho) d\rho}{(s^2+\rho^2)^{1/2}}
\\
& =   \frac{s\xi(s,r)}{{(}s^2+r^2{)^{1/2}}} - 2s\int_{\frac{s^2-1}{2}}^r\frac{\rho\del_\rho\xi(s, \rho) d\rho}{(s^2+\rho^2)^{1/2}}.
\endaligned
\end{equation} 
In the latter inequality, we used the fact that $\del_\rho\xi(s, \rho)\leq 0$ (thanks to our assumption $\chi'\geq 0$) and $\frac{s^2-1}{2} \geq 1$. 
Next, we rely on the following observation.  
Setting $f(s, \rho) : = \frac{\rho}{{(}s^2+\rho^2{)^{1/2}}}, $
we have 
$\del_\rho f(s, \rho) = \frac{s^2}{(s^2+\rho^2)^{3/2}} \geq 0$
and thus 
$$
\aligned
& s\int_{\frac{s^2-1}{2}}^r\frac{\del_\rho\xi(s, \rho)\rho d\rho}{(s^2+\rho^2)^{1/2}}
= s\int_{\frac{s^2-1}{2}}^rf(s,\rho)\del_{\rho}\xi(s,\rho) \, d\rho = \left[\frac{s\xi(s, \rho)\rho}{{(}s^2+\rho^2{)^{1/2}}} \right]_{\frac{s^2-1}{2}}^r - s\int_{\frac{s^2-1}{2}}^r\xi(s, \rho)\del_{\rho}f(s, \rho) \, d\rho
\\
& =  s\xi(s,r)f(s,r) - sf(s,\rhoH(s)) - s\int_{\frac{s^2-1}{2}}^r\xi(s, \rho)\del_{\rho}f(s, \rho) \, d\rho
\\
& \geq  s\xi(s,r)f(s,r) - sf(s, \rhoH(s)) - s\int_{\frac{s^2-1}{2}}^r\del_{\rho}f(s, \rho) \, d\rho
=  s(\xi(s,r)-1)f(s,r)\geq s(\xi(s, \rho)-1).
\endaligned
$$
In the above inequalities we have $0\leq f(s, \rho)\leq 1$ and we conclude that 
$
\del_sT\leq \frac{s\xi(s,r)}{{(}s^2+r^2{ )^{1/2}}} + 2s(1-\xi(s,r)).
$

%-------------------------------------

On the other hand, in order to establish a lower bound we write
$$
\aligned
\del_s \Time(s,r) =
& \frac{2s}{s^2+1}
+ \Bigg[\frac{s\xi(s, \rho)}{(s^2+\rho^2)^{1/2}} \Bigg]_{\frac{s^2-1}{2}}^r
- s\int_{(s^2-1)/2}^r\frac{\del_\rho\xi(s, \rho)}{(s^2+\rho^2)^{1/2}} \, d\rho
- s\int_{(s^2-1)/2}^r\frac{\rho\del_\rho\xi(s, \rho)}{(s^2+\rho^2)^{1/2}}  \, d\rho
\\
& \geq  \frac{s\xi(s,r)}{(s^2 + r^2)^{1/2}} - s\int_{(s^2-1)/2}^r\frac{\del_\rho\xi(s, \rho)\rho}{(s^2+\rho^2)^{1/2}} \, d\rho.
\endaligned
$$
Observing that $f(s,\rho) = \frac{\rho}{{ (}s^2+\rho^2{ )^{1/2}}}$ is increasing with respect to $\rho$, we infer that 
$$
\aligned
\del_s \Time(s,r)
& \geq  \frac{s\xi(s,r)}{(s^2 + r^2)^{1/2}} - sf(s, \rhoH(s))\int_{(s^2-1)/2}^r\del_{\rho} \xi(s, \rho) \, d\rho
= \frac{s\xi(s,r)}{(s^2 + r^2)^{1/2}} + (1-\xi(s,r))s \, \frac{s^2-1}{s^2+1}
\\
&
\geq \frac{s\xi(s,r)}{(s^2 + r^2)^{1/2}}  + \frac{3}{5}s(1-\xi(s,r)).
\endaligned
$$

\hskip.3cm

\noindent{\bf  Exterior domain $r\geq \rhoE(s)$.} In this case, the foliation coefficient $\xi$ vanishes identically and we have
$
\del_s\Time(s,r) = \del_s\Time(s, \rhoE(s))$.
The relevant bounds here are precisely the ones established in the previous case {\sl evaluated at the boundary} $r =\rhoE(s)$. In other words, we find 
$\frac{3}{5}s\leq \del_s \Time(s, \rhoE(s)) \leq 2s$.
\end{proof}

%==============================================================================================

\section{Proof of a Sobolev inequality} 
\label{appendix-B} 

\begin{proof}[Proof of Lemma~\ref{lem 1 30 -10 -217}]
1. We begin by proving the following claim: for any function $u: \RR^3_+ \mapsto \RR$ 
one has the following $L^6$ Sobolev inequality (the implied constant being a universal constant): 
\begin{equation}\label{eq3-14-02-2021}
\|u\|_{L^6(\RR^3_+)} \lesssim \| \nabla u\|_{L^2(\RR^3_+)} \simeq \sum_{a=1,2,3} \| \del_{a} u\|_{L^2(\RR^3_+)}. 
\end{equation}
By density, it is sufficient to establish the result for functions with compact support. 
Integrating the identity $\del_a u^4 = 4 \, u^3 \del_a u$ with respect to one of the variables, say $x^a$, from $+ \infty$ 
we obtain
$u^4(x) \leq 4 \int_0^{+ \infty} |u|^3(\xti) \, | \del_a u| (\xti) \, d\xti^a$ for 
$x \in \RR^3_+$, 
where $\xti$ denotes $x$ with the component $x^a$ replaced by $\xti^a$. 
Then, introducing the functions $w^1(x) = w^1(x^2, x^3), \ldots$ by 
$$
\aligned
& w^1(x^2, x^3) := \sup_{x^1\geq 0} |u(x)|^2,  
\qquad
w^2(x^1, x^3) := \sup_{x^2 \geq 0} |u(x)|^2,
\qquad
w^3(x^1, x^2) := \sup_{x^3\geq 0} |u(x)|^2,
\endaligned
$$
we see that
$
|w^a(x)|^2 \leq 4 \int_0^{+ \infty} |u|^3(\xti) \, | \del_a u| (\xti)  \, d\xti^a.
$
Next, writing  
$\widehat{d{x}}^1 = dx^2 \, dx^3$, 
$\widehat{d{x}}^2 = dx^1dx^3$, and $\widehat{d{x}}^3 = dx^1dx^2$, 
we have (for $a=1,2,3$)
\begin{equation}\label{eq 1 01-11-217}
\aligned
\int_{x^b\geq 0,b\neq a} (w^a)^2 \, \widehat{dx}^a 
& \leq 
4 \int_{x^b\geq 0,b\neq a} \int_0^{+ \infty} |u^3| \, | \del_a u| \, d \xti^a \widehat{dx}^a 
= 4 \int_{\RR^3_+} |u^3| \, | \del_a u| \, dx
\\
& 
\leq  4 \|u^3\|_{L^2(\RR^3_+)} \, \| \del_a u\|_{L^2(\RR^3_+)}
\lesssim \|u\|_{L^6(\RR^3_+)}^3\|\del_au\|_{L^2(\RR^3_+)}. 
\endaligned
\end{equation}
We see that
$$
\Big|
\int_{x^1\geq 0} w^2(x^1, x^3) \,  w^3(x^1, x^2) \, dx^1 
\Big| 
\leq \|w^2(\cdot, x^3)\|_{L^2(\RR^+_{x^1})} \,  \|w^3(\cdot, x^2)\|_{L^2(\RR^+_{x^1})}
$$
and  
$$
\aligned
&
\Big|
\int_{x^1, x^2 \geq 0}w^1(x^2, x^3) \, w^2(x^1, x^3) \, w^3(x^1, x^2) \, dx^1dx^2
\Big|
\leq  \|w^2(\cdot, x^3)\|_{L^2(\RR^+_{x^1})}
\int_{x_2 \geq 0} |w^1(x^2, x^3)| \,  
\|w^3(\cdot, x^2)\|_{L^2(\RR^+_{x^1})} \, dx^2
\\
& \leq  \|w^2(\cdot, x^3)\|_{L^2(\RR^+_{x^1})} \, 
\|w^1(\cdot, x^3)\|_{L^2(\RR^+_{x^2})} \, 
\big\| \|w^3(\cdot, \cdot)\|_{L^2(\RR^+_{x^1})}^2 \big\|_{L^2(\RR^+_{x^2})}
= 
\|w^1(\cdot, x^3)\|_{L^2(\RR^+_{x^2})}
\|w^2(\cdot, x^3)\|_{L^2(\RR^+_{x^1})}
\|w^3\|_{L^2(\RR^+_{x^1, x^2})}, 
\endaligned
$$
so we have
$$
\aligned
& \Big|
\int_{\RR_+^3} w^1(x^2, x^3) \, w^2(x^1, x^3) \, w^3(x^1, x^2) \, dx^1dx^2 \, dx^3
\Big|
\leq  \|w^3 \|_{L^2(\RR^+_{x^1, x^2})}
\int_{x^3\geq 0} 
\|w^1(\cdot, x^3)\|_{L^2(\RR^+_{x^2})} 
\|w^2(\cdot, x^3)\|_{L^2(\RR^+_{x^1})}
\, dx^3
\\
& \leq   \|w^3\|_{L^2(\RR^+_{x^1, x^2})} 
\, \big\| \|w^1\|_{L^2(\RR^+_{x^2})}^2 \big\|_{L^2(\RR^+_{x^3})}
\,  \big\| \|w^2 \, \|_{L^2(\RR^+_{x^1})}^2 \big\|_{L^2(\RR^+_{x^3})}
= \|w^1\|_{L^2(\RR^+_{x^2, x^3})}
\|w^2 \, \|_{L^2(\RR^+_{x^1, x^3})}
\|w^3\|_{L^2(\RR^+_{x^1, x^2})}.
\endaligned
$$
Then, applying \eqref{eq 1 01-11-217} we find 
$$
\aligned
& \Big|
\int_{\RR_+^3}w^1(x^2, x^3) \, w^2(x^1, x^3) \,  w^3(x^1, x^2) \, dx^1dx^2 \, dx^3
\Big|
\lesssim
\|u\|_{L^6(\RR^3_+)}^{9/2} 
\big(\| \del_1 u\|_{L^2(\RR^3_+)} \| \del_2 u\|_{L^2(\RR^3_+)} \| \del_3 u\|_{L^2(\RR^3_+)} \big)^{1/2}. 
\endaligned
$$
Combining this result with the inequality 
$\| u \|_{L^6(\RR^3_+)}^6 = \int_{\RR^3_+} u^6 \, dx \leq  \int_{\RR^3_+} w^1(x)  \, w^2(x) \,  w^3(x) \,  dx$, 
we arrive at \eqref{eq3-14-02-2021}.

\vskip.15cm

%----------------------------------------------------------------------------- 

2. Fix a point $x_0\in \RR^3_+$ and consider the cube $C_{\rho, x_0}$. We have
$|u(x_0) - u(x)| \leq  \int_0^1\left|(x-x_0)\cdot \nabla u(x_0+(x-x_0)t)\right| \, dt$
for all $x \in C_{\rho, x_0}$, which leads us to
$$
\aligned
& \Big| u(x_0) - \rho^{-3} \int_{C_{\rho, x_0}} u(x) \, dx \Big| 
\leq  \rho^{-3} \int_{C_{\rho, x_0}} |u(x)-u(x_0)| \, dx
\\
& \leq  \rho^{-3} \int_{C_{\rho, x_0}} \int_0^{1} \left|(x-x_0)\cdot \nabla u(x_0+(x-x_0)t)\right| \, dt dx
= \rho^{-3} \int_0^{1} \int_{C_{\rho, x_0}} \left|(x-x_0)\cdot \nabla u(x_0+(x-x_0)t)\right| \, dt dx
\endaligned
$$
and, therefore, 
\begin{equation}
\aligned
\Big| u(x_0) - \rho^{-3} \int_{C_{\rho, x_0}} u(x) \, dx \Big|  
& \lesssim   \rho^{-2} \int_0^{1} \int_{C_{\rho, x_0}} \left| \nabla u(x_0+(x-x_0)t)\right| \, dx \ dt
= \rho^{-2} \int_0^1t^{-3} \int_{C_{t\rho,0}} | \nabla u(x_0+y)|dy\ dt. 
\endaligned
\end{equation}
Then, by the Cauchy-Schwarz inequality, 
$$
\aligned
\int_{C_{t\rho,0}} | \nabla u(x_0+y)|dy
& \leq 
(t\rho)^{5/2} \| \nabla u(x_0+ \cdot)\|_{L^6(C_{t\rho,0})}
\leq (t\rho)^{5/2}
\| \nabla u\|_{L^6(C_{\rho, x_0})}
\endaligned
$$
and for all $t \in [0,1]$ 
\begin{equation}\label{eq 3 01-11-217}
\Big|
u(x_0) - \rho^{-3} \int_{C_{\rho, x_0}} u(x) \, dx
\Big|
\lesssim
\rho^{1/2} \| \nabla u\|_{L^6(C_{\rho, x_0})}.
\end{equation}
In the same manner, for the point $x_1 := (x_0^1+ \rho, x_0^2 + \rho, x_0^3+ \rho)$
by applying the same arguments but changing the sign of $\rho$
(using $C_{- \rho, x_1} = C_{\rho, x_0}$), we obtain 
\begin{equation}\label{eq 4 01-11-217-d}
\Big|
u(x_1) - \rho^{-3} \int_{C_{\rho, x_0}} u(x) \, dx
\Big|
\lesssim
\rho^{1/2} \, \| \nabla u\|_{L^6(C_{\rho, x_0})}.
\end{equation}
Combining \eqref{eq 3 01-11-217} and \eqref{eq 4 01-11-217-d} together, 
we arrive at the inequality 
$|u(x_0) - u(x_1)|
\lesssim
\rho^{1/2} \| \nabla u\|_{L^6(C_{\rho, x_0})}$ with $|x_0-x_1|\simeq \rho$. 

Next, we introduce a smooth cut-off function $\chi: \RR \mapsto \RR$,  
satisfying $\chi(x) = 0$ for $x \leq 0$ and $\chi(x) = 1$ for $x \geq 1$, and we define the auxiliary function 
\begin{equation}\label{eq:9144}
v_{x_0}(x) := \left(1- \chi(\rho^{-2} |x-x_0|^2)\right) u(x),
\qquad x \in C_{\rho, x_0}. 
\end{equation}
Since $x_1 = (x_0^1+ \rho, x_0^2 + \rho, x_0^3 + \rho)$, we find 
$
v_{x_0}(x_0) =u(x_0)$
and $v_{x_0}(x_1)=0, 
$
as well as 
$
|v_{x_0}(x_0) - v_{x_0}(x_1)|
\lesssim { \rho^{1/2}}
\| \nabla v_{x_0} \|_{L^6(C_{\rho, x_0})}.
$
The function $\del_\alpha v_{x_0}$ is regular and is compactly supported in $\RR^3_+$, 
thus we conclude that 
$$
\aligned
|v_{x_0}(x_0) - v_{x_0}(x_1)|
&
\lesssim
\sum_{1 \leq |I| \leq 2} \| \del^Iv_{x_0} \|_{L^2(\RR^3_+)} 
\lesssim
\sum_{1 \leq |I| \leq 2} \| \del^Iv_{x_0} \|_{L^2(C_{\rho, x_0})},
\endaligned
$$
which establishes the desired result since, in view of \eqref{eq:9144}, the norm of $v$ is bounded by the norm of $u$, namely 
$$
\sum_{1 \leq |I| \leq  2} \| \del^Iv_{x_0} \|_{L^2(C_{\rho, x_0})}
\lesssim (1+\rho^{-2}) \,  
\sum_{|I| \leq 2} \| \del^Iu \|_{L^2(C_{\rho, x_0})}.  
\qedhere
$$ 
\end{proof}  

%==============================================================================================

\section{Proof of calculus rules}
\label{appendix-calculus}

\begin{proof}[Proof of Proposition~\ref{prop1-02-02-2020}]  The following identity is immediate:
\begin{equation}\label{eq1-28-05-2021}
\delsN_a u = t^{-1}L_au - (x^a/r)(r-t)^{-1}\del_t u.
\end{equation}
Recalling Proposition~\ref{prop--fund-order}, we only need to consider the class of ordered operators. 
We are going to use decompositions of operators 
$Z = \del^IL^J\Omega^K$ involving $Z_i = \del^{I_i}L^{J_i}\Omega^{K_i}$ with 
$I_1+I_2=I$, $J_1+J_2=J$, and $K_1+K_2=K$. For this it is convenient to introduce the notation $Z_1 \odot Z_2= Z$, and similarly when three operators are required. 
We write  
%---------
\begin{equation}\label{eq1-09-06-2021}
Z \delsN_au = \sum_{Z_1 \odot Z_2 = Z}\!\!\!\!\!\!
Z_1(t^{-1})Z_2L_au 
- \sum_{Z_1 \odot Z_2\odot Z_3 = Z}\!\!\!\!\!\!\!\!\!\!\!\!
Z_1(x^a/r)Z_2((r-t)t^{-1})Z_3(\del_tu).
\end{equation}
%-----------------
For the first term in the right-hand side, we clearly have $
|Z_1(t^{-1}) \, |\lesssim t^{-1}\simeq r^{-1}\text{ in }\Mnear_{ [s_0, + \infty)}.
$
Furthermore, we find 
$$
Z_2L_au = \del^{I_2}L^{J_2}L_a\Omega^{K_2}u + \del^{I_2}L^{J_2}([\Omega^{K_2},L_a]u)
= \del^{I_2}L^{J_2}L_a\Omega^{K_2}u + \sum_{|K'|<|K|}  \sum_b  \Lambda^{Kb}_{aK'}
\del^{I_2}L^{J_2}L_b\Omega^{K'}u, 
$$
where \eqref{eq 4' comm} was used and $\Lambda^{Kb}_{aK'}$ are constants. 
We can now focus our attention
on $\del^{I_2}L^{J_2}L_a\Omega^{K_2}u$ with $|I_2| \leq|I|,|J_2| \leq|J|$
and $|K_2| \leq |K|$. We write $L_bL^{J_2'} = L^{J_2}L_a$ and by recalling \eqref{eq1-28-05-2021},
$$
\aligned
t^{-1}\del^{I_2}L^{J_2}L_a\Omega^{K_2}u 
& = t^{-1}\del^{I_2}L_bL^{J_2'}\Omega^{K_2}u 
= t^{-1}L_b\del^{I_2}L^{J_2'}\Omega^{K_2}u + t^{-1}[\del^{I_2},L_b](L^{J_2'}\Omega^{K_2}u)
\\
& = \delsN_b \del^{I_2}L^{J_2'}\Omega^{K_2}u + (x^a/r)(r/t-1)\del_t\del^{I_2}L^{J_2'}\Omega^{K_2}u + t^{-1}\sum_{|I_2'|=|I_2|\geq 1}\Lambda_{bI_2'}^{I_2} \del^{I_2'}L^{J_2'}\Omega^{K_2}u, 
\endaligned
$$
where \eqref{eq 1 comm} was used and $\Lambda_{bI_2'}^{I_2}$ are constants. Consequently, $Z_1(t^{-1})Z_2L_au $ is bounded as claimed in \eqref{eq2-03-02-2020}.
The second term in the right-hand side of \eqref{eq1-09-06-2021} is bounded (thanks to the last point of Lemma~\ref{lem 1 homo-ext}, \eqref{lem2-commu-ext} and \eqref{equa2-2-juin})  as follows: 
$$
|Z_1(x^a/r)Z_2((r-t)t^{-1})Z_3(\del_tu) \, |\lesssim \frac{|r-t|+1}{t}|\del u|_{p,k}
\lesssim \frac{|r-t|+1}{r}\sum_{\ord(Z)\leq p\atop \rank(Z)\leq k} \sum_\alpha |\del_{\alpha} Z u|.
$$
We have thus established  \eqref{eq2-03-02-2020}.  On the other hand, to handle \eqref{eq3-07-02-2020} we rely on the identity 
$$
\delsN_a u - \delsME_a u = (x^a/r)\Big(1-\frac{\xi(s,r)r}{(s^2+r^2)^{1/2}} \Big)\del_t u = (x^a/r)\frac{s^2+(1-\xi^2(s,r))r^2}{\xi(s,r)r(s^2+r^2)^{1/2} + (s^2+r^2)} \del_t u, 
$$
that is, $|\delsN_au-\delsME_a u| \leq \zeta^2|\del_t u| \leq \zeta \, | \del_t u|$
and, therefore, 
$|\delsN_au| \leq |\delsME_au| + \zeta \, |\del_t u|$. 
Finally, recalling also Lemma~\ref{Lem1-05-May-2020}, we have $\frac{|r-t|+1}{r}\leq \zeta^2$ and the desired result is established.
\end{proof}

%----------------------------------------------------------------------

\begin{proof}[Proof of Proposition~\ref{prop1-07-02-2020}] 

\begin{subequations}
Here, we rely on the identities 
\begin{equation}\label{eq2-26-05-2021}
\del_t\delsN_a u = \delsN_a\del_t u =  (x^a/r)(1-r/t) \del_t \, \del_t + t^{-1}L_a\del_t u, 
\end{equation}
\begin{equation}\label{eq1-07-02-2020}
\delsN_a\delsN_bu =  t^{-1}L_a\delsN_bu + (x^a/r)(1-r/t)\del_t\delsN_b u. 
\end{equation}
For any ordered operator $Z=\del^IL^J\Omega^K$ satisfying $\ord(Z)\leq p$ and $\rank(Z)\leq k$ we write
$$
| Z \del_t \delsN_au| 
\lesssim \big|\ Z \big((x^a/r)(1-r/t) \del_t \, \del_t u\big)\big| + \big| Z\big(t^{-1}L_a\del_t u\big)\big|. 
$$
Observe that $x^a/r$ is homogeneous of degree zero and $|Z(t^{-1}) \, |\lesssim r^{-1}$ in $\Mnear_{ [s_0, + \infty)}$ so that, in view of 
\eqref{eq2-07-02-2020}, \eqref{eq1-06-02-2020} is established for $|\del_t\delsN_au|_{p,k}$ (and we return to this inequality later in this proof). To handle \eqref{eq5-07-02-2020}, we use
\eqref{eq2-07-02-2020}.
Regarding \eqref{eq1-07-02-2020}, we note that
$$
\aligned
Z\big(\delsN_a\delsN_bu\big) = Z\big(t^{-1}L_a\delsN_bu\big) 
+ Z\big((x^a/r)(1-r/t)\del_t\delsN_b u\big).
\endaligned
$$
The first term in the right-hand side is bounded as 
$\big| Z \big(t^{-1}L_a\delsN_bu\big)\big| \lesssim    r^{-1} |\delsN u|_{p+1,k+1}$. 
For the second term, we rely on \eqref{eq2-07-02-2020}  and \eqref{eq1-06-02-2020} on $|\del_t\delsN_a u|$ (already established) and get 
$$
\aligned
\big| Z\big((x^a/r)(1-r/t)\del_t\delsN_b u\big)\big|
& \lesssim    \frac{|r-t|+1}{r} |\del_{ t}\delsN_{ b} u|_{p,k}
\lesssim \frac{|r-t|}{r}|\del_{ t}\delsN_{ b} u|_{p,k} + r^{-1}|\delsN u|_{p+1,k}
\\
& \lesssim  \frac{|r-t|^2}{r^2} |\del \del u|_{p,k} +  \frac{|r-t|}{r^2} |\del u|_{p+1,k+1} + r^{-1}|\delsN u|_{p+1,k}, 
\endaligned
$$
which implies \eqref{eq5-07-02-2020}.
It remains to handle the bound for $|\del_a\delsN_bu|_{p,k}$ and 
we observe that $\del_a\delsN_bu = \delsN_a\delsN_b u - (x^a/r)\del_t\delsN_b u$ and \eqref{eq5-07-02-2020} together with \eqref{eq1-06-02-2020} on $|\del_t\delsN_a u|_{p,k}$, we conclude with \eqref{eq1-06-02-2020} for $|\del_a\delsN_bu|_{p,k}$.
Finally, \eqref{eq4-07-02-2020} and \eqref{eq6-07-02-2020} are deduced from Lemma~\ref{lme1-26-05-2021}. 
\end{subequations}
\end{proof}

%----------------------------------------------------------- 

\begin{proof}[Proof of Proposition~\ref{prop1-10-02-2020}] 

Let us derive first the bounds in $\Mnear_{[s_0,s_1]}$ { with $\LOmega = L_a$}. Using the decomposition 
$L_a u = t\delsN_a u + (x^a/r)(r-t)\del_t u$
and recalling Lemmas~\ref{lem 1 homo-ext} { and~\ref{lem1-31-05-2021}}, in $\MMEnear_{[s_0,s_1]}$ we find 
$$
| Z L_au| \leq |\del^IL^J\Omega^K(t\delsN_au)| + | Z((x^a/r)(r-t) \del_t u)| 
\lesssim t \, | \delsN u|_{p,k} + (|r-t|+1) \, |\del u|_{p,k}.
$$
Furthermore, for $L_aL_bu$ we have 
$
| Z L_aL_bu| \lesssim |L_bu|_{p+1,k+1} \lesssim t \, | \delsN u|_{p+1,k+1} + (|r-t|+1) \, |\del u|_{p+1,k+1}.
$
On the other hand, using the notion of homogeneity we have
$$ 
L_aL_bu 
= r\big((t/r)\del_t + (x^a/r)\del_a\big) L_bu
$$ 
and, therefore, we have 
$|L_aL_bu|_{p,k} \lesssim t|L_b u|_{p+1,k}$ in $\Mnear_{[s_0,s_1]}$. 
Next, we claim that the following inequality holds: 
\begin{equation}\label{eq1-23-06-2021}
|\Omega_{ab} u|_{p,k}\lesssim |L u|_{p,k} \qquad \text{ in }\Mnear_{[s_0,s_1]}.
\end{equation}
This is true since $\Omega_{ab} = (x^a/t)L_b - (x^b/t)L_a$ and $|x^a/t|_{p,k}\lesssim 1$ in $\Mnear_{[s_0,s_1]}$ (which can be checked by induction). This shows the desired bounds on $|\Omega_{ab}u|$, { $|\Omega_{ab}L_c|, |L_c\Omega_{ab}|$} and $|\Omega_{ab}\Omega_{cd}u|$.
On the other hand, concerning the bound in $\Mfar_{[s_0,s_1]}$, we only need to write  
$$
r^{-1}L_au = (x^a/r)\del_t + (t/r)\del_a,\qquad r^{-1}\Omega_{ab}u =  (x^a/r)\del_b - (x^b/r)\del_a.
$$
In both right-hand sides, the coefficients are homogeneous of degree zero. In view of Lemma~\ref{lem 1 homo-ext}, the last 
inequality \eqref{eq1-24-06-2021} is  thus also established.
\end{proof}

%---------------------------------------------------------------------------------------

\begin{proof}[Proof of Lemma~\ref{lemma-611}] 
\begin{subequations}
Observe first that
\begin{equation}\label{eq2-01-06-2021}
[\del^IL^J\Omega^K,L_a] = \del^IL^J\big([\Omega^K,L_a]\big) + \del^I\big([L^J,L_a]\big)\Omega^K + [\del^I,L_a]L^J\Omega^K
=: T_1 + T_2 + T_3. 
\end{equation}
The first and third terms are easily decomposed thanks to \eqref{eq 4' comm} and \eqref{eq 1 comm}, as follows: 
$$
T_1  
\cong \sum_{|K'|<|K|} \sum_b \del^IL^JL_b\Omega^{K'} ,
\qquad\qquad
T_3 
\cong \sum_{|I'| = |I|{ \geq1}}\del^{I'}L^J\Omega^K, 
$$
which are included in the right-hand side of \eqref{eq1-01-06-2021}. 
Concerning $T_2$, by induction we easily check that 
\begin{equation}\label{eq1-02-06-2021}
[L^J,L_a]\cong \sum_{|J'|<|J|\atop c<d}L^{J'}\Omega_{cd} + \sum_{|J'|<|J|}L^{J'}. 
\end{equation}
It then follows that 
$
\del^I\big([L^J,L_a]\big)\Omega^K \cong \sum_{|J'|<|J|\atop c<d} \big(\del^IL^{J'}\Omega_{cd}\Omega^K + \del^IL^{J'}\Omega^K\big), 
$
which is also a term appearing in the right-hand side of \eqref{eq1-01-06-2021}. 
Finally, when $p=k$ the identity \eqref{eq2-01-06-2021} becomes
$
[\del^IL^J\Omega^K,L_a] = L^J\big([\Omega^K,L_a]\big) +  [L^J,L_a] \Omega^K, 
$
and in view of \eqref{eq 4' comm} and \eqref{eq1-02-06-2021}, we have thus established \eqref{eq3-02-06-2021}. 
\end{subequations}
\end{proof}

%-------------------------------------

\begin{proof}[Proof of Lemma~\ref{lem1-commu-ext}]
We rely on the identity \eqref{eq1-28-05-2021} and write
$$
\aligned
& \, [Z,\delsN_a\del_t]u 
= [Z,t^{-1}L_a\del_t]u + [Z,(x^a/r)(1-r/t)\del_t\del_t]u
\\
& =  t^{-1}[Z,L_a\del_t]u + (x^a/r)(1-r/t)[Z,\del_t\del_t]u
+ \sum_{Z_1 \odot Z_2=Z\atop \ord(Z_1)\geq 1}\big(Z_1(t^{-1})Z_2L_a\del_tu + Z_1\big((x^a/r)(1-r/t)\big)Z_2\del_t\del_tu\big)
\\
& =   t^{-1}[Z,L_a]\del_t u + t^{-1}L_a([Z,\del_t]u) + (x^a/r)(1-r/t)[Z,\del_t\del_t]u
+ \sum_{Z_1\odot Z_2=Z\atop \ord(Z_1)\geq 1}\big(Z_1(t^{-1})Z_2L_a\del_tu + Z_1\big((x^a/r)(1-r/t)\big)Z_2\del_t\del_tu\big).
\endaligned
$$ 
The first term in the right-hand side is bounded by $t^{-1}|\del u|_{p,k}$ thanks to \eqref{eq1-01-06-2021}. The second term is also bounded by $t^{-1}|\del u|_{p,k}$ via \eqref{eq 1 lem 1 depo-cH}. The third term is bounded by $\frac{|r-t|}{t}|\del\del u|_{p-1,k-1}$ by applying \eqref{eq 2 decompo-comm-H}. For the last two terms, we observe that
$$
\sum_{Z_1\odot Z_2=Z\atop \ord(Z) \geq 1}Z_1(t^{-1})Z_2L_a\del_tu 
= \sum_{Z_1 \odot Z_2=Z\atop \deg(Z_1) = 0, \, \rank(Z_1) \geq 1} Z_1(t^{-1})Z_2L_a\del_tu
+ \sum_{Z_1 \odot Z_2=Z\atop \deg(Z_1) \geq 1}Z_1(t^{-1}) Z_2 L_a\del_tu. 
$$
The first term in the right-hand side is bounded as follows: observe that $\rank(Z_2)\leq k-1$ so that this term is bounded by $t^{-1}|\del u|_{p,k}$. For the second term, we use that $\deg(Z_1) \geq 1$ or, equivalently, $|Z_1(t^{-1})| \lesssim t^{-2}$, so that 
the second term is bounded by
$$
t^{-2} \, \big| Z_2 L_a\del_tu\big|\lesssim t^{-1}|\del u|_{p,k}.
$$
Here we have used $L_a = t\del_a + x^a\del_t$ and we are near the light cone with  
$|I_2| \leq |I|-1$. A similar argument is applied to $\sum_{Z_1 \odot Z_2=Z\atop \ord(Z_1)\geq 1} Z_1\big((x^a/r)(1-r/t)\big)Z_2\del_t\del_tu$ and we find 
$$
\sum_{Z_1 \odot Z_2=Z\atop \ord(Z_1)\geq 1} \big|Z_1\big((x^a/r)(1-r/t)\big)Z_2\del_t\del_tu\big|
\lesssim \frac{|r-t|}{t}|\del\del u|_{p-1,k-1} + t^{-1}|\del u|_{p,k}.
$$
We have reached the conclusion \eqref{eq1-lem1-commu-ext}.
On the other hand, the proof of \eqref{eq2-lem1-commu-ext} is similar, and we rely on the decomposition
$$
\aligned
\delsN_a \delsN_b & =  \big((x^a/r)\del_a + \del_t\big)\big(t^{-1}L_b + (x^b/r)(1-r/t)\del_t\big)
\\
& =  t^{-1}(x^a/r)\del_aL_b + t^{-1}\del_tL_b - t^{-1}\delsN_b 
\\
& \quad 
+ (x^ax^b/r^2)(1-r/t)\del_a\del_t + (x^a/r)\del_a\big((x^b/r)(1-r/t)\big)\del_t
+(x^b/r)(1-r/t)\del_t\del_t + \del_t\big((x^b/r)(1-r/t)\big)\del_t
\\
& \cong
t^{-1} \Lambda_{1,ab}^{\alpha c} \del_\alpha L_c + (1-r/t)\Lambda_{2,ab}^{\alpha\beta}\del_\alpha\del_\beta + t^{-1}(1-r/t)\Lambda_{3,ab}^{\gamma}\del_{\gamma}
+t^{-1}\Lambda_{4,ab}^{\gamma}\del_{\gamma}, 
\endaligned
$$
where $\Lambda_{i,ab}$ are homogeneous expressions of degree zero. 
We can commute each term with $Z^I$ and the calculations are similar to the ones in the proof of \eqref{eq1-lem1-commu-ext}; we omit the details.
\end{proof}

%==============================================================================================

\section{Proof of pointwise decay properties of wave fields}
\label{appenDD} 

We provide here a proof of Proposition~\ref{Linfini wave} concerning the effect of a  source-term on the solutions to the wave equation and, specifically, we analyze the integral \eqref{Kirchhoff-source-basic}. For the convenience of the reader we recall here our notation. 
$$
\aligned
& \text{\bf Case 1:}  
&&\alpha_1 = -1+\upsilon ,
\quad
&& \alpha_2 = -1-\nu,
\quad
&& \alpha_3=-1+\mu,
\quad 
&& \upsilon + \mu < \nu,
\quad 
&& 0<\mu,\nu,\upsilon\leq 1/2.  
\\
& \text{\bf Case 2:}  
&&\alpha_1=0 ,
&&\alpha_2 = -2-\nu,
&& \alpha_3 = -1+\mu ,
&& 0 < \nu, \mu \leq 1/2. 
\\
&\text{\bf Case 3:} 
&&\alpha_1 = 0,
&& \alpha_2 = -2,
&& \alpha_3= -1-\mu,
&& \mu\in (0,1/2). 
\\
&\text{\bf  Case 4:} 
&& \alpha_1 = 0,
&& \alpha_2 = -2+\nu,
&& \alpha_3 = -1-\mu,
&&0<\nu< \mu < 1/2. 
\endaligned
$$ 

%--------------------------------------------------------------------------------------------------------------------------------------------------- 

\subsection{Bounds on the integral $I(\lambda;t,r)$}

Using  the change of variable $\omega  = \lambda^{-1}\rho^{-1/2}$ and observing that $\lambda\in[t^{-1},1] $ implies $\lambda^{-1}(r/t+1)-1 \geq 0$, we obtain  
\begin{equation}\label{eq1-06-01-2021}
\aligned
I(\lambda;t,r) & =  \frac{t}{r}\lambda^{\alpha_1+\alpha_2+\alpha_3 + 2}
\int_{|\lambda^{-1}(r/t-1)+1|}^{ |\lambda^{-1}(r/t+1)-1|}
\omega\big(1+\omega\big)^{\alpha_2}\big((\lambda t)^{-1} + |1-\omega|\big)^{\alpha_3} \, d\omega
\\
& \lesssim   \frac{t}{r}\lambda^{\alpha_1+\alpha_2+\alpha_3 + 2}
\int_{|\lambda^{-1}(r/t-1)+1|}^{\lambda^{-1}(r/t+1)-1}
\big(1+\omega\big)^{\alpha_2+1}\big((\lambda t)^{-1} + |1-\omega|\big)^{\alpha_3} \, d\omega
\\
& =   \frac{t}{r}\lambda^{\alpha_1+\alpha_2+\alpha_3 + 2}
\int_{|Z-1|}^{Y-1}
\big(1+\omega\big)^{\alpha_2+1}\big((\lambda t)^{-1} + |1-\omega|\big)^{\alpha_3} \, d\omega. 
\endaligned
\end{equation}
In order to analyze this expression further, we need to distinguish between several regimes, depending whether the expressions 
$
Z (\lambda; t, r) :=\lambda^{-1}(1 - r/t) 
$
and $
Y(\lambda; t, r) := \lambda^{-1}(1+ r/t)
$
belong to one of the following intervals: 
$$ 
Z \in (-\infty, -1] \cup [-1, 0] \cup [0, 1] \cup [1,2] \cup [2,4] \cup [4, +\infty), 
\qquad 
Y \in [0,2] \cup [2, 3] \cup [3, +\infty). 
$$

%----------------------------------

\paragraph{Regime I : $Z\leq -1$ and $Y\geq 3$.}   
% $\lambda^{-1}(r/t-1)+1\geq 2$ and $\lambda^{-1}(r/t+1)-1\geq 2$.} 
%
This is equivalent to
$
\lambda\leq  \frac{r-t}{t}$
and 
$ \lambda \leq \frac{r+t}{3t}.
$
In order to have $t^{-1}\leq \frac{r-t}{t}$ so that  this regime is non-empty, we require that 
$r-t\geq 1$.  
Then we have $1+\omega \lesssim |1-\omega| = \omega-1$ and 
\begin{equation}\label{eq4-08-01-2021}
I(\lambda;t,r)
\lesssim (t/r)\lambda^{\alpha_1+\alpha_2+\alpha_3+2} \int_{\lambda^{-1}(r/t-1)+1}^{\lambda^{-1}(r/t+1)-1}
(1+\omega)^{1+\alpha_2+\alpha_3} \, d\omega
\lesssim
\begin{cases}
|\nu-\mu|^{-1}(t/r)\lambda^{-1+\upsilon + \mu-\nu},\quad  &\text{Case 1,}
\\
(t/r)\lambda^{-1+\mu-\nu},\quad  &\text{Case 2},
\\
(t/r)\lambda^{-1-\mu},\quad &\text{Case 3},
\\
(t/r)\lambda^{-1-\mu+\nu},&\text{Case 4}.
\end{cases}
\end{equation}

\paragraph{Regime II : $-1 \leq Z  \leq 0$ and $Y\geq 3$.} 
% $1\leq \lambda^{-1}(r/t-1)+1\leq 2$ and $\lambda^{-1}(r/t+1)-1\geq 2$.}
This is equivalent to 
$
r\geq t$
and 
$  \lambda \geq\frac{r-t}{t}$
with 
$   \lambda\leq \frac{r+t}{3t}
$
and consequently, since   $\omega \geq 1$, we write $|\omega-1| = \omega-1$ and so
$$
\aligned
I(\lambda;t,r)
& \lesssim  (t/r)\lambda^{\alpha_1+\alpha_2+\alpha_3+2} \int_{\lambda^{-1}(r/t-1)+1}^{\lambda^{-1}(r/t+1)-1}
(1+\omega)^{1+\alpha_2}  \big((\lambda t)^{-1} + \omega -1\big)^{\alpha_3} \, d\omega.
\\
& \lesssim  (t/r)\lambda^{\alpha_1+\alpha_2+\alpha_3+2} 
\Big( 
\int_{\lambda^{-1}(r/t-1)+1}^2 + \int_2^{\lambda^{-1}(r/t+1)-1}
\Big)
(1+\omega)^{\alpha_2+1}  \big((\lambda t)^{-1} + \omega -1\big)^{\alpha_3} \, d\omega
\\
& \lesssim  (t/r)\lambda^{\alpha_2+\alpha_2+\alpha_3+2} 
\Big(\int_{\lambda^{-1}(r/t-1)+1}^2
\big((\lambda t)^{-1} + \omega -1\big)^{\alpha_3} \, d\omega
+  \int_2^{\lambda^{-1}(r/t+1)-1}
(1+\omega)^{1+\alpha_2+\alpha_3} d\omega\Big).
\endaligned
$$
This leads us to 
\begin{equation}\label{eq3-08-01-2021}
I(\lambda;t,r)\lesssim
\left\{
\aligned
&\big(\mu^{-1} + |\nu-\mu|^{-1}\big)(t/r)\lambda^{-1+\upsilon + \mu-\nu},\quad  &&\text{Case 1},
\\
&\mu^{-1}(t/r)\lambda^{-1+\mu-\nu}, &&\text{Case 2},
\\
&\mu^{-1}\Big(\frac{r-t+1}{t}\Big)^{-\mu}(t/r)\lambda^{-1} + (t/r)\lambda^{-1-\mu},&& \text{Case 3},
\\
&\mu^{-1}\Big(\frac{r-t+1}{t}\Big)^{-\mu}(t/r)\lambda^{-1+\nu} + (t/r)\lambda^{-1-\mu+\nu},
\quad && \text{Case 4}.
\endaligned
\right.
\end{equation}

\paragraph{Regime III :  $-1 \leq Z \leq 0$ and $Y \leq 3$.}  
% $1\leq \lambda^{-1}(r/t-1)+1\leq 2$ and $\lambda^{-1}(r/t+1)-1 \leq 2$.} 
This is equivalent to 
$
r\geq t,$
and $ \lambda \geq \frac{r-t}{t}$
with 
$  \lambda \geq \frac{r+t}{3t}.
$
Observing that $\omega \geq 1$, we find $|\omega-1| = \omega -1$ and then
$$
\aligned
I(\lambda;t,r)  & \lesssim  (t/r)\lambda^{\alpha_1+\alpha_2+\alpha_3+2} \int_{\lambda^{-1}(r/t-1)+1}^{\lambda^{-1}(r/t+1)-1}
(1+\omega)^{\alpha_2+1}  \big((\lambda t)^{-1} + \omega -1\big)^{\alpha_3} \, d\omega
\\
& \lesssim  (t/r)\lambda^{\alpha_1+\alpha_2+\alpha_3+2} \int_{\lambda^{-1}(r/t-1)+1}^{\lambda^{-1}(r/t+1)-1}
\big((\lambda t)^{-1} + \omega -1\big)^{\alpha_3} \, d\omega, 
\endaligned
$$
that is, by recalling that $\lambda \geq 1/3$,
\begin{equation}\label{eq3-06-01-2021}
I(\lambda;t,r)\lesssim 
\left\{
\aligned
&\mu^{-1}(t/r),\quad &&\text{Cases 1 and 2},
\\
&\mu^{-1}(t/r)\Big(\frac{r-t+1}{t}\Big)^{-\mu},\quad &&\text{Cases 3 and 4}.
\endaligned
\right. 
\end{equation}

\paragraph{Regime IV :  $0 \leq Z  \leq 1$ and $Y \geq 2$.} 
%  $0\leq 1-\lambda^{-1}(1-r/t) \leq 1$ and $\lambda^{-1}(r/t+1)-1\geq 1$.}
This is equivalent to saying
$
r\leq t$
with 
$ \frac{t-r}{t}\leq \lambda$ and 
$ \lambda\leq \frac{t+r}{2t}
$ 
and we find
$$
\aligned
I(\lambda;t,r) & \lesssim  
(t/r)\lambda^{\alpha_1+\alpha_2+\alpha_3+2}
\int_{1-\lambda^{-1}(1-r/t)}^{\lambda^{-1}(r/t+1)-1}
(1+\omega)^{\alpha_2+1}\big((\lambda t)^{-1} + |\omega-1|\big)^{\alpha_3} \, d\omega
\\
& =  (t/r)\lambda^{\alpha_1+\alpha_2+\alpha_3+2}\int_{1-\lambda^{-1}(1-r/t)}^1
+ \int_1^{\lambda^{-1}(r/t+1)-1}
(1+\omega)^{\alpha_2+1}\big((\lambda t)^{-1} + |\omega-1|\big)^{\alpha_3} \, d\omega
\\
& =:  I_1(\lambda;t,r) + I_2(\lambda;t,r). 
\endaligned
$$
For the term $I_1$, we observe that $(1+\omega)^{\alpha_2+1}\lesssim 1$, therefore 
$$
\aligned
I_1(\lambda ;t,r)\lesssim (t/r)\lambda^{\alpha_1+\alpha_2+\alpha_3+2}\int_{1-\lambda^{-1}(1-r/t)}^1
\big((\lambda t)^{-1} + 1 - \omega\big)^{\alpha_3} \, d\omega
\lesssim
\left\{
\aligned
&\mu^{-1}(t/r)\Big(\frac{t-r}{t}\Big)^{\mu}\lambda^{-1+\upsilon -\nu},\quad && \text{Case 1},
\\
&\mu^{-1}(t/r)\Big(\frac{t-r}{t}\Big)^{\mu}\lambda^{-1 -\nu},\quad && \text{Case 2},
\\
&\mu^{-1}(t/r)t^{\mu}\lambda^{-1},\quad &&\text{Case 3},
\\
&\mu^{-1}(t/r)t^{\mu}\lambda^{-1+\nu} ,\quad &&\text{Case 4}. 
\endaligned
\right.
\endaligned
$$
In the regime under consideration we have $\lambda \geq \frac{t-r}{t}$ and we conclude that
$$
I_1(\lambda;t,r)\lesssim 
\left\{
\aligned
&\mu^{-1}(t/r)\lambda^{-1+\upsilon +\mu -\nu},\quad && \text{Case 1},
\\
&\mu^{-1}(t/r)\lambda^{-1 +\mu -\nu},\quad && \text{Case 2},
\\
&\mu^{-1}(t/r)t^{\mu}\lambda^{-1},\quad &&\text{Case 3},
\\
&\mu^{-1}(t/r)t^{\mu}\lambda^{-1+\nu} ,\quad &&\text{Case 4}.
\endaligned
\right.
$$
%------------------ 
Concerning $I_2$, since $\omega\geq 1$ we have 
$$
I_2(\lambda;t,r)\lesssim (t/r)\lambda^{\alpha_1+\alpha_2+\alpha_3+2}\int_1^{\lambda^{-1}(r/t+1)-1}
(1+\omega)^{\alpha_2+1}\big((\lambda t)^{-1} + \omega -1\big)^{\alpha_3} \, d\omega.
$$
On one hand, when $\lambda^{-1}(r/t+1)-1\geq 2$ or, equivalently, $\lambda \leq \frac{t+r}{3t}$,  we write 
$$
\aligned
I_2(\lambda;t,r) & \lesssim   (t/r)\lambda^{\alpha_1+\alpha_2+\alpha_3+2}\int_1^2 + \int_2^{\lambda^{-1}(r/t+1)-1}
(1+\omega)^{\alpha_2+1}\big((\lambda t)^{-1} + \omega -1\big)^{\alpha_3} \, d\omega
\\
& \lesssim  (t/r)\lambda^{\alpha_1+\alpha_2+\alpha_3+2}\int_1^2\big((\lambda t)^{-1} + \omega -1\big)^{\alpha_3} \, d\omega
+  (t/r)\lambda^{\alpha_1+\alpha_2+\alpha_3+2}\int_2^{\lambda^{-1}(r/t+1)-1}
(1+\omega)^{1+\alpha_2+\alpha_3} \, d\omega, 
\endaligned
$$
and this leads us to
$$
I_2(\lambda;t,r)\lesssim
\begin{cases}
(\mu^{-1} + |\nu-\mu|^{-1})(t/r)\lambda^{-1+\upsilon + \mu-\nu},\quad & \text{Case 1},
\\
\mu^{-1}(t/r)\lambda^{-1+\mu-\nu},\quad & \text{Case 2},
\\
\mu^{-1}(t/r)t^{\mu}\lambda^{-1} ,&\text{Case 3,}
\\
\mu^{-1}(t/r)t^{\mu}\lambda^{-1+\nu}, &\text{Case 4}.
\end{cases}
$$
In Case 4 we used $\lambda \geq t^{-1}$. 
On other hand, when $\lambda^{-1}(r/t+1)-1\leq 2$ or, equivalently, $\lambda \geq \frac{t+r}{3t}$, we write 
$$
I_2(\lambda;t,r)\lesssim (t/r)\lambda^{\alpha_1+\alpha_2+\alpha_3+2}\int_1^{\lambda^{-1}(r/t+1)-1}
\big((\lambda t)^{-1} + \omega -1\big)^{\alpha_3} \, d\omega 
\lesssim 
\begin{cases} 
\mu^{-1}(t/r)\lambda^{-1+\upsilon+\mu-\nu}, \quad & \text{Case 1},
\\
\mu^{-1}(t/r)\lambda^{-1+\mu-\nu},\quad & \text{Case 2},
\\
\mu^{-1}(t/r)t^{\mu}\lambda^{-1}, \quad & \text{Case 3}, 
\\
\mu^{-1}(t/r)t^{\mu}\lambda^{-1+\nu}, &\text{Case 4}
\end{cases} 
$$
and we conclude that, in the region under consideration, 
\begin{equation}\label{eq4-06-01-2021}
I(\lambda;t,r)\lesssim
\left\{
\aligned 
&(\mu^{-1} + |\nu-\mu|^{-1})(t/r)\lambda^{-1+\upsilon + \mu-\nu},\quad && \text{Case 1},
\\
&\mu^{-1}(t/r)\lambda^{-1+\mu-\nu},\quad && \text{Case 2},
\\
&\mu^{-1}(t/r)t^{\mu}\lambda^{-1} ,&&\text{Case 3},
\\
&\mu^{-1}(t/r)t^{\mu}\lambda^{-1+\nu}, &&\text{Case 4}.
\endaligned
\right.
\end{equation}

\paragraph{Regime V : $0\leq Z \leq 1$ and $Y\leq 2$.} 
% $0\leq 1-\lambda^{-1}(1-r/t) \leq 1$ and $\lambda^{-1}(r/t+1)-1\leq 1$.}
In this regime we have 
$
r\leq t$
and  
$\frac{t-r}{t}\leq \lambda$
with  
$\frac{t+r}{2t}\leq \lambda
$ 
and we always have $\omega \leq 1$. Then we find 
$$
\aligned
I(\lambda;t,r)& \lesssim  (t/r)\lambda^{\alpha_1+\alpha_2+\alpha_3+2}\int_{1-\lambda^{-1}(1-r/t)}^{\lambda^{-1}(r/t+1)-1}
(1+\omega)^{\alpha_2+1}\big((\lambda t)^{-1} +1-\omega\big)^{\alpha_3} \, d\omega
\\
& \lesssim  (t/r)\lambda^{\alpha_1+\alpha_2+\alpha_3+2}\int_{1-\lambda^{-1}(1-r/t)}^{\lambda^{-1}(r/t+1)-1}
\big((\lambda t)^{-1} + 1-\omega \big)^{\alpha_3} \, d\omega.
\endaligned
$$
Since in this regime $1\geq \lambda \geq\frac{t+r}{2t}\geq 1/2$, we conclude that
\begin{equation}\label{eq1-07-01-2021}
I(\lambda;t,r)\lesssim
\left\{
\aligned
&\mu^{-1}(t/r)\Big(\frac{t-r+1}{t}\Big)^{\mu},\quad &&\text{Case 1},
\\
&\mu^{-1}(t/r)\Big(\frac{t-r+1}{t}\Big)^{\mu},\quad &&\text{Case 2},
\\
&\mu^{-1}(t/r)\Big(\lambda - \frac{t+r-1}{2t}\Big)^{-\mu},\quad &&\text{Cases 3 and 4}.
\endaligned
\right.
\end{equation}

\paragraph{Regime VI :  $1\leq Z \leq 2$ and $Y \geq 2$.} 
% $-1\leq 1-\lambda^{-1}(1 - r/t)\leq 0$ and $\lambda^{-1}(r/t+1)-1\geq 1$.}
We then obtain 
$
\frac{t-r}{2t}\leq \lambda\leq \frac{t-r}{t}$
and 
$ 
\lambda \leq \frac{t+r}{2t}, 
$
and in order for this regime to be non-empty, we must assume $\frac{t-r}{t}\geq t^{-1}$ or, equivalently, 
$t-r\geq 1$. Then we find 
$$
\aligned
I(\lambda;t,r)
& \lesssim  (t/r)\lambda^{\alpha_1+\alpha_2+\alpha_3+2}\int_{\lambda^{-1}(1-r/t)-1}^{\lambda^{-1}(r/t+1)-1}
(1+\omega)^{1+\alpha_2}\big((\lambda t)^{-1} + |1-\omega|\big)^{\alpha_3} \, d\omega
\\
& =(t/r)\lambda^{\alpha_1+\alpha_2+\alpha_3+2}\int_{\lambda^{-1}(1-r/t)-1}^1 + \int_1^{\lambda^{-1}(r/t+1)-1}
(1+\omega)^{1+\alpha_2}\big((\lambda t)^{-1} + |1-\omega|\big)^{\alpha_3} \, d\omega
\\
& =: I_1(\lambda;t,r)+I_2(\lambda;t,r).
\endaligned
$$
For $I_1$, we observe that $\omega\leq 1$, thus
$$
\aligned
I_1(\lambda;t,r) 
& =  (t/r)\lambda^{\alpha_1+\alpha_2+\alpha_3+2}\int_{\lambda^{-1}(1-r/t)-1}^1 
(1+\omega)^{1+\alpha_2}\big((\lambda t)^{-1} + 1-\omega\big)^{\alpha_3} \, d\omega
\\
& \lesssim   (t/r)\lambda^{\alpha_1+\alpha_2+\alpha_3+2}\int_{\lambda^{-1}(1-r/t)-1}^1 
\big((\lambda t)^{-1} + 1-\omega\big)^{\alpha_3} \, d\omega
\\
& \lesssim 
\left\{
\aligned 
& \mu^{-1}(t/r)\Big(\lambda - \frac{t-r-1}{2t}\Big)^{\mu}\lambda^{-1+\upsilon-\nu}\lesssim \mu^{-1}(t/r)\lambda^{-1+\upsilon+\mu-\nu},\quad &&\text{Case 1},
\\
& \mu^{-1}(t/r)\Big(\lambda - \frac{t-r-1}{2t}\Big)^{\mu}\lambda^{-1-\nu}\lesssim  \mu^{-1}(t/r)\lambda^{-1+\mu-\nu},\quad &&\text{Case 2},
\\
&\mu^{-1}(t/r)t^{\mu}\lambda^{-1}, &&\text{Case 3},
\\
&\mu^{-1}(t/r)t^{\mu}\lambda^{-1+\nu},&&\text{Case 4}.
\endaligned
\right.
\endaligned
$$
%Here we observeed that $(\lambda t)^{-1}\leq 1$ and $2\geq 2 - \lambda^{-1}(1-r/t)\geq 0$. 
%
For $I_2$, we have $\omega \geq 1$ and therefore 
$$
I_2(\lambda;t,r)\lesssim (t/r)\lambda^{\alpha_1+\alpha_2+\alpha_3+2}\int_1^{\lambda^{-1}(r/t+1)-1}
(1+\omega)^{1+\alpha_2}\big((\lambda t)^{-1} + \omega -1 \big)^{\alpha_3} \, d\omega
$$ 
\begin{itemize}

\item When $\lambda^{-1}(r/t+1)-1 \geq 2$ which is equivalent to saying $\lambda \leq \frac{t+r}{3t}$ we arrive at 
$$
\aligned
I_2(\lambda;t,r)& \lesssim  (t/r)\lambda^{\alpha_1+\alpha_2+\alpha_3+2}
\int_1^2 + \int_2^{\lambda^{-1}(r/t+1)-1}
(1+\omega)^{1+\alpha_2}\big((\lambda t)^{-1} + \omega -1 \big)^{\alpha_3} \, d\omega
\\
& \lesssim  (t/r)\lambda^{\alpha_1+\alpha_2+\alpha_3+2} \int_1^2 
\big((\lambda t)^{-1} + \omega -1 \big)^{\alpha_3} \, d\omega
+  \int_2^{\lambda^{-1}(r/t+1)-1}
(1+\omega)^{1+\alpha_2+\alpha_3} \, d\omega
\\
& \lesssim 
\left\{
\aligned
&\big(\mu^{-1}+ |\nu-\mu|^{-1}\big)(t/r)\lambda^{-1+\upsilon + \mu-\nu},\quad &&\text{Case 1},
\\
&\mu^{-1}(t/r)\lambda^{-1+\mu-\nu},\quad &&\text{Case 2},
\\
&\mu^{-1}(t/r)t^{\mu}\lambda^{-1},\quad &&\text{Case 3},
\\
&\mu^{-1}(t/r)t^{\mu}\lambda^{-1+\nu},\quad &&\text{Case 4},
\endaligned
\right.
\endaligned
$$

\item When $\lambda^{-1}(r/t+1)-1\leq 2$ which is equivalent to saying $\lambda \geq \frac{t+r}{3t}$, we find 
$$
\aligned
I_2(\lambda;t,r)& \lesssim  (t/r)\lambda^{\alpha_1+\alpha_2+\alpha_3+2}\int_1^{\lambda^{-1}(r/t+1)-1}
\big((\lambda t)^{-1} + \omega -1 \big)^{\alpha_2} \, d\omega
\lesssim
\left\{
\aligned
&\mu^{-1}(t/r)\lambda^{-1+\upsilon + \mu - \nu},\quad &&\text{Case 1},
\\
&\mu^{-1}(t/r)\lambda^{-1+\mu-\nu},\quad &&\text{Case 2},
\\
&\mu^{-1}(t/r)t^{\mu}\lambda^{-1},\quad &&\text{Case 3},
\\
&\mu^{-1}(t/r)t^{\mu}\lambda^{-1+\nu},\quad &&\text{Case 4}.
\endaligned
\right.
\endaligned
$$
\end{itemize}
Consequently we arrive at 
\begin{equation}\label{eq2-07-01-2021}
I(\lambda;t,r)\lesssim
\begin{cases}
\big(\mu^{-1}+ |\nu-\mu|^{-1}\big)(t/r)\lambda^{-1+\upsilon + \mu-\nu},\quad &\text{Case 1},
\\
\mu^{-1}(t/r)\lambda^{-1+\mu-\nu},\quad &\text{Case 2},
\\
\mu^{-1}(t/r)t^{\mu}\lambda^{-1},\quad &\text{Case 3},
\\
\mu^{-1}(t/r)t^{\mu}\lambda^{-1+\nu},\quad &\text{Case 4}.
\end{cases} 
\end{equation}

\paragraph{Regime VII :  $1\leq Z \leq 2$ and $Y\leq 2$.}   
In this regime we have 
$
\frac{t-r}{2t}\leq \lambda \leq \frac{t-r}{t}$
and 
$ \lambda \geq \frac{t+r}{2t}.
$
In order for this regime to be non-empty, we also assume that  $\frac{t-r}{t}\geq t^{-1}$ which is equivalent to $t-r\geq 1$.
Then, $\omega \leq 1$ and we have 
$$
\aligned
I(\lambda;t,r)\lesssim (t/r)\lambda^{\alpha_1+\alpha_2+\alpha_3+2}\int_{\lambda^{-1}(1-r/t)-1}^{\lambda^{-1}(r/t+1)-1}
(1+\omega)^{\alpha_2+1}\big((\lambda t)^{-1} + 1 - \omega\big)^{\alpha_3} \, d\omega. 
\endaligned
$$
Observing that
$
(\lambda t)^{-1} + 1-\omega \geq (\lambda t)^{-1} + 2 - \lambda^{-1}(r/t+1) 
= 2\lambda^{-1} \Big(\lambda - \frac{t+r-1}{2t}\Big)
$
and recalling that $\lambda \geq\frac{t+r}{2t}\geq 1/2$, we arrive at 
\begin{equation}\label{eq3-07-01-2021}
I(\lambda;t,r)\lesssim 
\begin{cases}
\Big(\lambda - \frac{t+r-1}{2t}\Big)^{-1+\mu},\quad &\text{Cases 1 and 2},
\\
\Big(\lambda - \frac{t+r-1}{2t}\Big)^{-1-\mu},\quad &\text{Cases 3 and 4}.
\end{cases}
\end{equation}

\paragraph{Regime VIII :  $Z \geq 2$ for a base point {\sl far} from the light cone.}  
In this regime we have 
$
\lambda \leq \frac{t-r}{2t}
$
and this implies that $\lambda^{-1}(r/t+1)-1 \geq \frac{t+3r}{t-r}\geq 1$ and $\omega\geq 1$. In order for the regime to be non-empty, we assume $\frac{t-r}{2t}\geq t^{-1}$ or, equivalently, $ t-r \geq 2$.
Then we find 
$$
\aligned
I(\lambda;t,r)& \lesssim  (t/r)\lambda^{\alpha_1+\alpha_2+\alpha_3+2}\int_{\lambda^{-1}(1-r/t)-1}^{\lambda^{-1}(r/t+1)-1}
(1+\omega)^{1+\alpha_2}\big((\lambda t)^{-1} + \omega - 1\big)^{\alpha_3} \, d\omega
\endaligned
$$
and we observe that
$
(\lambda t)^{-1} + \omega -1\geq 2\lambda^{-1}\Big(\frac{t-r+1}{2t} - \lambda\Big)$
and $(1+\omega)\geq \lambda^{-1}(1-r/t).
$
We arrive at
\begin{equation}\label{eq1-08-01-2021}
I(\lambda;t,r)\lesssim 
\begin{cases}
\lambda^{\upsilon}\Big(\frac{t-r}{t}\Big)^{-\nu}\Big(\frac{t-r+1}{2t} - \lambda\Big)^{-1+\mu},\quad &\text{Case 1},
\\
\Big(\frac{t-r}{t}\Big)^{-1-\nu}\Big(\frac{t-r+1}{2t}-\lambda\Big)^{-1+\mu},\quad &\text{Case 2},
\\
\Big(\frac{t-r}{t}\Big)^{-1}\Big(\frac{t-r+1}{2t}-\lambda\Big)^{-1-\mu},\quad &\text{Case 3},
\\
\Big(\frac{t-r}{t}\Big)^{-1+\nu}\Big(\frac{t-r+1}{2t}-\lambda\Big)^{-1-\mu},\quad &\text{Case 4}.
\end{cases}
\end{equation}
This bound is relevant in the sense that it removes the singularity arising from $(t/r)$ when $r \to  0^+$.
However, this bound cannot be used when $\frac{t-r}{t}\to 0$, that is,  when $(t,x)$ is near the light cone.

%------------------------------------------------

\paragraph{Regime IX : $2\leq Z \leq 4$ and $r\geq t/3$ for a base point {\sl near} the light cone.}  

We then have 
$
\frac{t-r}{4t}\leq \lambda \leq \frac{t-r}{2t}
$
and as in the previous regime, we need $t-r\geq 2$ to guarantee that our conditions are non-empty.
Then we find 
$$
\aligned
I(\lambda;t,r)& \lesssim  (t/r)\lambda^{\alpha_1+\alpha_2+\alpha_3+2}\int_{\lambda^{-1}(1-r/t)-1}^{\lambda^{-1}(r/t+1)-1}
(1+\omega)^{1+\alpha_2}\big((\lambda t)^{-1} + \omega - 1\big)^{\alpha_3} \, d\omega
\endaligned
$$
and we observe that $r\geq t/3$ implies 
$\frac{t-r}{2t}\leq \frac{t+r}{4t}$. 
Thus $\lambda \leq \frac{t+r}{4t}$ which implies $ \lambda^{-1}(r/t+1)-1 \geq 3$. 
We have 
$$
\aligned
I(\lambda;t,r)& \lesssim  \lambda^{\alpha_1+\alpha_2+\alpha_3+2}
\Big(
\int_{\lambda^{-1}(1-r/t)-1}^3
+ \int_3^{\lambda^{-1}(r/t+1)-1}
\Big)
(1+\omega)^{1+\alpha_2}\big((\lambda t)^{-1} + \omega - 1\big)^{\alpha_3} \, d\omega
\\
& \lesssim  \lambda^{\alpha_1+\alpha_2+\alpha_3+2}\int_{\lambda^{-1}(1-r/t)-1}^3
\big((\lambda t)^{-1} + \omega - 1\big)^{\alpha_3} \, d\omega
+
\lambda^{\alpha_1+\alpha_2+\alpha_3+2}\int_3^{\lambda^{-1}(r/t+1)-1} (1+\omega)^{1+\alpha_2+\alpha_3} \, d\omega.
\endaligned
$$ 
Then, we obtain 
\begin{equation}\label{eq2-08-01-2021}
I(\lambda;t,r)\lesssim 
\left\{
\aligned
&\big(\mu^{-1} + |\nu-\mu|^{-1}\big)\lambda^{-1+\upsilon + \mu - \nu},\quad &&\text{Case 1},
\\
&\mu^{-1}\lambda^{1+\mu-\nu},\quad &&\text{Case 2},
\\
&\mu^{-1}\lambda^{-1}\Big(\frac{t-r+1}{2t}-\lambda\Big)^{-\mu} + \lambda^{-1-\mu},\quad &&\text{Case 3},
\\
&\mu^{-1}\lambda^{-1+\nu}\Big(\frac{t-r+1}{2t}-\lambda\Big)^{-\mu} + \lambda^{-1+\nu-\mu},\quad &&\text{Case 4}.
\endaligned
\right.
\end{equation}

\paragraph{Regime X :  $Z \geq 4$ and $r\geq t/3$ for a base point {\sl near} the light cone.}  

In this regime we have 
$
\lambda \leq \frac{t-r}{4t}.
$
In order for this regime to be non-empty, we assume that $\frac{t-r}{4t}\geq t^{-1}$ which is equivalent to $t-r\geq 4$. We then find 
$$
\aligned
I(\lambda;t,r)
& \lesssim  (t/r)\lambda^{\alpha_1+\alpha_2+\alpha_3+2}\int_{\lambda^{-1}(1-r/t)-1}^{\lambda^{-1}(r/t+1)-1}
(1+\omega)^{1+\alpha_2}\big((\lambda t)^{-1} + \omega -1)^{\alpha_3} \, d\omega
\\
& \lesssim 
(t/r)\lambda^{\alpha_1+\alpha_2+\alpha_3+2}\int_{\lambda^{-1}(1-r/t)-1}^{\lambda^{-1}(r/t+1)-1}
(1+\omega)^{1+\alpha_2+\alpha_3} \, d\omega.
\endaligned
$$
This leads us to
\begin{equation}\label{eq5-08-01-2021}
I(\lambda;t,r)
\lesssim
\begin{cases}
|\nu-\mu|^{-1}\lambda^{-1+\upsilon+\mu-\nu},\quad &\text{Case 1},
\\
\lambda^{-1+\mu-\nu},\quad &\text{Case 2},
\\
\lambda^{-1-\mu},\quad &\text{Case 3},
\\
\lambda^{-1+\nu-\mu},\quad &\text{Case 4}.
\end{cases}
\end{equation}

%-----------------------------------------------------------------------------------------------------------------------------------------------------------

\subsection{Estimates for the solution to the wave equation}

Since we only need to consider the decay for large $t$ and/or large $r$, without loss of generality we always assume that $t\geq 6$, say. The estimate in $\{1\leq t\leq 6\}$ is easily checked by a direct calculation.

\paragraph{1. Estimate in $\{r\geq 2t\}$.} In this region we have $\frac{r-t}{t}\geq \frac{r+t}{3t}\geq  1$. Observe that $t\geq 6$ implies that $r\geq t+6 \geq t+1$. Then we find 
$$
[t^{-1},1] \subset [t^{-1},\frac{r+t}{3t}]\cap [t^{-1}, \frac{r-t}{t}]
$$
and we can control $I(\lambda;t,r)$ by \eqref{eq4-08-01-2021} and obtain
\begin{equation}\label{eq4-09-01-2021}
|u(t,x)|
\lesssim C_1 
\left\{
\aligned 
&|\nu-\mu-\upsilon|^{-1} |\nu-\mu|^{-1} \, (r+t)^{-1},\quad &&\text{Case 1}
\\
&
\left.
\aligned
&|\mu-\nu|^{-1} \, (r+t)^{-1}t^{\mu-\nu},\quad && \mu>\nu,
\\
& (r+t)^{-1}\ln (t+1),\quad &&\mu=\nu,
\\
&|\nu-\mu|^{-1} \, (r+t)^{-1},\quad && \mu<\nu,
\endaligned
\right\} &&\text{Case 2},
\\
&\mu^{-1} (r+t)^{-1} ,\quad &&\text{Case 3},
\\
&|\mu-\nu|^{-1}(t+r)^{-1},&&\text{Case 4}.
\endaligned
\right.
\end{equation}

\paragraph{2. Estimate in $\{t+1\leq r\leq 2t\}$.} In this region, we have $t^{-1} \leq \frac{r-t}{t}\leq \frac{r+t}{3t}\leq 1$  and 
$$
[t^{-1},1]\subset [t^{-1}, \frac{r-t}{t}]\cup [\frac{r-t}{t}, \frac{r+t}{3t}] \cup [\frac{r+t}{3t},1].
$$
On each sub-interval, we rely on \eqref{eq4-08-01-2021}, \eqref{eq3-08-01-2021}, and \eqref{eq3-06-01-2021}. A direct calculation shows that
\begin{equation}\label{eq1-27-06-2021}
|u(t,x)|
\lesssim C_1 
\left\{
\aligned 
&|\nu-\mu-\upsilon|^{-1} \big(\mu^{-1} + |\nu-\mu|^{-1}\big) (r+t)^{-1},\quad &&\text{Case 1}
\\
&
\left.
\aligned
&\mu^{-1}|\mu-\nu|^{-1}  (r+t)^{-1}t^{\mu-\nu},\quad && \mu>\nu,
\\
&\mu^{-1} (r+t)^{-1}\ln (t+1),\quad &&\mu=\nu,
\\
&\mu^{-1}|\nu-\mu|^{-1} (r+t)^{-1},\quad && \mu<\nu,
\endaligned
\right\} &&\text{Case 2},
\\
&\mu^{-1} (t+r)^{-1}\Big(1 +\crochet^{-\mu}\ln \Big(\frac{t}{\crochet}\Big)\Big),\quad &&\text{Case 3},
\\
&\big(|\mu-\nu|^{-1} + \mu^{-1}\nu^{-1}\crochet^{-\mu}t^{\nu}\big)  (t+r)^{-1} ,&&\text{Case 4}.
\endaligned
\right.
\end{equation}

\paragraph{3. Estimate in $\{t\leq r\leq t+1\}$.} In this region, we have $0\leq \frac{r-t}{t}\leq \frac{1}{t}\leq \frac{r+t}{3t}\leq 1$ and then
$$
[t^{-1},1] = [t^{-1}, \frac{r+t}{3t}]\cup [\frac{r+t}{3t},1].
$$
We rely on \eqref{eq3-08-01-2021} and \eqref{eq3-06-01-2021} and conclude that \eqref{eq1-27-06-2021} still holds in this case.

\paragraph{4. Estimate in $\{t-1\leq r\leq t\}$.} In this case, $0\leq \frac{t-r}{t}\leq \frac{1}{t}\leq \frac{t+r}{2t}\leq 1$. We observe that $t\geq 6$ implies $t/3\leq t-4\leq t-1$. Thus we have 
$$
[t^{-1},1] = [t^{-1},\frac{t+r}{2t}]\cup [\frac{t+r}{2t},1] 
$$
and on each sub-interval we rely on \eqref{eq4-06-01-2021} and \eqref{eq1-07-01-2021}, respectively. We obtain
\begin{equation}\label{eq3-09-01-2021}
|u(t,x)|
\lesssim C_1
\left\{
\aligned 
&|\nu-\mu-\upsilon|^{-1}\big(\mu^{-1}+|\mu-\nu|^{-1}\big) (r+t)^{-1},\quad &&\text{Case 1}
\\
&
\left.
\aligned
&\mu^{-1}|\mu-\nu|^{-1} (r+t)^{-1}t^{\mu-\nu},\quad && \mu>\nu,
\\
&\mu^{-1} (r+t)^{-1}\ln t,\quad &&\mu=\nu,
\\
&\mu^{-1}|\nu-\mu|^{-1} (r+t)^{-1},\quad && \mu<\nu,
\endaligned
\right\} &&\text{Case 2},
\\
&\mu^{-1}   (r+t)^{-1}\ln (t+1),\quad &&\text{Case 3},
\\
&\big(\mu^{-1}\nu^{-1} + |\mu-\nu|^{-1}\big) (t+r)^{-1}t^{\nu},&&\text{Case 4}.
\endaligned
\right.
\end{equation}

\paragraph{5. Estimate in $\{t-2\leq r\leq t-1\}$.} In this region we have $\frac{t-r}{2t}\leq t^{-1}\leq \frac{t-r}{t}\leq \frac{t+r}{2t}\leq 1$, therefore 
$$
[t^{-1},1] = [t^{-1}, \frac{t-r}{t}]\cup [\frac{t-r}{t}, \frac{t+r}{2t}]\cup [\frac{t+r}{2t}, 1].
$$
Then, we rely on \eqref{eq2-07-01-2021}, \eqref{eq4-06-01-2021}, and \eqref{eq1-07-01-2021} on each sub-interval. 
A direct calculation shows that (in fact, \eqref{eq2-07-01-2021}, \eqref{eq4-06-01-2021} give the same bound on $I(\lambda;t,r)$) \eqref{eq3-09-01-2021} holds for the present region.

\paragraph{6. Estimate in $\{t-4 \leq r\leq t-2\}$.} In this region, we have $\frac{t-r}{4t}\leq \frac{1}{t}\leq \frac{t-r}{2t}\leq \frac{t-r}{t}\leq \frac{t+r}{2t}\leq 1$, and then 
$$
[t^{-1},1] = [t^{-1}, \frac{t-r}{2t}]\cup[\frac{t-r}{2t}, \frac{t-r}{t}] \cup [\frac{t-r}{2t}, \frac{t+r}{2t}]\cup [\frac{t+r}{2t}, 1].
$$
On each sub-interval, we rely on \eqref{eq2-08-01-2021},  \eqref{eq2-07-01-2021}, \eqref{eq4-06-01-2021}, and \eqref{eq1-07-01-2021}. The integral on the first sub-interval for Cases 4 and 3 are critical so we write the argument in full details: 
$$
\aligned
\int_{t^{-1}}^{\frac{t-r}{2t}}I(\lambda;t,r) \, d\lambda 
& \lesssim
\begin{cases}
\mu^{-1} \int_{t^{-1}}^{\frac{t-r}{2t}}\lambda^{-1}\Big(\frac{t-r+1}{2t}-\lambda\Big)^{-\mu}d\lambda 
+ \int_{t^{-1}}^{\frac{t-r}{2t}}\lambda^{-1-\mu}d\lambda,
&\text{Case 3},
\\
\mu^{-1} \int_{t^{-1}}^{\frac{t-r}{2t}}\lambda^{-1+\nu}\Big(\frac{t-r+1}{2t}-\lambda\Big)^{-\mu}d\lambda 
+ \int_{t^{-1}}^{\frac{t-r}{2t}}\lambda^{-1-\mu+\nu}d\lambda,
&\text{Case 4},
\end{cases}
\\
& \lesssim 
\begin{cases}
\mu^{-1}t^{\mu}\int_{t^{-1}}^{\frac{t-r}{2t}}\lambda^{-1}d\lambda + \mu^{-1}t^{\mu},
&\text{Case 3},
\\
\mu^{-1}t^{\mu}\int_{t^{-1}}^{\frac{t-r}{2t}}\lambda^{-1+\nu}d\lambda + |\mu-\nu|^{-1}t^{\mu-\nu},
&\text{Case 4},
\end{cases} 
\hskip2.3cm
\lesssim
\begin{cases}
\mu^{-1}t^{\mu}\ln(1+t),
&\text{Case 3},
\\
\big(\mu^{-1}\nu^{-1} + |\mu-\nu|^{-1}\big)t^{\mu},
&\text{Case 4}.
\end{cases}
\endaligned
$$
This, together with similar estimates in the remaining sub-intervals (which we are omitted), shows that \eqref{eq3-09-01-2021} holds for the present region.

\paragraph{7. Estimate in $\{t/3<r\leq t-4\}$.} In this region, we have $t^{-1}\leq \frac{t-r}{4t}\leq \frac{t-r}{2t}\leq \frac{t-r}{t}\leq \frac{t+r}{2t}\leq 1$, and we especially observe that $t\geq 6$ implies $t/3\leq t-4$. Then, we have 
$$
[t^{-1},1] = [t^{-1},\frac{t-r}{4t}]\cup[\frac{t-r}{4t}, \frac{t-r}{2t}]\cup[\frac{t-r}{2t}, \frac{t-r}{t}] \cup [\frac{t-r}{2t}, \frac{t+r}{2t}]\cup [\frac{t+r}{2t}, 1].
$$
We rely on \eqref{eq5-08-01-2021}, \eqref{eq2-08-01-2021},  \eqref{eq2-07-01-2021}, \eqref{eq4-06-01-2021}, and \eqref{eq1-07-01-2021}. A calculation similar to what we did in the above case shows that \eqref{eq3-09-01-2021} also holds in this region.

\paragraph{8. Estimate in $\{0<r\leq t/3\}$.} In this region we have $t^{-1}\leq\frac{t-r}{2t}\leq \frac{t+r}{2t}\leq \frac{t-r}{t}\leq 1$, and then 
$$
[t^{-1},1] = [t^{-1},\frac{t-r}{2t}]\cup[\frac{t-r}{2t}, \frac{t+r}{2t}]\cup [\frac{t+r}{2t},\frac{t-r}{t}]\cup [\frac{t-r}{t},1].
$$
We rely on \eqref{eq1-08-01-2021}, \eqref{eq2-07-01-2021}, \eqref{eq3-07-01-2021}, and \eqref{eq1-07-01-2021}, and we observe that in the region under consideration
$
3/2\leq \frac{t-r}{t}\leq 1.
$
Let us provide the relevant calculation in $[t^{-1}, \frac{t-r}{2t}]$: 
$$
\int_{t^{-1}}^{\frac{t-r}{2t}}I(\lambda;t,r) \, d\lambda 
\lesssim
\left\{
\aligned
& \int_{t^{-1}}^\frac{t-r}{2t}\lambda^{\upsilon}\Big(\frac{t-r+1}{2t}-\lambda\Big)^{-1+\mu}d\lambda
\lesssim \mu^{-1}, &&\quad  \text{Case 1},
\\
& \int_{t^{-1}}^\frac{t-r}{2t}\Big(\frac{t-r+1}{2t}-\lambda\Big)^{-1+\mu}d\lambda
\lesssim \mu^{-1}, &&\quad  \text{Case 2},
\\
& \int_{t^{-1}}^\frac{t-r}{2t}\Big(\frac{t-r+1}{2t}-\lambda\Big)^{-1-\mu}d\lambda
\lesssim \mu^{-1}t^{\mu}, &&\quad  \text{Cases 3 and 4}.
\endaligned
\right.
$$
For the integral in $[\frac{t-r}{2t},\frac{t+r}{2t}]$, we need to deal with the singular factor $(t/r)$ in \eqref{eq2-07-01-2021}. Fortunately, this can be set off by the fact that the interval itself is small, saying, of length $(r/t)$. Recalling that $\frac{t-r}{2t}\geq 1/6$, we find 
$$
\int_{\frac{t-r}{2t}}^{\frac{t+r}{2t}}I(\lambda;t,r) \, d\lambda \lesssim
\left\{
\aligned
&\big(\mu^{-1} + |\nu-\mu|^{-1}\big)(t/r)\int_{\frac{t-r}{2t}}^{\frac{t+r}{2t}}d\lambda\lesssim 
\mu^{-1} + |\nu-\mu|^{-1},\quad &&\text{Case 1},
\\
&\mu^{-1}(t/r)\int_{\frac{t-r}{2t}}^{\frac{t+r}{2t}}d\lambda\lesssim \mu^{-1},\quad && \text{Case 2},
\\
&\mu^{-1}(t/r)t^{\mu}\int_{\frac{t-r}{2t}}^{\frac{t+r}{2t}}d\lambda\lesssim \mu^{-1}t^{\mu},\quad && \text{Cases 3 and 4}.
\endaligned
\right.
$$
On the other hand, the integral on $[\frac{t+r}{2t}, \frac{t-r}{t}]$ is straighforward, we only write the conclusion: 
$$
\int_{\frac{t+r}{2t}}^{\frac{t-r}{t}}I(\lambda;t,r) \, d\lambda \lesssim
\left\{
\aligned
&\mu^{-1}, \quad&&\text{Cases 1 and 2},
\\
&\mu^{-1}t^{\mu}, \quad&&\text{Cases 3 and 4}.
\endaligned
\right.
$$
Finally, the integral in $[\frac{t-r}{t},1]$ is bounded by $\mu^{-1}$ in each of the four cases. We only need to observe that this interval is of length $(r/t)$ which offsets the singular factor $(t/r)$ in \eqref{eq1-07-01-2021}. We conclude that \eqref{eq3-09-01-2021} also holds in the present region.

%==============================================================================================

\section{Method of characteristics}
\label{Annex-section-8}

\begin{proof}[Proof of Lemma~\ref{lem1-05-08-2020}]
The two parameterizations $(t,x)$ and $(s,x)$ are defined in  $\MME_{[s_0,s_1]}$, and satisfy $t = \Time(s,r)$ in \eqref{eq5-05-05-2020}. In view of $\del_tx^a = \del_s x^a =0$, the corresponding Jacobian reads 
$\del_ts = 1 / \del_s T$ and $\del_r s = - \del_r \Time / \del_s T$. We compute  
$$
\Pbf_H^{\N} \big( s \big) 
= \del_t s + \frac{4+\HN^{00}}{4-\HN^{00}} \del_rs 
= \frac{1}{\del_s T} \Big(1 - \frac{4+\HN^{00}}{4-\HN^{00}} \del_rT\Big), 
$$
in which $\del_sT>0$ and $0\leq \del_rT<1$.
By assumption, we have $\HN^{00} \leq0$ and $|\HN^{00}|\ll 1$ and, therefore, $\frac{4+\HN^{00}}{4-\HN^{00}} \leq 1$. 
Consequently, we have $\Pbf_H^{\N}(s) >0$, and the function $s$ is \textsl{strictly increasing along} the curve $\tau \mapsto \varphi_{t,x}(\tau)$. This establishes the statement (3). 
On the other hand, we have 
$
\Pbf_H^{\N}(t-r) =\frac{-2 \HN^{00}}{4-\HN^{00}} \geq 0,
$
so that the function $(t- r)$ is also \textsl{increasing along} $\varphi_{t,x}(\tau)$. 

Now consider an arbitrary point $(t,x)$ in the {\sl interior} of $\Mnear_{\ell,[s_0,s_1]}$ and a time $\tau< t$. At the point $\varphi_{t,x}(\tau)$, we have 
\begin{equation}\label{eq4-19-06-2020}
\tau^2 - \sum_a|\varphi^a(\tau;t,x)|^2 < t^2- r^2 =s^2,
\qquad \qquad 
\tau - \Big( \sum_a|\varphi^a(\tau;t,x)|^2 \Big)^{1/2} \leq t-r< 1, 
\end{equation}
so that the point $\varphi_{t,x}(\tau)$ belongs to $\MME_{s'}$ with $s'< s$. If $|t-\tau|$ sufficiently small, $\varphi_{t,x}(\tau)$ is contained in the interior of $\Mnear_{\ell,[s_0,s]}$. Here, our
second condition above is used in order to exclude that $\varphi_{t,x}$ might enter the region $\MH_{[s_0, s]}$ (where $t-r\geq 1$).

Extending the integral curve $\varphi_{t,x}$ backward in time, this curve eventually reaches the boundary of $\Mscr^{\near}_{\ell, [s_0,s_1]}$, which is composed of the components 
$$
\Mnear_{\ell,s}, \qquad \Lscr_{\ell, [s_0,s_1]},
\qquad
\Lscr_{[s_0,s_1]}, \qquad \Mnear_{\ell,s_0}. 
$$
We introduce the time 
$t_0 = \inf \big\{t_1 \, / \, \varphi_{t,x}([t_1,t))\in \textsl{ interior of }\Mscr^{\near}_{\ell, [s_0,s_1]} \big\}$
and, by continuity, we have $t_1<t$. Thanks to our earlier observation $\Pbf_H^{\N} \big( s \big) > 0$, 
we have $\varphi_{t,x}(t_0)\notin \MME_s$. From the second inequality in \eqref{eq4-19-06-2020}, 
we have $\varphi_{t,x}(t_0)\notin \{r=t-1\}$ and, therefore,
$$
\varphi_{t,x}(t_0)\in \Mnear_{\ell,s_0} \cup \Lscr_{\ell, [s_0,s_1]}. 
$$
The inequality $t_0\geq s_0$ holds since the region $\Mscr^{\near}_{\ell, [s_0,s_1]} $ lies in the future of the subset of $\MME_{s_0}$ within which $t\geq s_0$. 

When $(t,x)$ is on the boundary of $\Mscr^{\near}_{\ell, [s_0,s_1]}$, we only worry about the case where $(t,x)\in \Lscr_{[s_0,s_1]}$.  Recalling the property $\Pbf_H^{\N}(t-r) =\frac{-2 \HN^{00}}{4-\HN^{00}} \geq 0$ established above, we deduce that $\varphi_{t,x}(\tau)$ is still contained in $\Mscr^{\near}_{\ell, [s_0,s_1]}$ for $0<\tau\leq t$ with $t-\tau$ sufficiently small.  By extending the curve backward (with respect to the variable $\tau$), we meet $\Mnear_{\ell,s_0} \cup  \Lscr_{\ell, [s_0,s_1]}$. 
\end{proof}

%------------------------------------

\begin{proof}[Proof of Proposition~\ref{prop1-23-07-2020}]
We need to integrate \eqref{eq1-05-08-2020} and we present our argument for a general ordinary differential equation $u'(t) + P(t)u(t) = Q(t)$ posed on a real interval $[t_0,t_1]$ where $P,Q$ are continuous functions. 
By integration, we have 
$$
u(t) = u(t_0)e^{-\int_{t_0}^t P(\eta)d\eta} + \int_{t_0}^t Q(\tau) e^{-\int_{\tau}^t P(\eta)d\eta} d\tau.
$$
If we impose that $P \geq 0$ on $[t_0,t_1]$, we deduce that  $|u(t) \, |\leq |u(t_0)| + \int_{t_0}^t |Q(\tau)|d\tau$ and we can now apply this observation to \eqref{eq1-05-08-2020}. Our sign condition $\HN^{00}\leq 0$ implies $-\frac{2 \rho \HN^{00}}{(r-t+2)(4-\HN^{00})}\geq 0$ and, by Lemma~\ref{lem1-05-08-2020},
$$
\aligned
|(r-t+2)^\rho (\del_t-\del_r)(ru)| & \lesssim 
\sup_{\Omega_{s_0,s_1}^{\ell}} \Big( |(r-t+2)^\rho (\del_t-\del_r)(ru) \, |\Big)
\\
& \quad +  \int_{2}^t
\Big(
r(r-t+2)^\rho
\Big| -\Boxt_gu + \sum_{a<b}(r^{-1}\Omega_{ab})^2u + r^{-1}\HN^{00}X^{\N}[u] + \slashed{H}^{\N}[u]\Big|
\Big) \big|_{\varphi_{t,x}(\tau)} \, d\tau.
\endaligned
$$
Here, in view of the bounds \eqref{eq6-19-06-2020}, the desired inequality is established. 
\end{proof}

%=========================================================================

\section{Pointwise and energy estimates on the PDEs initial data}
\label{sec1-07-01-2022}

\subsection{Proof of Proposition \ref{lem1-07-01-2022}}

We begin with the following weighted Sobolev and Hardy inequalities. 

\begin{lemma}
Fix some parameter $\eta\geq 0$. 
For any function $u$ defined in $\RR^3$ and decaying sufficiently fast at infinity,
one has 
\begin{equation}\label{eq5-09-01-2022}
\la r\ra^{1+\eta} |u(x)|\lesssim \sum_{|I|+k\leq 2}\|\la r\ra^{\eta}\del^k_r\Omega^I u\|_{L^2(\RR^3)},
\end{equation}
\begin{equation}\label{eq6-09-01-2022}
\|\la r\ra^{-1 + \eta} u\|_{L^2(\RR^3)}\lesssim \|\la r\ra^{\eta}\del u\|_{L^2(\RR^3)}.
\end{equation}
\end{lemma}

\begin{proof} The estimate \eqref{eq5-09-01-2022} is a consequence of the  classical Sobolev inequality
$
|v|\lesssim \la r\ra^{-1}\sum_{|I|+k\leq 2}\|\del^k_r\Omega^I v\|_{L^2(\RR^3)}
$
in which $v  = \la r\ra^{\eta}u$. Concerning \eqref{eq6-09-01-2022}, it suffices to apply 
the classical proof of Hardy inequality
to the function $v = \la r\ra^{\eta} u$:
$$
r^{-2}(\la r\ra^{\eta}v^2)^2 = \del_a(x^a/r^2)(\la r\ra^{2\eta}v^2) 
= \del_a\big((x^a/r^2)\la r\ra^{2\eta}v^2\big)
- 2\eta\la r\ra^{2\eta-2}v^2 - 2\la r\ra^{\eta}r^{-1}v\,\la r\ra^{\eta}(x^a/r)\del_a v. 
$$
Integrating the above identity in the domain $\{\eps \leq |x|\leq R\}$ and letting $\eps\rightarrow 0^+$ and $R\to + \infty$, due to the fact that $\eta\geq 0$ we find
$$
\aligned
\| r^{-1}\la r\ra^{\eta}v\|_{L^2(\RR^3)}^2 
& = \int_{\RR^3}r^{-2}\la r\ra^{2\eta}v^2dx
= - 2\eta\int_{\RR^3} \la r\ra^{2\eta-2}v^2\,dx  -2\int_{\RR^3}\la r\ra^{\eta}r^{-1}v\,\la r\ra^{\eta}(x^a/r)\del_a v\,dx
\\
&\leq -2\int_{\RR^3}\la r\ra^{\eta}r^{-1}v\,\la r\ra^{\eta}(x^a/r)\del_a v\,dx
\leq 2 \, \| r^{-1}\la r\ra^{\eta}v\|_{L^2(\RR^3)}\|\la r\ra^{\eta}(x^a/r)\del_av\|_{L^2(\RR^3)}.
\qedhere
\endaligned
$$
%which leads to \eqref{eq6-09-01-2022}.
\end{proof}

\begin{proof}[Proof of Proposition \ref{lem1-07-01-2022}]

The bounds on $u_0$ and $\phi_0$ in \eqref{eq4-07-01-2022} are immediate.

\vskip.15cm

{\bf 1. Preliminary sup-norm estimate.} 
We observe that, for all $k + |K|\leq 2$ and $|J|\leq N-2$,   
$$
\aligned
\big\|\del_r^k\Omega^K \big(\la r\ra^{\kappa + |J|}\del_x^{J} \del_x u_0\big)\big\|_{L^2(\RR^3)} 
+ \big\|\del_r^k\Omega^K \big(\la r\ra^{\kappa + |J|}\del_x^{J} u_1\big)\big\|_{L^2(\RR^3)}
& \lesssim \epss,
\\
\big\|\del_r^k\Omega^K \big(\la r\ra^{\mu + |J|}\del_x^{J} \del_x \phi_0\big)\big\|_{L^2(\RR^3)} 
+ \big\|\del_r^k\Omega^K \big(\la r\ra^{\mu + |J|}\del_x^{J} \phi_0\big)\big\|_{L^2(\RR^3)} 
+ \big\|\del_r^k\Omega^K \big(\la r\ra^{\mu + |J|}\del_x^{J} \phi_1\big)\big\|_{L^2(\RR^3)}
& \lesssim \epss. 
\endaligned
$$
Then, applying \eqref{eq5-09-01-2022} with $\eta = \kappa$ and $\eta=\mu$ respectively, we find 
\begin{equation}\label{eq2-07-01-2022}
\la r\ra^{\kappa}\big(|\del_x^J \del_x u_0| + |\del_x^J u_1|\big) + \la r\ra^{\mu}\big(|\del_x^J \phi_0| + |\del_x^J \phi_1|\big)\leq C_N \, \eps \, \la r\ra^{-|J|-1}, 
\qquad\quad 
|J|\leq N-2, 
\end{equation}
where $C_N$ is a constant determined by $N$. The above estimate implies
$
\big|\del_r \del^Ju_0\big|\lesssim \la r\ra^{-\kappa-1 - |J|}\eps$,
which by integration from spacelike infinity leads us to 
$$
|\del^J u_0(x)| = \Big|\int_r^{+\infty}\del_r\del^Ju(\rho x/r) d\rho\Big| \lesssim \eps\int_r^{+\infty}\la \rho\ra^{-\kappa-1-|J|}d\rho. 
$$
In turn, we deduce that 
\begin{equation}\label{eq1-12-01-2022}
\la r\ra^{\kappa + |J|}|\del^Ju_0|\lesssim \eps,\qquad |J|\leq N-2.
\end{equation}

\vskip.3cm 

{\bf 2. Time-derivative of the scalar field.}
For the $L^2$ bound on $\overline{\phi}_1$, we recall \eqref{eq1-05-01-2022} and write 
$ 
\overline{\phi}_1 = \phi_1 + \text{h.o.t.}$, 
where each higher-order term contains at least one factor $\phi_1$ or $\del_x\phi_0$. 
Observe that thanks to \eqref{eq1-14-01-2022} the initial reference satisfies   
\begin{equation} \label{eq2-09-01-2022}
\| 
\la r\ra^{|I|} \, \del^I  h_{0ab}^{\star} \big\|_{L^\infty(\RR^3)} 
+ \| \la r\ra^{|J|+1} \, \del^J k_{0ab}^{\star}\big\|_{L^\infty(\RR^3)} 
\lesssim \epss, \qquad |I|\leq N+2,\quad|J|\leq N+1. 
\end{equation}
Recalling \eqref{eq2-09-01-2022} and \eqref{eq2-07-01-2022}, we obtain
\begin{equation}\label{eq2-08-01-2022}
\la r\ra^{\mu+|I|}|\del_x^I \overline{\phi}_1|
\lesssim \la r\ra^{\mu+|I|}|\del_x^I\phi_1| + \sum_{|J|\leq |I|}\big(\la r\ra^{\kappa+|J|-1}|\del_x^J u_0| + \la r\ra^{\mu+|J|}|\del_x^J\del_a\phi_0|\big). 
\end{equation}
In the above estimate, for the high-order terms we now only explain the treatment of the most critical term $u_0\del_a\phi$: 
$$
\la r\ra^{\mu + |I|}\del_x^I(u_0\del_a \phi) 
= \sum_{I_1+I_2=I}\la r\ra^{|I_1|}|\del_x^{I_1}u_0| \la r\ra^{\mu+|I_2|}|\del_x^{I_2}\del_a\phi_0|.
$$
When $|I_1|\leq N-2$, we apply \eqref{eq1-12-01-2022} and when $|I_2|\leq N-2$, we apply \eqref{eq2-07-01-2022} via the following relation:
$$
\la r\ra^{|I_1|}|\del_x^{I_1}u_0|\la r\ra^{\mu+|I_2|}|\del_x^{I_2}\del_a\phi_0|
\lesssim \la r\ra^{\mu - 2 + |I_1|}|\del_x^{I_1}u_0|
\lesssim \la r\ra^{\kappa + |I_1| - 1}|\del_x^{I_1}u_0|
$$ 
provided that $0\leq \mu-\kappa \leq 1$.  In both cases this term is bounded by the right-hand side of \eqref{eq2-08-01-2022}. Next, let us consider the $L^2$ norm of $\la r\ra^{\mu+|I|}|\del_x^I \overline{\phi}_1|$. For the first term of the sum in the right-hand side, we apply \eqref{eq6-09-01-2022} and obtain
$
\|\la r\ra^{\mu+|I|}\del_x^I\overline{\phi}_1\|_{L^2(\RR^3)}\lesssim \eps.
$

\vskip.3cm 

{\bf 3. Time-derivative of the metric.}
Similar arguments are now applied to deal with the components $\del_tg_{\alpha\beta}$ and we observe that, thanks to \eqref{eq4-16-12-2021},
$$
\aligned
u_{1ab}  &=  -2l_{0ab} + \del_au_{0b0} + \del_bu_{0a0} + \text{h.o.t.},
\qquad
\qquad
\qquad
u_{100} = 2\sum_a l_{0aa} {- w[g^{\star}]_0} + \text{h.o.t.},
\\
u_{1a0}  &=  \frac{1}{2}\del_au_{000} - \frac{1}{2}\sum_b\del_au_{0bb} + \sum_b\del_bu_{0ab} {- w[g^{\star}]_a} + \text{h.o.t.}.
\endaligned
$$
The linear terms can be bounded as expected. In each higher-order term, there exists at least one factor ${w[g^{\star}]_{\alpha}}, \del_a u_0, l_0, \phi_i$. In other words, there is no term that would be purely composed by $h^{\star}_0$ and $k^{\star}_0$. The most critical terms are the quadratic ones in the form $u_{0\alpha\beta} k^{\star}_{0ab}$. Thanks to \eqref{eq2-09-01-2022}, we have $|\del^I k_0^{\star}|\lesssim \epss r^{-1-|I|}$ and, by weighed Hardy's inequality, we find 
$$
\big\|\la r\ra^{\kappa+|I|}\del_x^I(u_{0\alpha\beta} k^{\star}_{0ab})\big\|_{L^2(\RR^3)}
\lesssim \epss\sum_{|J|\leq|I|}\| r^{\kappa+|J|-1}\del_x^J u_0\|_{L^2(\RR^3)}
\lesssim \epss \sum_{|J|\leq|I|}\|\la r\ra^{\kappa + |J|}\del\del_x^J u_0\|_{L^2(\RR^3)}\lesssim \epss\eps.
$$
For the remaining terms, we rely on \eqref{eq2-09-01-2022}, \eqref{eq1-07-01-2022}, and \eqref{eq2-07-01-2022} by choosing $c_0$ and $\epss$ to be sufficiently small.

%-------------------------------------------------------------------------------------------------------------------------

\subsection{Proof of Proposition \ref{prop1-07-01-2022}}

First of all, on the initial hypersurface $\{t=1\}$ we have 
$$
|Su| + |L_au| + |\Omega_{ab} u|\lesssim \la r\ra (|\del_xu| + |\del_tu|).
$$
Then for $\ord(\Gamma) = |I|\leq N$, by induction it follows that 
$$
\big\|\la r\ra^{\kappa}\Gamma \del u\|_{L^2(\RR^3)}
\lesssim 
\sum_{|J|+j\leq |I|} \Big( \big\|\la r\ra^{\kappa + j + |J|}\del_t^{j+1}\del_x^J u\big\|_{L^2(\RR^3)} +
\big\|\la r\ra^{\kappa + j + |J|}\del_t^j\del_x^J\del_a u\big\|_{L^2(\RR^3)}\Big).
$$
We thus need to bound the weighted $L^2$ norm of the right-hand side. This is achieved by induction on $j$, as follows. 
When $j= 0$, the argument is direct in view of \eqref{eq4a-07-01-2022} where $u_1 = \del_t u$. Suppose now that 
\begin{equation} 
\big\|\la r\ra^{\kappa + j + |J|}\del_t^{j+1}\del_x^J u\big\|_{L^2(\RR^3)} +
\big\|\la r\ra^{\kappa + j + |J|}\del_t^j\del_x^J\del_a u\big\|_{L^2(\RR^3)} \lesssim \eps
\end{equation}
for all $j\leq k$, and let us consider the case $j = k+1$. We only need to control 
$\big\|\la r\ra^{\kappa + k + 1 + |J|}\del_t^{k+2}\del_x^J u\big\|_{L^2(\RR^3)}$. To this end we differentiate the wave equation \eqref{eq 1 13-01-2019} and observe that
\begin{equation}\label{eq6-07-01-2022}
\del_t\del_t \del_t^k\del_x^J u (1,x) = \Delta_x \del_t^k\del_x^J u (1,x) + \text{h.o.t.}
\end{equation}
We claim that the quadratic terms in the right-hand side are bounded as expected
(and are at least better than the linear term). In fact, among these terms the most critical ones are $u\del\del h^{\star}$ (since the remaining terms contain at least one factor $\del u$ and
are bounded). Then, thanks to \eqref{eq2-09-01-2022} we have 
$
|\del_t^j \del_x^J \del\del h^{\star}|\lesssim \epss \la r\ra^{ -2 - j -|J|}$
for all $ j + |J|\leq N.
$
Based on this estimate, we thus obtain
$$
\la r\ra^{\kappa + k + |J|}\big|\del_t^k\del_x^J(u\del\del h^{\star})\big|\
\lesssim \epss \la r\ra^{\kappa-1}
\sum_{|J'|\leq|J|} \sum_{ k'\leq k}\big|\la r\ra^{k'+|J'|}\del_t^{k'}\del_x^{J'} u\big|.
$$
Applying the weighted Hardy inequality, we find 
$$
\|\la r\ra^{\kappa + k + |J|}\del_t^k\del_x^J(u\del\del h^{\star})\|_{L^2(\RR^3)}
\lesssim 
\epss\sum_{|J|\leq|I|} \sum_{ k'\leq k}\|\la r\ra^{\kappa + k' + |J'|}\del \del_t^{k'}\del_x^{J'} u\|_{L^2(\RR^3)}\lesssim \epss\eps.
$$
Then from \eqref{eq6-07-01-2022}, we obtain
$$
\aligned
\big\|\la r\ra^{\kappa + (k+1) + |J|}\del_t^{k+2}\del_x^J u\big\|_{L^2(\RR^3)}
& \lesssim  \sum_{a}\big\|\la r\ra^{\kappa + (k+1) + |J|}\del_t^k\del_x^J\del_a\del_a u\big\|_{L^2(\RR^3)} + \eps
\\
&= \sum_{a}\big\|\la r\ra^{\kappa + k + (1+|J|)}\del_t^k(\del_x^J\del_a)\del_a u\big\|_{L^2(\RR^3)} + \eps
\lesssim \eps, 
\endaligned
$$
which leads us to 
\begin{equation}\label{eq2-12-01-2022}
\|\la r \ra^{\kappa}\Gamma \del u(1,\cdot)\|_{L^2(\RR^3)}\lesssim \eps,\quad \ord(\Gamma)\leq N.
\end{equation}

A similar bound holds for the scalar field for all $\ord(Z)\leq N$: 
\begin{equation}\label{eq3-12-01-2022}
\big\|  \la r \ra^{\kappa}Z\del  \phi(1, \cdot) \big\|_{L^2(\RR^3)} + \|\la r\ra^{\kappa-1}Z \phi(1,\cdot)\|_{L^2(\RR^3)}\lesssim \eps.
\end{equation} 
Here we exclude $S$ since it does not commute with the Klein-Gordon operator and
therefore $\Gamma^I\phi$ need not be a solution to the Klein-Gordon equation. 
{ For the proof, in comparison to the treatment of the metric
the only difference is that, when we perform an induction on $j$ and due to the Klein-Gordon structure,  \eqref{eq6-07-01-2022} becomes 
$$
\del_t\del_t \del_t^k\del_x^J \phi (1,x) = \Delta_x \del_t^k\del_x^J \phi (1,x) - c^2\del_t^k\del_x^J\phi(1,x) + \text{h.o.t.}
$$
Consequently, we need the sharper bound \eqref{eq4b-07-01-2022}.}
Once \eqref{eq2-12-01-2022} and \eqref{eq3-12-01-2022} are established, we choose $\epss$ and $\eps$ sufficiently small so that the local-in-time solution extends to up to $\Mscr_{s_0}$. A weighted energy argument within the region $\{1\leq t\leq T(s_0,r)\}$ shows that the bounds on the high-order energy presented in \eqref{eq2-02-02-2022} are valid.
\end{proof}

%-----------------------------------------------------------

\subsection{Proof of Proposition \ref{prop1-12-01-2022} -- Kirchhoff formula in the homogeneous case}

We consider the solution by $u_{\init,\alpha\beta} = \Box^{-1}[u_{\alpha\beta}(1,x), \del_tu_{\alpha\beta}(1,x),0]$, and we recall Kirchhoff formula:
\begin{equation}\label{eq4-17-12-2021}
\aligned
& u_{\init,\alpha\beta}(t+1,x) 
= {1 \over 4\pi t^2} \int_{|x-y| =t} \Big(
u_{\alpha\beta}(1,y) +  \la\nabla_y u_{\alpha\beta}(1,y), y - x \ra \Big)d\sigma(y)
+
{1\over 4\pi t}\int_{|y-x|=t} \del_t u_{\alpha\beta}(1,y)  \, d\sigma(y). 
\endaligned
\end{equation} 
First of all, we establish the following result. In our application in the present paper we will be mainly interested in the regime where $\lambda$ is close to $1$.

\begin{proposition}\label{prop1-21-12-2021}
Provided  
$
|u_{\alpha\beta}(1,x)| + \la r\ra|\del_tu_{\alpha\beta}(1,x)|\lesssim \eps \, \la r\ra^{-\lambda}, 
$
one has 
\begin{equation}\label{eq2-17-12-2021}
|u_{\alpha\beta}(t,x)| \lesssim
\begin{cases}
(1-\lambda)^{-1}\la r+t \ra^{-\lambda}\vep, 
\qquad
&1/2<\lambda <1,
\\
(\lambda-1)^{-1}\la r+t\ra^{-1}\vep, &1<\lambda.
\end{cases}
\end{equation}
\end{proposition}

\begin{proof}
The result is checked from \eqref{eq4-17-12-2021}. A similar calculation as the one 
made below \eqref{eq--source-Kirchhoff} tells us that 
$$
\int_{|y-x|=t}f(y)d\sigma(y) = \int_{|y'|=1} f(x - ty') t^2d\sigma 
= t^2 \int_{\mathbb{S}^2} f(x - ty') d\sigma, 
$$
where $d\sigma$ is the Lebesgue measure on $\mathbb{S}^2$. Now if $|f(x)|\leq C_f \la |x|\ra^{-\alpha}$ with $\alpha\geq  0$, we can control the above expression. 

\begin{itemize}
\item 
When $x = 0$, a direct calculation shows that
%\begin{equation}\label{eq2-18-12-2021}
$\Big|\int_{|y-x|=t}f(y)d\sigma(y)\Big|\lesssim C_ft^{2-\alpha}.
$

\item When $x\neq 0$, it is convenient to introduce an adapted parameterization. Without loss of generality we let $x = (r,0,0)$ and the sphere $\{|y'| = 1\}$ is parameterized as follows.  

\end{itemize}
\noindent Indeed, we introduce two variables: 
\begin{itemize}

\item $\theta$ denotes the angle from $(1,0,0)$ to $y$ with $0\leq \theta\leq \pi$, and 

\item $\phi$ denotes the angle from the plane passing by the points $(1,0,0)$ and $(0,1,0)$ to the plane passing by the points $y$ and $(1,0,0)$, in which $0\leq \phi\leq 2 \pi$. 
\end{itemize} 
By elementary geometric and trigonometric arguments, we have 
$$
t^{-2}|x-ty'|^2 = |x/t - y'|^2 = (r/t)^2 + 1 - 2(r/t)\cos\theta,\qquad d\sigma =\sin\theta d\theta d\phi.
$$ 
Recalling that $\alpha\geq 0$, we have 
$$
\aligned
\Big|\int_{|y-x|=t}f(y)d\sigma(y)\Big|
& \lesssim  C_f t^{2-\alpha} \int_0^{\pi}\int_0^{2\pi} (|x/t-y'|^2+t^{-2})^{-\alpha/2}\,\sin\theta d\theta d\phi
\\
& \lesssim  C_f t^{2-\alpha} \int_0^{\pi} \big((r/t)^2 + 1 - 2(r/t)\cos\theta + t^{-2}\big)^{-\alpha/2}\,\sin\theta d\theta
\lesssim   C_f t^{3-\alpha}r^{-1} \int_{|1-r/t|^2}^{|1+r/t|^2}(\omega + t^{-2})^{-\alpha/2}d\omega.
\endaligned
$$
We distinguish between three case, the first two being trivial. 
\begin{itemize}

\item[$\bullet$] When $0< r/t<1/2$, one has $\omega + t^{-2}\geq 1/4$ and thus 
\begin{equation}\label{eq5-17-12-2021}
\Big|\int_{|y-x|=t}f(y)d\sigma(y)\Big|\lesssim C_ft^{3-\alpha}r^{-1}\int_{|1-r/t|^2}^{|1+r/t|^2}d\omega 
\lesssim C_ft^{2-\alpha}.
\end{equation}

\item[$\bullet$]When  $r/t\geq 3/2$, one has $\omega \geq 1/4$. One has $\omega\geq (r/t)^2/2$ on the interval $[|1-r/t|^2,|1+r/t|^2]$, and  thus
\begin{equation}\label{eq6-17-12-2021}
\aligned
&\Big|\int_{|y-x|=t}f(y)d\sigma(y)\Big|
\lesssim C_ft^{3-\alpha}r^{-1}\int_{|1-r/t|^2}^{|1+r/t|^2}(\omega+t^{-2})^{-\alpha/2}d\omega  
\\
&\lesssim C_f t^{3-\alpha}r^{-1} (r/t)^{-\alpha} \big((1+(r/t))^2-((r/t)-1)^2\big)\lesssim C_f t^2r^{-\alpha}.
\endaligned
\end{equation}

\item[$\bullet$] When $1/2 \leq r/t\leq 3/2$ we are in the critical case. We observe that $t\sim r$ and 
$$
\Big|\int_{|y-x|=t}f(y)d\sigma(y)\Big|\lesssim C_f t^{2-\alpha} \int_0^{25/4}(\omega + t^{-2})^{-\alpha/2}d\omega 
\lesssim 
\begin{cases}
C_f t^{2-\alpha}(2-\alpha)^{-1},\quad & 0\geq  \alpha/2>-1,
\\
C_f (\alpha-2)^{-1},\quad & \alpha/2<-1.
\end{cases} 
$$
\end{itemize}
In conclusion to the above cases, we have
\begin{equation}\label{eq3-18-12-2021}
\Big|\int_{|y-x|=t}f(y)d\sigma(y)\Big| 
\lesssim
\begin{cases}
C_f(2-\alpha)^{-1} t^2\la r+t\ra^{-\alpha},\quad & 0\leq \alpha<2,
\\
C_f(\alpha-2)^{-1} t^2\la r+t \ra^{-2} ,\quad &\alpha>2.
\end{cases}
\end{equation}
It remains to apply \eqref{eq4-17-12-2021} together with \eqref{eq3-18-12-2021}, and the desired bounds are obtained.
\end{proof}

%---------------------------------------------

\subsection{Proof of Proposition \ref{prop1-12-01-2022} --  completion of the proof}

Recalling \eqref{eq5-07-01-2022} and the weighted Klainerman-Sobolev's inequality (cf.~also \cite[Appendix C]{LR1} with $\kappa = 1/2+\gamma$ and $t=1$), we
obtain 
$
|\Gamma \del u(1,x)|\lesssim \eps \, \la r\ra^{-\kappa-3/2}$ for $\ord(\Gamma)\leq N-2$
(by omitting the estimates on commutators of generalized operators which have been treated in \cite{LR1}).
Applying similar arguments as we did for \eqref{eq1-12-01-2022}, we find 
\begin{equation}\label{eq4-12-01-2022}
\la r\ra|\del_t\Gamma u(1,x)| + |\Gamma u(1,x)|\lesssim \eps \, \la r\ra^{-\kappa-1/2},
\qquad 
\ord(\Gamma)\leq N-2.
\end{equation}
Observing that $\kappa+1/2>1$, we apply \eqref{prop1-21-12-2021} to the equation $\Box Z u = 0$ with $\lambda  = \kappa + 1/2>1$ and we use \eqref{eq4-12-01-2022}. The desired result is established.

%========================================================================================= 

\section{Estimates for sub-critical nonlinearities}
\label{appendix-EEE}

We are going to establish the desired estimates for sub-critical nonlinearities by relying only on the {\sl first three conditions} in \eqref{equa-new-conditions-hstar}, which are valid for, both, Class A and Class B metrics, namely 
\begin{equation}\label{equa-new-conditions-hstar-trois}
%%%%%%%%%%%%%%%%%%%%%     
\aligned
(C1) &&
|h^\star |_{N+1} 
& \lesssim \epss \, \la r+t\ra^{-\lambda}, 
\\
(C2) &&
|\del^{m} \del h^\star |_{N-m} 
& \lesssim \epss \, \la r+t\ra^{-1+\theta} \crochet^{-\kappa -m}, \quad 
&& m=0 \text{ and } 1, 
\\
(C3) &&
|\del^m \slashed \del h^{\star} |_{N-m} 
& \lesssim \epss \, \la r+t\ra^{-1-\kappa}\crochet^{-m}, \quad 
&& m=0 \text{ and } 1. 
\endaligned 
\end{equation} 

\begin{proof}[Proof of Lemma~\ref{prop1-23-05-2021}] 
{\bf Dealing with the term $ \Fbb^\star_{\alpha\beta}(u,g^{\star};\del g^{\star},\del g^{\star})$.} 
In view of the Poincar\'e-type estimate \eqref{eq1-12-05-2020} in combination with the Sobolev decay estimate 
and by using our condition (C1) above, 
we have 
$$
\aligned
\|s  \crochet^\kappa \zeta \, |\del h^{\star}|_p^2|u|_p\|_{L^2(\MME_s)} 
&\lesssim  \epss^2 \| r^{-2 +2\theta} s \, \crochet^{{-}\kappa} \zeta |u|_p\|_{L^2(\MME_s)}
\\
& \lesssim \epss^2 \| r^{-2+2\theta}s \,  \crochet^{-1+\kappa} \zeta |u|_p\|_{L^2(\MME_s)}
\lesssim  \delta^{-1} \epss^2 (\epss+C_1\eps) s^{-1-\delta}
\lesssim \epss (\epss+C_1\eps)  s^{-1-\delta}.
\endaligned
$$ 
Consequently, we find 
\begin{equation}\label{eq2-16-01-2020}
\|s \, \crochet^{\kappa} \zeta \, | \Fbb^\star_{\alpha\beta}(u,g^{\star};\del g^{\star},\del g^{\star}) |_N\|_{L^2(\MME_s)} \lesssim \epss (\epss + C_1\eps) s^{-1-\delta}. 
\end{equation}

\hskip.15cm 

%----------------

\noindent{\bf Dealing with the terms $\Bbb$.} 
Next, we consider the terms $\Bbb^{\star}[u]$ which satisfy \eqref{eq2-crossing} 
and we treat $\|s \, \crochet^{\kappa} \zeta  |\del h^{\star}|_p|u|_{p_1} |\del u|_{p_2} \|_{L^2(\Mscr_s)}$ first 
(with $p_1 + p_2 = N$). 

\begin{itemize} 

\item When $p_1\leq N-2$, we apply the pointwise decay \eqref{eq7-15-05-2020} enjoyed by $|u|_{p_1}$ and, thanks to the $L^2$ bound \eqref{eq7a-03-05-2020} and our condition (C2) above  we find 
$$
\aligned
\|s \, \crochet^{\kappa} \zeta  |\del h^{\star}|_p|u|_{p_1} |\del u|_{p_2} \|_{L^2(\MME_s)} 
& \lesssim 
\|s  \epss \, \la r+t\ra^{-1+\theta} \crochet^{-\kappa}
\, \delta^{-1} \, (\epss+C_1\eps)  \, s^{\delta} {1 \over r} \, \crochet^{-\kappa+1}
\, \crochet^{\kappa} \zeta |\del u|_{p_2} \|_{L^2(\MME_s)} 
\\
& \lesssim  \delta^{-1} \, (\epss+C_1\eps) \epss \, 
\|  
\la r+t\ra^{-1+\theta} 
\, s^{1+\delta} {1 \over r} \, \crochet^{-2\kappa+1}
\, \big( \crochet^{\kappa} \zeta |\del u|_{p_2}\big) \|_{L^2(\MME_s)} 
\\
& \lesssim  \delta^{-1} \, (\epss+C_1\eps) \epss \, 
\|  s^{-3+2\theta+\delta} \, \big( \crochet^{\kappa} \zeta |\del u|_{p_2}\big) \|_{L^2(\MME_s)} 
\lesssim  \delta^{-1} \epss  \, (\epss+C_1\eps) C_1 \eps \, s^{-3+2\theta+\delta}.   
\endaligned
$$

\item When $p_1\geq N-1 \geq 1$, then $p_2 \leq 2 \leq N-3$, by applying the condition (C2) for the reference  
and the pointwise decay \eqref{eq10-02-05-2020} to $|\del u|_{p_2}$ and, next, 
the Poincar\'e-type decay \eqref{eq1-12-05-2020} to $|u|_{p_1}$, we find 
$$
\aligned
\|s \, \crochet^{\kappa} \zeta  |\del h^{\star}|_p| u|_{p_1} |\del u|_{p_2} \|_{L^2(\MME_s)} 
& \lesssim  \epss (\epss+C_1\eps) C_1 \eps s^{-1-2 \lambda+\delta} \| \crochet^{-1+\kappa} |u|_{p_1} \|_{L^2(\MME_s)}
\\
& \lesssim  \epss (\epss+C_1\eps)^2 s^{-1-2 \lambda+2\delta}.
\endaligned
$$
\end{itemize}  
Hence, in both cases we arrive at
\begin{equation}\label{eq1-16-01-2020}
\aligned
\|s \, \crochet^{\kappa} \zeta  |\del h^{\star}|_p| u|_{p_1} |\del u|_{p_2} \|_{L^2(\MME_s)} 
\lesssim   \delta^{-1} \epss (\epss+C_1\eps)^2 s^{-1-\delta}.
\endaligned
\end{equation} 
For the term $|\del h^{\star}|_p^2|u|_{p_1} |u|_{p_2}$, without loss of generality we may assume that $p_1\leq N-2$,  hence $p_2 \geq 3\geq 1$. Recalling (C2) and \eqref{eq7-15-05-2020} and next using \eqref{eq1-12-05-2020}, we obtain  
$$
\aligned
\|s \, \crochet^{\kappa} \zeta  |\del h^{\star}|_p^2|u|_{p_1} |u|_{p_2} \|_{L^2(\MME_s)}
% &\lesssim \epss^2 \|s \, \crochet^{\kappa} \zeta \,  r^{-2-2 \lambda} |u|_{p_1} |u|_{p_2} \|_{L^2(\MME_s)}
&\lesssim \epss^2 \|s \, \crochet^{\kappa} \zeta \,  r^{-2+2 \theta}\crochet^{-2\kappa} |u|_{p_1} |u|_{p_2} \|_{L^2(\MME_s)}
\\
&  \lesssim  (\epss + C_1\eps) \epss^2\delta^{-1}
% \|s \, \crochet^{\kappa} \zeta \, r^{-2-2 \lambda} \, r^{-1} \crochet^{1-\kappa}s^{\delta} \, |u|_{p_2} \|_{L^2(\MME_s)}
\|s \, \crochet^{-\kappa} \zeta \, r^{-2+2 \theta} \, r^{-1} \crochet^{1-\kappa}s^{\delta} \, |u|_{p_2} \|_{L^2(\MME_s)}
\\
&  \lesssim (\epss + C_1\eps) \epss^2\delta^{-1} 
\| s^{1 + \delta} \, \zeta \, r^{-3+2 \theta} \, \crochet^{2-3\kappa} \, \big( \crochet^{\kappa-1}  |u|_{p_2} \big) \|_{L^2(\MME_s)}
\lesssim (\epss + C_1\eps)^2s^{-5+\delta+4\theta}
\endaligned
$$
% { where we used $\kappa + 2 \lambda\geq (3/2) \delta$ and 
with $ \delta^{-1}\epss\lesssim 1$.
Summarizing our estimates of the terms in \eqref{eq2-crossing}, we thus conclude that 
\begin{equation}\label{eq3-17-05-2020}
\|s \, \crochet^{\kappa} \zeta \, | \Bbb^\star_{\alpha\beta}[u]|_N\|_{L^2(\MME_s)} \lesssim (\epss + C_1\eps)^2s^{-1-\delta}.
\end{equation}

%--------------------

\noindent{\bf Dealing with the terms $\Cbb$.} 
In order to control the high-order terms $\Cbb^\star_{\alpha\beta}[u]$, we recall \eqref{eq3-crossing}. For the first term, we have the following bound when $\max(p_2,p_3) \leq N-3$:
$$
\aligned
\|s \, \crochet^{\kappa} \zeta |u|_{p_1} |\del u|_{p_2} |\del u |_{p_3} \|_{L^2(\MME_s)}
&\lesssim (\epss + C_1\eps)^2 \|s \, \crochet^{\kappa} \zeta \, r^{-2} \crochet^{-2 \kappa}s^{2 \delta} \,|u|_{p_1} \|_{L^2(\MME_s)}
\\
& \lesssim (\epss + C_1\eps)^2  s^{-3+2\delta} \| \crochet^{-1+\kappa} |u|_{p_1} \|_{L^2(\MME_s)}
\lesssim 
(\epss + C_1\eps)^2s^{-1-\delta}, 
\endaligned
$$
where we used \eqref{eq1-12-05-2020} together with \eqref{eq10-02-05-2020}.
%and { provided $\kappa \geq (3/2)\delta$} (for the second inequality above). 
When $\max(p_1,p_2)\leq N-3$, thanks to \eqref{eq7-15-05-2020}, \eqref{eq10-02-05-2020} and \eqref{eq7a-03-05-2020} we can write 
$$
\aligned
& \|s \, \crochet^{\kappa} \zeta |u|_{p_1} |\del u|_{p_2} |\del u |_{p_3} \|_{L^2(\MME_s)}
\lesssim  \delta^{-1} (\epss + C_1\eps)^2 \|s \, \crochet^{\kappa} \zeta \, r^{-2} 
\crochet^{1-2 \kappa}s^{2 \delta} \,|\del u|_{p_3} \|_{L^2(\MME_s)}
\\
& \lesssim (\epss + C_1\eps) \, s^{-3+2\delta} \|\zeta \crochet^{\kappa} |\del u|_{p_3} \|_{L^2(\MME_s)}
\lesssim (\epss + C_1\eps)^2 \, s^{-1-\delta}.
\endaligned
$$
%{ provided $\kappa \geq \delta/2$} (for the second inequality above). 
%
Next, for the second term in \eqref{eq3-crossing} for $\max(p_1,p_2)\leq N-3$, thanks to \eqref{eq7-15-05-2020} and \eqref{eq7a-03-05-2020} we find  
$$
\aligned
\|s \, \crochet^{\kappa} \zeta  |\del h^{\star}|_p  |u|_{p_1} |u|_{p_2} |\del u|_{p_3}  \|_{L^2(\MME_s)}
&
\lesssim \delta^{-2} \epss (\epss + C_1\eps)^2 \|s \, \crochet^{\kappa} \zeta r^{-1 +\theta}  r^{-2} \crochet^{2 (1-\kappa){-\kappa}} \,  s^{2 \delta} \, |\del u|_{p_3} \|_{L^2(\MME_s)}
\\
& \lesssim (\epss + C_1\eps)\, s^{-1+2\theta-6\kappa+2\delta} \|  \zeta \crochet^{\kappa} |\del u|_{p_3} \|_{L^2(\MME_s)}
\lesssim  (\epss + C_1\eps)^2s^{-1-\delta}.
\endaligned
$$
%{ provided $2 \kappa + \lambda \geq \delta$.}
%
For $\max(p_1,p_3)\leq N-3$, by applying \eqref{eq7-15-05-2020}, \eqref{eq10-02-05-2020} and \eqref{eq1-12-05-2020} we find 
$$
\aligned
& \|s \, \crochet^{\kappa} \zeta  |\del h^{\star}|_p  |u|_{p_1} |u|_{p_2} |\del u|_{p_3}  \|_{L^2(\MME_s)}
\lesssim  \delta^{-1} \epss (\epss + C_1\eps)^2 \|s \, \crochet^{\kappa} \zeta r^{-1+\theta}  
r^{-2} \crochet^{1- 3\kappa} s^{2 \delta} \, |u|_{p_2} \|_{L^2(\MME_s)}
\\
%----------------- 
& \lesssim  (\epss + C_1\eps)\, s^{-1+2\theta-6\kappa+2\delta} \| \crochet^{-1+\kappa} |u|_{p_2} \|_{L^2(\MME_s)}
\lesssim (\epss + C_1\eps)^2s^{-1-\delta}.
\endaligned
$$ 
On the other hand, let us consider the third term in \eqref{eq3-crossing}.
In the case $\max(p_1,p_2,p_3)\leq N-3$ 
we have 
$$
\aligned
\|s \, \crochet^{\kappa} \zeta  |\del u|_{p_1} |\del u|_{p_2} |u|_{p_3} |u|_{p_4} \|_{L^2(\MME_s)}
&
\lesssim  \delta^{-1} (\epss + C_1\eps)^3 \|s \, \crochet^{\kappa} \zeta r^{-2} \crochet^{-2 \kappa} \, r^{-1} \crochet^{1-\kappa}s^{3\delta} \, |u|_{p_4} \|_{L^2(\MME_s)}
\\
& \lesssim   (\epss + C_1\eps)^2 \, s^{-1-6\kappa+3\delta} \| \crochet^{-1+\kappa} |u|_{p_4} \|_{L^2(\MME_s)}
\lesssim (\epss + C_1\eps)^2s^{-1-\delta}. 
\endaligned
$$ 
For the case $\max(p_1,p_3,p_4) \leq N-3$, we have 
$$
\aligned
\|s \, \crochet^{\kappa} \zeta  |\del u|_{p_1} |\del u|_{p_2} |u|_{p_3} |u|_{p_4} \|_{L^2(\MME_s)}
&
\lesssim  (\epss + C_1\eps)\, \| s \, \crochet^{\kappa} \zeta r^{-3} \crochet^{2(1- \kappa)-\kappa} \, s^{3\delta} \, |\del u|_{p_2} \|_{L^2(\MME_s)}
\\
& \lesssim   
\delta^{-2}(C_1\eps )^3 \, s^{-1-6\kappa +3\delta} \| \crochet^{\kappa} \zeta \, | \del u|_{p_2} \|_{L^2(\MME_s)}
\lesssim  (C_1\eps)^2 s^{-1-\delta},
\endaligned
$$
{ since $\kappa \geq (2/3)\delta$.}
We thus conclude that  
$\|s \, \crochet^{\kappa} \zeta \, | \Cbb^\star_{\alpha\beta}[u]|_p\|_{L^2(\MME_s)} 
\lesssim (\epss + C_1\eps)^2 \, s^{-1-\delta}$.
\end{proof}

%--------------------------------------------------------------------------------------------------------------------------------------------------  

\begin{proof}[Proof of Lemma~\ref{prop2-23-05-2021}]       We begin by observing that 
\begin{equation}\label{eq5-17-05-2020}
\|s \, \crochet^{\kappa} \zeta | \phi^2 |_p\|_{L^2(\MME_s)} 
+ \|s \, \crochet^{\kappa} \zeta \, | \del_{\alpha} \phi \del_{\beta} \phi|_p\|_{L^2(\MME_s)} 
\lesssim 
(\epss + C_1\eps)^2s^{-1-\delta},
\end{equation} 
and we only write the relevant bound for $\del_{\alpha} \phi \del_{\beta} \phi$ since $\phi$ enjoys better $L^2$ and decay bounds. Recalling Lemma~\ref{lemma-111}, 
we have 
% \eqref{eq9-15-05-2020}, 
$$
\aligned
&
\|s \, \crochet^{\kappa} \zeta  |\del_{\alpha} \phi \del_{\beta}  \phi |_p\|_{L^2(\MME_s)}
\leq  \sum_{p_1+p_2=p} \|s \, \crochet^{\kappa} \zeta \, | \del \phi |_{p_1} |\del \phi |_{p_2} \|_{L^2(\MME_s)}
\\
& \lesssim 
(\epss + C_1\eps)s^{1+2\delta}\| r^{-1-\lambda}\crochet^{\kappa-2\mu} \crochet^\mu\zeta| \del\phi|_p\|_{L^2(\MME_s)}
+ (\epss + C_1\eps)s^{1+2\delta}\| \crochet^{1-2\mu + \kappa}r^{-2} \,\crochet^\mu\zeta|\del \phi |_p \|_{L^2(\MME_s)}
\\
& \lesssim (\epss + C_1\eps)\, s^{-1-2\min(\lambda,\mu)+2\delta} \, \Fenergy_{\kappa,c}^{\ME,p}(s, \phi)
\lesssim (\epss + C_1\eps)^2s^{-2\min(\lambda,\mu)+3\delta}. 
\endaligned
$$
Here, we use $ \min(\lambda,\kappa,\mu)\geq 1/2+ 2 \delta$
{(as well as $N\geq 9$ in order to guarantee $[N/2]\geq N-5$ for \eqref{eq9-15-05-2020}).} 
Next, recalling Lemma~\ref{lem-small} and the condition \eqref{equa-31-12-20} we have 
$
|h^{\star}|_p\lesssim \epss \ll 1, 
$
and therefore 
\begin{equation}\label{eq1-19-11-2020}
\aligned
&\sum_{\alpha, \beta} |2 \, T_{\alpha\beta} - Tg_{\alpha\beta} |_p
\lesssim \sum_{p_1+p_2=p} |g|_{p_1}\big( |\del \phi \del\phi|_{p_2} + |\phi^2|_{p_2}\big)
\\
&
\lesssim 
|g|\big( |\del \phi \del\phi|_p + |\phi^2|_p\big)
+\sum_{p_1+p_2=p, p_1\geq 1}|g|_{p_1}\big( |\del \phi \del\phi|_{p_2} + |\phi^2|_{p_2}\big)
\\
& \lesssim (1+|h^{\star}| + |u|) \big(|\del\phi\del\phi|_p + |\phi^2|_p\big)
+\sum_{p_1+p_2=p, p_1\geq 1}(|h^{\star}|_{p_1}+|u|_{p_1})\big( |\del \phi \del\phi|_{p_2} + |\phi^2|_{p_2}\big)
\\
& \lesssim (1+|h^{\star}|_p)\big( |\del \phi \del\phi|_p + |\phi^2|_p\big) 
+ \sum_{p_1+p_2=p}|u|_{p_1}\big( |\del \phi \del\phi|_{p_2} + |\phi^2|_{p_2}\big), 
\endaligned
\end{equation}
where we used $|g| \lesssim  1+|u| + |h^{\star}|$ and, when $p_1\geq 1$,
$|g|_{p_1}\lesssim |h^{\star}|_{p_1}+|u|_{p_1}$.
For the first term in the right-hand side of \eqref{eq1-19-11-2020}, recalling \eqref{eq5-17-05-2020} we find 
$$
\big\|s\crochet^{\kappa}\zeta(1+|h^{\star}|_p) \, |\del\phi\del\phi|_p\|_{L^2(\MME_s)} + \big\|s\crochet^{\kappa}\zeta(1+|h^{\star}|_p) \, |\phi^2|_p\big\|_{\MME_s} 
\lesssim (\epss + C_1\eps)^2s^{-1-\delta}.
$$
For the second term in the right-hand side of \eqref{eq1-19-11-2020}, for $p_1\leq N-3$ we have 
$$
\aligned
& \|s\crochet^{\kappa}\zeta|u|_{p_1}\big( |\del \phi \del\phi|_{p_2} + |\phi^2|_{p_2}\big)\|_{L^2(\MME_s)}
\lesssim \delta^{-1} (\epss + C_1\eps) \, \| s\crochet^{\kappa}\zeta\, r^{-1}\crochet^{1-\kappa}s^{\delta}\big( |\del \phi \del\phi|_{p_2} + |\phi^2|_{p_2}\big)\|_{L^2(\MME_s)}
\\
& \lesssim
\|s\crochet^{\kappa}\zeta\big( |\del \phi \del\phi|_{p_2} + |\phi^2|_{p_2}\big)\|_{L^2(\MME_s)}
\lesssim (\epss + C_1\eps)^2 s^{-1-\delta}, 
\endaligned
$$
where \eqref{eq5-17-05-2020} was used.
When $p_2\leq N-3$, we recall \eqref{eq11a-02-05-2020} and \eqref{eq1-18-05-2020} (with $N\geq 12$, say) and obtain 
$$
|\del \phi\del\phi|_{p_2} + |\phi^2|_{p_2}\lesssim |\del \phi|_{N-3}|\del \phi|_{N-8} + |\phi|_{ N-3}|\phi|_{N-8}
\lesssim (\epss + C_1\eps)^2 \, r^{-2}\crochet^{-2\mu}s^{1+2\delta}.
$$
We then deduce that 
$$
\aligned
\big\|s\crochet^{\kappa} \zeta |u|_{p_1}\big( |\del \phi \del\phi|_{p_2} + |\phi^2|_{p_2}\big)\big\|_{L^2(\MME_s)}
& \lesssim 
(\epss + C_1\eps)^2 s^{2 + 2\delta}\| r^{-2}\crochet^{1-2\mu} \, \crochet^{-1+\kappa}|u|_p\|_{L^2(\MME_s)}
\\
& \lesssim  (\epss + C_1\eps)s^{-2+2\delta}\big(
\Fenergy_{\kappa}^{\ME,p}(s,u) + \Fenergy_{\kappa}^0(s,u) \big)
\lesssim (\epss + C_1\eps)^2s^{-1-\delta}.  \qquad \qedhere
\endaligned 
$$
\end{proof} 

\end{document}